%% file: main.tex
\documentclass[12pt]{report}
\usepackage[a4paper,width=150mm,top=25mm,bottom=25mm]{geometry}

\counterwithout{figure}{chapter}

\usepackage{authblk} 
\usepackage[usenames, dvipsnames]{color}    
\usepackage{graphics}   
\usepackage{graphicx}            
\usepackage{amsmath,amssymb}
\usepackage[format=hang,justification = justified]{caption}   
\usepackage{subcaption} 
\usepackage[square, comma, sort&compress, numbers]{natbib}
\usepackage[linktocpage, colorlinks, citecolor=blue, linkcolor=blue, urlcolor=blue]{hyperref} 
\usepackage{enumerate}  
\usepackage{float}
\usepackage[T1]{fontenc}
\usepackage{lmodern}
\usepackage{calc}
\usepackage{OurCommands} 
\usepackage[hang,flushmargin]{footmisc} 
\usepackage{yfonts} 
\usepackage{pdfpages}
\usepackage{hhline}
\usepackage[title]{appendix}
\usepackage{subdepth} 
\usepackage{soul} 
\usepackage{amsthm}
\newtheorem{theorem}{Theorem}
\newtheorem{lemma}{Lemma}
\newtheorem{principle}{Principle}
\theoremstyle{definition}
\newtheorem{definition}{Definition}
\usepackage{IEEEtrantools}
\usepackage{feynmp}
\usepackage{braket}
\usepackage{simplewick}
\usetikzlibrary{arrows,positioning,decorations.markings,decorations.pathmorphing,calc}

\usepackage{multirow}
\usepackage{ragged2e}
\usepackage{tabularx}
\usepackage{setspace}
\usepackage{titlesec}
\usepackage{hyperref}

\usepackage[nopostdot, style=tree]{glossaries}
\makenoidxglossaries
\input{IntoAndOutro/1.Glossary}

\renewcommand*{\glossarymark}[1]{}

\usepackage[shortlabels]{enumitem}
\usepackage{xspace}
\usepackage{xcolor}
\usepackage[normalem]{ulem}

\definecolor{mygreen}{RGB}{63, 136, 39}

\begin{document}


\include{CoverPages/TitlePageETHZ}

\newpage
\thispagestyle{empty}
\mbox{}
\pagenumbering{gobble} 
\clearpage
\newpage

\include{IntoAndOutro/0.Abstract}
\newpage

\newpage
\thispagestyle{plain}
\mbox{}
\newpage

\include{IntoAndOutro/0.Zusammenfassung}

\newpage

\newpage
\thispagestyle{plain}
\mbox{}
\newpage

\include{IntoAndOutro/1.Acknowledgements}

\newpage

\newpage
\thispagestyle{plain}
\mbox{}
\newpage

\include{IntoAndOutro/0.Conventions}
\newpage

\newpage
\thispagestyle{plain}
\mbox{}
\newpage

\renewcommand{\contentsname}{Table of Contents}
\tableofcontents

\include{MainBody}

\newpage
\include{IntoAndOutro/1.Appendix}

\newpage
\thispagestyle{plain} 
\mbox{}
\printnoidxglossary[title=\huge{Glossary}]
\addcontentsline{toc}{chapter}{Glossary}

\newpage
\bibliographystyle{utcaps}
\bibliography{References}\addcontentsline{toc}{chapter}{Bibliography}

\end{document}

%% file: IntoAndOutro/1.Glossary.tex
\newglossaryentry{inertial frame of reference}{
    name={inertial frame of reference},
    description={A coordinate system, in which Newton's first law of motion holds.}
}

\newglossaryentry{spacetime}{
    name={spacetime},
    description={Spacetime is the collection of all possible events, represented by a sooth four-dimensional manifold endowed with a physical metric.}
}

\newglossaryentry{matter fields}{
    name={matter fields},
    description={Fields describing the non-gravitational laws of physics. They form a locally conserved energy-momentum tensor and are minimally and universally coupled to the physical metric of spacetime.}
}

\newglossaryentry{non-minimal fields}{
    name={non-minimal fields},
    description={Additional fields in the gravitational action on top of the physical metric, that couple non-minimally to the physical metric. In a metric theory of gravity, non-minimal fields exclusively appear in the gravitational action and do not couple to matter fields. In comparison to matter fields, there is no conserved energy-momentum tensor associated to non-minimal fields.}
}

\newglossaryentry{freely falling}{
    name={freely falling},
    description={Free fall is the motion of a test body when gravity is the only force acting on it. In gravity theories that satisfy the weak equivalence principle, freely falling frames are local inertial frames. However, in the presence of an inhomogeneous gravitational field, this notion of inertial frame is not uniquely defined throughout spacetime anymore.}
}

\newglossaryentry{physical observable}{
    name={physical observable},
    description={A measurable property of a physical system as an outcome of an experiment.}
}

\newglossaryentry{straight lines}{
    name={straight lines},
    description={The worldlines of free test-particles.}
}

\newglossaryentry{ghost}{
    name={ghost},
    description={A propagating degree of freedom with the wrong sign of the kinetic term. Ghosts are associated to Ostrogradsky instabilities arising in Lagrangian theories with higher order time derivatives that are non-degenerate. Such theories are unstable due to an unbound Hamiltonian from below.}
}

\newglossaryentry{worldline}{
    name={worldline},
    description={The trajectory of any physical test-particle in spacetime.}
}

\newglossaryentry{test particle}{
    name={test particle},
    description={A physical object with negligible self-gravitational energy that is also small enough to neglect any inhomogeneities of external fields.}
}

\newglossaryentry{effective field theory}{
    name={effective field theory},
    description={An effective field theory describes physical phenomena at a given length scale (or energy scale) by including only the relevant degrees of freedom, ignoring the precise structure of a theory that would be necessary to capture physics at shorter distances (or higher energies).}
}

\newglossaryentry{memory}{
    name={memory},
    description={A component of a gravitational polarization that induces a permanent, non-oscillatory offset between two test-masses in the idealized response of a gravitational wave detector.}
}

\newglossaryentry{tensor memory}
{
    name=tensor memory,
    description={(tensor/vector/scalar) Memory associated to a tensor/vector/scalar gravitational polarization.},
    parent=memory
}

\newglossaryentry{null memory}
{
    name=null memory,
    description={Memory sourced by the energy-momentum of radiation.},
    parent=memory
}

\newglossaryentry{ordinary memory}
{
    name=ordinary memory,
    description={Memory sourced by massive bodies that are unbound in their initial or final states (or both)},
    parent=memory
}

\newglossaryentry{waves}
{
    name=waves,
    description={Field-perturbations with characteristic high frequency $f_H$.}
}

\newglossaryentry{gravitational waves}
{
    name=gravitational waves,
    description={Waves of the physical metric perturbation.},
    parent=waves
}

\newglossaryentry{gravitational field}
{
    name=gravitational field,
    description={A tensor field (other than the physical metric) that is non-minimally coupled to the physical metric. In a metric theory of gravity, a gravitational field does not couple directly to matter fields, as only the physical metric does.}
}

\newglossaryentry{radiation}
{
    name=radiation,
    description={Field-perturbation that escape to null infinity.}
}

\newglossaryentry{gravitational radiation}
{
    name=gravitational radiation,
    description={Radiation of the physical metric perturbation.},
    parent=radiation
}

\newglossaryentry{limit to null infinity}
{
    name=limit to null infinity,
    description={Leading-order behavior in the limit $r\rightarrow \infty$ at fixed $u=t-r$, in source centered coordinates $\{t,x,y,z\}$, with $r=\sqrt{x^2+y^2+z^2}$.}
}

\newglossaryentry{radiative degrees of freedom}
{
    name=radiative degrees of freedom,
    description={Gauge-invariant perturbations given by a propagating solutions to wave-type equations in the radiation zone outside of any source.}
}

\newglossaryentry{propagating degrees of freedom}
{
    name=propagating degrees of freedom,
    description={Gauge-invariant perturbations given by a propagating solutions to wave-type equations.}
}

\newglossaryentry{non-dynamical degrees of freedom}
{
    name=non-dynamical degrees of freedom,
    description={Gauge-invariant perturbations given by a solutions to a Laplace-type equation.}
}

\newglossaryentry{unphysical degrees of freedom}
{
    name=unphysical degrees of freedom,
    description={Apparent degrees of freedom that are only an accident of a redundancy of description which are not part of any physical observable.}
}

\newglossaryentry{polarization}
{
    name=polarization,
    description={Property of a wave-like solution associated to a propagating degree of freedom that specifies the geometrical orientation of the oscillation, distinguishing it from other radiative modes.}
}

\newglossaryentry{tensor polarizations}
{
    name=tensor polarizations,
    description={(tensor/vector/scalar) Modes that transform like a tensor, a vector or a scalar under rotations about the direction of propagation.},
    parent=polarization
}

\newglossaryentry{gravitational polarizations}
{
    name=gravitational polarizations,
    description={Polarization modes of the physical metric perturbation far away from the source.},
    parent=polarization
}

\newglossaryentry{physical metric}
{
    name=physical metric,
    description={The metric that minimally couples to matter. In a metric theory of gravity, the physical metric is the only gravitational field that couples directly to matter.}
}

%% file: CoverPages/TitlePageETHZ.tex
\begin{titlepage}
   \begin{center}

        DISS. ETH NO. 29961

       \vspace*{1.5cm}
       \LARGE
       \textit{\textbf{Probing Gravity}}

        \vspace{0.5cm}

        \large
       \textit{Fundamental Aspects of Metric Theories and their Implications for Tests of General Relativity}

        \vspace{1.5cm}
       \normalsize
       A thesis submitted to attain the degree of

        \vspace{0.5cm}
        DOCTOR OF SCIENCES

        \vspace{0.5cm}
        (Dr. sc. ETH Z\"urich)
            
       \vspace{1.5cm}

       \normalsize Presented by

        \vspace{0.5cm}
        \large
       \textbf{JANN ZOSSO}

        \vspace{0.5cm}
       \normalsize
        \textit{Master of Science in Physics}
        
        \vspace{0.3cm}
        \textit{EPFL}
        
        \vspace{0.5cm} born on \textit{July 28$^\text{th}$, 1992}

        \vspace{1.5cm}
        \normalsize
        Accepted on the recommendation of
            
       \vspace{0.5cm}
        \textit{Prof. Dr. Lavinia Heisenberg}

        \vspace{0.3cm}
         \textit{Prof. Dr. Philippe Jetzer}

          \vspace{0.3cm}
          \textit{Prof. Dr. Luca Amendola}

           \vspace{0.3cm}
         \textit{Prof. Dr. Camille Bonvin}

         \vspace{0.3cm}
          \textit{Prof. Dr. Leonardo Senatore}

         \vspace{1cm}
      2024
            
   \end{center}
\end{titlepage}

%% file: IntoAndOutro/0.Abstract.tex
\newpage
\pagenumbering{roman}
\chapter*{Abstract}\addcontentsline{toc}{chapter}{Abstract}

\sloppy
\setstretch{1.0}
\small
Guided by the Einstein equivalence principle that identifies the phenomenon of gravitation as a manifestation of the dynamics of spacetime in contrast to a localizable force, we review and explore its consequences on formulating a theory of gravity. The resulting space of metric theories of gravity may address open conceptual and observational puzzles through a wealth of effects beyond general relativity, whose traces can be searched for within today's and tomorrow's gravitational testing grounds.

Above all, we offer a generic metric theory generalization of Isaacson's approach to the leading-order field equations of physical perturbations with a well-defined notion of energy-momentum carried by the gravitational waves. Within this framework, we identify the backreaction of the Isaacson energy-momentum flux onto the background spacetime with the displacement memory effect that induces a permanent distortion of space after the passage of a gravitational wave. This effect is a well-known prediction of GR whose dominant contribution captures its inherent non-linear nature, manifest in the ability of gravity to gravitate. However, the novel interpretation of memory as naturally arising within the Isaacson approach to gravitational waves comes with two main advantages. Firstly, it allows for a unified understanding of both the null and the ordinary memory effect, which are respectively sourced by unbound energy fluxes that do and do not reach asymptotic null infinity. Secondly, and most importantly, this approach allows for a consistent derivation of the memory formula for a large class of metric theories with considerable lessons to be learned for upcoming future measurements of the memory effect. Being sensitive not only to additional gravitational polarizations in the detector response but to any additional radiative degrees of freedom, memory may provide a valuable consistency test for beyond GR signatures in future gravitational wave observations.

Valuable probes of the theory of gravity are also found in the field of cosmology that equally promise a sharp increase in constrainability in the near future. Already current data are beginning to require a departure from the cosmological standard model with increasing statistical significance, due to the mismatch between the values of standard cosmological observables inferred from partially independent observations. In this context, we formulate a simple set of necessary conditions that a large class of late-time departures from today's standard model need to satisfy in order to tackle two of the most significant cosmological tensions simultaneously. Because our analytic work remains largely model agnostic, the results represent a general guideline on the search for a consistent resolution of the tensions and can in particular be applied to gain an intuition on the resulting constraints on the theory space beyond GR.

Finally, we analyze the theoretical consistency of beyond GR metric theories within their interpretation as quantum effective field theories. In doing so, we correct previous beliefs and show radiative stability for a model of luminal Horndeski and generalized Proca theory. However, we equally draw attention to the unsolved challenges of such a quantum field theory perspective of gravity theories and end the monograph with a speculation on a possible alternative take on the quantization of well-defined gravitational degrees of freedom within the philosophy of the Isaacson framework. 

\setstretch{1.2}
\normalsize

%% file: IntoAndOutro/0.Zusammenfassung.tex
\newpage

\chapter*{Zusammenfassung}\addcontentsline{toc}{chapter}{Zusammenfassung}

\setstretch{1.0}
\small
Auf der Grundlage des Einstein’schen Äquivalenzprinzips, welches die Gravitation als eine Manifestation der Dynamik der Raumzeit im Gegensatz zu einer lokalisierbaren Kraft identifiziert, werden die darauffolgenden Auswirkungen auf eine Formulierung einer Gravitationstheorie untersucht. Der resultierende Raum metrischer Theorien der Gravitation kann offene konzeptionelle und experimentelle Rätsel durch eine Vielzahl von Effekten jenseits der allgemeinen Relativitätstheorie angehen, deren Spuren in den heutigen und zukünftigen empirischen Datensätze gesucht werden können.

Insbesondere wird eine Verallgemeinerung von Isaacson’s Ansatz für die Feldgleichungen physikalischer Störungen führender Ordnung mit einem gut definierten Begriff des Energie-Impuls-Tensors von den Gravitationswellen, auf allgemeine metrische Theorien erarbeitet. In diesem Rahmen identifizieren wir die Rückwirkung des Isaacson-Energie-Impuls-Flusses auf die Hintergrund-Raumzeit als Ursache des Memory-Effektes, welcher eine permanente Verzerrung des Raums nach dem Durchgang einer Gravitationswelle beschreibt. Dieser Effekt ist eine allgemein bekannte Vorhersage der allgemeinen Relativitätstheorie, deren dominanter Beitrag die inhärente Nichtlinearität der Gravitation erfasst. Die neuartige Interpretation des Memory-Effektes als natürliche Konsequenz des Isaacson-Ansatzes zur Definition von Gravitationswellen, hat jedoch zwei wesentliche Vorteile. Erstens ermöglicht sie ein einheitliches Verständnis sowohl des sogenannten null- als auch des gewöhnlichen Memory-Effektes, die einerseits von lichtartigen und andererseits von massehaltigen ungebundenen Energieflüssen herrühren. Zweitens, und das ist der wichtigste Punkt, ermöglicht dieser Ansatz eine konsistente Ableitung des Memory-Effektes für ein weitgefasstes Spektrum an metrischer Gravitationstheorien, wobei erhebliche Erkenntnisse für bevorstehende Messungen des Memory-Effektes und die darauf basierenden Überprüfungen der allgemeinen Relativitätstheorie zu gewinnen sind.

Die Kosmologie stellt ebenfalls ein wertvolles Testfeld der Gravitationstheorie dar. Insbesondere weisen bereits heutige Datensätze zunehmend auf eine mögliche Notwendigkeit einer Abweichung vom aktuellen kosmologischen Standardmodell hin. Dies aufgrund einer zunehmenden statistischen Signifikanz von empirischen Diskrepanzen, besser bekannt als kosmologische Spannungen. In diesem Zusammenhang wird eine Methode vorgestellt, welche die Formulierung notwendigen Bedingungen ermöglicht um zwei der bedeutendsten kosmologischen Spannungen gleichzeitig anzugehen. Da diese analytische Herleitung weitgehend modellunabhängig ist, stellen die Ergebnisse einen allgemeinen Leitfaden für die Suche nach einer konsistenten Lösung dar, vor allem auch im Zusammenhang der möglichen alternativen metrische Gravitationstheorien.

Schliesslich wird die theoretische Konsistenz von allgemeinen metrischen Theorien im Rahmen ihrer Interpretation als quanteneffektive Feldtheorien untersucht. Dabei werden frühere Annahmen der Unstabilität bezüglich zwei wichtiger Ableitungs-Interaktionen basierten Theorien widerlegt. Es wird jedoch ebenfalls auf ungelösten Herausforderungen einer solchen quantenfeldtheoretischen Perspektive von Gravitationstheorien hingewiesen, inklusive einer Spekulation über eine mögliche alternative Herangehensweise an die Quantisierung gravitationeller Freiheitsgrade.

\setstretch{1.2}
\normalsize

%% file: IntoAndOutro/1.Acknowledgements.tex
\chapter*{\huge Acknowledgements}\addcontentsline{toc}{chapter}{Acknowledgements}

\small
First of all, I would like to express my inmost appreciation to my advisor, Prof. Lavinia Heisenberg, for her pertinent support both at the academic, as well as the personal level. Her exceptional mentoring qualities together with her modern and hierarchy free style of leadership unquestionably had a big share in converting my doctorate to an equally successful and enjoyable journey. Moreover, I'm especially thankful to my second supervisor, Prof. Philippe Jetzer, the chair of the doctoral exam, Prof. Lenonardo Degiorgi, and the additional examination committee, including Prof. Luca Amendola, Prof. Camille Bonvin and Prof. Leonardo Senatore, for their time, their beneficial advice and their willingness of providing valuable feedback to my work.
Additionally, I would like to extend my sincere thanks to Prof. Renato Renner, the lecturer of two courses I was a teaching assistant for, whose rigorous but refreshing understanding of fundamental physics represented a wealthy source of inspiration.

My deep gratitude also goes to Prof. Nicolás Yunes, for a very instructive and inspiring collaboration and for his pleasing hospitality, a gratitude I want to extend to all members of the Yunes Gravity Theory Group. A crucial part of this endeavor would therefore not have been possible without the generous support from an EHT Zürich Doc.Mobility fellowship that paved the way for my six moth visit at UIUC.

Special acknowledgements should also go to my various collaborators, my office mates and my colleagues at ITP and beyond, in particular Dr. Fabio D'Ambrosio, Dr. Andrea Giusti, Dr. Francesco Gozzini, Dr. Henri Inchausp\'{e}, Dr. Laura Johnson, David Maibach, Nadine Nussbaumer, Dr. Michael Ruf, Dr. Héctor Villarrubia-Rojo, Guangzi Xu and Stefan Zentarra for helpful insights, very valuable discussions, relaxing lunch breaks and in general a very memorable time at Hönggerberg. In this context, I want to especially also mention Dr. Shubhanshu Tiwari, who sparked my interest in the memory effect.

Lastly, I would be remiss if I did not mention my friends and family who supported me throughout the years of my doctorate. This in particular includes my Acky flat mates, with whom I would also want to live through the next pandemic, my sport buddies that I especially got to know through the incredibly diverse offers of the ASVZ, family friends in Chicago and Vancouver for their warm-hearted accommodation during my time abroad and all long-term friends, the reunions with whom always feel as if the last one was only yesterday. Above all, however, it is most precious to feel the unconditional support of my family, my sister Milena and my parents Myriam and André, providing a safe harbor as the foundation to explore the world together with my partner Zita, the wonder in my life I owe so much to.

%% file: IntoAndOutro/0.Conventions.tex
\chapter*{Conventions and Notation}\addcontentsline{toc}{chapter}{Conventions and Notation}

\paragraph{Units:} If not explicitly stated otherwise, we use natural units, in which $\hbar=c=1$.

\paragraph{Spacetime and Metric Signature:} Spacetime $(\M,\underline{g})$ is described by a four-dimensional, oriented and differentiable topological manifold $\M$ equipped with a metric $\underline{g}$ of signature $(-,+,+,+)$. Generically, underlined objects $\underline T$ denote tensor fields.

\paragraph{Indices:} Spacetime indices are denoted by Greek letters, $\alpha,\,\beta,\,...=0,\,1,\,2,\,3$, while spacial indices are denoted by Latin letters, $i,\,j,\,...=1,\,2,\,3$.


\paragraph{Symmetric and Antisymmetric parts:} We denote symmetrization and antisymmetrization by parentheses $(\;)$ and square brackets $[\;]$ around multiple indices
\begin{equation*}
    T_{(\alpha_1 ... \alpha_p)}=\frac{1}{p!}\sum_{\sigma}T_{\alpha_{\sigma(1)} ... \alpha_{\sigma(p)}}\,,\quad T_{[\alpha_1 ... \alpha_p]}=\frac{1}{p!}\sum_{\sigma} \text{sgn}(\sigma)\,T_{\alpha_{\sigma(1)} ... \alpha_{\sigma(p)}}\,.
\end{equation*}

\paragraph{Levi-Civita Symbol:} The totally antisymmetric symbol $\epsilon^{\mu\nu\rho\sigma}$ has $\epsilon^{0123}=+1$.

\paragraph{Derivatives and Connections:} Partial derivatives of the coordinates are denoted as $\partial_\mu$ or $_{,\mu}$. For instance, applied on a function $f$, we write $\frac{\partial f}{\partial x^\mu}=\partial_\mu f= f_{,\mu}\,.$

The components of the Levi-Civita connection, called Christoffel symbols, are given by
\begin{equation*}
    \Gamma^\lambda_{\mu\nu}=\frac{1}{2}g^{\lambda\rho}\left(g_{\mu\rho,\nu}+g_{\nu\rho,\mu}-g_{\mu\nu,\rho}\right)\,.
\end{equation*}
The associated metric-compatible and torsion-free covariant derivative is denoted as $\nabla_\mu$ or $_{;\mu}$. More general connection-coefficients are denoted by $\,\DGamma^{\,\lambda}_{\,\mu\nu}$, with associated covariant derivative $\DGrad_\mu$.

\paragraph{Curvature tensors:} The components of the Riemann curvature tensor are
\begin{equation*}
    R\ud{\mu}{\nu\rho\sigma}=\Gamma^\mu_{\nu\sigma,\rho}-\Gamma^\mu_{\nu\rho,\sigma}+\Gamma^\mu_{\lambda\rho}\Gamma^\lambda_{\nu\sigma}+\Gamma^\mu_{\lambda\sigma}\Gamma^\lambda_{\nu\rho}\,.
\end{equation*}
The Ricci tensor and Ricci scalar are respectively $R_{\mu\nu}=R\ud{\alpha}{\mu\alpha\nu}$ and $R=g^{\mu\nu}R_{\mu\nu}$.


%% file: MainBody.tex
\pagenumbering{gobble} 
\pagenumbering{arabic} 

\chapter*{\huge Introduction}\addcontentsline{toc}{chapter}{Introduction}

Our current understanding of the phenomenon of gravitation rests upon the revolution in physics that lead to the formulation of general relativity (GR) \cite{Einstein:1915EE,Hilbert:1915A,Einstein:1916GrundlagenGR,Weinberg1972,misner_gravitation_1973,WaldBook,landau_classical_2003,Flanagan:2005yc,maggiore2008gravitational,zee2013einstein,poisson2014gravity,Blau2017,guidry2019modern,Will:2018bme,carroll2019spacetime,Renner2020,YunesColemanMiller:2021lky,Jetzer:2022bme}. From a purely empirical point of view, general relativity accumulated over more than a hundred years a rock-solid experimental ground on a broad band of scales, with tests ranging from high-precision laboratory experiments, the direct observation of gravitational waves all the way to probes on cosmological scales \cite{Dyson1920gg,Clemence:1947uu,Pound:1959aa,Schiff:1960ddd,Kundig:1963kkl,Dicke:1964pna,Shapiro:1964kk,Nordtvedt:1968qs,Nordtvedt:1968first,Nordtvedt:1968:later,Greenstein:1971ff,Weinberg1972,misner_gravitation_1973,Taylor:1982ApJ,Bertotti:2003rm,Ciufolini:2004rq,Amendola:2004wa,Williams:2004qba,Weisberg:2004hi,Mattingly:2005re,Williams:2005rv,Kapner:2006si,Turyshev:2008dr,Merritt:2009ex,Everitt:2011hp,Dossett:2011tn,Hui:2012jb,Chatziioannou:2012rf,Yunes:2013dva,Will:2014kxa,poisson2014gravity,Wex:2014nva,Berti:2015itd,Yunes:2016jcc,Kostelecky:2016kfm,LIGOScientific:2016lio,Blau2017,Asmodelle:2017sxn,Sakstein:2017bws,Will:2018bme,LIGOScientific:2018dkp,Pardo:2018ipy,Ishak:2018his,Barack:2018yly,Kase:2018aps,carroll2019spacetime,Nair:2019iur,LIGOScientific:2019fpa,LIGOScientific:2020tif,Carson:2020rea,LIGOScientific:2021sio,Krishnendu:2021fga,Perkins:2021mhb,Durrer:2022fpc,Castello:2022uuu,Abidi:2022zyd,Jetzer:2022bme}.

Despite its extraordinary success and conceptual beauty, there are at least two reasons to believe that GR is not the ultimate theory of gravitation. Already quite early on, it became clear that it's apparent incompatibility with the rules of the quantum world [Part~\ref{Part: Quantum Gravity}] that governs physics at the smallest scales still leaves a lot of work for future generations of theoretical physicist's 
\cite{Rayski:1978jda,Isham:1992ms,Kiefer:2004xyv,Rickles:2006ee,Strominger:2009aj,Anderson:2010xm,Lindesay:2013iba,Ashtekar:2014ife,Giddings:2022jda,DeWitt:1957obj,DiMauro:2021mcu,zee2013einstein,Green:1987sp,Green:1987mn,Polchinski:1998rq,Polchinski:1998rr,Weinberg:2000cr,Zwiebach:2004tj,Mukhi:2011zz,Rovelli:1997yv,Gambini:2011zz,Ashtekar:2017yom,Ashtekar:2021kfp,Maldacena:1997re,Polchinski:2010hw,Hubeny:2014bla,Penedones:2016voo,Penrose:1964wq,Hawking:1967ju,misner_gravitation_1973,Birrell:1982ix,WaldBook,Fulling:1989nb,Wald:1995yp,Ford:1997hb,Mukhanov:2007zz,Parker:2009uva,carroll2019spacetime}. These tensions, together with the existence of singularities in the solutions of GR indicating the eventual breakdown of its description of nature, triggered the searches for a high-energy ultraviolet (UV) completion of the theory. Although so far the regime of quantum gravity remained out of empirical reach, one might still hope that eventually the unknown UV physics will leave its observational footprints, especially in the strong field regime. This in particular includes the presence of additional non-minimally coupled fields describing new degrees of freedom in the gravity sector at low energies, generically arising from string theory compactifications \cite{Zwiebach:1985uq,Gross:1986mw,Gross:1986iv,Moura:2006pz,Cano:2021rey}, as well as particular scenarios of other quantum gravity attempts \cite{Taveras:2008yf,Mercuri:2009zt}.

On the other hand, the quantum nature of matter also poses a serious theoretical challenge to operators influencing the long distances in the so-called infrared (IR) through the puzzle around the apparent absence of gravitating vacuum energy, known as the cosmological constant (CC) problem [Sec.~\ref{sSec: The CC Problem}], \cite{Weinberg:1988cp}. This theoretical issue is complemented with today's major open questions in cosmology [Part~\ref{Part: Cosmological Testing Ground}],\cite{Weinberg1972,Bertschinger:1993xt,Peebles:1994xt,Coles:1995bd,Liddle:2000cg,landau_classical_2003,Mukhanov:2005sc,Weinberg2008Cosmology,zee2013einstein,maggiore2018gravitationalV2,dodelson2020modern,Abdalla:2022yfr,Peebles:2022akh}. The current GR based understanding of the evolution of the universe in particular requires a postulation of the unknown components of dark energy causing the late time acceleration \cite{SupernovaSearchTeam:1998fmf,SupernovaCosmologyProject:1998vns,Astier:2012ba} that is intimately intertwined with the CC problem, the introduction of dark matter components \cite{Zwicky1933dd,Trimble:1987ee,Bertone:2016nfn,zee2013einstein,dodelson2020modern} foremost required for a coherent formation of large-scale structures, as well as a mechanism for viable cosmological initial conditions \cite{Guth:1980zm,Starobinsky:1980te,Sato:1980yn,Mukhanov:1981xt,Battefeld:2014uga,Brandenberger:2016vhg,Ijjas:2018qbo,Liddle:2000cg,Tsujikawa:2003jp,Cheung:2007st,Gorbunov:2011zzc,Rubakov:2017xzr,Vazquez:2018qdg} (see also \cite{Weinberg1972,Weinberg2008Cosmology,zee2013einstein,maggiore2018gravitationalV2,dodelson2020modern}). In addition, there exists an increasing significance of observational tensions within the GR-based cosmological standard model \cite{Zhao:2017cud,Riess:2019qba, Knox:2019rjx, DiValentino:2020vvd, DiValentino:2020zio,DiValentino:2021izs, Perivolaropoulos:2021jda,Abdalla:2022yfr,Peebles:2022akh,Hu:2023jqc}. While the open puzzles in cosmology might as well find their resolution in a better understanding and extension of the matter sector, including physics beyond the current standard model of particle physics, in this work, we choose to mostly focus on the equally exciting possibility of finding answers in beyond GR effects.

Indeed, both the unknown within the UV and the IR limits of GR drive a widespread search for a potential generalization of the current theory of gravity that might leave its traces in today's and near-future experiments (see \cite{Weinberg1972,Vilenkin:1985md,Copeland:2006wr,Nojiri:2006ri,Nojiri:2010wj,Clifton:2011jh,Hinterbichler:2011tt,Faraoni2011,Yunes:2013dva,deRham:2014zqa,Berti:2015itd,Bamba:2015uma,Cai:2015emx,Nojiri:2017ncd,Heisenberg:2018mxx,Heisenberg:2018vsk,DiValentino:2021izs,CANTATA:2021ktz,poisson2014gravity,papantonopoulos2014EntireBook,Joyce:2016vqv,PetrovKopeikinLompayTekin+2017,Amendola:2018ltt,Will:2018bme,carroll2019spacetime,BeltranJimenez:2019tme,YunesColemanMiller:2021lky,Shankaranarayanan:2022wbx,Heisenberg:2023lru} for reviews). In this context, the advent of the gravitational wave (GW) physics era rung in through the first direct measurements of gravitational waves \cite{LIGOScientific:2016aoc,LIGOScientific:2018mvr,LIGOScientific:2021usb,LIGOScientific:2021djp,KAGRA:2023pio}, represents a unique opportunity to probe the phenomenon of gravitation in new regimes [Part~\ref{Part: Gravitational Wave Testing Ground}], complementing local weak-field experiments as well as the long-range cosmological testing grounds. Based on the bright prospect of upcoming gravitational radiation observatories, GW science is believed to represent one of the most promising future research directions in physics and cosmology and, in particular, the quest towards a deeper understanding of the underlying theory of gravitation.


This promise rests on the ever-increasing number of gravitational wave observatories that will cover a wide range of GW frequencies, originating from all types of astrophysical and cosmic events violent enough to noticeably shake the fabric of spacetime. While the existing ground-based LIGO-Virgo-KAGRA detector network, soon to be joined by LIGO-India \cite{Saleem:2021iwi} is primarily sensitive to transient signals from compact binaries at a frequency range of $10^1$ - $10^3$ Hz, ongoing Pulsar Timing Array (PTA) experiments, listening to cosmological frequencies of the order of $10^{-9}$ - $10^{-3}$ Hz, have recently reported the first detection of a stochastic gravitational wave background  \cite{NANOGrav:2023gor,EPTA:2023fyk,Reardon:2023gzh,Xu:2023wog}. Planned space-based missions \cite{Taiji,TianQin}, in particular the Laser Interferometer Space Antenna (LISA) \cite{Armano:2016gg,LISA}, will try to fill the gap in between at $10^{-3}$ - $10^0$ Hz, such that together with envisioned 3rd-generation ground-based interferometers \cite{Punturo:2010zz,Maggiore:2019uih,Reitze:2019iox,Evans:2021gyd} the future GW detector network will be highly sensitive to a wide range of transient and continuous GW sources. These are not only expected to include a diverse bouquet of binary system types, but also supernova core collapse, rapidly rotating asymmetric neutron stars all the way to potential relics from the very early universe and yet unknown phenomena deep inside the cosmic flow (see \cite{maggiore2008gravitational,Cutler:2002me,Creighton:2011zz}).

However, in order to consistently put general relativity on trial against current and future data and better understand its limits and characteristics, it is first of all indispensable to describe a well-defined theory space beyond it [Part~\ref{Part: EFT of Gravity}], a task which is not always performed with equal care. Indeed, while it is possible to probe GR through blind parameter extensions, such general null tests might fall short due to their inability of capturing the complexity of realistic effects reflected in variations of multiple dependent beyond GR parameters. Put in other words, a lack of a second hypothesis could induce a fundamental bias towards GR that is important to prevent \cite{Yunes:2009ke,Chatziioannou:2012rf,Endlich:2017tqa}.
Conversely, without an alternative model to test for, it is also hard to exclude statistical flukes at the root of potential signatures that depart from the expectation \cite{Abdalla:2022yfr}.
Moreover, the interpretation of experimental results often reside on fundamental assumptions that can be broken though uninformed deformations of the theory parameters, rendering constraints inconsistent. The study of concrete alternative theories and their underlying principles is therefore decisive. This is true both for the analysis of full-fledged non-linear effects in individual gravity theories, as well as for the identification of theory agnostic smoking gun signals beyond GR. Ultimately, a systematic exploration of a viable space of gravity theories will also be rewarded through a better understanding of general relativity itself.

In this work, we choose to put the study of gravity theories on a firm footing by assuming the statements of the Einstein equivalence principle (EEP) [Principle~\ref{Principle:EEP}] at the root of the understanding of gravitation as a phenomenon of spacetime curvature within the initial formulation of GR. The present monograph will therefore start by offering a rather detailed review of the rationale behind this principle in Chapters~\ref{Sec:Theories of Gravity} and \ref{Sec:The Generalization to Gravity}.
Interestingly, the assumption of the EEP does not directly imply the theory of general relativity but leaves room for a large class of gravity theories known as metric theories [Def.~\ref{DefMetricTheory}], 
\cite{Dicke:1964pna,misner_gravitation_1973,poisson2014gravity,papantonopoulos2014modifications,Will:2014kxa,Will:2018bme,YunesColemanMiller:2021lky}. The principle characteristic of metric theories of gravity is their minimal and universal coupling [Principle~\ref{Principle:Universal and Minimal Coupling}] of matter to a unique physical metric that locally reduces to flat Minkowski spacetime, thus ensuring the postulates of the EEP. In Chapter~\ref{Sec:General Relativity} GR is then understood as the special metric theory with a minimal amount of propagating degrees of freedom [Theorem~\ref{Thm:LovelockTheorem}], the concept of which will be thoroughly analyzed in a perturbative approach in Chapter~\ref{Sec:PropagatingDOFs}.




The reason behind insisting on the EEP and an associated universal and minimal coupling of matter to a physical metric is not only based on empirical evidence, but stems from the fundamental requirement of describing dynamical spacetime as a self-sufficient concept, whose properties can be probed and studied independently of the composition and type of measurement devices \cite{Will:2018bme}. As it is sometimes forgotten, any empirical assessment of the phenomenon of gravitation is based on the study of matter, such that the coupling between the gravitational fields with matter represents a major ingredient of a gravity theory, influencing any interpretation of observations. Furthermore, a restriction to metric theories will allow for a viable definition of the notions of gravitational waves on arbitrary background spacetimes pioneered by Isaacson in the case of GR \cite{Isaacson_PhysRev.166.1263,Isaacson_PhysRev.166.1272,misner_gravitation_1973,Flanagan:2005yc,maggiore2008gravitational}, that will turn out to be crucial for the discussion of propagating degrees of freedom.



Chapter~\ref{Sec:The Theory Space Beyond GR} concludes Part~\ref{Part: EFT of Gravity} with a classification of concrete metric theories as effective theories of gravity in terms of their number and type of additional propagating degrees of freedom in the gravitational sector. In this task, Ostrogradski instabilities [Theorem~\ref{Thm:OstrogradskyTheorem}] provides a general guideline for model building and will further draw a decisive distinction between different types of gravity theories.



An introduction to Part~\ref{Part: Gravitational Wave Testing Ground} on the gravitational wave testing ground will be offered in Chapter~\ref{Sec:Radiation andGWs in Gravity}, detailing the notion of gravitational radiation in asymptotically flat space-times, together with a discussion about the concepts of gravitational wave speed and gravitational polarizations defined as the six radiative degrees of freedom that govern the physical effects of gravitational waves in metric theories. This will provide the final basis to describe the first main result of this work in Chapter~\ref{Sec:GWMemory}. Primarily, the Isaacson approach to gravitational waves will be shown to naturally describe the advent of a propagating low-frequency perturbation, which in the limit to null infinity gives rise to a gravitational memory contribution. The phenomenon of memory describing the scars in the fabric of spacetime left behind after a burst of gravitational waves is a prediction of GR \cite{Zeldovich:1974gvh,Christodoulou:1991cr,Ludvigsen:1989cr,Blanchet:1992br,Thorne:1992sdb,PhysRevD.44.R2945,Favata:2008yd,Favata:2009ii,Favata:2010zu,Bieri:2013ada,Strominger:2014pwa,Garfinkle:2022dnm} that is part of the hope of future GW observations to provide new insights into the workings of gravity \cite{vanHaasteren:2009fy,Johnson:2018xly,Yang:2018ceq,Favata:2009ii,Islo:2019qht,Burko:2020gse,Islam:2021old,Sun:2022pvh,LISA:2022kgy,Gasparotto:2023fcg,Ghosh:2023rbe,Goncharov:2023woe,Lasky:2016knh,Boersma:2020gxx,Grant:2022bla,Hubner:2019sly,Ebersold:2020zah,Hubner:2021amk}. In this context, the novel Isaacson approach to understanding gravitational displacement memory will allow for a natural generalization of the memory effect beyond GR. Especially, a general result for the functional form of memory in the tensor polarization will be proven [Theorem~\ref{Theorem1}], that is believed to entail important consequences for future memory based tests of GR.


Next, Chapter~\ref{Sec:CosmoNutshell} will switch gears to the second main topic in Part~\ref{Part: Cosmological Testing Ground} of this thesis and introduce the current standard model of cosmology. The thereby established concepts will allow for the presentation of a model agnostic approach in Chapter~\ref{Sec:CsomoTensions} that is able to formulate simple but effective guiding principles on the search for new physics. Applied to the context of current cosmological tensions mentioned above, we will be able to offer clear-cut analytic constraints that apply to almost any departures from the cosmological standard model, and will draw first connections to concrete metric theories of gravity. 

Part~\ref{Part: Quantum Gravity}, the third major subject on the viability of gravity theories in their contact with the quantum world, will be initialized in Chapter~\ref{Sec:Gravity and Quantum Physics} with a specific focus on a possible formulation of GR as an effective quantum theory of fields 
\cite{Feynman:1963ax,Weinberg:1964ew,Weinberg:1965rz,DeWitt:1967ub,Deser:1969wk,BOULWARE1975,PhysRev.96.1683,tHooft:1974toh,Donoghue:1993eb,Donoghue:1994dn,Dunbar:1994bn,Donoghue:1995cz,Feynman:1996kb,Weinberg:1995mt,Bjerrum-Bohr:2002gqz,Khriplovich:2002bt,Burgess:2003jk,maggiore2008gravitational,zee_quantum_2010,Donoghue:2012zc,zee2013einstein,PetrovKopeikinLompayTekin+2017} and an explicit connection to a common type of metric theories beyond GR. In this context, Chapter~\ref{Sec.Quantum Stability} offers an analysis of the quantum viability of two specific metric theories of gravity. Finally, Chapter~\ref{Sec:Challenges of the Quantum EFT of Gravity} will be devoted to the challenges that a quantum field theory viewpoint on gravity theories still has to face and ends the document with a speculation on a possible alternative approach inspired by the insights of the EEP and the Isaacson approach to the dynamical degrees of freedom of GR.


On top of the main thread delineated above, this monograph is also an attempt to providing a comprehensive introduction to the theory of gravity that nevertheless offers the explicit and detailed treatment of various important and interesting concepts that are not elaborated on in most introductory texts. As such, we tried to answer questions on various subtle points regarding the theoretical framework of physics on which we gained some insight throughout the doctoral studies and gathered the results of these efforts to create a document that we personally would have appreciated at an earlier stage of the research career. The emphasis should however be put here on the word ``attempt'', but we hope to at least partially have reached this goal. 
Questions to which we tempted to provide a more complete answer include: 
\begin{itemize}
    \item How does the concept of infinitesimal tangent vectors relate to the definition of a line element [Sec.~\ref{sSec:The Notion of Spacetime}];
    \item What is the difference between the notions of geodesics, autoparallels and straight lines [Sec.~\ref{sSec:Special Relativity}];
    \item What are the geometric objects of torsion and non-metricity and what is their role in metric theories of gravity [App.~\ref{sApp: connection and curvature} and \ref{sApp: Metric and Riemannian G}],[Sec.~\ref{ssSec:ExtraDOFs as a unique signature}];
    \item What is the fundamental difference between a local and a global symmetry [Sec.~\ref{Sec:Theories of Gravity}],[App.~\ref{App: Symmetires in Physics}];
    \item Related to this is the notion of gauge freedom and its promotion to a gauge symmetry [Sec.~\ref{Sec:Theories of Gravity}],[App.~\ref{App: Symmetires in Physics} and \ref{sApp: Spacetime Gaugefreedom and symmetries}]; 
    \item How is the freedom of diffeomorphic transformations precisely related to the freedom of coordinate transformations [App.~\ref{sApp:DiffsAndLieDer}]; 
    \item What is the physical metric of a gravity theory, and why is every metric theory defined through a Jordan frame [Sec.~\ref{sSec:Metric Theories} and \ref{ssSec: A Exact Theories}];  
    \item What is the distinction between matter fields and fields in the gravity sector [Secs.~\ref{sSec:Metric Theories} and \ref{sSec:Covariant Consrevation}]; 
    \item In what sense is a description in the Einstein frame equivalent to a description in the Jordan frame, and why should one be careful when using the Einstein frame of a theory [App.~\ref{ssSec: A Exact Theories}]; 
    \item Related to this is the distinction between a conformal transformation and a Weyl rescaling [App.~\ref{ssSec: A Exact Theories}];
    \item Also related is the fundamental absence of fifth forces in metric theories and the status of screening mechanisms [Secs.~\ref{sSec:Metric Theories} and \ref{ssSec:Screening}];
    \item What is the relation of infinitesimal one parameter families of diffeomorphisms, Lie derivatives and infinitesimal coordinate transformations [App.~\ref{sApp:DiffsAndLieDer}]; 
    \item How are perturbations of a theory on a manifold well-defined and how does their gauge freedom arise [Sec.~\ref{sSec:PerturbationTheory}]; 
    \item What is the distinction between unphysical gauge degrees of freedom, physical non-dynamical degrees of freedom and physical dynamical degrees of freedom [Sec.~\ref{sSec:GWs in GR}];
    \item How is the notion of dynamical degrees of freedom distinct from the measurable gravitational polarizations in metric theories of gravity [Secs.~\ref{ssSec:WavesInMetricTheoriesOfGravity}, \ref{sSec:GWPolGen} and \ref{sSec:GWPolExample}]; 
    \item Related to gravitational polarizations is the definition of the notions of scalar vector and tensor memory [Sec.~\ref{Sec:GWMemory}]; 
    \item How are gravitational waves defined as a physical concept with well-defined energy-momentum tensor [Secs.~\ref{ssSec:IsaacsonInGR} and \ref{ssSec:IsaacsonGeneral}];
    \item What is the difference between the notion of gravitational waves and gravitational radiation [Sec.~\ref{sSec:Asymptotic Flatness}]; 
    \item What justifies the use of the TT gauge in describing gravitational waves of GR [Secs.~\ref{ssSec:Local Wave Equation in GR}, \ref{ssSec:GaugeInvariantDecomposition} and \ref{sSec:GWPolExample}]; 
     \item How is the local spacial velocity of gravitational waves defined, and when can it depart from the speed of light [Secs~\ref{sSec:Special Relativity} and \ref{sSec:GWPropagation}].
    \item Why do current GW detectors only measure deviations in \textit{spacial} proper distances [Secs.~\ref{sSec:Special Relativity}, \ref{sSec:Metric Theories} and \ref{ssSec:The Physical Effects of Gravitational Waves}]; 
    \item When and how are spacial proper distances well-defined [Sec.~\ref{sSec:Special Relativity}]; 
    \item How are spacial distances in cosmology defined [Sec.~\ref{sSec:HomIsoUniverse}]; 
    \item What is the status of theories that taken at face value possess an Ostrogradsky instability, and how can they still be used [Secs.~\ref{sSec:OstrogradskyTheorem} and \ref{ssSec: B Perturbative Theories}];  
    \item Related to this are two distinct notions of well-posedness of a theory [Sec.~\ref{ssSec:WellPosedness}].

\end{itemize}

\part{Effective Theories of Gravity}\label{Part: EFT of Gravity}

\small
\noindent
\emph{\ul{Personal Contribution and References}}\\ 
\footnotesize 
\textit{Chapter~\ref{Sec:PropagatingDOFs} is based on \textbf{L. Heisenberg, N. Yunes, J. Zosso, 2023} \cite{Heisenberg:2023prj}, in particular Sec.~\ref{sSec:Isaacson and GWs Generalization Beyond GR}. Parts of the following treatment are also inspired from \cite{Weinberg1972,misner_gravitation_1973,WaldBook,Flanagan:2005yc,maggiore2008gravitational,poisson2014gravity,papantonopoulos2014modifications,zee2013einstein,Blau2017,Will:2018bme,carroll2019spacetime,Renner2020,Jetzer:2022bme}.}
\normalsize

\vspace{5mm}

\noindent
\textbf{Summary of Part I}\\ 
\noindent
We want to start by reviewing how modern gravity theories, including general relativity, emerge from the attempt of making Newtons theory of gravity compatible with the principles of special relativity (SR). This naturally leads to generalizing the Minkowski metric of SR to a dynamical object that captures the phenomenon of gravitation and the formulation of a key principle of theories of gravitation. This principle, generally known as Einstein equivalence principle, will subsequently guide us in constructing and classifying gravity theories beyond GR. At the same time, these considerations will make it clear that general relativity remains quite special among the large space of effective field theories of gravity. 
The notion of uniqueness of GR will also give an opportunity to introduce the concept of dynamical degrees of freedom.


\chapter{Special Relativity on a Manifold}\label{Sec:Theories of Gravity}

The initial formulation of special relativity \cite{Einstein:1905ve} was based on three key principles:
\begin{principle}\ul{Principles of Special Relativity}.\label{Principle:SR}
\begin{enumerate}[I.]
\item \ul{Principle of relativity}: In every inertial frame of reference, the non-gravitational physical laws are the same.
    \item \ul{Universality of the speed of light}: In every inertial frame of reference, the propagation speed of light, and any massless test particle, in empty space is given by the same constant $c$.
    \item \ul{Maximality of the speed of light}: In empty space, the speed of any massive test particle relative to any inertial frame is always less than the speed of light.
\end{enumerate}
\end{principle}
Here, an \textit{\gls{inertial frame of reference}}\footnote{All blue colored italic words are defined in the glossary in the Appendices.} is defined as a coordinate system, in which a \textit{\gls{test particle}} exhibits uniform motion in a straight line, whenever the net force on the mass is zero. In other words, it is a reference frame that satisfies Newtons first law. To each inertial frame, we can associate an inertial observer, whose trajectory is identified with the one of a free massive test particle that is at rest at the origin of the corresponding coordinate system. Note that inertial reference frames are only defined relative to each other and are related through a constant relative velocity. Moreover, for the moment, we explicitly exclude any gravitational effects. 

\section{The Notion of Spacetime}\label{sSec:The Notion of Spacetime}

It was later understood \cite{Poincare:1907ve,Minkowski:1909ve}, that the Principles~\ref{Principle:SR} of SR are naturally incorporated in the concept of a Minkowski spacetime $\mathbb{R}\times\mathbb{R}^3$, endowed with a corresponding Minkowski metric. We should therefore pause here for a moment and properly introduce the notion of \textit{\gls{spacetime}}. Intuitively, spacetime describes a ''container'' where all ``events'' that we can physically measure happen and where different events are related to each other through the laws of physics. Here, an event very generally corresponds to a possible interaction of matter that for a specific observer can be labeled by some time and place, which in particular also encompasses the simple manifestation of the location of a particle that in general can be deduced through a certain interaction. 

However, this intuitive picture needs to be rendered more precise in at least two aspects (see also \cite{misner_gravitation_1973}). First of all, it is important to realize that any probe of spacetime is fundamentally based on the study of test matter, whose influence on the space-time itself is assumed to be negligible. This implies that instead of some sort of container, spacetime itself can be defined as the collection of all possible events. Moreover, this set of all possible events is to be distinguished from the coordinate grid a specific (inertial) observer is using to study spacetime through the use of clocks and rulers to describe, for instance, the motion of test-particles. The set of all possible events is defined before the introduction of any specific coordinate system, which fundamentally depends on a given observer.




\paragraph{Spacetime as a Manifold.} Mathematically, spacetime is therefore a set of points with a certain structure, where the set of points may be labeled by an observer. It so happens, that this structure is conveniently given by assuming that spacetime is described through a differential manifold endowed with a metric providing a notion of distance and (hypothetical) causality between events. In other words, the mathematical framework of differential geometry include a convenient set of assumptions to serve as a model of spacetime.\footnote{Observe, however, that certain assumptions, in particular the requirement of differentiability or smoothness is of pure convenience and free of any empirical relation. Indeed, based on quantum mechanics, the definition of a manifold as a collection of events would presumably in the contrary yield a rather discontinuous notion of spacetime (we will come back to this point in Part~\ref{Part: Quantum Gravity}). However, on large enough scales, the assumption of smoothness is a very reasonable and practical one.} In these terms, Minkowski spacetime is therefore a very special spacetime with fixed metric and fixed topology. 

Of course, using the machinery of differential geometry to describe Minkowski spacetime is a bit of an overkill, and rests upon a considerable amount of hindsight. Indeed, Minkowski spacetime is often simply thought of being equivalent to a vector space. However, especially as concerns non-inertial effects, differential geometry actually naturally arises within special relativity itself. In the following, we will indeed offer a study of Minkowski spacetime within this more general framework that will turn out to be of great value.
In particular, this will allow the natural connection of the mathematical entities of differential geometry on a manifold with basic notions of physics. The considerations in this chapter will then serve as a smooth transition to the description of gravity theories in Chapter~\ref{Sec:The Generalization to Gravity}. A key role in this transition will be held by the trajectories of test particles through spacetime, called \textit{\gls{worldline}}, defined as a history of events of the manifestation of the location of a particle. More precisely, of particular interest will be worldlines of \textit{free} test particles, the \textit{\gls{straight lines}} of spacetime and their interplay between the notions of \textit{geodesics} and \textit{autoparallels} (both defined below).

In the main text, however, we want to focus on the physical aspects and will only provide a minimal introduction to the framework of differential manifolds, mostly considering its mathematical foundation as a prerequisite. However, in Appendix~\ref{App:DiffGeo} we offer a concise summary of the most important concepts and objects of differential pseudo-Riemannian geometry. Underlined mathematical definitions that we do not define in the main text are introduced in Appendix~\ref{App:DiffGeo}.

\paragraph{The Metric and Geodesics.}

Nevertheless, we want to at least provide a minimal introduction into the mathematical notation we employ. Let's therefore consider a spacetime given by a four-dimensional, differential \ul{manifold}, endowed with a \ul{metric}. One of the most basic notions in every spacetime are parameterized \ul{curves}, some of which will be associated to the worldlines of test particles. In some coordinate representation we will denote a curve as $x^\mu(\lambda)$, for some parameter $\lambda$ along the curve, where $\mu=0,1,2,3$ labels the coordinate system $x^\mu=(x^0,x^1,x^2,x^3)$. Note that a coordinate system or \ul{chart} is defined by a set of four scalar fields that attach a unique label to each point in spacetime\footnote{In particular, the coordinates $x^\mu$ should not be confused with components of a vector field.}. The \ul{tangent vectors} of all curves, defined through directional derivatives $d/d\lambda$ along the curves, then provide the notion of a \ul{tangent space} at each point in spacetime. More precisely, the tangent vectors $\underline{v}$\footnote{We underline tensorial quantities in order to distinguish them from their components.} can be characterized through their components expanded in a \ul{coordinate induced basis} $\underline{\partial}_\mu$
\begin{equation}\label{eq:Tangent Vector of Curve}
    \underline{v} = \dot x^\mu\underline{\partial}_\mu\,,
\end{equation}
where
\begin{equation}
    \dot x^\mu\equiv \frac{dx^\mu(\lambda)}{d\lambda}\,.
\end{equation}
The metric $\underline{g}$ of the spacetime, a non-degenerate and symmetric $\binom{0}{2}$-tensor with components $g_{\mu\nu}$ in a given coordinate system, then captures the notion of distances by providing a physical magnitude of vectors in terms of a norm
\begin{equation}\label{eq:Norm of Vector}
    \underline{g}(\underline{v},\underline{v})=g_{\mu\nu}\dot x^\mu \dot x^\nu\,.
\end{equation}

However, note that a norm of a vector on a general manifold does not define a distance between two spacetime points. This is because in general, a vector is only defined on the tangent space of a single point on the manifold and cannot connect two different points. In order to talk about a proper spacetime distance it is therefore useful to introduce the notion of \ul{infinitesimal tangent vectors} of a curve $x^\mu(\lambda)$, the components of which are defined as 
\begin{equation}\label{eq:Infinitesimal Tankgent vector}
    dx^\mu\equiv \dot x^\mu\,d\lambda\,.
\end{equation}
Such infinitesimal tangent vectors, as opposed to arbitrary vectors, connect two different but \ul{neighboring points} on the manifold. Thus, their norm defines an infinitesimal distance $ds$, termed \ul{line element} that is given by
\begin{equation}\label{eq:LineElementDef}
   \boxed{ ds^2\equiv  g_{\mu\nu} dx^\mu dx^\nu\,.}
\end{equation}

This infinitesimal distance on a curve can then be used to define the natural concept of distance on a manifold, which is given by the proper spacetime length of a curve known as \ul{arc-length} between two points $x^\mu_i=x^\mu(\lambda_i)$ and $x^\mu_f=x^\mu(\lambda_f)$, through
\begin{equation}\label{eq:Length of Curve}
    L\equiv \int_{x^\mu(\lambda)} ds =\int_{\lambda_i}^{\lambda_f}   d\lambda\,\sqrt{|g_{\mu\nu} \dot x^\mu \dot x^\nu|}\,.
\end{equation}
A \ul{geodesic} of a metric between two points $x^\mu_i$ and $x^\mu_f$ in spacetime is then defined as the curve, that extremizes the length $L$. Geodesics are therefore solutions of the \ul{geodesic equation}
\begin{equation}\label{eq:GeodesicEq}
    \boxed{\frac{d\dot x^\mu}{d\lambda}+\frac{1}{2}g^{\mu\nu}\left(g_{\alpha\nu,\beta}+g_{\beta\nu,\alpha}-g_{\alpha\beta,\nu}\right)\, \dot x^\alpha \dot x^\beta=0\,.}
\end{equation}
for any affine parametrization $\lambda$, such that the first integral satisfies
\begin{equation}
    g_{\mu\nu} \dot x^\mu \dot x^\nu=C= \text{const.}
\end{equation}

Furthermore, within a specific coordinate induced basis, the basis vectors of the tangent space $\underline{\partial}_\mu$ depend on the coordinates, such that also the components of tangent vectors will transform under a change of coordinates $x'^\mu=x'^\mu(x)$. According to the chain rule of derivatives (see Appendix~\ref{sApp:ManifoldCurvesTangent} for more details) the transformation of vector components reads
\begin{equation}\label{eq:VecTransformation}
    \dot x'^\mu=\frac{\partial x'^\mu}{\partial x^\nu}\dot x^\nu\,,
\end{equation}
while, for the metric, the transformation is given by
\begin{equation}\label{eq:GeneralCoordinateTransformation}
    g'_{\mu\nu}(x')=\frac{\partial x^\alpha}{\partial x'^\mu}\frac{\partial x^\beta}{\partial x'^\nu}\,g_{\alpha\beta}(x)\,.
\end{equation}

\paragraph{Gauge Freedom.}

In a theory on a manifold, such as special relativity, the mathematical objects defined above, for example the tangent vectors [Eq.~\eqref{eq:Tangent Vector of Curve}], their norm [Eq.~\eqref{eq:Norm of Vector}] and the length of a curve [Eq.~\eqref{eq:Length of Curve}], are associated to physical quantities and should therefore not depend on our choice of description of them. Indeed, it is a fundamental assumption in (classical) natural sciences, that well-defined measurable quantities, hence any \textit{\gls{physical observable}}, should not depend on the observer, or any individual choice he or she makes. Such freedom in the description in theory of physics is known as \textit{gauge freedom} (see also Appendix~\ref{App: Symmetires in Physics}). Indeed, any respectable theory of nature should not depend on particular choices of description of an observer.

An example of a gauge freedom is the choice of parametrization of a curve and indeed, the length of a curve defined in Eq.~\eqref{eq:Length of Curve} is independent of the choice of parametrization of the curve, as it should if we want to employ it as a well-defined physical quantity.
Another important example of gauge freedom is the choice of a coordinate system described above. For a theory defined on a manifold, the freedom of coordinate transformations can formally be described in terms of the freedom of performing diffeomorphic transformations, as discussed in Appendix~\ref{sApp: Spacetime Gaugefreedom and symmetries}. In fact, the tensor fields defined on a manifold, their norm, as well as the arc-length, are well-defined quantities, even before introducing any coordinates on the manifold. Describing Minkowski spacetime as a differentiable manifold with a special metric can therefore be viewed as a convenient way of manifesting the gauge freedom of coordinate transformations.

Moreover, any such gauge freedom in the description can be turned into a \textit{gauge symmetry} of the theory (see Appendix~\ref{App: Symmetires in Physics}). In other words, for any gauge freedom, one can find a formulation of the theory that is manifestly invariant under the gauge transformations. 
This is in particular also true for the gauge freedom of coordinate transformations. Any theory can be formulated in a coordinate transformation invariant way \cite{Weinberg1972,misner_gravitation_1973,carroll2019spacetime}, where a theory is defined as symmetric under coordinate transformations so long as its action is invariant. We refer the reader to Appendix~\ref{sApp: Gauge Symmetries and Proper Symmetries} for more details on these statements.

However, in practice, it is often convenient to work in a specific chart. Even more so in the theory of special relativity, which is a theory on a manifold with a given fixed metric. In other words, the metric as a tensor field on the manifold is fixed a priori and does not come with its own dynamical equations of motion. This also implies the existence of an a priori preferred choice of coordinate system for the description of special relativity. 
The strategy for the following subsection will thus be to identify the preferred chart in SR corresponding to the inertial frames of references, where we can link the mathematical concepts of spacetime to physical quantities and then generalized again to a description in arbitrary coordinates, a description that is indeed invariant under diffeomorphic transformations. 


\section{The Spacetime of Special Relativity}\label{sSec:Special Relativity}

\paragraph{Inertial Frames and Poincar\'e Invariance.}
We now want to show that within the mathematical framework described above (see also Appendix~\ref{App:DiffGeo}), the theory of special relativity and its principles are very naturally described in Minkowski spacetime $\mathbb{R}\times\mathbb{R}^3$. As a little warning, this first part will be a bit pedantic, as it will almost from scratch attempt to associate intuitive physical concepts such as trajectories and velocities with a description in Minkowski spacetime. However, we hope that such a careful introduction will eventually pay off in later chapters as a conceptual guideline.

The \textit{Minkowski metric} associate to Minkowski spacetime can be defined as a metric for which there exists a global coordinate systems $y^\mu=(t,x,y,z)$ called \textit{Minkowski coordinates}, in which the metric components reduce to
\begin{equation}\label{eq:MinkowskiMetric}
    g_{\mu\nu}(y)=\eta_{\mu\nu}=\text{diag}(-1,1,1,1)\,,
\end{equation}
at every spacetime point. Note that these coordinates span the entire spacetime. This special chart then naturally corresponds to the inertial reference frames of special relativity. In such coordinates, the associated line element, hence the physical infinitesimal spacetime distance $ds$ between nearby points with coordinates $y^\mu$ and $y^\mu+dy^\mu$, is given by
\begin{equation}\label{eq:SR}
    ds^2=\eta_{\mu\nu}dy^\mu dy^\nu=-dt^2+dx^2+dy^2+dz^2\, .
\end{equation}

First of all, we will derive a natural description of the worldlines of free physical test-particles in spacetime by associating them with a curve $y^\mu(\lambda)$. In Euclidean space within the Newtonian picture, the trajectories of free massive particles correspond to \textit{straight lines}, a nomenclature we will retain in Minkowski spacetime (see also the discussion on flatness and geodesic deviation below). In order to find out the description of a straight line in Minkowski spacetime of special relativity, it is a good starting point to consider an inertial observer $O$ and associate to that observer a coordinate system $y^\mu$, in the sense that the coordinate $t$ naturally corresponds to the time measured by the observer at rest at the origin in this coordinate system. More precisely, we associate a curve $y_O^\mu(\tau)$ to the observer, with $y_O^0=\tau=t$ and $y_O^i=0$, where $i=1,2,3$, whose only non-zero component of the tangent vector is in the $t$-direction $\dot y_O^\mu=t^\mu=\delta\ud{\mu}{0}$. Note that in this case, the parameter $\tau$, called \textit{proper time}, corresponds to the arc-length and satisfies
\begin{equation}\label{eq:Proper Time Definition}
    \eta_{\mu\nu}\, \dot y_O^\mu(\tau) \dot y_O^\nu(\tau)= -1\,.
\end{equation}
Moreover, we have that
\begin{equation}\label{eq:StraightLineInertialObserver}
   \frac{d \dot y_O^\mu}{d\tau}=\dot y_O^\nu\, \dot y_O\ud{\mu}{,\nu}=0\,.
\end{equation}
In other words, the tangent vector of an inertial observer does not change upon taking a directional derivative. 

Based on Eq.~\eqref{eq:Proper Time Definition}, we therefore postulate, that the equation of motion of any free test mass in an inertial frame of Minkowski spacetime reads\footnote{Note that this equation of motion is not invariant under general coordinate transformations. However, and crucially, this equation is invariant under the Lorentz transformations defined below and has thus the same form in every inertial frame.}
\begin{equation}\label{eq:StraightLineInertial}
   \frac{d \dot y^\mu}{d\lambda}=\dot y^\nu\, \dot y\ud{\mu}{,\nu}=0\,,
\end{equation}
for an affine parametrization, such that
\begin{equation}\label{eq:ParametrizationSpeed}
    \eta_{\mu\nu} \dot y^\mu \dot y^\nu= C\,,
\end{equation}
where $C$ is a constant. Within the Minkowski coordinates, such curves therefore correspond to the intuitive notion of ``straight''. Moreover, Eq.~\eqref{eq:Proper Time Definition} demands $C<0$ for any massive particle, which in turn implies the parametrization invariant requirement, that a worldline of a free massive particle satisfies
\begin{equation}\label{eq:Timelike Curve}
    ds^2<0\,.
\end{equation}
Any curve satisfying Eq.~\eqref{eq:Timelike Curve} at every point will be called a \textit{timelike} curve.
Moreover, in this context, Eq.~\eqref{eq:StraightLineInertial} is the relativistic version of the statement, that Newton's $1^\text{st}$ law for the motion of free test particles is satisfied in an inertial frame.

In order to connect the description in Minkowski spacetime to the Principles~\ref{Principle:SR} of special relativity, we also need to define a notion of spacial \textit{velocity}. As discussed above, any curve defines a natural vector at each point, the tangent vector with components $\dot y^\mu$ for some inertial observer $O$ also sometimes called $4$-velocity. However, as seen above, the norm of this vector depends on the parametrization of the curve and does therefore not have any physical significance. However, by interpreting the component $\dot y^0$ as a measure of time\footnote{Strictly speaking, for a general spacetime, a measure of time and space in the sense of connecting two events in spacetime is only given by the infinitesimal version of the tangent vector. However, since in the definition of the spacial velocity the parametrization dependent length of the components is effectively cancelled out this does not play a role here.} of the observer $O$ while $\dot y^i(\lambda_0)$ corresponds to a measure of spacial distance in the direction $i$, intuitively, the velocity of a physical test particle with worldline $y^\mu(\lambda)$, at some instant $\lambda=\lambda_0$ as measured by the inertial observer $O$ can locally be defined as
\begin{equation}\label{eq:definition Speed}
    v^i\equiv \frac{\dot y^i(\lambda_0)}{\dot y^0(\lambda_0)}\,,
\end{equation}
where we assume $\dot y^0>0$. Note that, decisively, this definition is independent of the parametrization of the worldline of the particle and therefore represents a physical quantity as measured from the perspective of the specific observer $O$.

The above definition of spacial velocity in turn allows us to draw the connection to the second principle of special relativity, which demands that in an inertial frame and at every instant, the Euclidean norm of the velocity of a light signal, or more generally a massless particle, satisfies (recall that we work in natural units, in which $c=1$)
\begin{equation}
    \lVert v_\text{light} \rVert \equiv \frac{\sqrt{\dot y^i\dot y^i}}{\dot y^0}=c=1\,,
\end{equation}
Minkowski spacetime is therefore constructed, such that tangent vectors of worldlines of massless particles are null $\eta_{\mu\nu}\dot y^\mu\dot y^\nu=-(y^0)^2+y^iy^i=0$, or equivalently $ds^2=0$. Moreover, it is natural to postulate, that a free massless particle also satisfies Eq.~\eqref{eq:StraightLineInertial} with $C=0$ with respect to the inertial observer $O$.

The $3^\text{rd}$ principle of special relativity, which demands that the maximum speed of propagation corresponds to the speed of light, then imposes, that any massive test particles, not only the free ones, follow timelike worldlines, satisfying $ds^2<0$ at every point, while \textit{spacelike} separated events $ds^2>0$ are therefore not causally connected. Observe that this classification into timelike, null and spacelike curves is coordinate invariant.

It remains to address the 1$^\text{st}$ of the SR Principles~\ref{Principle:SR}, more precisely the statement that there is more than one inertial frame of reference\footnote{Strictly speaking, inertial frames are even only defined with respect to each other.} for each of which the structure above needs to be preserved. A transformation from one inertial frame associated to coordinates $y^\mu$ and a second inertial frame $y'^\mu$ should therefore especially preserve the special form of the Minkowski metric \eqref{eq:MinkowskiMetric}, in order to ensure that the speed of light is the same in all inertial frames. Thus, we demand that
\begin{equation}\label{eq:PoincareCondition}
    \frac{\partial y^\alpha}{\partial y'^\mu}\frac{\partial y^\beta}{\partial y'^\nu}\eta_{\alpha\beta}=g'_{\mu\nu}(y')\stackrel{!}{=}\eta_{\mu\nu}\,,
\end{equation}
Very generally, two inertial observers are thus related through an at most linear coordinate transformation
\begin{equation}\label{eq:PoincareT}
    y'^\mu=(\Lambda^{-1})\ud{\mu}{\nu}y^\nu+a^\mu\,,
\end{equation}
with the additional requirement
\begin{equation}\label{eq:LorentzT}
    \Lambda\ud{\alpha}{\mu}\Lambda\ud{\beta}{\nu}\eta_{\alpha\beta}=\eta_{\mu\nu}\,,
\end{equation}
where both $\Lambda\ud{\mu}{\nu}$ and $a^\mu$ are independent of the coordinates.
The transformations in Eq.~\eqref{eq:PoincareT} together with Eq.~\eqref{eq:LorentzT} are known as \textit{Poincar\'e transformations}, while the purely linear piece $\Lambda\ud{\mu}{\nu}$ correspond to \textit{Lorentz transformations}. It is readily verified, that in the slow motion limit in which $c\rightarrow \infty$, these transformations reduce to the Galilean transformations, while the equations of motion for free test particles in Eq.\eqref{eq:StraightLineInertial} reduce to Newtons equation of a straight line in Euclidean space, therefore further validating the construction above.

The condition in Eq.~\eqref{eq:PoincareCondition} promotes the coordinate transformations in Eq.~\eqref{eq:PoincareT} to symmetries of the spacetime and, in particular, implies that the form of the line element in Minkowski coordinates remains the same
\begin{equation}\label{eq:LorentzSymmetryLineElement}
    ds^2=\eta_{\mu\nu}dy^\mu dy^\nu=\eta_{\mu\nu}dy'^\mu dy'^\nu\, .
\end{equation}
Note the crucial difference between the invariance of the line element under general coordinate transformations, which is in a sense a trivial statement, and Eq.~\eqref{eq:LorentzSymmetryLineElement}, called an \ul{isometry}, which is a profound statement about a metric of spacetime that remains unchanged under a particular subset of coordinate transformations. Such an invariance of the underlying structure of spacetime is a manifestation of a global (or proper) symmetry of the theory, with a far-reaching connection to conservation laws through Noether's $1^\text{st}$ theorem \cite{Noether:1918zz}. In Appendix~\ref{App: Symmetires in Physics} we review the Noether theorems, as well as the distinction between global and local symmetries and in Sec.~\ref{sSec:Covariant Consrevation} below, we will further discuss the implications of isometries and their associated conserved quantities, in particular the energy momentum tensor. But first, we will promote the insights gained in inertial frames to a formulation in arbitrary coordinates and in particular consider the generalization of the equations of a free particle.

\paragraph{Non-Inertial Frames and Acceleration.} 
The inertial frames discussed above represent preferred coordinate systems of Minkowski spacetime, such that for most situations it is useful to discuss special relativity within such inertial coordinates. However, as already discussed, general coordinate transformations are a gauge freedom of the theory, such that nothing prevents us from using a different coordinate representation. Indeed, in certain scenarios, other coordinates may be even more useful, for example, if the physical problem possesses a spherical symmetry or if one wants to study accelerated observers and apparent forces in non-inertial frames. Of course, as the reader might already know, formulating special relativity in arbitrary coordinates is ultimately interesting for our purposes as it represents a big leap towards unifying the theory of gravity with the principles of special relativity. It is nevertheless an interesting exercise to realize how much of the formalism that is usually only associated to gravity theories already serves in pure special relativity.

Let's therefore derive the motion of particles in special relativity in general coordinates by transforming inertial coordinates $y^\mu$ to arbitrary coordinates $ x^\mu=x^\mu(y)$. As we have already seen in Eq.~\eqref{eq:VecTransformation}, the components of the tangent vector of a curve $x^\mu(\lambda)$ in a coordinate induced basis transforms as
\begin{equation}
    \dot y^\mu=\frac{\partial y^\mu}{\partial x^\nu}\dot x^\nu\,.
\end{equation}
Hence, the worldline of free test particles in inertial frames in Eq.~\eqref{eq:StraightLineInertial} is modified as
\begin{equation}
    \frac{d\dot y^\mu}{d\lambda}=\frac{d}{d\lambda}\left(\frac{\partial y^\mu}{\partial x^\alpha}\dot x^\alpha\right)=\frac{\partial y^\mu}{\partial x^\alpha}\frac{d\dot x^\alpha}{d\lambda}+\frac{\partial^2 y^\mu}{\partial x^\alpha\partial x^\beta}\dot x^\alpha\dot x^\beta=0\,.
\end{equation}
In a general frame of reference, the equation of motion for a free test mass therefore reads
\begin{equation}\label{eq:StraightLineArbitrary}
   \frac{d\dot x^\mu}{d\lambda}+\Gamma^\mu_{\alpha\beta}\dot x^\alpha\dot x^\beta=\dot x^\alpha\left(\dot x\ud{\mu}{,\alpha}+\Gamma^\mu_{\alpha\beta} \dot x^\beta\right)=\dot x^\alpha \dot x\ud{\mu}{;\alpha}=\dot x^\alpha \nabla_\alpha\dot x^\mu=0\,,
\end{equation}
where we have identified the so called \ul{Christoffel symbols}
\begin{equation}\label{eq:ChristoffelFromCoordTransfo}
    \boxed{\Gamma^\mu_{\alpha\beta}\equiv\frac{\partial x^\mu}{\partial y^\rho}\frac{\partial^2 y^\rho}{\partial x^\alpha\partial x^\beta}\,,}
\end{equation}
as well as the \ul{covariant derivative} 
\begin{equation}
    \boxed{\nabla_\alpha\dot x^\mu\equiv\dot x\ud{\mu}{;\alpha}\equiv\dot x\ud{\mu}{,\alpha}+\Gamma^\mu_{\alpha\beta} \dot x^\beta\,.}
\end{equation}
Eq.~\eqref{eq:StraightLineArbitrary} therefore determines a ``straight line'', defined as the worldline of a free particle, in arbitrary coordinates. If such general coordinates are interpreted as the non-inertial frame of reference of an observer, then the additional term $\Gamma^\mu_{\alpha\beta}\dot x^\alpha\dot x^\beta$ in Eq.~\eqref{eq:StraightLineArbitrary} can be interpreted as an acceleration, associated to an apparent force, as seen by the non-inertial observer.

We now want to show that these worldlines of free test particles naturally coincide with the \textit{geodesics} of the spacetime (defined above) and simultaneously with the notion of \textit{autoparallels} of the \ul{Levi-Civita connection} (to be defined below). To achieve this, we first observe that in general frames, the components of the metric now depend on the coordinates
\begin{equation}\label{eq:metric Transform}
    g_{\mu \nu}(x)=\eta_{\alpha\beta}\frac{\partial y^\alpha}{\partial x^\mu}\frac{\partial y^\beta}{\partial x^\nu}\,,
\end{equation}
implying that in a general chart the derivative of the metric is non-vanishing. In fact, through Eq.~\eqref{eq:metric Transform} the derivative of the metric is intimately related to the Christoffel symbols via
\begin{equation}\label{eq:DMetricToChristoffel}
    g_{\mu\nu,\rho}(x)=\eta_{\alpha\beta}\left(\frac{\partial^2 y^\alpha}{\partial x^\rho\partial x^\mu}\frac{\partial y^\beta}{\partial x^\nu}+\frac{\partial y^\alpha}{\partial x^\mu}\frac{\partial^2 y^\beta}{\partial x^\rho\partial x^\nu}\right)=g_{\mu\lambda}\Gamma^\lambda_{\nu\rho}+g_{\nu\lambda}\Gamma^\lambda_{\mu\rho}\,.
\end{equation}
This equation can then be inverted to give
\begin{equation}\label{eq:Christoffel}
    \boxed{\Gamma^\lambda_{\mu\nu}=\frac{1}{2}g^{\lambda\rho}\left(g_{\mu\rho,\nu}+g_{\nu\rho,\mu}-g_{\mu\nu,\rho}\right)\,.}
\end{equation}
Using this result to compare Eqs.~\eqref{eq:GeodesicEq} and \eqref{eq:StraightLineArbitrary}, we can conclude that as expected the ``straight lines'' of a spacetime, hence the worldlines of free particles, coincide with the geodesics of the spacetime.\footnote{In fact, we could have concluded this already by comparing Eq.~\eqref{eq:StraightLineInertial} to the geodesic equation in Eq.\eqref{eq:GeodesicEq} in an inertial frame. Since the condition in Eq.~\eqref{eq:Timelike Curve} is invariant under coordinate transformations, this conclusion holds in any frame of reference.} In other words, free particles move through spacetime along extremal curves.

On the other hand, the advent of the Christoffel symbols in arbitrary coordinates can alternatively also be understood at the level of the coordinate induced basis $\underline{\partial}_\mu$ of the vector spaces, in which we expand generic vectors as $\underline{V}=V^\mu\underline{\partial}_\mu$. Indeed, from the point of view of the coordinate dependence of the basis vectors the appearance of the Christoffel symbols above stems from the fact that the vector space basis changes from point to point in spacetime, in the sense that the derivative of basis vectors is non-vanishing
\begin{equation}
    \underline{\partial}_{\mu,\nu}\equiv \Gamma^\lambda_{\mu\nu}\,\underline{\partial}_\lambda\,.
\end{equation}
More precisely, consider a curve $x^\mu(\lambda)$ and a vector field $\underline{V}(\lambda)$ defined at each point of the curve. Then both the vector components as well as the basis depend on the curve parameter $\lambda$, such that
\begin{equation}\label{eq:DefCovariantDerivativeD}
    \frac{d\underline{V}}{d\lambda}=\frac{dV^\alpha}{d\lambda}\underline{\partial}_\alpha+V^\alpha\frac{d}{d\lambda}\underline{\partial}_\alpha=\left(\frac{dV^\alpha}{d\lambda}+V^\mu\dot x^\nu\Gamma^\alpha_{\mu\nu}\right)\underline{\partial}_\alpha\equiv\frac{DV^\alpha}{d\lambda}\underline{\partial}_\alpha\,.
\end{equation}
In other words, in order to compute a sensible change in direction of a vector component along a curve, different tangent spaces must be related to each other via the concept of a \ul{connection}, in this case, the Levi-Civita connection whose coordinate representation of the coefficients is given by the Christoffel symbols (see Appendix~\ref{sApp: connection and curvature}). A vector is said to be \ul{parallel transported} along a curve, if its covariant derivative vanishes
\begin{equation}\label{eq:ParallelTransportOfVector}
    \boxed{\frac{DV^\mu}{d\lambda}=\dot x^\nu V\ud{\mu}{;\nu}=0\,.}
\end{equation}
Furthermore, if the tangent vector of a curve is itself parallel transported along the curve, we talk about an \ul{autoparallel}. 

Comparing again Eqs.~\eqref{eq:StraightLineArbitrary} and~\eqref{eq:ParallelTransportOfVector}, we further conclude that in this terminology, the straight lines of free particles also correspond to the autoparallels associated with the Levi-Civita connection. In other words, the autoparallels with respect to the Christoffel symbols coincide with the geodesics of the spacetime, which both correspond to the timelike trajectories of free, massive test particles. It is also very natural to extend this results to massless particles as well, hence, to world lines that are null. 

However, while geodesics have an immediate physical correspondence to the worldlines of test particles, the notion of parallel transportation can more directly be linked to the physical operation of moving a physical vector in spacetime. A sensible physical vector is for instance provided by a gyroscope, whose movements in spacetime without applying any forces can be considered as corresponding to the parallel transport in Minkowski spacetime. Since within special relativity such parallel transport is trivial, in the sense that it is path independent, the connection coefficients in an arbitrary frame will precisely correspond to the Christoffel symbols arising due to the coordinate dependence of the tangent space basis. Thus, curves whose tangent vector is given by parallel transporting the spin direction of a free test gyroscope will correspond to the autoparallels of the Levi-Civita connection.

In conclusion, we can formulate the \textit{Principle of Geodesic Motion} in Minkowski Spacetime, that reads:

\begin{principle}\ul{Principle of Geodesic Motion}.\label{Principle:GeodesicMotionSR}
The timelike and null geodesics of a spacetime that correspond to autoparallels of the Levi-Civita connection, are both equivalent to the physical worldlines of free test particles, which in turn coincide with the curves defined through the parallel transport of the tangent vector set by free test gyroscopes. These curves are defined as the ``straight lines'' of spacetime.
\end{principle}

\paragraph{Flatness and Geodesic Deviation.}

In the following, we want to introduce a few advanced notions of Minkowski spacetime as a special manifold with a particular fixed metric that will become important in later chapters. In Sec.~\ref{sSec: Theories of Minkowski Spacetime} we will then turn again to a description of physics by theories defined through the underlying structure of Minkowski spacetime.

From a highbrow point of view, the specialness of Minkowski spacetime is connected to the statement that through its invariance under the ten dimensional Poincar\'e transformations it corresponds to a \textit{maximally symmetric} spacetime \cite{Weinberg1972,zee2013einstein,carroll2019spacetime}. Locally, a maximally symmetric spacetime has the largest possible number of independent \ul{Killing vector fields}. For a $d$-dimensional manifold this maximal number is given by the number of independent components of the symmetric metric
\begin{equation}\label{eq:NumerofMaximallySymmetricKVF}
    \frac{d(d+1)}{2}\,,
\end{equation}
which for $d=4$ equals ten.
For $d>2$, maximal symmetry completely determines the metric $g_{\mu\nu}$ of a spacetime \cite{Weinberg1972} up to a constant curvature scalar
\begin{equation}\label{eq:CurvatureScalarK}
    K\equiv \frac{R}{d(d-1)}\,,
\end{equation}
with
\begin{equation}
    R_{\mu\nu\rho\sigma}=K(g_{\nu\rho}g_{\mu\sigma}-g_{\nu\sigma}g_{\mu\rho})\,,
\end{equation}
where $R$ is the \ul{Ricci scalar} and $R_{\mu\nu\rho\sigma}$ the \ul{Riemann curvature} tensor associated to the Levi-Civita tensor (see Appendix~\ref{App:DiffGeo}). The sign of the curvature $K$ divides maximally symmetric spacetimes into three types: Minkowski spacetime with $K=0$, de-Sitter (dS) spacetime with $K>0$ and Anti-de-Sitter (AdS) spacetime with $K<0$.

Thus, in contrast to the other two maximally symmetric spacetimes, Minkowski spacetime is flat, in the sense that the Riemann curvature tensor associated to the Levi-Civita connection vanishes everywhere
\begin{eqnarray}\label{eq:VanishingRiemann}
    R_{\mu\nu\rho\sigma} = 0\,.
\end{eqnarray}
As we will discover, deep down, this result is connected with the fact that the spacetime of special relativity equipped with an a priori Minkowski metric allows for the existence of global inertial frames. 

The corresponding geometric meaning of flatness, is best captured by the notion of parallel geodesics and geodesic deviation. Indeed, one of Euclid's fundamental axioms asserts that two straight lines never intersect when they are initially parallel. Having identified test particle trajectories with straight lines that correspond to geodesics of a spacetime, geodesic deviation also provides the physical interpretation of flatness.

Consider therefore two affinely parameterized and nearby geodesics $y_1^\mu(\lambda)$ and $y_2^\mu(\lambda)$ in a given inertial frame, separated by an infinitesimal vector $\delta y^\mu(\lambda)$\footnote{Note that here the infinitesimality is not given in terms of the parameter $\lambda$ along the geodesic, but rather between two different geodesics, which is why the infinitesimal vector is not denoted by $d y^\mu(\lambda)$. Below, we will introduce the notion of geodesic congruence that will clarify this point.},
such that $y_2^\mu(\lambda)=y_1^\mu(\lambda)+\delta y^\mu(\lambda)$.\footnote{More precisely, starting with two nearby points $p$ and $q$ separated by the infinitesimal vector $\delta y^\mu$ we consider two non-intersecting geodesics that pass through $p$ and $q$ respectively, parameterized by $\lambda_1$ and $\lambda_2$, such that $y_2^\mu(0)=y_1^\mu(0)+\delta y^\mu$. The separation vector along the geodesics can then be defined by $\delta y^\mu(\lambda)\equiv y_2^\mu(\lambda)-y_1^\mu(\lambda)$, where we adjust the geodesic parameters such that $\lambda_1=\lambda_2=\lambda$, assuming that the separation remains of infinitesimal norm $|\delta y^\mu|=\sqrt{|\eta_{\mu\nu}\delta y^\mu\delta y^\nu}|<<1$.} In an inertial frame, the tangent vectors of both geodesics satisfy the equation of a straight line given by Eq.~\eqref{eq:StraightLineInertial}
\begin{equation}
    \frac{d\dot y^\mu_{1,2}}{d \lambda}=0\,.
\end{equation}
Taking the difference between these two equations yields an evolution equation for the separation vector
\begin{equation}\label{eq:GeodesicDeviation Equation Minkowski coords}
    \frac{d^2\delta y^\mu}{d \lambda^2}=0\,,
\end{equation}
which precisely implies that two particle trajectories that are initially parallel, will never intersect. In other words, there is no \textit{geodesic deviation} in Minkowski spacetime. Observe that this result hols for any type of geodesic.

The same statement can also be made in arbitrary coordinates. Just as before, we simply perform a change of coordinates from $y^\mu$ to $x^\mu(y)$ with associated coordinate induced tangent space basis $\partial_\mu$, which yields non-trivial metric components $g_{\mu\nu}(x)$, with associated connection coefficients $\Gamma^\lambda_{\mu\nu}(x)$ defined through Eq.~\eqref{eq:Christoffel}. The separation $\delta y^\mu$ transforms as a regular vector component in the coordinate induced basis $\delta y^\mu=\frac{\partial y^\mu}{\partial x^\nu}\delta x^\mu$, such that the condition for vanishing geodesic deviation simply becomes
\begin{equation}\label{eq:GeodesicDeviationFlat}
    \boxed{\frac{D^2\delta x^\mu}{d \lambda^2}=0\,.}
\end{equation}
Note that this equation is now a coordinate invariant expression of the statement that a spacetime is flat. In Sec.~\ref{sSec:Metric Theories} we will understand that this equation as a definition of the flatness of Minkowski spacetime is intimately connected to Eq.~\eqref{eq:VanishingRiemann} above.

\paragraph{Simultaneity and Spacial Proper Distance.}
In Minkowski spacetime, spacial physical distances are globally well-defined because it is possible to introduce a notion of global simultaneity. This notion coincides with the a priori coordinate dependent simultaneity provided by equal time slices within Minkowski coordinates.\footnote{Note that although defined globally, such a notion of simultaneity is still tied to particular inertial observers and therefore not unique, in contrast to a Newtonian spacetime with a globally defined time.} Indeed, while in a general coordinate system the split between the temporal and the spacial part of tensor fields seems arbitrary, this arbitrariness is broken in Minkowski spacetime by the existence of inertial observers that provide a preferred notion of time in terms of their proper time $\tau$. However, for later use, it is worth examining how that notion of simultaneity can be fundamentally defined in more general coordinate systems. 
Such a careful consideration of simultaneity is important, as it allows a proper definition of purely \textit{spacial} distances, a concept which ultimately makes only sense for spacetime events that can be regarded as simultaneous with respect to a specific observer.

Simultaneity with respect to an observer $A$ of two spacelike separated neighboring events $p$ and $q$ can best be constructed by considering the geodesic of the physical observer $A$ that passes through $p$ as well as the geodesic of an observer $B$ that passes through $q$. There exists then a preferred coordinate system $(t,x^i)$ that is given by $t=\tau$, where $\tau$ is the proper time of observer $A$ and the coordinates $x^i=z^i$ are chosen such that $A$ is at the origin and the location of $B$ in the coordinate system does not change over time. In other words, the spacial coordinates are defined by the second physical observer $B$, whose geodesic we parametrize by the same proper time $\tau$.  Note that by construction, the line element of the metric of this (local) coordinate system takes the simple form
\begin{equation}\label{eq:SynchronousGauge}
    ds^2=-d\tau^2+g_{ij}dz^idz^j\,.
\end{equation}
In particular, any time-space components $g_{0i}$ vanish, while $g_{00}=-1$. Such a coordinate system is known as \textit{Gaussian normal coordinates} or \textit{synchronous coordinates} (see e.g. \cite{WaldBook,landau_classical_2003}) and also corresponds to the idea of a \textit{comoving coordinate system} that we will encounter in Sec.~\ref{sSec:HomIsoUniverse}.
In such a coordinate system $\{\tau,z^i\}$, it is in a sense trivial to identify simultaneous events by considering events that are labeled by the same proper time $\tau=\text{constant}$. In other words, all points on spacial slices of constant $\tau$ are defined as simultaneous events for the physical observers with proper time $\tau$. Thus, of course in particular the Minkowski coordinates of Minkowski spacetime are synchronous coordinate systems in which simultaneity is trivially defined.

Moreover, having defined two simultaneous events $p$ and $q$ it is now sensible to ask the question about the physical or proper \textit{spacial} distance $\ell$ between the two events. In particular for two nearby events separated by an infinitesimal coordinate distance $dz^i$\footnote{Note that here we restrict to an infinitesimal distance in order to also in more general situations unambiguously being able to talk about a vector that connects two spacetime points.} the infinitesimal \textit{spacial} proper distance $d\ell$ between the two events is intuitively given by
\begin{equation}\label{eq:SpacialDistanceDefSimp}
    d\ell^2=g_{ij}dz^idz^j\,.
\end{equation}

It is however important to realize, that these intuitive statements only hold in the special chart described above. It is therefore useful to provide a practical meaning of the notion of simultaneity introduced here that is independent of any coordinate system \cite{zee2013einstein}. Namely, operationally, the simultaneity between an event $p$ on the geodesic of $A$ and a neighboring point $q$ on a curve $B$ at fixed spacial coordinate $dz^i$ can be determined by sending a light signal from $A$ to $B$ and back to observer $A$. The duration of this process $T$ can be measured by $A$ in terms of its proper time. Then, the two events $p$ and $q$ that can very generally be defined as simultaneous are on the one hand the instant of arrival of the light ray at the observer $B$ that defines $q$, while the corresponding $p$ is given by the instant, when half of the proper time interval of $A$ is elapsed, hence $T/2\equiv d\tau$. Knowing that in vacuum light always travels at the speed $c=1$, the associated proper distance is then given by
\begin{equation}\label{eq:GeneralFormulaSpacialDistance}
    d\ell=c\frac{T}{2}=d\tau\,.
\end{equation}
In this very general and coordinate invariant setup, the coordinate dependent result in Eq.~\eqref{eq:SpacialDistanceDefSimp} can then easily be derived by demanding that for light $ds=0$, where the line element is given by the specific form in Eq.~\eqref{eq:SynchronousGauge}.

This rather cumbersome construction allows however the derivation of a general formula for local proper distance $d\ell$ between two geodesics separated by a coordinate distance $dx^i$ and of duration $d\tau$ short enough to neglect any variations of the components of the metric in a completely arbitrary coordinate system $\{t,x^i\}$. In such a chart the local line element has the general form
\begin{equation}
ds^2=g_{00}dt^2+2g_{0i}dtdx^i+g_{ij}dx^idx^j\,.
\end{equation}
In this case, Eq.~\eqref{eq:SpacialDistanceDefSimp} generalizes to (see e.g. \cite{zee2013einstein})
\begin{equation}\label{eq:ProperDistanceGeneral}
    d\ell^2=-g_{00}\left(\frac{1}{2}(dt_+-dt_-)\right)^2=\left(g_{ij}-\frac{g_{0i}g_{0j}}{g_{00}}\right)dx^idx^j\,,
\end{equation}
where $dt_-$ and $dt_+$ label the instants of sending and receiving the light signal respectively, assuming that the event $p$ is characterized by $t=0$.
This follows from the general formula in Eq.~\eqref{eq:GeneralFormulaSpacialDistance} by observing that now 
\begin{equation}
    \frac{T}{2}=d\tau=\frac{\sqrt{-g_{00}}}{2}(dt_+-dt_-)\,,
\end{equation}
while through the equation of null rays $ds^2=0=g_{00}dt^2+2g_{0i}dtdx^i+g_{ij}dx^idx^j$
one obtains the relation
\begin{equation}
    dt_{\pm}=\frac{1}{g_{00}}\left(-g_{0i}dx^i\pm\sqrt{(g_{0i}dx^i)^2-g_{00}g_{ij}dx^idx^j}\right)\,.
\end{equation}


\paragraph{Congruence of Timelike Geodesics and Spacial Geodesic Deviation.}

Observe that the above setup for defining spacial distances with two nearby geodesics represents the exact same situation considered when defining geodesic deviation one paragraph before, with the additional condition that the geodesics be timelike geodesics associated to physical observers. Indeed, the geodesic deviation of timelike coordinates is closely related to the existence of the synchronous coordinate system constructed above. Mathematically, these coordinates naturally correspond to a so-called \textit{congruence} of timelike geodesics $x^\mu(\tau,z^i)$ \cite{WaldBook,zee2013einstein,carroll2019spacetime}, that is, a series of nearby timelike geodesics that do not intersect, parameterized by a proper time $\tau$ and three additional (spacelike) parameters $z^i$. Such a congruence of timelike geodesics can be used to coordniatize the spacetime patch, precisely resulting in the synchonous chart $\{\tau,z^i\}$ described above.

In particular, two nearby geodesics can be viewed as forming part of the subset of a two-parameter family of geodesics $x^\mu(\tau,z)$. In this language, the infinitesimal separation vector $\delta x^\mu$ between the nearby geodesics is naturally given by the infinitesimal spacelike vector
\begin{equation}
    \delta x^\mu\equiv \frac{\partial}{\partial z} x^\mu(\tau,z) \,dz\,.
\end{equation}
Thus, this vector represents one natural basis vector of the synchronous coordinate system. And because the same is true for the infinitesimal tangent vector [Eq.~\eqref{eq:Infinitesimal Tankgent vector}] $dx^\mu=\dot x^\mu \,d\tau$, where
\begin{equation}
    \dot x^\mu\equiv \frac{\partial}{\partial \tau} x^\mu(\tau,z)\,,
\end{equation}
the separation vector can more formally be defined through the condition that the \ul{Lie Bracket} between them vanishes
\begin{equation}\label{eq:Condition Deviation Vector}
    [\underline{d x},\underline{\delta x}]^\mu=0\,.
\end{equation}

As we explicitly prove in Appendix~\ref{sApp:SpacialGeodesicDeviation} this implies that for timelike geodesics of a spacetime with Levi-Civita connection, the component of the separation vector in the direction of the geodesics 
\begin{equation}\label{eq:condition deviation}
    dx_\mu \delta x^\mu = \text{constant}\,,
\end{equation}
remains constant along the geodesics and can therefore without loss of generality be set to zero
\begin{equation}\label{eq:Condition temporal Deviation Vector}
    \boxed{d x_\mu\delta x^\mu=0\,.}
\end{equation}
Hence, the two vectors can be chosen to be orthogonal. In the synchronous chart, this therefore implies that we can without loss of generally choose $\delta x^0=0$ and concentrate on the spacial components $\delta x^i$ of the deviation vector only.
This statement has the profound implication that when measuring the geodesic deviation of timelike geodesics one is very generally probing the \textit{spacial} proper distance between the geodesics, where the concept of spacial proper distance was derived above. Indeed, this is true as long as the operational process of determining the proper spacial distance, hence sending a light-signal between the two geodesics, is short enough compared to the timescale of geodesic deviation.

As a little preview, these considerations will become crucial when defining the response of an idealized detector to gravitational waves in Sec.~\ref{ssSec:The Physical Effects of Gravitational Waves}. Indeed, while strictly speaking the above considerations so far are based on the existence of inertial observers in SR that naturally follow geodesics, the exact same conclusions will also hold locally in a spacetime with arbitrary metric.
As we will see below, in this case, the notion of inertial observers can locally be replaced by the concept of freely falling observers. Such freely falling observers can indeed be viewed as a set of locally defined inertial observers. In particular, freely falling observers locally also introduce a preferred notion of time and therefore also of space for any metric, such that it makes sense to talk about a proper \textit{spacial} distance.


\section{The Theories of Minkowski Spacetime}\label{sSec: Theories of Minkowski Spacetime}

It is time to return to physics and study in more detail how non-gravitational theories are described within the framework of Minkowski spacetime introduced above. This will in particular be important for the formulation of the ''matter sector'' of gravitational theories.

\paragraph{The Energy-Momentum Tensor.}

As discussed in the previous Section~\ref{sSec:Special Relativity}, the tangent vectors of timelike curves in a Minkowski frame, parameterized by the proper time $\tau$ represent the so called $4$-velocity $u^\mu(\tau)=\dot y^\mu(\tau)$ of a particle with mass and satisfies $u^\mu u_\mu=-1$ by the definition of proper time. For a particle with mass $m$, we can furthermore define an energy-momentum $4$-vector 
\begin{equation}
    p^\mu\equiv m\,u^\mu\,.
\end{equation}
The \textit{energy} of a particle with 4-velocity $u^\mu$ as measured by an inertial observer $O$ with 4-velocity $u_O^\mu$ who is at the location of the particle\footnote{Note that in Minkowski spacetime we can also define the energy of a particle from the perspective of an observer far away from the particle, since parallel transport is trivial.} is then given by \cite{WaldBook}
\begin{equation}\label{eq:EnergyDefMinkowski} 
    E=-p_\mu u_O^\mu\,.
\end{equation}
For a particle at rest with respect to the inertial observer, we recover $E=m c^2$, where we have restored the units in $c$. 

We will however also be interested in continuous distributions of matter, including matter fields. In order to make the transition from the point particle discussion above, it is enlightening to consider the energy-momentum density and the current of a system of $n$ particles, analogue to the intuitive notion of a charge and current density. A first crucial observation is that a naive definition of the energy-momentum density 
\begin{equation}
    T^\mu=\sum_n p_n^\mu(t)\,\delta^3(y-y_n(t))\,,
\end{equation}
in some inertial frame does not define a proper (Lorentz) tensor \cite{Weinberg1972}. In other words, it does not define an (inertial) observer independent object. Rather, an energy-momentum density can only consistently be described as a component of a more general object in direct connection with the associated current by defining a $\binom{2}{0}$-tensor called \textit{energy-momentum tensor} \cite{Weinberg1972,misner_gravitation_1973}
\begin{equation}\label{eq:Particle EMT}
    T^{\mu\nu}\equiv  \sum_n\,p_n^\mu(t)\, \frac{y_n^\nu(t)}{dt}\delta^4(y-y_n(t))=\sum_n \int d\tau \,p_n^\mu\, u_n^\nu\delta^4(y-y_n(\tau))\,,
\end{equation}
where $\delta^4(y-y_n(\tau))$ now indeed defines a scalar.
Note that this tensor is symmetric. Moreover, by straightforward computation it can be shown that this energy momentum tensor is conserved for free particles with constant $p^\mu_n$
\begin{equation}\label{eq:ConservationEMTEnsor}
    \partial_\mu T^{\mu\nu}=0\,.
\end{equation}
This is a crucial result as it implies the conservation of energy, momentum and angular momentum. For instance, from Eq.~\eqref{eq:EnergyDefMinkowski} the energy $4$-current density in the inertial frame with $u_O^\mu=t^\mu$, where $t^\mu\equiv \delta\ud{\mu}{0}$ is given by
\begin{equation}\label{eq:Energy4Current}
    J^\nu=t_\mu T^{\mu\nu}\,,
\end{equation}
Which is conserved due to Eq.~\eqref{eq:ConservationEMTEnsor}
\begin{equation}\label{eq:ConservationECurrent}
    \partial_\nu J^\nu=0\,.
\end{equation}

Locally, through Gauss's theorem, this conservation implies that the energy current across the tree dimensional boundaries $\partial V$ of a local spacetime volume $V$ is conserved \cite{misner_gravitation_1973,WaldBook,carroll2019spacetime}
\begin{equation}
    \int_{\partial V} J^\mu n_\mu dS=0\,,
\end{equation}
where $n^\mu$ is the unit normal to the boundary surface $\partial V$.
In other words, the same amount of energy that flows into the spacetime volume $U$ needs to flow out again. Analog considerations also hold for momentum and angular momentum currents (see e.g. \cite{Weinberg1972,misner_gravitation_1973}).

\begin{figure}[H]
    \centering
    \includegraphics[width=0.9\textwidth]{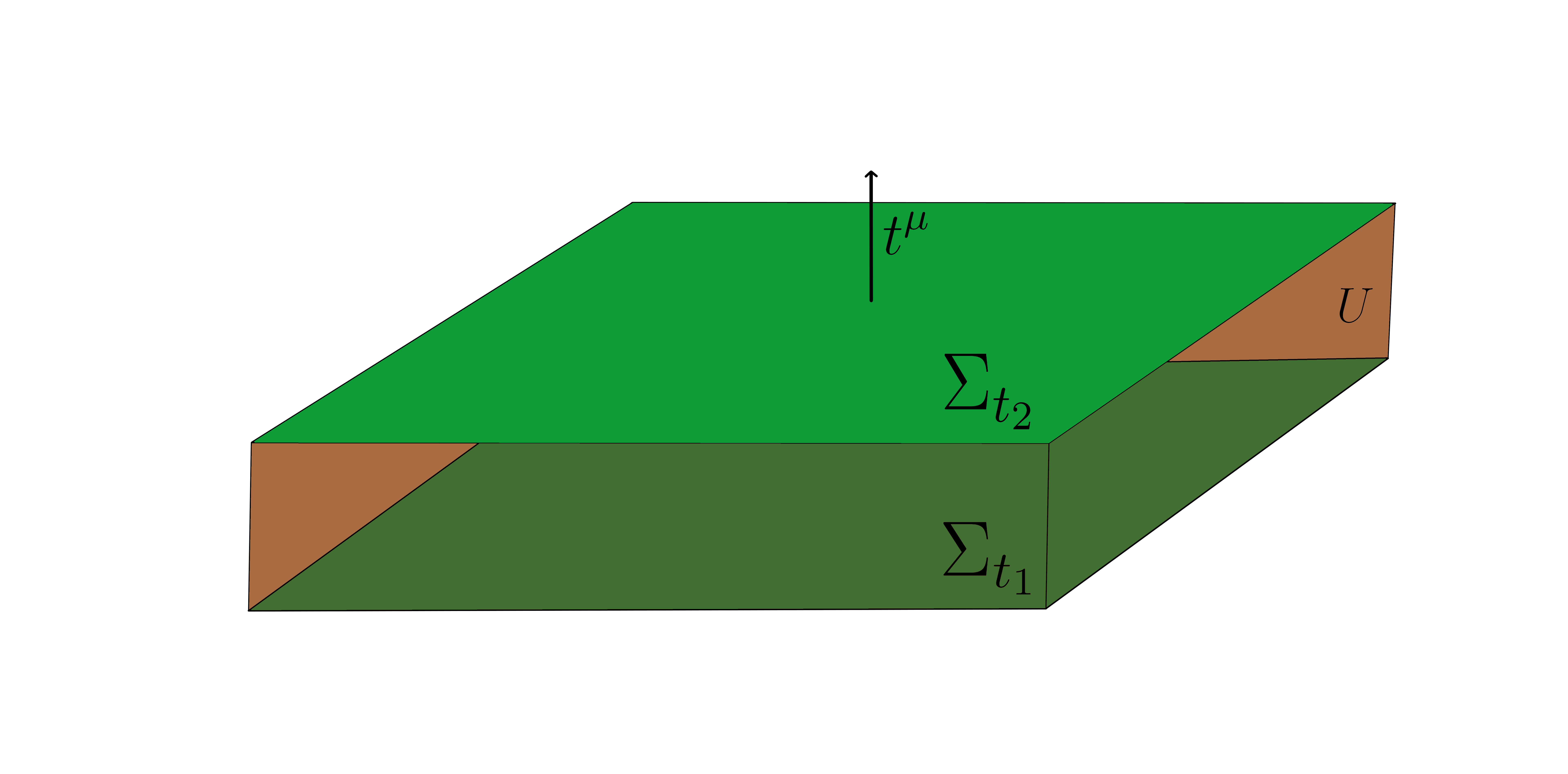}
    \caption{\small Spacial slices of Minkowski spacetime parameterized by a global time $t$ of some inertial frame. $U$ represents the spacetime volume delimited by the slices $\Sigma_{t_1}$ and $\Sigma_{t_2}$.}
    \label{fig:GaussLawMinkowski} 
\end{figure}

With a conserved current at hand, a more global statement across the entire Minkowski spacetime can also be made. Indeed, the conservation of the energy current in Eq.~\eqref{eq:ConservationECurrent} implies the existence of a \textit{conserved charge} \begin{equation}
    E(t)\equiv \int_{\Sigma_t} d^3x J^\nu t_\nu \,.
\end{equation}
Here, $\Sigma_t$ represents a spacial slice parametrized by the global time $t$ of the inertial frame (see Fig.~\ref{fig:GaussLawMinkowski}). Note that $t^\mu$ at the same time defines the energy current in Eq.~\eqref{eq:Energy4Current} and represents the unit normal to the constant time slices $\Sigma_t$.
The energy of the system is then independent of time, or equivalently the choice of the spacial slice $\Sigma_t$, since by Gauss's law
\begin{equation}\label{eq:Energyconservation Minkowski}
    E(t_2)-E(t_1)= \int_{\partial U} d^3x \,J^\nu t_\nu= \int_U d^4x \,\partial_\mu J^\mu=0 \,,
\end{equation}
provided that the energy is localized enough such that the integrals at spacial infinity vanish. As depicted in Fig.~\ref{fig:GaussLawMinkowski}, $U$ is the global spacetime volume that extends over the entire space with spacial boundaries $\Sigma_{t_1}$ and $\Sigma_{t_2}$, while $\partial U$ is its total boundary.

\paragraph{Matter Fields and the Matter Action.}

In order to consider more generic non-gravitational theories in Minkowski Spacetime, we will from now on, for simplicity, assume all non-gravitational physics to be describable by a set of \textit{\gls{matter fields}} $\Psi_\text{m}$. This shall include the point particle considerations above, any continuous matter distributions, for example perfect fluids, as well as any matter fields. The prime example of non-gravitational physics we are thinking of here is classical electrodynamics, but in principle this should include \textit{any} non-gravitational physics. For the sake of simplicity, we will however disregard any potential subtleties regarding spinors and quantized theories in general.

By definition, any Poincar\'e invariant theory of matter fields can be described within inertial frames of Minkowski spacetime. By changing between different inertial frames through Poincar\'e transformations in Eq.~\eqref{eq:PoincareT}, the components of the matter fields as tensor fields transform under the associated Lorentz transformations. For instance, a vector field such as the potential $A_\mu$ of electromagnetism transforms under such changes of frames as
\begin{equation}
    A'_\mu=A_\nu\,\Lambda\ud{\nu}{\mu}\,.
\end{equation}
The components of a vector with such a transformation law under Poincar\'e transformations is known as \textit{Lorentz vector}, which simply reflects its definition as a proper tensor field.
Furthermore, we will also assume that any such matter theory can be described through a so-called \textit{matter action} 
\begin{equation}\label{eq:MatterActionInertial}
    \mathcal{S}_\text{m}=\int d^4x\, \mathcal{L}_\text{m}\,,
\end{equation}
where $\mathcal{L}_\text{m}$ is the corresponding \textit{matter Lagrangian} density that depends on the fields and its derivatives.
This action conveniently determines the equations of motion of the matter fields through extermization (see also Appendix~\ref{sApp: Gauge Freedom})
\begin{equation}\label{eq:EOMMatterFields}
    \boxed{\frac{\delta S_\text{m}}{\delta \Psi_\text{m}}\equiv\mathcal{J}_\text{m}=0\,.}
\end{equation}

On the other hand, just as before, any theory of matter fields on Minkowski spacetime can also be formulated in arbitrary coordinates. In practice, the discussion in Sec.~\ref{sSec:Special Relativity} implies that the description in arbitrary coordinates of a theory in Minkowski spacetime can be achieved by starting from a formulation of the non-gravitational laws in an inertial frame of Minkowski spacetime and replace
\begin{equation}\label{eq: Minimal Coupling Procedure}
    \eta_{\mu\nu}\rightarrow g_{\mu\nu}\,,\quad \partial_\mu\rightarrow \nabla_\mu\,.
\end{equation}
In particular, written in arbitrary coordinates the matter action
\begin{equation}\label{eq:MatterAction}
    \boxed{\mathcal{S}_\text{m}=\int \sqrt{-g}\,d^4x\, \mathcal{L}_\text{m}\,,}
\end{equation}
is naturally invariant under general coordinate transformations upon the introduction of a more careful definition of the measure of integration on a manifold (see e.g. \cite{misner_gravitation_1973,WaldBook,carroll2019spacetime})
\begin{equation}\label{eq: Minimal Coupling Procedure 2}
    d^4x\rightarrow \sqrt{-g} \,d^4x\,,
\end{equation}
with $g\equiv \det g_{\mu\nu}$ the \ul{determinant of the metric}.

From the discussion of the energy-momentum density of a multi-particle system above, it should be clear that a central object to define for a theory on Minkowski spacetime is its total energy-momentum tensor.
A very convenient definition of a proper tensor describing the energy-momentum content is in fact provided by the variation of its action
\begin{equation}\label{eq:DefEnergyMomentumTensor}
    \boxed{T^{\mu\nu}\equiv \frac{-2}{\sqrt{-g}}\,\frac{\delta S_\text{m}}{\delta g_{\mu\nu}}\,,}
\end{equation}
by temporarily treating the Minkowski metric in arbitrary coordinates $g_{\mu\nu}$ as an independent field. As we will show in Sec.~\ref{sSec:Covariant Consrevation}, this tensor indeed defines an energy momentum tensor, as it is guaranteed to satisfy a conservation equation. Observe as well that by definition it represents a symmetric tensor. While such a definition of the energy-momentum tensor of might seem unusual for a theory on Minkowski spacetime, the object in Eq.~\eqref{eq:DefEnergyMomentumTensor} recovers the particle energy-momentum defined in Eq.~\eqref{eq:Particle EMT}, and is also equivalent to the expressions associated with a more direct application of the $1^\text{st}$ Noether Theorem~\ref{Thm:NoetherTheorem} to formulate conserved energy and momentum currents within a Minkowski frame. However, as we show in Appendix~\ref{sApp: Noethers Theorem} Noether currents in general, and therefore also the Noether energy-momentum tensor associated to the translation invariance of Minkowski spacetime, are not uniquely defined. This implies that a direct implementation of the Noether theorem may require a so called Belinfante improvement procedure \cite{Belinfante_1940Phy449B,Blaschke:2016ohs} in order to recover the energy-momentum tensor defined in Eq.~\eqref{eq:DefEnergyMomentumTensor} with all its desired properties. In Sec.~\ref{sSec:Covariant Consrevation} we will also provide an understanding of the energy-momentum conservation in a covariant language in association with the isometries of the Minkowski metric already mentioned above.


\chapter{The Generalization to Gravity}\label{Sec:The Generalization to Gravity}


\section{The Equivalence Principle}\label{sSec:Equivalence Principle}

So far, we explicitly excluded any gravitational effects. Unlike Maxwell's theory of classical Electrodynamics for instance, Newtons formulation of the laws of gravity is not invariant under Lorentz transformations and suffers from acausality due to instantaneous action. It turns out, however, that apparently the gravitational force is to be treated differently than other fundamental forces of nature, such that it is not a simple matter of making the theory of gravity compatible with Lorentz invariance (see however Sec.~\ref{sSec:Quantization of Gravity}). The reason is the so-called universal coupling of gravity, expressed in terms of equivalence principles. 


\paragraph{The Weak Equivalence Principle.} A first version of the equivalence principle is famously attributed to Galileo Galilei and Isaac Newton, who experimentally determined that the acceleration of a test mass in a homogeneous gravitational field is independent of its internal structure and physical properties, including its mass. Isaac Newton concertized this idea in his \textit{Philosophiae Naturalis Principia Mathematica} \cite{Newton:1686} by providing the underlying theoretical description. Newtons second law
\begin{equation}
    \vec{a}=\frac{\vec{F}}{m_I} \,,
\end{equation}
 suggests that in general, the acceleration of an object in an inertial frame depends on its inertial mass $m_I$, as well as possibly other intrinsic properties that determine the strength of the force. For instance, if the force is attributed to an electric field, the acceleration also depends on the charge of the object. The same could be true for the gravitational force, which Newton determined to be
 \begin{equation}
     \vec{F}_{G}=m_G\,\vec{g} \,,
 \end{equation}
where $m_G$ is the gravitational mass of the test object and where
 \begin{equation}
     \vec{g}=G\,m_i \frac{\vec{x}_i}{|\vec x_i|} \,,
 \end{equation}
 with Newtons constant $G$ and $x_i$ denoting the position of masses $m_i$ creating the gravitational field. If the gravitational mass $m_G$ would be different from the inertial mass $m_I$ of the test object, the gravitational force would be conceptually on the same footing as the electric force. However, experiments demand to very high precision that in fact, the inertial and gravitational masses are equivalent\footnote{More precisely, it suffices for the inertial and gravitational masses to be proportional to each other. Any constant proportional factor could be absorbed by the definition of Newtons constant.}
 \begin{equation}
     m_I=m_G\,,
 \end{equation}
 and thus, the acceleration of a test mass subject to a gravitational force only, is independent of its mass or any other internal structure. In other words, the coupling of gravity seems universal, in the sense that it acts in the same way on all test masses. On that aspect, the gravitational force behaves like an apparent force arising in non-inertial reference frames. These considerations are traditionally known as the \textit{Weak Equivalence Principle} (WEP).

\paragraph{The Einstein Equivalence Principle.} It was Albert Einstein \cite{Einstein:1907ve}, who realized the deep implications of this experimental result on the notion of the Principles~\ref{Principle:SR} of special relativity, discussed in Sec.~\ref{sSec:Special Relativity}. The WEP implies that locally, in a \textit{\gls{freely falling}} frame of a gravitational field, the motion of test masses will be indistinguishable from a configuration without the gravitational field. By ``locally'' we mean here local enough, such that inhomogeneities in the gravitational field can be neglected. 

Based on this result, Einstein postulated, that in fact no local measurement of a freely falling observer can detect the existence of an external gravitational field, in the sense that in any freely falling frame, any physical experiment involving only non-gravitational masses and energy will have the same outcome. The existence of ``small enough'' scales in order to satisfy the locality criteria, is guaranteed on very general grounds, due to the weakness of gravity compared to the other fundamental forces of nature, such that gravitational tidal forces can be neglected for short enough but still reasonably large characteristic scales of an experiment. Thus, the so-called \textit{Einstein Equivalence Principle} (EEP) asserts that (see e.g. \cite{Weinberg1972,WaldBook,misner_gravitation_1973,zee2013einstein,poisson2014gravity,Will:2018bme,carroll2019spacetime,Jetzer:2022bme}):

\begin{principle}\ul{Einstein Equivalence Principle}.\label{Principle:EEP}
Within any gravitational field, locally, the principles of special relativity hold with the same non-gravitational physical laws in all freely falling frames of reference. 
\end{principle}

In particular, this implies that in a small enough region of spacetime, in which the gravitational field can be considered approximately homogeneous and static, different freely falling observers are inertial observers related through Lorentz transformations. Note that it indeed makes sense to talk about inertial reference frames in that context, since two freely falling observers in a homogeneous gravitational field are related through a constant relative velocity due to the weak equivalence principle. In a gravitational field, the EEP therefore strongly suggests considering freely falling frames as ``local inertial frames''\footnote{It should be clarified, however, that in this context, inertial frames are defined as inertial with respect to all \textit{non-gravitational} forces. In other words, the equivalence principle suggests to treat gravity apart from conventional forces and define local acceleration as a measurement of the departure from free motion due to non-gravitational forces only.}.

The key implication here, however, is that unlike inertial frames in Minkowski spacetime, beyond the local approximation of a homogeneous gravitational field, different freely falling frames are no longer related to each other through Lorentz transformations. In other words, adding gravity to the picture, as compared to other forces or interactions, \textit{the notion of inertial observers is not uniquely defined anymore}.\footnote{In fact, in the theories of gravity discussed below, the relative velocity between observers at different spacetime points is not defined at all.} This due to the far-reaching consequence of the equivalence principle and the experimental fact, that in general, gravitational fields are not homogeneous. Indeed, any freely falling frame defines a set of inertial observers of special relativity, which are however not inertial among each other\footnote{For instance, within a Newtonian picture, two distant observers falling towards the center of earth would observe an acceleration between them.}. In other words, in the presence of gravity, the inertial observers of special relativity can only be defined in a restricted region of spacetime, associated to every local freely falling frame. Thus, while any gravitational effects can locally be turned off by choosing an appropriate frame, the true, or physical, gravitation lies in the non-local effects between different freely falling frames as determined by the inhomogeneities in the gravitational field. 

\paragraph{A Locally Lorentz Invariant Formulation of Gravity.} Therefore, given that:
\begin{enumerate}[(i)]
    \item SR can naturally be formulated as a theory on a four-dimensional flat manifold with associated Minkowski metric, as discussed in Secs.~\ref{sSec:Special Relativity} and \ref{sSec: Theories of Minkowski Spacetime};
    \item The EEP implies that locally, the effects of a gravitational field are not distinguishable from the effects of accelerated reference frames, which can be described through a metric $g_{\mu\nu}$ in generalized coordinates. In particular, the equations of motion of a free test mass in non-inertial reference frames, are determined by the geodesic equation Eq.~\eqref{eq:GeodesicEq};
    \item Different freely falling local inertial frames are not inertial to each other;
\end{enumerate}
this strongly suggests identifying the gravitational field with a general metric $g_{\mu\nu}(x)$ that describes a spacetime, in which the metric cannot be reduced to the Minkowski metric on a global level. As we will see below, this implies that the true effects of gravity can naturally be described as a manifestation of spacetime curvature.

On the other hand, the EEP is incorporated in the framework of viewing spacetime as a differential manifold through the existence of so called \ul{Riemann normal coordinates} $y^\mu$ at every point on the manifold, for which
\begin{equation}\label{eq:Riemann Normal Coordinates}
    g_{\mu\nu}(y)=\eta_{\mu\nu}\,,\qquad g_{\mu\nu,\alpha}(y)=0 \,,
\end{equation}
while in general
\begin{equation}\label{eq:Riemann Normal Coordinates 22}
  g_{\mu\nu,\alpha\beta}(y)\neq 0 \,.
\end{equation}
These normal coordinates at a given event can in fact be extended along an entire geodesic as we show in Appendix.~\ref{sApp: Normal Coordinates}, a construction known as \ul{Fermi normal coordinates}. The existence of such normal coordinates for every timelike geodesic are thus the mathematical manifestation of the postulated existence of freely falling frames within the EEP, since in a local enough region, the Metric is well approximated by the Minkowski metric and local effects of gravity will remain entirely negligible. It should however be mentioned, that the existence of Riemann normal coordinates and therefore also of Fermi normal coordinates in only ensured if the connection of the manifold is torsion-less and metric [App.~\ref{sApp: Normal Coordinates}]. This comment will become clear below.


In light of the overwhelming empirical evidence for the Einstein equivalence principle as well as local Poincar\'e invariance (see \cite{Mattingly:2005re,Will:2014kxa,Will:2018bme,Jetzer:2022bme} and references therein), we will choose to base the framework for studying theories of gravity on the considerations above. 
As an outlook, continuing on this route the main result will be the postulation of the principles of universal and minimal coupling to a physical metric, the meaning of which will be carved out in the next section. In the following, the theory of general relativity is then identified as a very special theory among the theory space delineated by the equivalence principle. The reasoning behind such a strict focus on the equivalence principle is based on the expectation that any future potentially more complete theory of fundamental physics will necessarily need to incorporate it in the limit of scales that are probed today. It is therefore interesting to study the conceptual implications of the EEP, as well as explore the room of theoretical possibilities that it leaves open.


\section{Metric Theories of Gravity}\label{sSec:Metric Theories}

Building up on the Einstein equivalence principle and the formulation of special relativity of Sec.~\ref{sSec:Special Relativity}, the arena for theories of gravitation is provided by a \textit{pseudo-Riemannian manifold} $(\mathscr M,g)$, defined as a differentiable manifold endowed with a symmetric metric tensor $g$ that is everywhere non-degenerate and smooth.\footnote{As already mentioned, in the Appendix~\ref{App:DiffGeo} we offer a concise summary of pseudo-Riemannian differential geometry.} Moreover, instead of assuming a priori a particular metric as in special relativity, the metric is treated as a dynamical object, subject to a set of field equations that determine its evolution as sourced by the matter content in the spacetime. This last statement can be regarded as the key novelty of modern theories of gravitation, which is based on the conviction that in a fundamental theory of physics we should impose as little structure as possible by hand.\footnote{In practice, however, it is often necessary to add additional structure in the form of a concrete background solution of spacetime or assume a certain asymptotic behavior of spacetime.}

\paragraph{True Effects of Gravitation.} Before talking about the dynamics of spacetime, we will first consider in more detail the physical effects\footnote{We employ here the adjective ``physical'', to refer to genuine, measurable effects.} of gravitation as implied by the equivalence principle. These fundamental notions underlying many gravitational experiments are independent of the specific equations of motion of the theory, and will therefore represent the foundation of any theory of gravity that we will consider. Moreover, as it is the case with every interaction, gravitation can only be probed through the coupling of the corresponding field with matter, such that this interplay is of critical importance to any theory of gravity. 


As discussed in the previous section~\ref{sSec:Equivalence Principle}, the key implication of the equivalence principle is that in a gravitational field, the notion of inertial frames is not uniquely defined anymore. Every freely falling frame associated to free particle motion defines a set of inertial frames, which are however not compatible with each other due to inhomogeneities in the gravitational field. The conclusion is that the physical effects of gravitation are to be found precisely within these incompatibilities of inertial frames. In Sec.~\ref{sSec:Special Relativity} we identified the worldlines of free particles with geodesics of a spacetime and also introduced the notion of vanishing geodesic deviation in special relativity. The physical effects of gravitation are therefore precisely expected to show up in the study of geodesic deviation of a general spacetime.

Let's therefore consider two affinely parameterized and nearby geodesics $x_1^\mu(\lambda)$ and $x_2^\mu(\lambda)$, such that $x_2^\mu(\lambda)= x_1^\mu(\lambda)+\delta x^\mu(\lambda)$, and [Eq.~\eqref{eq:Condition Deviation Vector}]
\begin{equation}
    [\underline{\delta x},\underline{\dot{x}}_1]=0\,,
\end{equation}
with $[.,.]$ the \ul{Lie brackets} and $\dot{x}_1^\mu$ the tangent vector of the geodesic.\footnote{Recall that a more rigorous definition of the separation vector requires the existence of a family of geodesics parameterized by $\lambda$, in which case $\delta x^\mu$ represents the tangent vector to the curves of constant $\lambda$ (see e.g. \cite{zee2013einstein,carroll2019spacetime}).} In a general coordinate system, the tangent vectors of the two geodesics individually satisfy the geodesic equation [Eq.~\eqref{eq:GeodesicEq}], which corresponds to a straight line of free test particles. Taking the difference between the two equations to first order in $\delta x^\mu$ and its derivative, again yields an equation for the separation vector that this time takes the form (see e.g. \cite{maggiore2008gravitational,zee2013einstein,Hodgkinson:1972jn})
\begin{equation}\label{eq:GeodesicDeviation}
    \boxed{\frac{D^2\delta x^\mu}{d\lambda^2}=-\left(\Gamma^\mu_{\nu\sigma,\rho}-\Gamma^\mu_{\nu\rho,\sigma}+\Gamma^\mu_{\lambda\rho}\Gamma^\lambda_{\nu\sigma}+\Gamma^\mu_{\lambda\sigma}\Gamma^\lambda_{\nu\rho}\right)\delta x^\rho \dot x_1^\nu\dot x_1^\sigma\,,}
\end{equation}
where $\Gamma^\mu_{\nu\sigma}$ represent the \ul{Christoffel symbols} constructed out of the metric defined in Eq.~\eqref{eq:Christoffel}. 
Comparing this to the corresponding equation in general coordinates of flat Minkowski spacetime [Eq.~\eqref{eq:GeodesicDeviationFlat}], it is evident, that the right-hand-side represent the anticipated departure from flatness that we want to attribute to physical gravitational effects, in particular the tidal forces within an inhomogeneous gravitational field. 

For timelike geodesics and by explicitly choosing freely falling coordinates for one of the geodesics, we can make it explicit that these effects are no mere coordinate artifact, but correspond to a physical effect of curved spacetimes. Consider therefore Fermi normal coordinates $\{t,y^i\}$ (defined in the Appendix~\ref{sApp: Normal Coordinates}) associated to the first geodesic $y_1^\mu(\lambda)$, such that $y_1^i(\lambda)=0$ and $g_{\mu\nu}(y_1)=\eta_{\mu\nu}$ and $g_{\mu\nu,\alpha}(y_1)=0$ for any point along the geodesic. We then expand the metric up to second order in spacial coordinates $y^i$ around the origin given by the geodesic
\begin{equation}
    g_{\mu\nu}(y)\simeq\eta_{\mu\nu}+N_{\mu\nu \,i j}\,y^i y^j\,,
\end{equation}
where the expansion coefficients $N_{\mu\nu \,i j}$ are explicitly given in Eq.~\eqref{FermiNormalCoords}, but their explicit form does not matter at this stage.
The only information we use for now is that the Christoffel symbols evaluated on the first geodesic vanish in Fermi normal coordinates, while however their spacial derivatives do not
\begin{equation}
    \Gamma^\mu_{\nu\sigma,i}\neq 0\,.
\end{equation}
Thus, in Fermi normal coordinates, the geodesic deviation equation reduces to
\begin{equation}\label{eq:GeodesicDeviationFermiNormalCoords}
    \boxed{\frac{d^2\delta y^\mu}{d\lambda^2}=-\left(\Gamma^\mu_{\nu\sigma,\rho}-\Gamma^\mu_{\nu\rho,\sigma}\right)\delta y^\rho \dot y_1^\nu\dot y_1^\sigma\,,}
\end{equation}
and therefore evidently does not reduce to the geodesic deviation equation in Minkowski spacetime established in Eq.~\eqref{eq:GeodesicDeviation Equation Minkowski coords}.

Although this is not evident from our derivation, the quantity in brackets in Eq.~\eqref{eq:GeodesicDeviation} indeed correspond to well-defined components of a tensor field\footnote{See e.g. \cite{carroll2019spacetime} for an explicitly covariant derivation.}, namely of the \ul{Riemann curvature tensor}
\begin{equation}\label{eq:Riemann Tensor}
    \boxed{R\ud{\mu}{\nu\rho\sigma}\equiv\Gamma^\mu_{\nu\sigma,\rho}-\Gamma^\mu_{\nu\rho,\sigma}+\Gamma^\mu_{\lambda\rho}\Gamma^\lambda_{\nu\sigma}+\Gamma^\mu_{\lambda\sigma}\Gamma^\lambda_{\nu\rho}\,,}
\end{equation}
associated to the Christoffel symbols. This quantity build out of the metric that we associate to the gravitational field, therefore determines the physical, or measurable, notion of curvature of spacetime attributed to the tidal stress induced in a body due to gravitational forces. Under the assumption that free test masses follow geodesics of spacetime, any measurement of tidal forces can therefore be viewed as experimental evidence for the curvature of spacetime.

At this point it should be noted, however, that in differential geometry, a more general $\binom{1}{3}$-curvature tensor $\underline{\mathcal{R}}$ can be introduced, which is not a priori related to any geodesic deviation. As we outline in the Appendix \ref{App:DiffGeo}, a curvature tensor is first of all associated to a general \ul{affine connection} of a spacetime with coefficients $\,\DGamma^{\,\mu}_{\,\alpha\beta}$ and associated covariant derivative $\DGrad_\mu$, which must not be related to any metric. Simply put, the components of the curvature tensor $\mathcal{R}\ud{\mu}{\nu\rho\sigma}$ at a certain location in spacetime are proportional to the change in direction $\Delta v^\mu$ after parallel transporting a vector $v^\mu$ around an infinitesimal closed loop of area\footnote{However, without a metric, we only have a relative notion of ``area''.} $a^{\sigma\rho}$ (see e.g. \cite{misner_gravitation_1973,zee2013einstein})
\begin{equation}\label{eq:Parallel Transport around Closed Curve}
    \Delta v^\mu=\mathcal{R}\ud{\mu}{\nu\rho\sigma} v^\nu a^{\sigma\rho}\,,
\end{equation}
where $\mathcal{R}\ud{\mu}{\nu\rho\sigma}$ is given by Eq.~\eqref{eq:Riemann Tensor} with the Christoffel symbols replaced by the general connection coefficients $\,\DGamma^{\,\mu}_{\,\alpha\beta}$. Thus, the change in the vector does not depend on the shape of the curve, but only on the area it encloses in a given plane. 
Observe as well that this definition does not require a metric on the manifold. It is only the specific parallel transport operation to connect different tangent spaces, corresponding to a freedom of choice one has in the framework of differential geometry, that determines the general curvature tensor $\underline{\mathcal{R}}$.

From that point of view, the equivalence principle therefore provides us with a physical choice of connection, namely the \ul{Levi-Civita connection} with associated Christoffel symbols $\Gamma^\mu_{\alpha\beta}$ with associated covariant derivative $\nabla_\mu$, that naturally arises in non-inertial frames. In this case, the abstract curvature tensor $\mathcal{R}\ud{\mu}{\nu\rho\sigma}$ is equal to the quantity $R\ud{\mu}{\nu\rho\sigma}$ that appears in the geodesic deviation. This should not entirely come as a surprise, since as already mentioned in Sec.~\ref{sSec:Equivalence Principle} above, it is the choice of a Levi-Civita connection that implies at every point on the manifold the existence of Fermi normal coordinates corresponding to the freely falling frames. Moreover, such a choice of a Levi-Civita connection is in fact unique, as asserted by the fundamental theorem of Riemannian geometry \cite{levi1917nozione,zbMATH06520113,zbMATH00052737}. The Levi-Civita connection is uniquely determined by the two conditions on the Christoffel symbols in Eqs.~\eqref{eq:NoTorsionA} and~\eqref{eq:NoNonMetricityA}, that is, the connection is symmetric $\Gamma^\mu_{\alpha\beta}=\Gamma^\mu_{\beta\alpha}$ and the covariant derivative on the metric vanishes $\nabla_\mu g_{\alpha\beta}=0$, which respectively imply a vanishing \ul{torsion} and \ul{non-metricity}. These two conditions directly imply the relation of the Christoffel symbols to the metric given in Eq.~\eqref{eq:Christoffel}. Below, we will further comment on the choice of connection.

Interestingly, the geometric interpretation of curvature in Eq.~\eqref{eq:Parallel Transport around Closed Curve} provides an alternative way to measure physical curvature, namely by parallel transporting a physical vector along a closed spacial path within a global coordinate system and comparing its change in direction to a vector that remained at the same spacial location. Indeed, note that while the definition in Eq.~\eqref{eq:Parallel Transport around Closed Curve} is only valid for an infinitesimal curve, the result can be generalized to arbitrary curves by constructing a large closed curve out of patches of infinitesimal curves, while noting that the contributions of internal lines will cancel. In other words, the resulting change in direction after parallel transporting a vector along a closed curve is still zero if and only if the curvature tensor vanishes everywhere. Moreover, the result can be compared to the expected shift in direction within a given metric. 

In Sec.~\ref{sSec:Special Relativity} we already considered a free test gyroscope to provide a physical vector that defines trivial parallel transport in Minkowski spacetime. In the present context, a gyroscope therefore provides the natural trivial parallel transport in any local Minkowski patch. By the equivalence principle, we therefore expect that a gyroscope also serves as an operational definition of parallel transport with respect to the Levi-Civita connection in a general spacetime. In fact, such an experiment has been performed by measuring a precision gyroscope moving in an orbit around the earth, thus measuring the curved metric through geodetic and frame-dragging precession \cite{Everitt:2011hp}.\footnote{See e.g. \cite{zee2013einstein,Jetzer:2022bme} for an explicit calculation.}

\paragraph{Universal and Minimal Metric-Coupling.}

Based on the Principle~\ref{Principle:GeodesicMotionSR} of geodesic motion in Minkowski spacetime, formulated in Sec.~\ref{sSec:Special Relativity}, the EEP~\ref{Principle:EEP} therefore leads us to construct a gravity theory with a spacetime, in which test particles follow the geodesics of a general metric and the spin vector of test gyroscopes are parallel transported with respect to the Levi-Civita connection of spacetime that depends on the metric and a derivative thereof. Locally, such gravitational effects cannot be distinguished from inertial effects in accelerated frames of references. In turn, the effects intrinsic to gravity are determined by the Riemann curvature tensor that depends on second derivatives of the metric, and manifest themselves for instance in geodesic deviation and the parallel transport of a gyroscope around a closed loop. In this framework, the metric of spacetime therefore incorporates all these physical effects of motion associated to the gravitational field and in the following, we will call that metric the \textit{physical metric}. 

Note, however, that essentially, these conclusions could have been drawn merely from the weak equivalence principle. The Einstein equivalence principle goes much further by postulating that not only the physical motion of test particles and vectors are governed by a unique physical metric, but any local law of physics that is based on Poincar\'e invariance and the Minkowski metric. In other words, all non-gravitational physics on Minkowski spacetime described in Sec.~\ref{sSec: Theories of Minkowski Spacetime} that is captured by a collective set of matter-fields $\Psi_\text{m}$, can be promoted to include gravitational effects by considering their formulation in generic coordinates after the effective replacements in Eq.~\eqref{eq: Minimal Coupling Procedure}. True gravitational effects, that cannot be mimicked by accelerated observers, then enter by treating the metric as a dynamical object that is determined through its own equations of motion, instead of assuming that spacetime is a priori Minkowski.

Using the practical procedure of Eq.~\eqref{eq: Minimal Coupling Procedure}\footnote{See the discussion in \cite{Weinberg1972,WaldBook,misner_gravitation_1973} regarding possible ambiguities and their resolution regarding this procedure.} to include gravity has the far-reaching consequence that the gravitational field, through its geometric interpretation, is acting on all matter in exactly the same way. That is to say, all matter fields $\Psi_m$ couple in a \textit{universal} way to a single gravitational field, the physical metric \cite{poisson2014gravity,Will:2018bme}. Furthermore, the coupling is \textit{minimal} \cite{Dicke:1964pna}, in the sense that the coupling only occurs through the metric itself and the Christoffel symbols within the covariant derivative. This in particular excludes couplings to the Riemann curvature tensor and any of its contractions involving more than one metric-derivatives.

Since the notion of universal and minimal coupling will be central in the following, we want to elaborate a bit more about this statement. Exactly as it was already the case for the motion of test particles, such a universal and minimal coupling to the physical metric and its single derivative, together with the existence of a Levi-Civita connection, ensure that in each event in spacetime there exist a local freely falling frame, corresponding to the normal coordinates introduced in Eq.~\eqref{eq:Riemann Normal Coordinates}, in which the equations of motion of all non-gravitational physics reduce to the ones constructed in Minkowski spacetime. In fact, it can be argued \cite{Dicke:1964pna,Will:2018bme} that the Einstein equivalence principle inevitably implies a unique description of gravity in terms of a minimal and universal coupling to a physical metric.
While it is perhaps debatable whether an entirely different description without referring to any manifold and metric for instance is ruled out completely, it is certain that within the framework of differential geometry, universal and minimal coupling is imminent. 

Indeed, both a direct coupling to other gravitational fields that are not the physical metric, as well as a coupling to higher order derivatives of the metric, would generally violate the equivalence principle. This is so, because otherwise, provided that a non-trivial gravitational field is present, there is no guarantee that one recovers the same laws of special relativity of matter fields in \textit{every} freely falling frame. The explicit presence of an additional gravitational field or a curvature component within the matter equations of motion would give rise to an additional force \textit{at a single spacetime point} in freely falling frames, also called ``fifth force'', that would influence the local physics of \textit{non-gravitating} matter fields, depending on the external gravitational field.\footnote{It is important to recognize, that the Einstein equivalence principle explicitly excludes experiments in which the self-gravitation of the energy of matter becomes non-negligible.} 

To be more concrete, so long as minimal coupling holds, the effect of the Riemann tensor is only felt by extended objects (or a system of multiple test masses) through so-called tidal forces. This is because if their interaction with gravity is only governed by the metric and the Christoffel symbol (through the covariant derivative) the local effect of gravity vanishes identically in any freely falling frame [Eq.~\eqref{eq:Freely falling frame}]. Gravitational effects can then fundamentally only be felt by extended objects. More precisely, for a test object of typical size $d$, the tidal effects would by dimensional analysis scale as $d/D$, where 
\begin{equation}\label{eq:characteristic spacetime dimension of curvature}
    D^{-2}\sim |R_{\mu\nu\rho\sigma}| \,,
\end{equation} 
represents the characteristic spacetime dimension of the external gravitational field.
Hence, for a small enough object in a small enough region, these effects can be made as negligible as desired, independently of the value of the curvature, as required by the equivalence principle. On the other hand, a violation of minimal coupling would entail an effect of curvature on the matter equations of motion at a single location, no matter the size of the object, and can therefore \textit{not} be made arbitrary small (see also \cite{Weinberg1972,misner_gravitation_1973}). Coming back to our dimensional analysis, such a coupling to curvature would necessitate the introduction of an additional ad hoc length scale $l_0$. This scale could of course still be chosen such that $l_0/D\ll 1$ whenever curvature is not too large, in order to recover constraints in weak gravitational fields, but not on any scale in principle (see e.g. \cite{Gonner:1976gq}).

The same is true for the presence of additional fields in the gravitational sector that might couple to the matter fields. If such a coupling would exist, then the EEP is violated whenever this additional gravitational field is non-trivial and induces a localized gravitational effect on the matter equations of motion. Indeed, Fermi normal coordinates of freely falling frames only ensure that the physical metric reduces to the Minkowski metric up to second order in derivatives. Again, the numerical value of the additional gravitational fields coupling to matter could always be chosen such that current experimental bounds are satisfied. 

However, on a conceptual level, precisely the condition that special relativity is recovered on \textit{all} experimentally accessible scales is fundamental. It is important to realize that this requirement goes beyond the rejection of ad hoc scales to match current observations.
Namely, it is this crucial property of the gravitational field, which allows one to talk about the physical metric as defining the notion of \textit{spacetime}, rather than being a field within spacetime. This is because a universal and minimal coupling ensures that any measurement of the properties of spacetime do by principle not depend on the location, as well as the composition and type of the (non-gravitational) measurement device  \cite{Will:2018bme}. For instance, this ensures that the concept of proper time or proper distance are true, observer independent characteristics of spacetime. These are assumptions at the basis of most empirical probes of spacetime, such that universal and minimal coupling can be regarded as a requirement of any testable theory of dynamical spacetime. When breaking these properties, substantial care is needed when interpreting experiments.

We therefore postulate, that the Einstein equivalence Principle~\ref{Principle:EEP} leads to the following concrete principle for theories of gravitation:

\begin{principle}\ul{Principle of Universal and Minimal Coupling}.\label{Principle:Universal and Minimal Coupling}
    Spacetime is endowed with a physical metric tensor $g_{\mu\nu}$, the world lines of test particles are the geodesics of that metric and the gravitational coupling of all matter fields $\Psi_m$ arises only through the physical metric $g_{\mu\nu}$ and its first derivative $g_{\mu\nu,\alpha}$, as it arises in a non-inertial frame in the absence of any gravitational field.
\end{principle}

\paragraph{Metric Theories.} Theories of gravitation that obey the principle of universal and minimal coupling to a physical metric are known as \textit{metric theories} \cite{Dicke:1964pna,poisson2014gravity,papantonopoulos2014modifications,Will:2018bme,YunesColemanMiller:2021lky}. However, note that this principle does not exclude additional ``gravitational'' fields in the theory. Indeed, so far we only discussed the coupling of gravity to the matter fields that describe all non-gravitational physics, but never considered the dynamics of the gravitational field itself. The EEP, together with the empirical evidence for the WEP and local Lorentz invariance of matter fields, do not constrain the gravitational sector, which therefore may contain additional effects, in particular also including Lorentz symmetry violations.

As concerns the terminology, we will however reserve the term ``gravitational'' for the physical metric and its physical effects through its coupling to matter and term any additional fields in the gravity sector as \textit{\gls{non-minimal fields}}, while collectively denoting them as $\Psi$. Here, a non-minimal coupling is literally defined as any coupling to the physical metric that is \textit{not} minimal in the sense of Principle~\ref{Principle:Universal and Minimal Coupling}. A decisive implication of the principle of universal coupling is, however, that none of the non-minimal fields associated to the gravity sector couple directly to matter. However, these potential additional fields in the theory, can of course still entail physical effects, although not through direct influence of non-gravitational matter, but precisely through their non-minimal coupling to the physical metric.\footnote{We want to note that from this perspective, additional non-minimal fields influence how ordinary matter sources the gravitational field described by the physical metric and therefore also might play a natural role as a dark matter candidate.} As we will show below, the main distinction between non-minimal fields and the matter fields defined in Sec.~\ref{sSec: Theories of Minkowski Spacetime} is that there is no locally conserved energy momentum tensor associated to the collection of non-minimal fields. To be more precise, no conserved energy momentum tensor that is independent of any higher-order metric curvature invariants can be defined. Therefore, they are distinct from the known non-gravitational physics described through matter fields, from which we assume that they make up all known non-gravitational mass and energy.

It is sometimes mentioned that such a strict distinction between non-minimal an matter fields is challenged by the expectation that quantum corrections of matter with gravitons inevitably lead to non-minimal couplings (see e.g. \cite{Padmanabhan:2004xk,papantonopoulos2014modifications}). This would however also directly imply that quantum effects spoil the equivalence principle, with deep implications also for GR and our fundamental understanding of GR. Thus, a clear distinction between non-minimal and matter fields should hold, at least on the same level as the concept of matter with an associated well-defined energy momentum-tensor holds in the current understanding of physics. Moreover, as we will discuss in Chapter~\ref{Sec.Quantum Stability}, quantum corrections are by no means to be treated on the same footing as classical operators, such that a distinction is tenable even in the presence of such corrections. Moreover, without a doubt, the reconciliation of gravity with quantum theory still lies in deep shadows, such that no hasty conclusions should be drawn. It could even be interesting to trace the implications of the principle of equivalence down to quantum scales (see also Chapter.~\ref{Sec:Challenges of the Quantum EFT of Gravity}).


It is now time to also address the dynamical side of gravity. We will do so by assuming that the theory can be formulated in terms of an action. This will allow us to provide a rigorous but still very general definition of dynamical metric theories of gravity.

\paragraph{Dynamics from an Action.} 

In order to formulate a dynamical theory of gravity, we will build on the considerations above and assume a four-dimensional differentiable Manifold with a Lorentzian metric and a Levi-Civita connection. Moreover, we require the theory to be local in order to respect causality. Based on the formulation of non-gravitational theories in non-inertial frames in Sec.~\ref{sSec: Theories of Minkowski Spacetime} and the connection of this formulation to gravity through the EEP, we furthermore choose a so-called general covariant formulation of the theory. This implies that the action should be a scalar under general coordinate transformations, hence diffeomorphism invariant (see Appendix~\ref{sApp:DiffsAndLieDer}). 
We will further require that all fields on the manifold are dynamical, in the sense that each field appearing in the action comes with its own equations of motion that determine its value. In particular, this applies also any additional non-minimal fields in the gravity sector on top of the physical metric. This is because we want to get rid of as much a priori structure on the spacetime as possible, in order to avoid any poorly motivated assumptions that might be too restrictive.

A dynamical formulation of gravity, should answer the question of how the energy-momentum density of matter fields, the source of gravity, generates the gravitational field. In particular, the non-minimal couplings of additional non-minimal fields in the gravity sector will precisely influence how matter generates the gravitational field and how it evolves and therefore have an indirect physical effect of any measurement of gravity and spacetime. Yet, as discussed above, according to the Principle~\ref{Principle:Universal and Minimal Coupling} of universal and minimal coupling, it is imperative that only the physical metric talks directly to the matter fields in order to preserve the notion of spacetime that can be probed independently of the measurement device. 

To formulate an action that should be dimensionless in natural units, we also need to set conventions regarding the dimensions of the fields. In general, the metric $g_{\mu\nu}$, as well as any other non-minimal fields $\Psi$ will be defined as dimensionless fields in terms of energy dimensions, hence $[g_{\mu\nu}]=E^0$ and $[\Psi]=E^0$. Note that for any derivative we have $[\partial_\mu]=E$, while $[d^4x]=E^{-4}$. For the leading order terms with two powers of derivatives this necessitates the introduction of a dimensionful bare gravitational constant $G$, with $[G]=E^{-2}$ that is conventionally introduced in the gravitational action through the combination $\kappa_0\equiv 8\pi G$ and is ultimately fixed through the requirement of an appropriate Newtonian limit. 

The above discussion culminates into the following definition of a metric theory of gravity:
\begin{definition} \textit{Metric Theory of Gravity.}\label{DefMetricTheory}
    Let $\mathscr M$ be a four-dimensional oriented and differentiable pseudo-Riemannian manifold equipped with a Lorentzian metric $g_{\mu\nu}$ and an associated Levi-Civita connection. A metric theory is a local and diffeomorphism invariant Lagrangian theory on $\mathscr M$ described by an action of the general form
    \begin{equation}\label{eq:ActionMetricTheory}
        \boxed{S=S_\text{G}+S_\text{m}=\int d^4x\,\sqrt{-g} \left(\frac{1}{2\kappa_0}\mathcal{L}_\text{G}[g,\Psi]+\mathcal{L}^{\myst{min}}_\text{m}[g,\Psi_\text{m}]\right)\,,}
    \end{equation}
    consisting of a matter Lagrangian $\mathcal{L}^{\myst{min}}_\text{m}$ minimally coupled to the metric $g$ only, and a gravitational Lagrangian $\mathcal{L}_\text{G}$ covariantly depending on the metric, as well as possibly on a set of additional non-minimal fields $\Psi$.
\end{definition}
As a heads-up, in this work we will generally restrict our attention to non-minimal fields in the action in the form of $k$-form fields (see Appendix~\ref{app:ExampleNullMemoryKForm} for a definition). These encompass the most important cases of the components of scalar and vector fields. Note that such a restriction immediately implies a limitation to bosonic fields. Moreover, a restriction to $k$-form fields also means that for simplicity, we will focus on theories with Abelian gauge groups only.

It is convenient to introduce a general symbol for the gravitational metric equations of a generic metric theory of gravity
\begin{equation}\label{eq:DefGenEinsteinTensor}
    \boxed{\mathcal{G}_{\mu\nu}\equiv \frac{1}{\sqrt{-g}}\,\frac{\delta S_\text{G}}{\delta g^{\mu\nu}}\,.}
\end{equation}
Following the variational principle (see e.g. \cite{Weinberg1972} or App.~\ref{App: Symmetires in Physics}, as well as a comment in Sec.~\ref{sSec:LovelockTheorem}), while recalling the energy momentum tensor of matter fields $T_{\mu\nu}$ defined in Eq.~\eqref{eq:DefEnergyMomentumTensor}, a generic metric theory of gravity is therefore governed by a set of \textit{metric equations of motion}
\begin{equation}\label{eq:EOM MetricTheory}
    \mathcal{G}_{\mu\nu}=\kappa_0 T_{\mu\nu}\,,
\end{equation}
together with the field equations of dynamical non-minimal fields and the matter equations of motion [Eq.~\eqref{eq:EOMMatterFields}] that we will collectively denote by
\begin{align}
    \frac{\delta S_\text{G}}{\delta \Psi} &\equiv\mathcal{J}=0\,,\label{eq:EOM MetricTheoryAdditionalfields}\\
    \frac{\delta S_\text{m}}{\delta \Psi_\text{m}} &\equiv\mathcal{J}_\text{m}=0\label{eq:EOMMatterFieldEquations}\,.
\end{align}
Of course, specific matter, as well as the non-minimal fields might also explicitly be sourced by some particular charge configuration. However, such charged entities are also themselves described through the collective concept of matter and non-minimal fields, such that these source terms do not appear explicitly.\footnote{In fact, explicit sources of non-minimal fields are often not considered, and non-minimal fields only arise through their coupling with the physical metric, or their own non-linearity.}

According to the Definition~\ref{DefMetricTheory} above, general metric theories of gravity therefore differ by their field content in the gravitational sector, as well as by the exact form of the action, which translates in a difference in the equations of motion in Eqs.~\eqref{eq:EOM MetricTheory} and \eqref{eq:EOM MetricTheoryAdditionalfields}. While this still leaves a lot of freedom, a key result of the above definition is that for any such metric theory, the energy-momentum tensor of matter fields in Eq.~\eqref{eq:DefEnergyMomentumTensor} is locally conserved, a statement we now want to elaborate on. 

\section{Conservation of Energy-Momentum}\label{sSec:Covariant Consrevation}


\paragraph{Local Conservation of the Energy-Momentum Tensor.}  Indeed, the local conservation of the energy-momentum tensor of matter fields follows directly from minimal and universal coupling together with the symmetry of coordinate invariance of the action in Eq.~\eqref{eq:ActionMetricTheory} (see also \cite{Trautman_1963,Weinberg1972,WaldBook,papantonopoulos2014modifications,Will:2018bme,carroll2019spacetime}).
Let's therefore consider an infinitesimal coordinate transformation (see App.~\ref{sApp:DiffsAndLieDer})
\begin{equation}\label{eq:Infinitesimal coordinate Transf}
    x^\mu\rightarrow x^\mu+\xi^\mu\,,
\end{equation}
generated by a vector field $\xi^\mu(x)$. Under this transformation, all the fields in the action, that is the metric $g_{\mu\nu}$, the dynamical non-minimal fields $\Psi$, as well as the dynamical matter fields $\Psi_\text{m}$, will transform. However, note that both pieces in the action, $S_\text{G}$ and $S_\text{m}$, are separately coordinate invariant, such that we can individually consider both pieces. The profound split between the gravitational Lagrangian and a universally and minimally coupled matter Lagrangian then further implies that the variation of $S_\text{m}$ under Eq.~\eqref{eq:Infinitesimal coordinate Transf} does not involve any variation due to the non-minimal fields, while $S_\text{G}$ in turn does not depend on the matter fields. Postponing a discussion on the variation of the gravitational action, consider the variation of the matter action that therefore reads \cite{carroll2019spacetime}
\begin{equation}\label{eq:variationOfMatterAction}
    \bar{\delta}S_\text{m}=\int d^4x\,\frac{\delta S_\text{m}}{\delta g_{\mu\nu}}\,\bar{\delta}g_{\mu\nu}+\int d^4x\,\frac{\delta S_\text{m}}{\delta \Psi_\text{m}}\,\bar{\delta}\Psi_\text{m}=0\,,
\end{equation}
where $\bar{\delta}$, denoting the \textit{total variation} defined in Appendix~\ref{App: Symmetires in Physics}, arises due to the integration over the coordinates in the action. The total variation of the metric and the matter fields, on the other hand, corresponds to the \ul{Lie derivative} (see App.~\ref{sApp:DiffsAndLieDer} and \ref{sApp: Metric and Riemannian G})
\begin{align}
    \bar{\delta}g_{\mu\nu}&=-\mathcal{L}_\xi g_{\mu\nu}=-\xi^\alpha\partial_\alpha g_{\mu\nu}-(\partial_\mu \xi^\alpha)g_{\alpha\nu}-(\partial_\nu \xi^\alpha)g_{\mu\alpha}=-2\nabla_{(\mu}\xi_{\nu)}\,,\label{eq:LieDerivativeMetric}\\
    \bar{\delta}\Psi_\text{m}&=-\mathcal{L}_\xi \Psi_\text{m}\,.
\end{align}

To continue, observe that the first term of Eq.~\eqref{eq:variationOfMatterAction} involves the definition of the energy-momentum tensor in Eq.~\eqref{eq:DefEnergyMomentumTensor}, whereas the second term contains the vanishing equations of motion of the matter fields Eq.~\eqref{eq:EOMMatterFields}. We can thus write
\begin{equation}\label{eq:variationOfMatterActionII}
    \bar{\delta}S_\text{m}=\int \sqrt{-g}d^4x\,T^{\mu\nu}\,\nabla_{\mu}\xi_{\nu}=0\,,
\end{equation}
where we dropped the symmetrization due to the symmetry of $T^{\mu\nu}$. 
Integrating by parts by assuming that the generating vector field $\xi^\mu$ vanishes at infinity, we finally obtain
\begin{equation}\label{eq:variationOfMatterActionIII}
    \int \sqrt{-g}d^4x\,\nabla_{\mu}T^{\mu\nu}\,\xi_{\nu}=0\,.
\end{equation}
Since this has to hold for any coordinate transformation, we conclude that indeed the diffemomorphism invariance of the matter action implies 
\begin{equation}\label{eq:CovariantConservationEMTensor}
    \boxed{\nabla_\mu T^{\mu\nu}=0\,.}
\end{equation}

This equation has the crucial implication that energy-momentum of matter fields is locally conserved (see e.g. \cite{misner_gravitation_1973}). This can also be understood by choosing local Riemann normal coordinates $y^\mu$, in which the covariant conservation equation in Eq.~\eqref{eq:CovariantConservationEMTensor} becomes
\begin{equation}\label{eq:ConservationEMTensorLocal}
  \pd_\mu T^{\mu\nu}(y)=0\,.
\end{equation}
For instance, based on this equation it is possible to locally define a conserved energy current density as in Sec.~\ref{sSec: Theories of Minkowski Spacetime}. Observe, however, that in contrast to Minkowski space, such conservation statements can in this general case only be made within the validity of the local Riemann normal coordinate frame (we will come back to this point below).

We want to emphasize the importance of the principle of universal and minimal coupling in obtaining this fundamental result. First of all, universal coupling implies that the second term in Eq.~\eqref{eq:variationOfMatterAction} corresponds to the vanishing equations of motion of matter fields. If there were no clean separation between the matter and the gravity sector of the action, the result in Eq.~\eqref{eq:CovariantConservationEMTensor} could not have been reached. On the other hand, the minimal coupling requirement assures that Eq.~\eqref{eq:ConservationEMTensorLocal} can truly be interpreted as an equation of energy-momentum conservation. Indeed, minimal coupling implies that the operator $T_{\mu\nu}$ defined in Eq.~\eqref{eq:DefEnergyMomentumTensor} only depends on gravitational fields through the physical metric and its first derivative, such that in local Riemann coordinates it is \textit{solely} determined through the matter fields themselves. If this were not the case, the operator would non-trivially depend on the external gravitational field, to which no local energy-momentum can be assigned. Thus, universal coupling not only dictates how matter should move in spacetime, but also influences the way energy-momentum density curves spacetime through an unambiguous definition of the matter energy-momentum tensor. Furthermore, the above result shows how the Principle~\ref{Principle:Universal and Minimal Coupling} of universal and minimal coupling draws a clear line between matter fields and non-minimal fields, for which no equivalent conservation equation of their ``energy-momentum tensor'' can be found. As we will further discuss in Part~\ref{Part: Cosmological Testing Ground}, the properties of non-minimal fields may be fitting to describe the dark sector within the model of cosmology. 

Finally, observe that as a consequence of the invariance of the matter action in Eq.~\eqref{eq:MatterAction} under coordinate transformations, the covariant conservation equation of the energy-momentum tensor [Eq.~\eqref{eq:CovariantConservationEMTensor}] is an instance of Noether's second Theorem~\ref{Thm:NoetherTheorem} that we discuss in Appendix~\ref{sApp: Noethers Theorem}. Indeed, the invariance of the matter action under generic coordinate transformations is a local or gauge symmetry of the formulation in arbitrary coordinate frames and in this context, the result in Eq.~\eqref{eq:variationOfMatterAction} can be viewed as a ``Bianchi identity'' \cite{Will:2018bme}.

\paragraph{The Contracted Bianchi Identities.} In the light of the comments above, it is instructive to also consider the variation of the gravitational action under Eq.~\eqref{eq:Infinitesimal coordinate Transf}. In this case, the second Noether Theorem~\ref{Thm:NoetherTheorem} applied to the symmetry of diffeomorphism of the gravitational part $S_\text{G}$ of the action will also lead to a set of Bianchi identities that are however distinct in nature from the local energy-momentum conservation. More concretely, the diffeomorphism invariance of $S_\text{G}$ implies, analogous to the considerations above, that
\begin{equation}\label{eq:variationOfGravAction}
    \bar{\delta}S_\text{G}=\int d^4x\,\frac{\delta S_\text{G}}{\delta g^{\mu\nu}}\,\bar{\delta}g^{\mu\nu}+\int d^4x\,\frac{\delta S_\text{G}}{\delta \Psi}\,\bar{\delta}\Psi=0\,,
\end{equation}
where again $\bar{\delta}$ denotes the total variation. Since crucially only the physical metric appears in both parts of the general action in Eq.~\eqref{eq:ActionMetricTheory}, the second term in Eq.~\eqref{eq:variationOfGravAction} this time involves the vanishing equations of motion of the non-minimal fields in Eq.~\eqref{eq:EOM MetricTheoryAdditionalfields}, such that Eq.~\eqref{eq:variationOfGravAction} becomes
\begin{equation}\label{eq:variationOfGravActionII}
    \bar{\delta}S_\text{G}=-2\int \sqrt{-g}d^4x\,\mathcal{G}_{\mu\nu}\,\nabla^{\mu}\xi^{\nu}=0\,,
\end{equation}
where we have used Eqs.~\eqref{eq:LieDerivativeMetric} and \eqref{eq:DefGenEinsteinTensor}, together with the symmetry of $\mathcal{G}_{\mu\nu}$.
An integration by parts then implies
\begin{equation}\label{eq:ContractedBianchiIdentity}
    \boxed{\nabla^\mu \mathcal{G}_{\mu\nu}=0\,,}
\end{equation}
known as \ul{contracted Bianchi identities}.

It is often stated, that the contracted Bianchi identities [Eq.~\eqref{eq:ContractedBianchiIdentity}] together with the metric field equations [Eq.~\eqref{eq:EOM MetricTheory}] imply the conservation of the energy-momentum tensor. From the above argument, it should be clear that such a statement is not entirely correct. Rather, both the Bianchi identities and the covariant conservation of the energy-momentum tensor are consequences of the universal and minimal coupling principle, the diffeomorphism invariant formulation and the equations of motion of non-minimal and matter fields. Moreover, neither depend on the metric equations of motion. In this sense, both equations are on the same footing and hold by their own right.
This subtlety in the origin of the covariant conservation of the energy-momentum tensor will play an important role when defining the energy-momentum carried by gravitational waves in later chapters. 

Moreover, note that Eq.~\eqref{eq:ContractedBianchiIdentity} cannot be considered as a statement of energy-momentum conservation, neither of the non-minimal fields nor of the physical metric. In fact, a proper definition of an energy-momentum tensor defining local conservation for the gravitational fields is a long-standing question that even puzzled Einstein back in the time \cite{Einstein:1916GrundlagenGR,misner_gravitation_1973,WaldBook,landau_classical_2003}. The difficulty in defining such a notion of localized energy-momentum of the physical metric can be tracked back to the EEP, implying that there are no true local effects of gravity. In Sec.~\eqref{ssSec:IsaacsonInGR} we will however discuss a well-defined notion of energy-momentum of the gravitational field.


\paragraph{Isometries and Conservation Laws} 

This is also a good moment to put the above statements of local energy-momentum conservation in contrast to the discussion in Sec.~\ref{sSec: Theories of Minkowski Spacetime} on Minkowski spacetime and further stress that the covariant conservation equation Eq.~\eqref{eq:CovariantConservationEMTensor} on its own merely ensures local energy-momentum conservation of the matter fields in a small enough region, in which gravity can be neglected. As soon as gravitational effects enter, however, in other words on scales where the inhomogeneities of the gravitational field captured by the non-trivial spacetime metric become important, the energy of matter fields can no longer be considered to be conserved because of the presence of the non-vanishing connection coefficients in the covariant conservation. In other words, matter fields can ``loose'' energy to the gravitational field, the local energy-momentum content of which is however not defined.  Eq.~\eqref{eq:CovariantConservationEMTensor} therefore implies that, in general, there is in fact no energy-momentum conservation in theories of gravity \cite{Blau2017,carroll2019spacetime}.

For a theory defined on Minkowski spacetime that admits global Minkowski charts, on the other hand, Eq.~\eqref{eq:CovariantConservationEMTensor} directly implies the ``proper'' conservation of the energy-momentum tensor in such inertial frames of Minkowski spacetime [Eq.~\eqref{eq:ConservationEMTEnsor}] with all its consequential implications on global energy conservation discussed thereafter.
More precisely, the global energy-momentum conservation on the entire Minkowski spacetime is distinct from the more general statement in Eq.~\eqref{eq:CovariantConservationEMTensor}, in that Eq.~\eqref{eq:ConservationEMTEnsor} is \textit{not} a consequence of diffeomorphism invariance. Rather, energy-momentum conservation across the spacetime follows from the special structure of Minkowski spacetime. In other words, it is the invariance of Minkowski spacetime under time and space translations which ensures conservation of energy-momentum. As such, it is the global symmetries associated to the $1^\text{st}$ Noether theorem which imply ``proper'' conservation laws. In contrast, recall that Eq.~\eqref{eq:CovariantConservationEMTensor} is a result based on the $2^\text{nd}$ Noether theorem.


To resolve this potential confusion, it is instructive to leave the inertial Minkowski frames and state global energy-momentum conservation in a covariant language \cite{Blau2017,carroll2019spacetime}. In this case it becomes apparent that Eq.~\eqref{eq:CovariantConservationEMTensor} alone does not imply any true conservation law but decisively requires the presence of global or ``proper'' symmetries of the spacetime metric known as \ul{isometries}. Indeed, in a covariant language, the presence of a global spacetime symmetry under a specific diffeomorphism is shown in Appendix~\ref{sApp: Spacetime Gaugefreedom and symmetries} to be associated to the existence of a \ul{Killing vector field} $K^\mu$, along which the Lie derivative of the metric vanishes
\begin{equation}\label{eq:KVFDef}
    \mathcal{L}_K g_{\mu\nu}=2\,\nabla_{(\mu}K_{\nu)}=0\,.
\end{equation}
Together with a conserved energy-momentum tensor, the existence of Killing vector fields then allow the definition of covariantly conserved currents
\begin{equation}
    J^\nu\equiv K_\mu T^{\mu\nu}\,,
\end{equation}
that satisfy
\begin{equation}
    \nabla_\nu J^\nu= \nabla_{(\nu} K_{\mu)} \,T^{\mu\nu}+K_\mu \,\nabla_\nu T^{\mu\nu}=0\,,
\end{equation}
due to Eqs.~\eqref{eq:KVFDef} and \eqref{eq:CovariantConservationEMTensor}. Only based on such a conserved current, an associated conserved charge can be defined. For instance, the time translation vector $t^\mu$ in the example in Sec.~\ref{sSec: Theories of Minkowski Spacetime} is nothing but a Killing vector associated to the symmetry of time translations of Minkowski spacetime, which implies the conservation of energy.
Moreover, given a Killing vector field $K^\mu$ and a geodesic $x^\mu(\lambda)$, the quantity $Q_K(\lambda)\equiv K^\mu\dot x^\nu g_{\mu\nu}$ is conserved along the geodesic. Indeed,
\begin{equation}
    \frac{dQ_K}{d\lambda}=K_\mu\frac{D\dot x^\mu}{d\lambda}+\dot x^\mu\frac{D K_\mu}{d\lambda}=0+\dot x^\mu\dot x^\nu\nabla_{(\nu} K_{\mu)}=0\,.
\end{equation}
In summary, Eq.~\eqref{eq:CovariantConservationEMTensor} only implies proper energy-momentum conservation if the spacetime in question possesses isometries, associated to a global symmetry under coordinate transformations, that imply the existence of Killing vector fields of the physical metric.

The clarifications above are mainly necessary due to the subtleties in the relation between a local symmetry and its global counterpart that also enter the distinction between the 1$^\text{st}$ and $2^\text{nd}$ Noether Theorems~\ref{Thm:NoetherTheorem}. Coordinate transformations exemplify, that a local gauge symmetry does not imply the invariance under any global subgroup. In other words, the invariance under generic coordinate transformations of a particular formulation of a theory, that in this case necessitates the presence of a generic metric $g_{\mu\nu}$ that transforms under coordinate changes, does not imply the existence of the global symmetry indicated by the condition 
\begin{equation}
    \bar{\delta}g_{\mu\nu}=-\mathcal{L}_K g_{\mu\nu}=2\,\nabla_{(\mu}K_{\nu)}=0\,.
\end{equation}
Indeed, this condition on the existence of KVFs is only satisfied in SR due to the special properties of the Minkowski metric.
This should be contrasted to the case of $U(1)$ gauge symmetric formulations in the presence of a vector potential $A_\mu$. In this case the local $U(1)$ gauge symmetry under the transformation $A_\mu\rightarrow A_\mu +\pd_\mu \Lambda$ always implies the presence of a proper global symmetry associated to charge conservation due to the simple fact that for any constant gauge parameter $\Lambda$, the total variation (or ``Lie derivative'') of the gauge field identically vanishes 
\begin{equation}
    \bar{\delta}A_\mu=\partial_\mu\Lambda\equiv 0\,,
\end{equation}
as discussed in more detail in Appendix~\ref{sApp: Gauge Symmetries and Proper Symmetries}.

\paragraph{Local Flatness and the Minkowski Metric.}

To close this chapter, we want to draw here as a little side note the (perhaps obvious) attention to the difference between the existence of Riemann normal coordinates and the definition of locally flat spacetimes. 
As discussed above, a manifold endowed with a metric and an associated Levi-Civita connection admits at every point normal coordinates (see also Appendix~\ref{sApp: Normal Coordinates}), in which 
\begin{equation}\label{eq:Freely falling frame}
    g_{\mu\nu}(y)=\eta_{\mu\nu}\,,\quad g_{\mu\nu,\alpha}(y)=0\,,
\end{equation}
up to second order at that point. In that sense, such a general spacetime naturally recovers the laws of special relativity in local enough regions.

While this is true for any metric, one should contrast such a local Minkowski form of the metric with the notion of local flatness:
\begin{definition} \textit{Flat Spacetime.}\label{DefFlatness}
    A spacetime is called flat, if for every point in the spacetime there exists a coordinate induced chart $y^\mu$, in which $g_{\mu\nu}(y)=\eta_{\mu\nu}$ for all points in the chart.
\end{definition}
\noindent
In contrast to the existence of normal coordinates, this notion of flatness is much stronger, as it can be shown that a spacetime is flat in the above sense if and only if the Riemann curvature tensor associated to the Levi-Civita connection vanishes $R\ud{\alpha}{\beta\mu\nu}=0$ \cite{misner_gravitation_1973,Renner2020}.\footnote{This statement is of course still closely connected to the fact that Riemman normal coordinates can also only be found if the torsion and non-metricity vanish.} Note that Def.~\ref{DefFlatness} is in fact only distinct from a Minkowski spacetime, defined as $\mathbb{R}\times\mathbb{R}^3$, due to potential topological effects that might prevent an extension of a Minkowski coordinate patch on the entire spacetime. As we will see below, flat spacetimes, and therefore also Minkowski spacetime, indeed represents a special solution to the dynamical equations for the metric.

This in particular clarifies a statement made back in Sec.~\ref{sSec:Special Relativity}. The condition $R\ud{\alpha}{\beta\mu\nu}=0$ on flatness [Eq.\eqref{eq:VanishingRiemann}] is directly connected with the existence of a global notion of inertial frames given by the Minkowski chart in Def.~\ref{DefFlatness}.



\chapter{General Relativity is still Special}\label{Sec:General Relativity}

We will now show, that in the space of metric theories of Definition~\ref{DefMetricTheory}, general relativity is the unique leading order theory that involves only the physical metric and no other non-minimal fields. This statement is due to a theorem by Lovelock \cite{Lovelock1969ArRMA,Navarro:2010zm,Charmousis:2014mia} (see also \cite{misner_gravitation_1973}) which considerably constraints the number and form of the $\binom{0}{2}$-tensors  $\mathcal{G}_{\mu\nu}$ that can appear on the left-hand side of the metric field equations given in Eq.~\eqref{eq:EOM MetricTheory}. Moreover, GR is the only known metric theory that satisfies a strong equivalence principle that goes beyond the Einstein equivalence Principle~\ref{Principle:EEP} at the basis of general metric theories of gravity.

\section{The Lovelock Theorem}\label{sSec:LovelockTheorem}

Recall from the discussion in Secs.~\ref{sSec:Metric Theories} and \ref{sSec:Covariant Consrevation} that the gravity part of the metric equations of motion $\mathcal{G}_{\mu\nu}$ needs to be symmetric and covariant, as well as divergence-free, hence $\nabla^\mu\mathcal{G}_{\mu\nu}=0$. This last requirement follows from the Principle~\ref{Principle:Universal and Minimal Coupling} of universal and minimal coupling, which requires the clear split between a gravitational and a matter action within Eq.~\eqref{eq:ActionMetricTheory} and diffeomorphism invariance. For a theory that only involves the physical metric in the gravitational part $S_\text{G}$, the gravity metric equation tensor, which in this case we will denote by $G_{\mu\nu}$, only depends on that metric and its derivatives.

Moreover, very generally, dimensional analysis imposes that the leading order theory at low energies involves as few derivatives as possible \cite{Weinberg1972}. More precisely, the leading order expression in $G_{\mu\nu}$ should only involve up to two powers of derivatives in each term.\footnote{In Chapter~\ref{Sec:The Theory Space Beyond GR} we will understand that this restriction to the lowest order terms is up to a subtlety fundamentally bound to the assumption of only propagating two degrees of freedom, a concept that we will properly introduce in Chapter~\ref{Sec:PropagatingDOFs}.} This restriction should be contrasted with a similar requirement on operators being \textit{second order in derivatives} that we will employ below. This latter statement will mean that there are only two derivative operators acting on each field, while however the number of derivative operators remains unconstrained. 

Together with the additional assumptions that went into the Definition~\ref{DefMetricTheory} of metric theories of gravity, the Lovelock theorem reads:
\begin{theorem}\label{Thm:LovelockTheorem} \textbf{The Lovelock Theorem} \cite{Lovelock1969ArRMA}.
Let $\mathscr M$ be a four-dimensional manifold endowed with a metric $g_{\mu\nu}$ and a Levi-Civita connection. The only $\binom{0}{2}$-tensor on $\mathscr M$ other than the metric itself, whose components are functions of the metric only, and that is
    \begin{itemize}
        \item symmetric,
        \item covariant,
        \item divergence-free,
        \item only involves up to two powers of derivative operators,
    \end{itemize}
is the \ul{Einstein tensor}
\begin{equation}\label{eq:EinsteinTensor}
    G_{\mu \nu}\equiv R_{\mu\nu}-\frac{1}{2}g_{\mu\nu}\,R\,.
\end{equation}
\end{theorem}

The Lovelock theorem directly implies that the unique leading order metric field equations [Eq.~\eqref{eq:EOM MetricTheory}] for such a metric theory with no additional non-minimal fields are given by 
\begin{equation}\label{eq:EinsteinFieldEquations}
    \boxed{G_{\mu\nu}+\Lambda g_{\mu\nu}=\kappa_0\,T_{\mu\nu}\,,}
\end{equation}
for some constant $\Lambda$ of energy dimension $[\Lambda]=E^2$, generally known as \textit{cosmological constant}. These field equations precisely correspond to the \textit{Einstein field equations} formulated by Einstein in 1915 \cite{Einstein:1915EE,Einstein:1916GrundlagenGR}.

For a Riemannian manifold without a boundary\footnote{For spacetime manifolds with boundaries, the variational problem of general relativity is not well posed \cite{Einstein:1915HamiltonP,Barth_1985} and counter terms need to be introduced. While there is no unique choice of counter terms, for non-null boundaries, there is a popular covariant counter term known as Gibbons-Hawking-York boundary term (see e.g. \cite{Hawking:1979ig}, that can be associated to the computation of BH entropy).} the Lovelock Theorem~\ref{Thm:LovelockTheorem} can also be formulated in terms of all possible diffeomorphism invariant and local metric scalars in the gravitational action $S_\text{G}$ that lead to field equations involving only up to two powers of derivatives. These are a simple constant $\Lambda$, the Ricci scalar $R$ as well as a combination known as \ul{Gauss-Bonnet curvature scalar}
\begin{equation}\label{eq:GBScalar}
    \mathcal{G}_{\myst{GB}}\equiv -\tilde R\ud{\mu\nu}{\rho\sigma} \tilde R\ud{\rho\sigma}{\mu\nu} =R^{\mu\nu\rho\sigma}R_{\mu\nu\rho\sigma}-4R^{\mu\nu}R_{\mu\nu}+R^2\,,
\end{equation}
where the Hodge dual of the Riemann tensor is defined as
\begin{equation}\label{eq:HodgeDualRiemannTensor}
    \tilde R\ud{\mu\nu}{\rho\sigma}\equiv\frac{1}{2}\epsilon^{\mu\nu\alpha\beta}R_{\alpha\beta\rho\sigma}\,,
\end{equation}
and where $\epsilon^{\mu\nu\alpha\beta}$ denotes the \textit{Levi-Civita tensor} of the physical metric that is related to the Levi-Civita symbol through the square root of the determinant (see e.g. \cite{carroll2019spacetime}).
In four dimensions, however, the Gauss-Bonnet scalar is a total derivative, in other words a purely topological term, that does not contribute to the equations of motion. Thus, the unique leading order gravitational action can be written as
\begin{equation}\label{eq:EinsteinHilbertAction}
        \boxed{S^{\myst{EH}}_G=\frac{1}{2\kappa_0}\int d^4x\sqrt{-g}\, \left[R-2\Lambda\right]\, ,}
\end{equation}
which corresponds to the \textit{Einstein-Hilbert action} \cite{Hilbert:1915A} that recovers the Einstein field equations [Eq.~\eqref{eq:EinsteinFieldEquations}] \cite{Weinberg1972}. For later use, note that while naively the action in Eq.~\eqref{eq:EinsteinHilbertAction} involves second-order derivatives of the metric, the corresponding equations of motion still remain at second-order in derivatives per field because the second derivatives in the Lagrangian appear linearly. In other words, upon a total derivative, the action can be rewritten in terms of purely first-order derivative terms.


\section{The Strong Equivalence Principle}\label{sSec:Strong Equivalence Principle}

The theory of general relativity also seems to be unique in the application of its equations of motion to the problem of motion of extended objects \cite{Einstein:1938yz,Damour:1986ny} (see also \cite{Hui:2009kc} for a nice review). Namely, it can be shown that in the limit in which tidal effects are negligible, the motion of extended, self-gravitating objects in an external gravitational field move just like point-like test-particles. In other words, up to tidal effects, all objects, regardless of their internal structure move along geodesics of the physical metric, a statement known as the \textit{effacement principle} of GR.

In the context of equivalence principles discussed in Se.~\ref{sSec:Equivalence Principle}, this result naturally leads to the conjecture that in GR, the Einstein equivalence Principle~\ref{Principle:EEP} that holds for all non-gravitational experiments can locally be extended to self-gravitating bodies and experiments involving gravitational forces. In other words, in a local freely falling frame in which inhomogeneities of external gravitational fields can be neglected, but that is big enough to encompass an extended system of gravitating matter along with its associated gravitational fields, the influence of the external gravitational field cannot be measured in any way. This statement is known as the \textit{strong equivalence principle} (SEP) (see e.g. \cite{poisson2014gravity,Will:2014kxa,Will:2018bme,Jetzer:2022bme}).

Arguably, GR is the only metric theory, in which the SEP holds \cite{Hui:2009kc,poisson2014gravity,Will:2014kxa,Will:2018bme}.\footnote{The SEP also holds in Nordstrom's, experimentally falsified, theory of gravity in which the gravitational field is associated to a single scalar field \cite{Deruelle:2011wu}, that is however not a metric theory of gravity.} This statement resides on the consideration that the presence of any form of non-minimal fields, which influence the way in which matter generates the gravitational field given by the physical metric, inevitably spoil the SEP as soon as the gravitational binding energy becomes non-negligible. This general behavior is known as the \textit{Nordtvedt effect} \cite{Nordtvedt:1968first,Nordtvedt:1968qs}.

In many local probes, in particular classical Cavendish experiments, such an associated violation of the SEP is however unobservably small due to the negligible fraction of gravitational binding energy \cite{Will:2018bme}. However, for larger systems, the effect can be tested for and, up to screening effects discussed in Sec.~\ref{ssSec:Screening} below, is able to put bounds on the space of metric theories beyond GR \cite{Will:2014kxa,Will:2018bme}. Concretely, such tests of the Nordtvedt effect can be carried out for example through lunar laser ranging of the earth moon system \cite{Williams:2005rv} (see also \cite{Hui:2012jb,Sakstein:2017bws} for test of the SEP through astrophysical black holes).

\section{Uniqueness in Dynamical Perturbations}\label{sSec:Uniqueness in Propagating DOFs}

Finally, as a smooth transition to the next chapter, we want to already anticipate the upcoming discussion and state that:
Up to perturbative corrections discussed in Sec.~\ref{ssSec: B Perturbative Theories}, the uniqueness of GR manifests itself in the statement that general relativity is the only dynamical metric theory of gravity with merely two propagating tensor degrees of freedom. This statement immediately requires a definition of the notion of ``propagating degrees of freedom'' which, especially in the context of gravity, is a subtle but crucial point that we will address in the next Chapter~\ref{Sec:PropagatingDOFs}. It will be worthwhile to spend some time in carefully understanding this concept, as the number of propagating degrees of freedom is an excellent tool to classify dynamical metric theories in Chapter~\ref{Sec:The Theory Space Beyond GR} and most importantly, the uniqueness of GR in that respect will allow for the formulation of clear-cut smoking gun signals for beyond GR effects that we will study in Part.~\ref{Part: Gravitational Wave Testing Ground}. 

Furthermore, the statement of uniqueness in terms of propagating degrees of freedom is closely related to uniqueness theorems of GR that can be formulated within a quantum effective field theory approach that we will discuss in Part~\ref{Part: Quantum Gravity}. Indeed, upon quantization, propagating degrees of freedom of tensor fields give rise to the notion of particles with spins depending on the behavior of the field under spacial rotations (see e.g. \cite{carroll2019spacetime}). According to Wigners little group classification \cite{Bargmann:1948ck} particle states are associated to unitary representations of the Poincare group through the irreducible representations of the stabilizer subgroup that leave a reduced form of the particle four-momentum invariant. In the massless case, the little group roughly corresponds to the group $SO(2)$ (see e.g. \cite{Weinberg:1995mt,Maggiore:2005qv,Schwartz:2014sze}) whose irreducible representations are all one dimensional. In parity preserving theories, however, the states come in pairs related by parity that must correspond to the same particle, which therefore possesses two degrees of freedom.\footnote{From this viewpoint one can understand that a description of such a particle via a rank two Lorentz tensor inevitably requires the introduction of a gauge redundancy that in the classical picture below will correspond to the gauge freedom introduced by diffeomorphism invariance (see also \cite{Schwartz:2014sze}).}  In this context, GR naturally arises as the unique description of a Lorentz invariant massless spin 2 particle with two degrees of freedom \cite{Weinberg:1964ew,Weinberg:1965rz,Deser:1969wk,BOULWARE1975,PhysRev.96.1683,Feynman:1996kb,maggiore2008gravitational}.
This GR uniqueness result in the quantum formulation can formally be understood from the requirement of unitarity together with Lorentz invariance of the $S$-matrix of scattering theory \cite{Weinberg:1964ew,Weinberg:1965rz}, while also at the level of scattering amplitudes GR can be proven to be the unique gauge-invariant theory of interacting, massless, spin-2 particles with second-order equations of motion \cite{BOULWARE1975,Krasnov:2014eza,Rodina:2016jyz}.

However, the existing quantum description of gravity does not come without its own conceptual difficulties (see Chapter~\ref{Sec:Challenges of the Quantum EFT of Gravity} and also \cite{WaldBook,Weinberg:1988cp}). In this work, we choose to postpone any quantum considerations and their related unsolved problems until Part~\ref{Part: Quantum Gravity} and first analyze the dynamical waves of gravity and their subtleties in a purely classical context. This will be the subject of the next Chapter. 


\newpage
\thispagestyle{plain} 
\mbox{}


\chapter{Propagating Degrees of Freedom}\label{Sec:PropagatingDOFs}

The concept of dynamical or \textit{\gls{propagating degrees of freedom}} (DOFs), although fundamental, is rather subtle, and it seems hard to give a precise but still practical general definition. Ultimately, the number of dynamical DOFs is related to the amount of Cauchy data needed to evolve a system starting from a given initial condition. For a theory defined on a Riemannian manifold, such a counting of required initial data can formally be related to the counting of available constraints within the generalized Hamiltonian formalism \cite{Dirac:1950pj,Anderson:1951ta,Lee:1990nz} (see also \cite{Dirac:1964tt,Golovnev:2022rui}) that is based on an Arnowitt, Deser and Misner (ADM) decomposition of spacetime \cite{ADM:1959zz,Arnowitt:1962hi}. A Hamiltonian formulation is a priori convenient for the counting of degrees of freedom due to the associated first derivative order nature of the equations, which allows to relate one Cauchy condition for every independent canonical variable. Yet, such an analysis of the full non-linear theory is often rather complex. Moreover, the actual evolution of a realistic system can only be performed numerically (see also Sec.~\ref{ssSec:WellPosedness}).

The notion of propagating degrees of freedom becomes however approachable in the realm of perturbation theory.
Indeed, very generally physics is concerned with the description of the processes and interactions in nature which are highly complex. The tool of perturbation theory that allows the determination of approximate solutions as small departures from a simpler (usually more symmetrical) exact solution is therefore indispensable in many respects. In this context, the components of the perturbations of the various tensor fields of the theory can be used to identify and describe the dynamical degrees of freedom of a theory.

However, in general, the components of the perturbation fields do not only describe physical propagating DOFs. Rather, there also exist \textit{\gls{non-dynamical degrees of freedom}} that are fully constrained but still physical and pure gauge modes of \textit{\gls{unphysical degrees of freedom}} that drop out of any physical observable. First of all, unphysical degrees of freedom are present if there exist any redundancy in the description of the perturbations. These can typically be eliminated through gauge-fixing or through a reformulation in terms of gauge-invariant variables. 
The remaining physical perturbative variables satisfy a given set of equations of motion on the background solution that can be treated analytically. Heuristically, the type of these equations of motion then determines whether the corresponding physical solutions corresponds to propagating or non-dynamical DOFs, where we distinguish between \textit{Laplace-type} equations, whose general solution does not introduce any additional time dependence such that the solution is completely tied to its source, and equations of the \textit{wave-type}, where this is not the case and the solution can acquire a self-sufficient propagation that is in principle independent of the source. 
As the names suggest, the prototypical equations of motion of both types are the purely spacial Laplace equation and the standard wave equation. The independent oscillation modes of the associated wave-like solutions, known as \textit{\gls{polarization}}s of the waves in a given field theory, are then directly connected to the number of propagating DOFs of the field.

Note, however, that in the context of metric theories of gravity defined on a manifold, the notion of propagation inevitably requires the introduction of a particular foliation. It will therefore be eminent to carefully single out the physical, coordinate independent information of the phenomenon of propagating waves.
Indeed, the notion of propagating waves is fundamentally only well-defined if the amplitude of the wave is small compared to some general ``background’’, and must therefore be considered within a perturbative setting. An intuitive picture for this statement is provided by the propagating waves of some given property of a fluid medium (see also \cite{misner_gravitation_1973}). In that case, the concept of waves is only meaningful if the waves are not too large in amplitude compared to other (possibly static) variations of that property within the fluid medium defining the background. On the other hand, the concept of waves that naturally arises in any field theory often assumes an implicit background, such as in the case of electromagnetic waves, where in the simplest standard description one assumes a vanishing electromagnetic field as a vacuum background. However, such an assumption simply conceals the necessity of a perturbative treatment since in the general case of an arbitrary electromagnetic field in particular in the presence of sources, electromagnetic waves can again only properly be described as small perturbations on a certain background value of the field.\footnote{Another option would be to consider the concept of radiation as energy that is inevitably carried away to infinity from a localized source as discussed for instance in \cite{Jackson:1998nia,griffiths_2017,YunesColemanMiller:2021lky,DAmbrosio:2022clk}. We will come back to this point in Sec.~\ref{sSec:Asymptotic Flatness} below, and give a clear distinction between the notion of waves and radiation.}  Even more so in metric theories of gravity without any a priori background, an explicit perturbative formulation is inevitable for the description of spacetime waves.

Moreover, there is a second fundamental assumption on the concept of waves that is often not stated explicitly, which is the existence of a clear separation in scales of variations of the waves compared to the background. In many contexts of physics, there exists a natural static background, such that also this second assumption becomes obsolete. Yet again, metric theories of gravity do not offer a given static background.
In describing physical degrees of freedom in perturbation theory of metric theories of gravity, it will thus be crucial to clearly state both the assumptions of small perturbations together with a clear separation of scales of variation [Sec.~\ref{sSec:GWs in GR}]. 

In the formalism that we are about to introduce, that is based on the above two points, in principle an arbitrary background spacetime can and will be considered. Nevertheless, as we will show, the description of physical DOFs including the wave solutions can locally be drastically simplified. Namely, outside any highly varying matter source, the physical, gauge invariant modes of any metric theory of gravity can locally be analyzed in Riemann normal coordinates for the physical metric and classified according to their type of equation of motion they satisfy. Thus, in such a setup, the number of propagating degrees of freedom of a theory can in principle be defined through the number of independent gauge-invariant modes given as solutions to a wave equation on an arbitrary slowly varying background. Since any metric theory of gravity can be put in the setup described above, such a counting might in principle\footnote{The practicality of this approach in counting propagating DOFs in more complex situations still needs to be investigated.} provide an unambiguous analytic method of determining the number of dynamical DOFs.

This rather powerful method should however be clearly distinguished from the pragmatic practice of choosing a particular background solution, often taken to be a Minkowski or cosmological solution. While often being ideal for a pertinent description of dynamical degrees of freedom, the counting of the number of dynamical degrees of freedom cannot be performed on a given background solution as this may conceal the existence of additional in many cases unhealthy (in the sense of destabilizing) DOFs.

In the following, we will first introduce the formulation of perturbation theory of metric theories at the basis of the description of waves. Subsequently, we will start by analyzing the case of GR and carefully outline the assumptions that go in the description of the associated dynamical degrees of freedom. The approach is then generalized to arbitrary dynamical metric theories of gravity.

\section{Perturbation Theory}\label{sSec:PerturbationTheory}

Intuitively, perturbation theory of metric theories of gravity assumes the existence of an exact but in principle arbitrary solution $\bar{g}_{\mu\nu}$, $\bar{\Psi}$ and $\bar{\Psi}_\text{m}$ that solves the system of Eqs.~\eqref{eq:EOM MetricTheory}, \eqref{eq:EOM MetricTheoryAdditionalfields} and \eqref{eq:EOMMatterFieldEquations} in some chart
\begin{equation}\label{eq:Background Equations}
    \mathcal{G}_{\mu\nu}[\bar{g},\bar{\Psi}]=\kappa_0\,T_{\mu\nu}[\bar{g},\bar{\Psi}_\text{m}]\,,\quad \mathcal{J}[\bar{g},\bar{\Psi}]=0\,,\quad \mathcal{J}_\text{m}[\bar{g},\bar{\Psi}_\text{m}]=0\,,
\end{equation}
and considers small perturbations of that system in order to obtain approximate solutions
\begin{equation}\label{eq:PerturbationSplit}
    g_{\mu\nu}=\bar{g}_{\mu\nu}+\delta g_{\mu\nu}\,,\quad \Psi=\bar\Psi+\delta\Psi\,,\quad \Psi_\text{m}=\bar\Psi_\text{m}+\delta\Psi_\text{m}\,,
\end{equation}
to more complex situations that cannot be solved exactly. Here we assume that\footnote{Locally, we can always choose a coordinate system in which the diagonal elements of $\bar{g}_{\mu\nu}$ are of $\mathcal{O}(1)$, while we for simplicity assume that this is also the case for all other fields.}
\begin{equation}
    |\bar{g}_{\mu\nu}|\,,\; |\bar{\Psi}|\,,\; |\bar{\Psi}_\text{m}| \sim 1 \,,\quad
     |\delta g_{\mu\nu}|\,,\; |\delta\Psi|\,,\; |\delta\Psi_\text{m}| \ll 1\,,
\end{equation}
and that the perturbations can be chosen small enough, such an evolution of the perturbed system remains close to the original solution, known as dynamical stability. In this case, the field equations of the true perturbed spacetime can be expanded in powers of the perturbations
\begin{align}
    \sum_{i=1}^\infty\frac{1}{i!}\,\phantom{}_{\mys{(i)}}\mathcal{G}_{\mu\nu}[\delta g,\delta\Psi]&=\kappa_0\, \sum_{i=1}^\infty\frac{1}{i!}\,\phantom{}_{\mys{(i)}}T_{\mu\nu}[\delta g,\delta\Psi_\text{m}]\,,\label{eq:FEPert}\\
     \sum_{i=1}^\infty\frac{1}{i!}\,\phantom{}_{\mys{(i)}}\mathcal{J}[\delta g,\delta\Psi]&=0\,,\label{eq:FEPertNonMinimal}\\
    \sum_{i=1}^\infty\frac{1}{i!}\,\phantom{}_{\mys{(i)}}\mathcal{J}_\text{m}[\delta g,\delta\Psi_\text{m}]&=0\,,
\end{align}
where $\phantom{}_{\mys{(i)}}O$ denotes the $i$th order in the perturbative expansion of the operator $O$. Furthermore, we omit the explicit dependence on the exact solutions, and we used Eq.~\eqref{eq:Background Equations} for the zeroth order $\phantom{}_{\mys{(0)}}O=O$.

However, for a theory defined on a manifold we should be careful in particular in writing equations like the split in Eq.~\eqref{eq:PerturbationSplit} between an exact solution and a perturbation in terms of components of tensors, since as discussed in length in Appendix~\ref{App:DiffGeo} it is important that we compare tensors on the same manifold and at equivalent spacetime points. In fact, a more formally correct treatment of perturbation theory of space-time manifolds requires the introduction of a one-parameter family of spacetimes\footnote{The individual spacetime manifolds are diffeomorphic to each other such that alternatively one could also consider a one-parameter family of structure on a single manifold.} $(\M_\epsilon,\Psi(\epsilon))$, where now for brevity all fields on the manifold including the metric are grouped within $\Psi$, with the exact solution $(\M,\bar{\Psi})$ corresponding to $\epsilon=0$ \cite{Stewart:1974uz} (see also \cite{Mukhanov:1990me,WaldBook,carroll2019spacetime}). 

Here, the original manifold $\M$ and $\M_\epsilon$ are related by a one parameter family of diffeomorphisms $\Phi_\epsilon$. This map identifies each point $p$ on $\M$ with a point $q$ on $\M_\epsilon$. The exact one-parameter family of solutions (defined on $\M_\epsilon$) can then be replaced by an approximate linear solution through \cite{Stewart:1974uz}
\begin{equation}
    \Psi(\epsilon)= \Phi_{\epsilon*}\bar\Psi+\delta\Psi+\mathcal{O}(\epsilon^2)
\end{equation}
where $\Phi_{\epsilon*}$ defines the pushforward (see Appendix~\ref{sApp:DiffsAndLieDer}). 

Therefore, to first order in $\epsilon$, the perturbations can now correctly be defined as the difference on $\M_\epsilon$ between the new one-parameter family of solutions and the pushed-forward of the original exact solution\footnote{Here, the restriction to $\mathcal{O}(\epsilon)$ is implicit.}
\begin{equation}\label{eq:SplitPerturbationWithPushforward2}
    \delta\Psi\equiv \Psi(\epsilon) -\Phi_{\epsilon*}\bar\Psi\,,
\end{equation}
or equivalently on the original manifold $\M$ as the difference between the original solution $\bar\Psi$ and the pullback of $\Psi(\epsilon)$ 
\begin{equation}\label{eq:SplitPerturbationWithPushforward}
    \delta\Psi\equiv \Phi^*_{\epsilon}\Psi(\epsilon) -\bar\Psi\,.
\end{equation}
While both options are equivalent we will opt here to follow the simplified and clear treatment in \cite{carroll2019spacetime} and mainly consider the latter. Note that a given perturbative solution corresponds to some particular value of $\epsilon$, for which one can expand all quantities and equations in the order parameter $\epsilon$.

In choosing the identification of spacetime points between $\M$ and $\M_\epsilon$, there is an intrinsic freedom directly related to the diffeomorphism invariance of a theory. This difference between two maps $\Phi_\epsilon$ and $\tilde{\Phi}_\epsilon$ is captured to first order in $\epsilon$ by some small one parameter family of diffeomorphisms $\delta\Phi_\epsilon$ on the background manifold that is generated by an associated small vector field $\xi^\mu$ on $\M$, with $|\xi^\mu|={\cal{O}}(\epsilon)$, such that $\tilde{\Phi}_\epsilon=\Phi_\epsilon \circ \delta\Phi_\epsilon$ \cite{Stewart:1974uz,carroll2019spacetime}.\footnote{The parameter of the background diffeomorphism is the same as the parameter of the one-parameter family of solutions. This can be justified, by requiring that a change in the description of a small deviation from the original solution should be restricted to an equally small transformation. A more formal justification requires the additional structure introduced in \cite{Stewart:1974uz}.} The newly defined perturbation then reads
\begin{align}\label{eq:GaugeFreedomPush forward}
    \delta\tilde{\Psi}&=\tilde{\Phi}^*_{\epsilon}\Psi(\epsilon)-\bar\Psi=(\Phi_\epsilon \circ \delta\Phi_\epsilon)^*\Psi(\epsilon)-\bar\Psi=\delta\Phi_\epsilon^*(\Phi_\epsilon^*\Psi(\epsilon))-\bar\Psi\nonumber\\
    &=\delta\Phi_\epsilon^*\delta\Psi+\delta\Phi_\epsilon^*\bar\Psi-\bar\Psi\,,
\end{align}
where in the second-to-last equality we have used the fact the pullback of a composition of diffeomorphisms is given by the composition of the pullbacks in the opposite order and in the last equality we plugged in the relation Eq.~\eqref{eq:SplitPerturbationWithPushforward} using that the pullback of the sum of two tensors is the sum of the pullbacks. All expressions in the equations above are evaluated at some $p\in\M$. Furthermore, to first order in $\epsilon$ we have
\begin{equation}
    \delta\Phi_\epsilon^*\delta\Psi=\delta\Psi\,,
\end{equation}
such that the difference between the two perturbations up to $\mathcal{O}(\epsilon^2)$ is given by
\begin{equation}\label{eq:GaugeFreedomPull back}
    \boxed{\delta\tilde{\Psi}-\delta\Psi=\epsilon\left(\frac{\delta\Phi_\epsilon^*\bar\Psi-\bar\Psi}{\epsilon}\right)=\mathcal{L}_\xi\bar{\Psi}\,,}
\end{equation}
where in the last equality we have used Eq.~\eqref{eq:LieDerivativeDefCompA} and the order in $\epsilon$ of $\mathcal{L}_\xi\bar{\Psi}$ is implicit in the smallness of $\xi^\mu$. Thus, the gauge freedom to linear order in perturbations is entirely captured by a Lie derivative of the original exact solution.

This statement can alternatively also be understood in a less formal picture of small coordinate transformations, although in this case one has to pay attention at which point the fields are evaluated at. Indeed, the gauge freedom can be understood as the possibility of creating fake perturbations through infinitesimal coordinate transformations 
\begin{equation}\label{eq:infcoordtransf}
    x^\mu\rightarrow x'^\mu=x^\mu+\xi^\mu\,,
\end{equation}
with $|\xi^\mu|=\mathcal{O}(\epsilon)$. These fake perturbations are measured by the difference of the transformed value of the fields of the exact solution $\bar\Psi'$ and the original $\bar\Psi$ but evaluated at the same point $x'^\mu$ that corresponds to the equivalent of $q\in\M_\epsilon$ in the active picture of diffeomorphisms
\begin{equation}
    \bar\Psi'(x')-\bar\Psi(x')=-\mathcal{L}_\xi\bar{\Psi}+\mathcal{O}(\epsilon^2)\,,
\end{equation}
where here we have used Eq.~\eqref{eq:LieDerivativeDefCompCoords2A}. Note that this equation is however not entirely equivalent to Eq.~\eqref{eq:GaugeFreedomPull back}, as also indicated by the minus sign. This is because the components of a tensor field after a coordinate transformation in Eq.~\eqref{eq:infcoordtransf} are equivalent to the components of the pushforward and not the pullback with the identification 
\begin{equation}
    x'^\mu_p\leftrightarrow x^\mu_q=x^\mu_{\delta\Phi(p)}\,,
\end{equation}
as shown in Eq.~\eqref{eq:EquivalenceDiffCordA}.

In order to obtain the equivalent of Eq.~\eqref{eq:LieDerivativeDefCompCoords2A} in terms of coordinate transformations, one would need to consider the alternative transformation (see also \cite{carroll2019spacetime})
\begin{equation}\label{eq:infcoordtransf2}
    x^\mu\rightarrow \tilde x^\mu=x^\mu-\xi^\mu\,,
\end{equation}
such that one can define the difference between $\tilde{\bar\Psi}$ and $\bar\Psi$ at the original location $x^\mu$ corresponding to $p\in\M$
\begin{equation}
    \tilde{\bar\Psi}(x)-\bar\Psi(x)=\mathcal{L}_\xi\bar{\Psi}+\mathcal{O}(\epsilon^2)\,,
\end{equation}
using Eq.~\eqref{eq:LieDerivativeDefCompCoords2A} and where the transformed tensor components $\tilde{\bar\Psi}(x)$ are defined in Eq.~\eqref{eq:DefTildeTransform}.

In particular, according to the general formula in Eq.~\eqref{eq:FormulaLieDerivativeGeneral} the linearized gauge freedom in perturbations for metric perturbations therefore reads 
\begin{align}
    \delta g_{\mu\nu}\rightarrow\delta g_{\mu\nu}+\mathcal{L}_\xi \bar{g}_{\mu\nu}&=\delta g_{\mu\nu}+\xi^\alpha\bar\nabla_\alpha \bar{g}_{\mu\nu}+(\bar\nabla_\mu \xi^\alpha)\bar{g}_{\alpha\nu}+(\bar\nabla_\nu \xi^\alpha)\bar{g}_{\mu\alpha}\,,\nonumber\\
    &=\delta g_{\mu\nu}+2\bar\nabla_{(\mu}\xi_{\nu)}\,,\label{eq:GaugeTransformationMetric}
\end{align}
whereas for a vector field and a scalar field we generally have that
\begin{align}
    \delta\Psi^\mu&\rightarrow\delta\Psi^\mu+\xi^\alpha\bar\nabla_\alpha \bar\Psi^\mu-\bar\Psi^\alpha\bar\nabla_\alpha \xi^\mu\,,\label{eq:GaugeTransformationVector}\\
    \delta\Psi&\rightarrow\delta\Psi+ \xi^\alpha\bar\nabla_\alpha \bar\Psi\,.\label{eq:GaugeTransformationScalar}
\end{align}
The last expression makes it apparent that linearized gauge transformations do not simply correspond to infinitesimal coordinate transformations, as it is often erroneously stated, simply because also scalars are subject to the gauge freedom.

In many textbooks and reviews of general relativity, gravitational waves are introduced at this point within the framework of linearized perturbation theory on a fixed background solution. However, as discussed in the introduction to this chapter, this misses out on an additional crucial assumption o the definition of waves, namely the existence of a clear separation of typical scales of variation. Very importantly, such parametric separation of physical scales not only leads to a well-defined notion of dynamical waves propagating on a background, but technically also renders the split in Eq.~\eqref{eq:SplitPerturbationWithPushforward} meaningful in concrete physical scenarios. Indeed, while Eq.~\eqref{eq:SplitPerturbationWithPushforward} as a mathematical tool is uniquely determined for a given one-parameter family of spacetimes, a physical situation only depends on one specific value of the small parameter $\epsilon$, such that without any additional assumption on the scales of variation, such a separation cannot be determined by local physical measurements \cite{Flanagan:2005yc}. Furthermore, the additional assumption will also be at the root of a meaningful definition of the energy-momentum associated with gravitational waves.

For simplicity of exposure, we will now first introduce the treatment of propagating degrees of freedom associated to well-defined gravitational waves with such an additional assumption as pioneered by Isaacson in the case of pure GR. In a second step, the considerations are then generalized to arbitrary metric theories of gravity in Sec.~\ref{sSec:Isaacson and GWs Generalization Beyond GR}.

As an outlook for later chapters, accounting for a separation in typical scales of variation also introduces an additional small parameter that in particular implies that Eqs.~\eqref{eq:FEPert} can in principle not simply be solved order by order in the label $(i)$. This will result in the fact that a purely linear treatment of gravitational waves even in the flat background spacetime approximation is in principle erroneous, as one misses an additional non-negligible contribution that is known as the GW memory effect \cite{Christodoulou:1991cr,Heisenberg:2023prj} that will be the subject of Chapter~\ref{Sec:GWMemory}.

\section{Propagating DOFs of GR}\label{sSec:GWs in GR}

\subsection{The Isaacson Approach}\label{ssSec:IsaacsonInGR}

Here we offer a review of the arguments originally brought forth by Isaacson \cite{Isaacson_PhysRev.166.1263,Isaacson_PhysRev.166.1272} (see also \cite{misner_gravitation_1973,Flanagan:2005yc,maggiore2008gravitational}) within GR, with the subtle difference that we explicitly work in a perturbation theory framework exposed in Sec.~\ref{sSec:PerturbationTheory} and therefore assume the existence of a (known) exact solution $\bar{g}_{\mu\nu}$ to the Einstein equations. This difference will be crucial in the following when identifying the memory component arising in the equations in Part~\ref{Part: Gravitational Wave Testing Ground}.
Consider therefore general relativity, hence the metric theory without any additional non-minimal fields on top of the Lorentzian metric $g_{\mu\nu}$ that is governed by the Einstein equations Eq.~\eqref{eq:EinsteinFieldEquations}, where for simplicity we set any cosmological constant $\Lambda$ to zero
\begin{equation}\label{eq:EinsteinEquationsWithoutCC}
    R_{\mu\nu}-\frac{1}{2}g_{\mu\nu} R=\kappa_0\,T_{\mu\nu}\,.
\end{equation}

\paragraph{The Isaacson Assumptions.} 
As discussed, within the Isaacson approach the notion of gravitational waves propagating on a background spacetime is given physical meaning by making two central assumptions: 
\begin{enumerate}[(1)]
    \item an exact solution $\bar{g}_{\mu\nu}$ is perturbed as in Eq.~\eqref {eq:PerturbationSplit}
    \begin{equation}\label{eq:PerturbationIsaacsonSplitGR}
        g_{\mu\nu}=\bar{g}_{\mu\nu}+\delta g_{\mu\nu}\,;
    \end{equation}
    \item there exists a clear separation of characteristic scales of variation that allows to separate the metric into a slowly varying background component $L$ and a highly varying piece $H$ 
    \begin{equation}\label{eq:IsaacsonSplitGR}
        g_{\mu\nu}=g^L_{\mu\nu}+\delta g^H_{\mu\nu}\,.
    \end{equation}
\end{enumerate}

Here we should pause for a moment and elaborate on a natural and practical way to single out the slowly varying part of any expression by performing an average $\langle...\rangle$ over a spacetime region of interest, with averaging kernel of characteristic scale in between of $L$ and $H$. The slowly varying contribution of any operator $O$ is then given by
\begin{equation}
    \left[O\right]^L\equiv\langle O\rangle\,,
\end{equation}
while the corresponding highly varying part simply reads 
\begin{equation}
    \left[O\right]^H\equiv O-\langle O\rangle\,.
\end{equation} 
A concrete covariant definition of such an averaging was given for instance in \cite{Brill:1964zz}. However, for our purposes, the exact averaging scheme is not important, so long as it satisfies the following set of properties to leading order
\cite{Isaacson_PhysRev.166.1272,Brill:1964zz,misner_gravitation_1973,Flanagan:2005yc,Zalaletdinov:2004wd,maggiore2008gravitational,Stein:2010pn}:
\begin{enumerate}[(I)]
    \item the average of an odd number of highly varying quantities vanishes;
    \item total covariant background derivatives of tensors average out to zero;
    \item as a corollary of the above, integration by parts of covariant derivatives are allowed.
\end{enumerate}

Combining the two assumptions in Eqs.~\eqref{eq:PerturbationIsaacsonSplitGR} and \eqref{eq:IsaacsonSplitGR} while further assuming that the exact solution is either entirely static or slowly varying, hence $\bar{g}_{\mu\nu}=\bar{g}^L_{\mu\nu}$, we can use the average defined above to obtain a general split of the approximate solution of the metric of the form
\begin{equation}\label{eq:IsaacsonSplitGeneral}
    \boxed{g_{\mu\nu}=\bar{g}_{\mu\nu}+\delta g_{\mu\nu}^L+\delta g_{\mu\nu}^H\,,}
\end{equation}
where the perturbations of the exact solutions are split according to their scales of variation
\begin{equation}
    \delta g_{\mu\nu}=\delta g_{\mu\nu}^L+\delta g_{\mu\nu}^H\,,
\end{equation}
while the slowly varying background is composed out of 
\begin{equation}
    g_{\mu\nu}^L=\langle g_{\mu\nu}\rangle=\bar{g}_{\mu\nu}+\delta g_{\mu\nu}^L\,.
\end{equation}

Formulated in other words, the Isaacson approach is the assumption of a physical situation in which there exists a perturbation $\delta g^H_{\mu\nu}$ of the metric that possesses a clearly distinct typical scale of variation compared to a slowly evolving background $g^L_{\mu\nu}$. In this case only, the notion of \textit{\gls{gravitational waves}} is well-defined. But of course, as discussed in Sec.~\ref{sSec:PerturbationTheory} above, not all independent components of the perturbation fields have physical meaning and the true gravitational waves with associated propagating degrees of freedom are the propagating gauge invariant modes within $\delta g^H_{\mu\nu}$. We will come back to this point below.

Concretely, the assumed physical separation of scales can for instance be formulated in terms of characteristic frequency dependence of the fields, where in this case we demand a clear distinction
\begin{equation}\label{eq:SmallParam1GR}
    f_{L}\ll f_{H}\,,
\end{equation} 
between a slowly varying background of frequencies lower than $f_L$ and high-frequency perturbations of typical frequency $f_H$. In Eq.~\eqref{eq:IsaacsonSplitGR} a super- or subscript $L$ and $H$ then indicates the dependence of the field components on the low or high frequencies respectively. Alternatively, the separation of scales can also be given in terms of scales of \textit{spacial} variations $L_L$ and $L_H$, instead of the temporal variations. In this case, instead of Eq.~\eqref{eq:SmallParam1GR} one demands  
\begin{equation}\label{eq:SmallParam1GRLength}
    L_H\ll L_B\,,
\end{equation}
also known as a \textit{short-wave expansion}, where in this case the scale $L_H$ is associated with the characteristic wavelength of the wavelike perturbation. 

Observe that demanding $L_H\ll L_B$ is in principle distinct from the condition on the frequencies in Eq.~\eqref{eq:SmallParam1GR}, because, while $L_H$ and $f_{H}$ are naturally related through the dispersion relation of the high-frequency wave, this is a priori not the case for the variations $L_L$ and $f_L$ of the background. The notion of slow variations in time or in space at the level of the background are in principle unrelated. However, the two choices are still interchangeable in the sense that the conclusions remain the equivalent, with the only difference that the distinction between the slowly varying background and the wave perturbations is drawn at a different level. It is interesting to note that from the point of view of current gravitational wave detectors on Earth, it is actually the condition in the frequency scales Eq.~\eqref{eq:SmallParam1GR} that dominates the distinction between gravitational waves and the background \cite{maggiore2008gravitational}.

In general, it is however advisable to assume both Eq.~\eqref{eq:SmallParam1GR} and Eq.~\eqref{eq:SmallParam1GRLength}, as we will see below. But for simplicity of exposure in this manuscript we choose to work with the condition in frequencies Eq.~\eqref{eq:SmallParam1GR} in the derivation of the equations. Moreover, for definiteness, we assume that the amplitudes of the (physical) perturbations in particular the high frequency perturbations are of the order of some small parameter $\alpha\ll1$ compared to the background assumed to be of order ${\cal{O}}(1)$
\begin{equation}\label{eq:SmallParam2GR}
    |\delta g_{\mu\nu}| = {\cal{O}}(\alpha)\,.
\end{equation}
Thus, in conclusion, the assumptions underlying the Isaacson approach are the existence of two small parameters, namely:
\begin{enumerate}[(1)]
    \item the amplitude of the perturbations
    \begin{equation}\label{eq:SmallParameterGRA2}
        \alpha\ll 1\,;
    \end{equation}
    \item the ratio of characteristic scales of frequencies
    \begin{equation}\label{eq:SmallParameterGRA1}
       \frac{f_L}{f_H}\ll 1\,.
    \end{equation}
\end{enumerate}
The existence of the additional small parameter is the reason for which the equations of the perturbations in Eq.~\eqref{eq:FEPert} cannot blindly be solved order by order in perturbations. Indeed, a derivative operator acting on a low- or high-frequency field posses a distinct order of magnitude that needs to be taken into account \cite{Isaacson_PhysRev.166.1263,misner_gravitation_1973}
\begin{align}
    \partial g^L_{\mu\nu}\leq\mathcal{O}(f_L)\,,\label{eq:OrderDerivativeL}\\
    \partial \delta g^H_{\mu\nu}= \mathcal{O}(\alpha f_H)\,.\label{eq:OrderDerivativeH}
\end{align}

\paragraph{Gauge Freedom and High Frequency Perturbations as Lorentz Tensors.}

It is interesting to further analyze the gauge freedom given by coordinate transformations in the light of the additional Isaacson split in Eq.~\eqref{eq:IsaacsonSplitGR}. This will reveal the true power of the Isaacson assumption.  

Namely, the physical split between a slowly varying background and a high-frequency perturbation always allows finding a local coordinate system, in which the background is flat Minkowski space $g^L_{\mu\nu}\simeq \eta_{\mu\nu}$
on top of which we still have the high-frequency perturbations (see also \cite{Flanagan:2005yc}). More precisely, consider an expansion of a general coordinate transformation $x'^\mu(x)$ to second order in $x/L_L$ or $xf_L$ around a given point
\begin{equation}\label{eq:ExpandedCoordinateTransf}
    x'^\mu(x)\approx a^\mu +(\Lambda^{-1})\ud{\mu}{\nu}\,x^\mu+\xi^\mu(x)\,,
\end{equation}
with $\xi^\mu(x)\sim x^2$ that change the total metric through Eq.~\eqref{eq:GeneralCoordinateTransformation}. Given the presence of a parametric separation of physical scales one can then further restrict to transformations with $\xi^\mu(x)=\xi^\mu_L(x)$, hence transformations with the same characteristics as the slowly varying background. Without loss of generality, we assume that $|\Lambda|\lesssim 1$ with a similar condition on $|\xi|$ such that the background metric remains at $|g^L_{\mu\nu}|={\cal{O}}(1)$ and the transformation in Eq.~\eqref{eq:ExpandedCoordinateTransf} with $\xi^\mu(x)=\xi^\mu_L(x)$ does not affect the defining nature of $\delta g^H_{\mu\nu}$, hence its high-frequency and perturbative properties. On the other hand, within a small enough region, through Eq.~\eqref{eq:ExpandedCoordinateTransf} and by using the torsion and non-metricity freeness of the Levi-Civita connection, the chart can be chosen as the normal coordinates with respect to the background $g^L_{\mu\nu}$ for which the background metric indeed reduces to the Minkowski form up to second order in $x$, as discussed in Appendix~\ref{sApp: Normal Coordinates} (see also \cite{Flanagan:2005yc})
\begin{equation}\label{eq:BackgroundMInkowskiForm}
    \boxed{g^L_{\mu\nu}=\eta_{\mu\nu}+{\cal{O}}\left(\frac{x^2}{L_L^2},x^2f_L^2\right)\,.}
\end{equation}

Intuitively, it should be clear that in a physical situation in which there exists a parametric separation between a slowly moving background and a varying perturbation, one can find a small enough region of spacetime in which the background is static and shows negligible spacial variation but large enough to capture the dynamics of the highly varying perturbations. In such a region of spacetime one can then choose a coordinate system that is flat Minkowski space for the background on top of which we describe the high-frequency perturbations
\begin{equation}
    g_{\mu\nu}\simeq \eta_{\mu\nu}+\delta g^H_{\mu\nu}\,,
\end{equation}
We want to stress that, imperatively, such a formulation of normal coordinates for the background is only possible with the Isaacson assumption of a clear separation of scales. Without such an assumption, one could only obtain normal coordinates of the full metric $g_{\mu\nu}$ restricted to an in this case even smaller region about any spacetime point, on which also any highly varying component looks static and smooth.

Once such a normal coordinate system for the background is found, the high-frequency perturbations $\delta g^H_{\mu\nu}$ can be viewed as true Lorentz tensors of special relativity as defined in Sec.~\ref{sSec: Theories of Minkowski Spacetime} that under Poincar\'e transformations between different inertial frames given in Eq.~\eqref{eq:PoincareT} change through
\begin{equation}\label{eq:LorentzTransformationMetric}
    \delta g'^H_{\mu\nu}=\delta g^H_{\alpha\beta}\,\Lambda\ud{\alpha}{\mu}\Lambda\ud{\beta}{\nu}\,,
\end{equation} 
where $\Lambda\ud{\mu}{\nu}$ are Lorentz matrices satisfying
\begin{equation}
    \Lambda\ud{\alpha}{\mu}\Lambda\ud{\beta}{\nu}\eta_{\alpha\beta}=\eta_{\mu\nu}\,.
\end{equation}
Further, the gauge freedom of perturbations associated to infinitesimal coordinate transformations $x'^\mu=x^\mu+\xi^\mu_H(x)$ translate in a gauge freedom linearly only affecting the high-frequency perturbations according to Eq.~\eqref{eq:GaugeTransformationMetric} as
\begin{equation}
    \delta g^H_{\mu\nu}\rightarrow\delta g^H_{\mu\nu}-2\partial_{(\mu}\xi_{\nu)}\,.
\end{equation}

\paragraph{The Leading Order Equations of Motion.}

We now want to come back to the field equations of perturbations in Eq.~\eqref{eq:FEPert} which in the case of GR reads
\begin{align}\label{eq:FEPertGR}
    \sum_{i=1}^\infty\frac{1}{i!}\,\phantom{}_{\mys{(i)}}G_{\mu\nu}[\delta g^L,\delta g^H]=\kappa_0\, \delta T_{\mu\nu}\,,
\end{align}
and analyze their form to lowest order in our bivariate expansion.
For simplicity, we have here denoted the sum of perturbed energy momentum tensors as \begin{equation}
    \delta T_{\mu\nu}\equiv \sum_{i=1}^\infty\frac{1}{i!}\,\phantom{}_{\mys{(i)}}T_{\mu\nu}[\delta g^L,\delta g^H,\delta\Psi^L_\text{m},\delta\Psi^H_\text{m}]\,,
\end{equation}
where also the matter fields are split into their low-and high-frequency components.
For concreteness, we will take the typical amplitude of high-frequency perturbations to be
\begin{equation}\label{eq:SmallParam2GRH}
    |\delta g^H_{\mu\nu}| = {\cal{O}}(\alpha)\,,
\end{equation}
while, in principle, the low-frequency terms could be of a different amplitude that we will denote as
\begin{equation}\label{eq:SmallParam2GRL}
    |\delta g^L_{\mu\nu}| = {\cal{O}}(\beta)\,,
\end{equation}
with $\beta\lesssim \alpha$.

Given an explicit theory, in this case GR, the expansion on the left-hand side can be computed explicitly. For instance, the first order terms can be derived from the expression of the perturbed Riemann tensor
\begin{align}
    _{\mys{(1)}}R_{\mu\nu\rho\sigma}[\delta g]=&-\frac{1}{2}\Big(\bar{\nabla}_\sigma\bar\nabla_\mu \delta g_{\nu\rho}+\bar\nabla_\rho\bar\nabla_\nu \delta g_{\mu\sigma}-\bar\nabla_\sigma\bar\nabla_\nu \delta g_{\mu\rho}-\bar\nabla_\rho\bar\nabla_\mu \delta g_{\nu\sigma}\label{eq:RiemmanFirstOrder}\\
    &\;\;\;\quad+\bar R_{\mu\gamma\rho\sigma}[\bar{g}]\delta g\ud{\gamma}{\nu}-\bar R_{\nu\gamma\rho\sigma}[\bar{g}]\delta g\ud{\gamma}{\mu}\Big)\,.\nonumber
\end{align}
From this expression, we obtain the first order perturbation of the Ricci tensor
\begin{equation}\label{eq:RicciFirstOrder}
    \phantom{}_{\mys{(1)}}R_{\nu\sigma}[\delta g]=-\frac{1}{2}\bar{g}^{\mu\rho}\Big(\bar\nabla_\sigma\bar\nabla_\mu \delta g_{\nu\rho}+\bar\nabla_\rho\bar\nabla_\nu \delta g_{\mu\sigma}-\bar\nabla_\sigma\bar\nabla_\nu \delta g_{\mu\rho}-\bar\nabla_\rho\bar\nabla_\mu \delta g_{\nu\sigma}\Big)\,,
\end{equation}
and the Ricci scalar
\begin{equation}\label{eq:RicciScalarFirstOrder}
\begin{split}
    \phantom{}_{\mys{(1)}}R[\delta g]=\bar\nabla_\nu\bar\nabla_\sigma \delta g^{\nu\sigma}-\bar\nabla_\nu\bar\nabla^\nu \delta g\ud{\sigma}{\sigma}-\delta g^{\nu\sigma}\bar R_{\nu\sigma}\,.
\end{split}
\end{equation}
This can be used in order to estimate the size of the leading order operators in each term of Eq.~\eqref{eq:FEPertGR}. In particular, according to Eqs.~\eqref{eq:OrderDerivativeL}, \eqref{eq:OrderDerivativeH}, \eqref{eq:SmallParam2GRH} and \eqref{eq:SmallParam2GRL}, we have that
\begin{equation}\label{eq:OrderCountingGRLow}
    \phantom{}_{\mys{(1)}}G_{\mu\nu}[\delta g^L]={\cal{O}}(\beta f_L^2)\,,\quad  \phantom{}_{\mys{(1)}}G_{\mu\nu}[\delta g^H]={\cal{O}}(\alpha f_H^2)\,,
\end{equation}
due to the presence of two derivative operators in each term, either applying on a high- and low-frequency perturbation or on the exact background. As there is no other mass scale in the theory, dimensional analysis in fact forces this structure upon every operator in the expansion. Therefore, even without looking at the explicit expressions at higher order one deduces for the leading order behavior of any higher order component of order $i\geq 2$
\begin{align}
    \phantom{}_{\mys{(i)}}G_{\mu\nu}[\delta g^L,\delta g^H]&={\cal{O}}(\alpha^i f_H^2)\,.\label{eq:OrderCountingGRHigh1}
\end{align}

At this stage, it is important to realize, that the Isaacson split between a low and high-frequency part also allows to impose such a decomposition at the level of the equations. This can be viewed as performing a multiple-scale analysis of the physical problem at hand. Such a decomposition at the level of the equations can easily be obtained via a space-time average $\langle...\rangle$ introduced above. Recall that by definition, such an average over an operator that linearly depends on a high-frequency component vanishes. However, averaging over the entire series in Eq.~\eqref{eq:FEPertGR} to obtain a low- and high-frequency leading order equation leads to the following key observation: Already at second order in perturbation fields, the average $\langle\phantom{}_{\mys{(2)}}G_{\mu\nu}[\delta g^H]\rangle$ will contain contributions both at the level of $f_H$, as well as at the background scales $f_L$, because two high-wave-vector modes can combine to form a low-frequency contribution. 

Therefore, while at the high-frequency level, the leading order equation up to ${\cal{O}}(\alpha^2 f_H^2)$ clearly reads
\begin{equation}\label{eq:EOMIISGR}
    \boxed{\phantom{}_{\mys{(1)}}G_{\mu\nu}[\delta g^H]=\kappa_0\left[\delta T_{\mu\nu}\right]^H\,,}
\end{equation}
at the low-frequency level we have instead
\begin{equation}\label{eq:EOMISGR}
    \boxed{\phantom{}_{\mys{(1)}}G_{\mu\nu}[\delta g^L]=-\frac{1}{2}\,\big\langle\phantom{}_{\mys{(2)}}G_{\mu\nu}[\delta g^H]\big\rangle+\kappa_0\big\langle\delta T_{\mu\nu}\big\rangle\,,}
\end{equation}
up to corrections of order ${\cal{O}}(\alpha^3 f_H^2)$ and ${\cal{O}}(\beta^2 f_L^2)$. Observe that this last equation relates the small scale $\beta$ to the original two small quantities defined in Eqs.~\eqref{eq:SmallParameterGRA1} and \eqref{eq:SmallParameterGRA2}. Concretely, in the absence of any matter perturbations, the scale of $\beta$ is determined through Eq.~\eqref{eq:EOMISGR} to be
\begin{equation}\label{eq:BetaParam}
    \boxed{\beta\sim \alpha^2\frac{f_H^2}{f_L^2}\,,}
\end{equation}
In this case, the requirement that $\beta \ll 1$ imposes a hierarchy between the two initial expansion parameters, namely\footnote{The original work by Isaacson \cite{Isaacson_PhysRev.166.1263} explicitly only considers the situation in which $\alpha\sim f_L/f_H$, reflecting the fact that the exact solution on top of which perturbations are defined was not properly subtracted.}
\begin{equation}
\label{eq:hierarchy-inequ}
    \alpha\ll \frac{f_L}{f_H}\,.
\end{equation}

Here, in the absence of matter perturbations the leading-order high-frequency equation [Eq.~\eqref{eq:EOMIISGR}] can be interpreted as a propagation equation for the leading-order gravitational waves $\delta g^H$. Including the matter perturbations promotes the equation to a sourced equation for the gravitational waves. On the other hand, Eq.~\eqref{eq:EOMISGR}, hence the leading-order, low-frequency equation can be viewed as a backreaction of the coarse-grained operator $\langle \phantom{}_{\mys{(2)}}G_{\mu\nu}[\delta g^H]\rangle$ of high frequency waves that gives rise to a perturbation of the background spacetime $g^L_{\mu\nu}$. Quite naturally, the right-hand side of Eq.~\eqref{eq:EOMISGR} can therefore be interpreted as the energy-momentum (pseudo)tensor of gravitational waves \cite{Isaacson_PhysRev.166.1263,Isaacson_PhysRev.166.1272,misner_gravitation_1973,Flanagan:2005yc,maggiore2008gravitational}
\begin{equation}\label{eq:DefPseudoEMTensorGR}
    \boxed{\phantom{}_{\mys{(2)}}t^{\myst{GR}}_{\mu\nu}[\delta g^H]\equiv -\frac{1}{2\kappa_0}\big\langle \phantom{}_{\mys{(2)}}G_{\mu\nu}[\delta g^H]\big\rangle\,.}
\end{equation}
Decisively, under the Isaacson assumptions, this expression, as well as all other terms in the leading order expansion, is gauge invariant under infinitesimal high-frequency coordinate transformations discussed above, up to higher order terms \cite{Isaacson_PhysRev.166.1263,maggiore2008gravitational}. Moreover, since covariant derivation and the average commute, the pseudotensor of GW energy is also covariantly conserved
\begin{equation}
    \bar\nabla^\mu\phantom{}_{\mys{(2)}}t^{\myst{GR}}_{\mu\nu}=0\,.
\end{equation}
Observe that this is true regardless of the form of the associated gravitational equations of motion. In the light of the statements back in Sec.~\ref{sSec:Covariant Consrevation} on the difficulty of defining a local notion of energy-momentum of the gravitational field, the Isaacson approach represents a way to precisely achieve this. Intuitively, in a situation where the notion of gravitational waves makes sense, a localized and physical energy-momentum content of the gravitational waves influencing the background spacetime can be defined through a coarse-graining over the small-scale details \cite{maggiore2008gravitational}.

In this section, now want to focus  on the propagation equation and the extraction of the propagating degrees of freedom. We will come back to the equally interesting low-frequency back-reaction equation in Part~\ref{Part: Gravitational Wave Testing Ground} of this manuscript.

\subsection{Local Wave Equation}\label{ssSec:Local Wave Equation in GR}

In Section~\ref{ssSec:IsaacsonInGR} above we showed that within an Isaacson approach to gravitational waves in GR, locally, we can \textit{always} choose coordinates in which the background spacetime is flat Minkowski spacetime that can be put in the Minkowski form for an inertial observer
\begin{equation}
    g^L_{\mu\nu}\simeq \eta_{\mu\nu}\,,
\end{equation}
whereas the high-frequency perturbations, that we will now denote as
\begin{equation}
    \delta g^H_{\mu\nu}=h_{\mu\nu}\,,
\end{equation}
define proper Lorentz tensors on that background. Moreover, the high-frequency perturbations admit a gauge freedom of the form
\begin{equation}
    h_{\mu\nu}\rightarrow h_{\mu\nu}+\mathcal{L}_{\xi^H} \eta_{\mu\nu}= h_{\mu\nu}+\partial_\mu\xi^H_\nu+\partial_\nu\xi^H_\mu\,,
\end{equation}
where $\xi^H$ defines an infinitesimal high-frequency transformation $\tilde{x}^\mu=x^\mu-\xi^\mu_H(x)$ of the Minkowski coordinates.

With all properties of a symmetric Lorentz tensor satisfied, it is tempting to analyze this perturbation field in analogy to the vector potential familiar in electrodynamics. From the one-dimensional $U(1)$ gauge invariance of the vector potential, we can therefore also expect that the four unphysical gauge artifacts are supplemented by four additional constraints from the equations of motion that impose four of the components to be non-dynamical, reducing the a priori ten degrees of freedom within the symmetric $h_{\mu\nu}$ down to the group-theoretically expected two propagating DOFs.

Indeed, it is a standard exercise to show this explicitly. As we will discuss in more detail in Part~\ref{Part: Gravitational Wave Testing Ground}, a neat analytic understanding of the Isaacson system above is provided by considering the perturbed action of the system. In this case, the relevant quantity is given by the second order Einstein-Hilbert gravitational action defined in Eq.~\eqref{eq:EinsteinHilbertAction} with vanishing cosmological constant, which can be written as
\begin{equation}\label{ActionGR2nd}
    _{\mys{(2)}}S^{\myst{EH}}_G=\frac{-1}{2\kappa_0}\int d^4x\,h^{\mu\nu}\mathcal{E}^{\alpha\beta}_{\mu\nu}h_{\alpha\beta}\,,
\end{equation}
where $\mathcal{E}_{\mu \nu}^{\alpha \beta}$ stands for the flat-space Lichnerowicz operator
\begin{equation}\label{eq:LichnerowiczOperator}
    \mathcal{E}^{\alpha\beta}_{\mu\nu}h_{\alpha\beta}\equiv-\frac{1}{4}\Big[\Box h_{\mu\nu}-2\partial_\alpha\partial_{(\mu}h\du{\nu)}{\alpha}+\partial_\mu\partial_\nu h^t-\eta_{\mu\nu}\left(\Box h^t-\partial_\alpha\partial_\beta h^{\alpha\beta}\right)\Big]\,,
\end{equation}
with $h^t\equiv \eta^{\mu\nu}h_{\mu\nu}$.
This operator in particular allows for a compact notation of the Fierz-Pauli Lagrangian, which up to integration by parts recovers the usual Fierz-Pauli combination
\begin{equation}
    h^{\mu\nu}\mathcal{E}^{\alpha\beta}_{\mu\nu}h_{\alpha\beta}\leftrightarrow\frac{1}{4}\Big[\partial_\mu h_{\alpha\beta}\partial^\mu h^{\alpha\beta}-\partial_\mu h^t\partial^\mu h^t+2\partial_\mu h^{\mu\nu}\partial_\nu h^t -2\partial_\mu h^{\mu\nu}\partial_\alpha h\ud{\alpha}{\nu}\Big]\,.
\end{equation}

In terms of the Lichnerowicz operator, the high-frequency propagation equation [Eq.~\eqref{eq:EOMIISGR}] becomes
\begin{equation}\label{eq:LinGRprop1}
    \mathcal{E}^{\alpha\beta}_{\mu\nu}h_{\alpha\beta} = \frac{\kappa_0}{2}\left[\delta T_{\mu\nu}\right]^H\,.
\end{equation}
At this stage we use the gauge freedom to choose the so-called \textit{harmonic gauge}
\begin{equation}
    \partial^\nu \bar{h}_{\mu\nu}=0\,,
\end{equation}
where we have defined
\begin{equation}\label{eq:HarmonicGaugeVariables}
    \bar{h}_{\mu\nu}\equiv h_{\mu\nu}-\frac{1}{2}\eta_{\mu\nu}\,h^t\,.
\end{equation}
The unfortunate standard notation $\bar{h}_{\mu\nu}$ should not be confused with the exact solution of the starting point of our perturbation approach. It is straightforward to show that such a gauge can always be chosen (see e.g. \cite{Weinberg1972,Flanagan:2005yc,maggiore2008gravitational}). Recall that such a gauge-fixing is required in order to get rid of the four unphysical degrees of freedom within the perturbation field associated to the four degrees of freedom in the gauge transformation in Eq.~\eqref{eq:GaugeTransformationMetric}.
For this choice, the six remaining independent components within $\bar{h}_{\mu\nu}$ satisfy a sourced wave equation, since Eq.~\eqref{eq:LinGRprop1} becomes
\begin{equation}\label{eq:LinGRprop2}
    \boxed{\Box \bar{h}_{\mu\nu}=-2\kappa_0 \left[\delta T_{\mu\nu}\right]^H\,.}
\end{equation}
This equation can be solved by standard Green's function methods as discussed in Sec.~\ref{sSec:GWGeneration}.

However, as we will explicitly see below, this by no means should be interpreted as six propagating degrees of freedom within GR. Indeed, the present approach obscures the distinction between non-dynamical and propagating DOFs (see Sec.~\ref{ssSec:GaugeInvariantDecomposition}) since there is a residual gauge freedom given by transformations that satisfy 
\begin{equation}
    \Box\xi^\mu_H=0\,,
\end{equation}
under which $\bar{h}_{\mu\nu}$ is not invariant. 
At this stage, in order to determine the propagating DOFs, the equations of motion of the specific theory need to be employed. However, for the sourced equation above, in general one cannot use the residual gauge freedom to explicitly set to zero components in the high-frequency perturbation. Within the present approach, this is only possible for components in $\bar{h}_{\mu\nu}$ satisfying Eq.~\eqref{eq:LinGRpropNoSource} outside any source \cite{maggiore2008gravitational}. To single out the true propagating degrees of freedom, we are therefore interested to consider regions in spacetime without any high-frequency matter source. Again, we will come back to the sourced equation in Part~\ref{Part: Gravitational Wave Testing Ground}.

Outside of any high-frequency content in the matter, a general solution to 
\begin{equation}\label{eq:LinGRpropNoSource}
    \Box \bar{h}_{\mu\nu}=0\,,
\end{equation}
is given by a superposition of plane waves.
For such solutions, the residual gauge freedom together with the equations of motion can completely fix the gauge by choosing the \textit{transverse-traceless} (TT) gauge, defined by \cite{maggiore2008gravitational,Flanagan:2005yc}
\begin{equation}\label{eq:TTgauge}
    h^{TT}_{0\nu}=0\,,\quad \delta^{ij}h^{TT}_{ij}=0 \,,\quad \partial^i h^{TT}_{ij}=0\,.
\end{equation}
Note that in this case 
\begin{equation}
    \bar{h}^{TT}_{\mu\nu}=h^{TT}_{\mu\nu}\,.
\end{equation}

We are thus left with two propagating degrees of freedom within $h^{TT}_{ij}$ defined as independent solutions of a propagating wave equation in the completely fixed TT gauge. In this sense, the Isaacson approach provides a well-defined method to analyze the propagating degrees of freedom of a theory, in this case GR, without loss of generality within this very convenient setup of a locally Minkowski background.

At this point, one might however wonder how the TT gauge-choice is justified and whether other possibilities would be permissible. Indeed, in that respect the direct approach taken above is rather unsatisfactory and as already mentioned is unable to draw a clear distinction between the unphysical gauge degrees of freedom and the physical but non-dynamical components within the tensor perturbations. In the end, only gauge-invariant modes can appear in physical observables, such as for instance in the response of a GW detector that we will treat in Part~\ref{Part: Gravitational Wave Testing Ground}. For this reason, as well as computational grounds in more complex multifield metric theories, it is of great value to make use of the symmetries of the local background and decompose metric perturbations into manifestly gauge invariant quantities.

\subsection{Gauge-Invariant Scalar-Vector-Tensor Decomposition}\label{ssSec:GaugeInvariantDecomposition}

A particularly useful way of investigating the number of physical degrees of freedom contained within metric perturbations was introduced by Bardeen in the context of cosmological perturbations \cite{Bardeen:1980kt} (see also \cite{Mukhanov:1990me,Bertschinger:1993xt,Flanagan:2005yc,poisson2014gravity,carroll2019spacetime}), which allows the direct identification of gauge invariant and thus physical components. To focus on an analysis of degrees of freedom stripped of any unnecessary clutter, we will again assume a coordinate system in which the slowly-varying background reduced to the Minkowski form on which we consider high-frequency perturbations $h_{\mu\nu}$ as in Sec.~\ref{ssSec:Local Wave Equation in GR} above. 

\paragraph{SVT Decomposition.} The starting point is an irreducible decomposition of the high-frequency perturbations according to their transformation properties under $SO(3)$ rotations (see \cite{Szapudi:2011iz} for a group theoretic account of cosmological perturbations)\footnote{It is interesting to note that an earlier gauge dependent approaches to cosmological perturbations by Lifshitz and Khalatnikov \cite{Lifshitz:1963ps} used an $SO(3)$ decomposition as well but only after first choosing a gauge in which perturbations were restricted to the spacial domain. When discussing GW polarizations within generic metric theories in Sec.~\ref{sSec:GWPolGen} we will again encounter both of these approaches.}

\begin{equation}\label{eq:hexpand}
h_{\mu\nu}=
    \left(\begin{array}{c|c c c} 
    	h_{00} &  & h_{0i}&\phantom{0} \\
    	\hline 
    	 &  &&\\
h_{i0} &  & h_{ii}& \\
 \phantom{0} &  & & \\
    \end{array}\right)
=
 \left(\begin{array}{c|c c c} 
    	S &  &V_i&\phantom{0} \\
    	\hline 
    	 &  &&\\
V_i &  & T\delta_{ij}+U_{ij}& \\
 \phantom{0} &  & & \\
    \end{array}\right)\,,
\end{equation}
Here $S$ and $T$ are $SO(3)$ scalars, $V_i$ is an $SO(3)$ vector and $U_{ij}$ is a traceless symmetric $SO(3)$ tensor, representing the $l=0,\,,1$ and $2$ irreducible representations (irreps) of $SO(3)$ of dimension $2l+1$. Note that this is an algebraic decomposition applicable to any symmetric tensor within a fixed inertial coordinate system of the background. Other inertial coordinate systems are related through Lorentz transformations.

To extract gauge invariant perturbations, however, a further decomposition of the $SO(3)$ irreps into the one dimensional irreps of the subgroup $SO(2)$ is required. These representations are labeled by an integer $\abs{m}\leq l$ and represent a rotation around an arbitrarily chosen direction. For a parity preserving theory, the two representations with $\pm m$ for $l>0$ are however grouped together. Indeed, without loss of generality (see e.g. \cite{poisson2014gravity}), any $SO(3)$ vector $V_i$ can further be Helmholtz-decomposed into its divergence-less and curl free parts
\begin{align}\label{eq:hexpand2}
V_i = V^T_i+\partial_i V^\parallel \,,
\end{align}
where $V^T_i$ is transverse, in the sense that
\begin{equation}
    \partial^i V^T_i=0\,.
\end{equation}
Similarly any symmetric-traceless tensor field $U_{ij}$ can uniquely be decomposed into a transverse-traceless, solenoidal and longitudinal part
\begin{align}\label{eq:hexpand3}
U_{ij}=h^{TT}_{ij}+2\partial_{(i} U^T_{j)}+\left(\partial_i\partial_j-\frac{1}{3}\delta_{ij}\Delta\right)U^\parallel\,,
\end{align}
where
\begin{equation}
    \partial^i h^{TT}_{ij}=0\,,\quad\delta^{ij} h^{TT}_{ij}=0\,,\quad\partial^i U^T_i=0
\end{equation}
where the notation $h^{TT}_{ij}$ should already ring a bell. In a specific normal background chart, we therefore uniquely decomposed the original ten-degrees of freedom of the high-frequency metric perturbations $h_{\mu\nu}$ into four scalars ($S$, $T$, $V^\parallel$ and $U^\parallel$) with $m=0$, two two-dimensional transverse spacial vectors ($V^T_i$ and $U^T_i$) with $m=\pm 1$ and a two-dimensional transverse and traceless tensor $h^{TT}_{ij}$ with $m=\pm 2$. Such a decomposition is therefore known as a scalar-vector-tensor (SVT) decomposition. Note that here the terminology ``scalar'', ``vector'' and ``tensor'' now refers to the behavior of the field perturbations under $SO(2)$ around an arbitrarily chosen direction as discussed above. The components are then longitudinal, respectively transverse with respect to said chosen direction.

We want to stress that the decomposition above is completely general. Indeed, such a decomposition into scalar, vector and tensor perturbations could be performed through an ADM analysis on a completely general background, in particular also in the case of an anisotropic background as shown for example in \cite{Pereira:2007yy}. However, if the background is invariant under $SO(3)$ spacial rotations as it is obviously the case for Minkowski spacetime the $SO(2)$ scalar, vector and tensor perturbations naturally decouple to linear order in any equation, which represents the main advantage of the approach.
Moreover, note that the further Helmholtz decomposition into the $SO(2)$ subgroup is not algebraic anymore and is actually only local in the Fourier domain \cite{Flanagan:2005yc,poisson2014gravity,carroll2019spacetime}. In other words, it is only well-defined for tensor fields defined on more than a single spacetime point. We will come back to that observation below.
 
\paragraph{Gauge Invariant Variables.} We are now in the position to discuss the gauge freedom in the metric perturbation given by Eq.~\eqref{eq:GaugeTransformationMetric}. 
\begin{equation}\label{GaugeTransf}
h_{\mu\nu}\rightarrow  h_{\mu\nu}+ \mathcal{L}_{\xi^H}\eta_{\mu\nu}= h_{\mu\nu}+ 2\partial_{(\mu}\xi^H_{\nu)}\,.
\end{equation}
Helmholtz-decomposing the vector $\xi^H_\mu$ as well into $\xi_0$ and $\xi_i=\xi^T_i+\partial_i \xi^\parallel$ one can verify that under the gauge freedom, the individual components transform as
\begin{subequations}
\begin{align}
S&\rightarrow S+ 2 \dot{\xi}_0 \,,& T&\rightarrow T+\frac{2}{3}\Delta \xi^\parallel \,,\\
 V^\parallel&\rightarrow V^\parallel +\xi_0+\dot{\xi}^\parallel\,, &  U^\parallel&\rightarrow U^\parallel+ 2 \xi^\parallel\,,\\
V^T_i&\rightarrow V^T_i + \dot{\xi}^T_i\,, & U^T_i&\rightarrow U^T_i+\xi^T_i\\
 h^{TT}_{ij}&\rightarrow h^{TT}_{ij}\,, & &
\end{align}
\end{subequations}
where $\Delta\equiv \partial_i\partial^i$ denotes the Laplace operator. Observe that the transverse-traceless part $h^{TT}_{ij}$ is already invariant. Moreover, one can define the additional gauge invariant quantities
\begin{subequations}\label{eq:gaugeInvariantMetricPerts1}
\begin{align}
\delta\Phi&\equiv S-2\dot{V}^\parallel+\ddot{U}^\parallel\\
\delta\Theta&\equiv T-\frac{1}{3}\Delta U^\parallel\\
\delta\Xi^T_i&\equiv V^T_i-\dot{U}^T_i\,.
\end{align}
\end{subequations}
Hence, as expected, four components in $h_{\mu\nu}$ are pure gauge artifacts, whereas there exist six physical gauge invariant degrees of freedom: two scalars one transverse vector and one transverse-traceless tensor.

As mentioned, at this point, it is the dynamics of a specific theory that decides how many and which of the physical DOFs are actually propagating. In the case of GR, the high frequency propagation equations [Eq.\eqref{eq:EOMIISGR}] in the absence of high-frequency sources
\begin{equation}
     \phantom{}_{\mys{(1)}}G_{\mu\nu}[\delta g^H]=0\,,
\end{equation}
can be reduced to the following form
\begin{equation}\label{linEE}
\Delta\delta\Phi=0\;\,,\quad \Delta\delta\Theta=0\;\,,\quad \Delta\delta\Xi_i=0\;\,,\quad \Box h^{TT}_{ij}=0\,.
\end{equation}
The explicit expressions in the presence of matter sources can for instance be found in \cite{Flanagan:2005yc}.

Thus, we immediately observe that only two of the six gauge invariant modes are true dynamical degrees of freedom satisfying a propagation equation. The four remaining DOFs satisfy a constraining Laplace equation lacking a time derivative, such that they are completely determined by boundary conditions as well as the energy-momentum content in the more general case. More precisely, only the spacial transverse-traceless part of the metric perturbations correspond therefore to what we defined as propagating degrees of freedom of the theory: the gauge-invariant perturbations that solve a wave equation. Thus, in GR, fundamentally only the transverse-traceless piece of the metric is dynamical, which justifies in retrospective the TT gauge introduced in Eq.~\eqref{eq:TTgauge}. 

\paragraph{Comments on the Different Approaches.} We want to stress again that the determination of propagating degrees of freedom given here crucially depends on the Isaacson approach that allows the description of high-frequency perturbations on a local Minkowski patch obtained through choosing Riemann normal coordinates on a completely arbitrary manifold. Note the difference between this approach and choosing a fixed Minkowski background from the start, in which case one could not claim a general result on the counting of propagating DOFs as the result must hold on arbitrary backgrounds. Within the example of GR above, it so happens that these two approaches coincide, but this will not be the case for more general metric theories. 

Moreover, we also want to contrast the manifestly gauge invariant identification of the propagating DOFs of this section with the TT gauge employed in Sec.~\ref{ssSec:Local Wave Equation in GR} within the manifestly local approach.
While in general the non-dynamical metric degrees of freedom need not vanish, in the absence of any source, the Laplace equations for well-behaved boundary conditions can be solved by
\begin{equation}
    \delta\Phi=\delta\Theta=\delta\Xi_i=0\,,
\end{equation}
such that only the TT modes remain non-zero. Such a solution would therefore precisely coincide with the TT gauge applied on the entire metric perturbations $h^{TT}_{\mu\nu}$ in Eq.~\eqref{eq:TTgauge}
\begin{equation}\label{eq:TTgaugeSEcond}
h^{TT}_{0\nu}=0\;\,,\quad \eta^{\mu\nu}h^{TT}_{\mu\nu}=0\;\,,\quad \partial^\mu h^{TT}_{\mu\nu}=0\,.
\end{equation} 
This coincides with the remark in Sec.~\ref{ssSec:Local Wave Equation in GR} the TT gauge is not valid in the presence of source terms.

Furthermore, despite the fact that the gauge invariant fields defined above are non-local in position space, the physical, in the sense of observable, information within the degrees of freedom of any metric theory are causal. This is explicit in the harmonic gauge variables defined in Eq.~\eqref{eq:HarmonicGaugeVariables}, which however lack the property of being gauge invariant. This trade-off in convenience of description is resolved by noting that the physical observables of gravity that we identified in Sec.~\eqref{sSec:Metric Theories} in the Riemann tensor arising in the geodesic deviation equation is both local and gauge-invariant. We will further elaborate on this in the context of perturbation theory in Sec.~\ref{sSec:GWObservation}. Thus, the metric components in any description we choose, be it manifestly local or gauge-invariant, must carry the physical and causal information relevant for observations. One could in fact also choose to exclusively work with manifestly gauge-invariant and local objects in terms of perturbations of curvature invariants only (see e.g.~ \cite{Koop_PhysRevD062002,Garfinkle:2022dnm}) which comes however with the drawback of higher complexity. In the end, all of these approaches are physically equivalent.

\section{The Generalization Beyond GR}\label{sSec:Isaacson and GWs Generalization Beyond GR}


We now want to generalize the above arguments and to more generic dynamical metric theories of gravity of Def.~\ref{DefMetricTheory}. The main steps will remain similar to the treatment within GR in Sec.~\ref{ssSec:IsaacsonInGR}, although we will highlight a few important technicalities. 
\subsection{The Generalized Isaacson Approach}\label{ssSec:IsaacsonGeneral}

Recall that a generic dynamical metric theory is defined by the existence of a physical metric $g_{\mu\nu}$ that couples minimally to matter and a set of additional dynamical non-minimal fields, which we will collectively refer to as $\Psi$. The Einstein equations are generalized to Eq.~\eqref{eq:EOM MetricTheory}
\begin{equation}
    \mathcal{G}_{\mu\nu}=\kappa_0 T_{\mu\nu}\,,
\end{equation}
while there are additional field equations for each dynamical non-minimal field that are schematically grouped in Eq.~\eqref{eq:EOM MetricTheoryAdditionalfields}
\begin{align}
    \mathcal{J}=0\,.
\end{align}

The starting point is of course again given by the Isaacson assumptions in Eqs.~\eqref{eq:SmallParameterGRA2} and \eqref{eq:SmallParameterGRA1} within the perturbative framework outlined in Sec.~\ref{sSec:PerturbationTheory} together with the general equations of the perturbations in Eqs~\eqref{eq:FEPert} and \eqref{eq:FEPertNonMinimal}. In particular, this assumes a well-defined split of any field into a dependence on the low- and high-frequency scales
\begin{equation}\label{eq:IsaacsonSplitGen}
   g_{\mu\nu}=g^L_{\mu\nu}+\delta g^H_{\mu\nu}\,,\quad \Psi=\Psi^L+\delta\Psi^H\,,
\end{equation}
where the background fields can further be decomposed into the exact solution about which we are expanding and the corresponding low-frequency perturbations
\begin{equation}\label{eq:IsaacsonSplitGen2}
     g^L_{\mu\nu}=\bar{g}_{\mu\nu}+\delta g^L_{\mu\nu}\,,\quad \Psi^L=\bar{\Psi}+\delta\Psi^L\,.
\end{equation}
For simplicity, we will assume that all high-frequency perturbations can be captured by the same small expansion parameter $\alpha$, such that
\begin{equation}\label{eq:SmallParam2}
    |\delta g^H_{\mu\nu}|\,,\;\;|\delta\Psi^H| =\cal{O}( \alpha)\,,
\end{equation}
where $\alpha \ll 1$ compared to the $\cal{O}(1)$ exact solution, although in practice of course the amplitudes of each perturbation might be different. Moreover, we again assume for concreteness a potentially different scale $\beta\lesssim\alpha$ for the amplitudes of the slowly-varying perturbations 
\begin{equation}
    |\delta g^L_{\mu\nu}|\,,\;\;|\delta \Psi^L| =\cal{O}( \beta)\,,
\end{equation}

By definition\footnote{Recall that the existence of normal coordinates is guaranteed in any theory with metric on a Riemannian manifold with torsion and non-metricity free connection.}, any metric theory of gravity also admits the existence of normal coordinates at each spacetime point, such that the arguments in Sec.~\ref{ssSec:IsaacsonInGR} on the existence of a local chart in which the slowly varying background metric $g^L_{\mu\nu}$ up to second order reduces to the Minkowski form, as in Eq.~\eqref{eq:BackgroundMInkowskiForm}. However, in general, we cannot assume anything about the local form of the additional non-minimal fields $\Psi$. For instance, the extra non-minimal fields could be such that they fundamentally break local Lorentz invariance, while the metric still reduced to Minkowski spacetime. Crucially, however, the principle of universal coupling restricts such fundamental violations of Lorentz invariance to the gravity sector. Thus, the equations governing the perturbation fields, including the ones of the metric perturbations, might be fundamentally changed due to the presence of the non-trivial local background of the non-minimal fields. However, generally, these equations will still admit a certain number of wave-type or Laplace-type solutions.

With this comment out of the way, we can therefore press forward and analyze the system of leading order equations of motion in the case of a generic metric theory. This is where a major difference to GR appears. Namely, a generic metric theory of gravity might involve additional fundamental energy scales, which complicates the order counting of the operators. These energy scales are typically of two kinds. First of all, the additional non-minimal fields might be massive or more generally admit a potential term that lacks any derivative operators. Moreover, there might be terms in the equations of motion with more than two derivative operators. 


Given the definition of the action of metric theories in Eq.~\eqref{eq:ActionMetricTheory} and the convention on dimensionalities laid out above that definition, each term in the equations of motion that has fewer than two derivative operators needs to be accompanied by some energy or mass scale that we will simply denote by $m$, while any term with more than two derivatives is divided by some energy scale $\lambda$. Thus, for a generic metric theory of gravity, the order counting of the first order derivative operators in Eq.~\eqref{eq:OrderCountingGRLow} needs to be generalized to 
\begin{equation}
   \phantom{}_{\mys{(1)}}\mathcal{G}_{\mu\nu}[\delta g^L,\delta\Psi^L]={\cal{O}}\left(\beta \frac{f_L^j m^k}{\lambda^l}\right)\,,\quad \phantom{}_{\mys{(1)}}\mathcal{G}_{\mu\nu}[\delta g^H,\delta\Psi^H]={\cal{O}}\left(\alpha \frac{f_H^j m^k}{\lambda^l}\right)\,,
\end{equation}
where the integer powers satisfy
\begin{equation}
    j+k-l=2\,,\qquad\text{with}\qquad j\,,\;k\,,\;l\geq 0\,.
\end{equation}
Similarly, the order of the higher perturbative order operators with $i\geq 2$ in Eq.~\eqref{eq:OrderCountingGRHigh1} becomes
\begin{align}
    \phantom{}_{\mys{(i)}}\mathcal{G}_{\mu\nu}[\delta g^L,\delta\Psi^L,\delta g^H,\delta\Psi^H]={\cal{O}}\left(\alpha^i \frac{f_H^j m^k}{\lambda^l}\right)\,,
\end{align}
with similar expressions for the perturbations of the equations of the non-minimal fields.

To formulate a sensible set of leading order low- and high-frequency equations of motion as in Eqs.~\eqref{eq:EOMIISGR} and \eqref{eq:EOMISGR} we need to ensure that the expansion in $\alpha$ of the higher order operators is not spoiled. In other words, we need to require that compared to the kinetic term of the leading order high-frequency operator, any higher order operator is subdominant
\begin{equation}
    \alpha\,f_H^2  \gg \alpha^{2} \frac{f_H^j m^k}{\lambda^l}\,,
\end{equation}
for all values of $j$, $k$ and $l$ in the permissible range.
This is achieved by demanding the following requirements
\begin{equation}
    m\lesssim f_H\lesssim \lambda\,.
\end{equation}
These conditions are in fact rooted in very physical assumptions. The requirement that $f_H\lesssim \lambda$ imposes that a metric theory with higher order derivative operators is only valid up to energy scales set by the parameters $\lambda$ for which the higher order operators do not dominate over the kinetic term. Note, however, that crucially, such non-linear terms can still become comparable to the kinetic term and considerably influence the dynamics. On the other hand, $m\lesssim f_H$ simply reflects the fact that the high-frequency fields should be at energy scales on which they are not dominated by the mass of the field, such that they can still be excited.

This is all we need in order to establish the analogue of Eqs.~\eqref{eq:EOMIISGR} and \eqref{eq:EOMISGR}, hence the leading order low- and high-frequency equations of motion within the Isaacson picture. Because of the Isaacson split between low- and high-frequency equations, which we insist is crucial in this case, we can consider the leading-order contributions of the low- and high-frequency equations separately and they read
\begin{align}
   \phantom{}_{\mys{(1)}}\mathcal{G}_{\mu\nu}[\delta g^H,\delta\Psi^H]&=\kappa_0\left[\delta T_{\mu\nu}\right]^H\,,\label{eq:EOMIIS2}\\
    \phantom{}_{\mys{(1)}}\mathcal{J}[\delta g^H,\delta\Psi^H]&=0\,,\label{eq:EOMIIS2Psi}\\
    _{\mys{(1)}}\mathcal{G}_{\mu\nu}[\delta g^L,\delta\Psi^L]&=-\frac{1}{2}\big\langle \phantom{}_{\mys{(2)}}\mathcal{G}_{\mu\nu}[\delta g^H,\delta\Psi^H]\big\rangle+\kappa_0\big\langle\delta T_{\mu\nu}\big\rangle\,,   \label{eq:EOMIS2}\\
    _{\mys{(1)}}\mathcal{J}[\delta g^H,\delta\Psi^H]&=-\frac{1}{2}\big\langle \phantom{}_{\mys{(2)}}\mathcal{J}[\delta g^H,\delta\Psi^H]\big\rangle\,.   \label{eq:EOMIS2Psi}
\end{align}

In analogy to the results in GR Eqs.~\eqref{eq:EOMIIS2} and \eqref{eq:EOMIIS2Psi} represent propagation equations for the high-frequency perturbations, while Eq.~\eqref{eq:EOMIS2} identifies the effective energy-momentum contribution of all the high-frequency perturbations
\begin{equation}\label{eq:DefPseudoEMTensorBeyondGR}
    \phantom{}_{\mys{(2)}}t_{\mu\nu}[\delta g^H,\delta\Psi^H]\equiv -\frac{1}{2\kappa_0}\big\langle \phantom{}_{\mys{(2)}}\mathcal{G}_{\mu\nu}[\delta g^H,\delta\Psi^H]\big\rangle\,.
\end{equation}
As in GR, to leading order, all terms in the equations above are gauge invariant and conserved. Moreover, a similar interpretation can be given to Eq.~\eqref{eq:EOMIS2Psi} although in practice this case is more subtle as we will discover in Part~\ref{Part: Gravitational Wave Testing Ground}.
Finally, observe that due to the assured presence of the non-negligible kinetic terms in each expression the relation $\beta\sim \alpha^2f_H^2/f_L^2$ in Eq.~\eqref{eq:BetaParam}, between the scale of amplitude of low- and high-frequency modes still holds.

\subsection{Waves in Metric Theories of Gravity}\label{ssSec:WavesInMetricTheoriesOfGravity}

In Sections~\ref{ssSec:Local Wave Equation in GR} and \ref{ssSec:GaugeInvariantDecomposition} above, we offered rather careful but straightforward deviation of the number of propagating degrees of freedom in GR and their association to the two transverse tensor modes in metric perturbations. This detailed discussion will now pay off, as we will be able to generalize this procedure to generic metric theories of gravity by merely dropping any input from the specific form of the Einstein equations.

In a generic metric theory of gravity, propagating degrees of freedom are still defined within the generalized Isaacson approach outlined above, as the gauge invariant field perturbations whose leading order propagation equations [Eqs.~\eqref{eq:EOMIIS2}and \eqref{eq:EOMIIS2Psi}] outside any high-frequency source on a local Minkowski patch of the background reduce to a wave-like equation. However, the presence of additional dynamical fields adds a certain amount of complexity that needs to be dealt with.

In principle, one can still follow the route in Sec.~\ref{ssSec:Local Wave Equation in GR} of completely gauge fixing with the aid of the equations of motion. Yet, in general, as indicated in Eqs.~\eqref{eq:EOMIIS2} and \eqref{eq:EOMIIS2Psi} the different propagation equations of the high-frequency perturbations might be coupled. In order to impose the analogue of the TT gauge on the metric perturbations together with potential additional convenient gauge choices of the non-minimal fields, one needs to find an appropriate field redefinition that decouples the perturbation variables. An explicit example of such a field redefinition will be given in Sec.~\ref{sSec:GWPolExample} where we will consider an explicit example of a rather general metric theory beyond GR.


\paragraph{SVT Decomposition of Fields.} Alternatively, one might also resort to the second approach introduced in Sec.~\ref{ssSec:GaugeInvariantDecomposition} of a scalar-vector-tensor decomposition into gauge-invariant variables. However, in contrast to GR, the rotational invariance of the general background in Riemann normal coordinates might not be guaranteed due to the presence of additional non-minimal fields, implying that the scalar, vector and tensor sectors do not automatically decouple. Nevertheless, this approach is still useful for an explicit description of dynamical and non-dynamical degrees of freedom, in particular when considering an explicit background solution as we will do in Sec.~\ref{sSec:GWPolExample}. Here, however, we want to remain on a general background and make a few general statements.

Regardless of the symmetries of the background in more general metric theories of gravity, the decomposition of the metric perturbations remains exactly the same, namely [Eqs.~\eqref{eq:hexpand}, \eqref{eq:hexpand2} and \eqref{eq:hexpand3}]
\begin{equation}\label{eq:hexpandBGR}
h_{\mu\nu}=
 \left(\begin{array}{c|c c c} 
    	S &  &V^T_i+\partial_i V^\parallel &\phantom{0} \\
    	\hline 
    	 &  &&\\
V^T_i+\partial_i V^\parallel  &  & T\delta_{ij}+h^{TT}_{ij}+2\partial_{(i} U^T_{j)}+\left(\partial_i\partial_j-\frac{1}{3}\delta_{ij}\Delta\right)U^\parallel& \\
 \phantom{0} &  & & \\
    \end{array}\right)\,,
\end{equation}
with the associated gauge invariant variables [Eq.~\eqref{eq:gaugeInvariantMetricPerts1}]
\begin{align}\label{eq:gaugeinvBGR}
\boxed{\delta\Phi= S-2\dot{V}^\parallel+\ddot{U}^\parallel\;,\quad\delta\Theta= T-\frac{1}{3}\partial^2U^\parallel\;,\quad\delta\Xi^T_i\equiv V^T_i-\dot{U}^T_i\;,\quad  h^{TT}_{ij}\,,}
\end{align}
representing six physical degrees of freedom, with
\begin{align}\label{eq:TransverseConditions}
\partial^i h^{TT}_{ij}=0\;,\quad\delta^{ij} h^{TT}_{ij}=0\;,\quad\partial^i \delta\Xi^T_i=0\,.
\end{align}

Recall that within GR it were the vacuum equations of motion, which at this point implied that only the TT-part satisfies a wave equation while all other gauge invariant degrees of freedom do not propagate. Rather, the remaining physical DOFs were part of constraint equations and could be set so zero in the absence of high-frequency sources. In a general metric theory of gravity, on the other hand, in principle all the six independent physical modes \eqref{eq:gaugeinvBGR} can be dynamical. With regard to the Helmholtz-decomposition, these correspond to two scalar, two vector and two tensor modes with respect to rotations about the direction of propagation within the chosen chart.

Similarly, all other non-minimal fields can be decomposed into $SO(2)$ scalars vectors and tensors. This of course implies that we restrict ourselves in this work to bosonic fields only, as already mentioned in Sec.~\ref{sSec:Metric Theories}. Indeed, fermionic fields do not usually play the role of a massless force carrier in known theories. Nonetheless, in principle, it could still be interesting to enrich metric theories with fermionic fields in the gravitational sector, an investigation we leave for future work. Moreover, no known consistent theories of massless perturbative fields with $SO(3)$ label $l>2$ are known (see e.g. \cite{maggiore2008gravitational,Schwartz:2014sze}). We will therefore restrict ourselves to considering fields with $l=0\,,\;1\,,\;2$ that are decomposed into $SO(2)$ scalar, vector and tensor representations of $m=0\,,\;\pm 1\,,\;\pm 2$ only.

Depending on whether the additional non-minimal fields in the metric theory of gravity are true dynamical fields or mere \textit{auxiliary fields} without any propagating degrees of freedom, their gauge-invariant perturbations in an SVT decomposition will mix with the gauge invariant DOFs of the physical metric. However, as long as the spacial rotational invariance of the background in local Minkowski coordinates is not broken by the presence of the non-minimal fields, which is for instance assured if the additional non-minimal fields are represented by scalar fields the scalar, the vector and the tensor sector at linear order are each automatically decoupled from one another.

\paragraph{Faithful Representation and Gravitational Waves.} At this point, one should keep in mind the distinction between the tensor fields defined on the manifold of spacetime that enter the definition of a metric theory at the level of its action in Eq.~\eqref{eq:ActionMetricTheory} and their associated perturbations that lead to a description of propagating degrees of freedom. Indeed, there are in principle countless ways of representing a given metric theory at the level of the action through field redefinitions and the introduction of auxiliary fields. In the next Chapter~\ref{Sec:The Theory Space Beyond GR} we will encounter explicit examples of formulations of theories that hide additional propagating degrees of freedom within higher order interaction terms. On the other hand, the number and nature of propagating degrees of freedom are physical properties that are invariant under any field redefinitions and therefore capture the true character of a theory. Of course, a given theory will still also depend on the exact form of interactions between the physical degrees of freedom, but the number and nature of propagating DOFs nevertheless represents a very useful tool for the classification of metric theories of gravity.

In the light of this discussion, it will be useful to define the notion of a \textit{faithful representation} of the metric theories of gravity
\begin{definition} \textit{A Faithful Representation of a Metric Theory of Gravity.}\label{DefFaithfulRep}
    A faithful representation of a metric theory of gravity of Definition~\ref{DefMetricTheory} is a formulation of the theory in which each propagating degree of freedom can naturally be associated to a field in the action. 
\end{definition}
Here, the term ``natural'' refers to a given free-field or leading order description. For example, a faithful representation requires that any massless scalar or vector DOF is associated to a corresponding field in the action and any massless transverse traceless tensor DOF should be associated to a symmetric tensor field, while a massive scalar and two vector DOFs could also be grouped in the description of a massive vector field. Note that this definition in particular excludes the presence of hidden degrees of freedom, but still allows for the presence of auxiliary fields that do not give rise to dynamical DOFs.

In a faithful representation of a metric theory of gravity, the physical metric therefore always only introduces the two degrees of freedom known from GR, while additional propagating DOFs need to be associated to extra non-minimal fields in the gravity sector. It is however important to realize, that such a faithful description does not imply that no other gauge invariant degree of freedom in the perturbations of the physical metric in Eq.~\eqref{eq:gaugeinvBGR} are dynamical. Indeed, depending on the coupling of the non-minimal field with the metric, the additional propagating DOF of the non-minimal field can, but must not, excite the physical modes within the metric. This is important because due to the Principle~\ref{Principle:Universal and Minimal Coupling} of universal and minimal coupling only the perturbations of the physical metric directly interact with the matter used in today's and future GW experiments and therefore only excitations of the physical metric are directly detectable.

From this point of view, we should make a distinction between the perturbations of the physical metric $g_{\mu\nu}$ and other field perturbations. While we therefore in general define propagating solutions of high-frequency perturbations as \textit{\gls{waves}}, we want to reserve the terminology \textit{\gls{gravitational waves}} for the high-frequency perturbations of the physical metric. More precisely, as we will describe in Sec~\ref{sSec:GWObservation}, the up to six propagating gauge invariant degrees of freedom of the physical metric will figure in the local response to gravitational waves governed by the geodesic deviation equation which will define the notion of six possible gravitational polarizations. The distinction between waves and gravitational waves is then equivalent to the distinction between the number of propagating degrees of freedom in a given metric theory, whose number is a priori not bound from above, and the concept of gravitational polarizations of which there can only be up to six within any metric theory of gravity. Note that both the number of propagating degrees of freedom and the number of gravitational polarizations of a theory do not depend on their description in the action.


\chapter{The Theory Space Beyond GR}\label{Sec:The Theory Space Beyond GR}

This final chapter of Part~\ref{Part: EFT of Gravity} will provide an overview of the most popular theories of gravity beyond GR within the framework of metric theories. The concrete models introduced here will play an important role in the subsequent parts of this manuscript. But first, we want to offer a deeper analysis of why the realm of metric theories provides an ideal framework to describe the space of effective (field) theories of gravity that can be tested for with current and near future experiments. For now, we will understand here the term ``effective theory'' in a purely classical sense, in which we are looking for an effective description of potential departures from GR that might be present for us to discover at the scales that we can currently probe. Moreover, we will also draw a clear distinction between higher-derivative metric theories that avoid Ostrogradsky ghost-like instabilities with and without additional perturbative constraints. The differences between these two classes are significant, and it is worth spelling them out in some detail in order to understand different strategies in testing the theory of general relativity and the searches for effects beyond it. 


\section{Effective Metric Theories of Gravity}\label{sSec:Why and how go beyond GR}

As presented in the introduction, the reasons for investigating theories beyond general relativity are manifold, but so are the number of proposed alternatives. So far, it seems that there exist no clear guidance nor hint towards a particularly preferred direction for the road beyond GR. This is mainly due to the unfortunate situation that while on a theoretical level many open questions remain, there is at present no clear-cut evidence of any empirical data that goes beyond the current standard theory. 

In the light of this situation, we want to argue that a good approach to take on the search for such observational departures from GR and a better understanding of the phenomenon of gravitation, is to consider the framework of metric theories of gravity as an \textit{\gls{effective field theory}} (EFT) description of any beyond GR effect that might be waiting for us to discover. The term ``effective field theory'' should here simply be understood as the expectation that whatever the future theory of gravity might look like concretely, possibly based on an entirely different basis of description as GR, its modifying effects on scales that we can probe with current experiments, be it at large or small scales, can be captured by a certain type of metric theory of gravity.

Note that this notion of EFT is broader than the widely used concept of what we will call \textit{quantum effective field theory} (qEFT) \cite{Weinberg:1978kz,Gasser:1983yg,Gasser:1984gg,Arzt:1992wz,Burgess:1992gx,Polchinski:1992ed,Cao1993,Weinberg:1995mt,Weinberg:2008hq,Burgess:2006bm,Burgess:2007pt,Davidson:2020gsx,Georgi:1993hh,Donoghue:1994dn,zee_quantum_2010,Donoghue:2012zc,Endlich:2017tqa} that is mainly based on Wilson's work on renormalization group (RG) methods \cite{Wilson:1973jj} in the context of critical phenomena. More specifically, the notion of qEFTs in particular also aims at the characterization of quantum (loop) corrections of a theory that in principle inevitably generate all possible interactions, as we will further discuss below and in particular also in Part~\ref{Part: Quantum Gravity}. Yet, although the precise notion of an EFT was developed in the context of quantum field theories, and many of the EFT tools are based on this mathematical framework, the philosophy behind effective (field) theories can, or even must, also be applied in a purely classical context (see e.g. \cite{Goldberger:2004jt,Goldberger:2007hy,Baumann:2010tm,Carrasco:2012cv,Carrasco:2013mua,Porto:2016pyg,Levi:2018nxp,Davidson:2020gsx}). For example, the description of fluids through the Navier-Stokes equations fundamentally only makes sense as an effective description that proves very useful despite the ignorance of all small-scale details.\footnote{See also \cite{Dubovsky:2011sj} for an EFT reformulation of hydrodynamics} In fact, any modern theory of physics can be regarded as an effective theory, a statement at the core of the possibility to construct meaningful descriptions of nature, despite our highly incomplete knowledge of it. 

Broadly speaking, we therefore aim for an optimal description of physical phenomena within a particular range of scales appropriate for current observational probes. Thus, schematically, an EFT is only valid up to an energy scale $\Lambda$ known as ``cutoff''. For energies $E$ below the cutoff, such a theory is typically organized in the order of increasing number of derivatives controlled by the expansion parameter $E/\Lambda$. Above the cutoff, however, the EFT is not valid anymore and physics is assumed to be governed by a UV completion that often involves the introduction of new degrees of freedom.\footnote{A low energy EFT can typically be constructed from a UV theory by ``integrating out'' the degrees of freedom relevant at the high energy scales \cite{Weinberg:1995mt}.}


As already discussed in the introduction, the consideration of a concrete theory space beyond GR to which a more fundamental description of nature might reduce to at our energy scales is important, since mere null-tests of parameterized beyond GR effects might miss out on more complex signatures and would not provide any conceptual guiding principle to advance the theoretical understanding.
The question remains as to why the effective theory space should be given by the metric theories of gravity in Definition~\ref{DefMetricTheory}? In our opinion, there are at least two major reasons:

\subsection{Metric Theories as Viable Theories of Spacetime}


As presented in Chapter~\ref{Sec:The Generalization to Gravity}, the overwhelming evidence for the weak equivalence principle, local Lorentz invariance and the resulting Einstein equivalence Principle~\ref{Principle:EEP} naturally leads to the conjecture of the Principle~\ref{Principle:Universal and Minimal Coupling} of universal and minimal coupling at the basis of the framework of metric theories of gravity. Of course, non-minimal or non-universal couplings to matter might still appear on yet unexplored scales and their study indeed bear interesting testable effects (see e.g. \cite{Gonner:1976gq,Moraes:2017zgm,Bonvin:2018ckp}). 

However, as discussed, the EEP is at the root of fundamental axioms that give meaning to many of the empirical probes and is therefore hard to disregard without radically challenging the foundation of all of modern physics. Namely, the EEP assumptions that enter the definition of metric theories assures that spacetime described through a physical metric is a self-sufficient concept whose observations do not depend on the precise experimental device and setup. This in particular concerns the interpretation of the experimental observations of the geodesic deviation in Eq.~\eqref{eq:GeodesicDeviation} that as we will discuss in detail in Sec.~\ref{sSec:GWObservation} is at the basis of all current gravitational wave experiments. The same is true for local energy-momentum conservation of matter fields, that also only fundamentally emerges in action based covariant gravity theories on a manifold as a consequence of the EEP and universal and minimal coupling.

And while GR, through the Lovelock Theorem~\ref{Thm:LovelockTheorem}, stands out as the simplest realization of a metric theory of gravity by being the unique leading order metric theory in four spacetime dimensions, build solely out of the metric, the guidance of the EEP does not directly limply GR but precisely results in the broader framework of metric theories of gravity. Thus, the strong empirical evidence for the EEP and local matter physics leaves room for interesting phenomenology beyond GR. 

\subsection{Extra Propagating Degrees of Freedom as a Unique Signature Beyond GR}\label{ssSec:ExtraDOFs as a unique signature}

Based on the Lovelock Theorem~\ref{Thm:LovelockTheorem}, one of the simplest options for describing a theory beyond GR is by allowing for non-minimal fields on top of the physical metric in the gravitational action. As long as the additional non-minimal fields do not directly couple to matter, the resulting theory is a metric theory, which due to the presence of the additional fields naturally involves extra propagating degrees of freedom compared to GR. Indeed, in that respect, the framework of metric theories of gravity seems tailor-made for the consistent and viable description of additional propagating degrees of freedom without spoiling the fundamental assumption of gravity theory given by the EEP.

This conclusion, however, even holds when insisting on the restriction of only considering the physical metric as the gravitational field in the action. This is because although the last assumption in the Lovelock Theorem~\ref{Thm:LovelockTheorem}, which restricts the equations of motion of GR to only contain up to two derivative operators, is already well rooted in dimensional analysis arguments, it can be associated to a much deeper insight: Namely, this assumption is fundamentally at the root of the statement that GR only propagates two degrees of freedom that we carefully discussed in the previous Chapter~\ref{Sec:PropagatingDOFs}. In effect, it is generally expected that the introduction of operators with more powers of curvature invariants in the action leads to additional propagating DOFs in the theory due to the requirement of extra initial data \cite{Simon:1990PhysRevD41,papantonopoulos2014modifications}. While this statement needs to be refined in a crucial aspect that will be the subject of the next Section~\ref{sSec:OstrogradskyTheorem}, this means that a departure from GR by considering higher-order operators build solely out of the physical metric still generally leads to additional propagating DOFs in the theory. An explicit example of such a theory is $f(R)$ gravity, that we will introduce in Sec.~\ref{ssSec: A Exact Theories} below. In turn, in an alternative faithful description (recall Def.~\ref{DefFaithfulRep}) of the resulting theory, these extra DOFs might again be described through additional non-minimal fields in the action of the metric theory.

But not only that, it turns out that even the breaking of any of the additional assumptions behind the Lovelock theorem that also underline the Definition~\ref{DefMetricTheory} of metric theories, that is, the restriction to four spacetime dimensions, the choice of a Levi-Civita connection, locality and diffeomorphism invariance, in most cases can actually also be accounted for by the introduction of additional fields within an effective description of gravity through metric theories, as long as the EEP is respected. Such additional fields then again typically introduce additional propagating DOFs into the theory. To illustrate this we will go through each of the four additional assumptions of metric theories mentioned above (see also \cite{papantonopoulos2014modifications}):
\begin{itemize}
    \item[] \textit{Restriction to four spacetime dimensions}: While on a theoretical basis it is a rather straightforward task to change the number of spacetime dimensions, clearly no additional dimensions have been detected experimentally. Hence, it must be required that for any higher-dimensional theory, there exists a consistent reduction to an effective theory in four spacetime dimensions within which the additional dimensions cannot be felt directly but entail indirect effects. In fact, historically, considering theories with extra dimensions was one of the first ways of obtaining consistent theories beyond general relativity in four dimensions. There are essentially two possibilities. The first one is to compactify the extra dimensions such that they are neither visible on large scales, not can anything with reasonably low energy escape in them.\footnote{That this is a viable approach can be argued based on the Heisenberg uncertainty principle \cite{zee2013einstein}.} The associated restriction to four dimensions is known as a Kaluza-Klein reduction \cite{Appelquist:1987nr,Dereli:1990he,zee2013einstein,deRham:2014zqa} and naturally leads to the introduction of additional fields with associated DOFs in the theory. In particular, Kaluza-Klein reductions of higher dimensional Lovelock gravity \cite{Charmousis:2014mia} leads to Galileon and Horndeski type of theories, which we will discuss in Chapter~\ref{ssSec: A Exact Theories}. The other possibility, is to consider large extra dimensions but to include a four-dimensional brane-world onto which the matter fields are confined, as brought forth by the Dvali-Gabadadze-Porrati (DGP) model \cite{Dvali:2000hr,Dvali:2000rv,Dvali:2000xg,Lue:2005ya}. Such models generally lead to theories that give a mass to the graviton \cite{deRham:2014zqa} and therefore inevitably introduces additional propagating DOFs.


    
    \item[] \textit{Levi-Civita connection}: When choosing a different connection than the Levi-Civita connection, in other words, when considering a non-zero \ul{torsion} and/or \ul{non-metricity} (see Appendix.~\ref{App:DiffGeo}), we need to distinguish two cases: 
    \begin{enumerate}[(i)]
    \item Matter fields couple to the torsion or the non-metricity. In this case, the principle of minimal coupling and therefore also the Einstein equivalence principle are violated. For instance, in this case the world lines of (certain) free test particles, hence the straight lines, might not correspond to geodesics of the physical metric anymore but rather to the autoparallels of the total connection that are distinct from the geodesics. Moreover, recall that with respect to a connection with non-vanishing torsion or non-metricity, the existence of Riemann normal coordinates is not guaranteed and one has to resort to non-coordinate basis to describe local Minkowski physics (see e.g. \cite{carroll2019spacetime} for an introduction to non-coordinate basis). Such theories have been studied intensively with possibly interesting results \cite{Cartan:1922prj,Cartan:1923prj,Einstein:1925tt,Trautman:1972prj,Blagojevic:2012bc,Katanaev:2013cqa,Cai:2015emx,Koivisto:2018aip,BeltranJimenez:2019bnx,BeltranJimenez:2019tme,Bahamonde:2021gfp,Heisenberg:2023lru,Heisenberg:2023wgk}, but as discussed, we will not consider such cases of EEP breaking any further. 
    \item If the Einstein equivalence principle is not violated, which means that all matter fields retain their minimal and universal coupling, then the introduction of torsion or non-metricity can be captured through the framework of metric theories of gravity. This is because the torsion and non-metricity are themselves proper tensor components, as opposed to the Levi-Civita part of the connection, the Christoffel symbols. Hence, their effect can in principle be captured by introducing additional tensor fields to the theory \cite{carroll2019spacetime}.
    \end{enumerate}
    \item[] \textit{Diffeomorphism invariance}: As already discussed, theories on a differential manifold come with a fundamental gauge freedom of diffeomorphic transformations (see also App.~\ref{sApp: Spacetime Gaugefreedom and symmetries}). Moreover, in principle, any gauge freedom can be turned into a gauge symmetry (see in particular App.~\ref{App: Symmetires in Physics}), which in particular also includes diffeomorphism invariance. However, in some cases the breaking of general coordinate invariance is an intrinsic feature of the theory, as it is the case for instance for ``massive'' gravity theories. From a (quantum) field theory point of view, defined on a fixed Minkowski spacetime (see Sec.~\ref{sSec:Quantization of Gravity}), considering a theory of a massive spin 2 particle instead of a massless one, is very natural. Here, the word ``massive'' refers to an altered propagation equation of field excitations associated to the gravitational field. Neglecting any matter interactions at first order in perturbation theory, it is straight forward to write down a corresponding theory \cite{Fierz:1939ix}. However, such a theory is not invariant under linearized coordinate transformations, such that also its fully non-linear counterpart of massive gravity therefore naturally breaks general covariance, or coordinate invariance (see  \cite{PhysRevD.33.3613,Hinterbichler:2011tt,deRham:2014zqa} and references therein). Yet, a diffeomorphism invariant formulation of such theories can be found by introducing additional fields through the St\"ukelberg trick \cite{Stueckelberg:1900zz,GREEN1991462,Siegel:1993sk,Arkani-Hamed:2002bjr,Ruegg:2003ps}. 

    \item[] \textit{Locality}: Fundamental non-locality is generally not desired due to inconsistencies with causality and instabilities. However, at an effective level, non-local terms in the action might appear with potential interesting consequences for cosmology \cite{Deser:2013uya,zee2013einstein,Belgacem:2020pdz}. Yet, in principle, any effective action can be rewritten in a local form by introducing auxiliary fields \cite{Nojiri:2007uq,Jhingan:2008ym,Koshelev:2008ie,Deser:2013uya}. In this case, however, one needs to be careful about over-counting the number of propagating degrees of freedom \cite{Belgacem:2020pdz}.
\end{itemize}

The above discussion, backed up by the quantum field theoretic arguments mentioned in Sec.~\ref{sSec:Uniqueness in Propagating DOFs}, renders the conjecture plausible, that GR is the unique consistent effective description of gravity in four space-time dimensions that only propagates two tensor degrees of freedom. Up to a caveat discussed in Sec.~\ref{sSec:OstrogradskyTheorem} below, this conjecture can immediately be turned into the statement that additional propagating DOFs not only represent a smoking gun signature beyond GR, but provide a unique opportunity to look for effects beyond the current standard description as very likely a deviation from GR introduces additional degrees of freedom. In turn, this provides a strong argument for the consideration of metric theories of gravity as a natural framework of effectively but consistently describing such addition propagating DOFs.

\section{Ostrogradsky Instabilities}\label{sSec:OstrogradskyTheorem}

We now want to refine the statement made above, that the introduction of additional powers of curvature invariants in the gravitational action leads to additional propagating DOFs in the theory. First of all, when introducing operators with additional powers of derivatives one generally needs to be careful, since in many cases, in particular if the equations of motion involve more than two derivative operators per field, the associated DOFs are unhealthy \textit{\gls{ghost}} instabilities which render a theory untenable. It is therefore imperative to avoid such instabilities when considering the theory space beyond the leading order in curvature terms, which can essentially be done in two ways. The first is to only introduce higher curvature operators that introduce healthy new propagating DOFs. For instance, this is prominently the case for $f(R)$ gravity that we will introduce below. The second option is to introduce additional constraints that assure that the presence of higher order curvature terms do not change the number of DOFs of the gravity theory (see e.g. \cite{Endlich:2017tqa}). In this case, the resulting perturbative high-energy corrections can be regarded as intrinsic to GR and can be associated to corrections that might be expected from high-energy quantum physics (see Sec.~\ref{ssSec: B Perturbative Theories}).

In fact, the potential instability of theories that include higher powers in derivatives in the Lagrangian plays a central role in formulating viable metric theories beyond GR, also when explicitly considering additional non-minimal fields, and will decisively structure the associated theory space. The nature of these instabilities is the essence of the Ostrogradsky Theorem~\ref{Thm:OstrogradskyTheorem} that we now want to discuss in some detail. As it was the case when introducing the concept of dynamical DOFs, the general statement of the theorem requires the use of the Hamiltonian formalism within an ADM decomposition of spacetime, that subsequently can be mapped to an analytically tractable perturbative setting.


\begin{theorem}\label{Thm:OstrogradskyTheorem}  \textbf{The Ostrogradsky Theorem} \cite{Ostrogradsky:1850fid,Woodard:2015zca}. If a Lagrangian theory with second order time derivatives or higher in its Lagrangian is non-degenerate, then the associated Hamiltonian of the theory is unbounded from below due to the existence of a linear instability.
\end{theorem}

Here, ``non-degeneracy'' refers to the ability to reexpress the highest time derivative operator in terms of canonical variables \cite{Chen:2012au}. 
Theories that posses an Ostrogradsky instability are of no use to describe physical systems as soon as interactions are turned on. This is because an unbounded Hamiltonian allows for an arbitrary fast decay of the vacuum via the creation of energy that can be compensated by negative energy states. Indeed, from a perturbation theory perspective, Ostrogradsky instabilities can be described through the notion of a ghost excitation with the wrong sign of the kinetic term \cite{Creminelli:2005qk,Deffayet:2005ys,papantonopoulos2014EntireBook,Salvio:2018crh,Ganz:2020skf}. Such instabilities are themselves extremely robust, since they are independent of the precise form of the interaction or the value of the coupling \cite{Eliezer:1989cr}.  The instability kicks in as soon as the ghost can be excited, is present both at the classical and the quantum level, and grows worse as more derivatives are added.

It is therefore imperative to avoid Ostrogradsky ghosts to formulate viable field theories of physics. Note that while the Ostrogradsky theorem primarily identifies unstable theories that should be disregarded, it also indicates how stable theories with higher order powers of derivatives in the action can be constructed. Indeed, in most cases, degenerate theories are stable \cite{Woodard:2006nt}. This is because in general, degenerate theories are guaranteed to involve additional constraints that reduce the phase space and therefore the number of degrees of freedom, which often kills any unwanted ghost excitation.
In general, there are two distinct possibilities in avoiding Ostrogradsky instabilities: 
\begin{enumerate}[(A)]
    \item The full theory does not suffer from any Ostrogradsky instability.
    \item The inclusion of additional constraints mitigates all ghost excitations.
\end{enumerate}

Theories of type (A) either do not involve any higher-order derivatives in the Lagrangian or do so in a degenerate way such that the full theory already incorporates constraints which exclude any ghost-like DOFs. The simplest possibility to avoid ghosts is to ensure that despite the presence of the higher-order derivative terms in the Lagrangian, the equations of motion still remain at second order in time derivatives per field operator. One example of such a theory is actually given by GR itself (recall Sec.~\ref{sSec:LovelockTheorem}). Other concrete examples and extensive constructions of such Ostrogradsky-stable higher-order derivative theories are discussed in Sec.~\ref{ssSec: A Exact Theories} below. 

However, while second-order equations of motion of the tensor fields generally imply the absence of ghost instabilities, this condition does not represent a necessary criterion, in particular as concerns multi-field theories \cite{deRham:2016wji}. This statement can be understood by again drawing the attention to the difference between the notion of tensor fields that appear in an action of a theory on a manifold and the associated field perturbations that can be used to describe the propagating degrees of freedom of a theory. In particular, the absence of ghosts refers to the perturbative level of the equations of motion, which essentially explains the loophole by which multi-field theories with equations of motion at higher-powers of derivatives per field can still remain stable. Note, however, that to conclude Ostrogradsky stability at the perturbative level, the absence of ghosts needs to be shown on arbitrary backgrounds which can however again be achieved in local Riemann normal coordinates in the Isaacson framework introduced in Chapter.~\ref{Sec:PropagatingDOFs}.

In contrast, theories, which taken at face value do involve an Ostrogradsky ghost, can be stabilized by the introduction of additional constraints that reduce the phase space of the theory \cite{Chen:2012au}. In particular, if there exists a Ostrogradsky-stable principal part of the theory, the higher order operators that would cause an instability can in principle still be considered as perturbations to the principal part, associated to a small expansion parameter. This is possible through the method of \textit{perturbative constraints} \cite{Eliezer:1989cr,Simon:1990PhysRevD41,Simon:1990jn,Yunes:2013dva}.\footnote{Note that, sometimes, this method is also called \textit{small-coupling approximation}. However, it is important to realize, that a smallness of a coupling alone, without the introduction of additional constraints, does not stabilize a theory against ghosts.} This method constructs perturbative solutions around the principal part that effectively throw away any instable branches. Moreover, the perturbative constraints can be applied to all orders in derivatives, which allows a systematic construction of higher order terms, as we will see explicitly in Sec.~\ref{ssSec: B Perturbative Theories}. Furthermore, in this approach, the number of degrees of freedom of the theory is not altered, and the constrained theory resembles the solutions of the principal part \cite{Simon:1990PhysRevD41}.

The two possibilities to avoid Ostrogradsky ghosts discussed above therefore divide the theory space of metric theories into two classes: 
\begin{enumerate}[(A)]
    \item Theories that are exact at the classical level.
    \item Perturbative theories that include higher-order corrections to an Ostrogradsky-stable principal part.
\end{enumerate}
In practice, only theories of type (A) come into question when considering long-range, or IR, modification in cosmological applications, which require more substantial departures from GR on cosmological scales. Such theories can be viewed as ``true'' classical theories of gravity that provide an alternative to GR. As discussed, the majority of such theories inevitably introduce new degrees of freedom that may result in larger modifications of gravity physics. 


On the other hand, theories of type (B) naturally capture observable (UV) and strong curvature corrections to an exact theory of type (A). While at first sight the introduction of seemingly ad-hoc constraints to render a theory stable might sound uninteresting, it is mainly the quantum world that motivates theories of type (B). Indeed, from a qEFT perspective it is generally expected, that given a fixed set of light degrees of freedom of an exact theory, the unknown UV physics gradually introduces correction effects as higher energies are probed, which can be parameterized by a set of local operators that only involve these low energy degrees of freedom theory (see e.g.\cite{Donoghue:1994dn,Weinberg:1995mt,Weinberg:2008hq,Endlich:2017tqa,Davidson:2020gsx}). Such an approach provides a natural separation of the still inaccessible and therefore uninteresting high energy contributions, from the in principle knowable low energy quantum effects, that can be captured by the introduction of a series of all possible higher order operators constructed out of the classical fields. And while the series of correcting operators is in principle infinite, the set of additional interactions needed to compute a physical observable to a given precision below a certain energy scale is always finite. The associated observable UV effects can then already be discussed from a purely classical point of view, precisely upon the use of the perturbative constraint techniques of type (B). We therefore postpone the discussion of such effects in an explicit quantum setting to Part~\ref{Part: Quantum Gravity}, and will retain for now a purely classical perspective. Two explicit examples of type (B) theories will be offered in Sec.~\ref{ssSec: B Perturbative Theories}.


\section{A Selective Overview of Metric Theories}\label{sSec:Overview of Metric Theories}

There exist a multitude of different metric theories of gravity (see e.g. \cite{Clifton:2011jh,Faraoni2011,Yunes:2013dva,papantonopoulos2014EntireBook,Berti:2015itd,Nojiri:2017ncd,Heisenberg:2018vsk,Will:2018bme,CANTATA:2021ktz} for a review). To a certain extent, the existence of extensive theory frameworks of a certain type, such as Horndeski theories (see below) allow for a unified description of a big portion of theory space. However, we will certainly not be able to cover all possibilities. We will rather restrict ourselves to the most widely known theories and only present the explicit action of the theories that will be used in the reminder of the manuscript. In all the metric theories below, $g_{\mu\nu}$ will denote the physical metric that is minimally coupled to matter.

\subsection{(A) Exact Theories}\label{ssSec: A Exact Theories}

In describing concrete examples of exact, hence Ostrogradski stable, metric theories, we will classify them according to the number and type of propagating DOFs. Interestingly, we will encounter theories that at first sight might look completely different, but in fact simply represent a subset of one-another, thus exemplifying the use of the notion of faithful representation introduced in Def.~\ref{DefFaithfulRep}. It therefore makes sense to classify the theories according to their physical properties in terms of propagating degrees of freedom instead of their tensor-field content that is description dependent.


\subsubsection{\ul{Scalar-Tensor Theories}}

Certainly, the most popular exact metric theories beyond GR are scalar-tensor (ST) theories that involve an additional non-minimal scalar field. Generalizations to multiscalar scenarios are of course always possible.

\paragraph{Generalized Brans-Dicke Gravity (gBD).}

Historically, one of the first alternative ST theory to GR considered was Brans-Dicke theory. The action of generalized Brans-Dicke gravity can be written as \cite{Brans:1961sx,Dicke:1961gz,Weinberg1972,poisson2014gravity,Will:2018bme,carroll2019spacetime}
\begin{equation}\label{ActionBD}
    S^{\myst{gBD}}=\frac{1}{2\kappa_0}\int \dd^4 x\sqrt{-g}\left(F(\Phi) R-\frac{\omega(\Phi)}{\Phi}g^{\mu\nu}\nabla_\mu\Phi\nabla_\nu\Phi-U(\Phi)\right)+S_\text{m}[g,\Psi_\text{m}]\,,
\end{equation}
with $\Phi$ a non-minimal scalar field and where $F(\Phi)$, $\omega(\Phi)$ and the potential $U(\Phi)$ are field dependent, smooth functionals. The corresponding equations of motion can for instance be found in \cite{poisson2014gravity,Will:2018bme,carroll2019spacetime}. This is a generalization of the original formulation of Brans-Dicke (BD) theory \cite{Brans:1961sx,Dicke:1961gz}, for which $F(\Phi)=\Phi$, $\omega(\Phi)=\omega=\text{const.}$ and $U(\Phi)=0$, such that 
\begin{equation}\label{ActionBD original}
    S^{\myst{BD}}=\frac{1}{2\kappa_0}\int \dd^4 x\sqrt{-g}\left(\Phi R-\frac{\omega}{\Phi}g^{\mu\nu}\nabla_\mu\Phi\nabla_\nu\Phi\right)+S_\text{m}[g,\Psi_\text{m}]\,.
\end{equation}
This theory reduces to GR in the limit $\omega\rightarrow\infty$. Note that therefore, in particular due to the variability of the coupling $\omega$, the generalized version in Eq.~\ref{ActionBD} has more freedom in the beyond GR effects to only dominate at certain scales. Moreover, through redefinitions of the scalar field, one may always fix one of the functionals $F$ or $\omega$ to a definite form. For example $F(\Phi)\rightarrow \Phi$, by redefining $\Phi\rightarrow F^{-1}(\Phi)$ with according modifications in $\omega$ and $U$.\footnote{Note that since matter fields are universally coupled to the physical metric only, such change of variables of non-minimal fields are generally allowed without altering the underlying assumptions.}

Due to the non-minimal coupling of the scalar field with the Ricci scalar, the theory can be thought of as locally redefining the bare newtons constant $G$ to the quantity $G_\text{eff}\equiv G/F(\Phi)$ that is locally measured, which therefore may depend on time and the position. This freedom and the associated connection to Mach's principle were the initial motivation for considering such a theory \cite{Weinberg1972}. 
Furthermore, note that since the gravitational Lagrangian only involves first derivatives of the scalar field and linear second derivatives for the metric (recall the discussion in Sec.~\ref{sSec:LovelockTheorem}), the theory is clearly of type (A) and propagates 3 degrees of freedom. Thus, the additional scalar field in the theory introduces precisely one additional DOF as compared to GR, such that the representation of the theory in Eq.~\ref{ActionBD} is faithful according to Def.~\ref{DefFaithfulRep}.

As an exact theory, gBD modifies gravity already in the weak field at large scales and consequently finds applications in cosmology \cite{Clifton:2011jh}. On the other hand, as concerns BH physics the theory is equivalent to GR and only non-BH compact objects, such as stellar objects and neutron stars (NS) can be modified within gGB theory \cite{Will1989ApJ,Damour:1996ke,Harada:1997mr,Harada:1998ge}. This is because the \textit{no-hair theorem} \cite{IsraelPhysRev:1967aa,Israel:1968bb,CarterPhysRevLett:1971,Hawking:1972aa,misner_gravitation_1973,POMazur_1982,PhysRevLett.34.905,Poisson:2009pwt} of stationary black hole solutions still applies \cite{Hawking:1972bb,Sotiriou:2011dz}.\footnote{However, this is not true for homogeneous, cosmological solutions to the scalar field equations \cite{Yunes:2013dva}.} Through modified NS solutions, the theory can however still leave its imprints in strong field processes, in particular in the presence of \textit{spontaneous scalarization} \cite{TDamour_1992,DamourPhysRevLett:1993}.

Finally, a comment on the so called ``Einstein frame'' and general redefinitions of the metric. The theory written in Eq.~\eqref{ActionBD} is traditionally known as the \textit{Jordan frame} formulation of gBD that represents the natural formulation of a metric theory with a physical metric that couples minimally and universally to matter fields and a non-minimal scalar field that only couples to the metric. The non-minimal coupling between the scalar field and the Ricci scalar can however be cast away by introducing a non-physical metric $\tilde{g}_{\mu\nu}$ that is related to the physical metric $g_{\mu\nu}$ through a \textit{Weyl transformation} (see e.g. \cite{poisson2014gravity,Will:2018bme,carroll2019spacetime}) 
\begin{equation}\label{eq:ConfTransfEinsteinFrame}
    g_{\mu\nu}(x)\rightarrow\tilde{g}_{\mu\nu}(x)=F(\Phi)\,g_{\mu\nu}(x)\,.
\end{equation}
In this context it is important to note the fundamental difference between such a Weyl rescaling and a conformal transformation
\begin{equation}
     g_{\mu\nu}(x)\rightarrow g'_{\mu\nu}(x')=\frac{\partial x^\alpha}{\partial x'^\mu}\frac{\partial x^\beta}{\partial x'^\nu}\,g_{\alpha\beta}(x)=F(\Phi)\,g_{\mu\nu}(x)\,.
\end{equation}
The latter is just a special case of a general coordinate transformation in Eq.~\eqref{eq:GeneralCoordinateTransformation} that represent by assumption a symmetry of metric theories of gravity, while the former is a replacement of the physical metric with a rescaled metric without any change of coordinates that is not a symmetry of the action (see Appendix~\ref{sApp: Symmetries} for a definition of the notion of symmetries).

The resulting action after performing the Weyl rescaling in Eq.~\eqref{eq:ConfTransfEinsteinFrame} is termed \textit{Einstein frame} formulation of the theory. This ``frame'' can sometimes be convenient for computation purposes, in particular since in the Einstein frame formulation the corresponding DOF excitations of the metric and the scalar field are already decoupled (see \cite{Heisenberg:2023prj} and Part~\ref{Part: Gravitational Wave Testing Ground} for more details on this comment). However, one should in general be cautious when performing redefinitions of the metric. In particular, the terminology of ``frame transformations'' in the context can be misleading. Indeed, as already mentioned, the gBD action in Eq.~\eqref{ActionBD} is \emph{not} invariant under Weyl transformations, as opposed to general coordinate transformations, and, therefore, the transformation in Eq.~\eqref{eq:ConfTransfEinsteinFrame} is not a symmetry of the theory. Certainly, Weyl transformations together with any other redefinition of the fields are still part of the gauge freedom of a theory.\footnote{Recall the distinction between the notions of gauge freedom and gauge symmetry discussed in Sec.~\ref{sSec:Special Relativity} and Appendix~\ref{App: Symmetires in Physics}.} However, not all such descriptions are equivalent, in the sense that key assumptions that went into the construction of the theory, which can influence the interpretation of experimental data for instance, may no longer hold. This is in particular true for the Weyl transformations of the physical metric in Eq.~\eqref{eq:ConfTransfEinsteinFrame}, after which the assumptions of universal and minimal coupling are broken. Thus, the transformed metric $\tilde{g}_{\mu\nu}$ can no longer be interpreted to describe an objective spacetime that locally recovers all Minkowskian non-gravitational physics. In other words, if the new metric $\tilde{g}_{\mu\nu}$ would be treated as the usual physical spacetime metric of the manifold implying the assumptions of the existence of Fermi normal coordinates corresponding to the freely falling frames of spacetime, the resulting theory would explicitly violate the EEP and no longer represent a mere reformulation of gBD. The formulation of a theory in the Einstein frame is therefore dangerous insofar as standard assumptions and intuitions on the new metric do no longer hold.

This generally applies to all field dependent redefinitions of the physical metric that are not explicit symmetry transformations of the matter action. In the case of Weyl transformations the implications on the matter sector remain however minimal, since the change in the metric simply corresponds to a field dependent rescaling of physical distances (see also \cite{Faraoni:2006fx}).\footnote{For instance, the matter action of an abelian vector field is invariant under Weyl rescalings.} It is therefore sometimes stated that the Jordan and Einstein frame metrics are equivalent up to rescaling of physical distances. Yet, in practice, one cannot artificially rescale physical distances depending on an unknown external field in order to recover results in agreement with the EEP, such that all observable effects should imperatively be computed within the original, physical Jordan frame spacetime metric \cite{poisson2014gravity,Will:2018bme}. This conclusion holds for all metric theories considered below.

\paragraph{Scalar Gauss-Bonnet Gravity (sGB).}
The theory of scalar Gauss-Bonnet gravity, also known as Einstein-dilaton-Gauss-Bonnet, is given by
\cite{Zwiebach:1985uq,Gross:1986iv,Boulware:1986dr,Moura:2006pz,Nojiri:2005vv,Nojiri:2006je,Pani:2009wy,Pani:2011xm}
\begin{equation}\label{eq:ActionsGB}
    S^{\myst{sGB}}=\frac{1}{2\kappa_0}\int \dd^4 x\sqrt{-g}\bigg( R-\frac{1}{2}g^{\mu\nu}\nabla_\mu\Phi\nabla_\nu\Phi
    +f(\Phi)\,\mathcal{G}_{\myst{GB}}\bigg)+S_\text{m}[g,\Psi_\text{m}]\,,
\end{equation}
where the Gauss-Bonnet curvature scalar $\mathcal{G}_{\myst{GB}}$ is defined in Eq.~\eqref{eq:GBScalar} and $f$ is an arbitrary (smooth) function. For constant values of $\Phi$ the theory reduces to GR because the Gauss-Bonnet term integrates to a boundary term. Note that the function $f$ necessarily involves a coupling $\epsilon^2\sim 1/\Lambda^2$ of dimension $[\epsilon^2]=E^{-2}$, such that at typical energy scales $E$ the interaction operator is in principle suppressed by a factor of $E^2/\Lambda^2$ compared to the kinetic terms. In the following, such mass scale factors that indicate the naive scale of cutoff of an EFT are to be understood implicitly whenever necessary.

The theory was first considered because of a Gauss-Bonnet curvature scalar coupling to the dilation arising in the context of low-energy effective string theory \cite{Zwiebach:1985uq,Gross:1986iv,Moura:2006pz,Nojiri:2017ncd}. Such a coupling also arises in the lowest order of the most general expansion of vacuum quantum operators of a metric coupled with a scalar \cite{Weinberg:2008hq} as we will discuss more closely in Sec.~\ref{ssSec: B Perturbative Theories} below. However, unlike other string-inspired gravity theories, sGB can still be treated as exact. This is because the resulting equations of motion remain at second-order in derivatives per field, which implies that the theory is degenerate and therefore Ostrogradsky-stable \cite{Nojiri:2017ncd}. Moreover, the theory only propagates three healthy DOFs and Eq.~\eqref{eq:ActionsGB} corresponds to a faithful representation.
Together with the existence of non-trivial black hole solutions that evade the no-hair theorem \cite{Kanti:1995vq,Pani:2009wy,Pani:2011xm,Maselli:2015tta,Blazquez-Salcedo:2016enn}, this makes the theory attractive for both cosmological applications, as well as strong field environments (see \cite{DeFelice:2010aj,Clifton:2011jh,Nojiri:2017ncd}, respectively \cite{Yunes:2013dva,Berti:2015itd,Silva:2017uqg,Elley:2022ept,Witek:2018dmd,Okounkova:2019zjf,Okounkova:2020rqw,East:2022rqi,Corman:2022xqg} and references therein). In Sec.~\ref{ssSec:WellPosedness} we will further comment on the existence of well-posed formulations of the theory that can be evolved numerically.

\paragraph{Double-Dual Riemann Gravity (ddR).}
There exist another Riemann curvature combination, for which a non-minimal derivative coupling to a scalar field preserves the structure of equations of motion with at most two derivatives per field, namely the double-dual Riemann tensor
\begin{align}\label{eq:ddR}
    L^{\mu\nu\alpha\beta}&\equiv\,\frac{1}{4}\epsilon^{\mu\nu\gamma\delta} R_{\gamma\delta\rho\sigma} \epsilon^{\rho\sigma\beta\alpha}\,,\\
    &=\,R^{\mu\nu\alpha\beta}+\big(R^{\mu\beta}g^{\nu\alpha}
    +R^{\nu\alpha}g^{\mu\beta}-R^{\mu\alpha}g^{\nu\beta}-R^{\nu\beta}g^{\mu\alpha}\big)+\frac{1}{2}R\big(g^{\mu\alpha}g^{\nu\beta}-g^{\mu\beta}g^{\nu\alpha}\big)\,.\nonumber
\end{align}
The associated gravitational action of so called double-dual Riemann gravity reads \cite{deRham:2011by,Charmousis:2011ea,Charmousis:2011bf}
\begin{equation}\label{ActionddR}
\begin{split}
    S_\text{G}^{\myst{ddR}}=\frac{1}{2\kappa_0}\int \dd^4 x\sqrt{-g}\bigg( R&+\nabla_\mu\Phi \nabla_\nu\Phi\,\Phi_{\alpha\beta}\,L^{\mu\nu\alpha\beta}-\frac{1}{2}g^{\mu\nu}\nabla_\mu\Phi\nabla_\nu\Phi\bigg)\,.
\end{split}
\end{equation}
Due to its degeneracy, this action can therefore also be regarded as exact and represents a faithful formulation.

\paragraph{f(R) Gravity.}
A very popular degenerate and thus Ostrogradsky stable departure from GR is also given by promoting the gravitational action to a general function $f$ of the Ricci scalar, known as $f(R)$ gravity \cite{Bergmann:1968aj,Ruzma:1969JETP,Buchdahl:1970MN,Sotiriou:2008rp,DeFelice:2010aj}
\begin{equation}\label{eq:Actionf(R)}
    S^{\myst{f(R)}}=\frac{1}{2\kappa_0}\int \dd^4 x\sqrt{-g}\,f(R)+S_\text{m}[g,\Psi_\text{m}]\,.
\end{equation}
This theory is indeed free of any Ostrogradsky instabilities \cite{MFerraris_1988,Woodard:2006nt,Sotiriou:2008rp,DeFelice:2010aj} and was successfully employed to construct alternative cosmological models (see \cite{Sotiriou:2008rp,DeFelice:2010aj,Clifton:2011jh,Faraoni2011,Nojiri:2017ncd} and references therein). However, $f(R)$ gravity is the first example of an unfaithful theory, as in general, the theory propagates an additional degree of freedom without explicitly introducing new non-minimal fields. 

The action in Eq.~\eqref{eq:Actionf(R)} is in fact equivalent to a particular subset of generalized Brans-Dicke gravity considered above, which makes it a scalar-tensor theory in disguise. Indeed, one can reformulate the theory by replacing the gravitational Lagrangian by $f(\Phi)+f'(\Phi)(R-\Phi)$, where $\Phi$ is a dynamical scalar. This theory is indeed equivalent to the action in Eq.~\eqref{eq:Actionf(R)}, since a variation with respect to the scalar field yields 
\begin{equation}
    f''(\Phi)(R-\Phi)=0\,,
\end{equation}
which implies $\Phi=R$ as long as $f''(\Phi)\neq 0$ \cite{OHanlon:1972xqa,Teyssandier:1983zz,Chiba:2003ir,Sotiriou:2008rp,DeFelice:2010aj,Will:2018bme}. Thus, $f(R)$ theories are but a subset of gBD gravity with\footnote{Note that one could further redefine a new scalar field $\tilde{\Phi}=f'(\Phi)$.}
\begin{equation}
    \omega(\Phi)=0\,,\quad U(\Phi)=\Phi f'(\Phi)-f(\Phi)\,.
\end{equation}

One can also define more general exact theories involving general functionals of the Gauss-Bonnet scalar in Eq.~\eqref{eq:GBScalar}, hence, $f(\mathcal{G}_{\myst{GB}})$ or even $f(R,\mathcal{G}_{\myst{GB}})$. The theory of $f(\mathcal{G}_{\myst{GB}})$ is however equivalent to the sGB gravity considered above, while $f(R,\mathcal{G}_{\myst{GB}})$ provide a mix between gBD and sGB with two scalar degrees of freedom \cite{DeFelice:2010aj}.

\paragraph{Horndeski Gravity.}
So far, we gathered a collection of scalar-tensor theories, which are united by their property of having equations of motion at second-order in derivatives per fields, thus evading any Ostrogradsky instabilities. One could therefore ask: what is the most general action of a scalar-tensor theory with this feature. The answer is given by the Horndeski action \cite{Horndeski:1974wa,Nicolis:2008in,Deffayet:2009wt,Deffayet:2009mn,Heisenberg:2018vsk,Kobayashi:2019hrl}
\begin{equation}\label{eq:ActionHorndeski}
    S^{\myst{H}}=\frac{1}{2\kappa_0}\int \dd^4 x\sqrt{-g}\left(\sum_{i=2}^5L^{\myst{H}}_i\right)+S_\text{m}[g,\Psi_\text{m}]\,,
\end{equation}
where
\begin{align}
        L^{\myst{H}}_2=&\,G_2(\Phi,X)\,,\\
        L^{\myst{H}}_3=&-G_3(\Phi,X)\Box\Phi\,,\\
        L^{\myst{H}}_4=&\,G_4(\Phi,X)\,R+G_{4X}\left[(\Box\Phi)^2-\Phi^{\mu\nu}\Phi_{\mu\nu}\right]\,,\\
        L^{\myst{H}}_5=&\,G_5(\Phi,X)\,G^{\mu\nu}\Phi_{\mu\nu}-\frac{G_{5X}}{6}\Big[(\Box\Phi)^3-3\,\Box\Phi\,\Phi^{\mu\nu}\Phi_{\mu\nu}+2\,\Phi_{\mu\nu}\Phi^{\nu\lambda}\Phi\du{\lambda}{\mu}\Big]\,,
\end{align}
with $\Phi_{\mu\nu}\equiv\nabla_{\mu}\nabla_{\nu}\Phi$, and where the $G_i$'s are arbitrary functionals of $\Phi$ and the kinetic combination $X\equiv - ({1}/{2}) \nabla_\mu\Phi\nabla^\mu\Phi$.\footnote{Note that up to integrations by parts, a term with $G_3(\Phi,X)=\Phi$ is equivalent to the kinetic term of the scalar, such that we specifically exclude such a term from $G_3$.} Moreover, we define $G_{iZ}\equiv \partial G_i/\partial Z$ for any operator $Z$ and recall the definition of the Einstein tensor $G_{\mu\nu}$ in Eq.~\eqref{eq:EinsteinTensor}.

As a little historical side-note, higher order derivative self interactions of the scalar field, also known as \textit{Galileon} interactions \cite{Nicolis:2008in}, naturally arise as the zero-helicity part of the graviton in higher dimensional models \cite{Dvali:2000hr} (see \cite{Hinterbichler:2011tt,deRham:2014zqa} for reviews). In Sec.~\ref{sSec:FlatSpaceGalileons} we will study Galileon theories in more detail, which will provide an understanding of the structure of the Lagrangian in Eq.~\eqref{eq:ActionHorndeski} through the construction of the corresponding most general Galileon theories in flat spacetime. A covariantization of these flat-space Galileon theories \cite{Deffayet:2009wt,Deffayet:2009mn} lead to a rediscovery of the work by Horndeski \cite{Horndeski:1974wa}.

As the most general scalar-tensor theory with second order equations of motion, the action in Eq.~\eqref{eq:ActionHorndeski} can actually be thought of as defining a large class of theories. In particular, it encompasses all the ST theories discussed above and inherits their applications to modifications in strong gravity regimes, as well as in cosmology \cite{Chow:2009fm,DeFelice:2010nf,Deffayet:2010qz,Appleby:2011aa,Deffayet:2011gz,Kobayashi:2011nu,Appleby:2012ba,Barreira:2012kk,Okada:2012mn,Bartolo:2013ws,Creminelli:2012my,Neveu:2013mfa,Barreira:2013jma,Barreira:2013eea,Gleyzes:2014dya,Heisenberg:2018vsk,Kobayashi:2019hrl}. The different subsets of gravity theories can be accessed through particular choices of the general functionals $G_i$. For example, Horndeski gravity reduces to BD theory for the choices
\begin{subequations}\label{eq:BDGs}
\begin{align}
    G_2&=\frac{2\omega}{\Phi} X\,\\
    G_4&=\Phi\,,\\
    G_i&=0 \;\;\text{otherwise}\,.
\end{align}
\end{subequations}
Moreover, as discussed above, $f(R)$ gravity is equivalent to a subset of gBG theory, and is thus also included in the Horndeski framework under the choices
\begin{subequations}\label{eq:f(R)s}
\begin{align}
    G_2&=f(\Phi)-\Phi f'(\Phi)\,\\
    G_4&=f'(\Phi)\,,\\
    G_i&=0 \;\;\text{otherwise}\,,
\end{align}
\end{subequations}
assuming that $f''(\Phi)\neq 0$.
On the other hand, and less trivially, sGB gravity can be obtained by choosing \cite{Kobayashi:2011nu,Kobayashi:2019hrl}
\begin{subequations}\label{eq:CorrespondencesGBHorndeski}
\begin{align}
        G_2&=X+8f^{(4)}(\Phi)X^2(3-\ln X)\,,\\ G_3&=4f^{(3)}(\Phi)X(7-3\ln X)\,,\\
        G_4&=1+4f^{(2)}(\Phi)X(2-\ln X)\,,\\
        G_5&=-f^{(1)}(\Phi)\ln X\,,
\end{align}
\end{subequations}
where $f^{(n)}(\Phi)\equiv\partial^n f/\partial\Phi^n$. Note that while this correspondence is not obvious at the level of the action, the resulting equations of motion are indeed equivalent. Similarly, choosing \cite{Narikawa:2013pjr,Kobayashi:2019hrl}
\begin{subequations}\label{eq:CorrespondenceddR}
    \begin{align}
        G_2&=X\,,\\
        G_5&=X\,,\\
        G_i&=0 \;\;\text{otherwise}\,,
    \end{align}
\end{subequations}
one recovers ddR gravity.

\paragraph{Degenerate Higher-Order Scalar-Tensor Gravity (DHOST).}
The Horndeski action represents the most general ST theory, with equations of motion at second-order in derivatives per fields. As already mentioned, a restriction to second-order EOMs of the fields ensures the absence of ghosts but is not, however, a necessary condition for a theory to be free of the Ostrogradsky instability. In particular for theories with multiple fields that can lead to a kinetic mixing of DOFs, the appearance of higher-order terms in the equations of motion of the unperturbed tensor fields must not imply the presence of a ghost excitation, as long as the system is still degenerate in its kinetic structure. In this case, the theory can be reformulated at the level of the perturbations to only involve second-order equations of motion. ST theories that use this loophole to go beyond the Horndeski framework are known as degenerate higher-order scalar-tensor theories (see \cite{deRham:2016wji,Heisenberg:2018vsk,Kobayashi:2019hrl} and references therein).

A first example of such theories and also the most relevant one can be obtained by applying an invertible disformal transformation through the replacement of the physical metric by\footnote{This transformation is invertible as long as $C(C-XC_X+2X^2D_X)\neq 0$ \cite{Kobayashi:2019hrl}.}
\begin{equation}\label{eq:DisformalTransf}
    g_{\mu\nu}\rightarrow \tilde{g}_{\mu\nu}=C(\Phi, X) g_{\mu\nu}+D(\Phi, X) \nabla_\mu\Phi\nabla_\nu\Phi\,.
\end{equation}
This leads to the so-called class Ia of quadratic DHOST theories that are by construction still degenerate despite the appearance of higher-order terms in the equations of motion. Note that, while at first sight the transformation in Eq.~\eqref{eq:DisformalTransf} appears like a gauge transformation (in the sense of Appendix~\ref{sApp: Gauge Freedom}), such that the resulting theory described by the new metric $\tilde{g}_{\mu\nu}$ should be equivalent to the original theory, a new theory can be obtained by performing the disformal transformation in the gravity sector only, while keeping a minimal and universal coupling to the new metric $\tilde{g}_{\mu\nu}$ within the matter Lagrangian\footnote{In other words, Horndeski theory is only equivalent to the Ia DHOST theory class with disformally coupled matter.} (recall the discussion above on the distinction between the Jordan and the Einstein frame). There exist a multitude of other possibilities to go beyond Horndeski, however, none of which seem to provide viable theories, at least as concerns applications to cosmology \cite{deRham:2016wji,Kobayashi:2019hrl}.

\subsubsection{\ul{Vector-Tensor Theories}}

An obvious generalization of the scalar-tensor theories presented above is to consider theories that also propagate vector DOFs. Naturally, such theories are described by the introduction of a non-minimal vector field in the gravity sector. As discussed, for parity preserving theories in four spacetime dimensions with local Lorentz invariance, massless vector DOFs naturally come in pairs of two as dictated by the Winger-classification of the solutions of relativistic wave-equations \cite{Bargmann:1948ck}. Such a description in terms of a vector field $A_\mu$ requires then the introduction of an internal gauge freedom
\begin{equation}\label{eq:GaugeTransformationVector2s}
    A_{\mu}\rightarrow A_\mu +\partial_\mu \Lambda\,,
\end{equation}
that is promoted to a symmetry of the action by only introducing the vector field in terms of the gauge invariant combination
\begin{equation}
    F_{\mu\nu}\equiv \nabla_\mu A_{\nu}-\nabla_\nu A_\mu=\partial_\mu A_\nu-\partial_\nu A_\mu\,,
\end{equation}
known as the \textit{field strength}.

On the other hand, the introduction of a mass term inevitably introduces an additional longitudinal scalar degree of freedom, which in the simplest case of an abelian vector field can be understood from the explicit breaking of the $U(1)$ gauge symmetry that was necessary in order to regulate the redundant excitations in the theory described in terms of a four-component vector field. This leads to a clear distinction between theories involving massive and massless non-minimal vector fields. While Horndeski theory represents an extensive set of scalar-tensor theories with second-order equations of motion, the equivalent construction for massive scalar-vector-tensor theories is therefore slightly richer in structure due to the necessity of introducing constraints for the vector field as well as for the metric. Below we will discuss a pure massless vector-tensor theory, while the massive case will be treated in the following scalar-vector-tensor paragraph.

\paragraph{Vector Horndeski Gravity (VH).}

Due to a no-go theorem for massless Galileon like vector interactions on flat spacetime \cite{Deffayet:2013tca}, the theory space of Ostrogradsky-stable vector-tensor theories described in terms of a metric and an abelian $U(1)$ gauge field is very restricted. There exists only one allowed non-minimal coupling to the double-dual Riemann tensor $L^{\mu\nu\alpha\beta}$ defined in Eq.~\eqref{eq:ddR}, which leads to the following most general vector-tensor theory with second order equations of motion \cite{Horndeski:1976gi,Barrow:2012ay}
\begin{equation}\label{eq:ActionVectorHorndeski}
    S^{\myst{VH}}=\frac{1}{2\kappa_0}\int \dd^4 x\sqrt{-g}\left(R+G_2(F,\tilde{F})+L^{\mu\nu\alpha\beta}F_{\mu\nu}F_{\alpha\beta}\right)\,.
\end{equation}
In the pure vector sector captured by the arbitrary function $G_2$, we defined the scalar quantities $F\equiv - ({1}/{4}) F^{\mu\nu}F_{\mu\nu}$ and $\tilde{F}\equiv F^{\mu\nu}\tilde{F}_{\mu\nu}$, with the Hodge dual
\begin{equation}\label{eq:HodgeDualVectorFieldStrength}
    \tilde{F}_{\mu\nu}\equiv\frac{1}{2}\epsilon_{\mu\nu\alpha\beta}F^{\alpha\beta}\,.
\end{equation}

This action can be regarded as an exact theory. Nevertheless, its applications to cosmology are limited due to the natural breaking of isotropy for non-trivial vector field background, as well as the general suppression of massless vector fields in an expanding universe. A generalization to non-abelian fields may circumvent such constraints (see e.g. \cite{Maleknejad:2011sq,Davydov:2015epx,Caldwell:2016sut,BeltranJimenez:2018ymu} and references therein). In particular, in the case of an internal non-abelian $SO(3)$ symmetry, homogeneity at the background level can be restored in a so called ``triad configuration'', for which the background consists of three orthogonal vector fields of identical values. 

\subsubsection{\ul{Scalar-Vector-Tensor Theories}}

\paragraph{Scalar-Vector Heisenberg-Horndeski Gravity (SVHH).}

Combining the scalar and vector Horndeski theories discussed above, one can construct a scalar-vector-tensor (SVT) theory that in addition to the actions given in Eqs.~\eqref{eq:ActionHorndeski} and \eqref{eq:ActionVectorHorndeski} include higher derivative scalar-vector interactions with second-order equations of motion \cite{Heisenberg:2018acv}
\begin{equation}\label{eq:ActionSVH}
    S^{\myst{SVH}}=\frac{1}{2\kappa_0}\int \dd^4 x\sqrt{-g}\left(\sum_{i=3}^5L^{\myst{H}}_i+\sum_{i=2}^4L_i\right)+S_\text{m}[g,\Psi_\text{m}]\,,
\end{equation}
where
\begin{align}
        L_2=&\,G_2(\Phi,X,Y,F,\tilde{F})\,,\\
        L_3=&\Big[\hat{G}_3(\Phi,X)\,g_{\alpha\beta}+\doublehat{G}_3(\Phi,X)\nabla_\alpha\Phi\nabla_\beta\Phi\Big]\tilde{F}^{\mu\alpha}\tilde{F}^{\nu\beta}\Phi_{\mu\nu}\,,\\
        L_4=&\,\hat{G}_4(\Phi,X)L^{\mu\nu\alpha\beta}F_{\mu\nu}F_{\alpha\beta}+\left[\doublehat{G}_4(\Phi)+\frac{1}{2}\hat{G}_{4X}\right]\tilde{F}^{\mu\alpha}\tilde{F}^{\nu\beta}\Phi_{\mu\nu}\Phi_{\alpha\beta}\,.
\end{align}
Recall that
\begin{equation}
    \Phi_{\mu\nu}\equiv\nabla_{\mu}\nabla_{\nu}\Phi\,,\; X\equiv - ({1}/{2}) \nabla_\mu\Phi\nabla^\mu\Phi\,,\; F\equiv - ({1}/{4}) F^{\mu\nu}F_{\mu\nu}\,,\;\tilde{F}\equiv F^{\mu\nu}\tilde{F}_{\mu\nu}\,,
\end{equation}
with $\tilde{F}_{\mu\nu}$ the Hodge dual in Eq.~\eqref{eq:HodgeDualVectorFieldStrength}. Furthermore, we have defined the additional mixed quantity $Y\equiv\nabla_\mu\Phi\nabla_\nu\Phi F^{\mu\alpha}F\ud{\nu}{\alpha}$.
The corresponding equations of motion associated with the action presented above can, for instance, be found in the Appendix of \cite{Heisenberg:2018mxx,Kobayashi:2011nu}.

This SVT theory propagates a scalar, two massless vector and two massless tensor DOFs. Imposing $G_{2\Phi^2}(\Phi,0,0,0,0)= 0$ also renders the scalar DOF massless and makes the theory the most general massless SVT of its kind with explicit second order equations of motion. For a constant scalar field, the theory reduces to the vector Horndeski gravity in Eq.~\eqref{eq:ActionVectorHorndeski}, while $\nabla_\mu A_\nu=0$ recovers scalar Horndeski theory of Eq.~\eqref{eq:ActionHorndeski}.

\paragraph{Generalized Proca Gravity (GP).}

An even richer structure of exact theories can be obtained by considering massive vector excitations. As discussed, through local Lorentz invariance, this naturally leads to an SVT theory. Using a description in terms of a single massive non-minimal vector field, the following action again captures all possible terms with second-order equations of motion \cite{Heisenberg:2014rta,Allys:2015sht,BeltranJimenez:2016rff}
\begin{equation}\label{eq:ActionGenProca}
    S^{\myst{GP}}=\frac{1}{2\kappa_0}\int \dd^4 x\sqrt{-g}\left(\sum_{i=2}^6L^{\myst{GP}}_i\right)+S_\text{m}[g,\Psi_\text{m}]\,,
\end{equation}
where
\begin{align}
        L^{\myst{GP}}_2=&\,G_2(A_\mu,F_{\mu\nu},\tilde{F}_{\mu\nu})\,,\\
        L^{\myst{GP}}_3=&\,G_3(Z)\nabla_\mu A^\mu\,,\\
        L^{\myst{GP}}_4=&\,G_4(Z)\,R+G_{4X}\left[(\nabla_\mu A^\mu)^2-\nabla_\mu A_\nu\nabla^\nu A^\mu\right]\,,\\
        L^{\myst{GP}}_5=&\,G_5(Z)\,G^{\mu\nu}\nabla_\mu A_\nu-\frac{G_{5Z}}{6}\Big[(\nabla_\mu A^\mu)^3-3\,\nabla_\mu A^\mu\,\nabla_\alpha A_\beta \nabla^\beta A^\alpha
        \\
        &+2\nabla_\mu A_\nu \nabla^\rho A^\mu\nabla^\nu A_\gamma-\hat{G}_5(Z)\tilde{F}^{\mu\rho}\tilde{F}\ud{\nu}{\rho}\nabla_\mu A_\nu\,\Big]\notag\,,\\
        L^{\myst{GP}}_6=&\,G_6(Z)\,L^{\mu\nu\rho\sigma}\nabla_\mu A_\nu \nabla_\rho A_\sigma-\frac{G_{6Z}}{2} \tilde{F}^{\mu\nu}\tilde{F}^{\rho\sigma}\nabla_\mu A_\rho \nabla_\nu A_\sigma\,,
\end{align}
with $Z\equiv - ({1}/{2}) A_\mu A^\mu$. Again, the quantity $L^{\mu\nu\alpha\beta}$ represents the double-dual Riemann tensor defined in Eq.~\eqref{eq:ddR}. Just as it was the case for Horndeski theory, this metric theory of gravity naturally arises as a covariant generalization of the most general Lorentz invariant flat-space massive vector theory with second order equations of motion \cite{Heisenberg:2014rta,Allys:2015sht,BeltranJimenez:2016rff}.
In Sec.~\ref{Sec:GenProca Quantum Stability} we will further analyze this flat-space counterpart of GP theory and especially also understand its unique structure that ensures the second order nature of the equations of motion. 

Similar to the SVHH theory above, this SVT propagates five DOFs, one scalar, two vectors and two tensors, with the difference that the vector modes are massive, and the scalar is hidden in the field $A_\mu$ which does not involve any gauge invariance.\footnote{An additional constraint imposed by the equations of motion ensures that the vector field indeed only describes three propagating DOFs.} A more faithful representation can be recovered through the St\"ukelberg mechanism \cite{Stueckelberg:1900zz,GREEN1991462,Siegel:1993sk,Arkani-Hamed:2002bjr,Ruegg:2003ps} that in this case explicitly introduces a scalar field into the theory by adding a redundancy in the description in the form of a gauge symmetry. This is obtained by effectively replacing the vector field by
\begin{equation}\label{eq:StuckelbergReplacementA}
    A_\mu\rightarrow A_\mu +\frac{1}{m}\nabla_\mu\Phi\,,
\end{equation}
for some mass scale $m$ that can be fixed by canonical normalization of the kinetic term of the scalar. It is important to note that the replacement in Eq.~\eqref{eq:StuckelbergReplacementA} does not represent a decomposition of $A_\mu$ into its transverse and longitudinal parts (see also \cite{Hinterbichler:2011tt}). Rather, it reformulates the theory by adding a redundancy in the description in terms of a new scalar field $\Phi$ and a gauge symmetry under the transformation
\begin{equation}
    \Phi\rightarrow \Phi + m \alpha\,, \quad A_\mu\rightarrow A_\mu -\nabla_\mu \alpha\,.
\end{equation}
The replacement in Eq.~\eqref{eq:StuckelbergReplacementA} is such that the field strength of the vector field remains untouched, while it naturally introduces a ``covariant derivative'' of the form $D_\mu\Phi\equiv\partial_\mu\phi+mA_\mu$, such that from Eq.~\eqref{eq:StuckelbergReplacementA} the new theory is defined through the replacements\footnote{In fact, the St\"ukelberg replacement is very closely related to an explicit reintroduction of an eaten Goldstone boson in the context of spontaneous symmetry breaking.}
\begin{equation}\label{eq:EffectiveStuckelbergRep}
    A_\mu\rightarrow\frac{1}{m}D_\mu\Phi\,,\quad F_{\mu\nu}\rightarrow F_{\mu\nu}\quad\text{and}\quad \tilde{F}_{\mu\nu}\rightarrow \tilde{F}_{\mu\nu}\,.
\end{equation}
For a unitary gauge choice of $\alpha=-\Phi/m$ that sets $\Phi=0$, we clearly recover the GP theory, such that the two descriptions are indeed equivalent.

As in the scalar Horndeski case, additional exact theories can be constructed from a massive vector field and a metric. For instance, using the disformal transformation trick discussed for the scalar Horndeski theories above, one can construct additional ghost free beyond GP interactions (see also \cite{Heisenberg:2016eld,Domenech:2018vqj}). Interestingly, it is also possible to construct a massive gravity inspired infinite tower of massive vector interactions that only propagate three healthy degrees of freedom despite higher-order equations of motion in the scalar-vector sector \cite{deRham:2020yet}. 
Obviously, one can also generalize the GP action by introducing yet another scalar DOF captured by an explicit scalar field, as for instance considered in \cite{Heisenberg:2018acv,Heisenberg:2018mxx}. The resulting SVT theory thus propagates six degrees of freedom in total. Finally, as in the massless case, also ``non-abelian'' multi Proca theories represent an attractive generalization with a multitude of interesting phenomenology in cosmology and gravitational wave signals (see e.g. \cite{Bento:1992wy,Golovnev:2008cf,Esposito-Farese:2009wbc,Allys:2016kbq,Rodriguez:2017wkg,BeltranJimenez:2016afo,BeltranJimenez:2018ymu}).

While gravity theories with non-minimal vector fields could at first sight not seem ideal candidates for applications to cosmology, for a massive vector field, there are various background configurations that are still compatible with a homogeneous and isotropic background (see e.g. \cite{Heisenberg:2018vsk}). Concrete cosmological applications of Proca theories and their generalizations can for instance be found in \cite{Boehmer:2007qa,Golovnev:2008cf,Jimenez:2013qsa,BeltranJimenez:2013fca,Tasinato:2013oja,Hull:2014bga,Khosravi:2014mua,Tasinato:2014eka,Hull:2015uwa,Jimenez:2015fva,Jimenez:2016opp,Heisenberg:2016eld,Kimura:2016rzw,Heisenberg:2016lux,Jimenez:2016upj,Allys:2016kbq,Lagos:2016wyv,DeFelice:2016yws,DeFelice:2016uil,Heisenberg:2016wtr,Emami:2016ldl,Rodriguez:2017wkg,deFelice:2017paw,Heisenberg:2018acv,Petrov:2018xtx,ErrastiDiez:2019trb}.
On the other hand, vector theories might of course also represent prime candidates to describe slight departures from the basic paradigms of cosmology.

\paragraph{Einstein-\AE{}ther Gravity (E\AE{}).} At this point we also want to mention Einstein-\AE{}ther gravity \cite{Jacobson:2000xp,Jacobson:2004ts,Eling:2005zq,Jacobson:2007veq,Bonvin:2007ap,Yagi:2013ava,Will:2018bme}
\begin{equation}\label{eq:Action EA}
    S^{\myst{\AE{}}}=\frac{1}{2\kappa_0}\int \dd^4 x\sqrt{-g}\left(R-K\ud{\alpha\beta}{\mu\nu}\nabla_\alpha A^\mu\nabla_\beta A^\nu\right)+\lambda (2Z-1)+S_\text{m}[g,\Psi_\text{m}]\,,
\end{equation}
where
\begin{equation}
    K\ud{\alpha\beta}{\mu\nu} = c_1\, g_{\alpha\beta}g_{\mu\nu}+c_2 \,\delta^\alpha_\mu\delta^\beta_\nu+c_3 \,\delta^\alpha_\nu\delta^\beta_\mu-c_4\,A^\alpha A^\beta g_{\mu\nu}\,,
\end{equation}
and in particular contrast it to the GP theory discussed above. Here $c_i$ define dimensionless coupling constants and $\lambda $ is a scalar field that serves as a Lagrange multiplier. E\AE{} metric gravity is defined as the most general action up to two powers of derivative operators constructed out of a physical metric $g_{\mu\nu}$ and a vector field $A_\mu$, where in contrast to GP gravity the vector field is \textit{a priori} constrained to represent a unit time-like vector field. In practice, this last constraint is incorporated through the Lagrange multiplier term in the action of the form $\lambda (g_{\mu\nu}A^\mu A^\nu + 1)$. Moreover, note that the omission of a term in the action of the form $R_{\rho\sigma} A^\rho A^\sigma$ is justified as up to total derivatives it can be expressed in terms of the difference of the $c_3$ and $c_2$ terms.

Just as GP gravity, the theory involves five propagating degrees of freedom, one scalar, two vector and two tensor DOFs. The fundamental difference lies however in the explicit local Lorentz breaking of the theory that in the case of E\AE{} gravity is introduced ``by hand''. In contrast, GP gravity can only spontaneously break local Lorentz invariance in the gravitational sector through special solutions of the vector field. In this context, we also want to point out that due to the additional constraint on the vector field, E\AE{} gravity is able to incorporate operators in the action that were not allowed in the GP case, since GP merely relies on internal constraints. Moreover, E\AE{} gravity explicitly restricts its construction to only involve up to two derivative operators in the action, which similarly to GP gravity could in principle however be extended to higher powers of derivatives without altering the number of propagating degrees of freedom.

\subsection{Screening}\label{ssSec:Screening}

The set of metric theories introduced above all exhibit departures from general relativity through the presence of additional non-minimal degrees of freedom that modify the equations of motion of the physical metric. The fact that general relativity has already been tested to high accuracy, especially in the weak field regime and on solar system scales, makes it therefore advantageous for models to exhibit a more or less natural way of recovering GR in these regimes. Such behaviors are known as \textit{screening mechanisms}, of which we will now offer a brief overview (see also \cite{Brax:2013ida,Deffayet:2015rzg,papantonopoulos2014modifications} for a review).

But first, we want to address a common misconception on the necessity of screening mechanisms. Namely, it is often stated that screening is necessary in order to comply with constraints on so called ``fifth force'' experiments. However, in the case of metric theories this statement is inaccurate, since, as already discussed, strictly speaking and by definition, metric theories of gravity do not give rise to any additional forces of nature that can locally act on test particles. In other words, even though the solution for the physical metric as sourced by some mass distribution might depart from the GR solution due to the presence of non-minimal fields in the metric equations of motion, the Einstein equivalence Principle~\ref{Principle:EEP} is still satisfied, and no local experiment will be able to detect any additional force. For metric theories of gravity, screening is therefore only necessary if the predictions for experiments involving self-gravitating objects would contradict measurements. Thus, in particular, screening might be necessary in the context of tests of the strong equivalence principle discussed in Sec.~\ref{sSec:Strong Equivalence Principle} through the Nordtvedt effect \cite{Nordtvedt:1968first,Nordtvedt:1968qs}.

This misconception mainly arises, because screening and equivalence principle tests are commonly treated in terms of an unphysical metric in the so-called Einstein frame (see discussion above) that facilitates computations as the non-linear mixing between the metric and the non-minimal fields are transformed away. In such a formulation, it appears as if a non-minimal field would actually directly couple to the energy-momentum tensor of matter. It is, however, imperative that such a coupling only arises due to a non-trivial redefinition of the physical metric and should by no means be taken too literal. In particular the natural freely falling frames of typical experiments are given by the normal coordinates with respect to the physical metric. In that sense, it is much safer to state results in terms of the physical point of view, in which non-minimal fields are not directly ``sourced'' by matter, but only indirectly through the solution of the physical metric.

\paragraph{Vainshtein Screening.} 

One of the most interesting screening mechanisms in the \textit{Vainshtein screening} \cite{Vainshtein:1972sx,Arkani-Hamed:2002bjr,Deffayet:2001uk,Babichev:2009jt,Babichev:2013usa,Heisenberg:2018vsk,Kobayashi:2019hrl} that is naturally present in theories with non-linear derivative self-interactions and therefore applies to the Horndeski type theories discussed above, in particular also GP gravity \cite{DeFelice:2016cri}. This screening mechanism, originally found as a solution to a puzzle regarding the massless limit of massive graviton theories, relies on the presence of an additional length scale in the theory, known as the \textit{Vainshtein radius}, that captures the scale at which the non-linear derivative interactions become important. As soon as the non-linear terms are non-negligible as compared to the kinetic term, the non-linear couplings between the physical metric and the non-minimal fields are naturally suppressed. This can be understood by noting that non-linear derivative interactions provide corrections to the leading kinetic term that, through canonical normalization to obtain a new effective kinetic behavior, translate into an effective suppression of the non-minimal couplings.

Thus, while on distance scales above the Vainshtein radius the influence of the non-minimal fields is unconstrained, their effect on smaller scales on modifying the physical metric is naturally suppressed due to the presence of the non-linear derivative interactions. It is therefore interesting to note that this effect does not rely on the suppression of the presence of a non-trivial non-minimal field on small scales, but on the contrary on the high non-linearity of the solution. On top of being rather natural, this comes with certain advantages compared to other screening mechanisms, in particular the chameleon mechanism.

\paragraph{Chameleon Mechanism.}

In contrast to the Vainshtein mechanism, the chameleon mechanism \cite{Mota:2003tc,Khoury:2003rn,Khoury:2003aq,Cembranos:2005fi,Faraoni:2009km,Khoury:2013yya} relies on a non-trivial potential of the non-minimal fields that essentially introduces a matter characteristic dependence of the range of a non-minimal field. For instance, dense regions in matter give rise to a suppression of the non-minimal field due to the advent of a large effective mass. This mechanism is in particular naturally associated to $f(R)$ theories \cite{Hu:2007nk,Capozziello:2007eu,Cognola:2007zu}. 

Interestingly, as shown in \cite{Hui:2009kc}, while the chameleon mechanism would effectively screen any strong equivalence principle violations of gravitating bodies within dense regions, it could on the other hand lead to large violations of the SEP for screened objects in an unscreened external gravitational field, provided there exists a natural background value of the non-minimal field. In other words, screened self-gravitating objects in an otherwise unscreened environment would not follow the geodesics of the physical metric compared to test-particles and unscreened bodies due to their local suppression of the background non-minimal field. Such an effect is not present for Vainstein screening.

\paragraph{Spontaneous Scalarization.} 

While the two screening mechanisms above mainly aim at allowing for non-trivial effects beyond GR on cosmological scales, there also exist mechanisms that favor an appearance of GR deviations in the strong gravity regime only. The oldest such mechanism is known as \textit{spontaneous scalarization} \cite{Damour:1993hw,Damour:1996ke}, that dynamically drives a non-minimal field into a non-trivial configuration. We already
mentioned this effect in the previous section. Interestingly, a similar effect was also conceived in the cosmological context, known as \textit{symmetron mechanism} \cite{Pietroni:2005pv,Olive:2007aj,Hinterbichler:2010es}.

\subsection{(B) Perturbative Theories}\label{ssSec: B Perturbative Theories}

We will now turn our attention to concrete examples of theories of type $(B)$. Recall that these represent theories that contain higher order operators that can only consistently be included if one imposes additional constraints on them in order to prevent the introduction of extra dynamical and in particular the unhealthy ghost-degrees of freedom. In practice, this can be done by imposing the equations of motion of a ghost-free principal part as a starting point of a perturbative series \cite{Simon:1990PhysRevD41,Burgess:2003jk,Weinberg:2008hq,Endlich:2017tqa}. Such perturbative theories in particular capture corrections naturally expected from a quantum UV completions of given exact theories. From this point of view, type (B) theories are not a new class on their own but capture possible quantum corrections to all theories of type (A). However, here the quantum origin of the perturbations should only be viewed as an underlying motivation. At this stage, we will therefore refer to these types of theories, which can in principle be treated on a purely classical level, as \textit{perturbative effective field theories} (pEFT). In Part~\ref{Part: Quantum Gravity} we will then make a more direct connection to quantum EFTs. 

Note, however, that here the terminology ``perturbative'' is not to be confused with the perturbation theory introduced in Sec.~\ref{sSec:PerturbationTheory}. Rather, while the linearized equations of motion of the DOFs of the principal part of a pEFT are still to be understood as computations about an exact but in principle arbitrary background solution as in Sec.~\ref{sSec:PerturbationTheory}, the equations of motion including the additional ``perturbative'' EFT terms are then to be solved as an additional ``perturbation'', order by order about the principal part as a result of the additional constraints that need to be imposed. This subtle distinction is clearer in the explicit perturbative quantum correction picture, where the equations of motion of the principal part naturally correspond to the dominating classical equations of motion. Moreover, while technically one should be able to consider any theory of type (A) as an exact principal part (see Part~\ref{Part: Quantum Gravity}), here we will only focus on theories, whose principal part is given by the leading order terms in an $E/\Lambda$ expansion of an exact theory, where $\Lambda$ characterizes a mass scale that represents the natural cutoff scale of the pEFT (not to be confused with the cosmological constant). 


\subsubsection{\ul{The perturbative EFT of GR}}

Let's start with the simplest option and consider a perturbative theory with GR as its principal part. As discussed, this boils down to adding extra terms to the action of GR that capture all possible higher order corrections to GR in a high-energy expansion, but treat them perturbatively, so as to retain the number of propagating DOFs of GR, hence two propagating tensor DOFs. This provides an effective theory that is in particular able to account for the most general corrections induced by a possible UV completion. Indeed, as we will discuss in more detail in Sec.~\ref{sSec:Quantization of Gravity}, this theory precisely corresponds to the quantum EFT expansion of GR, capturing first quantum corrections \cite{Donoghue:1993eb,Donoghue:1994dn,Bjerrum-Bohr:2002gqz}.

In the case of GR, the principal part of the action is given by the Einstein-Hilbert action in Eq.~\eqref{eq:EinsteinHilbertAction} that involves the Ricci scalar $R$, as well as a cosmological constant term $\Lambda$. This second term is however generally neglected due to the experimental evidence that locally it must be negligible and only becomes important on cosmological scales (see Part.~\ref{Part: Cosmological Testing Ground} and also e.g. \cite{Burgess:2003jk,Donoghue:2012zc,zee2013einstein}).\footnote{In the context of a qEFT however, such a suppression of a cosmological constant term represents one of the biggest puzzles of theoretical physics that we will come back to in Sec.~\ref{sSec: The CC Problem}. From that perspective, neglecting any CC contribution can also be viewed as sweeping unsolved issues under the rug, assuming that the puzzle will eventually be resolved, and press on to consider the tractable pieces.} Moreover, it is also common practice to consider vacuum equations of motion as a baseline for the expansion. In the case of GR, one therefore assumes the Einstein equations
\begin{equation}\label{eq:GR EE Vacuum}
    R_{\mu\nu}=R=0\,.
\end{equation}
This is justified in many concrete situations, in particular when considering the merger of a binary black hole system for example. However, it is important to keep in mind that the quantum effective theories that we will write down in this section crucially depend on these two assumptions and would need to be updated as soon as matter terms or a cosmological constant are present (see also \cite{Simon:1990PhysRevD41,Burgess:2003jk}).
Under these assumptions, we can then construct a perturbative effective field theory of GR by constructing all non-trivial higher order terms in a high-energy expansion as corrections to the Ricci scalar $R$, given the constraint imposed by the baseline equations of motion in Eq.~\eqref{eq:GR EE Vacuum} and up to topological (total derivative) terms.

Requiring explicit invariance under coordinate transformations of the action, the first such corrections naively enter at the quadratic order in curvature terms, hence at four powers of derivative operators.
\begin{equation}\label{eq:Terms Quadratic Gravity}
    R+\frac{1}{\MPl^2}\left(c_2 R^2 + \hat c_2 R_{\mu\nu}R^{\mu\nu}+\doublehat{c}_2 R_{\mu\nu\rho\sigma}R^{\mu\nu\rho\sigma}\right)\,,
\end{equation}
where we already omitted a fourth and a fifth contribution of the form $\tilde{R}_{\mu\nu\rho\sigma}R^{\mu\nu\rho\sigma}$ and  $\tilde{R}_{\mu\nu\rho\sigma}\tilde{R}^{\mu\nu\rho\sigma}$  that are total derivatives and thus vanishes up to boundary contributions, with $\tilde{R}_{\mu\nu\rho\sigma}$ the Hodge dual of the Riemann tensor defined in Eq.~\eqref{eq:HodgeDualRiemannTensor}. The additional factor of
\begin{equation}
    M_\text{P}=\frac{1}{2\kappa_0}\,,
\end{equation}
is required here on dimensional grounds, where the scale represents a natural first expectation in GR (see however comments below).
However, up to integrations by parts, these terms in fact all vanish due to Eq.~\eqref{eq:GR EE Vacuum}. Indeed, using the definition of the Gauss-Bonnet curvature scalar $\mathcal{G}_{\myst{GB}}$ given in Eq.~\eqref{eq:GBScalar}, that itself represents a discardable total derivative, one can reduce the quadratic correction $R_{\mu\nu\rho\sigma}R^{\mu\nu\rho\sigma}$ to a sum of terms that vanish under the constraint of the lowest order equations of motion $R_{\mu\nu}=R=0$. We want to mention at this point that in the light of the present considerations, an alternative to GR known as ``quadratic gravity'' (see e.g. \cite{Salvio:2018crh}), involving terms of the form $R_{\mu\nu}R^{\mu\nu}$ becomes obsolete as the terms either lead to propagating ghost degrees of freedom or vanish in the perturbative constraint approach described here. Terms of the form $R_{\mu\nu}R^{\mu\nu}$ could only possibly play a role in perturbative EFTs that consider a generalization of the leading order equations of motion in Eq.~\eqref{eq:GR EE Vacuum}.

The leading order corrections therefore appear at the level of terms involving six derivative operators, where, up to Bianchi identities, there exist two non-trivial independent terms \cite{Endlich:2017tqa,Carminati:1991,Fulling:1992vm}
\begin{equation}
    \mathcal{C}_3\equiv c_3 R_{\mu\nu\rho\sigma}R\ud{\mu\nu}{\alpha\beta}R^{\alpha\beta \rho\sigma}\,\quad  \tilde{\mathcal{C}}_3\equiv \tilde c_3\tilde{R}_{\mu\nu\rho\sigma}R\ud{\mu\nu}{\alpha\beta}R^{\alpha\beta \rho\sigma}\,,
\end{equation}
the first being parity even while the second is breaking parity. The effective perturbative action for the perturbative expansion of GR can therefore be written as 
\begin{equation}\label{eq:QEFT GR}
        S^{\myst{p}}=\frac{1}{2\kappa_0}\int d^4x\sqrt{-g}\, \left[R+\frac{\mathcal{C}_3+\tilde{\mathcal{C}}_3}{\MPl^4}+...\right]\,,
\end{equation}
where the dots indicate higher order terms with eight or more derivative operators, suppressed by even larger powers of the scale $\MPl$. Provided that GR is the unique viable exact metric theory of gravity that only propagates two tensor degrees of freedom (recall Sec.~\ref{sSec:Uniqueness in Propagating DOFs}) and that the perturbative extension above does not alter the number of dynamical DOFs, this pEFT expansion can be viewed as the most general covariant and Ostragradski stable effective description of two propagating tensor modes.

As mentioned, since the Planck mass $M_\text{P}$
represents a natural mass scale of the theory one usually expects that also the suppressing scale of higher order terms in the pEFT should correspond to the same order of magnitude, in which case the perturbative contributions would not be able to influence any plausible experiment in the near future. However, as many examples of pEFTs show, there could in principle exist new physical effects that appear at a much lower scale, and the cutoff of the pEFT of GR must in fact not be parametrically close to $M_\text{P}$. It is therefore worth to keep an open mind and test for such corrections even on distance scales on which gravity has already been tested with high precision laboratory experiments, since it is conceivable that quantum corrections to gravity remain unobservably small until the scale of spacetime curvature itself reaches a given threshold. These considerations represent the motivation to test for such effects in current gravitational wave experiments \cite{Endlich:2017tqa}.

\subsubsection{\ul{Scalar Tensor Theories}}

It is interesting to perform the same exercise as in GR but starting from the simplest possible principal part of a scalar-tensor theory, namely the Ricci scalar supplemented by a canonically normalized scalar field $\Phi$ with a general potential $U(\Phi)$. In this case the equations of motion of the principal part therefore simply read
\begin{equation}\label{eq:ST Vacuum EOM}
    R_{\mu\nu}=R=0\,, \quad \nabla_\mu\nabla^\mu\Phi + U(\Phi)=0\,.
\end{equation}
Using these equations, as well as the symmetries of the Riemann tensor, and up to integration by parts, the most general leading order correction of independent terms this time already appears at the four-derivative order and reads \cite{Weinberg:2008hq}
\begin{equation}\label{eq:QEFT ST}
       S^{\myst{p}}=\frac{1}{2\kappa_0}\int d^4x\sqrt{-g}\, \left[R-\frac{1}{2}\nabla_\mu\Phi\nabla^\mu\Phi-U(\Phi)+\frac{f(\Phi)\mathcal{C}_2+\hat f(\Phi)\hat{\mathcal{C}}_2 +\doublehat{f}(\Phi)\doublehat{\mathcal{C}}_2 }{\Lambda^2}+...\right]\,,
\end{equation}
with
\begin{align}
    \mathcal{C}_2&\equiv R_{\mu\nu\rho\sigma} R^{\mu\nu\rho\sigma}\,,&\hat{\mathcal{C}}_2&\equiv \tilde{R}_{\mu\nu\rho\sigma} R^{\mu\nu\rho\sigma}\,,&\doublehat{\mathcal{C}}_2&\equiv \left(\nabla_\mu\Phi\nabla^\mu\Phi\right)^2\,,
\end{align}
where again the Hodge dual of the Riemann tensor $\tilde{R}_{\mu\nu\rho\sigma}$ is defined in Eq.~\eqref{eq:HodgeDualRiemannTensor}. Note that this action reduces to GR for a constant scalar field, because both $\mathcal{C}_2$ and $\hat{\mathcal{C}}_2$ on their own are purely topological. 

Note in particular again the appearance of the parity odd term $\hat{\mathcal{C}}_2$, which is in fact often considered on its own. More precisely, with the particular choices of $U(\Phi)=0$ and $\hat f(\Phi)=\Phi$, such that $\Phi$ as a pseudo scalar could in principle compensate for the parity oddness of $\hat{\mathcal{C}}_2$, the resulting theory carries the name of dynamical Chern-Simons (dCS) gravity \cite{Jackiw:2003pm,Alexander:2009tp}
\begin{equation}\label{eq:ActiondCS}
    S^{\myst{dCS}}=\frac{1}{2\kappa_0}\int \dd^4 x\sqrt{-g}\bigg[ R -\frac{1}{2}g^{\mu\nu}\nabla_\mu\Phi\nabla_\nu\Phi+\Phi\,R\du{\mu\nu\rho\sigma} {}\tilde R^{\nu\mu\rho\sigma}\bigg]\,.
\end{equation}
In recent years, dCS gravity has received some attention, mainly because its rotating black holes have a nontrivial (pseudo)scalar profile \cite{Yunes:2009hc}. Unlike sGB gravity, however, the dCS interaction when taken at face value propagates a ghost degree of freedom~\cite{Motohashi:2011ds}. However, the theory can still consistently be considered under the perturbative approach discussed here~\cite{Yunes:2013dva}. 

Interestingly, the term $\mathcal{C}_2$ implies that also scalar Gauss-Bonnet theory already considered in Eq.~\eqref{eq:ActionsGB} is part of that perturbative action. Thus, sGB can both be considered as an exact theory on its own, but also features in the perturbative expansion of scalar-tensor theories.

\subsection{Well Posedness}\label{ssSec:WellPosedness}

We want to close this chapter with a comment on the notion of \textit{well-posedness} that will illustrate the practical differences between theories of type (A) and (B) considered above. Here, well-posedness of a theory is defined as the property that there exists a formulation of the theory, in which its partial differential field equations have a well-posed initial value problem and can be meaningfully evolved in numerical simulations, which is guaranteed if the system is strongly hyperbolic (see e.g. \cite{Alcubierre:2008jj,Baumgarte:2010ndz,Ripley:2022cdh,deRham:2023ngf}). While well-posedness cannot be regarded as a fundamental feature of a theory, the existence of a well-posed formulation is essential for numerical simulations, since without it, it is unclear what a numerical discretization of the system represents in the continuum limit.

In particular, the construction of a well posed formulation of GR allowed for the first stable numerical simulations of binary black hole systems \cite{Pretorius:2005gq,Campanelli:2005dd,Baker:2005vv,Pretorius2009}. Similarly, also well posed formulations of beyond GR theories are known with first successful numerical evolution's \cite{Noakes:1983xd,Delsate:2014hba,Papallo:2017qvl,Witek:2018dmd,Okounkova:2019zjf,Kovacs:2020ywu,Kovacs:2020pns,Okounkova:2020rqw,Held:2021pht,AresteSalo:2022hua,East:2022rqi,Corman:2022xqg,Ripley:2022cdh,deRham:2023ngf}. These include Einstein-\AE{}ther gravity but in particular also theories with higher order derivative interaction terms, such as Horndeski theory and dCS gravity. Recall that Horndeski gravity in particular also includes sGB gravity.

However, there is a fundamental difference between the well posed formulations of Horndeski theory and dCS, that reflects their classification into theories of type (A) and type (B). While both theories find their well-posed formulations in the so called \textit{weakly-coupled regime} (see e.g. \cite{Kovacs:2020ywu,Kovacs:2020pns,Ripley:2022cdh}) that reflects the need of a natural cutoff scale due to the presence of higher order derivative operators, Horndeski gravity, as a theory of type (A), admits a well posed formulation of its \emph{exact} equations of motion, while for dCS one needs to resort to a \emph{perturbative} (or \emph{order-reduction}) notion of well-posedness (see \cite{Ripley:2022cdh}). More precisely, within the perturbative scheme, a well posed formulation can only be found by reducing the principal part of the equations of motion to that of GR, which might not come as a surprise since only the perturbative equations of motion of dCS remain stable under Ostrogradsky ghosts. On the other hand, for Horndeski theory it is possible to construct a well posed formulation of the complete set of equations of motion, thus precisely reflecting the above distinction between theories of type (A) and (B).

\newpage
\thispagestyle{plain} 
\mbox{}



\part{The Gravitational Radiation Testing Ground}\label{Part: Gravitational Wave Testing Ground}

\small
\noindent
\emph{\ul{Personal Contribution and References}}\\ 
\footnotesize 
\textit{Chapters~\ref{Sec:Radiation andGWs in Gravity} and \ref{Sec:GWMemory} are based on \textbf{L. Heisenberg, N. Yunes, J. Zosso, 2023} \cite{Heisenberg:2023prj} and \textbf{L. Heisenberg, G. Xu, J. Zosso} \cite{Heisenberg:2024cc}, in particular Secs.~\ref{sSec:GWObservation}, \ref{sSec:GWPolExample}, \ref{sSec:GWMemoryinGR}, \ref{sSec:GW Memory beyond GR} and \ref{sec:SVTMem}. Parts of the following treatment are also inspired from \cite{misner_gravitation_1973,Flanagan:2005yc,maggiore2008gravitational,Creighton:2011zz,poisson2014gravity,guidry2019modern,YunesColemanMiller:2021lky,Jetzer:2022bme}.}
\normalsize

\vspace{5mm}

\noindent
\textbf{Summary of Part II}\\ 
\noindent
The metric theory space beyond GR introduced in the previous part can nowadays be directly tested against observations of gravitational radiation that we conferred about in the introduction. Concrete tests of gravity that can be carried out with gravitational wave data are manifold and range from basic consistency trials, including residual noise tests, waveform self-consistency checks and constraints on parameterized deviations from GR, to more specialized tests, for instance probes of the no-hair conjecture aiming at the quasi normal modes of the ringdown signal or tests targeting the number of large extra dimensions, all the way to direct comparison of beyond GR templates with the data, just to name a few. In this work, we will however put the focus on the implications of the fundamental aspects of gravity theory of the previous part on a specific set of GW based probes of gravitation.

For this, we will start by introducing the concept of radiation in asymptotically flat spacetimes and the basics of GW generation, followed by a slight detour of defining GW velocity in the Isaacson approach in connection with the existing stringent constraints on propagation speed. Subsequently, a thorough description of the physical response within today's GW detectors will be offered. This last point will closely be intertwined with the concept of gravitational polarizations, for which once again a restriction to metric theories of gravity seems essential. In turn, these considerations will introduce all the formalism needed to finally arrive at a well-defined description of the memory effect in metric theories of gravity, with an outlook on future memory based tests of GR.

\chapter{Radiation and Waves in Gravity}\label{Sec:Radiation andGWs in Gravity}

In Chapter~\ref{Sec:PropagatingDOFs} we carefully defined the concept of waves as gauge-invariant high-frequency perturbations with typical scale of variation $f_H$ that propagate on a slowly varying background that varies on scales lower than $f_L$ by assuming a so-called Isaacson split $f_L\ll f_H$ between the background and the high-frequency perturbations. Recall that the Isaacson split within the framework of perturbation theory around a slowly varying exact solution to the field equations $\{\bar{g}_{\mu\nu},\bar{\Psi}\}$, decomposes the perturbed approximate metric, together with all additional fields, at each spacetime point into three pieces as [Eq.~\eqref{eq:IsaacsonSplitGeneral}]
\begin{equation}\label{eq:IsaacsonExpansion}
    g_{\mu\nu}=\bar{g}_{\mu\nu}+\delta g^L_{\mu\nu}+\delta g^H_{\mu\nu}\,,\quad \Psi=\bar\Psi+\delta\Psi^L+\delta\Psi^H\,,
\end{equation}
where $\delta g^L_{\mu\nu}$ and $\delta\Psi^L$ represent low-frequency perturbations and $\delta g^H_{\mu\nu}\equiv h_{\mu\nu}$ and $\delta\Psi^H$ contain the information on high-frequency (gravitational) waves. 

Further, the Isaacson assumptions allow the formulation of a local chart, in which the low frequency background reduces to the Minkowski form [Eq.~\eqref{eq:BackgroundMInkowskiForm}]
\begin{eqnarray}
    g^L_{\mu\nu}=\bar{g}_{\mu\nu}+\delta g^L_{\mu\nu}\simeq \eta_{\mu\nu}\,,
\end{eqnarray}
while the high-frequency perturbations can be viewed as proper Lorentz vectors on that chart. Even though the existence of such a coordinate system is extremely useful for theoretical arguments in particular in connection with the definition of propagating degrees of freedom, in a general situation it is unrealistic to construct such a chart in practice. 

We therefore need to be more modest and add the further well motivated assumption that any source of gravity is sufficiently confided to a finite location, thereby allowing an expansion in the distance from the source that will be parameterized by some source centered radial coordinate $r$. The standard description of gravitational wave observations indeed relies on this additional assumption known as the concept of an \textit{asymptotically flat spacetime} that we will now introduce [Definition~\ref{DefAsymptoticallyFlat}], representing a good approximation for the realistic situation of observing gravitational waves from very distant sources. Foremost, this context provides a natural exact solution in the perturbative expansion in Eq.~\eqref{eq:IsaacsonExpansion}, namely a flat space vacuum solution $\{\bar{g}_{\mu\nu}=\eta_{\mu\nu},\bar\Psi\}$, while the perturbations in the far field limit correspond to the $\sim 1/r$ corrections, which in this case in principle not only describe the high-frequency waves but also the low-frequency perturbations $\delta g^L_{\mu\nu}$. We will therefore introduce the total metric perturbation at $\mathcal{O}(1/r)$
\begin{equation}\label{eq:TotalRadiationPerturbations}
    H_{\mu\nu} \equiv \delta g^L_{\mu\nu}+\delta g^H_{\mu\nu}\,,\quad \delta\Psi\equiv \delta\Psi^L+\delta\Psi^H  \,,
\end{equation}
that represent a perturbation (in the sense of Sec.~\ref{sSec:PerturbationTheory}) to the exact solution naturally provided by the asymptotically flat assumption.

These considerations therefore contrast the notion of \textit{\gls{waves}}, defined above as the physical and propagating high-frequency perturbations of fields, to the concept of what we will call \textit{\gls{radiation}} \cite{Isaacson_PhysRev.166.1263}. In general, radiation is defined as propagating modes to which it is possible to associate a power that is irreversibly carried away from a localized source to infinity. This captures the key characteristic of radiation as being energy which decouples completely from its origin and is lost in the system. In particular, this definition applies both to the familiar electromagnetic radiation within Maxwell's theory and to the gravitational case treated here \cite{DAmbrosio:2022clk}.

Radiation is therefore naturally described through the leading order field perturbations introduced above in Eq.~\eqref{eq:TotalRadiationPerturbations} as only terms at $\mathcal{O}(1/r)$ have a chance of describing a non-vanishing luminosity \cite{YunesColemanMiller:2021lky,DAmbrosio:2022clk}.
However, in the same way as not all components of the high-frequency perturbations represent waves, not all components in $ H_{\mu\nu}$ and $\delta\Psi$ are associated to radiation. Heuristically, this is because only the propagating degrees of freedom, hence the DOFs satisfying a wave-type equation that gain an independent evolution from the source, are able to carry away energy from the system. In that sense, the concept of radiation provides an alternative way of identifying the dynamical degrees of freedom of a theory. Indeed, throughout this chapter, the distinction between waves and radiation will remain secondary and will only become important in Chapter~\ref{Sec:GWMemory}. Moreover, below, we will also further analyze the distinction between the non-dynamical and dynamical components within the $\mathcal{O}(1/r)$ perturbations.

\section{Asymptotic Flatness}\label{sSec:Asymptotic Flatness}

The phenomenon of radiation in physics, although very intuitive, comes with certain subtleties which surface as soon as one tries to formulate a consistent definition, in particular in the context of gravity. This led to the conception of the notion of asymptotic flatness that we now want to introduce. Yet, we will refrain here from giving a precise introduction of the mathematical framework built around the pioneering works of Bondi, Metzner and Sachs (BMS) \cite{Bondi:1960jsa,Bondi:1962px,Sachs:1962wk}, Newman and Penrose (NP) \cite{Newman:1962,Penrose:1963,Newman:1963,Penrose:1965,Newman:1968}, as well as Geroch \cite{Geroch:1977jn}, Ashtekar \cite{Ashtekar:1981bq,Ashtekar:2014zsa} and many others \cite{WaldBook} and content ourselves with a minimal definition of asymptotic flatness which will be enough for most of the treatment considered here. For more details, we refer the interested reader to the following review \cite{DAmbrosio:2022clk}. 

All we will require for asymptotic flatness, is that any source is localized enough, such that in can be described in source centered coordinates $\{t,x,y,z\}$, known as asymptotic rest frame \cite{misner_gravitation_1973,Thorne:1980ru}.\footnote{We will generally neglect any self-induced accelerations of the source, such as black-hole remnant kicks, which would require a more rigorous definition of BMS rest frames \cite{Mitman:2022kwt}.} It will also be useful to define the radial coordinate
\begin{equation}\label{Eq:RadialSourceCenteredcoord}
    r\equiv\sqrt{x^2+y^2+z^2}\,,
\end{equation}
in spherical source-centered coordinates $\{t,r,\theta,\phi\}$.
Additionally, in the far-field limit, or \textit{radiation-zone}, corresponding to the large $r$ limit,
the background values of the fields admit a Minkowski background $\{\eta_{\mu\nu},\bar\Psi\}$. This background is perturbed in the radiation-zone by leading order corrections $H_{\mu\nu}$ and $\delta\Psi$ in the large $r$ expansion.
Depending on the situation, additional requirements can be demanded, such as the preservation of local Lorentz invariance of the solution $\bar{\Psi}$.

\begin{definition} \textit{Asymptotically Flat Spacetime.}\label{DefAsymptoticallyFlat}
    A spacetime of a metric theory of gravity in Definition~\ref{DefMetricTheory} is asymptotically flat, if any gravitational source is sufficiently localized, such that there exists a source centered chart $\{t,x,y,z\}$, with $r$ the associated radial source-centered coordinate defined in Eq.~\eqref{Eq:RadialSourceCenteredcoord}, for which the physical metric and the non-minimal fields take the form
        \begin{equation}\label{eq:perturbationsAsymptoticMetric}
             g_{\mu\nu}=\eta_{\mu\nu}+H_{\mu\nu}+\mathcal{O}\left(\frac{1}{r^2}\right)\,,
         \end{equation}
        and
        \begin{equation}\label{eq:perturbationsAsymptoticNonMinimal}
            \Psi=\bar{\Psi}+\delta\Psi +\mathcal{O}\left(\frac{1}{r^2}\right)\,,
        \end{equation}
        where the set $\{\eta_{\mu\nu},\bar\Psi\}$ represents and exact solution to the vacuum field equations with $\eta_{\mu\nu}$ the Minkowski metric.
\end{definition}

Heuristically, \textit{null} radiation, hence radiation from massless fields, can then be separated from the Coulombic pieces by defining a more sophisticated far field limit from a localized source. Indeed, through a convenient change of coordinates to the asymptotic retarded time of massless radiation
\begin{equation}\label{eq:Definition Asymptotic Retarded Time Massless}
    u\equiv t-r\,,
\end{equation}
one can define a \textit{\gls{limit to null infinity}} at large $r$ but at a fixed asymptotic retarded time $u$
\begin{equation}\label{eq:Limit to Null Infinity}
    r\rightarrow\infty\,,\quad u=\text{ const.}\,,
\end{equation}
up to the first nontrivial terms in the $r$ expansion.\footnote{The precise formulation of asymptotic flatness is in fact defined though the existence of an actual null boundary of the spacetime within a conformal completion that is reached in the limit to null infinity.} This novel limit separates null radiation from other far field contributions at spacial infinity. See also Fig.~\ref{fig:Asymptotically flat spacetime} below for a visual representation of asymptotic flatness and the role of the light-cone coordinate $u$.

In the following, we want to offer a quick analysis of the three stages of radiation from a localized source: the generation, the propagation and finally its detection. However, we do not have the ambition to self-sufficiently describe each stage and will mostly concentrate on introducing all relevant concepts for the discussion of gravitational wave memory in generic metric theories of gravity.

\section{Gravitational Wave Generation}\label{sSec:GWGeneration}

Computing the generation of gravitational radiation for realistic sources such as compact binary coalescence's (CBCs) is a very hard problem that can only be tackled numerically by the evolution of cleverly rewritten well posed Einstein equations and the extraction of the leading order physical modes far away from the source \cite{Pretorius:2005gq,Campanelli:2005dd,Baker:2005vv,Pretorius2009} (recall Sec.~\ref{ssSec:WellPosedness}). However, for our purposes, it will mostly be enough to know that in the radiation zone, the propagating information admits a special form whose time dependence is governed by an asymptotic retarded time, which in the massless and Lorentz preserving case is given by Eq.~\eqref{eq:Definition Asymptotic Retarded Time Massless}. It is therefore enlightening to at lest sketch the derivation of this asymptotic form within GR in the analytically tractable simplified situation of perturbative sources.

Moreover, we want to already mention here that gravitational waves are generically emitted by the bulk motion of its source, and therefore phase-coherently. This means that at each time instance, a rather narrow band of characteristic frequency can be associated to the gravitational radiation. In other words, in a first approximation a source only produces radiation at a certain scale $f_H$ which we will define to be our high-frequency scale and hence radiation is only found in the high-frequency perturbation within Eq.~\eqref{eq:TotalRadiationPerturbations}, with a given small amplitude $| h_{\mu\nu}|, |\delta\Psi^H|\sim \alpha\ll 1$. To linear order in $\alpha$, the concepts of waves and radiation introduced above therefore coincide, which is also the reason why such a distinction is generally not made. This will, however, not be true anymore as soon as we leave the linear approximation in $\alpha$ in Chapter~\ref{Sec:GWMemory}.

\subsection{Perturbative Sources in General Relativity}\label{eq:Perturbative Sources in General relavivity}

\paragraph{Solving the Inhomogeneous Wave Equation.}

Given the above comment, to consider the perturbative generation of radiation in general relativity, we can therefore to first order without loss of generality solve the high-frequency equation Eq.~\eqref{eq:LinGRprop2} derived in Sec.~\ref{ssSec:Local Wave Equation in GR} that we reproduce here for convenience
\begin{equation}\label{eq:LinGRprop2Second}
  \Box \bar{h}_{\mu\nu}=-2\kappa_0 \left[\delta T_{\mu\nu}\right]^H\,.
\end{equation}
Recall that the high-frequency field perturbations $\bar{h}_{\mu\nu}$ defined in Eq.~\eqref{eq:HarmonicGaugeVariables} already satisfy the Lorenz gauge
\begin{equation}
    \partial^\nu \bar{h}_{\mu\nu}=0\,,
\end{equation}
which is consistent with the conservation of the energy momentum tensor. Moreover, we want to point out the unfortunate clash of notation and hope that is clear in this context that $\bar{h}_{\mu\nu}$ does \textit{not} represent a background field, but a redefinition of the high-frequency field $\delta g^H_{\mu\nu}=h_{\mu\nu}$.

Equation~\eqref{eq:LinGRprop2Second} can then be solved by the method of Green's function
\begin{equation}
    \bar{h}_{\mu\nu}(x)=-2\kappa_0 \int d^4 x' \,G(x,x')\,\delta T_{\mu\nu}^H(x')\,,
\end{equation}
where the appropriate retarded Green's function is well known (see e.g. \cite{Jackson:1998nia})
\begin{equation}
    G(x,x')=\frac{\delta(t'-[t-|\mathbf{x}-\mathbf{x}'|])}{4\pi |\mathbf{x}-\mathbf{x}'|}\,.
\end{equation}
Here the quantity
\begin{equation}
    t_\text{ret}\equiv t-\frac{1}{c}|\mathbf{x}-\mathbf{x}'|\,,
\end{equation}
defines the \textit{retarded time}, taking into account the finite speed of propagation of information from the source coordinate $\mathbf{x}'$ to the event at $\mathbf{x}$. The solution to Eq.~\eqref{eq:LinGRprop2Second} can therefore be written as
\begin{equation}
    \boxed{\bar{h}_{\mu\nu}(t,\mathbf{x})=-\frac{\kappa_0}{2\pi} \int d^3 x' \,\frac{\delta T_{\mu\nu}^H(t_\text{ret},\mathbf{x}’)}{|\mathbf{x}-\mathbf{x}'|}\,.}
\end{equation}

For a confined high-frequency source the limit of large $r$ can be used to expand
\begin{equation}
|\mathbf{x}-\mathbf{x}'|=r-\mathbf{x}'\cdot\mathbf{n}+\mathcal{O}(1/r)\,.
\end{equation}
Note, however, that for a source of arbitrary velocities, it is not consistent to simply replace the leading order term $|\mathbf{x}-\mathbf{x}'|\sim r$ into the argument of the energy-momentum tensor, as the corresponding error precisely scales with the typical rate of change of the source (see e.g. \cite{Flanagan:2005yc,maggiore2008gravitational}). In the general case of arbitrary relativistic motions, one can nevertheless show that through a Fourier transform of the source the solution can be put into the form \cite{maggiore2008gravitational}
\begin{equation}\label{eq:GeneralWaveSolutionGR}
\bar{h}_{\mu\nu}(u,\mathbf{x})=\frac{1}{r}\, A_{\mu\nu}(u,\Omega)+\mathcal{O}(1/r^2)\,,
\end{equation}
where $u= t-r$ is the asymptotic retarded time coordinate defined in Eq.~\eqref{eq:Definition Asymptotic Retarded Time Massless}.


Recall (Sec.~\ref{sSec:GWs in GR}), however, that in GR only the spacial, transverse-traceless modes of the metric perturbations propagate. In other words, at this stage a priori the quantity $\bar{h}_{\mu\nu}$ also contains pure gauge as well as non-propagating degrees of freedom on top of the two propagating TT DOFs. Yet, as discussed, only the propagating modes are part of the radiation. At leading order in $1/r$ in the radiation zone at each point $\mathbf{x}$, these physical modes in the general solution of Eq.~\eqref{eq:GeneralWaveSolutionGR} can conveniently be extracted by an algebraic projection onto the traceless transverse space of the vector $\mathbf{n}$ that indicates the direction of propagation. More precisely, we define
\begin{equation}\label{eq: Def Direction of Propagation n 2}
    n_i\equiv \frac{x_i}{r}=\partial_i r\,,
\end{equation}
as the unit $n_in^i=1$ and radial source centered vector in the associated spacial Cartesian coordinate system. A specific spacial direction is usually parameterized by two angles $\Omega=\{\theta,\phi\}$ that represent the standard spherical angles of the source centered coordinate system, such that 
\begin{equation}\label{eq:Def Direction of Propagation n}
    \mathbf{n}(\Omega)=(\sin\theta \cos\phi,\,\sin\theta \sin\phi,\,\cos\theta )\,.
\end{equation}

The transverse space of a given direction $\mathbf{n}(\Omega)$ is then conveniently coordinatized by the two additional spacial vectors\footnote{This basis is explicitly constructed starting from an arbitrary Cartesian reference frame by first performing a rotation of $\phi$ around the $z$-axis in order to align the $x$-axis with the projection of $\mathbf{n}$ onto the $x$-$y$-plane, followed by a rotation of $\theta$ around the new $y$-axis to align the $z$-axis with the direction of travel $\mathbf{n}$ (see e.g. \cite{poisson2014gravity}).}
\begin{subequations}\label{eq:Def Transverse Vectors u v}
\begin{align}
\mathbf{u}(\Omega)&=(\cos\theta \cos\phi,\,\cos\theta \sin\phi,\,-\sin\theta )\,,\\
\mathbf{v}(\Omega)&=(-\sin\phi,\,\cos\phi,\,0)\,,
\end{align}
\end{subequations}
that together with $\mathbf{n}$ define an orthonormal spacial basis satisfying the completeness relation
\begin{equation}\label{eq:polcompleteness}
\delta_{ij}=n_in_j+u_iu_j+v_iv_j\,.
\end{equation} 

With such a basis at hand, one can define the transverse projector
\begin{equation}\label{eq:TransverseProjector}
\perp_{ij}\,\equiv \delta_{ij}-n_in_j=u_iu_j+v_iv_j\,,
\end{equation}
as well as the transverse-traceless projector
\begin{equation}\label{eq:Projectors}
    \boxed{\perp_{ijab}\,\equiv \,\perp_{ia}\perp_{jb}-\frac{1}{2}\perp_{ij}\perp_{ab}\,,}
\end{equation}
that is transverse to $n_i$ on all of its indices and traceless with respect to the $(ij)$ and $(ab)$ spacial indices. 

This is the TT projector that we were looking for, with which one can project any symmetric spacial tensor satisfying a homogeneous wave equation onto its TT part (see also \cite{maggiore2008gravitational}). Thus, in particular
\begin{equation}
    \perp_{ijab}(\Omega)\,\bar{h}^{ab}(x)=\perp_{ijab}(\Omega)\,h^{ab}(x) =h^{TT}_{ij}(x)\,,
\end{equation}
and the final solution of the gravitational waves produced by the perturbative source reads
 \begin{equation}\label{eq: Asymptotic Form GW GR}
     \boxed{h^{TT}_{ij}(t,\mathbf{x})=\frac{1}{r} \perp_{ijab}(\Omega) \,A_{ab}(t-r,\Omega)\equiv \frac{1}{r}  \,A^\text{TT}_{ab}(t-r,\Omega)\,.}
 \end{equation}
Note, however, that in performing such projections is important that $\bar{h}_{\mu\nu}$ already satisfies the Lorenz gauge as otherwise the corresponding equations would not reduce to a simple d'Alembert operator, which is decisive for the validity of such an algebraic projection \cite{maggiore2008gravitational}. 

\paragraph{Massless and Lorentz Preserving Radiation.} Therefore, indeed, the time dependence of the $\mathcal{O}(1/r)$ radiative perturbations is given by the asymptotic retarded time $u=t-r$ [Eq.~\eqref{eq:Definition Asymptotic Retarded Time Massless}]. This is important, because the special asymptotic form in Eq.~\eqref{eq: Asymptotic Form GW GR} directly implies that despite the fact that it represents a solution to a sourced wave equation, the asymptotic radiative modes $h^{TT}_{ij}$ explicitly satisfy the homogeneous one 
\begin{equation}
    \Box h^{TT}_{ij}=0\,.
\end{equation}
In fact this very generally applies to any type of massless and local Lorentz preserving radiation. To see this, consider a propagating degree of freedom that in the source centered coordinates admits the following asymptotic form
\begin{equation}\label{eq:GeneralFormRadiation}
    \boxed{\delta \Psi(u,r,\Omega)\sim \frac{1}{r}A(u,\Omega)}
\end{equation}
This directly implies that to first order in $r$, the solution satisfies
\begin{equation}\label{eq:IdentityRadiation}
    \partial_i \delta \Psi=-n_i \delta\dot\Psi\,,
\end{equation}
where $n_i$ was defined in Eq.~\eqref{eq: Def Direction of Propagation n 2}. In the radiation zone on the asymptotic Minkowski background, these asymptotic modes then indeed satisfy the homogeneous wave equation
\begin{equation}
    \Box \delta \Psi=-\delta\ddot\Psi +\pd_i\pd^i\delta\Psi=0\,.
\end{equation}

\paragraph{The Quadrupole Formula of GR.} If in addition one assumes non-relativistic motion for the source components with typical velocity $v$, the solution in Eq.~\eqref{eq: Asymptotic Form GW GR} can further be expanded in powers of $v/c$ and the emission of radiation will be dominated by the lowest multipole moments \cite{misner_gravitation_1973,Flanagan:2005yc,maggiore2008gravitational,Creighton:2011zz,poisson2014gravity,YunesColemanMiller:2021lky,Jetzer:2022bme}. Such an expansion is known as a Post Newtonian (PN) expansion of general relativity, that applied to situations beyond the perturbative regime represents a powerful analytic tool to approximate radiative solutions (see e.g. \cite{Blanchet:2013haa}). 

In general relativity, the leading order multipole moment is the quadrupole moment 
\begin{equation}\label{eq:DefQuandrupoleMoment First}
S_{ij}\equiv \int d^3x\, \delta T^H_{00}\, x_i x_j \,.
\end{equation}
This is because the monopole and the dipole are associated to the total mass and total momentum of the source that are approximately conserved up to radiation reaction. In other words, the variation of these moments is highly constraint such that their contribution to a high-frequency source is negligible.  Similarly, angular momentum conservation also constrains the lowest order current multipole moments.
However, in GR, the leading order radiation is more precisely not dominated by the quadrupole moment in Eq.~\eqref{eq:DefQuandrupoleMoment First}, but rather its traceless part.

This fact is clearly indicated by considering that due to Birkhoff's theorem \cite{Birkhoff:2011zz}, under a collapse or expansion of a non-rotating spherically symmetric and confined energy configuration, the far gravitational field remains invariant.\footnote{Note that this is not true for spherically symmetric configurations with non-zero mass current.} Yet, such a configuration would change the moment defined in Eq.~\eqref{eq:DefQuandrupoleMoment First} \cite{YunesColemanMiller:2021lky}. More precisely, the trace of $S_{ij}$ that is a scalar under rotation would naively vary under such a spherically symmetric collapse or expansion. Therefore, by consistency, the trace of the quadruple moment is not allowed to produce radiation but only its traceless part
\begin{equation}
    Q_{ij}\equiv S_{ij} -\frac{1}{3}\delta_{ij} S\ud{k}{k}\,,
\end{equation}
that transforms under rotations as a spin-2 operator. 

This statement is of course consistent with the finding that in GR only the TT part of the perturbation contains radiative degrees of freedom. Indeed, TT projection in Eq.~\eqref{eq: Asymptotic Form GW GR} naturally sets the trace to zero. In summary, to lowest order one therefore obtains the famous quadrupole formula of perturbative radiation of non-relativistic sources \cite{misner_gravitation_1973,Flanagan:2005yc,maggiore2008gravitational,Creighton:2011zz,poisson2014gravity,YunesColemanMiller:2021lky,Jetzer:2022bme}
\begin{equation}\label{eq:Quadrupole Formula}
    \boxed{h_{ij}^{TT}\simeq \frac{\kappa_0}{2\pi r}\,\ddot{Q}^{TT}_{ij}\,,}
\end{equation}
where
\begin{equation}
    Q^{TT}_{ij}=\perp_{ijab}\int d^3x \,\delta T^H_{00}(t,\mathbf{x})\,\left(x_ax_b-\frac{1}{3}\delta_{ab}r^2\right)\,.
\end{equation}
Here $\delta T^H_{00}$ to lowest order in $v/c$ represents the mass density but generally also includes any kinetic energy as well as gravitational binding energy.

\paragraph{Dimensional Estimates.} The quadrupole formula in Eq.~\eqref{eq:Quadrupole Formula} nicely serves for preliminary dimensional estimates and indicates that the typical (dimensionless) amplitude of gravitational waves scales as
\begin{equation}
    h\sim \frac{G}{r}\frac{\partial^2(Md^2)}{\partial t^2}\,,
\end{equation}
with $M$ the typical mass of the source, $d$ the typical size and $G$ Newtons constant. In order to recover factors of $c$ in physical units, it is useful to remember that the combination $GM/c^2$ and $GM/c^3$ have units of distance and time, respectively. Moreover, through Kepler's third law 
\begin{equation}\label{eq:Kepplers Law}
    f^2 d^3\sim G M\,,
\end{equation}
the typical size of the object can be replaced by a typical orbital frequency $f$, such that 
\begin{equation}
    h\sim \frac{G}{c^4}\,\frac{f^2Md^2}{r}\sim\frac{G^{5/3}}{c^4}\, \frac{M^{5/3}f^{2/3}}{r}\sim \frac{GM}{r}\left(fMG\right)^{2/3}\,.
\end{equation}
Notice the extremely small numerical factor in front of the expressions.

Numerically, we have
\begin{equation}
    h\sim 10^{-22}\left(\frac{\text{Mpc}}{r}\right)\left(\frac{M}{\text{M}_\odot}\right)^{5/3} \left(\frac{f}{\text{s}^{-1}}\right)^{2/3}\,,
\end{equation}
where $\text{M}_\odot$ denote solar masses.
For a source of a typical double neutron star system of 3 solar masses, a period of $0.01$ s and at a distance of a $100$ Mpc, the scale is then expected to be of the order of $h\sim10^{-22}$, while for a binary black hole coalescence the estimation is increased to $h\sim 10^{-21}$. Thus, even for such extremely violent events that can invert several solar masses of energy into gravitational waves, this is a tiny number. In order to get a feeling, a fractional change in spacetime distance (see Sec.~\ref{ssSec:The Physical Effects of Gravitational Waves}) due to the presence of a GW of that order corresponds to determining the distance to Alpha Centauri to a precision given by the width of a human hair. Such a measured fractional change in proper distance is frequently refereed to as the fractional strain of spacetime due to the deforming gravitational wave, a nomenclature taken over from the theory of elastic solids. In this language, the above discussion translates into the statement that spacetime is an extremely stiff medium.

Kepler's orbital law in Eq.~\eqref{eq:Kepplers Law} also allows for a convenient estimate of the frequency at merger or $f_\text{max}$ of quasi-circular equal mass binary black hole coalescence's. Namely, at merger, the typical size of the system $d$, the separation between the binaries, is of the order of the Schwarzschild radius $d\sim r_\text{S}=2GM$, such that
\begin{equation}\label{eq:frequency estimate}
    f_\text{max}\sim \sqrt{\frac{GM}{d^3}}\sim \frac{10^{-1}}{GM}\sim 10^4\,\left(\frac{\text{M}_\odot}{M}\right) \,\text{Hz}\,,
\end{equation}
since
\begin{equation}
    \frac{c^3}{G}\sim 10^{35}\,\frac{[\text{kg}]}{\text{[s]}}\sim 10^{5}\,\frac{[\text{M}_\odot]}{\text{[s]}}\,.
\end{equation}
For instance, ground-based detectors therefore mostly target equal mass mergers of $M\sim 10^2\,\text{M}_\odot$ at $f_\text{max}\sim 10^2$ Hz, while space-based observatories are sensitive to systems of total mass $M\sim 10^5\,\text{M}_\odot$ at $f_\text{max}\sim 10^{-1}$ Hz (recall the introduction).

\paragraph{SVT Decomposition and Coulombic Contributions.} Just as in Sec.~\ref{sSec:GWs in GR} also the perturbative emission of radiation can be considered in at least two alternative approaches, one being manifestly local and based on explicitly choosing the Lorenz gauge (employed above), and the other being a manifestly gauge invariant approach within an SVT decomposition. Recall that an SVT decomposition is in particular useful to explicitly also describe the non-dynamical degrees of freedom. Indeed, the projection onto the TT part above in describing the asymptotic high-frequency perturbation variable at $\mathcal{O}(1/r)$ is strictly speaking only justified with the knowledge that all non-dynamical Coulombic pieces associated to the gravitational potentials are to be found within the low-frequency perturbation, due to asymptotic conservation laws \cite{misner_gravitation_1973,Flanagan:2005yc}.

It is illuminating to discuss these statements within the SVT approach outlined in Sections~\ref{ssSec:GaugeInvariantDecomposition} and \ref{ssSec:WavesInMetricTheoriesOfGravity} of the total radiation. This framework will prove conceptually appealing, in particular also for the discussion of the physical response to radiation in Sec.~\ref{sSec:GWObservation}, as well as for the description of radiation in metric theories beyond GR. Consider therefore an SVT decomposition of the total perturbations of the physical metric $H_{\mu\nu}$ as in Eq.~\eqref{eq:hexpandBGR} but this time based on the asymptotically Minkowski background
\begin{equation}\label{eq:hexpandSecond}
H_{\mu\nu}=
 \left(\begin{array}{c|c c c} 
    	S &  &V^{T}_i+\partial_i V^{\parallel} &\phantom{0} \\
    	\hline 
    	 &  &&\\
V^{T}_i+\partial_i V^{\parallel}  &  & T\delta_{ij}+H^{TT}_{ij}+2\partial_{(i} U^{T}_{j)}+\left(\partial_i\partial_j-\frac{1}{3}\delta_{ij}\Delta\right)U^{\parallel}& \\
 \phantom{0} &  & & \\
    \end{array}\right)\,,
\end{equation}
where recall that in this case all perturbation also includes the low-frequency part. Six gauge invariant variables
\begin{align}\label{eq:gaugeinvBGRSecond}
\delta\Phi\;,\quad \delta\Theta\;,\quad\delta\Xi^{T}_i\;,\quad  H^{TT}_{ij}\,,
\end{align}
that satisfy
\begin{align}
\partial^i H^{TT}_{ij}=0\;,\quad\delta^{ij} H^{TT}_{ij}=0\;,\quad\partial^i \delta\Xi^{T}_i=0\,,
\end{align}
can then be identified as in Eq.~\eqref{eq:gaugeinvBGR}. A similar SVT decomposition also holds for the corresponding perturbed energy momentum tensor of matter fields, together with the identification of corresponding gauge invariant variables (see e.g. \cite{Mukhanov:1990me,Flanagan:2005yc}).

In GR, the linearized equations of motion [Eq.~\eqref{linEE}] then reduce to a set of sourced Laplace equations for the variables $\delta\Phi$, $\delta\Theta$ and $\delta\Xi^{T}_i$, while $H^{TT}_{ij}$ is the only variable that satisfies a sourced wave equation. While therefore only the TT component describe asymptotic radiation, the remaining gauge invariant variables describe $\mathcal{O}(1/r)$ contributions that directly associated to the total mass and total angular momentum of the system, whose time evolution is completely tied to the properties of the source (see also \cite{misner_gravitation_1973,Flanagan:2005yc,zee2013einstein}). 

Therefore, there are in fact non-zero non-propagating components within $H_{\mu\nu}$, as one could of course have guessed from black hole solutions for instance, but they are uninteresting insofar as their variations are highly constrained such that they do neither contribute to any asymptotic energy-momentum flux, nor do they induce any geodesic deviation governed by the leading order asymptotic Riemann tensor, as we will discuss in Sec.~\ref{sSec:GWObservation}. Heuristically, these quasi static contributions in the low-frequency perturbations that are entirely determined by the properties of the source are negligible in the radiation zone limit as soon as one applies either a temporal or spacial derivative on them, as in general a spacial derivative of a $1/r$ perturbation only survives if the derivative is hitting the exponential of a plane wave. 

A more precise description of the non-dynamical degrees of freedom, including their radiation-reaction, can be given in terms of the BMS formalism already alluded to above \cite{Bondi:1960jsa,Bondi:1962px,Sachs:1962wk,Strominger:2017zoo,GomezLopez:2017kcw,Compere:2019sm,Compere:2019gft}. More precisely, within asymptotic retarded coordinates [Eq.~\eqref{eq:Definition Asymptotic Retarded Time Massless}], coordinate gauge freedom can be used to choose a judicious \textit{Bondi gauge} form of the metric to eliminate all unphysical degrees of freedom while describing all physical dynamical and non-dynamical degrees of freedom within asymptotically flat spacetimes. In Sec.~\ref{MatchToAsymptoticsBD} we will get a taste of the use of such a Bondi form of the metric, in particular in connection with the associated formulation of the BMS balance laws.

\subsection{Radiation Emission Beyond GR}

Exactly as in GR, the full-fledged generation of radiation in metric theories beyond GR can only be addressed numerically for well posed formulations of the theory (recall Sec.~\ref{ssSec:WellPosedness}). Until now, NR simulations of realistic sources such as binary coalescences only exist for a handful of specific theories beyond GR \cite{Witek:2018dmd,Okounkova:2019zjf,Okounkova:2020rqw,Corman:2022xqg,East:2022rqi,AresteSalo:2022hua} and even fewer are able to evolve through merger. On the other hand, approximate waveforms can be modelled by combining PN efforts (see e.g. \cite{poisson2014gravity} for an explicit example in Brans-Dicke theory) with beyond GR ringdown computations \cite{Glampedakis:2019dqh,Wagle:2021tam,Chen:2021cts}. 

In the present work, we will, however, not be concerned with concrete solutions and only require the knowledge of the general form of asymptotic radiation. For massless and Lorentz preserving modes, this asymptotic structure will still be given by Eq.~\eqref{eq:GeneralFormRadiation}. Yet, an important aspect of beyond GR effects resides precisely in the possible modification of such a propagation behavior. This is why we will take a closer look at the propagation of waves in the next Sec.~\ref{sSec:GWPropagation} and generalize the possible asymptotic form of radiation in two important characteristics.
An explicit example of the description of radiating degrees of freedom together with their observational consequences will be given in Sec.~\ref{sSec:GWPolExample}, in both the localized gauge-fixing and the gauge-invariant SVT approaches.

Moreover, we also want to point out that obviously a second significant change in more general metric theories of gravity is given by the emission of extra radiation due to the potential presence of additional propagating degrees of freedom. In a faithful representation (Def.~\ref{DefFaithfulRep}) of such a theory, these additional DOFs are described by additional non-minimal fields $\Psi$ in the gravitational action that, depending on the precise form of the equations of motion and the properties of the source, can produce additional radiation of physical modes in the far field. 
In this context it is interesting to note that since the radiation of extra non-minimal fields, for instance an additional scalar field, can be sourced by the motion of an additional (scalar) charge carried by the source, such a radiation must not be restricted to a quadrupolar nature that dominates the gravitational radiation in the non-relativistic limit of GR discussed above. Indeed, just as in electrodynamics, the (approximate) conservation of a scalar charge of the source would only suppress the monopole term, such that dipole scalar emission can be present in principle. In Sec.~\ref{sSec:GWObservation} we will analyze the direct observational consequences of the radiation from extra non-minimal fields, which will lead to the generalization of the concept of directly observable gravitational polarizations.

\section{Gravitational Wave Propagation}\label{sSec:GWPropagation}

Once produced from a localized source as described in the previous section, radiation propagates towards a potential observer that can detect the ripples in spacetime. As mentioned, this propagation represents a further major stage in which radiation in GR can differ from radiation of more general metric theories \cite{Saltas:2014dha,Lombriser:2015sxa,Nishizawa:2017nef,Belgacem:2017ihm,Amendola:2017ovw,Belgacem:2018lbp}. Being mainly interested in the general form of asymptotic radiation [Sec.~\ref{ssSec:Description of Radiation General}], we will however primarily restrict our attention to an assessment of the velocity of propagating degrees of freedom in theories by employing the Isaacson approach introduced in Chapter~\ref{Sec:PropagatingDOFs}. More precisely, in Sec.~\ref{ssSec:SpeedofGR} we will offer a general argument for why gravitational waves in GR \textit{always} travel at the speed of light and discuss departures from that statement in more general metric theories of gravity in Sec.~\ref{ssSec:Speed Of Gravity Beyond GR}. Sec.~\ref{ssSec:PropagationSpeedConstraints} will then represent a slight detour and describe the existing constraint on the velocity of gravitational radiation in the context of propagation effects together with its stringent constraints on the theory space that will serve as a motivation to consider a particular subset of theories in Sec.~\ref{sSec:sec_HornSurv}.

\subsection{The Speed of Gravity in GR}\label{ssSec:SpeedofGR}

As discussed, we now want to provide a simple but surprisingly general argument for the statement that gravitational waves in GR always propagate at the speed of light. This argument will crucially depend on the assumptions of the Isaacson approach that is a prerequisite for a well-defined notion of gravitational waves in the first place (recall Sec.~\ref{ssSec:IsaacsonInGR}).

\paragraph{The General Local Argument.} 

As we derived explicitly in Sec.~\ref{ssSec:Local Wave Equation in GR} whenever gravitational waves can properly be defined as high-frequency perturbations, one can \textit{always} choose a local enough patch in which the low-frequency part of the metric reduced to the flat Minkowski metric [Eq.~\eqref{eq:BackgroundMInkowskiForm}]
\begin{equation}
    g_{\mu\nu}\simeq \eta_{\mu\nu}+\delta g^H_{\mu\nu}\,,
\end{equation}
Moreover, as it is also the case for matter waves, in particular electromagnetic waves, it is with respect to such inertial observers that one can unambiguously define a local notion of spacial velocity of gravitational waves (recall Eq.~\eqref{eq:definition Speed} for the local definition of spacial velocity of a particle with a given worldline). 

On this patch and outside any high-frequency sources the physical DOFs in $h^{TT}_{ij}$ satisfy a homogeneous massless wave equation 
\begin{equation}\label{eq:PlaneWaveEqMasslessTT}
    \Box h^{TT}_{ij}=-\ddot{h}^{TT}_{ij}+\Delta {h}^{TT}_{ij}=0\,.
\end{equation}
This simple statement already essentially proves that gravitational waves in GR \textit{always} travel luminally, hence with the speed of light, regardless of the precise solution of the general metric or the presence of any low-frequency matter, in other words no matter what the global background is. As we will further discuss below, this is a rather remarkable statement that ultimately follows from the Einstein equivalence Principle~\ref{Principle:EEP} ensuring the existence of local inertial observers and local Lorentz invariance. Crucially, in the case of GR, local Lorentz invariance is also ensured in the gravitational sector, regardless of the specific background.

It is a standard exercise to determine the speed of the waves based on Eq.~\eqref{eq:PlaneWaveEqMasslessTT} which we shall provide here for completeness. The simplest solution to Eq.~\eqref{eq:PlaneWaveEqMasslessTT} is given by a single plane wave that for concreteness we can choose to travel in the $z$ direction of the local Minkowski chart, that is described by the real part of
\begin{equation}\label{eq:PlaneWaveSolution}
    h^{TT}_{ij}=\mathcal{A}^{TT}_{ij} e^{ik_\mu x^\mu}= \mathcal{A}^{TT}_{ij} e^{-i(\omega t-k z)}\,,
\end{equation}
with wave frequency $\omega=k^0$ and $k=|\mathbf{k}|$ of the Fourier momentum vector $k^\mu$.
Equation~\eqref{eq:PlaneWaveEqMasslessTT} then implies that
\begin{equation}\label{eq:EquationMasslessWave}
   k_\mu k^\mu=-\omega^2+k^2=0\,,
\end{equation}
resulting in the following simple dispersion relation 
\begin{equation}
    \omega(k)=k\,.
\end{equation}
The velocity of the wave that carries physical information is then defined as the \textit{group velocity} or sound speed (see \cite{Jackson:1998nia} for a discussion of potential subtleties)
\begin{equation}\label{eq:GroupVelocity}
    v\equiv \frac{d\omega}{dk}\,,
\end{equation}
which in the case of a linear dispersion relation is equivalent to what is known as the \textit{phase velocity}
\begin{equation}\label{eq:PhaseVelocity}
    v_\text{ph}\equiv \frac{\omega}{k}\,.
\end{equation}
In this case, both velocities are as expected given by the speed of light in natural units $v=v_\text{ph}=1=c$.

\paragraph{Comparison to Electrodynamics.} 

It is interesting to put the general statement about the speed of gravitational waves in GR above into perspective and contrast it to the case of electromagnetic waves. In this case, local Lorentz invariance also ensures that in vacuum any massless propagating electromagnetic field travels at the speed of light $c$. Yet, as soon as a background medium spontaneously breaks Lorentz invariance, the local propagation speed can be modified. This is for example the case for light traveling through a dielectric material, for which the source-free Maxwell equations in physical units read \cite{Jackson:1998nia}
\begin{equation}\label{eq:MaxWellEquationsDielectric}
    -\ddot{E}+\frac{1}{\mu\,\epsilon}\Delta E=0\,,\qquad -\ddot{B}+\frac{1}{\mu\,\epsilon}\Delta B=0\,,
\end{equation}
where $\mu$ and $\epsilon$ are the permeability, respectively the permittivity of the medium, that are different from their vacuum values $\mu_0$ and $\epsilon_0$. Recall that in these units we have
\begin{equation}
    c=1/\sqrt{\epsilon_0\mu_0}\,.
\end{equation}

An associated plane-wave solution in the $z$ direction is again given by the real part of
\begin{equation}\label{eq:PlaneWaveSolution EM}
    E,B\sim e^{ik_\mu x^\mu}= e^{-i(\omega t-k z)}\,.
\end{equation}
Thus, this time Eq.~\eqref{eq:MaxWellEquationsDielectric} implies
\begin{equation}
    \omega^2=\frac{1}{\mu\,\epsilon} k^2\,,
\end{equation}
and the group velocity in Eq.~\eqref{eq:GroupVelocity} is given by
\begin{equation}\label{eq:GroupVelocityDielectric}
    v=\frac{d\omega}{dk}=\frac{1}{\sqrt{\mu\,\epsilon}}=\frac{c}{i_\text{ref}}\,,
\end{equation}
where $i_\text{ref}$ defines the refraction index
\begin{equation}
    i_\text{ref}\equiv \sqrt{\frac{\mu \,\epsilon}{\mu_0\epsilon_0}}\,.
\end{equation}
The planewave solution in Eq.~\eqref{eq:PlaneWaveSolution EM} thus becomes
\begin{equation}\label{eq:PlaneWaveLorentzViolating}
    E,B\sim e^{-i\omega( t-\frac{1}{v}z)}\,.
\end{equation}

Therefore, in contrast to the case of gravitational waves discussed above, a Lorentz breaking background medium can modify the speed of an electromagnetic wave. Moreover, observe that a priori there is no intrinsic restriction in Eq.~\eqref{eq:GroupVelocityDielectric} for the velocity to remain confined by $c$, reflecting the Lorentz violations.

\subsection{The Speed of Gravity beyond GR}\label{ssSec:Speed Of Gravity Beyond GR}

The argument for the universality of the speed of gravitational waves in GR above crucially relied on the fact that in GR there is a single gravitational field, the metric, that due to the Einstein equivalence Principle~\ref{Principle:EEP} fundamentally respects local Lorentz invariance. In other words, even for manifestly Lorentz breaking background solutions such as a cosmological spacetime (see Sec.~\ref{Sec:CosmoNutshell}), the local metric still reduces to the Minkowski metric. Moreover, the equations of motion of GR are such that its propagating degrees of freedom are massless and therefore intrinsically transverse. The speed of gravity in generic metric theories of gravity with additional dynamical DOFs can therefore differ from GR at least in two aspects: By introducing local Lorentz violations in the gravity sector and by introducing massive modes. In the following, we will discuss these two options more closely and give a certain number of concrete examples.

\subsubsection{\ul{Local Lorentz Breaking}}

In a generic metric theory of gravity, the EEP still holds. With the presence of extra non-minimal fields, however, this principle only assures local Lorentz invariance in the matter sector through the existence of Riemann normal coordinates of the physical metric and the Principle~\ref{Principle:Universal and Minimal Coupling} of minimal and universal coupling. The presence of additional non-minimal fields can however (fundamentally or spontaneously) break Lorentz symmetry in the gravity sector and therefore provide a background that can modify the propagating velocity of gravitational radiation through a detuning between the temporal and spacial derivatives. 

In order to illustrate this effect, we will take a look at two concrete classes of examples. The first one discusses theories that in an application to cosmology provide a natural cosmological background, which spontaneously breaks local Lorentz invariance. We will contrast this to the case of a manifestly Lorentz violating theory.

\paragraph{Horndeski Gravity.}

As a well known and rather general example of a metric theory of gravity, we want to discuss the propagation speed of gravitational radiation in Horndeski theory. While doing so, we will focus on the two tensorial TT DOFs that assuredly form part of the gravitational radiation that we can detect in typical GW experiments. We will come back to the detectability of additional modes in Sec.~\ref{sSec:GWObservation}. 

Recall that Horndeski gravity, introduced back in Sec.~\ref{ssSec: A Exact Theories}, is governed by the general action Eq.~\eqref{eq:ActionHorndeski} and represents the most general scalar-vector theory with equations of motions that remain at second order in derivatives per field. As an exact theory involving an additional scalar field, the EFT provides a natural cosmological background by assuming a rotationally invariant but spontaneously Lorentz symmetry breaking exact solution to the equations of motion with a characteristic frequency scale $f_L$\footnote{We are explicitly neglecting here any perturbations at the low-frequency background scales.} 
\begin{equation}\label{eq:HorndesiCosmologicalBackground}
    \Phi^L(x)=\bar{\phi}(t)\,,\qquad g^L_{\mu\nu}(x)=\bar g_{\mu\nu}(t)\,,
\end{equation}
where the homogeneous and isotropic cosmological background field configuration $\bar g_{\mu\nu}(t)$ is given in Eq.~\eqref{eq:FlatFLRWmetric}. Yet, in order to discuss the velocity of physical high-frequency perturbations on such a background, it suffices to choose a local enough chart in which the low-frequency part of the metric reduces to Minkowski spacetime and choose low-frequency Riemann normal coordinates, such that (see e.g. \cite{Flanagan:2005yc,Baldauf:2011bh,Dai:2015rda})
\begin{equation}
    \bar g_{\mu\nu}(y)= \eta_{\mu\nu}+\mathcal{O}(y^2 f_L^2)\,,
\end{equation}
on which we can describe the relevant high-frequency DOFs (recall the discussion in Sections.~\ref{ssSec:Local Wave Equation in GR} and~\ref{ssSec:WavesInMetricTheoriesOfGravity}). Indeed, the difference to GR is precisely that the background value of the scalar field in Eq.~\eqref{eq:HorndesiCosmologicalBackground} can in principle retain its Lorentz violating property on arbitrary small scales in freely falling frames. And it is precisely this Lorentz breaking background configuration of the non-minimally coupled scalar field that alters the GW propagation speed.

A direct computation of the equations of motion of the $TT$ part in the metric perturbations yields (see e.g. \cite{Kase:2018aps})
\begin{equation}\label{eq:PlaneWaveEqMasslessTT Gen}
   -\ddot{h}^{TT}_{ij}+c_T^2\,\Delta {h}^{TT}_{ij}=0\,.
\end{equation}
with
\begin{equation}
\boxed{c_T^2 = \frac{2 \bar G_4-2\bar X \bar G_{5,\Phi}-2\bar X \bar G_{5, X}\ddot{\bar{\phi}}}{4\,q_T}\,,} \label{eq:cTH}
\end{equation}
and where
\begin{equation}
    q_T=\frac14\left( 2( \bar G_4-2\bar X \bar G_{4,X})+2\bar X\,\bar G_{5,\Phi} \right)\label{eq:qTH}\,,
\end{equation}
is a coefficient associated to an effective gravitational coupling. Here, all functions are evaluated on the background configuration
\begin{equation}
   \bar G_i(\Phi,X)\equiv G_i(\bar\phi,\bar{X})\,,
\end{equation}
where 
\begin{equation}
    \bar X=\frac{1}{2}\dot{\bar{\phi}}^2\,.
\end{equation}

Following the discussion on plane wave solutions in Sec.~\ref{ssSec:SpeedofGR}, in particular Eq.~\eqref{eq:GroupVelocityDielectric} it follows immediately, that $c_T$ represents the velocity of the wave that for general functionals $G_4$ and $G_5$ is therefore indeed modified through the presence of a non-trivial Lorentz breaking non-minimal scalar background. Note that one could argue that for a cosmologically relevant theory, where the background value of the scalar field $\bar{\phi}(t)$ only evolves on cosmological timescales, on a local enough patch the background scalar field can be treated as a constant and thus in particular $\bar X\simeq 0$, implying $c_T\simeq 1$. Indeed, for Horndeski theory on a static Minkowski background, with $\bar{\phi}=\text{constant}$, local Lorentz symmetry is preserved and the GWs propagate luminally. 

Yet, even the tiniest amount of $\dot{\bar{\phi}}\neq 0$ with a change in the locally defined velocity will build up over time as the wave is traveling, even though locally such a change might not be detectable. In other words, in contrast to GR, the global propagation through the universe will still be modified in comparison the trajectory of light, which due to minimal and universal coupling only feels the background of the physical metric. We will come back to the assessment of such non-localized propagation effects in Sec.~\ref{ssSec:PropagationSpeedConstraints} below.

Thus, Horndeski gravity or other scalar-tensor theories with a nontrivial Lorentz breaking background value of the non-minimal scalar have the potential to fundamentally modify the speed of gravitational waves. However, this must not necessarily be the case for every value of the general functionals $G_i$. Concretely, given a time varying scalar background, one can thus ask the question, under what condition is the propagation still luminal, hence $c_T= 1$.
Using the background equations of motion this condition can be reduced to \cite{Kase:2018aps}
\begin{equation}
2G_{4,X}-2G_{5,\Phi}-\ddot{\bar{\phi}}\,G_{5,X}= 0\,.
\end{equation}
In other words, if one does not allow for a fine-tuning between different functionals\footnote{As shown in \cite{Ezquiaga:2017ekz}, a generalization to DHOST theories, also discussed in Sec.~\ref{ssSec: A Exact Theories}, could allow for a consistent cancellation of anomalous speed contributions without setting the functionals to zero.}, which is in general believed to be unstable \cite{Ezquiaga:2017ekz}, a luminal propagation of Horndeski theories therefore requires the constraints
\begin{equation}\label{eq:HG4G5c}
G_{4,X}=0 \qquad \text{and} \qquad G_5=\text{const.}\,.
\end{equation}
These constraints translate into a restriction of the luminal Horndeski Lagrangian in Eq.~\eqref{eq:ActionHorndeski} to the simple form
\begin{equation}\label{eq:Hsurv}
\boxed{L^{\myst{H}}_{\myst{luminal}}=G_2(\Phi,X)-G_3(\Phi,X)\Box\Phi+G_4(\Phi)\,R\,,}
\end{equation}
since $L^{\myst{H}}_5$ vanishes identically due to the Bianchi identity.

\paragraph{Generalized Proca Gravity.} 

While a single non-minimal field vector-tensor theories possess an intrinsic difficulty of modelling cosmological homogeneous and isotropic cosmological backgrounds, this is not the case for scalar-vector tensor theories, in particular the massive generalized Proca family described by the action in Eq.~\eqref{eq:ActionGenProca}. This gauge symmetry breaking theory contains a natural analogue cosmological background solution to Eq.~\eqref{eq:HorndesiCosmologicalBackground} of the form
\begin{equation}\label{eq:HGEnProcaCosmologicalBackground}
    A_\mu(x)=\bar{A}(t)\,,\qquad g_{\mu\nu}(x)=\bar g_{\mu\nu}(t)\,.
\end{equation}
with very similar GW speed equations as Horndeski gravity under the replacement $\dot{\bar{\phi}}(t)\rightarrow \bar{A}(t)$ (see \cite{DeFelice:2016uil}). However, note the crucial difference, that for a vector field even a constant value of the temporal component the background breaks local Lorentz symmetry and $c_T$ does not reduce to $c$ as it was the case for Horndeski theory. Hence, even a static ``Minkowski-like'' background solution of the non-minimal vector field can cause a modification of the local velocity of GWs and therefore induce a departure from the locally measurable speed of light.

Without fine-tuning, the luminality conditions remain [Eq.~\eqref{eq:HG4G5c}]
\begin{equation}\label{eq:HG4G5c Second}
G_{4,Z}=0 \qquad \text{and} \qquad G_{5,Z}=0\,.
\end{equation}
However, note that these conditions do not affect the additional Lagrangian $L^{\myst{GP}}_6$ that has no natural scalar Horndeski counterpart. Thus, the luminal Lagrangian is able to keep a larger structure
\begin{align}\label{eq:GenProcaLuminal}
\boxed{L^{\myst{GP}}_{\myst{luminal}}=L^{\myst{GP}}_2+L^{\myst{GP}}_3+G_4\,R+ L^{\myst{GP}}_6 \,,}
\end{align}
although it is worth noticing that in this case the non-minimal coupling to the Riemann tensor is lost completely.

\paragraph{Einstein-\AE{}ther Gravity.}

On the other hand, a second more direct option for metric theories to modify the speed of gravitational waves is through explicit Lorentz breaking. A well known example is Einstein-\AE{}ther, introduced in Eq.~\eqref{eq:Action EA}, that represents a scalar-vector-tensor theory with an a priori constraint on a non-minimal vector field to admit a Lorentz violating configuration. As studied in \cite{Jacobson:2004ts,Jacobson:2007veq} the theory admits a parameter space in which all physical gauge invariant modes are well-behaved and admit a linear dispersion relation of the form
\begin{equation}
    \omega(k) = v k\,,
\end{equation}
and thus in that respect behave just as the electromagnetic modes in a Lorentz violating medium discussed in Sec.~\ref{ssSec:SpeedofGR} above.

\subsubsection{\ul{Massive Degrees of Freedom}}

A second straightforward way in which the velocity of waves in metric theories beyond GR can differ is through the existence of massive modes. In this work, we will, however, not directly discuss massive gravity theories that involve massive tensorial degrees of freedom. Yet, we will still allow other non-minimal fields to describe massive DOFs, that might influence the perturbations of the physical metric through their non-minimal couplings.

The description of a massive but Lorentz preserving degree of freedom is fundamentally different from the Lorentz breaking case discussed above. A general massive but local Lorentz preserving mode on a Minkowski background satisfies the Lorentz invariant Klein-Gordon equation
\begin{equation}
    (\Box  -m^2)\delta\Psi=-\delta\ddot \Psi +\Delta \delta\Psi -m^2\delta\Psi=0\,.
\end{equation}
A plane wave solution, again for concreteness in the $z$ direction
\begin{equation}
    \delta\Psi\sim e^{ik_\mu x^\mu}=  e^{-i(\omega t-kz)}\,,
\end{equation}
is therefore characterized by the relation
\begin{equation}
    -k_\mu k^\mu=\omega^2-k^2=m^2\,.
\end{equation}
This results in the dispersion relation
\begin{equation}
    \omega(k)=\sqrt{k^2+m^2}\,,
\end{equation}
and in consequence leads to a group velocity [Eq.~\eqref{eq:GroupVelocity}] of the form
\begin{equation}\label{eq:SpeedMassiveDOF}
    v=\frac{d\omega}{dk}=\frac{k}{\sqrt{k^2+m^2}}=\sqrt{1-\frac{m^2}{\omega^2}}\,.
\end{equation}

Observe that this expression naturally preserves the Lorentz symmetric constraint $|v|<1$ in units of $c$. Moreover, in contrast to the Lorentz breaking case considered above, the group velocity is frequency dependent. Below, we will discuss the implication of this observation for the strategies of formulating constraints on such theories. Furthermore, the plane wave equation can therefore be written as
\begin{equation}\label{eq:PlaneWaveMassive}
    \boxed{\delta\Psi\sim e^{-i\omega(t-vz)}\,.}
\end{equation}
Note in particular the difference to the expression in the Lorentz violating case in Eq.~\eqref{eq:PlaneWaveLorentzViolating}. For consistency, we will generally assume a small enough mass such that the massive wave can still be treated as a radiation component that reaches an asymptotic region in the source centered coordinates \cite{poisson2014gravity}. This implies that any typical observer in the radiation zone is fundamentally not in the rest frame of the massive modes.

\subsection{Propagation Speed Constraints}\label{ssSec:PropagationSpeedConstraints}

\paragraph{Propagation Effects.}
Given the local statements on the velocity of waves for given inertial observers, one could ask about the propagation of such waves throughout an arbitrary spacetime background on scales larger than the local Minkowski patches. Within GR, high-frequency (or short wavelength) waves that locally travel on the Minkowski lightcone will always propagate along null geodesics of the entire arbitrary background spacetime (see also \cite{Isaacson_PhysRev.166.1263,misner_gravitation_1973,maggiore2008gravitational}).\footnote{This for instance also means that gravitational radiation also features gravitational lensing.} However, the non-localized propagation on general background spacetimes might induce physical effects on the wave that are not describable in the local frame.\footnote{For instance, in the case of a cosmological background, a global effect of the modified wave equation on a cosmological background implies that sub-Hubble (an expression explained in Part~\ref{Part: Cosmological Testing Ground}) waves decay as $1/a$, with $a$ the scale factor. Thus, any radiation component is subject to a fundamental gravitational redshift (see Sec.~\ref{sSec:HomIsoUniverse}).} 
As concerns the picture of radiation emitted from a localized source, such \textit{propagation effects} in particular in the cosmological setting, are often best studied by first considering an asymptotically flat limit around a source, which is a good approximation for small enough scales. Propagation effects can then be described to kick in on cosmological scales on top of such an asymptotically flat solution as the emitted radiation propagates further.

In more general metric theories of gravity, of course additional propagation effects might be considered based on the background solutions of the additional non-minimal fields. To give just one example, we want to mention \textit{gravitational wave birefringence} \cite{Grishchuk:1974ny,Yunes:2010yf,Yunes:2013dva,Creminelli:2014wna,Kostelecky:2016kfm,Nair:2019iur,Qiao:2019wsh,Zhao:2019xmm,Shao:2020shv,Yamada:2020zvt,Okounkova:2021xjv,Wang:2021gqm,ONeal-Ault:2021uwu,Zhao:2022pun} arising in parity violating backgrounds, that may lead to an asymmetry in both the propagation speeds and amplitudes of the left- and right-handed polarizations of the TT waves. Through a modification of the propagation speed, GW birefringence thus represents another potential source of Lorentz violations. Indeed, the fact that parity breaking also leads to a modification of the propagation speed can be understood from a fundamental relation between the local Lorentz and the parity symmetry \cite{Greenberg:2002uu}.

Here, we want to focus on the modifications of the local propagation speed. In terms of the general background, this implies that in contrast to luminal radiation, waves with an altered propagation speed do not propagate along the null cones of the general background metric. Rather, their propagation can be characterized by an alternative effective metric \cite{Bettoni:2016mij,Ezquiaga:2017ekz} that defines a different causal structure for the affected degrees of freedom.

\paragraph{Measurements of Propagation Speed.} 

While there exist multiple indirect probes of the speed of gravitational radiation \cite{Moore:2001bv,Yagi:2013qpa,Jimenez:2015bwa}, precise direct local measurements are tricky as they by definition rely on very short timescales, such as the difference of arrival time in different GW detectors. On the other hand, a non-local setup which would allow for a larger travel-time, increasing the precision seem unfeasible at first sight. First of all, this is due to a lack of any distant GW detector. More fundamentally, however, it is in principle relatively free of meaning to compute an ``averaged'' spacial velocity with respect to a given global coordinate system due to issues of defining simultaneous events and spacial proper distances (recall Sec.~\ref{sSec:Special Relativity}). Indeed, as already discussed, spacial velocities in generally curved spacetimes fundamentally only has a precise local meaning for a given observer.

However, one can imagine a special situation in which a non-local assessment of the luminosity of GW propagation is possible. Namely, by comparing the time of arrival of a GW signal and an electromagnetic signal emitted from the same distant source. In GR, under the assumption that light indeed travels along the light-cones of the physical metric of spacetime, gravitational and electromagnetic radiation both propagate along null geodesics in the short-wavelength limit. Thus, their path through curved spacetime is expected to be exactly the same, such that a difference in arrival time of two simultaneously emitted multi-messenger signals could be regarded as a clear indication of a departure from luminality of GWs. Of course, such a measurement relies on the non-verifiable assumption of simultaneous emission. However, even a rather large such uncertainty will be compensated by the large travel-time.

Precisely such a multimessenger event was recently observed, namely through the detection of a binary neutron star merger \cite{LIGOScientific:2017vwq} with associated electromagnetic counterpart in the form of a gamma-ray burst signal. By the mere fact that the optical signal was observed around $1.74$ s after the merger, together with very conservative assumptions on the distance and the not yet fully understood production of the gamma-ray burst which followed the binary NS coalescence, this single event is able to constrain the propagation speed of gravitational radiation to \cite{LIGOScientific:2017zic}
\begin{equation}\label{eq:Cconst}
-3\times 10^{-15}\leq c_{T}/c-1 \leq 7\times 10^{-16}\,.
\end{equation}
While the upper bound solely relies on the assumption that the gamma-ray burst did not occur before merger, the lower bound was obtained by setting the delay time between the emission of the two signals to $10$ s, even though most models expect a delay below $4$ s. Over a distance of $26$ Mpc, however, the constraint remains remarkable despite the conservative estimates. 

\paragraph{Implications on the Theory Space Beyond GR.} At first sight, this bound entails tremendous implication on non-minimally coupled additional gravitational degrees of freedom, in particular alternative dark energy models that involve a Lorentz breaking background configuration as discussed in Sec.~\ref{ssSec:Speed Of Gravity Beyond GR} above \cite{Bettoni:2016mij,Lombriser:2016yzn,Ezquiaga:2017ekz,Creminelli:2017sry,Sakstein:2017xjx,Baker:2017hug,Langlois:2017dyl,Heisenberg:2017qka,Amendola:2017orw,Akrami:2018yjz,Kase:2018aps}. For instance, luminal propagation strongly suggests that the theory space of Horndeski and generalized Proca theories, whenever employed in a cosmological setting, should be restricted to the Lagrangians in Eqs.~\eqref{eq:Hsurv} and \eqref{eq:GenProcaLuminal}.

At this point, it is important to stress, however, that the GW sound velocity defined through the group velocity in Eq.~\eqref{eq:GroupVelocity} of a gravitational EFT should more precisely be regarded as the low energy speed that dominates as long as higher order pEFT contributions remain sufficiently suppressed. This can for example be seen through an analysis of the retarded propagator \cite{Caldwell:1993xw,deRham:2019ctd}. As was pointed out in \cite{deRham:2018red}, irrelevant operators near the cutoff scale can significantly affect the speed of propagation of gravitational waves, thus unavoidably introducing a frequency dependence in the dispersion relation as the edge of validity of the EFT is approached. Moreover, assuming a Lorentz invariant UV-completion, one would naturally expect a luminal propagation at high enough energies, regardless of the details of the Lorentz-breaking background field configuration. 

It turns out, that cosmic EFT's such as Horndeski theories precisely break down at energies of the order of $100$ Hz or lower \cite{deRham:2018red}, which coincides with the frequency band of $10$ - $100$ Hz at which LIGO-Virgo constrains the speed of gravitational waves. Hence, it could technically be that the transition towards a Lorentz invariant UV completion of such a cosmological model happens before the LIGO band, such that  constraints in Eq.~\eqref{eq:Cconst} could be avoided. Stronger constraints will therefore be able to be posed with the planned LISA mission, sensitive to $10^{-3}$ - $10^0$ Hz GWs. It should be stressed, however, that such considerations above involve a lot of speculation and should themselves be taken with care. In fact a corresponding frequency dependence of the propagation speed entails further challenges. 

\paragraph{Constraints on Frequency Dependent Velocities.} Indeed, as soon as models with a running of the GW sound velocity with frequency are considered, the speed of gravity can be probed non-locally even without the existence of an optical counterpart. This brings us directly to observational constraints on massive DOFs as well, since in this case the velocity depends on the frequency as seen in Eq.~\eqref{eq:SpeedMassiveDOF}.

A frequency dependence of the sound speed can be probed non-locally by comparing different parts of an asymptotic radiative signal with distinct frequency content. For instance, for a typical CBC event, the low frequency early inspiral can be contrasted to the high frequency merger, where a potential mass of the DOFs associated to the measured gravitational waves would lead to a slower propagation of the lower frequencies and therefore distort the signal. More precisely, such a dispersion of GWs can be tested for through the gravitational phasing \cite{Will:1997bb} in order to formulate bounds on a potential mass of the graviton $m_g\leq1.27\times 10^{-23}$eV \cite{LIGOScientific:2021sio}. These represent dynamical tests that can be regarded as complementary to the much stronger bounds associated to cosmological constraints of $m_g\lesssim 10^{-30}$ eV \cite{Tolley:2017yje,DeFelice:2021trp}.

\subsection{Description of Asymptotic Radiation.}\label{ssSec:Description of Radiation General}

We want to end this section by using the above discussion on the propagation of (gravitational) waves on a local Minkowski background to given general expressions for the radiation, hence the $\mathcal{O}(1/r)$ propagating perturbations in the radiation zone of an asymptotically flat spacetime, that we will employ in this work. 


\paragraph{Radiation in GR.} 

The arguments for a luminal propagation of GWs in GR, given in Sec.~\ref{ssSec:SpeedofGR} for the locally defined gravitational waves also hold for the gravitational radiation in the setup of an asymptotically flat spacetime described at the beginning of this chapter. In other words, and as shown explicitly in Sec.~\ref{sSec:GWGeneration} for perturbative sources, the asymptotic propagating DOFs will always travel at the speed of light and the physical TT radiative modes can always be described in the far field limit as a superposition of plane waves. Moreover, due to linearity, and the fact that for an observation of radiation from a given astrophysical source the direction of the wave is very well-defined, it is sufficient to analyze individual species of plane waves that depend on a single frequency and a given propagation direction (see \cite{misner_gravitation_1973,maggiore2008gravitational}). This would be different when considering a stochastic GW background, for instance, where a more careful analysis of the superposition of waves is required. Moreover, in order to describe asymptotic outward radiation, a natural ``no-incoming radiation'' boundary condition is generally imposed.

Concretely, the propagating DOFs of GR can therefore without loss of generality be described through a single massless locally Lorentz invariant radially outward plane wave of the TT component of the metric perturbations with the general asymptotic form
\begin{equation}
    H_{ij}^{TT}(t,r,\Omega)=\Re\left[\frac{\mathcal{A}_{ij}^{TT}(\Omega)}{r}\,e^{i k_\mu x^\mu}\right]\,,
\end{equation}
where $\Re$ denotes the real part and the Fourier vector $k^\mu$ again satisfies Eq.~\eqref{eq:EquationMasslessWave}, indicating a luminal propagation.   
Imposing the wave to be in radially outward direction
\begin{equation}
    \mathbf{n}\equiv\frac{\mathbf{k}}{k}=\frac{\mathbf{x}}{r}\,,
\end{equation}
then assures that $\mathbf{n}$ coincides with Eq.~\eqref{eq:Def Direction of Propagation n} and 
\begin{equation}
     \boxed{H_{ij}^{TT}(u,r,\Omega)=\Re\left[\frac{\mathcal{A}_{ij}^{TT}(\Omega)}{r}\,e^{-i\omega u}\right]\,,}
\end{equation}
where $u=t-r$ is again the asymptotic retarded time. To first order in $1/r$, the propagating radiation satisfies [Eq.~\eqref{eq:IdentityRadiation}]
\begin{equation}\label{eq:IdentityRadiationSecond in}
    \partial^i H_{ij}^{TT}=-n^i  \,\dot H_{ij}^{TT}\,,
\end{equation}
reflecting the fact that a massless wave-equation is satisfied.

\paragraph{Radiation Beyond GR.} 
In more general metric theories of gravity, the additional propagating degrees of freedom can of course also be excited to be part of the asymptotic radiation. For simplicity, we will in the following however disregard the possibility of local Lorentz symmetry violations and therefore restrict ourselves to massless and massive locally Poincaré propagating DOFs. Following Eq.~\eqref{eq:PlaneWaveMassive}, each physical mode in Eq.~\eqref{eq:gaugeinvBGRSecond} of mass $m$ could therefore potentially be described by an asymptotic radially outward plane wave of the form
\begin{equation}\label{eq:GeneralPlaneWaveRadiationMassive}
    \boxed{\delta\Psi(t,r,\Omega)=\Re\left[\frac{\mathcal{A}(\Omega)}{r}\,e^{-i\omega(t-vr)}\right]\,,}
\end{equation}
with group velocity $v$ given in Eq.~\eqref{eq:SpeedMassiveDOF} and propagating along a radially outward direction $\mathbf{n}$.
Observe that therefore to first order in $1/r$ each propagating DOF satisfies (compare to Eq.~\eqref{eq:IdentityRadiationSecond in})
\begin{equation}\label{eq:IdentityRadiationSecond}
    \partial_i \delta\Psi=-v\,n_i  \,\delta \dot\Psi\,.
\end{equation}
The massless case is then simply given by the values $m=0$ and $v=1$.

\section{Gravitational Wave Observation}\label{sSec:GWObservation}

We now turn to the question of the experimental detection of gravitational radiation. First, Sec.~\ref{ssSec:The Physical Effects of Gravitational Waves} will provide a careful assessment of the observable effects of radiation that will be heavily based on the definition of metric theories of gravity and in particular the previous discussions in Secs.~\ref{sSec:Special Relativity} and \ref{sSec:Metric Theories}. This will directly lead to the identification of the six gravitational polarizations in Sec.~\ref{sSec:GWPolGen}, corresponding to the modes in the physical metric that govern an idealized GW detector response, more closely analyzed in Sec.~\ref{ssSec: GW Experiments}. Especially the careful description of gravitational polarizations, as well as the introduction into spin-weighted expansions will prove important for the subsequent Chapter~\ref{Sec:GWMemory} on the first main result of this work.

\subsection{The Physical Effects of Radiation}\label{ssSec:The Physical Effects of Gravitational Waves}

The starting point in discussing the observation of gravitational waves is to derive a general formula capturing the relevant physical effects of GWs. The framework of metric theories of gravity introduced in Chapter~\ref{Sec:The Generalization to Gravity} precisely provides a well-defined description of such a physical response through the geodesic deviation equation that in a generic coordinate system reads [Eq.~\eqref{eq:GeodesicDeviation}]
\begin{equation}\label{eq:GeodesicDeviation2}
    \frac{D^2\delta x^\mu}{d\lambda^2}=-R\ud{\mu}{\nu\rho\sigma}\,\delta x^\rho \,\dot x^\nu\dot \,x^\sigma\,,
\end{equation}
where $R\ud{\mu}{\nu\rho\sigma}$ is the Riemann tensor associated to the Levi-Civita connection and the physical metric $g_{\mu\nu}$ and $\delta x^\mu$ is the infinitesimal distance vector between two nearby geodesics $x^\mu(\lambda)$ and $x^\mu(\lambda)+\delta x^\mu(\lambda)$ at each value of $\lambda$, that satisfies
\begin{equation}\label{eq:ConditionDeviationVectorSecond}
    [\underline{\delta x},\underline{\dot{x}}]=0\,.
\end{equation}
Recall that the covariant derivative of the components of a vector field along a curve is given by [Eq.~\eqref{eq:DefCovariantDerivativeD}]
\begin{equation}
\frac{D\delta x^\mu}{D\lambda}\equiv \frac{d\delta x^\mu}{d\lambda}+\Gamma^\mu_{\nu\rho}\,\delta x^\nu\dot{x}^\rho\,,
\end{equation}
where $\Gamma^\mu_{\nu\rho}$ are the Christoffel symbols. Moreover, since $\delta x^\mu$ is an infinitesimal vector between two points on the manifold, its norm $g_{\mu\nu}\delta x^\mu\delta x^\nu$ corresponds to an infinitesimal measure of spacetime distance between the two geodesics. It is worth recalling, that the geodesic deviation equation is only valid up to first order in $\delta x^\mu$ and its derivative. More precisely, the equation is valid up to fractional errors of $\mathcal{O}(\delta x/L)$, where $L$ is the typical length-scale of variation of the curvature \cite{Flanagan:2005yc,maggiore2008gravitational}.

In the reminder of this subsection, we will simplify the geodesic deviation equation to a practical form and then compute the general response to gravitational radiation in an asymptotically flat spacetime. In contrast to the above sections, we will directly treat the most general case of an arbitrary metric theory and comment on the specific GR case on the way. 

\paragraph{Spacial Response on Timelike Geodesics.} 

The most natural way to physically measure a geodesic deviation is to consider the change in spacetime geodesics provided by two test masses (or physical observers) and hence to consider two timelike geodesics. In that case, as discussed at the end of Sec.~\ref{sSec:Special Relativity}, the geodesic deviation of a metric theory of gravity only contains information on the spacial separation between the geodesics. Indeed, as shown explicitly in Appendix~\ref{sApp:SpacialGeodesicDeviation}, the condition in Eq.~\eqref{eq:ConditionDeviationVectorSecond} implies that for a spacetime with vanishing torsion and non-metricity (implicit in our Definition~\ref{DefMetricTheory} of metric theories of gravity) the projection of the deviation vector onto the geodesic $\dot{x}_\mu\delta x^{\mu}$ remains constant along the geodesic and can therefore without loss of information be set to zero [Eq.~\eqref{eq:Condition temporal Deviation Vector}]
\begin{equation}
    \dot{x}_\mu\delta x^{\mu}=0\,.
\end{equation}
In that case, it makes sense to talk about a purely spacial deviation that therefore monitors a spacial proper distance between two simultaneous events of a given observer. Moreover, the simultaneity between events can operationally locally be determined through light-signal exchanges between the two physical observers defining the timelike geodesics.

In fact, the most practical method of determining the spacial proper distance between two events is precisely by measuring the light travel time between the two events by using the universality of the speed of light to convert a measure of proper time into a measure of proper distance as explicitly derived in Sec.~\ref{sSec:Special Relativity}. This is precisely the basics idea behind today's operational gravitational wave detectors. In the simplest case that we will treat here, a series of additional assumptions will allow us to considerably simplify the physical response to GW's even further.

It is important to realize that the physical response to gravitational fields, which in an idealized setup can be measured by the movement of test-masses given by the geodesic deviation in Eq.~\eqref{eq:GeodesicDeviation2}, is valid in all metric theories of gravity and is independent of the gravitational equations of motion. This is because the geodesic equation that is satisfied by any test-mass is a direct consequence of the minimal coupling to matter. Note however, that such a universal statement is lost in theories that would break the Einstein equivalence principle as not all test-masses are guaranteed to follow the geodesics of the physical metric, such that in principle the determination of the gravitational field, and therefore of spacetime would intrinsically depend on the experimental setup, in particular the nature of test-masses used to perform the experiment. 

\paragraph{Geodesic Deviation in Fermi Normal Coordinates.} 

In order to derive the classic result of the geodesic deviation equation relevant for the physical GW response in arbitrary metric theories of gravity, it is useful to consider a particular type of simplifying coordinates, the Fermi normal coordinates $\{t,y^i\}$ that we already encountered (see App.~\ref{sApp: Normal Coordinates}). These coordinates can be viewed as the closest that one can get to Minkowski coordinates in a general spacetime by considering the freely falling frame of one of the geodesics. In other words, Fermi normal coordinates describe Riemann normal coordinates along an entire timelike geodesic $y^\mu(\tau)$ that can be constructed for every metric theory of gravity with Levi-Civita connection. The precise form of the coordinates are given in Eq.~\eqref{FermiNormalCoords} but for our purposes we again only require knowing the general form of the metric, that reads
\begin{equation}\label{eq:FermiNormalCoordsSecond}
    g_{\mu\nu}(y)=\eta_{\mu\nu}+N_{\mu\nu \,i j}\,y^i y^j+\mathcal{O}\left(\frac{y^3}{D^3}\right)\,,
\end{equation}
expanded up to second order in spacial coordinates $y^i$ in the given flotation around the origin of the spacial grid that is set by the geodesic. Moreover, recall that the coefficients $N_{\mu\nu \,i j}$ are of the order of the spacetime curvature $D^{-2}\sim|R_{\mu\nu\rho\sigma}|$ [Eq.~\eqref{eq:characteristic spacetime dimension of curvature}] evaluated on the geodesic.

As already derived in Eq.~\eqref{eq:GeodesicDeviationFermiNormalCoords} in these coordinates up to the given error, the geodesic deviation simplifies to 
\begin{equation}\label{eq:GeodesicDeviationFermiNormalCoordsSecond}
    \frac{D^2\delta y^\mu}{d\tau^2}=-\left(\Gamma^\mu_{\nu\sigma,\rho}-\Gamma^\mu_{\nu\rho,\sigma}\right)\delta y^\rho \dot y^\nu\dot y^\sigma\,,
\end{equation}
with $\dot y^\mu$ the tangent vector of the first derivative, since the Christoffel symbols evaluated on $y^\mu(\tau)$ vanish and the Riemann tensor reduces to
\begin{equation}\label{FermiNormalRiemann}
R\ud{\mu}{\nu\rho\sigma}=\Gamma^\mu_{\nu\sigma,\rho}-\Gamma^\mu_{\nu\rho,\sigma}\,.
\end{equation}

For timelike geodesics the above expression is very useful as Eq.~\eqref{eq:FermiNormalCoordsSecond} implies that to first order in $y^i$, the deviation vector $\delta y^\mu$ directly measures proper spacetime distances in the sense that its (infinitesimal) norm is given by
\begin{equation}\label{eq:FormOfMetricFermiN}
  g_{\mu\nu}  \delta y^\mu\delta y^\nu=\eta_{\mu\nu}\delta y^\mu\delta y^\nu+\mathcal{O}\left(\frac{y^2}{D^2}\right)\,.
\end{equation}
This also means that to leading order the temporal components $\delta y^0$ are a direct measure of proper time
\begin{equation}
    d\tau^2=(\delta y^0)^2+\mathcal{O}\left(\frac{y^2}{D^2}\right)\,,
\end{equation}
while the spacial components $\delta y^i$ naturally correspond to a spacial proper distance defined in Eq.~\eqref{eq:ProperDistanceGeneral}
\begin{equation}
    d\ell^2=\delta_{ij}\delta y^i\delta y^j+\mathcal{O}\left(\frac{y^2}{D^2}\right)\,.
\end{equation}
Thus, to first order in the expansion to which we will restrict ourselves, the geodesic deviation measured by $\delta y^i$ in Fermi normal coordinates directly corresponds to the deviation in proper distance that we are after. To make this point clear, we will from now on write 
\begin{equation}
    \ell^i\equiv \delta y^i\,.
\end{equation}

Moreover, recall that in Fermi normal coordinates, by definition the tangent vector $\dot y^\mu$ of the first derivative only has a temporal component along the direction of proper time, hence
\begin{equation}
    \dot{y}^0=1\,,\qquad \dot{y}^i=0\,.
\end{equation}
Furthermore, note that only spacial derivatives acting on Christoffel symbols evaluated on the geodesic contribute, because the expansion to second order in Eq.~\eqref{eq:FormOfMetricFermiN} only involves the spacial coordinates. Therefore, only terms with at least two spacial derivatives of the metric do not identically vanish when evaluated on the geodesic at $y^i=0$, and thus
\begin{equation}\label{FermiNormalChris}
\partial_0\Gamma^\mu_{\rho\nu}=0\,.
\end{equation}
Finally, Eq.~\eqref{eq:FormOfMetricFermiN} also implies that any directional covariant derivative along the geodesic evaluated on the geodesic can be replaced by a time derivative and thus a derivative of proper time. Hence, in particular, we have
\begin{equation}
    \frac{D^2 \ell^i}{d\tau^2}= \ddot{\ell}^i\,.
\end{equation}

Gathering all of the above, as well as recalling that any deviation in proper time is trivial such that without loss of generality one can set $\delta \ell^0=0$, in Fermi normal coordinates the geodesic deviation equation to leading order can be written as an equation for the evolution of proper distance that reads
\begin{equation}\label{eq:GeodesicDeviationFermiNormalCoordsThird}
    \boxed{ \ddot{\ell}^i=-R\ud{i}{0j0}\, \ell^j\,, }
\end{equation}
where
\begin{equation}
    R\ud{i}{0j0}=\Gamma^i_{00,j}\,.
\end{equation}

Even though Equation~\eqref{eq:GeodesicDeviationFermiNormalCoordsThird} was derived in a particular coordinate system, it in fact represents the physical response to a non-trivial curvature on two timelike geodesics in any frame in which the assumptions that entered the formulation of the geodesic deviation equation hold and the spacetime region is localized enough. This is because it describes the evolution of an infinitesimal vector of proper distance due to the electric parity component of the Riemann tensor of the physical metric, which both are gauge invariant concepts. Moreover, we want to stress again that this result is valid for all metric theories of gravity with Levi-Civita connection that obey the Principle~\ref{Principle:Universal and Minimal Coupling} of universal and minimal coupling and in particular is independent of the equations of motion.

\paragraph{Spacial Response in the Radiation Zone.}

In order to describe the response given by Eq.~\eqref{eq:GeodesicDeviationFermiNormalCoordsThird} of an idealized GW detector consisting of two freely falling test masses, whose proper spacial distance is monitored through light-travel time measurements, we therefore need to evaluate the associated electric part of the Riemann tensor $R\ud{i}{0j0}$ in the appropriate limit of incoming gravitational radiation. As discussed, for simplicity we will restrict to the asymptotically flat case and assume the presence of radiation as propagating $\mathcal{O}(1/r)$ corrections that can formally be described as perturbations on top of a Minkowski background [Eqs.~\eqref{eq:perturbationsAsymptoticMetric} and \eqref{eq:perturbationsAsymptoticNonMinimal}] 
\begin{align}
    g_{\mu\nu}=\eta_{\mu\nu}+H_{\mu\nu}+\mathcal{O}(1/r^2)\,,\quad \Psi = \bar\Psi+\delta\Psi+\mathcal{O}(1/r^2)\,.
\end{align}
Observe that Eq.~\eqref{eq:GeodesicDeviationFermiNormalCoordsThird} of the physical response to radiation in a generic metric theory of gravity then directly implies that only the perturbations of the physical metric have a measurable impact on an idealized GW detector, as already mentioned on several occasions.

It remains to actually compute the Riemann tensor $R_{i0j0}$ to leading order in $H_{\mu\nu}$
\begin{equation}
    R_{\mu\nu\rho\sigma}=\phantom{}_{\mys{(1)}}R_{\mu\nu\rho\sigma}+\mathcal{O}(1/r^2)\,.
\end{equation}
The expression for the linearized Riemann tensor of a given perturbation and background variable was given in Eq.~\eqref{eq:RiemmanFirstOrder}, such that the electric parity components on a flat background reduce to
\begin{equation}\label{eq:linRiemann}
\phantom{}_{\mys{(1)}}R_{i0j0}[H]=-\frac{1}{2}\left(\partial_0\partial_0 H_{ij}+\partial_{i}\partial_j H_{00}-\partial_0\partial_i H_{0j}-\partial_0\partial_j H_{0i}\right)\,.
\end{equation}
Because $\phantom{}_{\mys{(1)}}R_{\mu\nu\rho\sigma}$ is a gauge invariant quantity, we can actually calculate it in any gauge we like. However, it is illuminating to confirm this statement by explicitly showing that it can be written in terms of gauge invariant variables of the metric perturbations, all of which contribute to the physical response.

Let's therefore consider the general SVT expansion of gravitational radiation given in Eq.~\eqref{eq:hexpandSecond} and evaluate the expression Eq.~\eqref{eq:linRiemann}
\begin{align}
\phantom{}_{\mys{(1)}}R_{i0j0}=-\frac{1}{2}&\Big(\partial_{0}\partial_{0}H^{TT}_{ij}-2 \partial_{0}\partial_{(i}[V^{T}-\dot{U}^{T}]_{j)}+\delta_{ij}\partial_{0}\partial_{0}\left[T-\frac{1}{3}\Delta U^{\parallel}\right]\nonumber\\
&+\partial_i\partial_j \left[S-2\dot{V}^{\parallel}+\ddot{U}^{\parallel}\right]\Big)\,.
\end{align}
Indeed, the expressions in the square brackets precisely correspond to the special combinations of gauge invariant variables identified back in Eq.~\eqref{eq:gaugeinvBGR} and the linearized Riemann tensor can be written as
\begin{align}\label{eq:PertRiemannFirst}
\boxed{\phantom{}_{\mys{(1)}}R_{i0j0}=-\frac{1}{2}\Big(\partial_{0}\partial_{0}H^{TT}_{ij}-2 \partial_{0}\partial_{(i}\delta\Xi^T_{j)}+\delta_{ij}\partial_{0}\partial_{0}\delta\Theta+\partial_i\partial_j \delta\Phi\Big)\,.}
\end{align}
Thus as anticipated, the leading order electric part of the Riemann tensor can indeed be written entirely in terms of gauge-invariant quantities. This represents the local combination of gauge-invariant variables of the perturbations of the physical metric that can be detected in a typical GW experiment.

Moreover, observe that due to the presence of the derivative operators in the physical response, effectively only radiative terms in the $H_{\mu\nu}$ perturbation contribute (recall the discussion at the end of Sec.~\ref{eq:Perturbative Sources in General relavivity}). This in particular also implies that we can assume that each gauge invariant component in Eq.~\eqref{eq:PertRiemannFirst} is composed of a superposition of (possibly massive) plane wave solutions that satisfy the relation in Eq.~\eqref{eq:IdentityRadiationSecond}. Note that by assumption, we disregard any Lorentz symmetry violating cases or more general equations of state of the waves. Effectively, this allows us to perform a replacement $\partial_i\rightarrow -n_i v\partial_0$, where the direction of propagation $n_i$ is equal for each plane wave, but the velocity $v$ can in principle represent an entire sum of different velocities. Thus, for instance we can rewrite
\begin{equation}
    \partial_i\partial_j\delta\Phi=n_in_j v^2_{\mys{\Phi}} \delta\Phi\,,
\end{equation}
where $v_{\mys{\Phi}} $ represents the group velocity of the radiative part of the variable $\delta\Phi$, or possibly a sum of velocities of potential superpositions.
Moreover, it will also be useful to redefine the gauge invariant scalar variables and replace $\delta\Phi$ in favor of a new variable
\begin{equation}
    \delta\Upsilon\equiv \delta\Theta+v_{\mys{\Phi}}^2\delta\Phi\,.
\end{equation}
in order to match their values to the polarization basis that we will use below.

Considering these remarks we are now in a position to write the perturbed Riemann tensor as
\begin{equation}\label{eq:linRiemann00}
\phantom{}_{\mys{(1)}}R_{i0j0}=-\frac{1}{2}\,\ddot{P}_{ij}\,,
\end{equation}
where
\begin{equation}\label{eq:General Formula ofGWStrain}
\boxed{P_{ij}=H^{TT}_{ij}+2v_{\mys{\Xi}} n_{(i} \delta\Xi^{T}_{j)}+\left[\delta_{ij}-n_in_j\right] \delta\Theta +n_in_j\delta\Upsilon\,.}
\end{equation}
Plugging this result into the geodesic deviation equation  [Eq.~\eqref{eq:GeodesicDeviationFermiNormalCoordsThird}] governing the physical response to the presence of gravitational radiation, the equation can now easily be integrated to first order in the proper distance displacement to yield
\begin{equation}\label{eq:changeinsi}
\ell_i=\ell_i^0+\frac{1}{2}P_{ij}\,\ell^j_0 \,,
\end{equation}
where $\ell_i^0$ defines an initial proper distance separation. A change in proper distance in a particular direction $\ell_i=\ell \,e_i$, with $e_i$ the Cartesian basis vectors, therefore reads
\begin{equation}\label{eq:changeins}
\boxed{\frac{\Delta \ell}{\ell_0}=\frac{1}{2}P_{ij}\,e^ie^j\,,} 
\end{equation}
with $\Delta \ell\equiv \ell-\ell_0$.

Equation~\eqref{eq:changeins}, frequently refereed to as the fractional strain of spacetime, in summary captures the physical effects of gravitational radiation, given the asymptotic gauge invariant wave modes that one can plug into Eq.~\eqref{eq:General Formula ofGWStrain}. We want to stress, again, that this response is solely purely spacial due to our (very natural) choice of monitoring the geodesic deviation between two timelike geodesics. Thus, as already discussed, gravitational radiation affects the proper distance that is naturally captured via light travel time measurements at the basis of today's interferometric GW detectors (see Sec.\ref{ssSec: GW Experiments} below). Observe, in particular, that this implies that the total change in proper distance $\Delta\ell$ scales with the corresponding rest length $\ell_0$ as clearly visible in Eq.~\eqref{eq:changeins}.


Moreover, one should keep in mind that this result is fundamentally based on the validity of the geodesic deviation equation [Eq.~\eqref{eq:GeodesicDeviation2}], which is only valid up to corrections of the order $\ell/L$, where $L$ represents the typical scale of length variation of the curvature component, that in this case corresponds to the typical wavelength of the radiation \cite{Flanagan:2005yc}. Thus, the requirement for Eq.~\eqref{eq:changeins} to be viable
\begin{equation}
    \ell\ll L\,,
\end{equation}
is satisfied, as long as the typical size of the detector is smaller than the size of the wavelength. Observe that this condition automatically ensures that the second type of errors given by the expansion of the Fermi normal coordinates (recall the discussion around Eq.~\eqref{eq:FermiNormalCoordsSecond}) with $\ell<y$ is automatically satisfied
\begin{equation}
    \ell\ll D\sim L/\sqrt{h}\,,
\end{equation}
since the Riemann tensor scales as $D^{-2}\sim h/L^2$. In other words, the curvature scale of the waves $D$ is much larger than the characteristic length-scale of variation $L$. As soon as the wavelength of the waves becomes comparable to the detector size, as it will be the case for the high frequency spectrum within the LISA space mission for instance, a more sophisticated analysis in the TT gauge is required (see e.g. \cite{maggiore2008gravitational}).

We also want to comment that a measurement of gravitational radiation through a monitoring of changes in spacial proper distances through light-travel time measurements as described in Eq.~\eqref{eq:changeins} implies that such detectors are directly measuring information about the phase of the wave. This is only possible because, as already mentioned, typical GWs are generated by the bulk motion of a system and are therefore emitted phase-coherently. This is in contrast to standard observations of electromagnetic waves, since photons usually originate from independent events of local charges within a larger source. As a consequence, the wavelength of GWs is typically larger or comparable to the size of the source and can therefore not be used for imaging and is closer in analogy to hearing sound. However, measuring the strain instead of an overall energy flux as in the electromagnetic case comes with the advantage that the radiative strain merely falls off as $1/r$ with the distance of the source while a flux of energy decays with $1/r^2$, thus compensating for the extreme weakness of the amplitude.\footnote{This fact, together with the weakness of interaction of gravitational waves with surrounding matter makes GW signals the optimal source for very early cosmic information and could allow us to possibly even look past the current horizon of direct information given by the cosmic microwave background (see Sec.~\ref{sSec:CMB}).} Moreover, improving the sensitivity of an instrument measuring the strain by a certain factor increases the number of potential sources by the volume and hence the factor cubed.

\subsection{Gravitational Polarizations}\label{sSec:GWPolGen}

With the spacial response relying on timelike geodesic deviation in Eq.~\eqref{eq:changeinsi} together with the response matrix in Eq.~\eqref{eq:General Formula ofGWStrain} at hand, we can now talk about gravitational \textit{\gls{polarization}s} of generic metric theories of gravity \cite{Eardley:1973zuo,Eardley:1973zzz} (see also \cite{poisson2014gravity,Will:2018bme}). But first, we will quickly review the case of GR.

Recall that for GR, only the spacial components $H^{TT}_{ij}$ of the gauge invariant metric perturbations propagate, hence $P_{ij}=H^{TT}_{ij}$. This TT tensor can naturally be expanded in a polarization basis
\begin{equation}\label{GWPolexpansion GR}
H^{TT}_{ij}=\sum_{\lambda} H_\lambda e^\lambda_{ij}\,,
\end{equation}
with
\begin{equation}\label{eq:Polarization Vectors Basis}
    e^{\lambda}_{ij}e_{\tilde{\lambda}}^{ij}=2\delta_{\lambda\tilde{\lambda}}\,.
\end{equation}

The most frequently used $+$/$\times$ polarization basis vectors read\footnote{Another option would be for example circularly polarized left- and right-handed modes \cite{carroll2019spacetime}.}
\begin{equation}
e^+_{ij}\equiv u_iu_j-v_iv_j\;,\qquad e^\times_{ij}\equiv u_iv_j+v_iu_j\,.
\end{equation}
Thus, any symmetric, transverse-traceless tensor can be expanded in such a tensorial basis constructed out of $\mathbf{u}$ and $\mathbf{v}$ as
\begin{equation}\label{eq:pluscrossex}
H^{TT}_{ij}=H_+\, e^+_{ij}+H_\times\, e^\times_{ij}\,.
\end{equation}

On the other hand, for a generic metric theory of gravity, the gauge invariant variables also involve temporal components of the metric, such that strictly speaking, gravitational polarizations of the metric are more fundamentally defined as an expansion of the spacial response metric $P_{ij}$ into an appropriate basis space. Indeed, observe that, similar to the TT fields, a transverse vector $V^T_i$ can always be decomposed in a polarization basis in the $\mathbf{u}$, $\mathbf{v}$ space as
\begin{equation}
    V^\text{T}_i=V_u e^u_i+V_v e^v_i\,,
\end{equation}
where simply
\begin{equation}
    e^u_i\equiv u_i\,,\quad e^v_i\equiv v_i\,,\quad e^\lambda_ie_{\tilde\lambda}^i=\delta_{\lambda\tilde\lambda}\,.
\end{equation}
Using this result, together with Eq.~\eqref{eq:pluscrossex} one can naturally define the following six polarization modes of the GW response matrix
\begin{subequations}\label{eq:GWgenPoldef}
\begin{align}
P_+&\equiv\frac{1}{2}e_+^{ij}\,H^{TT}_{ij}\,,& P_\times&\equiv\frac{1}{2}e_\times^{ij}\,H^{TT}_{ij}\,,& P_b&\equiv \delta\Theta\,,\\
P_{u}&\equiv e_u^i \,v_\Xi\,\delta\Xi^T_{i}\,,& P_{v}&\equiv e_v^i \,v_\Xi\,\delta\Xi^T_{i}\,,& P_l&\equiv \,\delta\Upsilon\,,
\end{align}
\end{subequations}
such that
\begin{align}\label{eq:Pijexpand}
\boxed{P_{ij}=e^+_{ij}\,P_++e^\times_{ij}\,P_\times+e^u_{ij}\,P_u+e^v_{ij}\,P_v+e^b_{ij}\,P_b+e^l_{ij}\,P_l\,,}
\end{align}
where
\begin{subequations}\label{eq:PolTensors}
\begin{align}
e^+_{ij}&=u_iu_j-v_iv_j\,,& e^\times_{ij}&=u_iv_j+v_iu_j\,,& e^b_{ij}&\equiv u_iu_j+v_iv_j\,,\\
e^u_{ij}&\equiv n_iu_j+u_in_j\,,& e^v_{ij}&\equiv n_iv_j+v_in_j\,,& e^l_{ij} &\equiv n_in_j\,.
\end{align}
\end{subequations}
These tensors define a complete polarization basis of a spacial symmetric tensor $P_{ij}$ and therefore also satisfy orthogonality relations of the form Eq.~\eqref{eq:Polarization Vectors Basis}. Moreover, each polarization mode can be extracted as
\begin{equation}
    P_\lambda=\frac{1}{2}P_{ij}e^{ij}_\lambda\,,
\end{equation}
except for the longitudinal polarization $\lambda=l$, for which we simply have $P_l=P_{ij}e^{ij}_l$.

Each of the polarization modes defined in Eq.~\eqref{eq:GWgenPoldef} is associated to one (possibly propagating) degree of freedom in the \textit{physical metric}. We emphasize here the physical metric, because as already mentioned several times, in a metric theory of gravity, only the perturbations of the physical metric can directly influence matter, and therefore the gravitational polarizations exclusively refer to the polarizations of the physical metric. Indeed, one should make a clear distinction between the notion of propagating DOFs of a metric theory and the gravitational polarizations discussed here. While there is no limit in the number of propagating degrees of freedom in a given metric theory, there are only up to six distinct gravitational polarizations. Moreover, not all the propagating DOFs in a theory are necessarily associated to a corresponding gravitational polarization mode. Quite the opposite, in many concrete examples, there exist more radiative DOFs than gravitational polarizations, while in other cases a greater number of polarizations are excited than there exist propagating DOFs in the theory. In this context, a faithful representation of a metric theory is very useful, as such a description allows considering the propagating DOFs as a truly distinct concept, which may or may not excite gravitational polarizations of the physical metric, depending on the non-minimal couplings of the field with the physical metric. These statements will be exemplified in a concrete set metric theory beyond GR in Sec.~\ref{sSec:GWPolExample} below.

\begin{figure}
\centering
\includegraphics[scale=0.38]{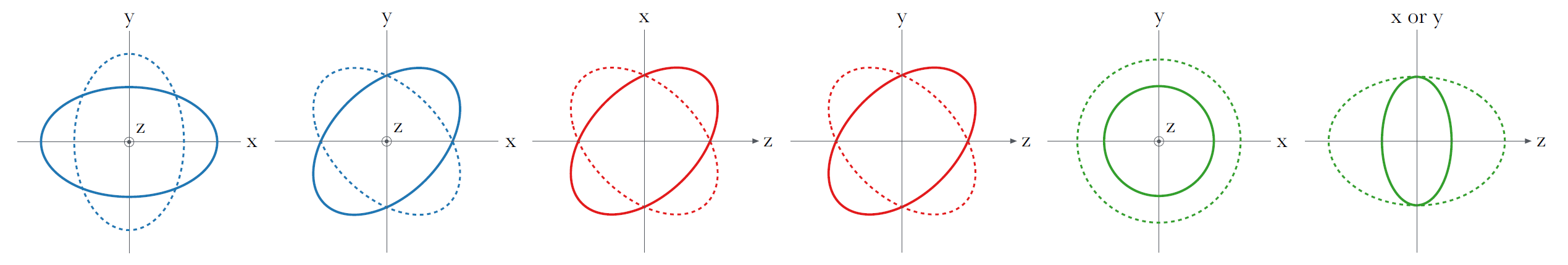}
\caption{\label{fig:GWPol}\small{The effect of the gravitational polarizations on the relative proper distance in the timelike geodesic deviation equation for gravitational radiation traveling in the $z$-direction. The modes are expanded in the polarization basis given in Eqs.~\eqref{eq:Pijexpand} and \eqref{eq:PolTensors}, with $\mathbf{n}=\mathbf{e}_z$, hence $\theta=0$, while the transverse axis are chosen such that $\mathbf{u}=\mathbf{e}_x$ and $\mathbf{v}=\mathbf{e}_y$, corresponding to the choice $\phi=0$. From left to right: plus and cross transverse-traceless tensor polarizations $P_+$, $P_\times$ (blue), two mixed vector polarizations $P_u$, $P_v$ (red) and two scalar polarizations (green) where the breathing mode $P_b$ is transverse and $P_l$ is purely longitudinal. (figure taken from \textit{T. Callister et al., (2017)} \cite{Callister:2017ocg})}}
\end{figure}

\paragraph{Pictorial Representation of Gravitational Polarizations.} 
It is instructive to picture the effects of the different polarization modes by plotting the relative variation of proper distance over time with respect to a given reference frame. A standard choice is to consider gravitational radiation travelling in the $z$-direction, hence $\mathbf{n}=\mathbf{e}_z$ corresponding to $\theta=0$, as well as choosing $\phi=0$ such that the transverse basis becomes $\mathbf{u}=\mathbf{e}_x$ and $\mathbf{v}=\mathbf{e}_y$. In this special frame, the spacial response matrix in terms of gravitational polarizations in Eq.~\eqref{eq:Pijexpand} reads
\begin{equation}\label{eq:ResponsezP}
P_{ij}=
\begin{pmatrix}
P_b+P_+ & P_\times & P_u\\
P_\times & P_b-P_+ & P_v\\
P_u & P_v & P_l
\end{pmatrix}\,.
\end{equation}
The result is plotted in Fig. \ref{fig:GWPol}.

At this point, a word of caution in the interpretation of Fig.~\ref{fig:GWPol} is in order. In the special frame of Fermi normal coordinates used in the derivation of the timelike geodesic deviation above, centered at the origin of the spacial basis vectors in Fig.~\ref{fig:GWPol}, it is possible to interpret the pictures as the deformation of a ring of test particles. Such deformations represent the movement of test particles with respect to the locally flat background in Fermi normal coordinates. However, what is physically represented is more precisely the change in proper spacial distance, that in Fermi normal coordinates directly correspond to the coordinate distances. Indeed, to talk about the movement of a test particle with respect to a fixed rigid ruler that itself does not stretch under the effect of GWs in practice only makes sense very locally. Thus, to avoid confusion, it is usually much safer to think about the deformations in Fig.~\ref{fig:GWPol} as changes in proper spacial distances, that can naturally be measured by light-travel time experiments as already discussed. In other words, the leading order effect is a stretch of the space in between two test masses, rather than some particular ``movement'' of test masses. This also directly explains why the effect scales with the initial distance between two test masses.

\paragraph{Spin-Weighted Functions.} 

There exist an additional intrinsic freedom in the description of gravitational polarizations that is concealed in the common representation in Eq.~\eqref{eq:ResponsezP}. Namely, the transverse $\mathbf{u}$ and $\mathbf{v}$ basis introduced in Eq.~\eqref{eq:Def Transverse Vectors u v} is not unique and is only defined up to a rotation $\psi$ along the longitudinal direction. Such a rotation according to the defining vector representation is given by
\begin{equation}\label{eq:rotaroundn}
u'_i=\cos\psi \, u_i-\sin\psi \, v_i\;,\quad v'_i=\cos\psi \, v_i+\sin\psi \, u_i\;,\quad n'_i=n_i\,.
\end{equation}
Observe that in the particular case of choosing the direction of propagation to coincide with the $z$-axis, such rotations are degenerate with rotations of $\phi$ of the source centered coordinate system. However, in general, the rotations about the axis of propagation represent a distinct local operation.

Indeed, the completeness relation in Eq.~\eqref{eq:polcompleteness}, and as a consequence also each term in the expansion in Eq.~\eqref{eq:Pijexpand}, are invariant under the rotations of the transverse basis in Eq.~\eqref{eq:rotaroundn}, such that these transformations represent a residual freedom of description of the SVT decomposition. However, this implies that the individual polarization modes defined in Eq.~\eqref{eq:GWgenPoldef} themselves do transform under such rotations, namely
\begin{subequations}\label{eq:TransformationsSpinWeightMetricPerturbations}
\begin{align}
P'_+&=\cos 2\psi \, P_+-\sin 2\psi \, P_\times\;,& P'_\times&=\cos 2\psi \, P_\times +\sin 2\psi \, P_+\,,\\
P'_u&=\cos\psi \, P_u-\sin\psi \, P_v\;,& P'_v&=\cos\psi \, P_v+\sin\psi \, P_u\,,\\
P'_b&=P_b\;,& P'_l&=P_l\,.
\end{align}
\end{subequations}
These transformations reveal the tensorial nature of each gravitational polarization modes. Namely, the $+$ and $\times$ modes are invariant under rotations of $\psi=180^{\circ}$ and are thus associated with a $SO(2)$ tensor irreducible representation labeled by $m=\pm 2$ (recall the discussion in Sec.~\ref{ssSec:GaugeInvariantDecomposition}\footnote{As already mentioned, from a field theoretic perspective the label $m$ corresponds to the helicity, which in the case of massless fields is a Lorentz invariant notion labeling irreps of the little group. Note however the subtleties discussed in \cite{Eardley:1973zuo}.}). On the other hand, the vector modes, corresponding to $m=\pm 1$, are invariant under rotations of $\psi=360^{\circ}$ around the axis of propagation as familiar from electromagnetic waves, while the scalar modes are invariant. These statements can also nicely be seen optically in Fig.~\ref{fig:GWPol}.

More generally, such an internal $U(1)$ freedom is inherent to any function $f(\theta,\phi)$ on the sphere. Under a rotation about $\mathbf{n}$, a function $f(\theta,\phi)$ can change its value through a phase, even though its argument (the point on the sphere) stays the same. In this context, it is useful to define functions $f_s$ with a definite value of so called \textit{spin-weight} $s$, such that the function transforms under such $U(1)$ rotations as
\begin{equation}\label{eq:SpinWeightTransfos}
    f_s(\theta,\phi)\rightarrow f'_s(\theta,\phi)=f_s(\theta,\phi)\,e^{is\psi}.
\end{equation}

One is therefore lead to define an alternative complex basis of the transverse space by defining the vector\footnote{Observe that this vector precisely corresponds to the components of the Newman-Penrose tetrad $\underline m=\frac{1}{\sqrt{2}}\left(\underline{\partial}_\theta+i\sin\theta\underline{\partial}_\phi\right)$ that parameterizes the 2-sphere metric through $2m_{(i}\bar{m}_{j)}$, invariant under the $U(1)$ transformation in Eq.~\eqref{eq:SpinweightTransformation m}, while the area element is given by $2m_{[i}\bar{m}_{j]}$ \cite{DAmbrosio:2022clk}.}
\begin{equation}\label{eq:Def m}
    m_i\equiv \frac{1}{\sqrt{2}}(u_i+iv_i)\,,
\end{equation}
alongside its complex conjugate
\begin{equation}\label{eq:Def mbar}
    \bar m_i= \frac{1}{\sqrt{2}}(u_i-iv_i)\,,
\end{equation}
where $m_im^i=0$ and whose components are of definite spin-weight $s=1$ and $s=-1$, respectively, as determined by their behavior under rotations about the longitudinal direction given in Eq.~\eqref{eq:rotaroundn}
\begin{equation}\label{eq:SpinweightTransformation m}
    m'_i\rightarrow e^{i\psi}m_i\,.
\end{equation}
One can then construct combinations of polarization modes of any given radially outward tensor radiation with a given spin-weight by simply contracting the tensors on the sphere with the appropriate combination of the basis vector $m_i$.

In particular, as concerns the response matrix $P_{ij}$, one can for instance isolate the dominant $+$ and $\times$ polarizations of GR and define a complex function $P_{\mys{-2}}$ of definite spinweight $s=-2$ through
\begin{equation}\label{eq:ScalarPt}
     P_{\mys{-2}}\equiv P^{ij}\bar{m}_i\bar{m}_j=\frac{1}{2}P^{ij}(e^+_{ij}-i\, e^\times_{ij})=P_+-iP_\times\,.
\end{equation}
Similar scalars of definite spin-weight can also be defined for the scalar and vector polarizations. More generally, given any spacial vector $V_i$ one can construct a function $V_{\mys{-1}}$ of spinweight $s=-1$ through
\begin{equation}\label{Scalara}
     V_{\mys{-1}}\equiv V^{i}\sqrt{2}\bar{m}_i=V_{u}-iV_{v}\,.
\end{equation}
Note here that $\bar m_i$ automatically selects the transverse part of the vector, while the combination $\bar{m}_i\bar{m}_j$ projects onto the $TT$ space. Similarly, the new basis vectors $m_i$ can also be used to describe the transverse projector defined in Eq.~\eqref{eq:TransverseProjector}
\begin{equation}
\perp_{ij}\equiv \delta_{ij}-n_in_j=u_iu_j+v_iv_j=m_i\bar m_j+\bar m_im_j\,.
\end{equation}

The definition of spin-weighted functions is important, as they allow for a consistent decomposition into spherical harmonics on the sphere, which in this case have to be generalized to so called \textit{spin-weighted spherical harmonics} (SWSH) introduced in Appendix~\ref{App:TTM Expansion}. Indeed, for instance the angular dependence of the spin-weight $s=-2$ scalar function $P_{\mys{-2}}$ defined in Eq.~\eqref{eq:ScalarPt} can naturally be expanded in terms of SWSH as
\begin{equation}
    P_{\mys{-2}}(\theta,\phi)=\sum_{l=2}^{\infty}\sum_{m=-l}^{l}\,P_{lm}\,_{\mys{-2}}Y_{lm}(\theta,\phi)\,.
\end{equation}
Note that the transformation of the SWSHs under rotations given in Eq.~\eqref{rotSWSH} ensures that the modes $P_{lm}$ transform in the usual way under a rotation $R$ of a given coordinate system, namely
\begin{equation}
    P_{lm}\rightarrow P'_{lm}=\sum_{m'=-l}^{l}P_{lm'}\,\mathfrak{D}^l_{m'm}(R^{-1})\,,
\end{equation}
with $\mathfrak{D}$ denoting the Wigner-D matrices also introduced in Appendix~\ref{App:TTM Expansion}. This is because the spin-weight ambiguity is taken care off by the SWSH and is one of the main reasons why a decomposition into SWSH is preferred over a decomposition in terms of standard spherical harmonics. In particular, for rotations around the $z$ axis parameterized by the angle $\phi$, the Wigner-D matrices have a particularly simple form
\begin{equation}
    \mathfrak{D}^l_{m'm}(0,0,\phi)=\delta_{m'm}e^{im\phi}\,,
\end{equation}
such that in this case the $s=-2$ modes simply transform as
\begin{equation}\label{phRotationsofModes}
    P_{lm}\rightarrow P'_{lm}=P_{lm}e^{-im\phi}\,.
\end{equation}

\paragraph{The Synchronous Gauge.} Finally, we want to mention that the above discussions can also be held in a particular gauge in which the perturbations of the physical metric $H_{\mu\nu}$ are chosen to be purely spacial $H_{00}=H_{0i}=0$, and thus
\begin{equation}\label{eq:spacialh}
H_{\mu\nu}=
    \left(\begin{array}{c|c c c} 
    	0 &  & 0&\phantom{0} \\
    	\hline 
    	 &  &&\\
0 &  & H_{ij}& \\
 \phantom{0} &  & & \\
    \end{array}\right)\,.
\end{equation}
It is in fact always possible to find such a gauge choice and the associated coordinate system precisely corresponds to the synchronous (or Gaussian normal) frame defined back in Eq.~\eqref{eq:SynchronousGauge}  (see also \cite{WaldBook,landau_classical_2003,carroll2019spacetime}). 

Equations~\eqref{eq:linRiemann} and \eqref{eq:linRiemann00} imply that in this particular chart the response matrix simply corresponds to the metric perturbations
\begin{equation}
    P_{ij}=H_{ij}\,.
\end{equation}
Thus, Eq.~\eqref{eq:Pijexpand} suggests that in this gauge we can expand the gravitational radiation into six polarizations modes
\begin{equation}\label{GWPolexpansion}
H_{ij}=\sum_{\lambda} H_\lambda e^\lambda_{ij}\,.
\end{equation}
where the polarization basis is given by Eq.~\eqref{eq:PolTensors} and where
\begin{equation}
    H_{\lambda}=P_{\lambda}\,.
\end{equation}
Often, gravitational polarizations are discussed in this particular gauge. It is however important to remember that choosing such a gauge is ultimately justified by the knowledge that only the fully gauge-invariant response in the geodesic deviation is restricted to the spacial space.

It is interesting to note that these special coordinates also allow for a simple re-derivation of the gravitational radiation response in terms of proper distance displacements in Eq.~\eqref{eq:changeins}, a derivation we will now offer explicitly.

First, we will show that in such a local chart where $H_{00}=H_{0i}=0$ the proper time $\tau$ of a test mass initially at rest is the same as the coordinate time $t$ up to irrelevant corrections. To this end, consider the geodesic equation of a test mass evaluated at $\tau=0$ for which $dx^i/d\tau=0$
\begin{equation}
\frac{d^2x^i}{d\tau^2}=-\Gamma^i_{00}\left(\frac{dx^0}{d\tau}\right)^2=0\,.
\end{equation}
This equation vanishes, because
\begin{equation}
\Gamma^i_{00}=\frac{1}{2}\left(2\partial_0H_{0i}-\partial_iH_{00}\right)=0\,,
\end{equation}
vanishes in the synchronous gauge. Thus, for a test mass initially at rest, $dx^i/d\tau$ remains zero at all times. This now directly implies that along such a time-like trajectory $x^\mu(\tau)$
\begin{equation}\label{eq:TimeCoordinateSynchronous gauge}
d\tau^2=dt^2(\tau)-\left(\delta_{ij}+h_{ij}\right)\frac{dx^i}{d\tau}\frac{dx^j}{d\tau}d\tau^2=dt^2\,,
\end{equation}
which is precisely what we wanted to show.

Note that in this respect, this chart therefore corresponds to the TT-gauge of GR in which test charges remain at rest even in the presence of GWs such that the coordinate distance between test masses remains constant too (see e.g. \cite{maggiore2008gravitational}). 
This is no surprise, as the synchronous coordinate system was precisely constructed by using a set (or congruence) of timelike geodesics as definitions of the coordinates. Hence, timelike geodesics (initially at rest in the frame) correspond to comoving observers whose coordinate values do not change over time. Moreover, the synchronous chart also explicitly uses the proper time of the comoving timelike geodesics as a time coordinate, which also explains the result in Eq.~\eqref{eq:TimeCoordinateSynchronous gauge}.

With this result at hand, one can now easily determine the physical effect of gravitational waves on the proper distance between for instance two simultaneous events $(\tau,0)$ and $(\tau,z^0_i)$, where $z_i=z e_i$ for some spacial basis vectors $e_i$.
This is because, as already discussed in Sec.~\ref{sSec:Special Relativity}, the spacial proper distance between the two events in synchronous coordinates is simply given by [Eq.~\eqref{eq:SpacialDistanceDefSimp}]
\begin{equation}
    \ell = \sqrt{g_{ij} z^iz^j}=z_0\sqrt{\left(\delta_{ij}+H_{ij}\right)e^ie^j}\simeq \ell_0\left(1+\frac{1}{2}H_{ij}e^ie^j\right)\,.
\end{equation}
This follows, because the initial coordinate separation in the absence of gravitational waves on the asymptotically flat spacetime corresponds to the initial proper distance $\ell_0=\delta_{ij}z^iz^j=z_0$. Therefore,
\begin{equation}
\frac{\Delta \ell}{\ell_0}=\frac{1}{2}H_{ij}e^ie^j\,,
\end{equation}
which, remembering that in synchronous coordinates $H_{ij}=P_{ij}$, precisely corresponds to \eqref{eq:changeins}.

While not being particularly enlightening as compared to the full treatment in terms of manifestly gauge invariant perturbations given above, the reformulations of the results in this particular gauge represents a nice consistency check. This discussion is in fact reminiscent of the distinction between an early, gauge dependent approach to the treatment of cosmological perturbations \cite{Lifshitz:1963ps} as opposed to the manifestly gauge invariant formulation first provided by Bardeen \cite{Bardeen:1980kt}, as already mentioned back in Sec.~\ref{ssSec:GaugeInvariantDecomposition}. In GR such a distinction with the associated subtleties usually does not come up as the transverse-traceless modes, the only dynamical gauge invariant quantities, are already purely spacial.


\subsection{Gravitational Wave Experiments}\label{ssSec: GW Experiments}

We will now apply the results of the previous subsection, in which we analyzed physical effects of gravitational radiation in generic metric theories of gravity, to the specific case of laser interferometers in the low wavelength regime. The section is then concluded by offering a brief overview over the ongoing and planned tests of additional gravitational polarizations

\paragraph{Quadrupole Detectors.} As we already stressed many times, the physical effects of gravitational radiation can fundamentally be detected by monitoring light-travel time changes in the timelike geodesic deviation equation as governed by the formula in Eq.~\eqref{eq:changeins}. 
A particularly smart way of measuring light travel time is to use the idea behind the laser interferometer employed by Michelson to measure the speed of light. The basic setup (see e.g. \cite{maggiore2008gravitational}) consists of two perpendicular detector arms\footnote{The specific angle between the arms is not an essential feature as we will see below, as long as the angle is not too small of course.} of equal rest-length $\ell_0$ with mirrors at each end. Detectors measuring differential arm motion are generally referred to as \textit{quadrupole detectors}. By the use of a beam-splitter, a coherent laser beam of frequency $\omega_\text{lb}$ is sent along each arm and after traveling once back and forth, the two beams meet again at the beam-splitter where the phase difference $\Delta \varphi_\text{lb}$ can be measured. In the absence of GWs, $\Delta \varphi_\text{lb}$ vanishes but as soon as the relative proper length of the two arms varies due to GWs as predicted by the strain equation Eq.~\eqref{eq:changeins} the different light travel times result in a measurable phase difference. In the limit where the arm-length of the quadrupole detector is small compared to the wavelength $L$ of the gravitational wave $\ell_0/L\ll 1$ or equivalently large compared to the frequency $f \ell_0\ll 1$ and by choosing a coordinate system in which the two detector arms determine the $x$ and $y$ axis we have \cite{maggiore2008gravitational}
\begin{equation}\label{phasedifference}
\Delta\varphi\simeq \omega_\text{lb}(s^x-s^y) =\omega_\text{lb} \ell_0\frac{1}{2}\left(e^x_ie^x_j-e^y_ie^y_j\right)P^{ij}\equiv\omega_\text{lb} \ell_0 P(t) \,.
\end{equation}
where according to Eq.~\eqref{eq:changeinsi} we have written
\begin{equation}
\ell^x=\ell_0\left(1+\frac{1}{2}P^{ij}e^x_ie^x_j\right)\;,\quad \ell^y=\ell_0\left(1+\frac{1}{2}P^{ij}e^y_ie^y_j\right)\,.
\end{equation}
Equation \eqref{phasedifference} defines the \textit{detector response function}
\begin{equation}
P(t)=\frac{1}{2}\left(e^x_ie^x_j-e^y_ie^y_j\right)P^{ij}\,.
\end{equation}
Note that the limit $f \ell^0\ll 1$ appropriate for ground based detectors is the same that is appropriate for the derivation of the geodesic deviation equation and therefore also of Eq.~\eqref{eq:changeins} in the first place (recall the discussion in Sec.~\ref{ssSec:The Physical Effects of Gravitational Waves}).

\paragraph{Detector Pattern Functions.} The response to each individual mode can now conveniently be determined via the expansion in gravitational polarization modes [Eq.~\eqref{eq:Pijexpand}]
\begin{equation}\label{eq:Signal in Detector}
\boxed{P(t)= F_+\,h_++F_\times\,h_\times+F_u\,h_u+F_v\,h_v+F_b\,h_b+F_l\,h_l\,.}
\end{equation}
Here, the quantities $F_\lambda(\Omega)$ called \textit{detector pattern functions} are defined as
\begin{equation}\label{eq:defDetectorPattern}
F_\lambda\equiv \frac{1}{2}\left(e^x_ie^x_j-e^y_ie^y_j\right)e_\lambda^{ij}\,,
\end{equation}
for each polarization tensor in Eq.~\eqref{eq:PolTensors}. The prefactor in this equation is usually termed \textit{detector tensor} and reflects the detector geometry. For example, if the detector arms would make an angle $\chi$ instead of being perpendicular, the detector tensor would simply be multiplied by a factor of $\sin\chi$ \cite{poisson2014gravity}. The detector pattern functions on the other hand include as well the information of the detector response to each polarization and therefore the directional dependence of the sensitivity of the detector to each of the modes. 

For each detector geometry they can be calculated once and for all by relating the reference frame of the detector to the coordinate system of the wave characterized by the direction of the source $\mathbf{N}=-\mathbf{n}$. Recall that back in Sec.~\ref{sSec:GWGeneration} we already did precisely that, but in this case by relating the natural Cartesian coordinate system of the source to the direction of propagation of the waves $\mathbf{n}$ defined in Eq.~\eqref{eq:Def Direction of Propagation n}, together with an associated tangent space $\{\mathbf{u},\mathbf{v}\}$ in Eq.~\eqref{eq:Def Transverse Vectors u v} at each point on the sphere. Thus, the result for the detector centered coordinate system with spherical coordinates $\{\theta,\phi\}$ will be exactly the same
\begin{subequations}\label{eq:vectorbasisp}
\begin{align}
&\mathbf{N}=(\sin\theta \cos\phi,\,\sin\theta \sin\phi,\,\cos\theta )\,,\\
&\mathbf{U}=(\cos\theta \cos\phi ,\,\cos\theta \sin\phi,\,-\sin\theta )\,,\\
&\mathbf{V}=(-\sin\phi, \cos\phi ,0)\,.
\end{align}
\end{subequations}

Recall, however, that there exists an additional freedom in rotating the coordinate system of the radiation around the direction of propagation given by the transformations in Eq.~\eqref{eq:rotaroundn}. It is useful to once explicitly account for this freedom, since different conventions are chosen in the literature. Thus, the basis in Eq.~\eqref{eq:vectorbasisp} can be generalized to a ``prime'' basis with the freedom of an additional angle $\psi$
\begin{subequations}\label{eq:vectorbasispGen}
\begin{align}
&\mathbf{N}'=(\sin\theta \cos\phi,\,\sin\theta \sin\phi,\,\cos\theta )=\mathbf{N}\\
&\mathbf{U}'=(\cos\theta \cos\phi \cos\psi+\sin\phi\sin\psi,\,\cos\theta \sin\phi \cos\psi-\cos\phi \sin\psi,\,-\sin\theta\cos\psi)\\
&\mathbf{V}'=(\cos\theta \cos\phi \sin\psi-\sin\phi\cos\psi,\,\cos\theta\sin\phi\sin\psi+ \cos\phi \cos\psi,\,-\sin\theta\sin\psi)\,.
\end{align}
\end{subequations}
Plugging this basis into Eq.~\eqref{eq:defDetectorPattern} by using Eq.~\eqref{eq:PolTensors} in terms of the general vectors in Eq.~\eqref{eq:vectorbasispGen} above, results in
\begin{subequations}\label{eq:DetectorPattern}
\begin{align}
F_+&=\frac{1}{2}\left(1+\cos^2\theta\right)\cos2\phi\cos2\psi+\cos\theta\sin2\phi\sin2\psi\\
F_\times&=\frac{1}{2}\left(1+\cos^2\theta\right)\cos2\phi\sin2\psi-\cos\theta\sin2\phi\cos2\psi\\
F_u&=\frac{1}{2}\sin2\theta\cos2\phi\cos\psi+\sin\theta\sin2\phi\sin\psi\\
F_v&=\frac{1}{2}\sin2\theta\cos2\phi\sin\psi-\sin\theta\sin2\phi\cos\psi\\
F_{b}&=-\frac{1}{2}\sin^2\theta\cos2\phi \label{DPb}\\
F_l&=\frac{1}{2}\sin^2\theta\cos2\phi\,.\label{DPl}
\end{align}
\end{subequations}
These are completely general results for the detection pattern functions of any signal arriving from a direction $\mathbf{N}$ given by the two angles $\theta$ and $\phi$ and the third angle $\psi$ representing a freedom in the description of the polarization basis. We want to stress again, however, that these results were obtained in the low frequency limit $f\ell^0\ll 1$ and a generic analysis would in particular require a more general version of Eq.~\eqref{phasedifference} (see \cite{Rakhmanov:2008is} for general GR results). Furthermore, for any realistic detector, the antenna pattern functions are actually functions of time given by the peculiar motion of the detector. For short transient signals such as CBCs, this rotation is however negligible. At the least for LISA, these two points will however become relevant. 

\begin{figure}
\centering
\includegraphics[scale=0.38]{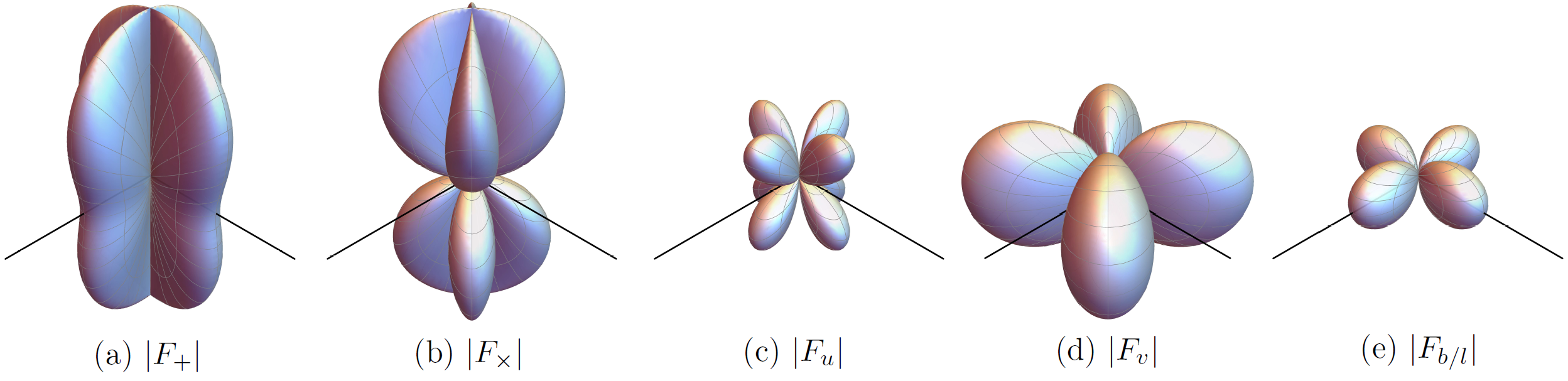}
\caption{\label{fig:DetectorP}\small{Absolute value of the detector pattern functions $\abs{F_\lambda}$ for each GW polarization in the low frequency limit as a function of the source direction $\mathbf{N}$ parametrized by the angles $\theta$, $\phi$ and where $\psi=0$. The radial distance represents the angular response to a unit-amplitude GW of an ideal interferometer with detector arms along the $x$ and $y$-axis, as indicated by the black lines. The response to the breathing and longitudinal scalar modes is identical. (Figure taken from \textit{M. Isi et al., (2017)} \cite{Isi:2017equ})}}
\end{figure}

Since the angle $\psi$ is a mere freedom of description, we choose to again simply set it zero $\psi=0$ for concreteness. The absolute value of the angular response of the detector to each polarization is then shown in Fig. \ref{fig:DetectorP}. For each mode, the interferometer has blind spots which represent directions for which the GW produces equal changes in proper distance to both arms. Moreover, the detector responds not equally strongly to all the polarizations. For instance, in average, the response to the plus and cross modes is significantly greater than compared to the scalar signals.
Observe that the detector response functions for the two scalars, hence the breathing [Eq.~\eqref{DPb}] and longitudinal [Eq.~\eqref{DPl}] modes are equal up to a sign $F_b=-F_l$ such that only the combination $P_b-P_l$ of the two polarizations can be measured. Other than that, and apart from certain blind spots, all polarizations modes can in principle be detected. However, in order to fully resolve the five distinct polarizations characterized by the five amplitudes $P_+$, $P_\times$, $P_u$, $P_v$ and $P_b-P_l$ in principle a total of five independent detectors are needed, provided that the direction of the source, hence the two angles $\theta$ and $\phi$ are determined independently either through an electromagnetic counterpart or by using the time delay of the signal between different detectors.

\paragraph{Searching for Additional Polarizations.} Similar to the radiation speed constraints considered in Sec.~\ref{ssSec:PropagationSpeedConstraints} above, the model agnostic search for additional gravitational polarizations as a smoking gun signal for beyond GR effects have a great potential of posing tight bounds on alternative theories \cite{Eardley:1973zuo,Eardley:1973zzz,Will:2014kxa}. However, due to the still limited operational ground-based detector network based on CBC observations, together with the fact that, choosing redundancy over diverse sensitivity, the two LIGO interferometers were constructed nearly coaligned, in order to produce almost identical signals, the present constraints on additional polarization modes are not as significant yet \cite{LIGOScientific:2017ycc,Isi:2017fbj,LIGOScientific:2018dkp,LIGOScientific:2021sio}.

On the other hand, long-duration, or persistent, signals will be able to probe more than a single point in the response patterns such that information about GW polarizations can be extracted without the need of additional detectors or even independent knowledge of the source location. This is in particular interesting for future detections of GWs emitted by rotating neutron stars \cite{Isi:2015cva,Isi:2017equ,LIGOScientific:2017ous} as well as similar persistent signals observed with the LISA mission \cite{Tinto:2010hz}. Furthermore, also the measurements of the stochastic GW background might be used to constrain additional polarization modes \cite{Nishizawa:2009bf,Nishizawa:2009jh,Nishizawa:2013eqa,LIGOScientific:2018czr,Callister:2017ocg}, in particular also in PTA experiments, that due to the increasing number of individual pulsars distributed in different directions are actually very well suited to measure the distinct polarization content \cite{Lee_2008,daSilvaAlves:2011fp,Chamberlin:2011ev}.

\section{SVT Example: DOFs and Polarizations}\label{sSec:GWPolExample}

It is time to finally discuss an explicit example of a metric theory of gravity beyond GR and analyze its content in propagating degrees of freedom and gravitational polarizations. More precisely, we will take a closer look at the scalar-vector Heisenberg-Horndeski (SVHH) theory that we introduced in Sec.~\ref{ssSec: A Exact Theories}. As discussed, this theory is a generalization of the well-known scalar Horndeski theory\footnote{See also \cite{Hou:2017bqj} for a discussion of the polarizations in scalar Horndeski theory.} and represents the most general action of two metric tensor modes, together with a single scalar and a pair of massless vector DOFs, described  with second order equations of motion, thus avoiding any Ostrogradsky instabilities (recall Sec.~\ref{sSec:OstrogradskyTheorem}). This theory therefore represents a rather large class of metric theories, encompassing many specific theories studied in different contexts. In its faithful representation (recall Def.~\ref{DefFaithfulRep}), the theory is described by the action in Eq.~\eqref{eq:ActionSVH}, with a physical metric $g_{\mu\nu}$, a scalar field $\Phi$ and a massless vector field $A_\mu$. Thus, the additional non-minimal fields involve $\Psi=\{\Phi,A_\mu\}$, while for simplicity, we will neglect here any explicit presence of matter fields $\Psi_\text{m}$. In particular, in order to describe two massless DOFs assuming local Lorentz invariance, the vector field introduces an additional gauge redundancy [Eq.~\eqref{eq:GaugeTransformationVector2s}]
\begin{equation}\label{eq:gaugeTrasf A}
    A_{\mu}\rightarrow A_{\mu}+\partial_\mu\Lambda\,,
\end{equation}
that is promoted to a symmetry of the action by only considering the gauge invariant combination of the field strength
\begin{equation}
    F_{\mu\nu}=\nabla_\mu A_\nu -\nabla_\nu A_\mu= \partial_\mu A_\nu -\partial_\nu A_\mu\,.
\end{equation}

In order to study the gravitational polarizations in this theory, we will assume that an isolated system produces radiation, whose physical DOFs asymptote to future null infinity with a $1/r$ falloff as described in Sec.~\ref{sSec:Asymptotic Flatness}. Therefore, we formally impose an exact time-independent background solution in a perturbation theory setting describe in Sec.~\ref{sSec:PerturbationTheory}, which is naturally selected by the asymptotic flatness condition to be
\begin{equation}\label{eq:BackgroundSolMetric SVH}
    \bar{g}_{\mu\nu}=\eta_{\mu\nu}\,,\quad \bar{\Psi}=\{\bar{\Phi},\bar{A}_\mu\}\,,
\end{equation}
where local Lorentz invariance in the asymptotic solution imposes
\begin{equation}\label{eq:BackgroundSolutionNonMinimalFields SVH}
    \bar{\Phi}=\text{constant}\,,\qquad \bar{A}_\mu=0\,.
\end{equation}
This background indeed solves the exact equations of motion Eq.~\ref{eq:Background Equations}
by imposing the conditions
\begin{equation}\label{eq:backgroundeqCond sVH}
    \bar{G}_2=\bar{G}_{2\Phi}=0\,,
\end{equation}
where here we define the notation $\bar{G}_i\equiv G_i(\bar{\Phi},0,...,0)$, hence, the functionals evaluated on the background. Observe that the conditions in Eq.~\eqref{eq:backgroundeqCond sVH} imply that for a nontrivial scalar background $\bar\Phi\neq 0$, a theory with a simple mass term potential for the scalar field $G_2=-X+M^2\Phi^2$ is not allowed, which simply corresponds to the fact that in this case the solution is driven to $\bar\Phi= 0$. A solution with $\bar\Phi\neq 0$ that still includes a mass term for the scalar perturbation can however still be obtained, for instance through a standard spontaneous symmetry breaking potential $G_2=-X+(\Phi^2-\bar\Phi^2)^2$, that naturally drives the background solution to a non-zero $\bar\Phi$.

We further assume that there exists a split of the $1/r$ perturbations into a slowly varying- and high-frequency components, as in Eqs.~\eqref{eq:IsaacsonSplitGen} and \eqref{eq:IsaacsonSplitGen2}
\begin{equation}
    g_{\mu\nu}=\eta_{\mu\nu}+\delta g_{\mu\nu}^L+\delta g_{\mu\nu}^H\,,\quad \Psi=\bar\Psi+\delta\Psi^L+\delta\Psi^H\,,
\end{equation}
and to a first approximation only consider the presence of high frequency fields of characteristic amplitude $\cal{O}(\alpha)$ in the radiation zone
\begin{equation}\label{eq:HighFreqFields}
    \delta g^H_{\mu\nu}=h_{\mu\nu}\,,\quad \delta\Psi^H=\{\varphi,a_{\mu}\}\,.
\end{equation}
This is an excellent first approximation, since as discussed, to linear order, GWs are generally produced phase coherently. As already mentioned, this implies that for the high-frequency field in Eq.~\eqref{eq:HighFreqFields}, the terms \textit{radiation} and \textit{wave} can be used interchangeably. Moreover, without loss of generality, we only need to focus on the dynamical degrees of freedom and can neglect any Coulombic potential terms in the perturbations. 
We further want to ensure a nonvanishing kinetic term for at least the tensor perturbations by imposing $\bar{G}_{4}\neq 0$. For a certain gravitational wave source within the SVHH theory, the scalar and vector waves might or might not be excited, depending on the concrete situation.

For pedagogical reasons, we will now tackle the task of describing the dynamical high-frequency DOFs in Eq.~\eqref{eq:HighFreqFields} and their connection with the gravitational polarizations that can be measured in a typical GW experiment in the two equivalent approaches presented in this work. That is, we will first solve the first order perturbation equations in a general SVT decomposition by explicitly identifying all gauge invariant variables. In a second round, we will then obtain the same results through manifestly local gauge-fixing procedures (see also \cite{Hou:2017bqj} for the polarization content in pure Horndeski theory). This exercise will in particular prove itself useful for the considerations Chapter~\ref{Sec:GWMemory}.

\subsection{Gauge Invariant Polarizations in SVHH Gravity}

The first order propagation equations of motion of the high-frequency radiation given in Eqs.~\eqref{eq:EOMIIS2} and \eqref{eq:EOMIIS2Psi} can be solved in an SVT decomposition described in Sections~\ref{ssSec:GaugeInvariantDecomposition} and \ref{ssSec:WavesInMetricTheoriesOfGravity} that to first order in perturbations will decompose into a set of scalar, vector and tensor equations. Moreover, and crucially, such a decomposition allows the identification of manifestly gauge invariant perturbations under the gauge freedom introduced in the perturbative treatment of theories on a manifold as described in detail in Sec.~\ref{sSec:PerturbationTheory}. Concretely, the metric perturbations in Eq.~\eqref{eq:HighFreqFields} on a given background are only defined up to the gauge transformations generated by a small high-frequency vector field $\xi^\mu$ as [Eq.~\eqref{eq:GaugeTransformationMetric}]
\begin{align}
    \delta g_{\mu\nu}\rightarrow\delta g_{\mu\nu}+\mathcal{L}_\xi \bar{g}_{\mu\nu}=\delta g_{\mu\nu}+2\bar\nabla_{(\mu}\xi_{\nu)}\,,\label{eq:GaugeTransformationMetric SVH}
\end{align}
while on the other hand, the perturbations of the vector and scalar fields in Eq.~\eqref{eq:HighFreqFields} generally transform as [Eqs.~\eqref{eq:GaugeTransformationVector} and \eqref{eq:GaugeTransformationScalar}]
\begin{align}
    a^\mu&\rightarrow a^\mu+\xi^\alpha\bar\nabla_\alpha \bar A^\mu-\bar A^\alpha\bar\nabla_\alpha \xi^\mu\,,\label{eq:GaugeTransformationVector SVH}\\
    \varphi&\rightarrow \varphi+ \xi^\alpha\bar\nabla_\alpha \bar\Phi\,.\label{eq:GaugeTransformationScalar SVH}
\end{align}

On the given background solution in Eqs.~\eqref{eq:BackgroundSolMetric SVH} and \eqref{eq:BackgroundSolutionNonMinimalFields SVH}, the metric perturbations therefore satisfy the familiar linearized gauge transformation
\begin{align}
    h_{\mu\nu}\rightarrow h_{\mu\nu}+2\partial_{(\mu}\xi_{\nu)}\,,\label{eq:GaugeTransformationMetric SVH 2}
\end{align}
whereas both the vector and the scalar waves in Eq.~\eqref{eq:HighFreqFields} do not transform at all
\begin{align}
    a^\mu&\rightarrow a^\mu\,,\label{eq:GaugeTransformationVector SVH 2}\\
    \varphi&\rightarrow\varphi\,.\label{eq:GaugeTransformationScalar SVH 2}
\end{align}
However, the perturbed vector field $a_\mu$ inherits the internal gauge freedom in Eq.~\eqref{eq:gaugeTrasf A} and hence is only defined up to the following internal transformations
\begin{equation}
    a_\mu\rightarrow a_\mu+\partial_\mu \Lambda
\end{equation}

The general split of the metric perturbations was already offered explicitly in Eq.~\eqref{eq:hexpandBGR} together with the identification of six gauge invariant modes described by the variables $\{\delta\Phi,\delta\Theta,\delta\Xi^T_i,h^{TT}_{ij}\}$ in Eq.~\eqref{eq:gaugeinvBGR}, satisfying [Eq.~\eqref{eq:TransverseConditions}]
\begin{align}\label{eq:TransverseConditions SVH}
\partial^i h^{TT}_{ij}=0\;,\quad\delta^{ij} h^{TT}_{ij}=0\;,\quad\partial^i \delta\Xi^T_i=0\,.
\end{align}
On the other hand, due to Eqs.~\eqref{eq:GaugeTransformationVector SVH 2} and \eqref{eq:GaugeTransformationScalar SVH 2}, the scalar $\varphi$ as well as all the components in the SVT decomposition of the vector field 
\begin{equation}\label{eq:aexpand SVH}
 a_\mu=
 \left(\begin{array}{c} 
    	a_0  \\
    	\hline 
        a^T_i+\partial_i a^\parallel \\
    \end{array}\right)\,,
\end{equation}
where $a^T_i$ is transverse
\begin{equation}
    \partial^ia^T_i=0\,.
\end{equation}
Thus, the internal gauge symmetry in the high-frequency sector 
\begin{subequations}
    \begin{align}
        a_0&\rightarrow a_0+\dot\Lambda\,,\\
        a^\parallel &\rightarrow a^\parallel+\Lambda\,,\\
        a^T_i &\rightarrow  a^T_i\,, 
    \end{align}
\end{subequations}
reduces the number of physical DOFs by one unit and the SVT scalars in the vector field can generally be described by a single manifestly gauge invariant quantity
\begin{equation}
    \delta\Omega \equiv a_0 -\dot a^\parallel\,.
\end{equation}

Outside any sources in our asymptotic limit, the first order high-frequency equations [Eqs.~\eqref{eq:EOMIIS2} and \eqref{eq:EOMIIS2Psi}] for the metric, can then be cast into a set of scalar vector and tensor equations in terms of manifestly gauge invariant quantities only
\begin{subequations}\label{eq:Metric SVH}
    \begin{align}
        \Delta (\delta\Theta +\sigma \varphi) = 0 \,,\label{eq:Th SVH} \\
        \Delta (\delta\Phi - \sigma \varphi) = 0 \,, \label{eq:Ph SVH} \\
         \Delta \delta\Xi^T_i = 0 \,, \label{eq:Xi SVH} \\
        \Box h^{TT}_{ij} = 0 \label{eq:hTT SVH}\,,
    \end{align}
\end{subequations}
where $\Delta$ and $\Box$ denote the flat-space Laplace and wave operators, while the corresponding equations for the non-minimal scalar and vector fields [\eqref{eq:EOMIIS2Psi}] become
\begin{subequations}\label{eq:Vector And Scalar SVH}
    \begin{align}
        (\Box -m^2)\varphi = 0 \,, \label{eq:varph SVH}\\
         \Delta \delta\Omega = 0 \,, \label{eq:Om SVH}\\
        \Box a^T_i= 0 \label{eq:aT SVH}\,.
    \end{align}
\end{subequations}
Here we have defined the variables
\begin{equation}\label{eq:DefSigma}
    \sigma \equiv \frac{\bar{G}_{4,\Phi}}{\bar{G}_4}\,,
\end{equation}
as well as the mass of the scalar field
\begin{align}\label{eq:DefMass SVH}
    m^2 \equiv \frac{\bar{G}_{2,\Phi\Phi}}{\bar{G}_{2,X}-2\bar{G}_{3,\Phi}+3(\bar{G}_{4,\Phi})^2 / \bar{G_4} } =\frac{\bar{G}_{2,\Phi\Phi}}{\bar{G}_{2,X}-2\bar{G}_{3,\Phi}+3\sigma^2\bar{G_4} }\,. 
\end{align}
assuming that $\bar{G}_{2,X}-2\bar{G}_{3,\Phi}+3(\bar{G}_{4,\Phi})^2 / \bar{G_4}\neq0$.

These equations beautifully display the number of propagating degrees of freedom in SVHH gravity, namely two massless tensor DOFs in Eq.~\eqref{eq:hTT SVH}, two massless vector DOFs in Eq.~\eqref{eq:aT SVH} and one potentially massive degree of freedom as a solution to the Klein-Gordon equation in Eq.~\eqref{eq:varph SVH}. This result could have been expected, since in a faithful representation of the theory, generally only the TT perturbations of the physical metric satisfies a dynamical equation as in GR, while the other DOFs correspond to the natural description in terms of additional non-minimal fields. Moreover, again as in GR, Eqs.~\eqref{eq:Xi SVH} and \eqref{eq:Om SVH} in the absence of any source imply that with sufficiently well-behaved boundary conditions that we assume, the equations of motion constrain the fields $\delta\Xi^T_i$ and $\delta\Omega$ is such a way that we can set them to zero. However, and this is the crucial difference, the non-minimal coupling of the scalar field to the physical metric lead to a coupling of the scalar DOF to the two gauge invariant metric scalar variables in Eqs.~\eqref{eq:Th SVH} and \eqref{eq:Ph SVH}, resulting in the natural solutions
\begin{equation}
    \delta\Theta=-\sigma \varphi\,,\qquad \delta\Phi=\sigma \varphi\,.
\end{equation}
Thus, in the light of the discussion in Sec.~\ref{sSec:GWPolGen}, the scalar DOF excites additional modes of the physical metric, that can in turn be detected in a typical GW experiment. Note that this is not the case for the vector field. Indeed, for the Lorentz preserving background equation in Eq.~\eqref{eq:BackgroundSolutionNonMinimalFields SVH}, a vector field is not able to excite any gravitational polarization modes and due to the Principle~\ref{Principle:Universal and Minimal Coupling} of universal and minimal coupling cannot be directly detected. This is a result that in fact also holds if the vector modes are massive (see e.g. \cite{Dong:2023xyb}), such that we conjecture the following general result: Vector polarizations only arise in local Lorentz breaking background configurations.

More concretely, the response matrix $P_{ij}$  [Eq.~\eqref{eq:General Formula ofGWStrain}], that governs the physical effect of gravitational radiation that can be measured as a change in proper distance as described in Eq.~\eqref{eq:changeins}, reads in the case of SVHH gravity
\begin{equation}\label{eq:General Formula ofGWStrain SVH}
\boxed{P_{ij}=h^{TT}_{ij}+\left[\delta_{ij}-n_in_j\right] \delta\Theta +n_in_j\delta\Upsilon\,,}
\end{equation}
where
\begin{subequations}
\begin{align}
    \delta\Theta&=-\sigma \varphi\,,\\
    \delta\Upsilon&=\delta\Theta+v^2\delta\Phi=\sigma\,\varphi\left(v^2-1\right)\,.
\end{align}
\end{subequations}
and $v\leq 1$ defines the group velocity [Eq.~\eqref{eq:SpeedMassiveDOF}] of the potentially massive scalar wave.
Thus, in SVHH gravity, the gravitational polarization modes defined in Eq.~\eqref{eq:GWgenPoldef} that span the response matrix in the polarization space [Eq.~\eqref{eq:Pijexpand}] are given by
\begin{align}\label{eq:Polarizations SVH}
P_{+/\times}&=\frac{1}{2}e_{+/\times}^{ij}\,h^{TT}_{ij}\,,& P_{u/v}&=0\,,& P_b&\equiv -\sigma \varphi\,,&P_l&\equiv \sigma\,\varphi\left(v^2-1\right)\,\,,
\end{align}
From this equation, we can immediately conclude three things:
\begin{enumerate}[1)]
    \item If $\sigma=0$, hence, the coefficient defined in Eq.~\eqref{eq:DefSigma}, that controls the non-minimal coupling of the scalar field to the Ricci scalar, vanishes, then the scalar-vector theory does not excite any extra gravitational polarizations and only the two TT polarizations of GR can be observed directly.
    \item If $\sigma\neq 0$, but  $m=0$, hence the mass of the scalar perturbation defined in Eq.~\eqref{eq:DefMass SVH} vanishes, which through Eq.~\eqref{eq:SpeedMassiveDOF} implies that $v=1$, then $P_l=0$ and only the transverse breathing polarization $P_b$ is excited. This coincides with the expectation, that massless radiation is always purely transverse.
    \item If both $\sigma \neq 0$ and $m\neq0$, hence $v\neq 1$ and $v\neq 0$, then the single scalar degree of freedom simultaneously excites two additional gravitational polarizations $P_b$ and $P_l$, where $P_l$ is enhanced for decreasing group velocities and becomes equal in amplitude to the transverse polarization in the rest-frame of  its group velocity. However, in general one cannot consider to be in such a rest-frame, given the assumed situation of a localized source that produces the radiation.
\end{enumerate}

We want to take advantage of this nice example to stress again the clear distinction between the concepts of dynamical degrees of freedom of a theory and its gravitational polarizations. A general SVHH gravity has five propagating DOFs that can be excited in a given event. And depending on the precise structure of the theory, these DOFs either can excite or cannot excite gravitational polarizations of the physical metric. For instance, the two TT DOFs always excite the two TT polarizations of the physical metric (this is the reason why in GR such a distinction is seldom made). As for the scalar DOFs, it can, depending on the precise nature of the theory, either excite no additional polarization (case 1)), excite only one additional transverse polarization (case 2)) or excite two additional gravitational polarization modes (case 3)).

On the other hand, the two vector DOFs never excite any additional gravitational polarizations. We postulate here that this is in fact a general result for any ghost-free massless Lorentz preserving vector degrees of freedom. This is because with the restriction of building general actions in terms of the field-strength of the vector field in order to explicitly preserve the invariance under the internal gauge symmetry, no non-trivial non-minimal coupling to the Ricci scalar is allowed, which is generally necessary to excite additional gravitational polarizations. Moreover, this also coincides with the intuition that a massless wave should only be able to excite purely transverse polarizations (as it is for instance the case for the scalar DOF discussed here). According to this argument, the gravitational vector polarizations $P_u$ and $P_v$ that incorporate a longitudinal component could therefore only be excited my massive vector waves.


\subsection{Concrete Metric Theories: Polarizations}\label{sSec: Concrete Metric theories GW pol}

It is also instructive to look at particular example theories that the space of SVHH gravity encompasses. As discussed in Sec.~\ref{ssSec: A Exact Theories}, Horndeski gravity reduces to various exact popular theories widely used in the literature. These include:

\paragraph{Brans-Dicke Gravity.} BD theory is obtained by choosing the functionals following values [Eq.~\eqref{eq:BDGs}] of the general functionals in Eq.~\eqref{eq:ActionSVH} 
\begin{subequations}\label{eq:BDGs SVH}
\begin{align}
    G_2&=\frac{2\omega}{\Phi} X\,\\
    G_4&=\Phi\,,\\
    G_i&=0 \;\;\text{otherwise}\,.
\end{align}
\end{subequations}
Thus, in BD theory, with a nontrivial scalar background $\bar\Phi\neq 0$ we have
\begin{equation}
\sigma=\frac{\bar{G}_{4\Phi}}{\bar{G}_{4}}=\frac{1}{\bar{\Phi}}\neq 0\,.
\end{equation}
Moreover, the mass of the scalar perturbation vanishes, since
\begin{align}
    m^2 \propto \bar{G}_{2,\Phi\Phi} =0 \,. 
\end{align}
Thus BD theory only excited the breathing mode polarization.

\paragraph{f(R) Gravity.} On the other hand, $f(R)$ gravity, with $f''(R)\neq 0$ is equivalent to choosing [Eq.~\eqref{eq:f(R)s}]
\begin{subequations}\label{eq:f(R)s SVH}
\begin{align}
    G_2&=f(\Phi)-\Phi f'(\Phi)\,\\
    G_4&=f'(\Phi)\,,\\
    G_i&=0 \;\;\text{otherwise}\,.
\end{align}
\end{subequations}
Thus, first of all, in order to satisfy the background equation conditions in Eq.~\eqref{eq:backgroundeqCond sVH} we need to impose in this case
\begin{equation}
    \bar\Phi=f(\bar\Phi)=0\,.
\end{equation}
Then, we obtain similar to the BD case
\begin{equation}
\sigma=\frac{\bar{G}_{4\Phi}}{\bar{G}_{4}}=\frac{f''(\bar\Phi)}{f'(\bar{\Phi})}\neq 0\,,
\end{equation}
while however this time in general
\begin{align}
    m^2= \frac{\bar{G}_{2,\Phi\Phi}\bar{G_4}}{3(\bar{G}_{4,\Phi})^2 }=\frac{f'(\bar\Phi)}{3f''(\bar\Phi)}\neq 0 \,. 
\end{align}
Hence, the single additional scalar DOF in $f(R)$ in general excites two additional gravitational polarizations. Moreover, note that for $f(R)$, the coefficient $\sigma$ is directly related to the mass through $\sigma m^2=1/3$. This implies that also the breathing modes in Eq.~\eqref{eq:Polarizations SVH} depends on the mass of the fields. This is therefore a special feature of $f(R)$ gravities, since for $\sigma$ independent of the mass, which is usually assumed, only the longitudinal polarization depends on the mass through the velocity of the field.

\paragraph{Scalar-Gauss-Bonnet Gravity.} Finally, sGB gravity can be obtained by choosing [Eq.~\eqref{eq:CorrespondencesGBHorndeski}]
\begin{subequations}\label{eq:CorrespondencesGBHorndeski SVH}
\begin{align}
        G_2&=X+8f^{(4)}(\Phi)X^2(3-\ln X)\,,\\ G_3&=4f^{(3)}(\Phi)X(7-3\ln X)\,,\\
        G_4&=1+4f^{(2)}(\Phi)X(2-\ln X)\,,\\
        G_5&=-f^{(1)}(\Phi)\ln X\,,
\end{align}
\end{subequations}
where $f^{(n)}(\Phi)\equiv\partial^n f/\partial\Phi^n$. Hence, with $\bar\Phi=\text{const.}$ one obtains the simple relations
\begin{equation}
    \sigma=m=0\,.
\end{equation}
Therefore the massless scalar DOF in sGB gravity does not excite any additional gravitational polarization in a typical GW response.

\subsection{Manifestly Local Polarizations in SVHH Gravity}

We now want to repeat the same exercise above, but instead of identifying manifestly gauge invariant but non-local perturbation variables, we want to use the local approach in the Lorentz and ultimately in the TT gauge. To address the leading-order wave propagation in this case, it is very useful to first expand the action in Eq.~\eqref{eq:ActionSVH} to second order in perturbations, which facilitates the determination of the physical dynamical degrees of freedom in the theory. Moreover, the action approach will also be useful in a later stage of this work.

The second-order action in SVT theory contains a kinetic term that couples the high-frequency metric and scalar perturbations $h_{\mu\nu}$ and $\varphi$. This term can, however, be removed through the field redefinition\footnote{Note that this redefinition is equivalent to a traditional transition from the Jordan to the Einstein ``frame'', that redefines the physical metric to an unphysical one. It is however much cleaner to make this redefinition only at the level of the perturbations, since in the end we will discover that the observationally relevant information in both the original and the redefined field perturbations coincide.}
\begin{equation}\label{RelationPhToh}
    \hat{h}_{\mu\nu}\equiv h_{\mu\nu}+\eta_{\mu\nu}\sigma\,\varphi\,,
\end{equation}
where $\sigma$ is the same background variable that was already defined in Eq.~\eqref{eq:DefSigma}. 
Moreover, the scalar and vector perturbation can be rescaled so that their kinetic terms in the second-order action are canonically normalized. The necessary rescaling is
\begin{equation}\label{RescaledPhi}
    \hat{\varphi}\equiv \rho\,\varphi\,,\quad \hat{a}_\mu\equiv \zeta\,a_\mu\,,
\end{equation}  
where
\begin{equation}
    \rho\equiv\sqrt{3\,\sigma^2+\frac{(\bar G_{2,X}-2\,\bar G_{3,\Phi})}{\bar G_4}}=\sqrt{\frac{\bar G_{2,\Phi\Phi}}{\bar{G}_4}}\,\frac{1}{m}\,,
\end{equation}
with $m$ defined in Eq.~\eqref{eq:DefMass SVH} and
\begin{equation}
    \zeta\equiv \sqrt{\frac{\bar G_{2,F}}{\bar G_4}}\,.
\end{equation}

We require here that the coefficients $\sigma$, $\rho$ and $\zeta$ are real and positive, which is also imposed by the positivity of the energy carried by the perturbations as we will see explicitly in Chapter~\ref{Sec:GWMemory}.
In terms of the new variables in Eqs.~\eqref{RelationPhToh} and \eqref{RescaledPhi}, the second-order action of SVT theory then simply reads
\begin{equation}\label{ActionSVT2nd}
    _{\mys{(2)}}S^{\myst{SVH}}=\frac{-1}{2\kappa_\text{eff}}\int\dd^4x\bigg[\hat{h}^{\mu\nu}\mathcal{E}^{\alpha\beta}_{\mu\nu}\hat{h}_{\alpha\beta}+\frac{1}{4}\hat{f}_{\mu\nu}\hat{f}^{\mu\nu}+\frac{1}{2}\left(\partial_\mu\hat{\varphi}\partial^\mu\hat{\varphi}-m^2\hat{\varphi}^2\right)\bigg]\,,
\end{equation}
where we define the field strength of the leading-order vector perturbation 
\begin{equation}
    \hat{f}_{\mu\nu}\equiv\partial_\mu \hat{a}_{\nu}-\partial_\nu \hat{a}_\mu\,,
\end{equation}
as well as an effective gravitational coupling
\begin{equation}\label{eq:kappaeff SVH}
    \kappa_\text{eff}\equiv 8\pi G_\text{eff}\,,\quad G_\text{eff}\equiv \frac{G}{\bar G_4}\,,
\end{equation}
where $G$ is the bare Newtons constant. Moreover, recall the definitions of the trace
\begin{equation}
    \hat{h}^t=\eta^{\mu\nu}\hat{h}_{\mu\nu}
\end{equation} 
and the flat-space, Lichnerowicz operator [Eq.~\eqref{eq:LichnerowiczOperator}]
\begin{equation}
    \mathcal{E}^{\alpha\beta}_{\mu\nu}\hat{h}_{\alpha\beta}=-\frac{1}{4}\Big[\Box \hat{h}_{\mu\nu}-2\partial_\alpha\partial_{(\mu}\hat{h}\du{\nu)}{\alpha}+\partial_\mu\partial_\nu \hat{h}^t-\eta_{\mu\nu}\left(\Box\hat{h}^t-\partial_\alpha\partial_\beta \hat{h}^{\alpha\beta}\right)\Big]\,.
\end{equation}
Observe that in terms of the new variables, the second-order action in Eq.~\eqref{ActionSVT2nd} is nothing but the linearized Einstein-Hilbert action with a sum of additional canonical fields. 

As already discussed above, in particular also in Sec.~\ref{sSec:PerturbationTheory}, the high-frequency perturbed values of fields on a manifold are subject to gauge redundancies that can be associated to infinitesimal coordinate transformations of the form $x^\mu\rightarrow \tilde x^\mu=x^\mu-\xi_H^\mu$, with $|\xi_H^\mu|\ll 1$. Concretely, for SVHH gravity with locally Lorentz preserving asymptotic background solutions only the metric perturbation transforms under this gauge symmetry as
\begin{equation}\label{eq:CoordGaugeTransf}
  h_{\mu\nu}\rightarrow h_{\mu\nu}-2\,\eta_{\alpha(\nu}\partial_{\mu)} \xi_H^\alpha\,.
\end{equation}
Note that this gauge freedom is entirely inherited by the redefined perturbation variable $\hat h_{\mu\nu}$.
On the other hand, the vector perturbation $a^\mu$ is subject to a different, internal gauge freedom 
\begin{equation}\label{eq:U(1)GaugeTransf}
  a_{\mu}\rightarrow a_{\mu}+\partial_\mu\Lambda^H\,.
\end{equation}
By performing suitable coordinate [Eq.~\eqref{eq:CoordGaugeTransf}] and $U(1)$ gauge transformations [Eq.~\eqref{eq:U(1)GaugeTransf}], in the radiation zone one can then impose at the level of the equations of motion the following TT gauge conditions\footnote{Note that these gauge conditions here are not to be confused with the stronger notion of \textit{TT-gauge} that impose a vanishing of all $\hat h_{0i}$ components that can only be imposed outside any source (see e.g. \cite{maggiore2008gravitational}). The notion of TT gauge presented here is in fact still compatible with a sourced equation, as long as the source is conserved and traceless, a fact we will use in subsequent chapters.}
\begin{equation}\label{eq:FirstOrderGaugeFixing}
    \partial_\mu \hat{h}^{\mu\nu}=0\,,\quad\hat{h}^t=0\quad \text{and}\quad \partial_\mu \hat{a}^{\mu}=0\,.
\end{equation}
In this gauge, it is no surprise that the leading-order wave propagation described by Eqs.~\eqref{eq:EOMIIS2} and \eqref{eq:EOMIIS2Psi} lead to decoupled wave equations for all the hatted perturbations
\begin{equation}\label{eq:FirstOrderProp}
    \Box \hat{h}_{\mu\nu}=0\,,\quad \Box \hat{a}_\mu=0\,,\quad (\Box-m^2) \hat{\varphi}=0\,.
\end{equation}

The solutions to these equations in principle represent the dynamical DOFs of the theory. Yet, in contrast to the manifestly gauge invariant approach discussed above, in these local equations it seems that also additional components of the fields satisfy a wave equation. However, just as in GR, there is in fact a residual gauge freedom left over after fixing Eq.~\eqref{eq:FirstOrderGaugeFixing}, given by transformations satisfying $\Box\xi_{H}^{\mu}=\partial_\mu\xi_H^\mu=0$ and $\Box\Lambda^H=0$ and not all remaining components are invariant under these additional transformations. In the radiation zone, this residual gauge freedom can be employed to single out the true dynamical DOFs, which in the case of the tensorial perturbation correspond to the $TT$ part, as already shown on several occasions, while for the vector perturbations the true DOFs are found in the transverse part. In Sec.~\ref{sSec:GWGeneration} we showed that for a superposition of plane waves in the radiation zone, these true dynamical DOFs can be singled out through a transverse [Eq.~\eqref{eq:TransverseProjector}] and transverse-traceless projection [Eq.~\eqref{eq:Projectors}]
\begin{equation}\label{eq:TT modes h and a SVH}
    \perp_{ijab}\hat{h}_{ab}=\hat{h}^\text{TT}_{ij}=h^\text{TT}_{ij}\,,\quad \hat{a}^\text{T}_{i}=\perp_{ij}\hat{a}_{j}\,.
\end{equation}
Decisively, the TT part of the original perturbation of the physical metric coincides with the TT part of the redefined field [Eq.~\eqref{RelationPhToh}]. Thus, the same is therefore also true for the associated polarization modes
\begin{equation}\label{eq:polmodes h SVH}
    \hat{h}_{+/\times}=\frac{1}{2}e^{ij}_{+/\times}\,\hat{h}_{ij}=h_{+/\times}=\frac{1}{2}e^{ij}_{+/\times}\,h_{ij}\,.
\end{equation}
Similarly, also the transverse vector modes can be described in terms of polarizations in the transverse $\{\mathbf{u},\mathbf{v}\}$ basis
\begin{equation}\label{eq:polmodes a SVH}
    \hat{a}_{u/v}=e^i_{u/v} \hat{a}_i\,.
\end{equation}
These polarization modes, together with the scalar perturbation $\hat{\varphi}$ represent the leading-order tensor, vector and scalar radiation respectively that in source centered coordinates $\{t,r,\theta,\phi\}$ in the radiation zone take the general form
\begin{align}
    &\left\{\hat{h}_{+/\times}\,,\;\hat{a}_{u/v}\,,\;\hat{\varphi} \right\} \sim \frac{1}{r} \left\{f^h_{+/\times}(t-r,\theta,\phi), f^a_{u/v}(t-r,\theta,\phi), f^\varphi(t-vr,\theta,\phi) \right\}\,.\label{eq:OutgoingPlaneWave}
\end{align}
for some functions $f^{h,a,\varphi}$.

We are now in a position to rederive the gravitational polarizations of this theory by evaluating the leading order Riemann tensor in Eq.~\eqref{eq:linRiemann} for the perturbations of the physical metric at hand. As already discussed, only the true radiative modes will contribute, such that using Eq.~\eqref{RelationPhToh} we can decompose the leading-order wave of the high-frequency perturbations of the physical metric as
\begin{equation}
    h_{ij}=\hat{h}_{ij}^\text{TT}-\delta_{ij}\frac{\sigma}{\rho}\,\hat{\varphi}\quad \text{and}\quad h_{00}=\frac{\sigma}{\rho}\,\hat{\varphi}\,.
\end{equation}
After imposing the falloff of Eq.~\eqref{eq:OutgoingPlaneWave}, and the corresponding replacements $\partial_i\rightarrow -n_i v \partial_0$, the electric part of the linearized Riemann tensor reads
\begin{equation}\label{ElectricRiemannLin}
\begin{split}
    \phantom{}_{\mys{(1)}}R_{i0j0}&=-\frac{1}{2}\left(\ddot{\hat{h}}^\text{TT}_{ij}-\left[\delta_{ij}-n_in_j\right]\frac{\sigma}{\rho}\,\ddot{\hat{\varphi}}+n_in_j(v^2-1)\frac{\sigma}{\rho}\,\ddot{\hat{\varphi}}\right)\\
    &=-\frac{1}{2}\left(e^+_{ij}\,\ddot{h}_++ e^\times_{ij}\,\ddot{h}_\times-e^b_{ij}\sigma\,\ddot{\varphi}+e^l_{ij}(v^2-1)\sigma\,\ddot{\varphi}\right)\,.
\end{split}
\end{equation}
By Eqs.~\eqref{eq:linRiemann00} and \eqref{eq:Pijexpand} we therefore precisely recover the result in Eq.~\eqref{eq:Polarizations SVH}. In terms of this derivation, the vector DOFs of the theory do not excite additional polarizations, as their non-minimal coupling in the action is not such that a redefinition of the fields is necessary in order to obtain variables for which one obtains a set of decoupled wave equations. On the other hand, the scalar sector recovers the well known results from scalar Horndeski theory \cite{Hou:2017bqj}.


\newpage
\thispagestyle{plain} 
\mbox{}


\chapter{Gravitational Wave Memory}\label{Sec:GWMemory}

So far, we only treated gravitational waves and radiation at the linear level, in other words to linear order in $\alpha$, where $\alpha$ characterizes the size of the GW perturbation amplitude. However, and especially in an asymptotically flat scenario far away from any matter source where perturbations are naturally characterized by a $\sim 1/r$ expansion, a purely linear treatment of gravitational waves does intrinsically not suffice \cite{Christodoulou:1991cr} (see also \cite{Heisenberg:2023prj}). This is due to the fact that radiation by definition carries energy, the source of gravity, all the way to the radiation zone. The resulting back-reaction of the energy and momentum of the waves induces a non-negligible contribution to the dynamics of spacetime, which in the case of gravitational radiation therefore fundamentally relies on the non-linearity of gravity. 

In GR, this effect is known to contribute to the gravitational wave \textit{\gls{memory}} \cite{Christodoulou:1991cr,Ludvigsen:1989cr,Blanchet:1992br,Thorne:1992sdb,PhysRevD.44.R2945} (see also \cite{Favata:2008yd,Favata:2009ii,Favata:2010zu,Bieri:2013ada,Garfinkle:2022dnm}). Memory, or more precisely \textit{displacement memory} \cite{Barnich:2009se,Pasterski:2015tva,Nichols:2017rqr,Nichols:2018qac,Compere:2019gft}, of gravitational radiation is defined as a permanent change in proper distance after the passage of a gravitational wave in the geodesic deviation response defined in Eq.~\eqref{eq:changeins}. However, not only the energy carried by radiation can induce such a lasting distortion of spacetime. Indeed, a displacement memory component was first discovered in the context of unbound massive components within the production of gravitational waves in GR \cite{Zeldovich:1974gvh,Turner:1977gvh,Braginsky:1985vlg,Braginsky:1987gvh}, such as in hyperbolic encounter binaries (see also \cite{Favata:2008yd,Favata:2010zu}). Thus, any unbound source of energy in an isolated system emitting gravitational radiation, ranging from supernovae ejection of matter or neutrinos \cite{Epstein:1978gvh,Burrows:1995bb,Ott:2008wt,Murphy:2009}, gamma-ray burst jets \cite{Sago:2004pn} or even CBC remnant kicks \cite{Merritt:2004xa,Gonzalez:2006md,Favata:2008ti} induces GW memory (see also \cite{Thorne:1992sdb,Bieri:2013ada,Garfinkle:2022dnm}). In this context, it can be useful to distinguish between so called \textit{\gls{null memory}} that is sourced by massless radiation that reaches null infinity of asymptotically flat spacetime and \textit{\gls{ordinary memory}} which encompasses all massive unbound objects \cite{Bieri:2013ada}.

\paragraph{Definition of Memory.} However, independently of the precise type, displacement memory can be defined in the geodesic deviation equation [Eq.~\eqref{eq:changeins}] as follows: In an idealized situation, by definition the difference in proper distance $\Delta\ell$ before the presence of any gravitational radiation at some initial proper time $\tau_0\rightarrow -\infty$ vanishes, hence $\Delta\ell(\tau_0)=\ell(\tau_0)-\ell_0=0$. As a burst of gravitational waves passes by, the difference in proper distance is starting to oscillate around the zero value defined by the initial time. GW displacement memory is then defined as a permanent change
\begin{equation}
    \Delta\ell(\tau_f)=\ell(\tau_f)-\ell_0\neq 0\,,
\end{equation}
for a time $\tau_f\rightarrow\infty$ well after the passage of a burst of gravitational radiation. In other words, GW memory is a modification of the rest proper length compared to an initial value, and in this sense permanently distorts spacetime.
Thus, any piece within the radiative response matrix $P_{ij}$ governed by the gravitational polarizations of the physical metric that induces such a permanent displacement will be called a memory component. 

In the light of the discussion in Sec.~\ref{sSec:GWPolGen}, on a very general basis, metric theories of gravity are expected to contain memory that can be associated with each of the six gravitational polarizations. This naturally leads to a distinction between scalar, vector and \gls{tensor memory}, where in this terminology, the terms ``scalar'', ``vector'' or ``tensor'' refer to the polarization type that induces a permanent displacement. This distinction in SVT memory should not be confused with the tensorial nature of the leading order waves that acts as a source of memory, which is not restricted in any way.

\paragraph{Computation of Memory.} Within GR, a derivation of the memory effect in asymptotically flat spacetimes is well understood in terms of a post-Newtonian expansion \cite{Blanchet:1992br} but in particular also through a deep connection of memory to the supertranslations of the asymptotic BMS group \cite{Bondi:1962px,Sachs:1962wk,Geroch:1977jn,Ashtekar:1981bq,Strominger:2014pwa,Strominger:2017zoo,Compere:2019sm}. In this context, the non-trivial energy carried by null radiation can neatly be described within the non-linear Newman-Penrose approach in terms of non-trivial BMS balance laws \cite{Christodoulou:1991cr,FrauendienerJ,Ashtekar:2014zsa,Compere:2019gft,DAmbrosio:2022clk}.

As concerns beyond GR theories, on the other hand, the memory effect has only been investigated in a handful of concrete theories. For instance, within a post-Newtonian expansion of Brans-Dicke theory [Eq.~\ref{ActionBD}], a new memory contribution originating from a dipole-dipole coupling was found~\cite{lang_compact_2014,lang_compact_2015,tahura_gravitational-wave_2021}. Moreover, also the BMS balance laws were recently derived in BD theory \cite{hou_gravitational_2021,tahura_brans-dicke_2021,hou_conserved_2021,hou_gravitational_2021_2} (see Sec.~\ref{MatchToAsymptoticsBD}), by showing that the theory retains the same asymptotic group structure as in GR, despite its altered peeling properties. As a consistency check, the resulting memory component was also matched to the earlier PN calculation \cite{tahura_gravitational-wave_2021}. Furthermore, different aspects of the scalar memory within BD theory were investigated in~\cite{du_gravitational_2016,koyama_testing_2020}. Similarly, also the BMS balance laws  of dynamical Chern-Simons gravity [Eq.~\eqref{eq:ActiondCS}] were established in \cite{hou_gravitational_2022,Hou:2021bxz}.

In the present work, we will take advantage of the careful definition of gravitational waves, gravitational radiation and dynamical degrees of freedom in the preceding chapters, in particular the Isaacson approach discussed in Sections~\ref{ssSec:IsaacsonInGR} and~\ref{ssSec:IsaacsonGeneral}, and present a novel consistent framework to compute and understand gravitational wave memory. This will allow in a first step to provide a unified description of null and ordinary memory in GR [Sec.~\ref{sSec:UnifiedTreatmentof Null and Ordinary Memory}]. Foremost, however, in Sections~\ref{sSec:GW Memory beyond GR} and \ref{sec:SVTMem} this new approach to null and ordinary displacement memory is shown to be readily generalizable to any metric theory of gravity beyond GR.

\section{Displacement Memory in GR}\label{sSec:GWMemoryinGR}

In this section, we will show that the Isaacson approach to defining gravitational waves outlined in Sections~\ref{ssSec:IsaacsonInGR} and~\ref{ssSec:IsaacsonGeneral} represents the ideal framework to investigate and compute gravitational wave memory effects in very general settings. To illustrate this, we will first analyze the simplest scenario and consider general relativity on an asymptotically flat spacetime [Def.~\ref{DefAsymptoticallyFlat}]. As discussed, this implies that formally we work in a perturbation theory setting around Minkowski spacetime and consider the radiation zone outside of any source. In other words, we consider the leading order in a $1/r$ expansion in source centered asymptotic Minkowski coordinates $\{t,r,\theta,\phi\}$. For the massless radiation of GR, it will further be essential to perform the correct radiation zone limit given by the \textit{limit to null infinity} defined as $r\rightarrow \infty$ at constant asymptotic retarded time $u=t-r$ (recall Sec.~\ref{sSec:Asymptotic Flatness}). As mentioned, the Isaacson approach will allow for a unified treatment of all types of memory, including unbound energy-momentum from radiation of massless as well as massive fields to individual localized matter junks.

\subsection{A Unified Treatment of Null and Ordinary Memory}\label{sSec:UnifiedTreatmentof Null and Ordinary Memory}

Thus, let's consider the system of leading order wave equations in the Isaacson picture we derived in Eqs.~\eqref{eq:EOMIISGR} and \eqref{eq:EOMISGR}
\begin{align}
    \phantom{}_{\mys{(1)}}G_{\mu\nu}[h]&=0\,,\label{eq:EOMIISGR2}\\
    \phantom{}_{\mys{(1)}}G_{\mu\nu}[\delta h]&=-\frac{1}{2}\,\big\langle\phantom{}_{\mys{(2)}}G_{\mu\nu}[h]\big\rangle+\kappa_0\big\langle\delta T_{\mu\nu}\big\rangle\,, \label{eq:EOMISGR2}
\end{align}
and solve them in the limit to null infinity, where we identified $\delta g^H_{\mu\nu}=h_{\mu\nu}$ and $\delta g^L_{\mu\nu}=\delta h_{\mu\nu}$ to comply with standard notation. 
Here, $\langle\delta T_{\mu\nu}\rangle$ represents contributions to the asymptotic energy momentum flux from unbound matter radiation or particles, while $\langle\phantom{}_{\mys{(2)}}G_{\mu\nu}[h]\rangle$ is associated to the coarse-grained Isaacson energy momentum tensor of the leading order high-frequency waves in Eq.~\eqref{eq:DefPseudoEMTensorGR}.
In this setup we therefore assume the presence of leading order high-frequency gravitational waves $h_{\mu\nu}$ of small amplitude $\mathcal{O}(\alpha)$ in the radiation zone as discussed in Sec.~\ref{sSec:GWGeneration} that satisfy a propagation equation [Eq.~\eqref{eq:EOMIISGR2}], as well as a potential presence of additional matter perturbations in the form of radiation or localized massive particles satisfying similar leading order propagation equations. 

Recall that within the Isaacson approach, the back-reaction of the energy-momentum carried by such unbound perturbations can be consistently discussed through the leading order low-frequency equation of the physical metric in Eq.~\eqref{eq:EOMISGR2}. As we will now show, the physical modes within the resulting low-frequency metric perturbation $\delta g^L_{\mu\nu}=\delta h_{\mu\nu}$ will precisely correspond to a memory contribution in any GW detector response. For this, we will require additional information of the energy-momentum tensor of perturbations in the radiation zone. Since the most interesting contribution to the memory will be the one that is sourced by the gravitational waves themselves, we will first consider the energy-momentum of the leading order high-frequency radiation, hence the leading order waves, in more detail.

\paragraph{Leading Order Wave.}  Recall, that based on the considerations in Sec.~\ref{ssSec:Local Wave Equation in GR} as well as the preceding Chapter~\ref{Sec:Radiation andGWs in Gravity}, the leading order propagation equation of the gravitational waves in the gauge in Eq.~\eqref{eq:FirstOrderGaugeFixing}
\begin{equation}\label{eq:FirstOrderGaugeFixingGR}
    h^t=\eta^{\mu\nu}h_{\mu\nu}=0\,,\qquad \partial_\mu h^{\mu\nu}=0\,,
\end{equation}
is given by a wave equation
\begin{equation}\label{eq:EOMIISGR3}
    \phantom{}_{\mys{(1)}}G_{\mu\nu}[h]=-\frac{1}{2}\Box h_{\mu\nu}=0\,.
\end{equation}
In the appropriate gauge, this equation can be reduced to two modes describing the dynamical DOFs of the theory as solutions to the wave equation
\begin{equation}
    \Box h_{+/\times} =0\,,
\end{equation}
where 
\begin{equation}
    h_{+/\times}=\frac{1}{2} e_{+/\times}^{ij}\,h_{ij}\,.
\end{equation}
In practice, these leading order waves are assumed to be known \textit{a priori} and could correspond to the standard waveforms of for instance CBC events.

\paragraph{Radiative Energy-Momentum.} However, as already mentioned, this is only half of the story, since Eq.~\eqref{eq:EOMIISGR3} only represents the leading order \textit{high-frequency} equation and there exists a second leading order equation for the low-frequency fields given in Eq.~\eqref{eq:EOMISGR}. As discussed, the low frequency perturbation components that are directly tied to the localized source are not relevant in the asymptotic regime. Yet, the presence of the high-frequency waves in the radiation zone inevitably represent another source of gravity, the energy-momentum carried by the gravitational radiation itself. Back in Sec.~\ref{ssSec:IsaacsonInGR} we already identified the pseudo energy-momentum tensor of gravitational waves as the quantity in Eq.~\eqref{eq:DefPseudoEMTensorGR}
\begin{equation}\label{eq:StressEnergyGRSecond2}
    \phantom{}_{\mys{(2)}}t^{\myst{GR}}_{\mu\nu}[h]\equiv -\frac{1}{2\kappa_0}\big\langle \phantom{}_{\mys{(2)}}G_{\mu\nu}[h]\big\rangle\,,
\end{equation}
which our asymptotic expansion takes the form
\begin{align}\label{eq:StressEnergyGRThird}
    \phantom{}_{\mys{(2)}}t_{\mu\nu}^{\myst{GR}}=\frac{1}{4\kappa_0}\Big\langle \partial_\mu h_{\alpha\beta}\partial_\nu h^{\alpha\beta}\Big\rangle=\frac{1}{4\kappa_0}\Big\langle \partial_\mu h^{TT}_{ij}\partial_\nu h_{TT}^{ij}\Big\rangle=\frac{1}{2\kappa_0}\,\Big\langle \dot{h}_+^2+\dot{h}_\times^2\Big\rangle\,l_\mu\,l_\nu\,.
\end{align}
where we have defined the null vector
\begin{equation}\label{eq:FirstDef l}
    l_\mu\equiv -\nabla_\mu t+\nabla_\mu r\,,
\end{equation}
with $\nabla_\mu r=\partial_\mu r=\delta_{\mu\ i} \,\partial_i r=\delta_{\mu\ i}\,n_i$.

In the first equality, we simply evaluated the Einstein tensor at second order in the perturbation variable\footnote{Note that compared to \cite{maggiore2008gravitational} for instance, we have factored out the $1/2!$ prefactor of the second order operator explicitly.} and imposed the gauge conditions in Eq.~\eqref{eq:FirstOrderGaugeFixingGR} while also performing integrations by parts that are allowed due to the averaging (recall the discussion in Sec.~\eqref{ssSec:IsaacsonInGR}). One can not stress enough that the Isaacson approach, hence the assumption of the existence of a clear separation of scales for the waves that also introduces the averaging in the definition of the energy momentum tensor of the waves, is crucial here. Only in this framework is it possible to consistently define a gauge invariant and conserved energy momentum tensor of gravitational radiation. Fundamentally, this is because due to the Einstein equivalence Principle~\ref{Principle:EEP}, in general it is not possible to define a local energy-momentum for a gravitational field. The Isaacson approach provides however a well-defined procedure to identify the energy carried by a wave by allowing for a consistent coarse-gaining procedure. 

That the energy momentum tensor in Eq.~\eqref{eq:StressEnergyGRSecond2} is gauge invariant also justifies the second equality in Eq.~\eqref{eq:StressEnergyGRThird}, implying that the energy momentum tensor can ultimately be written in terms of gauge-invariant high-frequency degrees of freedom only, which in GR are entirely given by the $TT$ part of the metric perturbations.\footnote{More fundamentally, if one would define the energy momentum tensor of the full metric perturbation, hence also involving the low-frequency pieces, still only the gauge-invariant radiative variables (here the $TT$ part) would contribute, as the non-radiative gauge invariant quantities drop out due to the differentiation's, just as it was the case in the electric part of the Riemann tensor in Eq.~\eqref{eq:linRiemann00}.} The third equality follows from
\begin{equation}
    h_{ij}^{TT}=e^+_{ij}\,h_++e^\times_{ij}\,h_\times\,,
\end{equation}
with Eq.~\eqref{eq:Polarization Vectors Basis}, such that
\begin{equation}
    \Big\langle \partial_\mu h^{TT}_{ij}\partial_\nu h_{TT}^{ij}\Big\rangle=2\, \Big\langle \partial_\mu h_+\partial_\nu h_++\partial_\mu h_\times\partial_\nu h_\times\Big\rangle\,.
\end{equation}
Moreover, on the spacial derivatives one can use the general form of asymptotic radiation given in Eq.~\eqref{eq:GeneralFormRadiation} implying to leading order in $1/r$
\begin{equation}
    \partial_i h_\lambda = -n_i \dot h_\lambda\,,
\end{equation}
to finally arrive at Eq.~\eqref{eq:StressEnergyGRThird}.

\paragraph{General Structure of Asymptotic Energy-Momentum.} 

At this point, we want to pause and derive a general structure of asymptotic energy-momentum tensors that will prove important in the following. For this, we consider a general asymptotic energy momentum tensor $T^\text{a}_{\mu\nu}$ that is conserved on the asymptotically flat spacetime 
\begin{equation}\label{eq:Conservation asymptotic EMT}
    \partial_\mu T_\text{a}^{\mu\nu}=0\,,
\end{equation}
and whose energy-momentum in source-centered coordinates $\{t,r,\Omega\}$ is transported by an asymptotic group velocity $v$ in the radial direction. 
In order to be as general as possible, we will leave the asymptotic speed $v$ as a general variable. The case of luminal gravitational waves will then simply follow by setting $v=1$.

We now want to show that such an asymptotic energy momentum tensor always has the form 
\begin{equation}\label{eq:General Form asymptotic EMT}
    \boxed{T^\text{a}_{\mu\nu}(u,r,\Omega)=T^\text{a}_{00}(u,r,\Omega)\,l_\mu l_\nu=\frac{1}{r^2}\,F(u,\Omega)\,l_\mu l_\nu\,,}
\end{equation}
for some function $F(u,\Omega)$ that is related to a purely radial outward energy flux. For arbitrary velocities, the vector $l_\mu$ already introduced in Eq.~\eqref{eq:FirstDef l} now generalizes to
\begin{equation}
    l_\mu\equiv -\nabla_\mu t+v\,\nabla_\mu r\,,
\end{equation}
while similarly, the asymptotic retarded time $u$ for general velocities $v$ becomes
\begin{equation}\label{eq:asymptoticRet Time v}
    u\equiv t-\frac{r}{v}\,.
\end{equation}

To see this, first recall that $T^\text{a}_{00}$ defines an energy density that carries energy with a velocity $v$ away from the source, such that by definition one can define an energy flux through (see also \cite{misner_gravitation_1973,maggiore2008gravitational})
\begin{equation}
    v\,T^\text{a}_{00}=\frac{1}{r^2}\frac{dE(u,\Omega)}{dud\Omega}\,,
\end{equation}
This corresponds to the typical inverse square law. In Eq.~\eqref{eq:General Form asymptotic EMT} we can therefore identify
\begin{equation}\label{eq:rel F to Flux}
    \boxed{F(u,\Omega)\equiv \frac{1}{v}\frac{dE}{dud\Omega}\,.}
\end{equation}
As a consequence of the form of arguments of $F(u,\Omega)=F(t-r/v,\Omega)$, to leading order in $1/r$, the energy density satisfies [Eq.~\eqref{eq:IdentityRadiation}]
\begin{equation}\label{eq:IdentityEMT deriv}
    \partial_i T^\text{a}_{00}=-\frac{n_i}{v}\partial_0 T^\text{a}_{00}\,.
\end{equation}

Equation.~\eqref{eq:rel F to Flux} can in fact be further justified through the following arguments. First of all, note that since $T^\text{a}_{00}$ defines an energy density, we can write
\begin{equation}\label{eq:DefEdV}
    E_{dV}=T^\text{a}_{00}\,dr \, r^2 d\Omega
\end{equation}
for an infinitesimal volume element\footnote{Recall that for high-frequency waves, we define a low-frequency energy momentum tensor at a given spacetime point through a coarse-graining average over spacetime.} $dV=drdA'=dr\, r^2 d\Omega$ at some fixed location in time and space in the radiation zone. Now, by assumption, the energy is emitted from the source with asymptotic velocity $v$, such that in a given direction $\Omega$ the energy $E$ in a cell of volume $dV$ is conserved as it propagates radially outward with $r=r_0+vt$. Thus, for a given point in the asymptotic spacetime, $E_{dV}(t,r,\Omega)$ changes over time (at fixed radius) in the same way as it will change as one reduces the radius (at fixed time) with velocity $v$. More precisely, we have that
\begin{equation}
    \frac{d E_{dV}}{d t}=-v \frac{d E_{dV}}{d r}\,.
\end{equation}
It then directly follows that the energy within $dV$ only depends on the particular combination given by the asymptotic retarded time in Eq.~\eqref{eq:asymptoticRet Time v} 
\begin{equation}
    E_{dV}(t,r,\Omega)=E_{dV}(u,\Omega)\,.
\end{equation}
From Eq.~\eqref{eq:DefEdV}, we therefore obtain
\begin{equation}
    T^\text{a}_{00}=\frac{1}{r^2}\frac{dE_{dV}}{drd\Omega}=-\frac{1}{r^2}\frac{1}{v}\frac{dE_{dV}(u,\Omega)}{dud\Omega}\,.
\end{equation}
Eq.~\eqref{eq:rel F to Flux} is then finally reached by changing perspective $dE=-dE_{dV}$ and considering an outward flowing positive energy flux, rather than an energy loss through the surface element $d\Omega$.

On the other hand, the time component of the conservation of the energy-momentum tensor in Eq.~\eqref{eq:Conservation asymptotic EMT} also implies that (see also \cite{maggiore2008gravitational})
\begin{equation}\label{eq:relation partial i EMT}
    \partial_i T_\text{a}^{i0}=-\partial_0 T_\text{a}^{00}=v\,n^i\,\partial_i T^\text{a}_{00}\,,
\end{equation}
where in the last equality we have employed Eq.~\eqref{eq:IdentityEMT deriv}. Using Stockes' theorem for a spherical shell in the radiation zone with unit normal $n_i$ for the outer boundary, we can integrate the above equation and obtain
\begin{equation}\label{eq:RelTi0}
    n_iT_\text{a}^{i0}=v\,T_\text{a}^{00}\,,
\end{equation}
such that by neglecting any transverse components of the energy momentum tensor we can conclude that 
\begin{equation}\label{eq:RelTi0 2}
    T_\text{a}^{i0}=v \,n^i \,T_\text{a}^{00}\,.
\end{equation}
Similarly, the spacial components of Eq.~\eqref{eq:Conservation asymptotic EMT} require
\begin{equation}
    \partial_i T_\text{a}^{ij}=-\partial_0 T_\text{a}^{0j}=-v\,n^j \,\partial_0 T_\text{a}^{00}=v^2\,n^in^j\,\partial_i T_\text{a}^{00}\,,
\end{equation}
such that
\begin{equation}\label{eq:RelTij}
    T_\text{a}^{ij}=v^2\,n^i n^j \,T_\text{a}^{00}\,.
\end{equation}
Combining Eqs.~\eqref{eq:RelTi0 2} and \eqref{eq:RelTij}, thus indeed results in the general formula in Eq.~\eqref{eq:General Form asymptotic EMT}.

\paragraph{Solving the Leading Order Low-Frequency Equation.}

We are now ready to explicitly solve the low-frequency back-reaction equation [Eq.~\eqref{eq:EOMISGR2}]
\begin{align}
    \phantom{}_{\mys{(1)}}G_{\mu\nu}[\delta h]&=\kappa_0T^a_{\mu\nu}\,. \label{eq:EOMISGR2Second}
\end{align}
More precisely, we will solve this equation very generically for any asymptotic energy-momentum tensor $T^a_{\mu\nu}(u,r,\Omega)$ with the properties defined above, in particular Eq.~\eqref{eq:General Form asymptotic EMT}.

First, note that the left-hand-side of this equation has precisely the same structure as the left-hand-side of the propagation equation in Eq.~\eqref{eq:EOMIISGR2} but for the low-frequency perturbation $\delta h_{\mu\nu}$. Thus, by choosing the same appropriate TT gauge choice as in Eq.~\eqref{eq:FirstOrderGaugeFixingGR} also for the low-frequency perturbation, the left-hand-side of the asymptotic back-reaction equation will also be of the form given in Eq.~\eqref{eq:EOMIISGR3}. Technically, the TT gauge choice in Eq.~\eqref{eq:FirstOrderGaugeFixingGR} can only be imposed if the asymptotic energy momentum tensor is traceless, which is not guaranteed. Yet, in this case one can perform a redefinition of the low-frequency field variable to the trace-reversed combination (recall Eq.~\eqref{eq:HarmonicGaugeVariables})
\begin{equation}
    \delta \bar{h}_{\mu\nu}\equiv \delta  h_{\mu\nu}-\frac{1}{2}\eta_{\mu\nu}\delta  h^{t}\,,
\end{equation}
for which $\phantom{}_{\mys{(1)}}G_{\mu\nu}[\delta \bar h]$ will again reduce to a massless wave equation without imposing tracelessness on $\delta h_{\mu\nu}$. This subtlety is only arising due to our tensor field description of gravitational waves, since in the end, the physical radiative TT modes in $ \delta \bar{h}_{\mu\nu}$ and  $\delta h_{\mu\nu}$ outside the source are equivalent. This parallels the discussion for the sourced propagation equation of the high-frequency field in Sec.~\ref{ssSec:Local Wave Equation in GR}. In the following we will therefore disregard this technical detail and write $\delta \bar h_{\mu\nu}=\delta h_{\mu\nu}$, where it is understood that technically only $\delta \bar h^\text{TT}_{\mu\nu}=\delta h^\text{TT}_{\mu\nu}$ is satisfied.

In the radiation zone, the leading order low-frequency equation that we want to solve has therefore the general form
\begin{equation}\label{eq:EOMISGRSecond}
    \Box \delta h_{\mu\nu}=-2\kappa_0 \,T^\text{a}_{\mu\nu}\,,
\end{equation}
where, as discussed, the energy momentum tensor stands for any type of asymptotic contributions from either the gravitational waves $\phantom{}_{\mys{(2)}}t^{\myst{GR}}_{\mu\nu}[h]$ or any perturbative unbound matter contribution $\langle\delta T_{\mu\nu}\rangle$.

We will now explicitly present the procedure to solve Eq.~\eqref{eq:EOMISGRSecond} for field points $(t,\vec{x})$ in the limit of outgoing null rays and show that the result in fact corresponds to a general formula for the tensor displacement memory of GR, that encompasses both null and ordinary memory. 
The basic structure of the arguments laid out below, if reduced to the memory sourced by gravitational waves, are very similar to an alternative understanding of memory arising within the Landau-Lifshitz approach to the Einstein equations in \cite{PhysRevD.44.R2945} (see also \cite{Favata:2008yd,Garfinkle:2022dnm}). However, we want to stress that the philosophy behind the Landau-Lifshitz approach is rather distinct and would in particular not allow for a straightforward generalization to metric theories beyond GR.

First of all, the sourced wave equation [Eq.~\eqref{eq:EOMISGRSecond}] can be solved through the standard retarded Green's function method
\begin{equation}\label{eq:GenSolutionWaveEq}
    \delta h_{\mu\nu}(x)=-2\kappa_0\int d^4 x' G(x-x')\,T^\text{a}_{\mu\nu}(x')\,,
\end{equation}
where
\begin{equation}\label{eq:GreensFunction}
    G(x-x')=-\frac{\delta (t_{\myst{ret}}-t')}{4\pi |\mathbf{x}-\mathbf{x}'|}\,,
\end{equation}
with 
\begin{equation}
    t_{\myst{ret}}\equiv t-\frac{|\mathbf{x}-\mathbf{x}'|}{c}\,,
\end{equation}
the retarded time of the low frequency radiation that is travelling at the speed of light. While Eq.~\eqref{eq:GenSolutionWaveEq} represents the general solution, one still needs to perform a limit to null infinity in order to find an expression that captures the observationally relevant radiative content. 

To do so, it is important to note that while the source $T^\text{a}_{\mu\nu}(x')$ can itself be constructed out of null waves within the radiation zone such that in principle both $|\mathbf{x}|$ and $|\mathbf{x}'|$ are large, there still exists a parametric hierarchy $|\mathbf{x}'|\ll |\mathbf{x}|$ in the integral in Eq.~\eqref{eq:GenSolutionWaveEq}. More precisely, we need to assume that in the past null cone of any point $(t,\vec{x})$ where the low-frequency perturbation is evaluated, the integrated source term is in fact only non-zero in regions that satisfy $|\mathbf{x}'|\ll |\mathbf{x}|$, an assumption whose consistency can be checked retrospectively (see e.g. \cite{Garfinkle:2022dnm}). In other words, we need to ensure that the solution $\delta h_{\mu\nu}(x)$ is still evaluated outside its own source. This is indeed satisfied for all known examples. Switching to spherical coordinates with $\mathbf{x}=r\mathbf{n}$ and $\mathbf{x}'=r'\mathbf{n}'$, this allows an expansion of the form
\begin{equation}\label{eq:largerRel}
    |\mathbf{x}-\mathbf{x}'|\simeq r \left(1+\frac{r'}{r}\mathbf{n}'\cdot \mathbf{n}\right)\,.
\end{equation}
Thus, in the limit to null infinity, the retarded time asymptotes to its well known value
\begin{equation}
     t_{\myst{ret}}\rightarrow u=t-r\,,
\end{equation}
which defines the asymptotic limit in Eq.~\eqref{eq:Limit to Null Infinity}. In order to perform such a limit to null infinity, we should therefore first switch to the appropriate asymptotic coordinates $\{u,r,\Omega=(\theta,\phi)\}$.\footnote{As it is customary in a large part of the gravitational wave community, we will however still use a Minkowski basis $\{t,x,y,z\}$ for the index structure of tensor components.}

On the other hand, we also want to transform to convenient asymptotic coordinates $\{u',r',\Omega'\}$ for the source. These are chosen by recalling that the source energy-momentum tensor instead is a natural function of the asymptotic retarded time 
\begin{equation}\label{eq:AsRetTimeMassive}
    u'=t'-\frac{r'}{v}\,,
\end{equation}
depending on the asymptotic group velocity $v$ of the entity that carries the energy. After the change of variables $\dd^4x'\rightarrow \dd u' r'^2\dd r'\dd^2\Omega'$, the special structure of the asymptotic energy-momentum tensor in Eq.~\eqref{eq:General Form asymptotic EMT} ensures that the only dependence on $r$ and $r'$ is within the Green's function, since Eq.~\eqref{eq:GenSolutionWaveEq} becomes 
\begin{equation}\label{eq:GenSolutionWaveEq2}
    \delta h_{\mu\nu}(x)=-2\kappa_0\int du'\int d^2\Omega '\int_0^\infty dr' G(x-x')\,F(u',\Omega')l'_\mu l'_\nu\,.
\end{equation}

For performing the limit to null infinity, we can therefore only concentrate on the retarded Green's function in Eq.~\eqref{eq:GreensFunction}, which to leading order in these asymptotic retarded coordinates together with the relation in Eq.~\eqref{eq:largerRel} becomes
\begin{equation}\label{eq:Greesfunction Ex}
    G(x-x')=-\frac{\mathcal{V}\,\delta(r'-\mathcal{V}(u-u'))}{4\pi r}
\end{equation}
where we used the identity
\begin{equation}
    \delta(f(x))=\sum_i\frac{\delta(x-x_i)}{|g'(x_i)|}\,,\quad \text{for all roots }x_i\text{ of } f(x)\,,
\end{equation}
and we have defined
\begin{equation}
   \mathcal{V}\equiv \frac{v}{1-v \,\mathbf{n}'\cdot\mathbf{n}}\,.
\end{equation}

Finally, recall that the physical modes of the gravitational radiation is captured by the propagating TT component of the metric perturbation $\delta \hat{h}_{ij}^\text{TT}=\delta h_{ij}^\text{TT}$ of the physical metric (recall Secs.~\ref{sSec:GWGeneration} and \ref{sSec:GWPolGen}). This means that the measurable effect of the tensor null memory of GR is given by a projection of the spatial components of Eq.~\eqref{eq:GenSolutionWaveEq2} onto its transverse-traceless part. We thus only need to consider projection of the components of the asymptotic energy momentum tensor in Eq.~\eqref{eq:General Form asymptotic EMT} onto the TT part of the low-frequency components in a given angular direction $\Omega$
\begin{equation}
    [T^\text{a}_{ij}(u',r',\Omega')]^\text{TT}=\frac{1}{r'^2}\,F(u',\Omega')\,\perp_{ijab}(\Omega) \,v^2\,n'_a n'_b\,,
\end{equation}
where recall that $n'_i=n_i(\Omega')$.
Traditionally, a superscript TT denotes a projection of the free indices with the transverse-traceless projection operator $\perp_{ijab}(\Omega)$ defined in Eq.~\eqref{eq:Projectors} with respect to a given direction $\Omega$. Below we will continue to use this notation.

Considering all of the above while remembering that $F(u',\Omega')$ is related to the energy flux through [Eq.~\eqref{eq:rel F to Flux}]
\begin{equation}
   F(u',\Omega')= \frac{1}{v} \frac{dE}{du'd\Omega'}\,,
\end{equation}
one can now simply plug the expression for the Green's function in Eq.~\eqref{eq:Greesfunction Ex} into Eq.~\eqref{eq:GenSolutionWaveEq}, impose $u'\leq u$ given by the retardation condition of the Green's function, kill the integration over $r'$ and obtain in the limit to null infinity
\begin{align}\label{eq:DispMemoryGR}
    \boxed{\delta h_{ij}^\text{TT}(u,r,\Omega)=\,\frac{\kappa_0}{2\pi r} \int_{-\infty}^udu'\int_{S^2} d^2\Omega'\,\frac{dE}{du'd\Omega'}\,\left[\frac{\perp_{ijab}(\Omega)\,v^2\,n'_a n'_b}{1-v\,\mathbf{n}'\cdot\mathbf{n}(\Omega)}\right]\,.}
\end{align}
Observe that crucially, the result is a well-defined quantity even in the massless case with $v=1$, since if $\mathbf{n}'=\mathbf{n}$ the numerator vanishes as well and the limit is well-defined. Moreover, the integral over asymptotic retarded time $u'$ of the source is bound by the asymptotic retarded time $u$ due to the delta function in Eq.~\eqref{eq:Greesfunction Ex}, which for $r'\geq 0$ selects $u\geq u'$, since $\mathcal{V}>0$ whenever $\mathbf{n}'\neq \mathbf{n}$ and $1\geq v >0$.

Equation~\eqref{eq:DispMemoryGR} represents the general formula for the propagating low-frequency perturbation that is sourced by unbound coarse-grained energy-momentum that escapes the otherwise localized source. Moreover, as promised, this low-frequency contribution precisely corresponds to a memory component in the detector response. This is because for a given anisotropic energy-loss of the source, after performing the angular integral, the expression contains a time integral of over a function of definite sign, which therefore inevitably induces a permanent distortion of proper distance, translating to a permanent displacement of the detector strain. In Sec.~\ref{NullMemory for GW Radiation in GR} below we will explicitly offer a derivation of how the angular integral can be performed in a spin-weighted spherical harmonics decomposition. 

It is however important to realize, that despite the fact that the above derivation of the memory formula solely relied on the backreaction of unbound energy-momentum content focusing on the asymptotic region, the presence of a source or interaction event from which the energy content is ``unbound'' is crucial. Indeed, in \cite{Tolish:2014bka} it was explicitly shown that there is no memory effect for null sources propagating on an unbound null-geodesic. This result can be understood as follows. So far, we have not talked about energy-momentum that is unbound in the limit to past infinity, which must however in principle also be considered. For instance, ordinary memory is also produced in a hyperbolic encounter of an initially unbound objects that gets deflected in an interaction event with another mass. The memory component vanishes however with a vanishing deflection. In that sense, memory arises only if there is an imbalance between unbound objects of past and future infinity either in number or in direction, as otherwise their memory contributions ``cancel-out''. The necessity of such an interaction event or source also justifies the nomenclature of a ``gravitational wave'' memory, as any such interaction event will also be accompanied by a burst of gravitational radiation.

Apart from these remarks, we want to stress that the formula in Eq.~\eqref{eq:DispMemoryGR} is very general and can be applied to gravitational waves emitted by the source, as well as matter waves or matter particles. To illustrate this, we will now consider the most important special cases and recover the formulas in the existing literature.

\paragraph{Gravitational Waves.} 
For the TT gravitational waves of GR, one can simply set $v=1$ to obtain the known non-linear memory formula \cite{Christodoulou:1991cr,PhysRevD.44.R2945,Favata:2010zu}
\begin{align}\label{NonLinDispMemoryGR}
   \delta h_{ij}^\text{TT}&=\frac{\kappa_0}{2\pi}\int_{-\infty}^u du' \int_{S^2} d^2\Omega'\,F_{\myst{GR}}(u',\Omega')\,\left[\frac{\perp_{ijab}(\Omega)\,n'_an'_b}{1-\mathbf{n}'\cdot\mathbf{n}(\Omega)}\right]\,,\nonumber\\
   &=\frac{4G}{r}\int_{-\infty}^u du' \int_{S^2}d^2\Omega'\,\frac{dE_{\myst{GR}}}{du'd\Omega'}\,\left[\frac{n'_in'_j}{1-\mathbf{n}'\cdot\mathbf{n}}\right]^\text{TT}\,,
\end{align}
where, as already mentioned, a superscript TT denotes a projection with $\perp_{ijab}(\Omega)$ onto the TT part, and [Eqs.~\eqref{eq:t00GR} and \eqref{eq:DefFlux GR}]
\begin{equation}\label{eq:DefFlux GR 2}
   F_{\myst{GR}}(u',\Omega')= \frac{dE_{\myst{GR}}}{dud\Omega}=\frac{r^2}{2\kappa_0}\,\big\langle \dot{h}_+^2+\dot{h}_\times^2\big\rangle\,.
\end{equation}
Recall that this energy flux of radiation at speed $c$ can also be related to the energy density
\begin{equation}\label{eq:t00GR}
    \phantom{}_{\mys{(2)}}t^{\myst{GR}}_{00}(u,r,\Omega)=\frac{1}{2\kappa_0}\,\big\langle \dot{h}_+^2+\dot{h}_\times^2\big\rangle
\end{equation}
through
\begin{equation}\label{eq:DefFlux GR}
    c \, \phantom{}_{\mys{(2)}}t^{\myst{GR}}_{00}(u,r,\Omega)=\frac{c}{r^2}F_{\myst{GR}}(u,\Omega)= \frac{1}{r^2} \frac{dE_{\myst{GR}}}{dud\Omega}\,.
\end{equation}
Note as well that the additional factors of $r$ in Eq.~\eqref{eq:DefFlux GR 2} are only necessary due to our definition of the polarization modes $h_{+/\times}$ that keep and explicit $r$ dependence. A similar formula would also hold for any massless radiation of matter fields, such as electromagnetic radiation.

\paragraph{Massive Point Particle.} The displacement memory formula in Eq.~\eqref{eq:DispMemoryGR} however also encompasses the case of a massive unbound particle of mass $M$ with a given asymptotic radial velocity $v_\text{p}$ in a given direction $\Omega_\text{p}$ in the source centered coordinate system. In this case, the trajectory of the particle $\mathbf{x}_\text{p}(t')=r_\text{p}(t')\mathbf{n}_\text{p}$, with $\mathbf{n}_\text{p}=\mathbf{n}(\Omega_\text{p})$
is governed by the equation
\begin{equation}\label{eq:EOMParticle}
    r_\text{p}(t')=v_\text{p} (t'-t_\text{p})=v_\text{p} \,u'+r'-v_\text{p} t_\text{p}\,,
\end{equation}
for some constant $t_\text{p}$ where we used Eq.~\eqref{eq:AsRetTimeMassive}. The energy momentum tensor of such a localized particle is given by [Eq.~\eqref{eq:Particle EMT}]
\begin{equation}
    \phantom{}_{\mys{(2)}}T_{ij}(x')=\frac{M}{\sqrt{1-v_\text{p}^2}} \,v_\text{p}^2\,n'_i n'_j \,\delta^3(\mathbf{x}'-\mathbf{x}_\text{p}(t'))\,.
\end{equation}
Here, we have dropped the explicit averaging over small spacetime scales for simplicity. However, outside of the point like approximation, a macroscopinc distribution of matter would naturally provide a rather smooth energy-momentum tensor that contributes to the background scales. On the other hand, also quantum particles would come with an effective ``size'' through the Heisenberg uncertainty principle and a definition of the corresponding energy would require an averaging over the associated scales (see also \cite{maggiore2008gravitational}). These inherent scales of the unbound energy-momentum would then translate into a characteristic rise-time of the associated memory that represents the natural high-frequency cutoff of the signal. Within the scope of this work, we will however content ourselves with the simple point-like approximation that is required to recover the existing literature.

Upon a change of coordinates to spherical coordinates as well as the asymptotic retarded time $u'$ the energy-momentum tensor also assumes the general form in Eq.~\eqref{eq:General Form asymptotic EMT}
\begin{equation}
    \phantom{}_{\mys{(2)}}T_{ij}(u',r',\Omega')=\frac{1}{r'^2} F(u',\Omega') \,v_\text{p}^2\,n_in_j\,,
\end{equation}
where
\begin{equation}
    F(u',\Omega')=\frac{M}{\sqrt{1-v_\text{p}^2}}\delta(u'-t_\text{p})\delta(\Omega'-\Omega_\text{p})\,,
\end{equation}
To obtain this result we have used the equations of motion in Eq.~\eqref{eq:EOMParticle} to set
\begin{equation}
    \delta(r'-r_\text{p}(u',r'))=\delta(r'-v_\text{p}u'-r'+v_\text{p} t_\text{p})=\frac{\delta(u'-t_\text{p})}{v_\text{p}}\,.
\end{equation}

Plugging this expression into the general memory formula in Eq.~\eqref{eq:DispMemoryGR} we recover the well known result for linear memory \cite{Braginsky:1987gvh,Thorne:1992sdb}
\begin{align}
    \delta h_{ij}^\text{TT}(u,r,\Omega)&=\,\frac{\kappa_0}{2\pi r} \frac{M}{\sqrt{1-v_\text{p}}}\,\Theta(u-t_\text{p})\left[\frac{\perp_{ijab}(\Omega)\,v_\text{p}^2\,n^a(\Omega_\text{p}) n^b(\Omega_\text{p})}{1-v_\text{p}\,n_k(\Omega_\text{p})n^k(\Omega)}\right]\,,\nonumber\\
    &=\,\frac{4G}{r} \frac{M}{\sqrt{1-v_\text{p}}}\,\Theta(u-t_\text{p})\left[\frac{v_\text{p}^2\,n^\text{p}_i n^\text{p}_j}{1-v_\text{p}\,\mathbf{n}_\text{p}\cdot\mathbf{n}}\right]^\text{TT}\,,\label{eq:Particle memory single}
\end{align}
where $\Theta$ is the Heaviside step function.
For a collection of particles of different masses $M_A$ and velocities $v_A$, the contribution will then just correspond to a sum of the above
\begin{align}\label{eq:Patricle contribution memory}
    \delta h_{ij}^\text{TT}(u,r,\Omega)=\frac{4G}{r} \sum_{A}\frac{\pm M_A}{\sqrt{1-v_A}}\,\Theta(u-t_\text{p})\left[\frac{v_A^2\,n^\text{p}_i n^\text{p}_j}{1-v_A\,\mathbf{n}_\text{p}\cdot\mathbf{n}}\right]^\text{TT}\,,
\end{align}
where incoming particles pick up an additional minus sign and it is understood that in this case the radial velocity is negative.

As remarked above, this formula should be thought of as a difference between (gravitationally) unbound objects before and after an interaction event. If there is no interaction at all, the initial and final contributions to the memory cancel each other out. Moreover, observe that the transition from considering massive particles and null matter, hence from ordinary to null memory, is fluent. The memory formula for the emission of a massless particle is given by Eq.~\ref{eq:Particle memory single} with $v_\text{p}=1$ and $E=M/\sqrt{1-v_\text{p}}$. This statement was also studied in an explicit example in \cite{Tolish:2014oda}. From the perspective of our unified derivation of null and ordinary memory in Sec.~\ref{sSec:UnifiedTreatmentof Null and Ordinary Memory}, this fact can of course be readily understood and will also apply to the emission of waves, as we will discuss in Sec.~\ref{sSec:Memory From Massive Fields}.

\subsection{Null Memory of Gravitational Radiation in GR}\label{NullMemory for GW Radiation in GR}

By far the most important \cite{Christodoulou:1991cr,Favata:2008ti} and most interesting memory contribution is however the null memory induced by the gravitational waves themselves [Eq.~\eqref{NonLinDispMemoryGR}] that can also be written as 
\begin{align}\label{NonLinDispMemoryGR 2s}
    \boxed{\delta h_{ij}^\text{TT}=\,\frac{\kappa_0}{2\pi r} \int_{S^2}\dd^2\Omega'\,\mathcal{F}_{\myst{GR}}(u,\Omega')\,\left[\frac{n'_in'_j}{1-\vec{n}'\cdot\vec{n}}\right]^\text{TT}\,,}
\end{align}
where recall that the superscript TT denotes a projection onto the TT component via $\perp_{ijab}(\Omega)$ defined in Eq.~\eqref{eq:Projectors} and where we define the energy per solid angle as
\begin{equation}\label{eq:EperSolidAngleGR}
    \mathcal{F}_{\myst{GR}}(u,\Omega')\equiv\int_{-\infty}^u\dd u'\,\frac{dE_{\myst{GR}}}{du'd\Omega'}=r^2\int_{-\infty}^u\dd u'\,t_{00}^{\myst{GR}}(u',r,\Omega')\,.
\end{equation}
We therefore want to analyze this contribution in more detail and especially further simplify the expression for practical use.

\paragraph{Spin-Weighted Spherical Harmonic Decomposition of Memory.}
It is particularly useful to decompose the GW memory solution into a spin-weighted spherical harmonics expansion (see also Appendix~\ref{App:TTM Expansion}). Such an expansion first requires the definition of the spin-weight $s=-2$ memory quantity (recall Eq.~\eqref{eq:ScalarPt})
\begin{equation}\label{eq:SW Memory quantity GR}
    \delta h(u,r,\Omega)\equiv \delta h_{ij}^\text{TT}\bar m^i\bar m^j=\delta h_+-i\delta h_\times=\sum_{lm}\,\delta h_{lm}(u,r)\,\,_{\mys{-2}}Y_{lm}(\Omega)\,,
\end{equation}
where the complex transverse vector $m_i$ of spin-weight $s=-1$ was defined in Eq.~\eqref{eq:Def mbar}. It is then useful to determine the value of the coefficients $\delta h_{lm}$ of the expansion in spin-weighted spherical harmonics of Eq.~\eqref{eq:SW Memory quantity GR} (see App.~\ref{App:TTM Expansion}). Writing the memory in this alternative form will ultimately also allow a direct comparison to the memory arising from BMS balance laws or within a systematic PN expansion, as we will see in a later section. 

The simplest way to obtain the spin-weighted spherical harmonic coefficients is to first expand the TT-projected term in the square brackets of Eq.~\eqref{NonLinDispMemoryGR 2s} as a geometric series and then transform the result to a symmetric trace-free (STF) basis, which can subsequently be related to the spin-weighted spherical harmonic expansion. In Appendix~\ref{App:TTM Expansion} we gathered a collection of formulas for different multipole expansions and the relations between them, and also introduce the notation we use for STF tensors. 

As we show explicitly in Appendix~\ref{DerivationEq} the transformation of the TT-projected square brackets in Eq.~\eqref{NonLinDispMemoryGR 2s} in terms of STF tensors results in the identity [see also Eq.~(2.34) in \cite{BlanchetPaper}]
\begin{equation}\label{eq:RelBlanchet}
   \left[\frac{n'_in'_j}{1-\vec{n}'\cdot\vec{n}}\right]^\text{TT}=\,\perp_{ijab}\sum_{l=2}^\infty \frac{2(2l+1)!!}{(l+2)!}\,n_{L-2}n'_{\langle abL-2\rangle}\,,
\end{equation}
such that 
\begin{equation}\label{NonLinDispMemoryST}
\begin{split}
    \delta h_{ij}^\text{TT}=\frac{\kappa_0}{2\pi r}&\perp_{ijab}\sum_{l=2}^\infty \frac{1}{l !}\,n_{L-2}\,\frac{2(2l+1)!!}{(l+1)(l+2)}\int_{S^2}\dd^2\Omega'\,\mathcal{F}_{\myst{GR}}(u,\Omega')\,n'_{\langle abL-2\rangle}\,.
\end{split}
\end{equation}
By comparing to the general STF multipole expansion of a rank-2 TT tensor written out in Eq.~\eqref{eq:AExpansionULVL}, we immediately see that the memory only contributes via the electric-parity multipole, namely
\begin{align}
    \delta U_{L}&=\frac{\kappa_0}{8\pi r}\frac{2(2l+1)!!}{(l+1)(l+2)}\int_{S^2}\dd^2\Omega'\,\mathcal{F}_{\myst{GR}}(u,\Omega')\,n'_{\langle L\rangle}\,,\\
    \delta V_{L}&=0\,,
\end{align}
where we have relabeled $ijL-2\rightarrow L$ through multi-index notation.

A change to the pure-spin TT harmonic basis using Eq.~\eqref{eq:AUVlmToULVL} as well as Eq.~\eqref{eq:AYlmToNL} then yields\footnote{The memory computed in a PN expansion assumes precisely this form, as explicitly shown in GR \cite{Blanchet:1992br} (see also \cite{Favata:2010zu}). However, we define the mass multipole without factoring out the $r^{-1}$ dependence. Note as well that we could have obtained Eq.~\eqref{ResU} more directly by using the identity in Eq.~\eqref{eq:AUlmCalc}.}
\begin{equation}\label{ResU}
    \delta U_{l m}=\frac{4\kappa_0}{r}\sqrt{\frac{(l-2)!}{2(l+2)!}}\int_{S^2}\dd^2 \Omega' \, \mathcal{F}_{\myst{GR}}(u,\Omega')\,Y^*_{l m}(\Omega')\,.
\end{equation}
This expression can finally be related to the spin-weighted spherical harmonic expansion through Eq.~\eqref{eq:AUVlmToHlm} to give
\begin{equation}\label{NonLinDispMemoryModesGR}
  \boxed{\delta h^{lm}_{\myst{GR}}(u,r)=\,\frac{1}{r}\sqrt{\frac{(l-2)!}{(l+2)!}}\int_{-\infty}^u\dd u'\int_{S^2}\dd^2\Omega'\,Y^*_{lm}(\Omega')\,r'^2\big\langle \dot{h}_+^2+\dot{h}_\times^2\big\rangle\,,}
\end{equation}
where [Eqs.~\eqref{eq:t00GR} and \eqref{eq:DefFlux GR}]
\begin{equation}
   r'^2\big\langle \dot{h}_+^2+\dot{h}_\times^2\big\rangle = 2\kappa_0 F_{\myst{GR}}(u',\Omega')\,,
\end{equation}
only depends on $u'$ and the angular coordinates.

Furthermore, the angular integral in this expression can be evaluated analytically as a sum of 3$j$ symbols by also expanding the leading-order waves in spin-weighted spherical harmonics
\begin{equation}\label{eq:Leading order waves strain}
    h(u',r',\Omega')=h_{ij}\bar{m}^i\bar{m}^j=h_+-ih_\times=\sum_{l=2}^{\infty}\sum_{m=-l}^l\,h_{lm}(u',r')\,_{\mys{-2}}Y_{lm}(\Omega')\,,
\end{equation}
and applying the identity in Eq.~\eqref{SWSHTrippleInt}, which involves three spin-weighted spherical harmonics. More precisely
\begin{equation}
   \big\langle \dot{h}_+^2+\dot{h}_\times^2\big\rangle=\big\langle |\dot h|^2\big\rangle=\sum_{l_1 m_1}\sum_{l_2 m_2}\big\langle \dot h_{l_1m_1}\dot h^*_{l_2m_2}\big\rangle\,_{\mys{-2}}Y_{l_1m_1}\,_{\mys{-2}}Y^*_{l_2m_2}\,,
\end{equation}
such that using the rule for complex conjugation of the SWSH 
\begin{equation}
    (-1)^{s+m}\, \phantom{}_{\mys{-s}}Y^*_{l-m}=\phantom{}_{\mys s}Y_{lm}\,,\label{CCSWSH H}
\end{equation}
one obtains
\begin{align}\label{NonLinDispMemoryModesGR 2}
  \delta h^{lm}=&\,\frac{1}{r}\sqrt{\frac{(l-2)!}{(l+2)!}}\int_{-\infty}^u\dd u'\sum_{l_1=2}^\infty\sum_{m_1=-l_1}^{l_1}\sum_{l_2=2}^\infty\sum_{m_2=-l_2}^{l_2}(-1)^{m+m_2}r'^2\big\langle \dot h_{l_1m_1}\dot h^*_{l_2m_2}\big\rangle\nonumber\\
  &\qquad\qquad\times \int_{S^2} \dd^2\Omega'\,\,_{\mys{-2}}Y_{l_1m_1}\,_{\mys{2}}Y_{l_2-m_2} \,Y_{l-m}\,.
\end{align}
The angular integral therefore indeed precisely has the form applicable to the relation in Eq.~\eqref{SWSHTrippleInt}, such that we finally arrive at the full expression of the memory component, given a spin-weighted mode-decomposition of the leading order high-frequency wave
\begin{align}\label{NonLinDispMemoryModesGR 3}
  \delta h^{lm}=&\,\frac{1}{r}\sqrt{\frac{(l-2)!}{(l+2)!}}\int_{-\infty}^u\dd u'\sum_{l_1=2}^\infty\sum_{m_1=-l_1}^{l_1}\sum_{l_2=2}^\infty\sum_{m_2=-l_2}^{l_2}(-1)^{m+m_2}r'^2\big\langle \dot h_{l_1m_1}\dot h^*_{l_2m_2}\big\rangle\nonumber\\
  &\times \sqrt{\frac{(2l_1+1)(2l_2+1)(2l+1)}{4\pi}}
  \footnotesize{
\begin{pmatrix}
l_1 & l_2 & l\\
m_1 & -m_2 & -m
\end{pmatrix}
\begin{pmatrix}
l_1 & l_2 & l\\
2 & -2 & 0
\end{pmatrix}}\,,
\end{align}
where the expression of the $3-j$ symbols is only non-zero for
\begin{equation}\label{eq:SelectionRule M}
    m=m_2-m_1\,,
\end{equation}
and
\begin{equation}\label{eq:SelectionRule L}
    |l_1-l_2|\leq l\leq l_1+l_2\,.
\end{equation}

\paragraph{Leading Order Memory for a Non-Precessing CBC.} Assuming a particular source of GWs, namely a non-precessing CBC event, one can obtain a simple leading order expression of the general formula for memory in Eq.~\eqref{NonLinDispMemoryModesGR 3} that allows for additional insight \cite{Favata:2008yd,Favata:2009ii,Favata:2010zu}. This is because for a non-precessing CBC and choosing a coordinate system in which the binary lies in the $x-y$ plane, we have
\begin{enumerate}[(i)]
    \item The modes of the high-frequency wave satisfy
    \begin{equation}\label{eq:SymmOrbitPlaneModes}
        h_{lm}(t)=(-1)^l\,h^*_{l-m}(t)\,;
\end{equation}
\item  The leading order terms of the high-frequency modes are proportional to 
\begin{equation}
    h_{lm}\propto e^{-im\phi(t)}\,,
\end{equation}
where $\phi(t)$ is the leading orbital phase, that to a first approximation coincides with the angle of the spherical coordinate system.
\item The leading order modes of the high-frequency waves are $h_{22}$ and $h_{2-2}$.
\end{enumerate} 
The first condition (i) above follows from the fact that for a non-precessing binary systems in a frame where the orbital plane coincides with the $x-y$ plane, the high-frequency waves enjoy a symmetry under reflection across the orbital plane 
\begin{equation}
    h(t,\theta,\phi)=h^*(t,\pi-\theta,\phi).
\end{equation}
Using the identity in Eq.~\eqref{CCSWSHandthshift}, we therefore have
\begin{equation}
 \sum_{lm}h_{lm} \,_{\scriptscriptstyle-2}Y_{lm}(\theta,\phi)=\sum_{lm}h^*_{lm} \,_{\scriptscriptstyle-2}Y^*_{lm}(\pi-\theta,\phi)=\sum_{lm} (-1)^l h^*_{l-m}\, _{\scriptscriptstyle-2}Y_{lm}(\theta,\phi)\,,
\end{equation}
where in the last equality, we have also relabeled $m\rightarrow -m$, which implies Eq.~\eqref{eq:SymmOrbitPlaneModes}. 

On the other hand, strictly speaking the condition (ii) is only valid in the quasi-circular inspiral phase in the given coordinate system, but remains a good approximation throughout merger as well. More precisely, in the inspiral phase of a non-precessing CBC the SWSH modes can be decomposed as
\begin{equation}\label{decompAph}
    h_{lm}=A_{lm}e^{-i\phi_{lm}}\,,
\end{equation}
where the phases of the modes up to higher post-Newtonian corrections satisfy \cite{Boyle:2014ioa,CalderonBustillo:2015lrg,Varma:2018mmi,Barkett:2019tus}
\begin{equation}\label{PhaseOrbPhaseRel}
    \phi_{lm}\simeq m\phi(t) \,.
\end{equation}
Heuristically, this relation can be understood by comparing Eq.~\eqref{decompAph} with Eq.~\eqref{phRotationsofModes}, together with the fact that for a circular and stable binary system, rotations by $\phi$ around the $z$ axis are the passive transformation counterpart of advancing in the orbital phase $\phi(t)$.\footnote{The orbital phase $\phi(t)$ in a binary system is defined as the angle traced out by the evolution of the lighter object $m_2$ in a coordinate system centered at $m_1$. Defining a center of mass frame whose $x$-axis at $\phi(t)=0$ points from the lighter to the heavier object, $\phi(t)$ is equivalent to the angle traced out by $m_1$ in this coordinate system. Note that if $m_1=m_2$ there is an ambiguity of $\pi$ in defining $\phi(t)$ which is reflected in the fact that in this case modes with odd index $m$ vanish up to a certain approximation.} Observe that this approximate relation therefore decisively relies on the nice rotation properties of $h_{lm}$ ensured by an expansion in SWSH.

Similarly, explicit post-Newtonian computations show that the modes $h_{22}$ and $h_{2-2}$ indeed capture the dominant quadrupole radiation, in the sense that at the dominant order, it gives the only time-varying contribution to the waveform \cite{Kidder:2007rt,Creighton:2011zz,Faye:2012we,Faye:2012xt}.


The conditions (i-iii) above translate into the observation that only the leading order spin-weighted memory modes are given by
\begin{equation}
    \delta h_{20}=\frac{r}{7}\sqrt{\frac{5}{6\pi}}  \int_{-\infty}^u\dd u'\big\langle |\dot h_{22}|^2\big\rangle\,,
\end{equation}
and
\begin{equation}
    \delta h_{40}=\frac{r}{1260}\sqrt{\frac{5}{2\pi}}  \int_{-\infty}^u\dd u'\big\langle |\dot h_{22}|^2\big\rangle\,.
\end{equation}
This follows because, the selection rule in Eq.~\eqref{eq:SelectionRule M}, together with the condition (ii), imply that the leading order memory modes will be proportional to
\begin{equation}
    \delta h_{lm}\propto \big\langle e^{-i(m_1-m_2)\phi(t)}\big\rangle= \big\langle e^{im\phi(t)}\big\rangle\,,
\end{equation}
such that only the $m=0$ modes are non-oscillatory. Hence, all modes with $m\neq 0$ will be highly suppressed by the averaging over the high-frequency regime. Furthermore, the condition (iii) further implies that $l\leq 4$ due to the selection rule in Eq.~\eqref{eq:SelectionRule L}. And finally, condition (i) then ensures that $\delta h_{30}=0$, as the $m_{1,2}=\pm 2$ contributions will cancel each other out.

Plugging the results into the expansion in Eq.~\eqref{eq:SW Memory quantity GR} finally yields
\begin{align}
    \delta h_+(u,r,\theta,\phi)&=\frac{r}{192\pi}\,\sin^2\theta(17+\cos^2\theta)\,\int_{-\infty}^u\dd u'\big\langle |\dot h_{22}|^2\big\rangle\,, \label{MemQuadrupole}\\
    \delta h_\times(u,r,\theta,\phi)&=0\,.
\end{align}
Thus, notably, in the given coordinate system, the memory signal only has a single polarization. This can be understood retrospectively by observing that
\begin{equation}
    \Re  [\delta h] = \delta h_+\,,
\end{equation}
where $\Re$ selects the real part, since $\delta h_{ij}$ and $e_{+/\times}^{ij}$ are both real. On the other hand, if only the memory modes $\delta h_{lm}$ with $m=0$ and even $l$ contribute, then the entire memory component is real
\begin{equation}
    \Re[\delta h]=\delta h_+=\sum_{l} \Re\left[\delta h_{l0}\,_{\mys{-2}}Y_{l0}\right]=\sum_{l}  \delta h_{l0}\,_{\mys{-2}}Y_{l0}=\delta h\,,
\end{equation}
where we have used Eqs.~\eqref{CCSWSHandthshift} and \eqref{eq:SymmOrbitPlaneModes}.

Moreover, observe that Eq.~\eqref{MemQuadrupole} also implies that edge-on systems with $\theta=\pi/2$ are most optimal for memory detection, while the memory effect in the above approximation vanishes in the face-on limit $\theta=0$.

\paragraph{Example Waveform for a Non-Precessing CBC.}

\begin{figure}[H]
\centering
\includegraphics[scale=0.208]{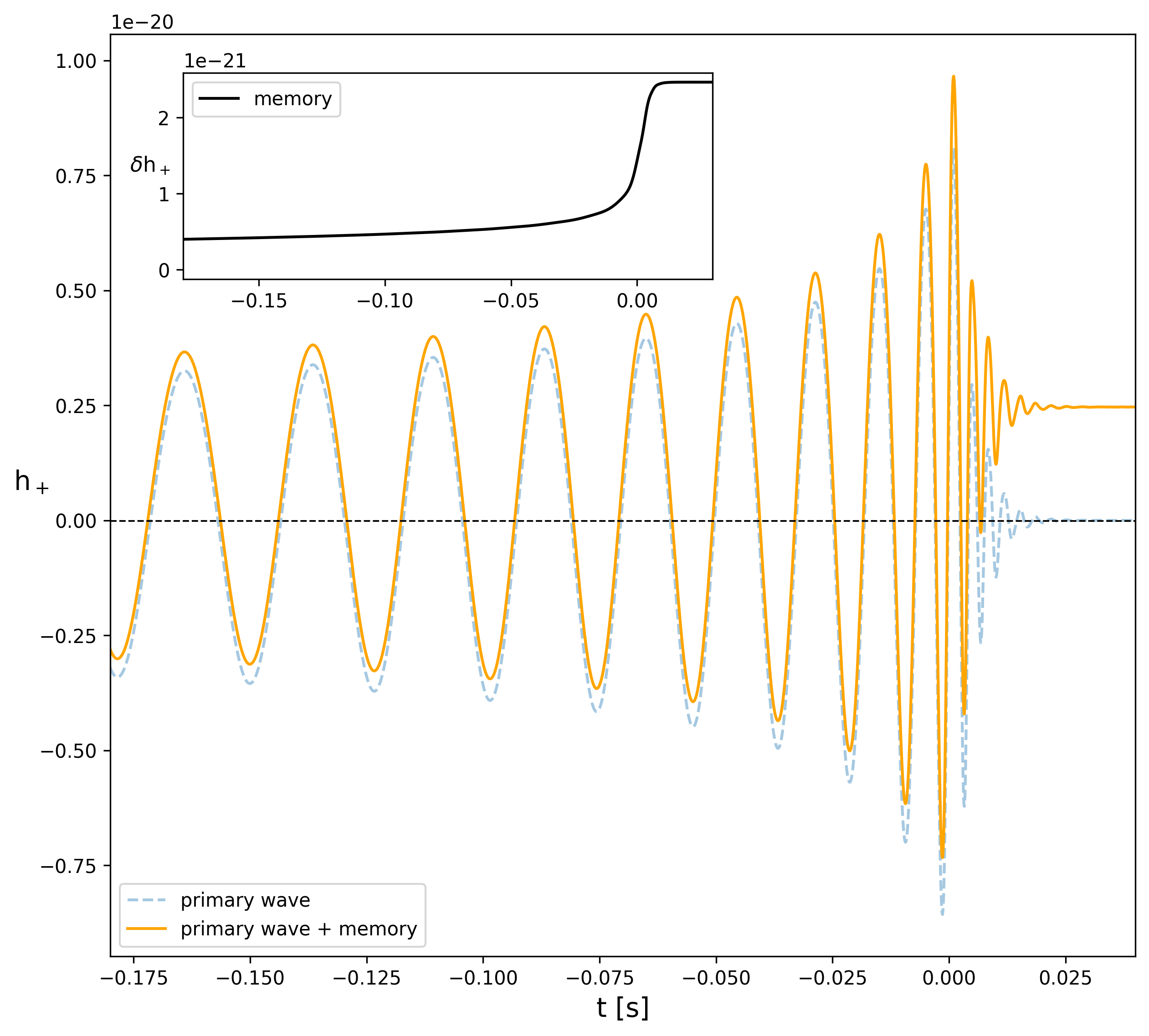}
\caption{\label{fig:MemoryWaveform}\small{The $+$ polarization of the waveform strain $h_+$ [Eq.~\eqref{eq:Leading order waves strain}] without memory (dashed blue) and $h_++\delta h_+$ [Eq.~\eqref{eq:SW Memory quantity GR}] with the memory correction (solid orange), from a non-spinning equal mass binary black hole coalescence of total mass $M_\text{tot}=60\;\text{M}_\odot$, at a luminosity distance $d_L = 20$ Mpc (see Eq.~\eqref{eq:LuminosityDistanceSecond} for a definition) and as viewed from an inclination angle $\theta=\pi/2$ (edge on), where the binary inspiral is lying in the $x$-$y$ plane of the source centered coordinate system. The waveform corresponds to the NR surrogate approximant \cite{Varma:2018mmi}, while the associated memory signal is computed through the publicly available {\fontfamily{qcr}\selectfont python} package in \cite{Talbot:2018sgr}. Top left panel: Strain of the displacement memory $\delta h_+$ in the $+$ polarization.}}
\end{figure}

Especially for precessing binaries with spin, however, the subdominant source modes become important, such that \eqref{MemQuadrupole} has to be updated to include higher order contributions \cite{Talbot:2018sgr}. Through Eqs.~\ref{NonLinDispMemoryModesGR 3} and \ref{eq:SW Memory quantity GR} one can evaluate the memory correction for a favorite waveform model of the primary wave by plugging in the SWSH modes of the model and performing a numerical time integration. In Fig.~\ref{fig:MemoryWaveform} we show the result of such a computation for an example waveform.

The example in Fig.~\ref{fig:MemoryWaveform} illustrates that the memory effect, although naively corresponding to a second order effect, induces a rather non-negligible correction that can plainly be seen by eye. This can partially be understood from Eq.~\eqref{eq:BetaParam}
\begin{equation}
   \beta\sim \alpha^2\frac{f_H^2}{f_L^2}\,,
\end{equation}
which implies that the null memory is enhanced by an additional factor of $f_H^2/f_L^2$ compared to usual second order terms at $\mathcal{O}(\alpha^2)$ that indeed remain negligible. This parallels the observation in \cite{Favata:2008yd}, that, while the hereditary time integral of oscillatory corrections scales with the orbital timescale, the memory scales with the radiation-reaction timescale instead. However, note that an estimate of $\beta$ as given above does not represent a faithful estimate of the amplitude of the memory. Indeed, if the $1/r$ scaling of the memory signal is properly taken into account, its amplitude is boosted up to roughly $10\%$ of the primary wave oscillatory signal as shown in \cite{Thorne:1992sdb}. This estimate is nicely confirmed in Fig.~\ref{fig:MemoryWaveform}.

\subsection{Future Prospects for Memory Detection}\label{ssSec:FutureProspects MemoryDetection}

Excitingly, while the memory effect has not yet been observed, the prospects of its detection are positive and a first observation is believed to be around the corner. Indeed, single event detections are expected both with next generation ground-based observatories \cite{Grant:2022bla,Johnson:2018xly,Islam:2021old,Goncharov:2023woe} such as the Einstein Telescope \cite{Punturo:2010zz,Maggiore:2019uih} and Cosmic Explorer \cite{Reitze:2019iox,Evans:2021gyd}, as well as with space borne experiments \cite{Favata:2009ii,Islo:2019qht,Burko:2020gse,Islam:2021old,Sun:2022pvh,LISA:2022kgy,Gasparotto:2023fcg,Ghosh:2023rbe,Goncharov:2023woe}, in particular the Laser Interferometer Space Antenna (LISA) mission \cite{LISA}. Current GW detectors might have a chance to find evidence of memory through stacking of events \cite{Lasky:2016knh,Boersma:2020gxx,Grant:2022bla} but so far no detection of the memory effect was reported in past observation runs \cite{Hubner:2019sly,Ebersold:2020zah,Hubner:2021amk}, consistent with earlier forecasts \cite{Favata:2009ii,Johnson:2018xly,Yang:2018ceq}. PTA observations might also be able to eventually capture a memory signal, although LISA is expected to be faster in doing so \cite{vanHaasteren:2009fy,Islo:2019qht,NANOGrav:2019vto}.

\paragraph{Characteristics of the Memory Signal.} For a deeper understanding of future prospects of the detectability of the signal in GR, as well as potential future applications, it is instructive to consider some of the important characteristics of the tensor null memory signal. Firstly, from our derivation of the memory signal through the Isaacson approach presented in Sec.~\ref{ssSec:IsaacsonInGR} it is clear that the memory effect as a correction to the $\mathcal{O}(\alpha)$ gravitational waves is fundamentally different from its high-frequency source. Namely, the memory signal as a propagating low-frequency perturbation sourced by the coarse-grained radiative energy-momentum is a component of the low-frequency background of characteristic frequency $f_L$, and therefore parametrically separated from the high-frequency signal of $f_H$ in time-frequency space according to Eq.~\eqref{eq:hierarchy-inequ}. This fact provides a key principle for the extraction of the memory signal as clearly distinguishable from its high-frequency counterpart. 

However, given that realistic detectors are only sensitive to a bounded frequency interval, the separation in characteristic frequencies between the memory signal and its primary wave ought not to be too large, as otherwise a simultaneous detection in a single type of detector would fail. Luckily, the parametric separation for the realistic scenario of equal mass CBCs of total mass $M_\text{tot}$ for example is minimal, as can already be seen through an estimation of the maximal frequencies of typical events. Indeed, recall that the high-frequency scale $f_H$, which in this case can be approximated with the frequency at merger, is estimated through Eq.~\eqref{eq:frequency estimate}
\begin{equation}
    f_H\sim 10^4\left(\frac{\text{M}_\odot}{M_\text{tot}}\right) \,\text{Hz}\,,
\end{equation}
respectively corresponding to $f_H \approx 10^2$ Hz and $f_H \approx 10^{-1}$ Hz for ground- and space-based detectors sensitive to $10^2 \; \text{M}_\odot$ and $10^5\; \text{M}_\odot$ (total) mass binaries. On the other hand, the characteristic frequency of the memory can be estimated from the inverse of the rise time of the memory at merger $t_\text{rt}$ for the given total mass of the binary, which from Fig.~\ref{fig:MemoryWaveform} can be approximated as
\begin{equation}
\frac{t_\text{rt}}{M_\text{tot}}\sim \frac{0.05 \,\text{s}}{60 \, \text{M}_\odot}\sim 10^{-3}\text{ to } 10^{-2} \;\frac{\text{s}}{ \text{M}_\odot}\,,
\end{equation}
such that
\begin{equation}
    f_L\sim 10^{3}\text{ to } 10^{2}  \left(\frac{\text{M}_\odot}{M_\text{tot}}\right) \,\text{Hz}\,.
\end{equation}
Hence, we have that $f_L \approx 10$ to $1$ Hz and $f_L \approx 10^{-2}$ to $10^{-3}$ Hz for ground- and space-based detectors, respectively. 

\paragraph{Detector Sensitivity to Memory.} Therefore, a memory signal from a given CBC can in principle be in-band of the same detector targeting the high-frequency primary wave. The question is however still whether a given detector is sensitive enough, in other words has a low enough noise level, that a memory signal can be extracted. Moreover, since a tensor null memory signal always comes with its high-frequency counterpart, the memory effect will also need to be distinguished from the primary signal. As discussed, this might well be achieved through a separation in time-frequency space. 

In this context, it is however important to realize, that the amplitude of a signal in the time series is not the decisive factor for the detectability of a signal. Rather, a given signal needs to be compared to the noise level of the detector. More precisely, in the time series of a detector output $s(t)$ there will be a noise component $n(t)$, that we assume here to be stationary. On the other hand, for a given detector in a metric theory of gravity, the relevant time series of the signal in a given direction and a given distance is provided by the projection of each of the six gravitational polarizations with the corresponding detector pattern functions that we defined in Eq.~\eqref{eq:Signal in Detector}
\begin{equation}\label{eq:Signal in Detector Mem}
P(t)= F_+\,h_++F_\times\,h_\times+F_u\,h_u+F_v\,h_v+F_b\,h_b+F_l\,h_l\,,
\end{equation}
where the detector pattern functions for a perpendicular quadrupole detector where explicitly computed in Eq.~\eqref{eq:DetectorPattern}. For GR, of course, only the first two tensor TT polarizations are present. Moreover, recall that concentrating on \textit{tensor} null memory precisely also means that all additional polarization contributions are neglected. The question of detectability of gravitational radiation is therefore more precisely stated in terms of the extraction of the detector signal $P(t)$, that includes a potential memory signal, from the noisy detector response
\begin{equation}
    s(t)=P(t)+n(t)\,.
\end{equation}

Now, the noise $n(t)$ whose characterization via the ensemble average\footnote{Since in practice one does not have access to multiple realizations of the noise, the ensemble average needs to be replaced by a time average of the stationary noise.} of its Fourier components
\begin{equation}
    \langle \tilde n(f)\tilde n^*(f')\rangle =\frac{1}{2}\delta(f-f')\,S_n(f)\,,
\end{equation}
has the crucial property of being diagonal in frequency space (see also Sec.~\ref{sSec:Initial Conditions} below). This is true as long as the noise is stationary, since different Fourier modes are uncorrelated. Here, $S_n(f)$ defines the so-called \textit{power spectral density} (PSD) of dimension Hz$^{-1}$, and we define the time-frequency Fourier transform of a time series $x(t)$ as
\begin{equation}
    \tilde x(f)\equiv \int_{-\infty}^\infty dt \,x(t)\,e^{-2\pi i\,ft}\,,
\end{equation}
with inverse Fourier transform
\begin{equation}
   x(t)= \int_{-\infty}^\infty dt \,\tilde x(f)\,e^{2\pi i\,ft}\,.
\end{equation}
Since the detector time-series are real, we have that $\tilde x(-f)=\tilde x^*(f)$ such that the PSD is an even function
\begin{equation}
    S_n(f)=S_n(-f)\,,
\end{equation}
therefore called a \textit{one-sided} PSD that allows the conversion of the full Fourier integrals to integrals over positive frequencies only.

Because of this property, the question of detectability of a signal in the detector response is conveniently addressed in Fourier space. More precisely, the problem of extracting a signal with known theoretical shape, from a noisy detector output has a known optimal solution through the constriction of a so-called \textit{Wiener optimal filter} \cite{Wiener:1949}. This allows the introduction of a \textit{signal-to-noise ratio} (SNR) $\rho_{\myst{SNR}}$ that measures the ratio between the filtered response in presence and absence of the true signal. An optimal filter is then such that it maximizes the signal-to-noise-ratio, a technique that is accordingly also known as \textit{matched filtering}. The final result for the SNR of the Wiener optimal filter constructed out of the Fourier transform of the signal template $\tilde P(f)$ and the noise PSD $S_n(f)$ is \cite{maggiore2008gravitational,Creighton:2011zz,Moore:2014lga,YunesColemanMiller:2021lky}
\begin{equation}\label{eq:SNR def}
    \rho_{\myst{SNR}}^2=\int_0^{\infty} df\,\frac{4\,|\tilde P(f)|}{S_n(f)}\,.
\end{equation}
A threshold of an SNR above unity, usually around $3$, is then chosen as a practical definition for detectability of a given signal.

\begin{figure}
    \centering
    \includegraphics[width=0.85\textwidth]{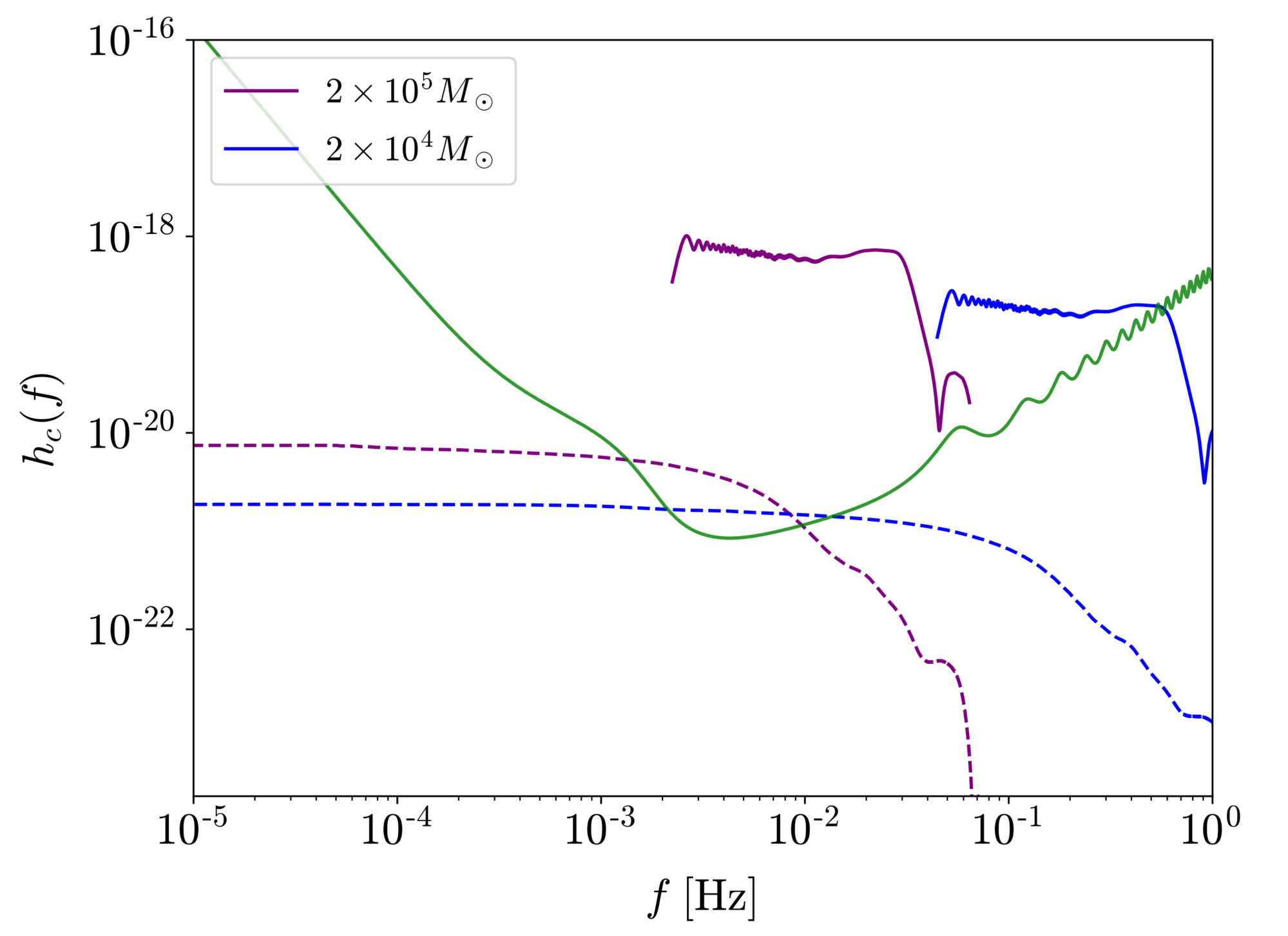}
    \caption{\small Preliminary assessment of the detectability of memory with LISA through a comparison of the characteristic strain of memory (dashed lines) in Eq.~\eqref{eq:cs of memory} with the sensitivity [Eq.~\eqref{eq:Noise amplitude based sensitivity}] of LISA (green solid line). The memory is shown for two representative equal mass and non-spinning binary black hole coalescence's of (purple) total mass $M_\text{tot}=2\cdot 10^4\;\text{M}_\odot$ at redshift $z=0.5$ (see Sec.~\ref{sSec:HomIsoUniverse} below for a definition of redshift and its relation to distance) and (blue) $M_\text{tot}=2\cdot 10^4\;\text{M}_\odot$ at $z=2$, as seen from a fixed direction $\theta = 40^{\circ}$ and $\phi=0$ with the binary lying in the $x$-$y$ plane of the source centered coordinate system. The solid lines correspond to the characteristic strain $h_c(f)\equiv 2f \,|\tilde h_+-i\tilde h_\times|$ of the high-frequency radiative signal emitted by the last 25 cycles before merger as modeled by the NR surrogate approximant \cite{Varma:2018mmi}, while the dashed lines represent $\delta h_c(f)$ of the associated memory signal of Eq.~\eqref{NonLinDispMemoryGR 2s} computed through the publicly available {\fontfamily{qcr}\selectfont python} package in \cite{Talbot:2018sgr}. (Figure taken from \textit{S. Gasparotto et al., (2023)} \cite{Gasparotto:2023fcg})}
    \label{fig:Memory With LISA}
\end{figure}

However, the quantification of the amplitude of gravitational radiation in the detector through the comparison of the Fourier transform of the detector signal $|\tilde P(f)|$ and the noise PSD does not capture the crucial property that while an instantaneous amplitude might reside orders of magnitude below the noise level, an integration of the SNR over time might lift the SNR to a detectable level. It is therefore useful to define an alternative characterization of the SNR that accounts for this effect. This is accomplished through the introduction of the so-called \textit{characteristic strain} $P_c(f)$ alongside its noise counterpart given by the \textit{noise amplitude} $P_n(f)$, defined as \cite{Moore:2014lga}
\begin{equation}
    P_c(f)\equiv 2\,f\,|\tilde P(f)|\,,\qquad P^2_n(f)\equiv f\,S_n(f)\,,
\end{equation}
where both the characteristic strain and the noise amplitude are dimensionless.
Observe that in terms of these variables, the SNR in Eq.~\eqref{eq:SNR def} can be written as
\begin{equation}
    \rho_{\myst{SNR}}^2=\int_{-\infty}^{\infty} d(\log f)\,\left(\frac{ P_c(f)}{P_n(f)}\right)^2\,.
\end{equation}
This implies that plotted in a log-log scale the area between the curves of the characteristic strain of the signal and the detector noise amplitude are directly related to the SNR. In other words, the height of the signal above the noise level can be used to gain an intuition of the corresponding SNR.

\paragraph{Detectability Estimates for Memory.} In principle, assessing detectability of the memory signal therefore requires the computation of the Fourier transform of the memory signal within a given detector and the evaluation of its characteristic strain. To obtain an estimate for the memory SNR within a given detector, it is however common practice to further define an SNR measure that is averaged over the sky localization as well as the polarizations. Accounting for the sky averages over the detector pattern functions $\langle F_\lambda \rangle$ (see e.g. \cite{Babak:2021mhe}) within the noise PSD defines the detector \textit{sensitivity}
\begin{equation}
    S_s(f)=\frac{S_n(f)}{\langle F_\lambda \rangle}\,.
\end{equation}
and equivalently the noise amplitude based sensitivity 
\begin{equation}\label{eq:Noise amplitude based sensitivity}
    P_s(f)=\sqrt{\frac{f S_n(f)}{\langle F_\lambda \rangle}}\,.
\end{equation}
This implicitly takes into account the projection in terms of detector pattern functions, such that for the computation of a sky-averaged SNR one can directly employ the polarization modes $\delta h_+$ and $\delta h_\times$ within the spin-weighted scalar quantity $\delta h= \delta h_+-ih_\times$ [Eq.~\eqref{eq:SW Memory quantity GR}].

In Fig.~\ref{fig:Memory With LISA} an example of such a preliminary estimate for the detectability of memory within LISA is plotted for two representative non-spinning equal mass binary black hole coalescence's. Obviously, in such an estimate for LISA, a sufficiently powerful global fit \cite{Cornish:2005qw,Vallisneri:2008ye,Littenberg:2023xpl} is assumed, that is able to simultaneously distinguish between the mixture of sources present in the detector. Observe that the scales of the maximal frequencies corresponding to $f_H$ and $f_L$ of the high-frequency primary signal and the memory respectively match the order of magnitude estimates provided above.

To gain an intuition for the memory signal and its characteristic strain in Fourier space shown in Fig.~\ref{fig:Memory With LISA}, it is useful to think of it as an approximation of a step function. Indeed, the Fourier transform of the Heaviside step function is well known and scales like the inverse of the frequency $\sim 1/f$. Therefore, in a first approximation, the characteristic strain of the memory effect
\begin{equation}\label{eq:cs of memory}
    \delta h_c(f)\equiv 2 f\,|\delta\tilde h|\,,
\end{equation}
is given by a horizontal line in frequency space and therefore corresponds to a signal of equal characteristic amplitude on the entire frequency span. However, and crucially, the departure from the step function captured by the scale of the rise time $t_\text{rs}$ of the memory signal effectively introduces a maximal frequency up to which the memory effect contributes and beyond which the characteristic strain drops sharply (see Fig.~\ref{fig:Memory With LISA} and also \cite{Favata:2009ii}). Thus, the detectability of memory in a first approximation is determined by the height of the characteristic strain related to the amplitude of the memory effect in terms of its saturation value, and the scale of its maximal frequency corresponding to $f_L$ that is determined through the inverse rise-time.


\section{Gravitational Wave Memory beyond GR}\label{sSec:GW Memory beyond GR}

\small{\textit{Selected parts of this section are taken over from the original work \cite{Heisenberg:2023prj} of the author. Based on this remark, we will refrain from introducing explicit quotation marks to indicate direct citations.}}
\\

\normalsize
\noindent
The Isaacson approach to understanding and computing displacement memory that we successfully applied in the previous section in the context of GR, can equally be used to examine memory within all types of metric theories, thanks to our generalization of the Isaacson approach presented in Sec.~\ref{ssSec:IsaacsonGeneral}. Based on these assumptions together with a notion of Lorentz-preserving asymptotically flat spacetimes discussed in Sec.~\ref{sSec:Asymptotic Flatness} we will in fact be able to prove a theorem for the functional form of null memory in metric theories of gravity. The most important conclusion of the result will be that the functional form of the tensor null memory is only modified through the presence of additional contributions to the null flux $F_{\Psi}(u',\Omega')$ from extra dynamical degrees of freedom that might be excited in a given GR emission event
\begin{align}\label{NonLinDispMemory Thm pre}
    \boxed{\delta  h_{ij}^{\text{TT}}(u,r,\Omega)=\,\frac{\kappa_\text{eff}}{2\pi r}\int_{-\infty}^u du' \int_{S^2} d^2\Omega'\,\left[F_{\myst{GR}}+F_{\Psi}\right](u',\Omega')\,\left[\frac{n'_in'_j}{1-\vec{n}'\cdot\vec{n}}\right]^\text{TT}\,,}
\end{align}
This is true, regardless of whether the extra DOFs do or do not excite additional polarizations of the physical metric. As such, gravitational wave memory might be used in the future to not only complement current searches for additional polarizations but extend them to any type of radiative modes beyond GR. More precisely, an independent extraction of the memory signal of a given GW event might be used as a universal tool for the search for the prime signature of beyond GR effects: the additional propagating degrees of freedom in the gravitational sector.

Indeed, the most important aspect of the result in Eq.~\eqref{NonLinDispMemory Thm pre} can be considered to be the statement that the functional form of the memory is \textit{only} modified by the presence of additional energy fluxes that \textit{must} be associated to additional degrees of freedom in the theory, which themselves will not contribute to the TT polarizations in the primary radiation. In other words, a modification of the theory that is not associated to the presence of additional degrees of freedom does not alter the functional form of memory. The potential implication of this result for future memory based tests of GR will further be discussed in  Sec.~\ref{eq:Summary and Outlook}.

On the other hand, Theorem~\ref{Theorem1} also serves as a guide for the computation of displacement memory in a wide class of metric theories of gravity. While a direct computation of memory in specific metric theories of gravity of course does not need to rely on the theorem, the formulation of this general result will allow us to explore the boundaries of the validity of the associated functional form of tensor null memory, thus identifying potentially interesting cases for a generalization of the results beyond the null memory.

For simplicity, the Theorem~\ref{Theorem1} will primarily focus on the tensor null memory sourced by gravitational and non-minimal null radiation. This is motivated by the expectation that for binary coalescence's the ordinary memory, associated with unbound massive objects of the system, will generally be subdominant, a statement that is confirmed both in GR \cite{Christodoulou:1991cr,Favata:2008ti} and in Brans-Dicke theory \cite{tahura_gravitational-wave_2021}. However, the statements of the theorem are by no means bound to tensor null memory and the massive case will be discussed in the context of a concrete example theory beyond GR in Sec.~\ref{sSec:Memory From Massive Fields}.

In the next subsection [Sec.~\ref{sec:StatementOfClaim}], we will offer an outline of the rationale behind the proof of the memory Theorem~\ref{Theorem1}, while the full proof can be found in the Appendix~\ref{App:ProofOfMemoryTheorem}. To do so, we will first introduce relevant technical tools as well as present Lemma~\ref{LemmaP} that lie at the core of the theorem. In the subsequent subsection [Sec.~\ref{sSec:Analysis of Theorem}] we will discuss the scope of the theorem in more detail, and in particular present its workings for the null memory of metric theories with an arbitrary number of additional non-minimal $k$-form fields (to be defined below). A later section [Sec.~\ref{sec:SVTMem}] will then be devoted to the concrete example of the SVHH gravity, whose degrees of freedom and polarization content we already examined back in Sec.~\ref{sSec:GWPolExample}. This will in particular also demonstrate the use of the theorem to obtain memory formulas for concrete theories beyond GR. However, it will also give us a chance to generalize the statements of Theorem~\ref{Theorem1} beyond the tensor null memory and also address the questions of memory sourced by massive fields as well as scalar and vector memory of different polarization type (recall the definitions in the introduction to this chapter). Finally, in Sec.~\ref{MatchToAsymptoticsBD} we will make contact with other works on memory beyond GR that have been carried out within the special case of Brans-Dicke theory, that will serve as a valuable consistency check of the results in Theorem~\ref{Theorem1}.

\subsection{A Theorem for Memory of Metric Theories}\label{sec:StatementOfClaim}

As mentioned, before stating the precise form of the theorem, we first want to present additional insight into the computation of the leading order evolution equations of the Isaacson approach, in particular the computation of the effective energy-momentum tensors that govern the low-frequency equation. For this we will introduce the so-called \textit{second-variation} approach that was already considered for instance in \cite{Maccallum:1973gf} (see also \cite{Stein:2010pn}).

\paragraph{The Second-Variation Approach.}\label{sec:IsaacsonSecondVariation}
The second-variation framework is primarily a tool to compute the energy-momentum tensor for gravitational fields through the variation of an effective action, in a similar way it is possible to do so for matter fields (recall Eq.~\eqref{eq:DefEnergyMomentumTensor}). As discussed, in general this is not possible for gravitational and non-minimal fields. More precisely, for the physical metric it is fundamentally not possible to define a local energy-momentum tensor due to the Einstein equivalence principle. The same is true for additional non-minimal fields in metric theories, as by definition their non-minimal coupling to the physical metric prevents from an unambiguous definition of a corresponding energy-momentum tensor (see Sec.~\ref{sSec:Covariant Consrevation}). 

However, recall that in the context of perturbation theory (see Sec.~\ref{sSec:PerturbationTheory}) together with Isaacson assumptions, one can unambiguously define the energy momentum tensor of high-frequency metric and non-minimal perturbations through a split of a parametric separation between high-frequency (short-wavelength) perturbations and slowly varying field components. In terms of the physical metric, as well as all other non-minimal fields $\Psi$, in our notation such a split reads [Eq.~\eqref{eq:IsaacsonSplitGen}]
\begin{equation}\label{eq:IsaacsonSplitGen Thm}
   g_{\mu\nu}=g^L_{\mu\nu}+\delta g^H_{\mu\nu}\,,\quad \Psi=\Psi^L+\delta\Psi^H\,,
\end{equation}
where the slowly-varying background fields $g^L_{\mu\nu}$ and $\Psi^L$ admit a further split into the exact solution of the perturbative approach and a corresponding low-frequency perturbation [Eq.~\eqref{eq:IsaacsonSplitGen2}]
\begin{equation}\label{eq:IsaacsonSplitGen2 Thm}
     g^L_{\mu\nu}=\bar{g}_{\mu\nu}+\delta g^L_{\mu\nu}\,,\quad \Psi^L=\bar{\Psi}+\delta\Psi^L\,.
\end{equation}
We want to remark at this point, that while we will mostly concentrate here on the gravitational action with the corresponding metric and non-minimal fields, the arguments here would equally go through when considering matter fields.

The second-variation approach then asserts that under such conditions, the coarse-grained energy-momentum tensors for the high-frequency perturbations can be computed through a variation of an effective action with respect to the slowly-varying metric component $g^L_{\mu\nu}$ that is temporarily treated as an independent field. This is much like the strategy of defining a well-defined energy-momentum tensor for matter fields on a Minkowski background through Eq.~\eqref{eq:DefEnergyMomentumTensor}.

Concretely, in the second-variation method the action $S$ of a given metric theory is first expanded to second order in high-frequency perturbation fields and is then promoted to an effective one by treating the slowly-varying background fields $g^L_{\mu\nu}$ and $\Psi^L$ and the high-frequency perturbations as independent fields
\begin{equation}\label{eq:second variation action}
    S_\text{eff}\equiv S_\text{eff}[g^L,\Psi^L]+\phantom{}_{\mys{(2)}}S_\text{eff}[\delta g^H,\delta\Psi^H]\,.
\end{equation}
In defining $S_\text{eff}$ we already omitted the piece $\phantom{}_{\mys{(1)}}S_\text{eff}[\delta g^H,\delta\Psi^H]$ that is linear in high-frequency fields, since it can be neglected without loss of generality as we will show below.
Moreover, recall that a subscript $\phantom{}_{\mys{(i)}}O[\delta p]$ denotes the $i$th order in the perturbative expansion of an operator $O$ evaluated on the perturbation fields $\delta p$. Thus, in Eq.~\eqref{eq:second variation action} the perturbed action is only evaluated with respect to the high-frequency perturbations $\delta g^H$ and $\delta\Psi^H$, while $g^L_{\mu\nu}$ and $\Psi^L$ serve as implicit background fields. Then, the leading order effective energy-momentum tensor of the high-frequency fields [Eq.~\eqref{eq:DefPseudoEMTensorBeyondGR}] can be computed in analogy to Eq.~\eqref{eq:DefEnergyMomentumTensor} as
\begin{equation}\label{eq:SecondOrderVariationEMTensor}
    \boxed{\frac{-2}{\sqrt{-\bar g^L}}\Bigg\langle\frac{\delta \phantom{}_{\mys{(2)}}S_\text{eff}}{\delta  g_L^{\mu\nu}}\Bigg\rangle\equiv \phantom{}_{\mys{(2)}}t_{\mu\nu}[\delta g^H,\delta\Psi^H]}
\end{equation}
upon averaging out the small scales and where a subsequent replacement of the background metric with its true value is understood. Note that to leading order, this effectively corresponds to a replacement $g^L_{\mu\nu}=\bar g_{\mu\nu}$, since the presence of any additional low-frequency perturbation on top of the two high-frequency perturbations would necessarily be of higher order.

However, the second-variation method not only serves as a definition of a high-frequency energy-momentum pseudo-tensor, but can in fact be used to derive the entire low-frequency metric equation [Eq.~\eqref{eq:EOMIS2}]
\begin{align}
   _{\mys{(1)}}\mathcal{G}_{\mu\nu}[\delta g^L,\delta\Psi^L]=\kappa_0\,\phantom{}_{\mys{(2)}}t_{\mu\nu}[\delta g^H,\delta\Psi^H]\,,   \label{eq:EOMIS2 Thm}
\end{align}
through a leading order variation of the effective action 
\begin{equation}\label{eq:MemEqAction}
       \left[\frac{\delta S_\text{eff}}{\delta  g_L^{\mu\nu}}\right]^L=0\,.
\end{equation}
This is because a variation of the zeroth-order action $S_\text{eff}[g^L,\Psi^L]$ in Eq.~\eqref{eq:second variation action} of course recovers the full metric field equations of the theory but in terms of the background fields $\mathcal{G}_{\mu\nu}[g^L,\Psi^L]$. Due to the split in Eq.~\eqref{eq:IsaacsonSplitGen2 Thm} the corresponding leading order term is then however provided by the first order of the equation operator in low-frequency perturbations, since by definition, the background solution $\{\bar g_{\mu\nu},\bar\Psi\}$ solves the equations of motion. Thus, the leading order term of the variation of the background action in Eq.~\eqref{eq:MemEqAction} indeed recovers the left-hand side of Eq.~\eqref{eq:EOMIS2 Thm}
\begin{equation}
    \mathcal{G}_{\mu\nu}[g^L,\Psi^L]=\underset{=0}{\underbrace{\mathcal{G}_{\mu\nu}[\bar g,\bar\Psi]}}+\phantom{}_{\mys{(1)}}\mathcal{G}_{\mu\nu}[\delta g^L,\delta\Psi^L]+...\,.
\end{equation}
On the other hand, the linear term of the second variation action $\phantom{}_{\mys{(1)}}S_\text{eff}[\delta g^H,\delta\Psi^H]$ is indeed irrelevant as any operator with only one instance of high-frequency perturbation fields will vanish upon a restriction to the low-frequency equations. From that point of view, the effective action of the second-variation approach in Eq.~\eqref{eq:second variation action} can be viewed as a gravitational action of the low-frequency fields, with $\phantom{}_{\mys{(2)}}S_\text{eff}[\delta g^H,\delta\Psi^H]$ playing the role of a matter action that provides the effective energy-momentum tensor of the dynamical equation [Eq.~\eqref{eq:EOMIS2 Thm}] of the a priori unknown slowly-varying background fields.

On the other hand, the effective action can also be used to derive the leading order high-frequency equations [Eqs.~\eqref{eq:EOMIIS2} and \eqref{eq:EOMIIS2Psi}]
\begin{align}\label{eq:EOMIIS2 Thm}
   \phantom{}_{\mys{(1)}}\mathcal{G}_{\mu\nu}[\delta g^H,\delta\Psi^H]=0\,,\quad \phantom{}_{\mys{(1)}}\mathcal{J}[\delta g^H,\delta\Psi^H]=0\,,
\end{align}
through
\begin{equation}\label{eq:PropEqAction}
       \left[\frac{\delta S_\text{eff}}{\delta \delta g_H^{\mu\nu}}\right]^H= \frac{\delta \phantom{}_{\mys{(2)}}S_\text{eff}}{\delta \delta g_H^{\mu\nu}}=0\,,\quad  \left[\frac{\delta S_\text{eff}}{\delta \delta\Psi^H}\right]^H=\frac{\delta \phantom{}_{\mys{(2)}}S_\text{eff}}{\delta \delta\Psi^H}=0\,.
\end{equation}
The correspondence between Eq.~\eqref{eq:PropEqAction} and Eq.~\eqref{eq:EOMIIS2 Thm} is ensured, because very generally a variation of a perturbed action with respect to a perturbation field yields the same equation that one obtains by perturbing the total field equations computed from the full action \cite{Maccallum:1973gf,Taub1971}.
Moreover, note that in this case we can again safely neglect any linear piece $\phantom{}_{\mys{(1)}}S[\delta g^H,\delta\Psi^H]$ in high-frequency perturbations, since a variation of this term would not contain any high-frequency components.\footnote{In fact, a variation of $\phantom{}_{\mys{(1)}}S_\text{eff}[\delta g^H,\delta\Psi^H]$ with respect to a high-frequency perturbation field would simply give back the corresponding background equation, and therefore, does not contain any additional information.}
In the logic alluded to above, the high-frequency propagation equations simply correspond to the equations of motion of the effective ``matter fields''.

\paragraph{The Philosophy of the Theorem and a First Lemma.} The second-variation approach discussed above allows for a crucial insight that will allow us to prove a general functional form of the displacement memory merely based on an assumption on the first order equations of motion of metric theories. Namely, the low-frequency (memory) equation [Eq.~\eqref{eq:EOMIS2 Thm}], in particular the coarse-grained energy momentum tensor of high-frequency fields only depends on the second-order effective action, which also governs the leading order high-frequency propagation equations [Eq.~\eqref{eq:EOMIIS2 Thm}]. This is a non-trivial statement to the extent that the energy-momentum tensor appearing in Eq.~\eqref{eq:EOMIS2 Thm} is a second order quantity in perturbations, which naively would not be captured by a second order action. However, the second variation method shows that the low-frequency, averaged, portion of the second-order perturbation equations of motion in fact still only depends on $\phantom{}_{\mys(2)}S_\text{eff}[\delta g^H,\delta\Psi^H]$ that naively should only govern the linear equations in high-frequency perturbations. This statement can readily be verified for concrete theories, in particular for GR.

For the memory theorem, we will concentrate ourselves in asymptotically flat spacetimes that admit a natural flat background Minkowski solution $\{\eta_{\mu\nu},\bar\Psi\}$ (see Sec.~\ref{sSec:Asymptotic Flatness}). Recall that by flat [Def.~\ref{DefFlatness}], one requires a background vacuum solution of the physical metric with vanishing curvature, given by the Minkowski metric. Based on our assumptions of a vanishing torsion and non-metricity in the connection, this implies the existence of a preferred set of asymptotic (source centered) Minkowski coordinates, in which the Minkowski metric has the Minkowski form $\eta_{\mu\nu}$ for which in particular the Christoffel symbols vanish. In the following, we will exclusively choose such a preferred Minkowski chart. Furthermore, we will require the asymptotic background to preserve Lorentz invariance also in the gravitational sector. To ensure this, we will simply assume that the background values of any non-minimal field and their derivatives vanish $\partial_\alpha\bar\Psi=\bar\Psi=0$, except for scalar fields, which are allowed to retain a non-zero constant asymptotic background value $\bar\Psi=$ constant.

As a preparation for the theorem, we now want to remark that in the limit to null infinity, we can formulate a slightly stronger statement than Eq.~\eqref{eq:SecondOrderVariationEMTensor}, namely, that the asymptotic energy momentum tensor of the high-frequency fields only depends on the \textit{flat} second-order effective action 
\begin{equation}\label{eq:flatSA}
    \phantom{}_{\mys(2)}S^{\myst{flat}}_\text{eff}[\delta g^H,\delta\Psi^H]\equiv \phantom{}_{\mys(2)}S_\text{eff}[\delta g^H,\delta\Psi^H]\big\lvert_{(g^L_{\mu\nu}=\eta_{\mu\nu},\Psi^L=\bar\Psi)}\;.
\end{equation}
In fact, we already computed the flat second order action for explicit examples, in particular GR [Eq.~\eqref{ActionGR2nd}] and SVHH gravity [Eq.~\eqref{ActionSVT2nd}].

While at first sight, this might seem like a trivial statement, one should recall that the \textit{effective} action is a priori defined with respect to an independent and arbitrary background metric and a replacement with the true background metric, in this case the asymptotic Minkowski metric is only executed after performing the variation. It could thus technically be that there is a non-trivial term in the full effective action, not present in its flat-space counterpart, that gives rise to a term with a non-trivial contribution in the limit to null infinity. However, we explicitly show that this is not the case by proving the Lemma \ref{LemmaP} in Appendix~\ref{App:Proof of Lemma}. Of course, while varying the effective action, the flat background fields are still considered as generic independent entities over which one can perform the variation. The statement is, however, that only the terms in the effective action contribute that survive a restriction to the flat background in Minkowski coordinates.
\begin{lemma}\label{LemmaP}
    In the limit to null infinity
    \begin{equation}\label{eq:EqLemma}
    \frac{-2}{\sqrt{-\eta}} \Bigg\langle\frac{\delta \phantom{}_{\mys{(2)}}S^{\myst{flat}}_{\text{eff}}}{\delta \eta^{\mu\nu}}\Bigg\rangle= \phantom{}_{\mys{(2)}}t_{\mu\nu}[\delta g^H,\delta\Psi^H]\,,
     \end{equation}
    and hence, the leading-order energy-momentum tensor [Eq.~\eqref{eq:SecondOrderVariationEMTensor}] only depends on the flat, second-order effective action defined in Eq.~\eqref{eq:flatSA}.
\end{lemma}

Moreover, as we have shown in Sec.~\ref{sSec:UnifiedTreatmentof Null and Ordinary Memory} above, very generally any asymptotic energy momentum tensor has the following form [Eq.~\eqref{eq:General Form asymptotic EMT}]
\begin{equation}\label{eq:General Form asymptotic EMT Thm}
   T^\text{a}_{\mu\nu}(u,r,\Omega)=T^\text{a}_{00}(u,r,\Omega)\,l_\mu l_\nu=\frac{1}{r^2}\,F(u,\Omega)\,l_\mu l_\nu\,,
\end{equation}
for a function $F(u,\Omega)$ related to a purely radial outward energy flux of radial asymptotic velocity $v$. For simplicity of the statement of the theorem, we will however restrict ourselves to purely massless gravitational degrees of freedom. As mentioned, the massive case will be further discussed in Sec.~\ref{sSec:Memory From Massive Fields} below. Recall that for null sources, the asymptotic retarded time $u$ and the vector $l_\mu$ are respectively given by
\begin{equation}\label{eq:asymptoticRet Time v Thm}
    u= t-r\,.
\end{equation}
and
\begin{equation}
    l_\mu= -\nabla_\mu t+\,\nabla_\mu r\,,
\end{equation}
with $\nabla_\mu r=\delta_{\mu i} n_i$.
The restriction to massless modes allows one to describe the energy moment tensor $\phantom{}_{\mys{(2)}}t_{\mu\nu}[\delta g^H,\Psi^H]$ of high-frequency perturbations in the limit to null infinity as a sum or superposition of asymptotic energy momentum fluxes of the form in Eq.~\eqref{eq:General Form asymptotic EMT Thm} 
\begin{equation}
    \phantom{}_{\mys{(2)}}t_{\mu\nu}[\delta g^H,\delta\Psi^H](u,r,\Omega)=\frac{1}{r^2}\sum_i F_i(u,\Omega)l_\mu l_\nu\,.
\end{equation}
Observe that since the asymptotic group velocity of massive fields would depend on the frequency of the waves, this would imply that the corresponding asymptotic energy momentum tensor also depends on the frequency content of the emitted waves. To postpone the treatment of these subtleties is the only reason we will for now concentrate on massless sources.   


The key to Theorem~\ref{Theorem1} then relies on the realization that based on a simple assumption on the form of the propagation equation of the high-frequency perturbations, namely that the TT modes of the theory satisfy a decoupled massless wave equation, the second-variation approach can be used in order to show that the leading order memory equation will still be of the form in Eq.~\eqref{eq:EOMISGRSecond} but with $T^\text{a}_{\mu\nu}$ replaced by a superposition of asymptotic energy-momentum tensors. Indeed, to leading order we can solve for each memory contribution in exactly the same way as explicitly shown in Sec.~\ref{sSec:UnifiedTreatmentof Null and Ordinary Memory} above. In particular, the computation of the final memory formula did not rely on any of the specifics of the unbound source of energy and momentum, except for the general structure in Eq.~\eqref{eq:General Form asymptotic EMT Thm}. The conclusion of the theorem will thus be that in a very broad class of metric theories of gravity, the tensor memory formula remains of the same functional as given in Eq.~\eqref{eq:DispMemoryGR} with the energy flux given by superposition of contributions, from which important conclusions can be drawn.

\paragraph{The Statement of the Theorem.} We are now ready to state the precise form of the theorem for the functional form of displacement tensor memory in metric theories of gravity. As already discussed, in formulating the theorem we will entirely focus on the gravitational part of the action. However, linearity implies that any memory contribution of potential unbound matter sources could also be considered in parallel.

\begin{theorem}\label{Theorem1} Consider a dynamical metric theory [Def.~\ref{DefMetricTheory}], for which
    \begin{enumerate}[(i)]
         \item the space-time is asymptotically flat as in Def.~\ref{DefAsymptoticallyFlat} with a background $\{\eta_{\mu\nu},\bar\Psi\}$ that solves the vacuum field equations and preserves local Lorentz invariance, with $\partial_\alpha\bar\Psi=0$ for scalar fields and $\bar\Psi=0$ for all other tensor fields and where $\eta_{\mu\nu}$ is the Minkowski metric.
        \item the assumptions of Sections~\ref{sSec:PerturbationTheory} and \ref{ssSec:IsaacsonGeneral} hold with the exact solution given by the Minkowski background introduced above. In particular the Eqs.~\eqref{eq:IsaacsonSplitGen}, \eqref{eq:IsaacsonSplitGen2} are satisfied, such that the leading-order, low-frequency metric equation can be written as [Eq.~\eqref{eq:EOMIS2}]
        \begin{equation}\label{eq:EOMISTh}
            _{\mys{(1)}}\mathcal{G}_{\mu\nu}[\delta g^L,\delta\Psi^L]=-\frac{1}{2}\big\langle \phantom{}_{\mys{(2)}}\mathcal{G}_{\mu\nu}[\delta g^H,\delta\Psi^H]\big\rangle\,,
        \end{equation}
        while the leading-order, high-frequency propagation equations [Eqs.~\eqref{eq:EOMIIS2} and \eqref{eq:EOMIIS2Psi}] are 
        \begin{align}\label{eq:EOMIISTh}
          \phantom{}_{\mys{(1)}}\mathcal{G}_{\mu\nu}[\delta g^H,\delta\Psi^H]=0\,, \qquad\phantom{}_{\mys{(1)}}\mathcal{J}[\delta g^H,\delta\Psi^H]=0\,;
        \end{align}

        \item in a faithful representation [Def.~\ref{DefFaithfulRep}], the physical metric $g_{\mu\nu}$ and the non-minimal fields $\Psi$ describe $2+N$ massless dynamical degrees of freedom that can potentially be excited as radiative modes. Moreover, there exists a set of leading-order high-frequency field perturbations 
        \begin{equation}
            \hat{h}_{\mu\nu}^H=W(\delta g^H,\delta\Psi^H)\,,\qquad \hat{\psi}^H=V(\delta\Psi^H)\,,
        \end{equation}
        for some functions $W$ and $V$, that describe $2+N$ propagating degrees of freedom, for which, in the limit to null infinity, the leading-order, high-frequency propagation equations [Eqs.~\eqref{eq:EOMIISTh}] reduce to a set of decoupled massless wave equations for the tensor field associated to the metric
        \begin{equation}\label{eq:ThWaveEq}
            \Box \hat{h}_{\mu\nu}^H=0\,,\quad \phantom{}_{\mys{(1)}}\mathcal{J}[\hat{\psi}^H]=0\,,
        \end{equation}
        upon imposing the Lorenz gauge as well as tracelessness
        \begin{equation}\label{eq:ThLorenzGauge}
        \partial^\mu\hat{h}_{\mu\nu}^H=0\,,\quad\eta^{\mu\nu}\hat{h}_{\mu\nu}^H=0\,.
        \end{equation}
    \end{enumerate}
    Then, in the limit to null infinity, the solution to the leading-order low-frequency metric equation [Eq.~\eqref{eq:EOMISTh}] gives a tensor memory component of the form
    \begin{align}\label{NonLinDispMemory Thm}
    \delta \hat h_{ij}^{\text{TT}\,L}(u,r,\Omega)=\,\frac{\kappa_\text{eff}}{2\pi r} \int_{S^2} d^2\Omega'\,\mathcal{F}(u,\Omega')\,\left[\frac{n'_in'_j}{1-\vec{n}'\cdot\vec{n}}\right]^\text{TT}\,,
\end{align}
where $\delta \hat h^L_{\mu\nu}=W(\delta g^L,\delta\Psi^L)$ satisfies the Lorenz gauge $\partial^\mu\delta \hat h^L_{\mu\nu}=0$, and where $\kappa_\text{eff}=\kappa_0\bar{A}$, with $\bar{A}$ a function that only depends on the Minkowski background. Moreover,
\begin{align}\label{eq:EperSolidAngleBeyondGR}
    \mathcal{F}(u,\Omega')=r^2\int_{-\infty}^u d u'\,\phantom{}_{\mys{(2)}}t_{00}[\hat h^H,\hat\psi^H](u',r,\Omega')\,,
\end{align}
where $\phantom{}_{\mys{(2)}}t_{\mu\nu}$ has the following properties:
    \begin{enumerate}
        \item[(a)] it is conserved: $\partial^\mu\phantom{}_{\mys{(2)}}t_{\mu\nu}=0$;
        \item[(b)] it can be written as a sum of terms:
        \begin{equation} 
         \phantom{}_{\mys{(2)}}t_{\mu\nu}[\hat h^H,\hat\psi^H]=\phantom{}_{\mys{(2)}}t^{\myst{GR}}_{\mu\nu}[\hat h^H]+ \phantom{}_{\mys{(2)}}t^{\hat{\psi}}_{\mu\nu}[\hat \psi^H]\,,
    \end{equation}
     where
     \begin{equation}
         \phantom{}_{\mys{(2)}}t^{\myst{GR}}_{\mu\nu}\propto\Big\langle \partial_\mu \hat h^H_{\alpha\beta}\partial_\nu \hat h^{H\alpha\beta}\Big\rangle\,;
     \end{equation}
     \item[(c)] it is invariant under infinitesimal coordinate transformations
     \begin{equation}
         x^\mu\rightarrow x '^\mu= x^\mu+\xi_H^{\mu}\,.
     \end{equation}
    \end{enumerate}
\end{theorem}

\subsection{Analysis of the Theorem}\label{sSec:Analysis of Theorem}

Let us begin an in-depth discussion of the theorem by making several technical remarks. 

\paragraph{Technical Remarks on the Theorem}\label{sec:ThmRemarks}

    First, let us stress that in assumption (iii) we only require the \textit{first-order propagation equations} to reduce to a set of decoupled second-order wave equations. One of the main results of the theorem is therefore that the decoupling between fields remains intact even at $\mathcal{O}(\alpha^2)$ at the low-frequency level.

    Furthermore, observe that assumption (iii) is quite generic. Indeed, in any theory satisfying second-order equations of motion, the first-order propagation equation will only involve two derivative operators. But as we will discuss below, this is even true for a large class of theories, whose field equations are higher-order in derivatives. Masslessness, together with local Lorentz invariance, would then actually ensure that the leading-order propagation equation generically take the form of a massless wave equation.

    Moreover, also a decoupling of the equations at first order in perturbations is quite generic. First, recall that a Minkowski background ensures that the tensor, vector and scalar sectors\footnote{The terms "tensor", "vector" and "scalar" refer here to the polarization type of each mode.} can always be decoupled at leading order in perturbations and each of these sectors can therefore be treated individually. Hence, potential couplings between perturbations at the level of the leading-order perturbation equations could only arise within each of these sectors. We are, however, not aware of any concrete massless theory that admits such a coupling of first-order perturbations. Indeed, explicit examples of such coupled equations typically only arise in theories that include an explicit mass term, such as, for instance, in massive bigravity models (see e.g. \cite{Comelli:2012db,DeFelice:2013nba}) or massive multi-Proca theories \cite{BeltranJimenez:2016afo}. It would, however, be interesting to explore null memory for such theories with coupled leading-order perturbation equations, a task we leave for future work.
   
    Also, any typical massless theory involving multiple interacting vector or scalar fields at the level of the full action, such as non-Abelian vector fields or typical scalar multifield models (see e.g. \cite{Dimakis:2019qfs}), naturally decouple to leading order in perturbations on a Minkowski background, and thus, still abide by the decoupling assumption in (iii).

    Finally, as mentioned in the proof of the theorem, the theorem also implies that if the leading-order propagation equations only involve up to two-derivative operators, which is generally expected for ghost-free theories, then also the low-frequency $\mathcal{O}(\alpha^2)$ term only involves two derivative operators. In particular, the latter directly implies that in any theory satisfying the assumptions of the theorem, the memory equation will only directly depend on terms in the action that involve two derivative operators. This is nicely exemplified in the SVHH theory result of the radiative energy-momentum tensor that we will consider below in Eq.~\eqref{eq:StressEnergyFourth}.

\paragraph{An Explicit Memory Formula for $k$-Form Fields.}

To continue the discussion of the theorem, we want to be more specific and consider a concrete type of possible additional gravitational fields, which covers a large class of theories considered in the literature. Namely, from now on, we will focus on dynamical metric theories whose additional gravitational fields $\Psi$ are $k$-form field potentials with an associated Abelian $U(1)$ gauge symmetry.

Recall that a differential $k$-form field $\Psi$ is a totally antisymmetric tensor field, which in a coordinate-induced basis can be written as 
\begin{equation}
    \Psi=\frac{1}{k!}\,\Psi_{\mu_1...\mu_k}\,dx^{\mu_1}\wedge...\wedge dx^{\mu_k}\,,
\end{equation}
with $k<d$ and where $"\wedge"$ denotes the exterior product. Such $k$-form fields naturally generalize $U(1)$ vector field potentials because their field strength $\mathscr{F}\equiv d\Psi$ is invariant under Abelian gauge transformations $\Psi\rightarrow \Psi + d\Lambda$, where $\Lambda$ is an arbitrary $(k-1)$-form and $d$ is the exterior derivative. See for instance \cite{Henneaux:1986ht,zee_quantum_2010} for a review of the topic.

In particular, this restriction implies that we focus on theories with Abelian gauge groups, but we want to remark that similar conclusions should also hold in the non-Abelian case. As already mentioned, a restriction to $k$-form fields also implies a limitation to bosonic fields. 


The collection of these additional dynamical $k$-form fields are assumed to describe $N$ additional propagating gravitational degrees of freedom. Thus, the theory admits $N$ independent and propagating solutions to the wave equations, characterized through $N$ modes in the canonically normalized, second-order action, which we will denote as $\hat{\psi}_\lambda$, where $\lambda=1,..,N$.  Note that in four spacetime dimensions, we only consider $k$-forms for $k<4$. A $0$-form field simply corresponds to a scalar field, while a 1-form field naturally describes an Abelian vector field, and, therefore, it carries two propagating degrees of freedom. A 2-form field, on the other hand, again only describes one dynamical mode equivalent to a scalar degree of freedom (see e.g. \cite{Heisenberg:2019akx}). A 3-form field will not contain any propagating modes in four dimensions simply because the components of the associate 4-form field strength are constant (see e.g. \cite{Bandos:2019wgy}).\footnote{However, a non-trivial coupling to the metric of such fields can for example lead to a dynamical contribution to the cosmological constant \cite{Duncan:1989ug} and may thus still have physical implications.}

In Appendix~\ref{app:ExampleNullMemoryKForm} we offer for this class of theories the explicit derivation of the form of the energy-momentum tensor, that results in an explicit formula of the associated tensor null-memory formula in a spin-weighted spherical harmonic composition. Provided that the assumptions of Theorem~\ref{Theorem1} hold, the end result is [Eq.~\eqref{NonLinDispMemoryGen2App}]
\begin{equation}\label{NonLinDispMemoryGen2}
    \begin{split}
     &\delta h_H^{l m}(u,r)=\frac{1}{r} \sqrt{\frac{(l-2)!}{(l+2)!}}\int_{S^2} d^2 \Omega'\,Y^*_{l m}(\Omega') \int_{-\infty}^{u}d u'\,r^2\bigg\langle | \dot{\hat{h}}_+|^2+| \dot{\hat{h}}_\times|^2+\sum_{\lambda=1}^N|\dot{\hat{\psi}}_\lambda|^2\bigg\rangle\,,
    \end{split}
\end{equation}
where $\hat{h}_{+,\times}=h_{+,\times}$ correspond to the polarization modes of the perturbations of the physical metric. Indeed, as accounted for in the statement of the Theorem~\ref{Theorem1}, for certain theories it is necessary to redefine the tensor perturbation variable $\hat{h}_{\mu\nu}^H=W(h^H,\Psi^H)$ to obtain a perturbation variable that satisfies a first-order wave equation in the appropriate gauge. However, the TT component of this redefined variable, and thus, also the polarization modes $h_{+,\times}$ always correspond to the TT component of the physical metric present in the detector response $\hat{h}^{\text{TT}H}_{\mu\nu}=h^{\text{TT}H}_{\mu\nu}$. The same is true for the memory component. The need for such a change of variables to decouple the leading-order equations is, typically, a sign of the presence of additional gravitational polarizations.

Let us end this paragraph by stressing that the tensor null-memory result in Eq.~\eqref{NonLinDispMemoryGen2} was obtained without any knowledge of the precise form of the Lagrangian, and it simply follows from Theorem~\ref{Theorem1} and the resulting solution of the memory-evolution equation. The coupling constants of a specific theory would then enter through a transformation from the canonically normalized modes $\hat{\psi}_\lambda$ to the physical modes of the theory (this point will further be discussed below). The expression in Eq.~\eqref{NonLinDispMemoryGen2} represents a generalization of the explicit SVT theory example that we will consider in Sec.~\ref{sec:SVTMem} and should be compared to the result in Eq.~\eqref{NonLinDispMemoryModesSVT}. 

\paragraph{Scope of the Memory Theorem.}

While in the above paragraph we simply assumed that Theorem~\ref{Theorem1} holds, we will now explore in greater detail the scope of the theorem and investigate which types of theories satisfy the assumptions of Theorem~\ref{Theorem1}. As discussed above, we will however restrict ourselves to dynamical metric theories of gravity that admit an arbitrary number of additional $k$-form fields in the gravitational sector. Such theories however still encompass a very large class of concrete metric theories considered in the literature.

First of all, the theorem clearly encompasses any covariantized version of massless $k$-form Galileon theories \cite{Deffayet:2010zh,Deffayet:2016von}, restricting the full equations of motion to second order. Such theories represent a natural generalization of the SVT class of theories with second-order equations of motion discussed in Sec.~\ref{ssSec: A Exact Theories}. In particular this includes Horndeski theory in Eq.~\eqref{eq:ActionHorndeski} as well as its SVHH generalization in Eq.~\eqref{eq:ActionSVH}, that include concrete beyond GR theories such as BD theory [Eq.~\eqref{ActionBD}], sGB gravity [Eq.~\eqref{eq:ActionsGB}], $f(R)$ gravity [Eq.~\eqref{eq:Actionf(R)}], and double-dual Riemann gravity [Eq.~\eqref{ActionddR}]. 

The memory formula in Eq.~\eqref{NonLinDispMemoryGen2}, however, is not restricted to theories that satisfy second-order equations of motion. A first interesting concrete example of a theory that does not fall under the class of covariantized $k$-form Galileon theories is dCS gravity, given in Eq.~\eqref{eq:ActiondCS}.
Just as in sGB gravity the dCS action near null infinity expanded to second-order in perturbations simply reduces to the GR one with a canonical scalar field, because by assumption of a Lorentz preserving asymptotically flat spacetime the background scalar value is a constant and the Pontryagin density is of higher order in this limit. Therefore, the \textit{linear-order}, high-frequency propagation equations are just given by two decoupled, second-order wave equations for the metric and the scalar field perturbations. In turn, this fact implies that for dCS gravity the tensor null-memory formula is given by Eq.~\eqref{NonLinDispMemoryGen2}, with $N=1$, $\Psi\to\Theta$ and corresponding leading-order wave mode $\hat\psi_\lambda\to\hat\vartheta=\vartheta$. For dCS gravity, this result is indeed confirmed by the explicit computation of the associated BMS balance laws in \cite{hou_gravitational_2022}.\footnote{In Sec.~\ref{MatchToAsymptoticsBD} below we will explicitly show how it is possible to derive a memory formula from BMS balance laws.}
Note, however, that while the dCS coupling does not enter explicitly into the memory equation, it still has an implicit effect through the dependence of the metric and scalar perturbations on the coupling. 

Recall, however, that outside of the limit to null infinity (with a non-trivial scalar background), dCS, when taken at face value, has in fact higher order equations of motion at the linearized level and propagate a ghost, that renders the theory untenable. As discussed in Chapter~\ref{Sec:The Theory Space Beyond GR}, this implies that dCS can only consistently be treated as a theory of type (B) considered in Sec.~\ref{ssSec: B Perturbative Theories} which require additional constraints to ensure a limited number of propagating degrees of freedom. These constraints are naturally imposed by assuming the equations of motion of the ghost-free principal part of the theory and treating any higher order terms as explicit perturbations. 

Thus, up to such corrections, also any theory of type (B) that involve higher powers of curvature invariants together with additional nonminimal couplings to other gauge-invariant, Abelian, $k$-form fields are expected to comply with the assumptions of Theorem~\ref{Theorem1}. This is because the massless, covariance and local-Lorentz invariance conditions ensure that the leading-order propagation equations of the principal parts of such theories still reduce to massless wave equations at null infinity. Note that as discussed above, as long as there are no additional spin 2 tensor fields also the decoupling of the leading order propagation equations between the tensor fields and all other non-minimal fields is ensured, due to the rotational invariance of the asymptotic Minkowski background.

\section{SVT Example: Displacement Memory}\label{sec:SVTMem}
\small{\textit{Selected parts of this section are taken over from the original work \cite{Heisenberg:2023prj} of the author. Based on this remark, we will refrain from introducing explicit quotation marks to indicate direct citations.}}
\\

\normalsize
\noindent
We now want to come back to the explicit example of an SVT metric theory beyond GR, the scalar-vector Heisenberg-Horndeski theory [Eq.~\eqref{eq:ActionSVH}], whose dynamical DOFs and gravitational polarizations we discussed in detail in Sec.~\ref{sSec:GWPolExample}. This will not only allow us to see the memory Theorem~\ref{Theorem1} in action, but will also allow us to discuss the concepts of scalar and vector memory, as well as memory arising from massive fields in more detail.
We will however first still stick to considering massless degrees of freedom only to match the explicit exposure of Theorem~\ref{Theorem1} above and will come back to the massive case in Sec.\ref{sSec:Memory From Massive Fields} below.

\subsection{Tensor Memory from Massless Fields}\label{eq:Memory From Massless Fields}

Let's therefore momentarily set the mass of the scalar field perturbation $m$ defined in Eq.~\eqref{eq:DefMass SVH} to zero, hence demand that
\begin{equation}\label{eq:restriction to massless scalar}
    \bar{G}_{2,\Phi\Phi}=0\,.
\end{equation}

\paragraph{Computing Memory from Theorem~\ref{Theorem1}.} Based on Theorem~\ref{Theorem1}, the precise memory formula for SVT gravity is given by Eq.~\eqref{NonLinDispMemory Thm}
\begin{align}\label{NonLinDispMemory SVH Gem^n}
    \delta  h_{ij}^{\text{TT}}(u,r,\Omega)=\,\frac{\kappa_\text{eff}}{2\pi r} \int_{S^2} d^2\Omega'\,\mathcal{F}_{\myst{SVH}}(u,\Omega')\,\left[\frac{n'_in'_j}{1-\vec{n}'\cdot\vec{n}}\right]^\text{TT}\,,
\end{align}
where
\begin{align}\label{eq:EperSolidAngle SVH}
    \mathcal{F}_{\myst{SVH}}(u,\Omega')=\int_{-\infty}^u d u'\,F_{\myst{SVH}}(u',\Omega')=r^2\int_{-\infty}^u d u'\,t^{\myst{SVH}}_{00}(u',r,\Omega')\,.
\end{align}
Thus, we can compute the displacement memory formula of SVHH by simply computing its asymptotic energy-momentum tensor as well as determining $\kappa_\text{eff}$.

Based on its second order action that we derived in Eq.~\eqref{ActionSVT2nd} this can in fact readily be done through the use of Lemma~\ref{LemmaP} and results in
\begin{equation}
    \phantom{}_{\mys{(2)}}t_{\mu\nu}^{\myst{SVH}}=t_{\mu\nu}^{\myst{GR}}+ t_{\mu\nu}^{\hat{a}}+ t_{\mu\nu}^{\hat{\varphi}}\,,
\end{equation}
where
\begin{subequations}
\begin{align}
    t_{\mu\nu}^{\myst{GR}}&=\frac{1}{4\kappa_\text{eff}}\Big\langle \partial_\mu\hat{h}_{\alpha\beta}\partial_\nu\hat{h}^{\alpha\beta}\Big\rangle\,,\\
    t_{\mu\nu}^{\hat{a}}&=\frac{1}{2\kappa_\text{eff}}\Big\langle\hat{f}_{\mu\alpha}\hat{f}\du{\nu}{\alpha}\Big\rangle=\frac{1}{2\kappa_\text{eff}}\Big\langle\partial_{\mu}\hat{a}_{\alpha}\partial_{\nu}\hat{a}^{\alpha}\Big\rangle\,,\\
    t_{\mu\nu}^{\hat{\varphi}}&=\frac{1}{2\kappa_\text{eff}}\Big\langle\partial_\mu\hat{\varphi}\partial_\nu\hat{\varphi}\Big\rangle\,,\label{eq: EMT scalar SVH}
\end{align}
\end{subequations}
and $\kappa_\text{eff}$ is given in Eq.~\eqref{eq:kappaeff SVH}.

Indeed, this result can readily be verified through an explicit computation of the averaged second order equations of motion $\langle \phantom{}_{\mys{(2)}}\mathcal{G}^{\myst{SVH}}_{\mu\nu}\phantom{}\rangle$ of SVHH gravity. We already want to remark at this point that the result of the energy momentum holds even if the scalar field perturbation is massive. This is because the average that allows for integrations by parts and the propagation equations of motion of the high-frequency fields [Eq.~\eqref{eq:FirstOrderProp}] cancel any additional contribution. Moreover, because of the spacetime averages (derivatives and averaging commute) and the wave equations, this total energy-momentum tensor is conserved as well as traceless
\begin{align}\label{eq:StressEnergyTot}
   \partial^\mu t_{\mu\nu}^{\myst{SVH}}=0\,,\quad\eta^{\mu\nu}  t_{\mu\nu}^{\myst{SVH}}=0\,.
\end{align}
Furthermore, gauge invariance can easily be checked (see e.g. \cite{maggiore2008gravitational}) such that the total stress-energy (pseudo)tensor only depends on the modes in Eqs.~\eqref{eq:TT modes h and a SVH}, \eqref{eq:polmodes h SVH}
and \eqref{eq:polmodes a SVH}, namely
\begin{align}\label{eq:StressEnergySecond}
    t_{\mu\nu}^{\myst{SVH}}=\frac{1}{4\kappa_\text{eff}}\Big\langle \partial_\mu\hat{h}^\text{ TT}_{ij}\partial_\nu\hat{h}_{\text{TT}}^{ij}+2\,\partial_\mu\hat{a}^\text{T}_i\partial_\nu\hat{a}_{\text{T}}^i+2\,\partial_\mu\hat{\varphi}\partial_\nu\hat{\varphi}\Big\rangle\,.
\end{align}
and thus
\begin{equation}\label{eq:EperSolidAngle}
    F_{\myst{SVH}}(u',\Omega')=\frac{dE_{\myst{SVH}}}{du'd\Omega'}=r^2\,t_{00}^{\myst{SVH}}(u',r,\Omega')=\frac{r^2}{2\kappa_\text{eff}}\,\Big\langle |\dot{\hat{h}}|^2+|\dot{\hat{a}}|^2+\dot{\hat{\varphi}}^2\Big\rangle\,.
\end{equation}

However, one has to remember at this point that the physical, in the sense of observationally relevant, modes are characterized in terms of the perturbations of the original fields that appear in the full action. Indeed, it is the scalar perturbation $\varphi$ that is associated with a potentially observable additional breathing mode, as shown in Eq.~\eqref{ElectricRiemannLin}. Thus, in terms of the physical modes, the radiative energy density becomes
\begin{align}
    t_{00}^{\scriptscriptstyle{\text{SVH}}}(u',r,\Omega')
    =\frac{1}{2\kappa_\text{eff}}\Bigg\langle |\dot{h}|^2+\zeta^2\,|\dot{a}|^2+\rho^2\,\dot{\varphi}^2\Bigg\rangle\,,\label{eq:StressEnergyFourth}
\end{align}
where we recall the definitions
\begin{equation}
    \rho=\sqrt{3\,\sigma^2+\frac{(\bar G_{2,X}-2\,\bar G_{3,\Phi})}{\bar G_4}}\,,\quad  \sigma= \frac{\bar G_{4\Phi}}{\bar G_{4}}\,\,,\quad\zeta\equiv \sqrt{\frac{\bar G_{2,F}}{\bar G_4}}\,.
\end{equation}

The tensor null memory for SVT gravity can thus be simply evaluated by inserting the expression for the radiative energy density in Eq.~\eqref{eq:StressEnergyFourth}
into the expression for the time-integrated energy flux in Eq.~\eqref{eq:EperSolidAngle SVH}. 
Observe that this tensor displacement memory is sourced by \textit{all} radiative degrees of freedom in the SVT theory, independent of whether these radiative modes excite additional gravitational polarizations in the physical metric or not. More precisely, a nonzero value of $\sigma$, which implies that the breathing mode is excited (recall the discussion in Sec.~\ref{sSec:GWPolExample}), only influences the value of the scalar prefactor $\rho$, but it does not determine whether the emitted scalar radiation provides an additional tensor memory source in principle. Moreover, the energy density emitted in vector modes backreacts to produce tensor memory, even though they are in no way connected to any gravitational polarizations of the physical metric.

Moreover, the displacement memory can for practical use also be given explicitly in terms of modes of a spin-weighted spherical harmonics expansion, as explicitly derived in Sec.~\ref{NullMemory for GW Radiation in GR} in terms of which the displacement memory is given by [Eq.~\eqref{NonLinDispMemoryModesGR}]
\begin{equation}\label{NonLinDispMemoryModesSVT}
  \boxed{\delta h^{lm}_{\myst{SVH}}=\,\frac{1}{r}\sqrt{\frac{(l-2)!}{(l+2)!}}\int_{S^2}d^2\Omega'\,Y^*_{lm}(\Omega')\int_{-\infty}^u d u'\,r^2\,\Bigg\langle |\dot{h}|^2+\zeta^2\,|\dot{a}|^2+\rho^2\,\dot{\varphi}^2\Bigg\rangle\,,}
\end{equation}
Recall that the angular integral in this expression can be evaluated analytically as a sum of 3$j$ symbols by expanding the leading-order waves in spin-weighted spherical harmonics and applying the identity in Eq.~\eqref{SWSHTrippleInt}.

\paragraph{Manifestly Local Derivation of Memory.} Of course, the memory equation for SVHH gravity could also have been computed directly without the use of Theorem~\ref{Theorem1}. It is perhaps instructive to quickly go through the necessary steps in such a direct computation, in particular for the transition to the massive mode case (see \cite{Heisenberg:2023prj} for more details). An explicit calculation is of course based on the leading order low-frequency tensor memory equation Eq.~\eqref{eq:EOMIS2} and its solution. In the manifestly local approach discussed here, and analogous to the high-frequency perturbations, this would require a field redefinition of the form
\begin{equation}\label{RelationPhTohMemory}
    \delta\hat{h}_{\mu\nu}\equiv \delta h_{\mu\nu}+\eta_{\mu\nu}\sigma\,\delta\varphi\,.
\end{equation}
in order to decouple the low-frequency perturbations. This transformation also explicitly features in the statement of the memory theorem. Moreover, thanks to the properties of the source term in Eq.~\eqref{eq:StressEnergyTot}, infinitesimal coordinate transformations at the low-frequency level $\xi_\mu^L$ can be used to once again impose the gauge conditions
\begin{equation}\label{eq:De-DonderSecond}
    \partial^\mu\delta\hat{h}_{\mu\nu}=0\,,\quad \eta^{\mu\nu}\delta\hat{h}_{\mu\nu}=0\,,
\end{equation}
such that the left-hand side of Eq.~\eqref{eq:EOMIS2} reduces to a wave equation. Hence, the metric equation [Eq.~\eqref{eq:EOMIS2}] in the asymptotic region of an asymptotically flat spacetime described by SVT gravity simply reads
\begin{equation}\label{eq:SecondOrderEOMMem}
   \Box\delta\hat{h}_{\mu\nu}=-2\kappa_\text{eff}\,t^{\myst{SVH}}_{\mu\nu}\,.
\end{equation}
This equation can then be solved as presented in Sec.~\ref{sSec:UnifiedTreatmentof Null and Ordinary Memory}. As already stressed on several occasions, the physically relevant modes of the ``hatted'' tensor variables are actually equivalent to the physical perturbations of the physical metric $\delta\hat{h}^\text{TT}_{ij}=\delta\hat{h}^\text{TT}_{ij}$. It is however important to note that the necessity for field redefinitions as well as the gauge conditions are a mere artifact of the manifestly local approach discussed here. Indeed, throughout this work and in particular in Sec.~\ref{sSec:GWPolExample} we already explicitly discussed the alternative manifestly gauge invariant approach within an SVT decomposition of the field perturbations, in which all the subtleties related to identification of physical modes are washed away. In the same way, also the computation of memory can be carried out without such subtleties, as we now want to briefly show.


\paragraph{Gauge Invariant Derivation of Memory.}

Recall that within an SVT decomposition, one can directly identify the manifestly gauge invariant degrees of freedom of a theory and determine which ones are propagating DOFs by examining their propagation equations. In the case of SVHH theory, the high frequency propagation equations were given in Eqs.~\eqref{eq:Metric SVH} and \eqref{eq:Vector And Scalar SVH}. Moreover, in a faithful description of the theory, if additional gravitational polarizations of the physical metric are excited, then there exist additional relations between those and the additional degrees of freedom in the theory.

The same SVT decomposition can be carried out for the low-frequency components such that one can write down the memory equation based on the leading order low-frequency equations of motion Eq.~\eqref{eq:EOMIS2} directly in terms of the physical variables as
\begin{align}\label{eq:backreaction}
    \Box \delta h^{TT}_{ij} = -2\kappa_{\myst{eff}} \perp_{ijab}  t^{ab}_{\myst{SVH}}[h^\text{TT}_{ij},a^\text{T}_i,\varphi] \,.
\end{align}
This sourced wave equation can then be solved in the exact same manner as we did in the manifestly local approach.

\subsection{Memory from Massive Fields}\label{sSec:Memory From Massive Fields}

Until now, we focused primarily focused on the memory beyond GR that itself arises as a TT polarization (tensor memory) and that is sourced by massless fields that propagate to null infinity (null memory). While the possibility of scalar and vector memory will be discussed in Sec.~\ref{sSec:Scalar and Vector Memory} below, we now want to catch up and discuss memory that can also be sourced by massive fields. As already remarked on several occasions, a restriction to null memory was only based of pure convenience to postpone certain subtleties in the presentation of the memory formula to this subsection here. Fundamentally, however, there is no restriction in applying our Isaacson approach to the computation of memory in the case of massive non-minimal fields.

Indeed, recall the general structure of the energy momentum tensor in Eq.~\eqref{eq:General Form asymptotic EMT} 
\begin{equation}\label{eq:General Form asymptotic EMT SVH}
   T^\text{a}_{\mu\nu}(u,r,\Omega)=T^\text{a}_{00}(u,r,\Omega)\,l_\mu l_\nu=\frac{1}{r^2}\,F(u,\Omega)\,l_\mu l_\nu\,,
\end{equation}
where $F(u,\Omega)$ is related to a purely radial outward energy flux of velocity $v$, where
\begin{equation}
    l_\mu\equiv -\nabla_\mu t+v\,\nabla_\mu r\,,
\end{equation}
and
\begin{equation}\label{eq:asymptoticRet Time v SVH}
    u\equiv t-\frac{r}{v}\,.
\end{equation}
This general form of an asymptotic energy-momentum tensor, that in particular also applies to the special case of massive particles, is also valid for the asymptotic energy-flux of massive fields. This can be readily confirmed by comparing to an explicit computation of the energy momentum tensor of the non-minimal scalar field in SVHH theory. From now on, we thus alleviate the restriction in Eq.~\eqref{eq:restriction to massless scalar} and consider a scalar perturbation $\varphi$ of non-zero mass $m$ defined in Eq.~\eqref{eq:DefMass SVH}. 

First of all, for a massive scalar field, the form of the asymptotic energy momentum tensor is in fact unaltered and still reads [Eq.~\eqref{eq: EMT scalar SVH}]
\begin{equation}
    t_{\mu\nu}^{\varphi}\propto\Big\langle\partial_\mu\varphi\partial_\nu\varphi\Big\rangle\,.\label{eq: EMT scalar SVH 2}
\end{equation}
As already remarked, this is because the average that allows for integrations by parts and the propagation equations of motion of the high-frequency fields [Eq.~\eqref{eq:FirstOrderProp}] cancel any additional contribution. However, what changes is the form of the asymptotic solution of the scalar field, which is now given by a superposition of plane-wave solutions to the Klein-Gordon equations that we described in Eq.~\eqref{eq:GeneralPlaneWaveRadiationMassive}. Especially, the time dependence of the asymptotic massive modes is not governed by the asymptotic retarded time in Eq.~\eqref{eq:asymptoticRet Time v SVH}, but rather by the combination $t-vr$ where in this case the group velocity $v$ is distinct from the phase velocity. Consequently, in contrast to Eq.~\eqref{eq:relation partial i EMT} the massive fields satisfy to leading order in $1/r$ the relation [Eq.~\eqref{eq:IdentityRadiationSecond}]
\begin{equation}
    \partial_i \varphi=-v_i\,\partial_0 \varphi\,.
\end{equation}
In combination with the explicit result in Eq.~\eqref{eq: EMT scalar SVH 2} together with a standard inverse square law this implies that the structure of the radiative energy-momentum tensor is precisely of the form in Eq.~\eqref{eq:General Form asymptotic EMT SVH}. Therefore, a corresponding memory equation can be solved in the same way as it was the case for unbound massless field sources.

However, there is a crucial difference between the massive and the massless case, namely that the group velocity of massive fields is frequency dependent. As a consequence, the general form in Eq.~\eqref{eq:General Form asymptotic EMT SVH} is different for each plane-wave contribution in our superposition of plane-waves. In other words, if the emission is not that of a monochromatic massive wave, the total asymptotic energy-momentum tensor is given by a sum of frequency dependent contributions. Of course, this does not influence the corresponding computation of the memory per se, since at the linear level in the radially outward case each individual plane-wave contribution can be considered separately. The difference is that the final result can not be reported in the same was as for the null memory in Eq.~\eqref{NonLinDispMemory Thm}, since the energy flux of the massive fields as well as the directional contribution in the angular brackets will further depend on the velocity of the source-waves and therefore their frequency. Hence, the memory contribution of a massive wave, which thus by definition is a contribution to the ordinary memory, is more precisely of the form
\begin{align}\label{NonLinDispMemory Thm massive}
    \boxed{\delta  h_{\varphi \,ij}^{\text{TT}}(u,r,\Omega)=\,\frac{\kappa_\text{eff}}{2\pi r} \int_0^\infty dv\int_{S^2} d^2\Omega'\,\mathcal{F}_{\varphi}(u,\Omega',v)\,\left[\frac{v^2\,n'_in'_j}{1-v\,\vec{n}'\cdot\vec{n}}\right]^\text{TT}\,.}
\end{align}
This is in complete analogy to the case of a collection of massive particles in Eq.~\eqref{eq:Patricle contribution memory} that is given by a sum of individual contributions.

As discussed, the transition between the massive and the massless case is however smooth and generally, the contribution to the ordinary memory from massive particles will only remain non-negligible if their velocity in the source-centered frame is not too far from luminality. One could therefore expect that the influence of the explicit frequency dependence of massive modes will not play a significant role in realistic scenarios, a statement which however needs to be investigated in more detail.


\subsection{Scalar and Vector Memory}\label{sSec:Scalar and Vector Memory}

We also want to briefly address here the question of whether scalar or vector memory contributions can arise within SVT theory, although we will content ourselves with a first glance at the problem. Experimentally, such contributions are not expected to play a major role, since already the measurement of an additional gravitational polarization mode in a GW detector response would signify a revolution in physics and is correspondingly based on current constraints expected to be hard to detect, let alone the memory contribution within such a scalar or vector gravitational polarization. 

From a theoretical point of view, the question is however still interesting. Within the Isaacson approach to memory presented in this work, a scalar or vector memory contribution could in principle arise through the leading order low-frequency equation of the additional non-minimal fields in the theory in Eq.~\eqref{eq:EOMIS2Psi}. Indeed, until now, we merely focused on the corresponding metric memory equation [Eq.~\eqref{eq:EOMIS2}] that naturally gives rise to a memory in the tensor TT polarization of the detector response.

Within the SVHH theory considered here, we could therefore ask whether there are nontrivial scalar or vector memory contributions $\delta\varphi$ and $\delta a_\mu$ arising from the corresponding low-frequency equations of the low-frequency perturbations of the non-minimal vector and scalar fields. For the vector perturbations, the answer can be given right away, since SVT gravity never excites any vector polarizations in the physical metric. By definition, this directly implies that SVT gravity will not contain any vector memory either. 

For the scalar leading order low-frequency equation, we obtain
\begin{align}\label{eq:ScalarMemoryEquation}
    \left(\Box-m^2\right)\delta\hat\varphi=\frac{1}{2}\,\varphi^2\Bigg[m^2&\Bigg\{\frac{2\bar G_{4,\Phi}\left(2\bar G_{2,X}-4\bar G_{3,\Phi}+3\bar G_{4,\Phi\Phi}-9 G_{4,\Phi}^2/\bar G_4\right)}{\bar G_4}\notag\\
    &+2\bar G_{3,\Phi\Phi}-\bar G_{2,\Phi X}\Bigg\}+ \bar G_{3,\Phi\Phi\Phi}\Bigg]\,.
\end{align}
Observe that for a scalar field with vanishing potential, the source term completely vanishes, and the scalar memory equation reduces to
\begin{align}\label{eq:ScalarMemoryEquation 2}
    \Box\delta\hat\varphi=0\,.
\end{align}
In this case, no additional scalar null memory is generated, implying that the only nontrivial null-memory component is the tensor null memory. On the other hand, at first sight we do not see any obstacle for (ordinary) scalar memory to arise for a scalar field with non-trivial potential, in particular a massive scalar field. Note however, that in the massive case a massive Klein-Gordon equation needs to be solved, for which we expect additional suppression.


\subsection{Concrete Metric Theories: Displacement Memory}\label{sSec:Concrete Theories}

We close this section by offering the explicit results for the tensor null memory of interesting subclasses of Horndeski gravity that we already discussed in Sec.~\ref{sSec:GWPolExample}.

\paragraph{Brans-Dicke Gravity.} 
Recall that Horndeski gravity reduces to BD theory [Eq.~\eqref{ActionBD original}] for the choices [Eq.~\eqref{eq:BDGs SVH}]
\begin{align}\label{eq:BDGs 2}
    G_2=\frac{2\omega}{\Phi} X\,, \qquad G_4=\Phi\,,\qquad G_i=0 \quad\;\text{otherwise}\,.
\end{align}
Inserting Eq.~\eqref{eq:BDGs 2} into Eq.~\eqref{eq:StressEnergyFourth}, the corresponding energy-momentum (pseudo)tensor sourcing the tensor memory therefore reads 
\begin{equation}\label{sourceTensorMemoryBD}
    t_{00}^{\myst{BD}}=\frac{1}{2\kappa^{\myst{BD}}_{\text{eff}}}\Bigg\langle |\dot{h}|^2+(2\omega+3)\left(\frac{\dot{\varphi}}{\bar{\Phi}}\right)^2\Bigg\rangle\,,
\end{equation}
where 
\begin{equation}
    \kappa^{\myst{BD}}_{\text{eff}}=\frac{\kappa_0}{\bar{\Phi}}\,,
\end{equation}
such that Eq.~\eqref{NonLinDispMemoryModesSVT} becomes
\begin{equation}\label{NonLinDispMemoryModesBD}
  \boxed{\delta h^{lm}_{\myst{BD}}=r\sqrt{\frac{(l-2)!}{(l+2)!}}\int_{S^2}\dd^2\Omega'\,Y^*_{lm}(\Omega')\int_{-\infty}^u\dd u'\Bigg\langle\,|\dot{h}|^2+(2\omega+3)\left(\frac{\dot\varphi}{\bar{\Phi}}\right)^2\Bigg\rangle\,.}
\end{equation}

Moreover, recall that in this theory 
\begin{equation}
\sigma=\frac{\bar G_{4,\Phi}}{\bar G_{4}}=\frac{1}{\bar{\Phi}}\neq 0\,,
\end{equation}
such that BD gravity has an additional breathing polarization. As discussed above, this fact only minimally modifies the memory formula of Eq.~\eqref{NonLinDispMemoryModesBD}, since $\omega$ is already constrained to be a large number (e.g.~$\omega > 4 \times 10^4$ due to constraints from the tracking of the Cassini spacecraft and the Shapiro time delay~\cite{Bertotti:2003rm}).
The existence of such an additional scalar polarization in BD theory is only relevant for memory inasmuch as it in principle allows for the possibility to also measure scalar memory, hence memory within the scalar polarization of the detector response. Yet, as already mentioned, in the massless case we do not find any analogous scalar null-memory component because there is no analogous null source for the scalar mode with trivial potential.

Considering the explicit result for the memory component of BD theory is also interesting because BD theory is one of the few theories beyond GR where memory was already computed using different techniques. Equation~\eqref{NonLinDispMemoryModesBD} thus represents a nice opportunity to cross-check our results. In Sec.~\ref{MatchToAsymptoticsBD} below, we will explicitly show that our result in Eq.~\eqref{NonLinDispMemoryModesBD}, precisely matches the memory extracted from the BMS balance laws in BD theory, which were previously computed in \cite{hou_gravitational_2021,tahura_brans-dicke_2021,hou_conserved_2021,hou_gravitational_2021_2}. In fact, we deliberately chose here to represent the memory contribution in terms of the SWSH modes, as it is in this form that memory can naturally be compared to the results from the asymptotic energy balance laws. In relating the BMS balance law results to our memory formula ind Sec.~\ref{MatchToAsymptoticsBD} will furthermore exemplify how the Isaacson approach can shed some light on the interpretation of memory as an independent low-frequency signal within the full non-linear approach to gravitational radiation. 

Moreover, our finding that there is no additional scalar null memory component in BD theory also agrees with the results in \cite{hou_gravitational_2021,tahura_brans-dicke_2021,hou_conserved_2021,hou_gravitational_2021_2} as they do not find a full BMS constraint for the scalar. In the terminology of \cite{tahura_brans-dicke_2021} this implies that non-trivial displacement contributions in the scalar polarization are no true memory components but represents rather more general \textit{persistent gravitational wave observables}. A permanent displacement in the scalar polarization contribution to the Riemann tensor therefore does not arise through an emission from any null radiation. However, a permanent scalar displacement can till potentially arise through other mechanisms.\footnote{See for instance \cite{du_gravitational_2016} where a non-vanishing shift in the scalar is shown to arise as a consequence of the no-hair theorem.} 

Finally, an explicit memory formula for BD theory was previously also reported in \cite{du_gravitational_2016}, which does however not agree with our result, in particular as concerns the form of the energy flux sourcing the memory. However, based on Theorem~\ref{Theorem1} together with the explicit cross-check against the BMS balance law computations, we are rather confident that Eq.~\eqref{NonLinDispMemoryModesBD} captures the correct memory contribution of BD theory.

\paragraph{f(R) Gravity.} On the other hand, recall that $f(R)$ gravity, with $f''(R)\neq 0$ is equivalent to choosing [Eq.~\eqref{eq:f(R)s SVH}]
\begin{align}\label{eq:f(R)s 2}
    G_2=f(\Phi)-\Phi f'(\Phi)\,,\qquad G_4&=f'(\Phi)\,,\qquad G_i=0 \quad\;\text{otherwise}\,.
\end{align}

Thus, we have that
\begin{equation}
    \rho=\sqrt{3}\sigma\,,\qquad \sigma= \frac{f''(\bar\Phi)}{f'(\bar\Phi)}\,,
\end{equation}
and therefore the radiative energy density governing the memory [Eq.~\eqref{eq:StressEnergyFourth}] becomes
\begin{align}\label{eq:StressEnergyFourth f(R)}
    t_{00}^{\myst{f(R)}}(u',r,\Omega')
    =\frac{1}{2\kappa^{\myst{f(R)}}_\text{eff}}\Bigg\langle |\dot{h}|^2+3\left(\frac{f''(\bar\Phi)}{f'(\bar\Phi)}\right)^2\,\dot{\varphi}^2\Bigg\rangle\,,
\end{align}
where 
\begin{equation}
    \kappa^{\myst{f(R)}}_{\text{eff}}=\frac{\kappa_0}{f'(\bar{\Phi})}\,.
\end{equation}

However, since the mass of the scalar field is in this case non-zero
\begin{equation}
    m^2=\frac{f'(\bar\Phi)}{3f''(\bar\Phi)}\,,
\end{equation}
the memory correction also depends on the frequency spectrum of the emitted scalar wave, as discussed above.

\paragraph{Scalar Gauss-Bonnet Gravity.}
On the other hand, recall that sGB theory can be obtained by choosing [Eq.~\eqref{eq:CorrespondencesGBHorndeski SVH}]
\begin{subequations}\label{eq:CorrespondencesGBHorndeski 2}
    \begin{align}
        G_2&=X+8f^{(4)}(\Phi)X^2(3-\ln X)\,,\\ G_3&=4f^{(3)}(\Phi)X(7-3\ln X)\,,\\
        G_4&=1+4f^{(2)}(\Phi)X(2-\ln X)\,,\\
        G_5&=-f^{(1)}(\Phi)\ln X\,,
    \end{align}
\end{subequations}
where $f^{(n)}(\Phi)\equiv\partial^n f/\partial\Phi^n$. 
Although at first glance the correspondence of sGB to Horndeski theory given by Eq.~\eqref{eq:CorrespondencesGBHorndeski 2} could therefore suggest that for nontrivial functions $f(\Phi)$ the sGB term could actually contribute nontrivially to the memory beyond the contribution from the kinetic term of the scalar field, this is not the case. A closer look reveals that
\begin{equation}
    3\frac{\bar G^2_{4,\Phi}}{\bar G^2_4}+\frac{(\bar G_{2,X}-2\,\bar G_{3,\Phi})}{\bar G_4}\Bigg\lvert_\text{sGB}=\lim_{X\rightarrow 0}\;1-4 w(X)\left(f^{(2)}(\varphi_0)+2f^{(4)}(\varphi_0)\right)X+...=1,
\end{equation}
where we defined $w(X)\equiv 2-\ln X$ as well as $f^{(N)}$ for the $N$th derivative of $f$. Hence, the higher-order sGB term does not modify the memory formula and the theory simply contributes through the canonical scalar term within Eq.~\eqref{eq:StressEnergyFourth} as
\begin{equation}\label{eq:sGBenergymomentum}
    t_{00}^{\myst{sGB}}
    =\frac{1}{2\kappa_0}\bigg\langle |\dot{ h}|^2+\,\dot{\varphi}^2\bigg\rangle\,.
\end{equation}
Indeed, by proving Theorem~\ref{Theorem1} we have shown that any term in the action involving more than two derivative operators will not modify the tensor memory in an explicit way.

Furthermore, since 
\begin{equation}
\sigma=\frac{\bar G_{4\Phi}}{\bar G_{4}}=0\,,
\end{equation}
recall that sGB gravity does not excite any additional scalar polarizations (breathing or longitudinal) within the physical metric, as opposed to the BD theory considered above. However, as discussed, the presence of the additional degree of freedom in the theory still explicitly modifies the tensor memory formula. Moreover, note that this also implies that sGB by definition only features tensor null memory.


\section{Memory from BMS Balance Laws}\label{MatchToAsymptoticsBD}
\small{\textit{Parts of this section are taken over from the original work \cite{Heisenberg:2023prj} of the author. Based on this remark, we will refrain from introducing explicit quotation marks to indicate direct citations.}}
\\

\normalsize
\noindent
The purpose of this section is two-fold. On the one hand, we would like to exemplify how the notion of an isolated memory component as the low-frequency signal of a gravitational wave arises in the context of the treatment of a fully non-linear treatment in asymptotically flat spacetimes. On the other hand, we want to explicitly make the connection between previous work on BMS balance laws and memory in Brans-Dicke theory and our general memory formula, as a valuable cross-check within this particular example.

Starting from the action of Brans-Dicke Gravity in Eq.~\eqref{ActionBD original}, together with a definition of asymptotic flatness that coincides with the one used in GR (see e.g. \cite{Geroch:1977jn,Ashtekar:1981bq,Ashtekar:2014zsa,WaldBook,DAmbrosio:2022clk})\footnote{Note, however, that, for instance, certain aspects of the peeling theorem need no longer hold when the field equations are not Einstein's.}, the authors in  \cite{hou_gravitational_2021,tahura_brans-dicke_2021,hou_conserved_2021,hou_gravitational_2021_2} arrive at the BMS supermomentum flux-balance law in asymptotic spherical Bondi coordinates $\{u,r,x^A\}$, $x^A=\{\theta,\phi\}$ [see for instance Eqs.~(10)--(12) in \cite{hou_gravitational_2021_2}, from which we also adopt the notation]
\begin{equation}\label{eq:BMSBalanceLawI}
   0=\frac{\varphi_0}{4\pi G}\int_{S^2}d^2\Omega\,\alpha \Bigg\{\Delta M+\int_{-\infty}^\infty\dd u'\Bigg[\frac{1}{2}N_{AB}N^{AB}+\mathcal{D}_A\mathcal{D}_BN^{AB}+(2\omega+3)\left(\frac{N}{\varphi_0}\right)^2\Bigg]\Bigg\}\,,
\end{equation}
where $\varphi_0$ is the asymptotic value of the scalar field, $\mathcal D_A$ is the covariant derivative on $S^2$, $\alpha=\alpha(\theta,\phi)$ is an arbitrary function on $S^2$ parametrizing supertranslations, and $M$ denotes the Bondi mass aspect that is associated to the total energy of the isolated system through Eq.~\eqref{eq:Supermomentum Charge E}, while 
\begin{equation}
    \Delta M\equiv M(u\rightarrow\infty)-M(u\rightarrow-\infty)\,.
\end{equation}
Moreover, as in \cite{hou_gravitational_2021_2} we write
\begin{equation}\label{eq:BondiNews}
     N_{AB}\equiv -\dot{\hat{c}}_{AB}\,, \quad N\equiv \dot\varphi_1\,,
\end{equation}
where $\hat{c}_{AB}$ is the symmetric, traceless and transverse shear tensor and $\varphi_1$ the component of the scalar that falls off as $\sim 1/r$.

In short (see e.g. \cite{Strominger:2017zoo,Compere:2019gft,DAmbrosio:2022clk}), the BMS flux balance laws are a statement of conservation of a charge
\begin{equation}\label{eq:Supermomentum Charge}
    \mathcal{Q}_{(\alpha)}(u)\propto\int_{S^2}d\Omega^2\,\alpha(\theta,\phi)\,M(u,\theta,\phi)
\end{equation}
associated to an asymptotic Killing vector field of so called \textit{supertranslations}
\begin{equation}
    \xi_{(\alpha)}=\alpha \partial_u\,.
\end{equation}
Indeed, the fixed structure of asymptotically flat spacetimes allows for the question of asymptotic isometries with associated KVFs and conserved charges of the underlying BMS group. However, in contrast to the statements back in Sec.~\ref{sSec: Theories of Minkowski Spacetime} on the conservation of energy in Minkowski spacetime, the asymptotic supermomentum charges are only conserved up to a flux term at the boundary of the asymptotically flat spacetime.

\begin{figure}
    \centering
    \includegraphics[width=0.56\textwidth]{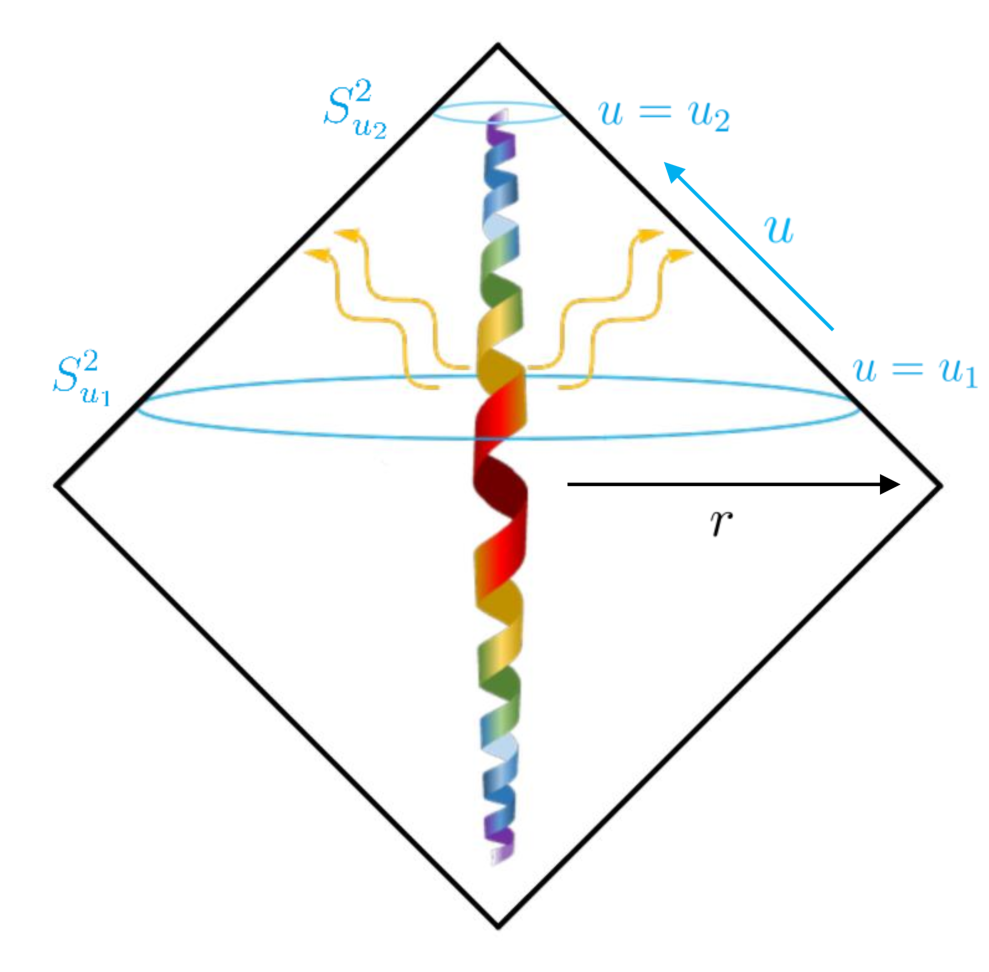}
    \caption{\small Penrose diagram representation of a conformally completed asymptotically flat spacetime in asymptotic light-cone coordinates $\{u,r,\theta,\phi\}$, with $u=t-r$ and where time $t$ flows vertically. A localized source at the center of the spacial coordinate system is represented to emit null radiation (yellow) towards asymptotic null infinity coordinatized by the asymptotic retarded time $u$. The angular coordinates are not shown, but the asymptotic two-spheres at the instances $u_2$ and $u_1$ are represented by circles (blue) that leave the 2D plane. (Figure adapted from \textit{F. D'Ambrosio, S. Fell, L. Heisenberg, D. Maibach, S. Zentarra, J. Zosso, (2022)} \cite{DAmbrosio:2022clk}.)}
    \label{fig:Asymptotically flat spacetime}
\end{figure}

More precisely, in analogy to Eq.~\eqref{eq:Energyconservation Minkowski}, Eq.~\eqref{eq:BMSBalanceLawI} states that the super momentum charge in Eq.~\eqref{eq:Supermomentum Charge} between an instant $u_2$ and $u_1$ on the asymptotic two-sphere is only conserved up to the flux crossing the boundary of the volume defined between the two spheres $S^2_{u_2}$ and $S^2_{u_1}$
\begin{equation}
    \mathcal{Q}_{(\alpha)}(u_2)-\mathcal{Q}_{(\alpha)}(u_1)=-\mathcal{F}_{(\alpha)}(u_1,u_2)\,,
\end{equation}
where
\begin{equation}
    \mathcal{F}_{(\alpha)}(u_1,u_2)\propto  \int_{u_1}^{u_2}\dd u'\int_{S^2}d^2\Omega\,\alpha\Bigg[\frac{1}{2}N_{AB}N^{AB}+\mathcal{D}_A\mathcal{D}_BN^{AB}+(2\omega+3)\left(\frac{N}{\varphi_0}\right)^2\Bigg]\,.
\end{equation}
These statements are visually represented in a Penrose diagram in Fig.~\ref{fig:Asymptotically flat spacetime}, where an isolated source is producing null radiation, causing a change in the asymptotic supermomentum charge over time. In fact, for $\alpha(\theta,\phi)=1$, the supermomentum charge corresponds to the energy of the isolated system
\begin{equation}\label{eq:Supermomentum Charge E}
    E(u)\propto\int_{S^2}d\Omega^2\,M(u,\theta,\phi)\,,
\end{equation}
and the balance law [Eq.~\eqref{eq:BMSBalanceLawI}] simply reduced to a statement of conservation of energy
\begin{equation}\label{eq:BMSBalanceLawI 2s}
   \Delta E \propto \int_{-\infty}^\infty\dd u'\int_{S^2}d^2\Omega\Bigg[\frac{1}{2}N_{AB}N^{AB}+(2\omega+3)\left(\frac{N}{\varphi_0}\right)^2\Bigg]\Bigg\}\,,
\end{equation}
where the right-hand side precisely corresponds to the asymptotic energy flux. Observe, however, that therefore for a non-trivial angular dependence of the supertranslation, there is an additional contribution $\mathcal{D}_A\mathcal{D}_BN^{AB}$ to the change in supermomentum charge. It turns out, that precisely this additional contribution can be associated to a memory effect as a permanent change in the asymptotic shear tensor that induced a permanent offset in the proper distance \cite{Strominger:2014pwa,Strominger:2017zoo,Compere:2019gft}.

The goal will now thus be to massage the BMS balance law in Eq.~\eqref{eq:BMSBalanceLawI} into a form, from which this null-memory component can be extracted. For this we first of all expand the shear on the asymptotic two-sphere as
\begin{equation}\label{eq:ShearExpand}
    \hat{c}_{AB}= \bar{c} \,\bar m_A\bar m_B+ c\, m_A m_B\,,
\end{equation}
where in this context an overbar denotes complex conjugation and where in spherical coordinates [Eq.~\eqref{eq:Def m}]
\begin{equation}
    m=\frac{1}{\sqrt{2}}\left(\partial_\theta+i\sin\theta\partial_\phi\right)\,.
\end{equation}
Using Eqs.~\eqref{eq:BondiNews} and \eqref{eq:ShearExpand}, as well as the definition of the angular derivative operator [Eq.~\eqref{eth}], which implies that we have 
\begin{equation}
    \mathcal{D}_A\mathcal{D}_B \hat{c}^{AB}=\frac{1}{2}\left(\eth^2 c+\bar{\eth}^2\bar{c}\right)\,,
\end{equation}
the flux-balance law in Eq.~\eqref{eq:BMSBalanceLawI} can be rewritten as
\begin{align}\label{BMSSupermomentumBalanceLawBD}
  \int_{S^2} d^2\Omega\,\alpha\, \Delta M=\frac{1}{4}\int_{-\infty}^\infty d u'\int_{S^2} d^2\Omega\,\alpha\Bigg[\,|\dot{c}|^2-\Re[\eth^2\dot{c}]+(2\omega+3)\left(\frac{\dot\varphi_1}{\varphi_0}\right)^2\Bigg]\,.
\end{align}

To single out the tensor null memory from the above relation, we can first set the subdominant, left-hand side to zero. In fact, this contribution to the permanent change in the shear is associated to the ordinary memory. We then rewrite the BMS supermomentum balance law in Eq.~\eqref{BMSSupermomentumBalanceLawBD} by moving the second to last term to the left, while carrying out the $u'$ integral to obtain
\begin{align}\label{BMSSupermomentumBalanceLawBD2}
  \int_{S^2} d^2\Omega\,\alpha\,\Re[\eth^2\Delta{c}]=\int_{-\infty}^\infty d u'\int_{S^2} d^2\Omega\,\alpha\Bigg[\,|\dot{c}|^2+(2\omega+3)\left(\frac{\dot\varphi_1}{\varphi_0}\right)^2\Bigg]\,.
\end{align}
Furthermore, expanding $c$ with spin-weight $s=-2$ on the left-hand side as
\begin{equation}
   c(u,\theta,\phi)=\sum_{l=2}^\infty \sum_{m=-l}^l c_{lm}(u)\,_{\scriptscriptstyle-2}Y_{lm}(\theta,\phi)\,,
\end{equation}
using the relation [Eq.~\eqref{eq:ASWSHid1}] 
\begin{equation}
    \eth^2\phantom{}_{\scriptscriptstyle-2}Y_{lm}=\sqrt{\frac{(l+2)!}{(l-2)!}}\,Y_{lm}\,,
\end{equation}
as well as choosing $\alpha(\theta,\phi)=Y^*_{lm}(\theta,\phi)$, we obtain
\begin{align}\label{BMSSupermomentumBalanceLawBD3}
\begin{split}
  \frac{1}{2}\big(\Delta c^{lm}+(-1)^m&\Delta\bar{c}^{l-m}\big)=\sqrt{\frac{(l-2)!}{(l+2)!}}\int_{S^2} d^2\Omega\,Y^*_{lm} \int_{-\infty}^\infty d u'\left[\,|\dot{c}|^2+(2\omega+3)\left(\frac{\dot\varphi_1}{\varphi_0}\right)^2\right]\,.
\end{split}
\end{align}

To continue, we want to note that the symmetric and traceless shear tensor on the transverse $2$-sphere in Eq.~\eqref{eq:ShearExpand} also naturally defines the rank-2 TT tensor
\begin{equation}
    \hat c_{ij}\equiv e_i^Ae_j^B \hat c_{AB}=c\, m_im_j+\bar{c} \,\bar{m}_i\bar{m}_j
\end{equation}
where $e_i^A$ represent the embedding of the unit $S^2$ basis 
\begin{equation}\label{eq:embedding}
    e^{i}_A=\frac{\partial n^i}{\partial x^A}
\end{equation} 
This basis change is such that
\begin{equation}\label{eq:embedding res1}
    e_{i}^Am_A=m_i\,,
\end{equation}
as well as
\begin{equation}\label{eq:embedding res2}
    e_{i}^Ae_{j}^B\, \gamma_{AB}=2m_{(i}\bar{m}_{j)}=\delta_{ij}-n_in_j\,,
\end{equation}
while
\begin{equation}
    \delta_{ij}e^{i}_Ae^{j}_B=\gamma_{AB}=2m_{(A}\bar{m}_{B)}\,.
\end{equation}

Thus, using the spin-2, TT tensor harmonic expansion in Eq.~\eqref{eq:AHlmToUlmVlm} of $\hat c_{ij}$, we can separate $c^{lm}$ into its electric- and magnetic-parity moments
\begin{equation}\label{eq:clmToUlmVlm}
    c^{lm}=\frac{1}{\sqrt{2}}\left[U_c^{lm}-iV_c^{lm}\right]\,.
\end{equation}
Observe that the left-hand side in Eq.~\eqref{BMSSupermomentumBalanceLawBD3} therefore precisely corresponds to the electric-parity part [Eq.~\eqref{eq:AUVlmToHlm}]. We thus finally arrive at
\begin{align}\label{BMSSupermomentumBalanceLawBD4}
 \boxed{\Delta U_c^{lm}=\sqrt{\frac{2(l-2)!}{(l+2)!}}\int_{S^2} d^2\Omega\,Y^*_{lm} \int_{-\infty}^\infty d u'\left[\,|\dot{c}|^2+(2\omega+3)\left(\frac{\dot\varphi_1}{\varphi_0}\right)^2\right]\,.}
\end{align}

From the balance laws, we can therefore single out the total tensor displacement memory, and hence, the lasting nonzero component within the electric-parity multipole of the shear $c$, which ultimately induces a lasting offset in the detector strain. Note, however, that the shear and the scalar field which enter the balance laws are the total shear and scalar field at $\mathcal{O}(r^{-1})$ within the full nonlinear theory, and therefore, in particular, they already contain all possible memory contributions.
Nevertheless, the result can be interpreted as a computation of the total memory offset after the passage of the gravitational waves. This is because the BD energy flux that enters the right-hand side of the Eq.~\eqref{BMSSupermomentumBalanceLawBD4}, vanishes as $u\rightarrow \pm\infty$, where, by assumption, no gravitational waves reach null infinity.

However, to actually use the BMS balance laws as a tool to compute the low-frequency displacement memory characterized by the measurable monotonically increasing and non-oscillatory, time-dependent raise of the memory, which is what gravitational wave detectors are sensitive to (recall Sec.~\ref{ssSec:FutureProspects MemoryDetection}), requires a slight reinterpretation of Eq.~\eqref{BMSSupermomentumBalanceLawBD4}. More precisely, a time dependent extraction of memory in fact requires an Isaacson-type to distinguish between a high- and low-frequency part of the shear and the scalar
\begin{equation}\label{BMSSupermomentumBalanceLawBD5pre}
    c=c^L+c^H\,,\quad \varphi_1= \varphi_1^L+\varphi_1^H\,,
\end{equation}
in order to gradually integrate over retarded time, while extracting the low-frequency part of the expression by averaging out the small scales.\footnote{Observe that in order to compute the full memory in Eq.~\eqref{BMSSupermomentumBalanceLawBD4}, such an averaging is irrelevant.} Only then will it be possible to connect the BMS balance-law result to the computation in Eq.~\eqref{sourceTensorMemoryBD}.

After averaging, any cross terms of the form ``$c^Lc^H$'' or ``$\varphi_1^L\varphi_1^H$'' on the right-hand side in Eq.~\eqref{BMSSupermomentumBalanceLawBD4} will vanish. Moreover, we assume that we can neglect any contribution of low-frequency components ``$c^Lc^L$'' or ``$\varphi_1^L\varphi_1^L$'' which can be interpreted as the 
``memory of the memory''. In other words, we assume that the source modes for the memory themselves have a negligible memory component, which is indeed a reasonable assumption \cite{Talbot:2018sgr}. Furthermore imposing $c(u\rightarrow-\infty)=0$, we therefore have
\begin{align}\label{BMSSupermomentumBalanceLawBD5}
  \boxed{c_L^{lm}(u)=\sqrt{\frac{(l-2)!}{(l+2)!}}\int_{S^2}\dd^2\Omega\,Y^*_{lm}\int_{-\infty}^u\dd u'\Bigg\langle\,|\dot{c}^H|^2+(2\omega+3)\left(\frac{\dot\varphi^H_1}{\varphi_0}\right)^2\Bigg\rangle\,,}
\end{align}
where $c_L^{lm}(u)$ is the resulting low-frequency correction to the shear, given high-frequency radiation modes $c^H$ and $\varphi_1^H$. Here we have used Eq.~\eqref{eq:clmToUlmVlm} with $\delta V_c^{lm}=0$ to rewrite Eq.~\eqref{BMSSupermomentumBalanceLawBD5pre} in terms of the shear.


As a last step before finally being able to compare results, we need to connect the perturbative shear and scalar field defined here with the perturbations used in the main text and ensure that these are indeed the same quantities. In the case of BD theory, the easiest way to establish this correspondence is to compare the corresponding leading $\mathcal{O}\left(r^{-1}\right)$ terms of the electric part of the Riemann tensor. In \cite{hou_gravitational_2021,tahura_brans-dicke_2021}, these terms were computed and found to be (see e.g. Eq.~(2.44) in \cite{hou_gravitational_2021})\footnote{Note, however, that the authors in \cite{hou_gravitational_2021,tahura_brans-dicke_2021} report the result in an orthonormal tetrad basis, instead of the spherical coordinates employed here.}
\begin{equation}
    R_{uAuB}=-\frac{1}{2r}\left(\ddot{\hat{c}}_{AB}-\gamma_{AB}\frac{\ddot{\varphi}_1}{\varphi_0}\right)+\mathcal{O}\left(\frac{1}{r^2}\right)\,.
\end{equation}
By using the embedding of the unit $S^2$ basis defined in Eq.~\eqref{eq:embedding}, we can convert the leading-order expression to a $\{t,x,y,z\}$ Minkowski basis, which yields
\begin{align}\label{RiemannAsymptotic}
    R_{0i0j}&=e_{i}^Ae_{j}^BR_{uAuB}=-\frac{e_{i}^Ae_{j}^B}{2r}\left(\ddot{\hat{c}}_{AB}-\gamma_{AB}\frac{\ddot{\varphi}_1}{\varphi_0}\right)\,,\notag\\
    &=-\frac{1}{2r}\bigg(\underset{= \frac{1}{2}e^+_{ij}\,(\ddot{c}+\ddot{\bar{c}})+\frac{i}{2}e^+_{ij}\,(\ddot{c}-\ddot{\bar{c}})}{\underbrace{m_im_j\,\ddot{c}+\bar m_i\bar m_j\,\ddot{\bar{c}}}}-(\delta_{ij}-n_in_j)\frac{\ddot{\varphi}_1}{\varphi_0}\bigg)\,,\notag\\
    &=-\frac{1}{2r}\left(e^+_{ij}\,\ddot c_++e^\times_{ij}\,\ddot c_\times-(\delta_{ij}-n_in_j)\frac{\ddot{\varphi}_1}{\varphi_0}\right)\,,\notag
\end{align}
where we used Eqs.~\eqref{eq:embedding res1} and \eqref{eq:embedding res2} and we defined 
\begin{equation}
    c_+\equiv \Re [c]\,,\qquad c_\times \equiv -\Im [c]\,.
\end{equation}
Hence, comparing to Eq.~\eqref{ElectricRiemannLin} with 
\begin{equation}
    \sigma=\frac{1}{\varphi_0}=\frac{1}{\bar\Phi}\,,
\end{equation}
we obtain the correspondence
\begin{subequations}
\begin{align}
    c^H_+(u,\Omega)&=\lim_{r\rightarrow\infty}rh_+(u,r,\Omega)\,,\\
    c^H_\times(u,\Omega)&=\lim_{r\rightarrow\infty}rh_\times(u,r,\Omega)\,,\\
   \varphi^H_1(u,\Omega)&=\lim_{r\rightarrow\infty}r\varphi(u,r,\Omega)\,,
\end{align} 
\end{subequations}
while therefore as well
\begin{equation}
   c_L^{lm}=\lim_{r\rightarrow\infty}r\delta h^{lm}\,,
\end{equation}
such that Eq.~\eqref{BMSSupermomentumBalanceLawBD5} indeed corresponds to the result in Eq.~\eqref{NonLinDispMemoryModesBD}. This represents a powerful crosscheck of our results.

\section{Summary and Outlook}\label{eq:Summary and Outlook}

In summary, the Isaacson approach to gravitational waves, whose generalization to metric theories of gravity beyond GR was presented in Chapter~\ref{Sec:PropagatingDOFs}, provides a powerful and conceptually sound approach to investigating the memory effect in generic theories of gravitation. These considerations culminated in the Theorem \ref{Theorem1} for the functional form of the dominant tensor null memory. The essence of the theorem states that very generically, for dynamical metric theories of gravity defined in Definition~\ref{DefMetricTheory} with a viable EFT expansion, the tensor null memory is of the form in Eq.~\eqref{NonLinDispMemory Thm}. In other words, null memory is modified in comparison to GR in two ways: (I) through contributions to the energy fluxes at null infinity of additional, massless, dynamical degrees of freedom in the theory; and (II) through modifications in the generation and propagation of the leading-order tensor perturbations. 

This simple result could have interesting implications, as we will now discuss in more detail. First of all, as presented in Sec.~\ref{ssSec:FutureProspects MemoryDetection}, 
planned space-based and next-generation ground-based detectors are expected to provide the first direct measurements of the tensor memory effect in the near future. These observations of gravitational wave memory may play an important role in establishing a better understanding of gravity and constraining modifications of GR. This is because tensor memory is a very special, nonlinear correction to the gravitational wave response. Indeed, the memory's dominant null component can be understood as being sourced by the leading-order wave front itself. A detection of memory would therefore represent a first direct measurement of the ``ability of gravity to gravitate'', reflecting its inherent non-linear nature. Furthermore, concerning CBCs the memory signal is mainly sensitive to the merger of an event, and thus precisely targets the phase of strongest gravity where beyond GR effects might emerge. In this context, the computation of the explicit formula for null memory in the most general, massless SVT theory with second-order equations of motion, together with its subsequent generalizations, represents a significant step towards memory based modeled searches of GR deviations in the fully non-linear regime that require a parameterization for the beyond-GR effects.

But foremost, one of the main discoveries presented above, namely the simple but very generic conclusion that the functional form of the tensor null memory of dynamical metric theories of gravity is merely modified from the GR expectation through the presence of additional null fluxes of extra non-minimal fields, already provides an important hint towards an exciting future application of memory observations. Namely, the memory effect could potentially be exploited to develop a largely model-insensitive test of GR of perhaps the most straightforward manifestation of new physics [Chapter~\ref{Sec:The Theory Space Beyond GR}]: the existence of additional scalar, vectorial or tensorial propagating degrees of freedom. Since memory is sensitive to any kind of energy-momentum emitted from the source, such a test would not only complement ongoing searches for additional gravitational polarizations [Secs.~\ref{sSec:GWPolGen} and \ref{ssSec: GW Experiments}], but also extend the sensitivity of such tests to scenarios in which additional gravitational fields do not excite any other polarization modes of the physical metric.

More precisely, a model independent extraction of both the primary TT high-frequency signal, as well as the memory signal could in principle be used as a cross-check between the computed and measured displacement memory, the failure of which would necessarily point towards the presence of additional degrees of freedom.
The above statement involves of course many idealizations, and, for instance, necessitate additional independent information on the inclination angle in order to extrapolate the angular distribution over the sky of the primary signal that could be obtained through an electromagnetic counterpart or the detection with multiple interferometers. Nevertheless, such a universal consistency check might provide an interesting new handle, for instance in determining whether potential beyond GR signals within gravitational wave observations really do originate from the presence of additional non-minimal degrees of freedom in the phenomenon of gravitation.

On the other hand, the present work may also serve as a basis for future advances on the theoretical side. For instance, a definition of the memory-evolution equation as the leading-order low-frequency equation of motion in the Isaacson picture, is expected to allow for an even broader study of memory beyond GR. Indeed, in principle, the Isaacson approach could also be applied to the field equations of theories which we explicitly disregarded in this work, such as theories with non-dynamical field content or local Lorentz breaking. Note that such a local Lorentz symmetry breaking can either occur explicitly as in Einstein-\AE{}ther theories discussed in Eq.~\eqref{eq:Action EA} or spontaneously in the asymptotic background solution in theories with additional non-minimal massive vector fields such as generalized Proca [Eq.~\eqref{eq:ActionGenProca}]. In such a general setting, one could also think about solving the resulting memory-evolution equation not in the vicinity of null infinity, but in a different appropriate limit. 
This last point could potentially also be explored for the study of the tensor memory in metric theories with broken diffeomorphism invariance, such as massive gravity theories \cite{deRham:2014zqa}.

Another avenue for future work would be to investigate a possible generalization of the BMS balance laws to a wide class of theories, based on the results obtained in this paper. In light of the close connection between BMS balance laws and tensor null memory established in Sec.~\ref{MatchToAsymptoticsBD} in the case of BD theory, our results strongly suggest that the approach of \cite{hou_gravitational_2021,tahura_brans-dicke_2021,hou_conserved_2021,hou_gravitational_2021_2} for BD gravity may be readily generalized to the asymptotic structure of any dynamical metric theories. It would be interesting to explore this conjecture in detail, especially with regard to the ordinary memory, which we have explicitly excluded in the discussion in Sec.~\ref{MatchToAsymptoticsBD}, as well as what concerns scalar or vector null memory, briefly discussed in Sec.~\ref{sSec:Scalar and Vector Memory}. 


\newpage
\thispagestyle{plain} 
\mbox{}


\part{The Cosmological Testing Ground}\label{Part: Cosmological Testing Ground}

\small
\noindent
\emph{\ul{Personal Contribution and References}}\\ 
\footnotesize 
\textit{Parts of Chapter~\ref{Sec:CosmoNutshell} are inspired from the treatment in \cite{Weinberg1972,Weinberg2008Cosmology,zee2013einstein,maggiore2018gravitationalV2,dodelson2020modern}. Chapter~\ref{Sec:CsomoTensions} is based on \textbf{L. Heisenberg, H. Villarrubia-Rojo, J. Zosso, (2022), (2023)} \cite{Heisenberg:2022gqk,Heisenberg:2022lob}. While H.V. conceived the ideas and performed initial computations, including the numerical simulations, J.Z. contributed by writing all
Mathematica notebooks of the project in parallel to the python based code of H.V., verifying and partially
extending early results. All authors contributed to the conceptual developments, the discussion of the results and the final manuscripts. L. H. supervised the projects.}
\normalsize

\vspace{5mm}

\noindent
\textbf{Summary of Part III}\\ 
\noindent
Describing spacetime as a dynamical concept within metric theories of gravity especially also allows for an understanding of the astonishing finding that the spacetime structure of the universe has not always been as it is today. The study of this evolution represents another major gravitational testing ground, as it poses a serious challenge to the underlying theoretical description, hopefully providing a guideline on the search for new physics. While the dark sector, as well as the mechanism for initial conditions of the cosmological standard model remains poorly understood, recent cosmological tensions might even indicate the necessity for a departure from the current general relativity based model itself with increasing statistical significance.

After introducing today's cosmological standard model, we will present a largely model independent approach that is well suited for an analytical study of cosmological tensions. This will allow for the formulation of necessary conditions that a large class of standard model departures need to satisfy in order to consistently alleviate the tensions. Such general constraints can be viewed as first guiding principles towards formulating viable alternative theories, in particular with implications on the metric theory space.

\chapter{Cosmology in a Nutshell}\label{Sec:CosmoNutshell}

In this chapter, we provide a lightning overview of the current GR based cosmological standard model and introduce all necessary concepts for the subsequent chapter on the exciting prospect of learning about new physics through current tensions in cosmological observables. For a more in depth treatment of the rich subject of cosmology, we refer the reader to the reviews in \cite{Weinberg1972,Bertschinger:1993xt,Peebles:1994xt,Coles:1995bd,Liddle:2000cg,landau_classical_2003,Mukhanov:2005sc,Weinberg2008Cosmology,Gorbunov:2011zzc,zee2013einstein,maggiore2018gravitationalV2,dodelson2020modern}.


\section{A Homogeneous and Isotropic Universe}\label{sSec:HomIsoUniverse}

\subsection{The Cosmic Metric and its Evolution Equations}

\paragraph{The Spacetime Geometry.} 
The philosophical hypothesis of anti-anthropocentrism, backed up by experimental evidence, suggest that on the largest scales, spacial slices of the universe we live in are homogeneous and isotropic. More precisely, the expectation that there are no preferred spacial locations in the universe, together with the observation that on cosmic scales the universe is spherically symmetric about us, requires that on average, the cosmos is isotropic about every point \cite{Weinberg1972,Weinberg2008Cosmology}. This is the essence of the so-called \textit{Cosmological Principle} (see below). 

Note, however, that a priori, the above arguments for the cosmological principle do not explicitly imply homogeneity and isotropy at \textit{all times}, although this additional assumption is in most cases implicit. 
Moreover, we want to stress that isotropy is first of all an assumption that is based on empirical data, in particular of the measured cosmic microwave background to be introduced below. This experimental evidence is however only provided up to a fundamental dipole subtractions that might solely be due to our own peculiar velocity with respect to a cosmic rest-frame, but might in principle also hide a fundamental cosmic anisotropy \cite{Secrest:2022uvx,Peebles:2022akh}. 
Furthermore, in a scenario of a universe with a beginning, pure theoretical considerations based on the assumption of random initial conditions would rather expect a universe with multiple causal patches, as we will discuss in more detail below. Nevertheless, the assumption of isotropy seems consistent with the largest parts of today's data and represents a decisive basis of the current standard model. This allows for the definition of freely falling observers, the \textit{comoving observers}, that move together with the average velocity of free-falling matter in the universe. It is with respect to such ``typical'' observers that the universe is assumed to be isotropic.

Mathematically, homogeneity and isotropy impose very tight constraints on the functional form of the spacetime metric describing the universe on cosmic scales, which is assumed to represent a \textit{background metric} that can be used as an exact solution about which perturbation variables can be considered (recall Sec.~\ref{sSec:PerturbationTheory}). In fact, for a spacial metric in three space dimensions, the requirements of the invariance of the metric under rotations and translations, in other words the existence of six Killing vector fields associated to the six independent isometries (see Sec.~\ref{sSec:Special Relativity} and App.~\ref{sApp: Spacetime Gaugefreedom and symmetries}) implies that the metric is maximally symmetric since for $d=3$, the maximal number of KVFs is [Eq.~\eqref{eq:NumerofMaximallySymmetricKVF}]
\begin{equation}
    \frac{d(d+1)}{2}=6\,.
\end{equation}
Thus, the spacial metric can only represent three different types of spaces: flat Euclidean space $\mathbb{R}^3$, the sphere $S^3$ and its negatively curved counterpart the hyperboloid $H^3$ (see e.g. \cite{Weinberg1972,zee2013einstein,carroll2019spacetime}). Therefore, a mathematically precise formulation of the cosmological principle reads:
\begin{principle}\ul{Cosmological Principle}.\label{Principle:Cosmological}
The background metric describing the observable universe can be foliated into maximally symmetric spacetime slices.
\end{principle}
This is completely analogous to the notion of maximally symmetric spacetimes involving ten Killing vector fields in four spacetime dimensions with the three possibilities of Minkowski, de-Sitter and Anti-de-Sitter spacetimes (recall the discussion in Sec.~\ref{sSec:Special Relativity}). The difference, is that spacial homogeneity and isotropy only determines the symmetries of the spacial metric and leaves room for a breaking of time translation invariance and three ``Lorentz boosts''. One of the key findings of modern cosmology is in fact the breaking of time translations, describing a non-trivially evolving universe.

However, the extensions of such maximally symmetric spaces to a corresponding spacetime is still highly restricted. Indeed, it can be shown \cite{Weinberg1972} that the only possibility is a metric that admits the existence of coordinates $(t,x^i)$ in which the line element (defined in Eq.~\eqref{eq:LineElementDef}) takes the form known as Friedmann-Lemaitre-Robertson-Walker (FLRW) metric solution
\begin{equation}\label{eq:LineElementHomIsoGenX}
    ds^2=  - dt^2 + a^2(t) \left(\delta^{ij} +k \,\frac{x^ix^j}{1-k\mathbf{x}^2}\right)dx^idx^j\,.
\end{equation}
where $a(t)$ is the so-called \textit{scale factor} for which we assume that $a(t)>0$.\footnote{Naively, a value of $a(t)=0$ would describe the vanishing of space associated to a ``beginning of the universe'' commonly known as ``Big Bang''.} Moreover, the constant $k$ controls the sign of the curvature scalar defined in Eq.~\eqref{eq:CurvatureScalarK}, and hence determines the type of the space
\begin{equation}
		k=\left\{\begin{array}{l}
		\displaystyle
			+1\qquad\text{sphere}\\[8pt]
			\displaystyle
			-1\qquad\text{hyperboloid}\\[8pt]
			\displaystyle
			\phantom{+}0\qquad\text{Euclidean}\,.
		\end{array}\right.
	\end{equation}

Observe that in these coordinates, the Christoffel symbol $\Gamma^i_{00}=0$ vanishes.
This implies that free test particles following the geodesics of Eq.~\eqref{eq:GeodesicEq} that are initially at rest, remain at rest.\footnote{In this aspect, this coordinate system is equivalent to the TT gauge discussed in Part~\ref{Part: Gravitational Wave Testing Ground}.} Hence, the coordinates $x^i$ follow the motion of an expected mean motion of typical freely falling observers associated to the comoving observers introduced above and are thus known as \textit{comoving coordinates}. Moreover, because $g_{tt}=-1$, the cosmic time $t$ also corresponds to the proper time of such comoving observers. Indeed, the FLRW coordinates precisely correspond to a set of \textit{global} synchronous coordinates (recall Sec.~\ref{sSec:Special Relativity}).

Through a change of variables to spherical coordinates $(t,r,\theta,\phi)$ for which
\begin{equation}
    d\mathbf{x}^2=dr^2+r^2d\Omega^2\,\quad \text{where}\quad d\Omega^2\equiv d\theta^2+sin^2\theta\,d\phi^2\,,
\end{equation}
the line element in Eq.~\eqref{eq:LineElementHomIsoGenX} becomes
\begin{equation}\label{eq:LineElementHomIsoGenR}
    \boxed{ds^2=  - dt^2 + a^2(t) \left(\frac{dr^2}{1-kr^2} +r^2 d\Omega^2\right)\,.}
\end{equation}
Since current observations constrain the universe to be approximately flat, we will in the following for simplicity choose $k=0$ and therefore only consider flat universes. In this case, the background metric simply becomes
\begin{equation}\label{eq:FlatFLRWmetric}
     ds^2=  - dt^2 + a^2(t)\, d\mathbf{x}^2=  - dt^2 + a^2(t) \left(dr^2 +r^2 d\Omega^2\right)\,.
\end{equation}
Moreover, note that for $k=0$, the overall normalization of the scale factor has no significance, as it corresponds to a simple rescaling of the coordinates $\mathbf{x}$. Hence, only ratios of the values of $a(t)$ matter. As it is custom, we choose a scale factor of unity $a(t_0)=1$ at the cosmic time today, conventionally denoted as $t_0$. We want to mention, however, that most of the subsequent discussion actually would also go through for arbitrary values of $k$ (see e.g. \cite{Weinberg2008Cosmology,carroll2019spacetime}).

\paragraph{A Dynamical Universe.} 
The physical meaning of the scale factor $a(t)$ can be enlightened by considering the notions of proper spacial distances in such cosmological spacetimes. However, the notion of proper \textit{spacial} distance in GR is only well-defined locally, due to the lack of a global notion of simultaneity. Indeed, locally, that is within a region of a non-evolving metric, simultaneity can for instance be established by bouncing light between two comoving observers as we already discussed in some detail in Sec.~\ref{sSec:Special Relativity}. The associated proper distance $d\ell$ in terms of comoving coordinates is given by Eq.~\eqref{eq:ProperDistanceGeneral}, which for synchronous coordinates simply corresponds to the spacial part of the metric and therefore reads
\begin{equation}\label{eq:ProperDistanceCosmologyLocal}
    d\ell^2=\left(g_{ij}-\frac{g_{0i}g_{0j}}{g_{00}}\right)dx^idx^j=a(t)^2d\mathbf{x}^2=a^2(t) \left(dr^2 +r^2 d\Omega^2\right)\,.
\end{equation}
Thus, one could define a set of physical spacial coordinates
\begin{equation}
    \mathbf{x}_{\text{ph}}\equiv a(t) \mathbf{x}\,,
\end{equation}
that more faithfully represent physical distances in a local sense. 

Moreover, it is tempting to integrate the relation in Eq.~\eqref{eq:ProperDistanceCosmologyLocal} and define a global notion of purely spacial ``proper distance'' between the origin and another comoving object at radial comoving coordinate $R$ at some instant $t$
\begin{equation}\label{eq:ProperDistanceCosmologyGlobal}
    \ell(R,t)=a(t) \int^R_0 dr =a(t) R\,.
\end{equation}
However, for a non-trivially time evolving universe there is not really a practical notion of such a physical spacial distance, in the sense of the absence of any realistic construction of simultaneous events that are essential for the definition of proper spacial distances (recall Sec.~\ref{sSec:Special Relativity}). In order to view Eq.~\eqref{eq:ProperDistanceCosmologyGlobal} as an actual proper distance between two galaxies, one would require a ``cosmic conspiracy'' \cite{Weinberg1972,zee2013einstein}, in which at the cosmic time $t$ a series of comoving observers are lined up between the origin and $R$, all of which performing local light signal travel time experiments between local neighbors. Nevertheless, the assumed existence of a preferred global cosmic time $t$ still justifies the conceptual interpretation of Eq.~\eqref{eq:ProperDistanceCosmologyGlobal} as a physical distance within these coordinates. In other words, such a definition of a global notion of spacial proper distances is rendered possible through the assumed existence of the global set of synchronous coordinates represented by the FLRW chart, which as we will see coincides with alternative and more pragmatic measures of distance in cosmology in the limit of small $R$.

Based on the above discussion, the scale factor $a(t)$ may effectively be viewed as controlling the physical ``size'' of the universe. A change in $a(t)$ represents an expansion or contraction of the universe as a whole that manifests itself as an increase of physical ''distances'' between comoving observers. 
This expansion history is conveniently captured by the fractional rate of change of $a(t)$ known as \textit{Hubble parameter} or Hubble function
\begin{equation}\label{eq:HubbleParameter}
    \boxed{H(t)\equiv \frac{\dot a(t)}{a(t)}\,.}
\end{equation}
Its value today defines the \textit{Hubble constant}
\begin{equation}\label{eq:HubbleConstant}
    H_0\equiv \frac{\dot a(t_0)}{a(t_0)}\,,
\end{equation}
that governs the series expansion of the scale factor for times not far in the past 
\begin{equation}\label{eq:SeriesExpansionScaleFactor}
    a(t)=a(t_0)[1+(t-t_0)H_0+...]\,.
\end{equation}
It is custom to also introduce an associated dimensionless quantity $h$ through
\begin{equation}\label{eq:Def little h}
    H_0\equiv  100\ h\ \text{km}\,\text{s}^{-1}\,\text{Mpc}^{-1}\,.
\end{equation}

\begin{figure}
    \centering
    \includegraphics[width=0.7\textwidth]{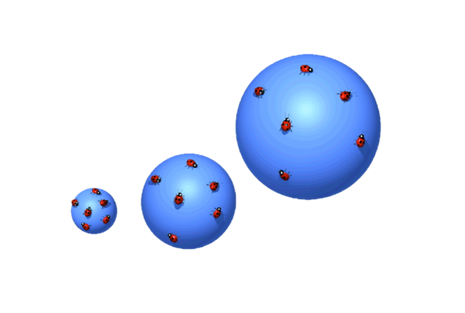}
    \caption{\small Analogy of an expanding universe: The ladybugs represent galaxies with their own peculiar velocity within an expanding universe represented by an inflating balloon. Regardless of their peculiar velocities, in the mean the physical distances between the galaxies increases and effectively, all galaxies are drawn away from each other. (Figure taken from \cite{LoebImage})}
    \label{fig:ExpandingUniverse}
\end{figure}

An intuition for the consequences of an expanding and contracting space can be gained from typical analogies, such as bugs walking on an inflating balloon in Figure~\ref{fig:ExpandingUniverse}.
It is however important to realize, that the universe is by no means expanding within ``something'' as one could erroneously conclude from such analogies in embedded spaces. Rather, the expansion happens intrinsic in a four dimensional spacetime and manifests itself through a change of the metric, providing a notion of spacetime distance between events of the spacetime manifold. As we will discuss below, observations confirm that our universe is currently expanding and will continue to do so in an accelerated manner. This also implies that reversing into the past, the scale factor continuously decreases until one reaches a potential beginning of the universe at a ``Big Bang'' as $a(t)\rightarrow 0$. Yet again, according to GR, such a beginning of the cosmos did not occur at a specific ``location'' with an explosion-like expansion in all directions. Instead, the entire infinite space is always present but compared to today, the physical distances between comoving points were much lower, indicating a highly increased density. In fact, within GR the actual limit of $a(t)\rightarrow 0$ that would naively describe the ``creation'' of space from nothing, is not describable, as the theory breaks down towards that singularity. 

\paragraph{Homogeneous and Isotropic Matter.}

The assumptions of homogeneity and isotropy also impose stringent constraints on the mean values of tensor fields describing matter in the universe \cite{Weinberg1972,Weinberg2008Cosmology}. In a $3+1$ slicing defined by the comoving coordinates, homogeneity implies that any scalar under spacial coordinate transformations can only depend on the cosmic time, while isotropy requires the components of any three-vector to vanish. Thus, the background mean values of any vector field components $V^\mu$, such as the current of baryons,  must at every point in a comoving reference frame read
\begin{equation}
    V^i=0\,,\quad V^0=v(t)\,,
\end{equation}
for some density function $v(t)$. Similarly, the energy-momentum tensor of all matter $T^{\mu\nu}$ must at every spacetime point in comoving coordinates have the form
\begin{equation}\label{eq:ComponentsEMTBackground}
    T^{00}=\rho(t)\,,\quad T^{0i}=0\,,\quad T^{ij}=p(t)\,a^{-2}(t)\delta^{ij}\,.
\end{equation}
Note that these are precisely the components of a perfect fluid 
\begin{equation}\label{eq:EnergyMomentumHomogeneousIsotropic}
    T^{\mu\nu}=(\rho+p)u^\mu u^\nu+p g^{\mu\nu}\,,
\end{equation}
evaluated in its rest frame with $u^{\mu}=(1,0,0,0)$, with $\rho$ the local energy density and $p$ the local pressure. Equation~\eqref{eq:ComponentsEMTBackground} therefore implies that the rest frame of the fluid must everywhere coincide with the comoving frame.

Minimal and universal coupling of the physical metric further implies the covariant conservation of the energy-momentum tensor in Eq.~\eqref{eq:CovariantConservationEMTensor} that results in the relation
\begin{equation}\label{eq:ConservationOfMatterCosmology}
    \dot\rho+3H(\rho+p)=0\,,
\end{equation}
where a dot denotes a derivative with respect to cosmic time $t$.
Together with a general equation of state
\begin{equation}\label{eq:EquationOfStateGeneral}
    p(t)=w(t)\rho(t)\,,
\end{equation}
Eq.~\eqref{eq:ConservationOfMatterCosmology} admits the general solution \cite{maggiore2018gravitationalV2}
\begin{equation}\label{eq:GeneralRhoNotConstOmega}
    \boxed{\rho(x)=\rho_0\,\exp\left\{-3\int^x_{0}dx'[1+w(x')]\right\}\,,}
\end{equation}
with $x\equiv\log a(t)$ and where $\rho_0$ represents the energy density today. For a constant $w$, this reduces to
\begin{equation}\label{eq:GeneralRhoConstOmega}
    \boxed{\rho(a)=\rho_0\,a^{-3(1+w)}\,.}
\end{equation}
In particular, there are three important limiting cases: Relativistic matter, hence, matter particles whose mass is negligible, which in cosmology is referred to as hot matter or \textit{radiation} ($r$), satisfies $p_r=\rho_r/3$; Non-relativistic, or cold matter on the other hand, mostly simply called \textit{matter} ($m$) is pressureless $p_m=0$; Moreover, introducing a \textit{cosmological constant} (CC) commonly denoted by ($\Lambda$) in the vacuum Einstein equations (c.f. Eq.~\eqref{eq:EinsteinFieldEquations}), can also be viewed as a fluid with energy momentum tensor (see Sec.~\ref{sSec:CCDarkEnergy})
\begin{equation}\label{eq:EMTforCC}
    T^{\mu\nu}_{\Lambda}=-\frac{\Lambda}{\kappa_0} g^{\mu\nu}\,,
\end{equation}
and negative pressure $p_\Lambda=-\rho_\Lambda$. For these three types of fluids, the energy density respectively satisfies
\begin{subequations}\label{eq:EnergyTypesUniverse}
\begin{align}
    &\text{cold matter}\,: & w &=0\,, &\rho_m&=\rho_{m0}\, a^{-3}\,,\\
    &\text{radiation}\,: & w &=\frac{1}{3}\,, &\rho_r&=\rho_{r0}\, a^{-4}\,,\\
    &\text{CC}\,: & w &=-1\,, &\rho_\Lambda&=\text{const}\,.
\end{align}
\end{subequations}
These results can be understood as follows: The local energy density of cold matter primarily results from its mass, which is diluted by a factor of $a^{-3}$ as the universe expands. On the other hand, for radiation or relativistic matter with $w=1/3$, the energy of each particle already redshifts away by $a^{-1}$ as computed explicitly below, which explains the factor of $a^{-4}$. 
The associated non-conservation of energy is a manifestation of the lack of timelike Killing vector fields in general FLRW solutions. Finally, while a cosmological constant as a geometric quantity has a priori nothing to do with ``matter'', in a quantum context, a cosmological constant can actually be associated to vacuum energy of matter, filling the entire spacetime with a constant energy density (see Sec.~\ref{sSec: The CC Problem}). Regardless of this association, the current cosmological standard model in fact requires the presence of a cosmological constant in the universe, or at least an energy content called dark energy that behaves very close to it.

\paragraph{The Friedmann Equation.}

The precise form of the scale factor and its evolution is finally dictated within general relativity by the Einstein field equations [Eq.~\eqref{eq:EinsteinFieldEquations}]
\begin{equation}
    G_{\mu\nu}+\Lambda g_{\mu\nu}=\kappa_0 T_{\mu\nu}\,.
\end{equation}
It is illuminating to temporarily allow for an arbitrary value of the curvature of space captured by the scalar $k$ within the general FLRW Ansatz in Eq.~\eqref{eq:LineElementHomIsoGenR}. In a comoving frame, the fundamental $\mu=\nu=0$ component is known as \textit{Friedmann equation} governing the expansion of the universe and reads
\begin{equation}\label{eq:FriedmannEquationGen}
    H^2+\frac{k}{a^2}=\frac{\kappa_0}{3}\rho\,.   
\end{equation}
where the cosmological constant contribution has been included in the total energy density $\rho$ according to Eq.~\eqref{eq:EMTforCC}. This equation can first of all be interpreted as the matter content of the universe dictating the value of the curvature of space $k$. Indeed, evaluated today, the Friedmann equation becomes
\begin{equation}
    \frac{3k}{\kappa_0a_0^2}=\rho_0-\rho_c\,.   
\end{equation}
where we defined the \textit{critical density}
\begin{equation}\label{eq:CriticalDensityToday}
    \rho_c\equiv\frac{3H_0^2}{\kappa_0}\,.
\end{equation}
Hence, whether $k=\pm 1$ or $k=0$ is determined if the total matter energy momentum density today $\rho_0$ (including the contribution from the cosmological constant) is greater than, less than or equal than the critical density $\rho_c$. Observations strongly support the assumption that indeed $\rho_0\simeq \rho_c$, such that we will again set $k=0$ in what follows.\footnote{From the perspective above, it is however rather unlikely that the universe should be spatially flat with $k=0$. We will discuss this point in greater detail below.}

Provided that there is no exchange of energy between different components, the total energy density content of the universe can be described by a sum of individual components labelled by an index $\lambda$
\begin{equation}\label{eq:SumOfRhos}
    \rho(t)=\sum_\lambda \rho_\lambda(t)\,.
\end{equation}
In this case, it is convenient to introduce the notion of energy fractions today of each species normalized by the critical density, called \textit{density parameters}
\begin{equation}\label{eq:DefDensityParameters}
    \Omega_\lambda\equiv \frac{\rho_{\lambda 0}}{\rho_0}\,,
\end{equation}
such that according to Eq.~\eqref{eq:GeneralRhoConstOmega} we have
\begin{equation}
    \rho_\lambda(a)=\rho_0\Omega_\lambda a^{-3(1+w_\lambda)}\,.
\end{equation}
It turns out that good approximation for the description of the energy content of the universe is to simply consider a sum of the limiting cases of non-relativistic cold matter, relativistic radiation and the cosmological constant introduced in Eqs.~\eqref{eq:EnergyTypesUniverse} above. In this case, the Friedmann equation [Eq.~\eqref{eq:FriedmannEquationGen}] can be written as
\begin{equation}\label{eq:FriedmannEq}
   \boxed{ H(t)=H_0\left(\Omega_r\,a^{-4}(t)+\Omega_m\,a^{-3}(t)+\Omega_{\Lambda}\right)^{1/2}\,,}
\end{equation}
together with the additional constraint from Eq.~\eqref{eq:SumOfRhos}
\begin{equation}\label{eq:ConstraintOnDensityParametersFirst}
    \boxed{1=\Omega_r+\Omega_m+\Omega_\Lambda\,.}
\end{equation}

Written in the form above, the Friedmann equation indicates that cold matter and radiation dominate earlier epochs of expansion as $a(t)$ becomes smaller. Eventually, however, as soon as the universe expanded enough to suppress the other contributions, a cosmological constant will take over. Within each of these epochs of \textit{radiation domination}, \textit{matter domination} and \textit{cosmological constant domination}, the Friedmann equation in Eq.~\eqref{eq:FriedmannEq} can be solved by effectively neglecting the remaining matter contributions. If $w\neq -1$, the solution reads
\begin{equation}\label{eq:GenSolutionScaleFactor}
    a_\lambda\propto t^{\frac{2}{3(1+w_\lambda)}}\,,
\end{equation}
while for the special case of $w=-1$ with a constant $\rho$, one obtains
\begin{equation}\label{eq:deSitterSolution}
    a_\Lambda\propto e^{Ht}\,,
\end{equation}
for $H$ a constant related to the cosmological constant through $H^2=\Lambda/3$. 

As an interesting side note, the solution of Eq.~\eqref{eq:deSitterSolution} together with Eq.~\eqref{eq:FlatFLRWmetric} describes a maximally symmetric \textit{spacetime}. In other words, the solution admits four more Killing vector fields than imposed by the cosmological principle and the spacetime corresponds to either dS or AdS, depending on the sign of the cosmological constant $\Lambda$. In contrast to the other cases described by Eq.~\eqref{eq:GenSolutionScaleFactor}, the existence of a timelike KVF for these solutions therefore allows for a definition of conserved energy according to the considerations in Secs.~\ref{sSec: Theories of Minkowski Spacetime} and \ref{sSec:Covariant Consrevation}. Note, however, that in a comoving frame with FLRW slicing, this fact is obscured.

\subsection{Distances and Horizons in Cosmology}

In order to experimentally evaluate whether our universe is evolving over time as suggested by GR through the FLRW solution introduced above, it is necessary to monitor the change in distance between us as observers and other freely moving objects in the universe. Since the definition of proper distance at a given instant between two simultaneous events in Eq.~\eqref{eq:ProperDistanceCosmologyGlobal} is not practical, a notion of distance closer to observations is therefore required. Since the different notions of ``distance'' in cosmology are sometimes subject to confusion, we want to address this question here in quite some detail. 

\paragraph{Light Travel Distances and Event Horizons.}

As today's cosmology is for the largest parts based on observations of electromagnetic signals reaching us from the cosmos\footnote{Excitingly, this will drastically change in the future due to observations of gravitational wave signals originating from events at cosmic distances.}, the basis of such pragmatic definitions of ``distances'', already alluded to above, is to give up the requirement of simultaneity within the definition of a spacial distance and instead consider the light travel-time between events at different times as a measure of distance. With such a pragmatic non-simultaneous notion of distance, observations at larger distances are equivalent to observations at earlier times in the past, such that variables of time can be viewed as variables of distance as well.

Observe that in comoving coordinates $(t,r,\theta,\phi)$, the distance between two comoving observers at say $r=0$ and $r=R$ does not change over time. This provides a relation between the comoving distance $R$ and the time $t_r-t_e$ it takes for a light signal to travel between $r=0$ and $r=R$ for any emission time $t_e$ and reception time $t_r$. More precisely, through the light-like condition $dt=a(t)dr$ in comoving coordinates, one can write
\begin{equation}\label{eq:ComovingDistanceFirstInt}
    R=\int^R_0 dr=\int^{t_r}_{t_e} \frac{dt'}{a(t')}\,.
\end{equation}
Viewed in this way, the coordinate distance $R$ depends on the $t_e$ and $t_r$, which introduces the notion of \textit{comoving distance} $R(t_r,t_e)$ between an event at an emission time $t_e$ at $r=0$, and an event at a receiving time $t_r>t_e$ at $r=R$
\begin{equation}\label{eq:ComovingDistanceFirst}
   \boxed{R(t_r,t_e)\equiv  \int^{t_r}_{t_e}\frac{dt'}{a(t')}\,.}
\end{equation}
This relation between the light travel-time and comoving distances is at the basis of all pragmatic concepts of distance in cosmology.

Yet, by definition, the comoving distance introduced above is a coordinate dependent notion. As an analogue to the proper spacial distance between two simultaneous events in Eq.~\eqref{eq:ProperDistanceCosmologyGlobal} we are therefore also interested in defining an associated proper physical length of the light-travel-time distance considered above. This can of course again been done on the basis of the universality of the speed of light $c$. More precisely, consider a comoving observer at $r=0$ that sends a light signal at $t_e$ to a distant comoving observer at $r=R$, who receives the signal at a cosmic time $t_r(R)$. But since the duration of this process in cosmic time $t_r(R)-t_e$ precisely corresponds to the proper time of the comoving observers, a coordinate independent notion of ``proper'' \textit{light travel distance} $\ell_l(d,t_e)$ between two events at different spacial locations $r=0$ and $r=R$ at two different times $t_e$ and $t_r$ can simply be obtained by multiplying the proper time $t_r(R)-t_e$ with the speed of light \cite{zee2013einstein}
\begin{equation}\label{eq:LightTravelTimeDist}
    \boxed{\ell_l(R,t_e)\equiv c \,[t_r(R)-t_e]\,,}
\end{equation}
where $t_r$ is related to a comoving distance $R$ through Eq.~\eqref{eq:ComovingDistanceFirst}. In comparison, the instantaneous proper distance $\ell(R,t)$ in Eq.~\eqref{eq:ProperDistanceCosmologyGlobal} was defined between two \textit{simultaneous} events separated by a comoving distance $R$.


It is instructive to evaluate the light travel-time distance in Eq.~\eqref{eq:LightTravelTimeDist} for the specific solutions of the scale factor corresponding to different types of matter given in Eqs.~\eqref{eq:GenSolutionScaleFactor} and \eqref{eq:deSitterSolution}. For a universe dominated by a positive cosmological constant\footnote{Note that here the time variable $t$ starts at $-\infty$ and reaches today's cosmic time at $t_0=0$.} $a(t)=e^{H t}$, with $H$ a positive constant, the integral of the comoving distance in Eq.~\eqref{eq:ComovingDistanceFirst} can readily be evaluated to give
\begin{equation}\label{eq:comovingDistanceCC}
    R(t_r,t_e)=\frac{1}{H}\left(e^{-H t_e}-e^{-H t_r}\right)\,.
\end{equation}
Interestingly, this solution is finite for large $t_r$
\begin{equation}\label{eq:RMaxDeSitter}
    R_\text{max}(t_e)=\lim_{t_r\rightarrow \infty} \,\frac{1}{H}\left(e^{-H t_e}-e^{-H t_r}\right) = \frac{e^{-H t_e}}{H}\,,
\end{equation}
which indicates that a signal send out today at $t_e$ will only reach a finite portion of the universe. Indeed, inverting Eq.~\eqref{eq:comovingDistanceCC} and plugging the result into the proper light travel distance in Eq.~\eqref{eq:LightTravelTimeDist} gives
\begin{equation}
    \ell_l(R,t_e)= -\frac{1}{H}\ln\left[1-e^{H t_e} H\,R\right]\,,
\end{equation}
which diverges when $R$ approaches $R_\text{max}$. In other words, the universe expands too fast for light to keep up, which defines a physical \textit{event horizon} of the spacetime known as \textit{de-Sitter horizon}. On the other hand, for $R\ll 1$, the physical light-travel distance coincides with the proper simultaneous spacial distance defined in Eq.~\eqref{eq:ProperDistanceCosmologyGlobal}
\begin{equation}
    \ell_l(R,t_e)=e^{H t_e} \,R+\mathcal{O}(R^2)\,.
\end{equation}

For a solution of the form Eqs.~\eqref{eq:GenSolutionScaleFactor}\footnote{Here, $t_0=1$, while the time variable starts at $t=0$.} $a(t)=t^\alpha$ for a constant 
\begin{equation}\label{eq:Alphacoeff}
    \alpha=\frac{2}{3(1+w_\lambda)}\,,
\end{equation}
on the other hand, the comoving distance becomes
\begin{equation}\label{eq:comovingDistancealpha}
    R(t_r,t_e)=\frac{1}{\alpha-1}(t_e^{1-\alpha}-t_r^{1-\alpha})\,.
\end{equation}
In this case we have to distinguish two cases. If $\alpha>1$, this expression again admits a finite limit at large $t_r$
\begin{equation}
    \alpha>1\,, \quad\Rightarrow \quad R_\text{max}(t_e)=\frac{t_e^{1-\alpha}}{\alpha-1}\,,
\end{equation}
and the universe comes with a future de-Sitter-like horizon. The presence of such a horizon is in fact tied to an accelerated expansion of the universe with $\ddot a>0$. In terms of the equation of state, according to Eq.~\eqref{eq:Alphacoeff}, the universe is therefore expanding in an accelerated manner as long as $w < -1/3$.
On the other hand, for $\alpha<1$, the comoving distance has no maximal value and a light signal send out today can in principle reach the infinite universe. We can again invert Eq.~\eqref{eq:comovingDistancealpha} and compute the associated light travel distance
\begin{equation}
    \ell_l(R,t_e)= \left[t_e^{1-\alpha}+R(1-\alpha)\right]^{\frac{1}{1-\alpha}}-t_e\,.
\end{equation}
This expression again diverges as $R\rightarrow R_\text{max}$ for $\alpha>1$ and coincides with the proper distance of Eq.~\eqref{eq:ProperDistanceCosmologyGlobal} in the small $R$ limit
\begin{equation}
    \ell_l(R,t_e)=t_e^\alpha R+\mathcal{O}(R^2)\,.
\end{equation}


Reciprocally, the cosmological principle implies that the event horizons $R_\text{max}$ described above also define the region from which no light signal will ever reach us as receivers of light. In particular, for an acceleratingly expanding universe with $a(t)=e^{H t}$, with $H$ a constant, Eq.~\eqref{eq:RMaxDeSitter} tells us that over time, the horizon radius is exponentially shrinking, such that eventually all comoving galaxies at fixed comoving distance will eventually pass out of our horizon and no signal from other galaxies will ever reach us again. Physically, a cosmic expansion is causing a redshift of light as we will explicitly show below, such that the ``passing out of our event horizon'' of a light source is equivalent to reaching an infinite redshift of the emitted light on our way to us. 

\paragraph{The Hubble Radius and the Hubble Horizon.}
Frequently, an analogy of the cosmological redshift to the familiar Doppler redshift is drawn, by imagining that the increasing redshift of distant galaxies is due to their increasing velocity with respect to us, caused by the cosmic expansion. However, it is important to note that such an analogy is on a practical level only accurate locally over times of negligible change in scale factor. Naively, based on the global physical spacial distance between simultaneous events introduced in Eq.~\eqref{eq:ProperDistanceCosmologyGlobal} one could still define a ``physical'' or ``proper'' velocity
\begin{equation}
    v_\text{p}(R,t)\equiv \frac{d}{dt}\ell(R,t)= H(t)\, \ell(R,t)\,,
\end{equation}
which for some time $t$ increases with comoving distance $R$. In particular, for an evolution dominated by a cosmological constant with constant Hubble parameter $H$, the associated de-Sitter horizon in Eq.~\eqref{eq:RMaxDeSitter} can be interpreted as the radius for which the physical velocity at $t_e$ exceeds the speed of light
\begin{equation}\label{eq:properVelocity}
    v_\text{p}(R_\text{max},t_e)=1\,,
\end{equation}
which implies a proper distance radius of
\begin{equation}
    \ell(R_\text{max},t_e)=\frac{1}{H}\,.
\end{equation}
Indeed, Eq.~\eqref{eq:RMaxDeSitter} together with Eq.~\eqref{eq:ProperDistanceCosmologyGlobal} imply that the de-Sitter horizon radius in terms of physical distance $\ell(R_\text{max},t_e)=a(t_e) R_\text{max}=e^{H t_e}e^{-H t_e}/H=1/H$ is set by the so called \textit{Hubble radius}
\begin{equation}\label{eq:HubbleRadius}
   \ell_H(t)\equiv c/H(t)\,.
\end{equation}

Sometimes, the Hubble radius, defining the physical distance radius at which the physical velocity exceeds the speed of light at some time $t_e$ in a general cosmology, is erroneously called a ''Hubble horizon''. However, this is only true by chance in the specific case of de-Sitter space. Indeed, as we have shown above, if the expansion is not accelerating, there is no physical event horizon and the Hubble radius defined at some instant $t_e$ has nothing to do with an actual event horizon. In general, while an event horizon is a global concept, that requires knowledge of the entire spacetime, the Hubble radius is only defined at a particular cosmic time and has no physical meaning as a horizon except in certain special cases.\footnote{However, the Hubble radius is a very useful scale in terms of cosmological perturbations discussed in Sec.~\ref{sSec:LCDM} below, since it discriminates two scales of radically different behaviors of perturbation modes.} The lack of physical meaning as a horizon of the Hubble radius is associated to the lack of physical meaning of the proper velocity defined in Eq.~\eqref{eq:properVelocity}. This quantity is not an actual velocity, defined as a rate of movement between the object and a local inertial frame. More precisely, the velocity of a distant galaxy compared to us has no physical meaning, unless we imagine the cosmic conspiracy described when defining the proper distance in Eq.~\eqref{eq:ProperDistanceCosmologyGlobal}. Moreover, a value of $v_p$ greater than the speed of light does by no means indicate a violation of special relativity, as this does not imply that information can be transported faster than the speed of light, since all worldlines of physical particles remain inside their local lightcone.

\paragraph{The Particle Horizon, Sound Horizon and Conformal Time.} Apart from the event horizons, or future horizons, defined above, a universe which has a start at some time $t=0$ exhibit another type of horizon called \textit{particle horizon} $r_C(t)$. This horizon is defined by the greatest value of comoving distance $R_\text{max}$ that any physical particle emitted at $r=0$ at time $t=0$ can reach in cosmic time $t$, which is simply given by the comoving radial distance $R(0,t)$ that light travels during that time period, thus
\begin{equation}\label{eq:ParticleHorizon}
    r_C(t)\equiv R(0,t)=\int_0^{t}\frac{dt'}{a(t')}\,.
\end{equation}
According to Eq.~\eqref{eq:FriedmannEq}, it is expected, that at early times the energy density is dominated by radiation, for which $a(t)\propto t^{1/2}$, such that $r_C(t)\propto 2t^{1/2}$.
This comoving distance coincides with the definition of so-called \textit{conformal time} $\eta(t)=r_C(t)$, which is the time variable satisfying $d\eta=dt/a(t)$, that renders the FLRW line element in Eq.~\eqref{eq:FlatFLRWmetric} conformally flat
\begin{equation}
    ds^2=a^2(\eta)\left[-d\eta^2+d\mathbf{x}^2\right]\,.
\end{equation}
Note that in terms of conformal time, light cones form angles of $45^\circ$ as one is  accustomed from Minkowski coordinates.
Furthermore, for a sound wave traveling at the sound speed $c_s$, for instance within the primordial plasma, the related concept of \textit{comoving sound horizon} can be defined
\begin{equation}\label{eq:SoundHorizonFirst}
    \boxed{r_\text{s}(t)\equiv \int_0^{t}c_\text{s}(t')\,\frac{dt'}{a(t')}\,.}
\end{equation}

\paragraph{A Different Perspective on the Comoving Distances.} In Eq.~\eqref{eq:LightTravelTimeDist} we defined a ``proper'' spacial distance in cosmology based on the light travel-time in terms of a coordinate independent geodesic length given by the proper time of comoving observers between two events labeled by an emission and reception time. Yet, except for theoretical considerations as given above, there is no actual need for such proper distances in cosmology, precisely because of the existence of preferred comoving coordinates that can be used in order to unambiguously label distances to objects in the universe according to the time the light signal spend to reach us. Because we obviously only observe the objects at today's cosmic time $t_0$ whose light signals were sent out in the past at $t<t_0$, it is custom to feed these assumptions into the general definition of the comoving distance in Eq.~\eqref{eq:comovingDistancealpha} and define a ``lookback'' comoving distance
\begin{equation}\label{eq:ComovingDistanceSecond}
   \boxed{d_C(t)\equiv  \int^{t_0}_t\frac{dt'}{a(t')}\,.}
\end{equation}
While the lookback comoving distance and the associated length measure in terms of cosmic time variables would in principle be enough to label the distance of all observed objects, different ways of determining distances in practice suggest the introduction of additional definitions, including the luminosity and angular diameter distances that we will introduce at the end of this section. But first, we need to introduce the cosmological redshift.

\paragraph{Cosmological Redshift.} 
It is very useful to define an alternative variable of ``time'' with a value of zero today, at $t_0$ about which one can naturally expand for close by observations. It turns out that such a variable is provided by the shift in frequency due to the cosmic background expansion that was already briefly mentioned above. We will now explicitly derive this shift in frequencies, which manifested in a shift of spectral lines in the light from distant galaxies, in fact, represents the main source of knowledge of the local evolution of $a(t)$. More precisely, a cosmological shift in frequency can be related to a change in the scale factor by considering two maxima of a light signal emitted at $t_e$ at a comoving distance $d_C(t_e)=R$, with an arrival time separated by $\Delta t_0$ and a time separation at emission of $\Delta t_e$. Indeed, since the comoving distance between comoving objects does not change over time we have
\begin{equation}
    R=  \int^{t_0}_{t_e}\frac{dt'}{a(t')}=\int^{t_0+\Delta t_0}_{t_e+\Delta t_e}\frac{dt'}{a(t')}\,.
\end{equation}
This implies that
\begin{equation}
    0=  \int^{t_0+\Delta t_0}_{t_0}\frac{dt'}{a(t')}-\int^{t_e+\Delta t_e}_{t_e}\frac{dt'}{a(t')}\simeq \frac{\Delta t_0}{a(t_0)}-\frac{\Delta t_e}{a(t_e)}\,,
\end{equation}
where in the last equality, we assumed that $a(t)$ is approximately constant over the time intervals $\Delta t_e$ and $\Delta t_0$.
Hence, in terms of frequencies $\nu_i=1/\Delta t_i$ we can define a quantity $z$ through
\begin{equation}
    1+z\equiv \frac{\nu_e}{\nu_0}=\frac{a(t_0)}{a(t_e)}\,,
\end{equation}
which is zero today, positive for a redshift $\nu_0<\nu_e$ when the scale factor is increasing $a(t_0)>a(t_e)$ and negative for a blueshift $\nu_0<\nu_e$. Since observationally, the universe is expanding, the factor $z$ is known as \textit{cosmological redshift}, and thus $z\geq 0$. 

If not stated otherwise, we will in the following label the emission time as $t<t_0$, where the cosmic time today satisfies $a(t_0)=1$. In this notation, the relation between redshift and scale factor reads
\begin{equation}
  a = \frac{1}{1+z} \,, \quad \Leftrightarrow  \quad z = \frac{1}{a}-1 \,.
\end{equation}
Moreover, in a purely expanding universe in which the scale factor $a(t)$ ever increases, the scale factor itself and consequently also the redshift can effectively replace the cosmic time variable in an unambiguous way. Therefore, in the following, we will often use the time $t$ and the redshift $z$ as interchangeable variables.

For instance, the comoving sound horizon defined in Eq.~\eqref{eq:SoundHorizonFirst}, as well as the comoving distance in Eq.~\eqref{eq:ComovingDistanceSecond} can respectively also be expressed in terms of redshifts as
\begin{equation}\label{eq:SoundHorizonSecond z}
		r_\text{s}(z) =\int^\infty_z \,c_\text{s}(z')\,\frac{d z'}{H(z')}\,,
\end{equation}
and
\begin{equation}\label{eq:ComovingDistanceThird}
   d_C(z)=\int^z_0\frac{dz'}{H(z')}\,.
\end{equation}
These relations simply follows from the changes of variables
\begin{equation}
   \int^{t}_0\frac{dt'}{a(t')}=\int^{a(t)}_{0}\frac{da'}{a'^2H(a')}=\int^\infty_z\frac{dz'}{H(z')}\,,
\end{equation}
respectively
\begin{equation}
   \int^{t_0}_t\frac{dt'}{a(t')}=\int^{1}_{a(t)}\frac{da'}{a'^2H(a')}=\int^z_0\frac{dz'}{H(z')}\,.
\end{equation}
One of the advantages of using the redshift as time variable is that for instance the expression of the comoving distance in Eq.~\eqref{eq:ComovingDistanceThird} is readily expanded in terms of small redshifts to give
\begin{equation}\label{eq:ComovingDistanceThirdExpanded}
   \boxed{d_C(z)=\frac{z}{H_0}+\mathcal{O}(z^2)\,.}
\end{equation}
Note that for nearby sources, hence at small redshift for which the scale factor is effectively unity
\begin{equation}
    a(t)=1+(t-t_0) H_0+\mathcal{O}(t^2)\,,
\end{equation}
the notions of comoving distance [Eq.~\eqref{eq:ComovingDistanceFirst}], proper distance [Eq.~\eqref{eq:ProperDistanceCosmologyGlobal}] and light travel distance [Eq.~\eqref{eq:LightTravelTimeDist}] coincide to first order
\begin{equation}\label{eq:DistanceRedshiftRelation}
    R=\ell=\ell_l\simeq t_0-t \equiv d =\frac{z}{H_0}\,.
\end{equation}
In this case, the local distance $d$, as well as the associated instantaneous radial velocity
\begin{equation}\label{eq:HubbleLaw}
    v=H_0 d = z\,,
\end{equation}
that provides the leading term of the proper velocity $v_p(t)\simeq v + \mathcal{O}(t) $ defined in Eq.~\eqref{eq:properVelocity}, are well-defined physical notions. This is in the sense that the distance $d$ is a proper spacial distance between two simultaneous spacelike separated events, while the velocity $v$ is an unambiguous relative velocity for which the cosmological redshift can be interpreted as arising from the Doppler effect associated to $v$. 

Note that the distance-redshift relation in Eq.~\eqref{eq:DistanceRedshiftRelation} indicates, that locally, it is possible to measure $H_0$ by ``simply'' measuring the redshift $z$ and distances $d$ to nearby comoving objects, without requiring any knowledge on the expansion history of $a(t)$. The velocity-distance relation in Eq.~\eqref{eq:HubbleLaw} is known as \textit{Hubble's law}, due to the first measurement of $H_0$ by Edwin Hubble in 1929 that started the journey of modern cosmology. This journey of understanding the evolution and content of a non-static universe resulted in the current $\Lambda$CDM standard model of cosmology that we will introduce in the next Section~\ref{sSec:LCDM}. As we will discover, the exact value of the Hubble constant is however still subject to vigorous debate, whose resolution might very well be one of the clues guiding us beyond the current horizon of knowledge.


\paragraph{Angular Diameter and Luminosity Distances.} 
One of the reasons why the value of $H_0$ is still up for debate is that despite the simple relation in Eq.~\eqref{eq:DistanceRedshiftRelation}, the measurement of distance to a given light source is by no means simple. While measurements of redshift are straightforward through the shift of spectral lines, the distance is mainly obtained through three different methods: 
\begin{enumerate}[(1)]
\item By exploiting the motion of the earth around the sun, geometric triangulations can be used in order to define distances of nearby object. 
\item For an extended object of known proper distance $s$ that today subtends a small angle $\theta$ in the sky, the distance is given by $d=s/\theta$. 
\item Knowing the absolute luminosity $L$ of a light source, the distance can be determined by measuring the apparent luminosity $l$ through $d=\sqrt{L/(4\pi l)}$. 
\end{enumerate}
While the method (1) is limited by the scale of motion of the earth, method (2) and (3) effectively introduce uncertainties through the assumed a priori knowledge of $s$ and $L$. Moreover, the above relations are only valid locally, in a static patch. Looking further into the past to observe more interesting objects, corrections from the cosmic expansion need to be taken into account. In this last paragraph of this section, we will introduce the notions of the so-called angular diameter and luminosity distances that naturally capture these expansion effects of the background universe.

For an extended light source at a comoving distance $R$ that emits light at $t_e$ that we observe today, the proper distance $s$ normal to the line of sight is equal to $s=a(t_e)\,R\,\theta$. The so-called \textit{angular diameter distance}\footnote{Note that this relation only holds in a Euclidean universe.}
\begin{equation}\label{eq:AngularDistanceSecond}
    d_A(t)\equiv a(t)\, d_C(t)
\end{equation}
is therefore introduced to preserve the usual relation of Euclidean geometry
\begin{equation}\label{eq:AngularDiameterDistanceDefAngle}
    \theta=s/d_A\,.
\end{equation}
On the other hand, the luminosity is defined as a power, in other words an energy per time. Cosmic expansion therefore affects the Luminosity distance relation by reducing the energy of the photons by a redshift factor $1/(1+z)$, while at the same time also reducing the rate of arrival of individual photons by the same factor. Note that the proper area of the sphere around the luminous object encompassing the earth is not modified, since $a(t_0)=1$. Therefore, a \textit{luminosity distance} 
\begin{equation}\label{eq:LuminosityDistanceSecond}
    d_L(z)\equiv (1+z)\,d_C(z)
\end{equation}
is introduced, in order to preserve the familiar relation
\begin{equation}\label{eq:apparentLuminosity}
    l=\frac{L}{4\pi d_L^2}\,.
\end{equation}

In summary, the lookback comoving distance [Eq.~\eqref{eq:ComovingDistanceSecond}], the angular diameter distance [Eq.~\eqref{eq:AngularDistanceSecond}] and luminosity distance [Eq.~\eqref{eq:LuminosityDistanceSecond}] are defied as
\begin{subequations}\label{eq:DistancesCosmology all}
\begin{align}
    &\text{comoving}\,: & d_C&=\int^{t_0}_t\frac{dt'}{a(t')}=\int^z_0\frac{dz'}{H(z')}\,,\\
    &\text{angular diameter}\,: & d_A&=a(t)\,d_C(t)=\frac{1}{1+z}d_C(z)\,,\\
    &\text{luminosity}\,: & d_L&=\frac{1}{a(t)}d_C(t)=(1+z)\,d_C(z)\,.
\end{align}
\end{subequations}

\section{The $\Lambda$CDM Cosmological Standard Model}\label{sSec:LCDM}

Based on a dynamical background spacetime discussed in the previous section, a multitude of breakthroughs over the last century were combined to form a current concordance model of cosmology describing the content and evolution of our universe that can account for almost all empirical observations to great precision with only a few free parameters. In this section, we offer a concise summary of this cosmological standard model.

\subsection{The Basics of $\Lambda$CDM}

\paragraph{An Expanding Background Universe.} As discussed above, direct measurements of the movements of distant objects first attributed to Hubble indicate that our universe is expanding. Projecting backwards, Hubble's measurements therefore suggest that the evolution of the cosmos must have started off in a very contracted and dense state. 
 A confirmation of the associated so called ``Big-Bang''-paradigm of an expanding universe was indeed subsequently provided by two important predictions such a dense beginning would entail on the matter content. 

First of all, the universe is filled with known matter, whose most important constituents can be grouped into baryons ($b$), which in cosmology refers to both nuclei and electrons, since the mass of the electrons remains largely negligible, photons ($\gamma$) and neutrinos ($\nu$). Quite amazingly, cosmology offers a study of the creation and abundances of light elements observed in the universe today. 

Indeed, the Big-Bang assumption implies the existence of an extremely dense and hot environment at early cosmic times, in which even nuclei could not be held together, and all particles highly interacted, forming an equilibrium described by a single \textit{temperature of the universe}. In an expanding universe, this equilibrium temperature decreases continuously, such that the temperature can also be used as a measure of cosmic time. A decreasing equilibrium temperature also implies that eventually the universe cooled well below the typical nuclear binding energies and light elements began to form. This process is known as \textit{Big Bang Nucleosynthesis} (BBN). 

On top of the formation of bound objects due to a drop in temperature, there is a second fundamental process happening in an expanding universe. Namely, as long as the interaction rate of a particle remains above the expansion rate of the universe, the equilibrium with the cosmic matter soup can be maintained, while otherwise if falls out of the equilibrium with the other species and ``freezes out''. A combination of these two processes together with knowledge of the conditions of the early universe allows computing the expected primordial abundances of all light elements. These computations can be compared to direct observations of the amount of light elements in the cosmos, confirming the model and constraining its parameters (see e.g. \cite{Weinberg2008Cosmology,dodelson2020modern}).

However, the most important confirmation of the assumption of an expanding universe was given by the observation of the relic electromagnetic radiation released from the hot primordial plasma when the expansion-driven decrease in temperature finally allowed the combination of the free electrons and protons to form neutral hydrogen. Since then, a process called \textit{decoupling}, these primordial photons travelled almost freely through spacetime, filling the entire universe with a very isotropic black-body spectrum at a temperature today of \cite{Fixsen:2009ug}
\begin{equation}
    T_0=2.72548\pm  0.00057 \text{ K}\,,
\end{equation}
known as the \textit{cosmic microwave background} (CMB). The name arises because the wavelengths of the cosmic photons nowadays lie in the microwave regime. CMB observations still represent the bedrock of modern cosmology, and in particular provides the strongest evidence for the fundamental assumption of a very homogeneous and isotropic universe at early times and at large scales today. Moreover, the CMB formation represents the furthest distance, or time that we are able to receive electromagnetic signals from, as decoupling precisely represents the moment in which the universe became transparent to light.\footnote{Note that this restriction does not apply to gravitational waves, which therefore represent a possibility to receive direct information from pre-CMB physics.}

\paragraph{The Cosmological Standard Model and its Energy Content.} Based on the assumptions of a homogeneous and isotropic background, together with the Big-Bang paradigm of an expanding universe, associated cosmological observations of the cosmic evolution can be fitted by a model with a given set of free parameters that crucially involve the density parameters of different types of Energy contents in Eq.~\eqref{eq:FriedmannEq}. Such inferences can then be compared with the expectation of direct observations of the known matter. The subsequent formulation of a precise GR-based cosmological standard model, called $\Lambda$CDM, came however with two tremendous and name-giving surprises, representing key open questions of modern physics: 
\begin{enumerate}[(i)]
    \item It requires the existence of so-called \textit{cold dark matter} (cdm) that mainly only interacts through the gravitational force, but actually dominates the current energy density of non-relativistic matter.
    \item Multiple independent measurements also point towards the existence of an energy component known as \textit{dark energy} that is at least close to admitting a constant energy density described by a cosmological constant $\Lambda$, as discussed above.
\end{enumerate}
The universe is therefore not only filled with known matter in the form of baryons, photons and neutrinos, but as it turns out for the most part with the unknown ingredients of cold dark matter and dark energy that in $\Lambda$CDM is modeled by a cosmological constant. 

More precisely, recall that based on homogeneity and isotropy, the evolution of the background spacetime of the flat $\Lambda$CDM universe is governed by the Hubble function in Eq.~\eqref{eq:FriedmannEq} determined by the Friedmann equation that in terms of redshift reads
\begin{align}\label{eq:HubbleParameterLCDMSecond}
		\boxed{H^2_\text{\tiny $\Lambda$CDM} = H_0^2\, \Big(\Omega_m (1+z)^{3} + \Omega_r (1+z)^{4} + \Omega_\Lambda\Big)\,.}
\end{align}	
As discussed the evolution is governed by the amount of relativistic matter, called radiation (r), clumping non-relativistic or cold matter (m) and the cosmological constant ($\Lambda$). Whether a given matter particle is relativistic or non-relativistic depends on the ratio between the temperature of the universe and the mass of the particle. However, for the relevant epochs of the universe all baryons behave non-relativistically, while photons only know the relativistic state. Moreover, in the standard model also the new ingredients of dark matter and dark energy are assumed to be permanently associated to cold matter and the CC. The only species that transitioned from hot to cold after matter-radiation equality, are cosmic neutrinos. Since the observation of neutrino oscillations \cite{Super-Kamiokande:1998kpq} it is known that the three flavors of neutrinos must have a sum of masses of at least 
\begin{equation}\label{eq:NeutrinoMasses}
   \sum_{i=1}^3 m_{\nu_i} \gtrsim \, 0.06 eV\,.
\end{equation}
While cosmic neutrinos have not been detected directly yet, their presence is strongly suggested by multiple cosmological observations (see e.g. \cite{Lesgourgues:2012uu,dodelson2020modern}). Their energy density lies however well below the ones of photons except after their non-relativistic transition deep in the matter dominated era and for the most part of the subsequent treatments, neutrinos will not play a significant role. Thus, if not stated otherwise, their energy densities will by default be neglected. 

In summary, we can therefore make the following associations
\begin{equation}
    \Omega_m=\Omega_{cdm}+\Omega_b\,,\qquad \Omega_r=\Omega_\gamma\,,
\end{equation}
where the corresponding $\Lambda$CDM values of the density parameters employed in this work are found in Table~\ref{table:PlankBestFit}. The standard $\Lambda$CDM background evolution started in a radiation dominated era filled with a primordial equilibrium plasma of photons and electrons and light elements formed during BBN that evolved in a spatially translational and rotational invariant spacetime according to Eq.~\eqref{eq:HubbleParameterLCDMSecond}. However, the background history of the universe is only half of the story and many of today's precision experiments of cosmology depend on the departures of homogeneity and isotropy in the form of cosmological perturbations.

\paragraph{Cosmological Perturbations.}

Matter in the early universe could only be distributed in a homogeneous and isotropic way up to a certain degree due to the large matter clumps in the form of galaxies and galaxy clusters observed today. Indeed, this is confirmed by the observation of tiny anisotropies imprinted in the CMB. These are traces of primordial matter fluctuations that, due to the attractive nature of gravity, eventually grew to the structure in the universe. However, before decoupling, the highly interactive baryon photon plasma prevented any baryonic matter perturbations to grow and after that, a universe consisting only of known interacting matter would not have had enough time to form the observed large galactic structures. This fact represents one of the most stringent clues for the existence of a solely gravitationally interacting and clumping cold dark matter that dominates the non-relativistic matter energy budget, whose perturbations were able to grow well before the production of the CMB. In general, the observation and study of the perturbed universe provide powerful probes of the underlying model. Here we will not be able to discuss cosmological perturbations in full detail, but will restrict the discussion to a pertinent qualitative understanding and refer to the excellent treatments in \cite{Bardeen:1980kt,Mukhanov:1990me,Ma:1995ey,Liddle:2000cg,Weinberg2008Cosmology,carroll2019spacetime,maggiore2018gravitationalV2,dodelson2020modern}.

In any metric theory of gravity, perturbations to a given background solution can be treated within the framework of a scalar-vector-tensor decomposition described in Sec.\ref{ssSec:GaugeInvariantDecomposition}.
Moreover, the CMB temperature fluctuation measurements tell us that at early times, perturbations to the FLRW background had to be of the order of $\delta\rho/\rho\sim 10^{-5}$, while at late times perturbations remain small on large scales, as we will understand below. Hence, only being interested in early times and large scales, it is an excellent approximation to only consider first order or linear perturbations. If the background is furthermore invariant under spacial rotations, the scalar, vector and tensor perturbation sectors within the SVT decomposition decouple from each other at linear order and can therefore be treated separately. Finally, in this chapter, we will entirely focus on the scalar sector, while leaving aside the tensor sector, which in GR corresponds to gravitational waves on a cosmological background. In GR based cosmology, the vector sector does generally not play a role, as vector perturbations simply decay as long as they are not sourced.

We therefore concentrate on the scalar sector of linear perturbations where there exist two independent metric perturbations and a number of matter perturbations including energy density and pressure perturbations for each matter species that must however not all be independent. The relation and evolution of such perturbations in the matter distributions and the metric are governed by the perturbed Einstein equations described in Sec.~\ref{ssSec:GaugeInvariantDecomposition}. The main difference compared to the treatment in previous chapters is that we assume an exact background solution, which is provided by the FLRW metric. In GR, the two metric scalar modes are therefore still non-dynamical but are this time inevitably sourced by the matter homogeneous and isotropic matter content (recall Sec.~\ref{sSec:HomIsoUniverse}). Moreover, the same gauge freedom in describing perturbations persists. While an explicitly gauge invariant formulation is possible as famously shown by Bardeen \cite{Bardeen:1980kt}, cosmologists still like to work in a definite gauge, usually chosen to be the \textit{conformal Newtonian gauge} (see e.g. \cite{maggiore2018gravitationalV2,dodelson2020modern}) that we will also adopt here. The scalar perturbation variables of Newtonian gauge actually correspond to the gauge-invariant Bardeen variables, but the interpretation of matter perturbations depends in principle on the different gauges. However, cosmological observables will in the end not depend on any gauge choices.

Furthermore, while the Einstein field equations suffice to describe an expanding universe in the perfect fluid approximation, which is a good approximation for the description of a highly interacting fluid in equilibrium or decoupled non-relativistic cold species, this is not the case anymore as soon as out of equilibrium effects need to be considered. Indeed, in general, the set of gravitational equations needs to be supplemented by the Boltzmann equations that describe the statistical behavior of thermodynamic systems not necessarily in equilibrium (see e.g. \cite{Weinberg2008Cosmology,dodelson2020modern}). The full Einstein-Boltzmann system of equations is quite complex and generally only admits numerical solutions. For us, it will however be sufficient to understand a set of key concepts and definitions, together with the underlying assumptions. Namely,
\begin{itemize}
    \item There is only one independent scalar perturbation that we will denote as $\phi$ that describes the gravitational potential on top of the homogeneous and isotropic FLRW background. This assumption holds as long as the anisotropic stress in the energy-momentum tensor of matter can be neglected, which is a good approximation for all relevant scales that we will be interested in (see e.g. \cite{maggiore2018gravitationalV2,dodelson2020modern}).
    \item For matter, the most relevant perturbations are the density perturbations and pressure perturbations
    \begin{equation}
        \rho(\mathbf{x},t)=\overline\rho(t)+\delta\rho(\mathbf{x},t)\,,\qquad p(\mathbf{x},t)=\overline p(t)+\delta p(\mathbf{x},t)\,,
    \end{equation}
    where $\overline\rho(t)$ and $\overline p(t)$ are the FLRW background values, now defined as spacial averages, which are independent of spacial positions under the assumption of the cosmological principle. All relevant fluids will be \textit{barotropic fluids}, defined by the fact that their pressure only depends on the density $p=p(\rho)$ even at the level of perturbations \cite{maggiore2018gravitationalV2}. This implies that the density and pressure perturbations of each species are related by their adiabatic sound speed $c_s$
    \begin{equation}
        \delta p=\frac{dp}{d\rho}\,\delta\rho\equiv c_{s}^2\,\delta\rho\,.
    \end{equation}
    It further follows that for the $\Lambda$CDM species introduced in Eq.~\eqref{eq:EnergyTypesUniverse} with constant equation of state $p=w\rho$, the sound speed squared is equal to the equation of state parameter
    \begin{equation}
        c_s^2=w\,.
    \end{equation}
    since at linear order $dp/d\rho=d\overline p/d\overline\rho=w$ and the total quantities satisfy \begin{equation}
        p=w \rho\,.
    \end{equation}
    However, for fluids with multiple species, the equation of state parameter and thus also the sound speed depend on time.
    Moreover, the perturbations are conveniently characterized by the density contrast or \textit{overdensities}\footnote{In first order perturbation equations we will for instance often write $\rho$ instead of the explicit background quantity $\overline \rho$ as it amounts to the same.}
\begin{equation}\label{eq:Overdensities def}
    \delta(\mathbf{x},t)\equiv \frac{\rho(\mathbf{x},t)}{\overline{\rho}(t)}-1=\frac{\delta\rho}{\rho}\,.
\end{equation}
\item In the $\Lambda$CDM model, in which dark energy is described by a cosmological constant, there are no dark energy perturbations since $w=c_s^2=-1$ (see \cite{maggiore2018gravitationalV2}). Therefore, at leading order only radiation, baryon and dark matter perturbations are present.
 \item During the early times of radiation domination, matter is dominated by the primordial baryon-photon plasma fluid, while cold dark matter evolves independently. While the relativistic photons alone satisfy $c_s=w=1/\sqrt{3}$, the combined baryon-photon fluid $\rho_{br}=\rho_r+\rho_b$ and $p_{br}=p_r=1/3\rho_r$ has a sound speed of
 \begin{equation}
     c_s^2=\frac{dp}{d\rho}=\frac{1}{3}\left(1+\frac{\partial\rho_b}{\partial\rho_r}\right)^{-1}=\frac{1}{3}\left(1+\frac{-3\rho_{b0}a^{-4}}{-4\rho_{r0}a^{-5}}\right)^{-1}=\frac{1}{3}\left(1+\frac{3\rho_{b}}{4\rho_{r}}\right)^{-1}\,,
 \end{equation}
 thus
\begin{equation}\label{eq:SoundspeedPrimordialFluid}
    c_s=\frac{1}{\sqrt{3(1+R_{br}(a))}}\,,
\end{equation}
where $R_{br}(a)$ is the \textit{baryon-to-photon energy ratio}
\begin{equation}
    R_{br}(a)\equiv \frac{3}{4}\frac{\rho_b(a)}{\rho_r(a)}=\frac{3}{4}\frac{\Omega_b}{\Omega_r}a\,.
\end{equation}
The baryons makes the fluid heavier and consequently lower the sound speed. The factor of $a$ in the baryon-to-photon energy ratio arises due to the difference in time dependent decay of the relativistic and non-relativistic species, and captures the fact that the influence of baryons increases over time as radiation decays faster.
    \item At late times, after the CMB photons decouple from the neutralized baryon fluid, baryons essentially follow the cold dark matter wells and the dominant non-relativistic matter can be described in terms of a total matter density
    \begin{equation}
        \rho_m(\mathbf{x},t)=\overline\rho_{m}(t)+\delta\rho_{cdm}(\mathbf{x},t)+\delta\rho_{b}(\mathbf{x},t)\,.
    \end{equation}
    \item While the adiabatic conditions on the sound speed discussed above naturally arise in a fluid in thermal equilibrium dominated by temperature fluctuations, the initial conditions are generally  assumed to be adiabatic from the start \cite{Weinberg2008Cosmology,maggiore2018gravitationalV2,dodelson2020modern}. This implies that the scalar perturbations arise from a single independent variable. Thus, in the following, we assume that we only need to specify the initial conditions of one quantity in the scalar sector.
    \item In cosmology, it is extremely useful to work in Fourier space, as the equations for linear perturbations on a solely time dependent background turn into a set of decoupled differential equations. In this case, each $\mathbf{k}$ mode evolves independently of all other modes. Moreover, the Fourier domain allows for a clear separation of perturbations according to their scale, which for instance at late times allows a discrimination between still-linear large scale modes and non-linear small scale perturbations. Since in cosmology only a portion of the entire universe is observable and quantities do not tend to zero at these boundaries, the Fourier transform cannot be defined as an integral over spacial infinity. Rather, one must consider a finite volume $V$, such that (see e.g. \cite{maggiore2018gravitationalV2})
    \begin{equation}\label{eq:FourierTransformDef}
    \tilde{f}(\mathbf{k})\equiv\frac{1}{\sqrt{V}}\int_Vd^3x\,f(\mathbf{x})\,e^{-i\mathbf{k}\cdot\mathbf{x}}\,,
    \end{equation}
    for any space dependent quantity $f$. Observe that this implies that for a dimensionless quantity $f$ (such as the metric perturbation $\phi$ or the overdensity $\rho$) the Fourier transform as dimensions of $k^{-3/2}$, hence $[\tilde f]=L^{3/2}$. The associated inverse Fourier transform is given by
    \begin{equation}
        f(\mathbf{x})=\sqrt{V}\int d^3k\, \tilde{f}(\mathbf{k})\,e^{i\mathbf{k}\cdot\mathbf{x}}\,,
    \end{equation}
    If $f(\mathbf{x})$ is real, then its Fourier transform satisfies $\tilde{f}^*(\mathbf{k})=\tilde{f}(-\mathbf{k})$.
    \item In Fourier space the equations naturally separate in two regimes, namely on the one hand the \textit{super-Hubble}\footnote{It is also common to call this the super-horizon regime. However, as discussed above, the Hubble radius can only be interpreted as a horizon in a universe that forever expands exponentially.} (super-H) regime in which the comoving scale $k$ of the perturbation is larger than the comoving Hubble radius (see Eq.~\eqref{eq:HubbleRadius})
    \begin{equation}
        \frac{\ell_H}{a} \ll \frac{1}{k}\quad\Leftrightarrow\quad k\ll aH\,.
    \end{equation}
    In this large scale limit essentially all $k$ dependent terms associated to pressure inducing oscillations can be dropped, and the evolution is only governed by the gravitational pull and the Hubble friction caused by the background expansion. On the other hand, on \textit{sub-Hubble} (sub-H) scales
    \begin{equation}
        \frac{\ell_H}{a} \gg \frac{1}{k}\quad\Leftrightarrow\quad k\gg aH\,,
    \end{equation}
    where the scalar solutions generally correspond to damped oscillations.
\end{itemize}

In the subsequent Sections, we will follow the history of the observable cosmos, highlighting key epochs in the evolution and discuss two central observational pillars in more detail: The model-fitting of the CMB powerspectrum and the local measurements of the clustering of matter on large scales. We will however start at the ``beginning'' by describing the cosmological initial conditions of the $\Lambda$CDM model that evolved to an expanding universe that we see today.

\subsection{Random Initial Conditions}\label{sSec:Initial Conditions}

As already mentioned, in this chapter we will focus on the scalar sector within an SVT decomposition of linear perturbations, as the scalars determine the observed anisotropies and inhomogeneities in the universe. Let's therefore consider a scalar variable such as the energy density $\rho(\mathbf{x},t)$ to be concrete, whose space dependence is introduced by the perturbations conveniently described in terms of density contrast $\delta(\mathbf{x},t)$ defined in Eq.~\eqref{eq:Overdensities def}. The following considerations hold for any perturbed quantity, in particular the scalar metric perturbations, but because of the various simplifying assumptions, standard cosmology actually only requires the formulation of initial conditions for a single quantity in the scalar sector.

On general grounds, it is expected, that the initial conditions at a given time\footnote{In this context, it will be most convenient to choose the scale factor $a$ as a measure of time.} $a_\text{in}$ at a given location are random. That is, the value of a given observable such as the energy density $\rho(\mathbf{x},a_\text{in})\equiv\rho_\text{in}(\mathbf{x})$ at a given location is stochastic, in the sense that it is a variable that is drawn from a certain distribution whose precise value cannot be determined deterministically. This assumption could simply represent our ignorance of the process leading to the initial conditions. 

\paragraph{Background Initial Conditions: The Horizon Problem.} At this stage, we should distinguish between the initial conditions for the background value $\overline{\rho}_\text{in}$ and the perturbations $\delta_\text{in}(\mathbf{x})$. Cosmology requires a mechanism that explains both the advent of a homogeneous and isotropic early universe that is however slightly departed from through primordial perturbations. Let's first analyze the homogeneous and isotropic initial conditions.

As already mentioned, the requirement of homogeneity and isotropy arises from the observation of a very rotationally invariant universe about us together with the assumptions of no preferred location, implying isotropy around any comoving observer resulting in homogeneity. For the background quantity $\overline{\rho}_\text{in}$, this then imposes an independence of the spacial location. Based on the expectation of random initial conditions, a mostly homogeneous and isotropic universe could be explained through a mechanism of natural alignment, such as through an establishment of a thermal equilibrium of initial values. The formation of such an equilibrium in turn requires causal contact between different patches of the universe.


However, in a universe with a starting point naively described as $a\rightarrow 0$, there was only a finite time for causal light cones to spread out on the comoving grid, described by the particle horizon introduced in Eq.~\eqref{eq:ParticleHorizon}. Now, we are not able to observe the universe at the earliest times, and over time the particle horizons keep growing. But for a universe starting in RD, even at the time of the CMB, which is so far our earliest probe, the particle horizons would on today's sky correspond to patches separated by angles of $\approx 1.2^\circ$ only (see e.g. \cite{dodelson2020modern}). Hence, since the temperature of the CMB depends on the energy density, one would actually expect random initial conditions to produces a very anisotropic background radiation. This discrepancy between the expectation of random initial conditions and the observed isotropy is known as the \textit{horizon problem}. Thus, while we do not expect to understand the precise value of $\overline{\rho}_\text{in}$, observations indicate a stage of evolution of the universe before $a_\text{in}$ that cannot be captured by Eq.~\eqref{eq:HubbleParameterLCDMSecond}. Any such earlier epoch will be considered as part of the mechanism of initial conditions that should therefore explain how the observable universe was in causal contact at $a_\text{in}$ (or at least at the time of the CMB).

\paragraph{Perturbation Initial Conditions: Cosmological Statistics.}
On the other hand, a mechanism for initial conditions should also provide an origin of the observed perturbations. The randomness hypothesis implies that at every location in space, the scalar perturbations, hence for instance the scalar gravitational potential $\phi_\text{in}(\mathbf{x})$ or the overdensity $\delta_\text{in}(\mathbf{x})$ or equivalently the total energy density $\rho_\text{in}(\mathbf{x})$ at time $a_\text{in}$ should be a stochastic variable. Indeed, within the leading hypothesis of the initial condition mechanism that we will describe below, the origin of cosmological perturbations are quantum fluctuations of a primordial field and therefore intrinsically random. Hence, also in the subsequent evolution, these scalar observables carry their stochastic past with them. For concreteness, we will in the following focus on the energy density $\rho(\mathbf{x})$ and the associated overdensities at a given time in order to discuss their statistical properties. As we will now explicitly show, the cosmological principle largely restricts the form of allowed perturbations on a homogeneous and isotropic background.

First, the central limit theorem implies that for a large number of random processes involved in the creation of the initial conditions the distribution from which the values are drawn should be at least very close to a Gaussian distribution, an assumption that has been verified observationally. 
We therefore suppose that at every location in space, the scalar perturbations, hence for instance the initial total energy density $\rho_\text{in}(\mathbf{x})$ at time $a_\text{in}$ is drawn from a Gaussian distribution. 
If we restrict our attention to a single point, then such a Gaussian distribution is entirely characterized by the mean value $\langle \rho(\mathbf{x})\rangle$ and its \textit{variance}
\begin{equation}
    \sigma^2= \big\langle (\rho(\mathbf{x})-\langle \rho(\mathbf{x}) \rangle)^2 \big\rangle=\langle \rho^2(\mathbf{x}) \rangle- \langle \rho(\mathbf{x}) \rangle^2 \,.
\end{equation}
However, we do not have access to multiple realizations of the universe and the only thing we can do is to measure quantities at different locations in space, assuming that this is equivalent to probing a representative set of the distribution. This is a good estimate of the ensemble average as long as the volume is big enough to ensure that the associated variance $\sigma^2$ tends to zero on large enough patches, which is the case due to the \textit{ergodic theorem} (see e.g. \cite{Weinberg2008Cosmology}). The ensemble average $\langle ...\rangle$ is in cosmology therefore replaced by a spacial average
\begin{equation}
    \boxed{\langle \rho(\mathbf{x})\rangle=\overline{\rho}(\mathbf{x})\equiv \frac{1}{V}\int_Vd^3x'\,\rho(\mathbf{x}+\mathbf{x}')\,.}
\end{equation}
This in particular implies that by definition
\begin{equation}
    \langle \delta(\mathbf{x})\rangle=0\,.
\end{equation}
Moreover, due to the cosmological principle the average of any cosmological quantity over the observable universe does not depend on the spacial location, thus simply $\langle \rho(\mathbf{x})\rangle=\overline{\rho}$, where $\overline{\rho}$ is the background FLRW value. As discussed, this translates into the expectation that the observable universe was in causal contact during the formation of initial conditions.

Within causally connected regions, the value of a quantity at different locations $\rho(\mathbf{x})$ and $\rho(\mathbf{x}')$ must not be independent of each other. The distribution is therefore more precisely given by an infinite dimensional Gaussian characterized by variances at every spacial point and the correlations between different locations. For a Gaussian process, this correlation is entirely captured by the \textit{two-point correlation function} 
\begin{equation}\label{eq:TwoPointCorrelationFunction}
    \xi_\rho(\mathbf{x},\mathbf{x}')\equiv \frac{\langle \rho(\mathbf{x})\rho(\mathbf{x}')\rangle}{\overline{\rho}^2}-1=\langle \delta(\mathbf{x})\delta(\mathbf{x}')\rangle
\end{equation}
In principle, to characterize the distribution of initial conditions therefore requires the knowledge of a variance at each spacial location together with the two-point functions in Eq.~\eqref{eq:TwoPointCorrelationFunction}. Fortunately, however, the cosmological principle actually severely restricts the form of the cosmic distribution. First of all, homogeneity implies that the variances do not depend on the spacial position and thus the distribution depends on one single variance.  Moreover, homogeneity and isotropy, hence invariance under spacial translations and rotations, also restricts the functional form of any two-point correlation function to only depend on the distance between the two points
\begin{equation}
    \xi(\mathbf{x},\mathbf{x}')=\xi(|\mathbf{x}-\mathbf{x}'|)\,.
\end{equation}

At this point we should mention that in particular for energy densities it is not possible to measure its value at a single point but rather over a volume $\Delta V$ of a certain size. Thus, the physical meaning of the two-point correlation function should be understood as providing a measure for the joint probability of measuring particular densities within two volumes $\Delta V_1$ and $\Delta V_2$ around $\mathbf{x}$ and $\mathbf{x}'$ respectively that are different from the expected mean density.
This physical interpretation implies that the two-point function measures the structure in the distribution of perturbations, and hence the clustering property of the observable. For example, the correlation function of dark matter energy densities  within the current cosmological model has a positive amplitude that decreases with the distance $R=|\mathbf{x}-\mathbf{x}'|$, recovering the perfectly homogeneous background on the largest scales. Over time, the amplitude of the correlation function increases as more and more structure forms. In particular, for a structure consisting of halos of a certain average comoving size $R$, the two-point function would indicate a correlation up to separations of size $R$, followed by a sudden drop on for larger distances.

On the other hand, the variance of overdensities given by $\xi(0)$ is of no particular use, as it is in general diverging due to a lack of small scale cutoff.\footnote{This statement depends on the initial conditions but is at least true in $\Lambda$CDM (see also \cite{Weinberg2008Cosmology}).} In order to make sense of the variance, one needs to take the remark on the measurability of a density function seriously and instead define an amplitude of energy fluctuations on a particular scale by smoothing out the overdensities over scales of comoving radius R. This effectively washes out all clustering on smaller scales and focuses on the scale $R$ and larger. The smoothed-out overdensities are defined by weighting the density contrasts by a window function of scale $R$
\begin{equation}
			\boxed{\delta_{R}(\mathbf{x})\equiv \int_V d^3x'\,\delta(\mathbf{x}')W_R\big(|\mathbf{x}-\mathbf{x}'|\big)\ ,}
\end{equation}
where the window function can be chosen as a \textit{tophat} of radius $R$ of the form
\begin{equation}
    W_R(x)=
				\left\{\begin{array}{ll}
					\displaystyle\frac{3}{4\pi R^3}\ , &\quad x<R\,,\\[8pt]
					0\ ,                               &\quad x\geq R\,.
				\end{array}\right.	
\end{equation}
Note that such smoothed-out density contrasts retain a zero mean
\begin{equation}
    \langle\delta_{R}\rangle=0\,.
\end{equation}
Moreover, because of the scale $R$ introduced by the smoothing, the interesting statistical quantity is now the variance 
\begin{equation}\label{eq:DefSigmaRFirst}
   \boxed{\sigma^2_R\equiv\langle \delta_{R}^2(\mathbf{x})\rangle\,,}
\end{equation}
that is again independent of $\mathbf{x}$ due to the homogeneity of the background. 


To go further, it is useful to consider also the statistical information within the Fourier transform of the random variables, which even more so provide information on the clustering at each given size. More precisely, computing the correlation functions between momentum modes yields
\begin{align}
    \langle\tilde \delta(\mathbf{k})\tilde \delta^*(\mathbf{k}')\rangle&=\frac{1}{V}\int_Vd^3x d^3x'\langle\delta(\mathbf{x})\delta(\mathbf{x}')\rangle\,e^{-i\mathbf{k}\cdot\mathbf{x}}e^{i\mathbf{k}'\cdot\mathbf{x}'}\,,\nonumber\\
    &=\frac{1}{V}\int_Vd^3x e^{-i(\mathbf{k}-\mathbf{k}')\cdot\mathbf{x}}\int d^3R\,\xi(|\mathbf{R}|)\,e^{-i\mathbf{k}'\cdot\mathbf{R}}\,,\nonumber\\
    &=\frac{1}{V}\int_Vd^3x (2\pi)^3\delta^3(\mathbf{k}-\mathbf{k}')\int d^3R\,\xi(R)\,e^{-i\mathbf{k}\cdot\mathbf{R}}\,,\nonumber\\
    &=\frac{1}{V}(2\pi)^3\delta^3(\mathbf{k}-\mathbf{k}')\,4\pi\int_0^{R_\text{max}}dR \,R^2\,\xi_\rho(R)\left(\frac{\sin(kR)}{kR}\right)\,,
\end{align}
where $k\equiv|\mathbf{k}|$, $\mathbf{R}\equiv\mathbf{x}-\mathbf{x}'$, $|\mathbf{R}|\equiv R$ and $R_\text{max}$ is the largest separation within the volume $V$. Thus, in contrast to the spacial variables, the Fourier space fluctuations of different values of momenta $\mathbf{k}$ are not correlated, and the distribution is characterized by the variances 
\begin{equation}\label{eq:Power Spectrum}
    \langle|\tilde \delta(\mathbf{k})|^2\rangle\equiv P_\rho(k)=4\pi\int_0^{R_\text{max}}dR \,R^2\,\xi_\rho(R)\left(\frac{\sin(kR)}{kR}\right)\,,
\end{equation}
that only depend on the modulus $k$ but are in principle distinct for every $k$. The variance of momentum space variables is known as the \textit{power spectrum} which measures the amplitude of clustering at each scale $k$. Note that with the conventions of the Fourier transform in Eq.~\eqref{eq:FourierTransformDef} the power spectrum has the dimensions of $k^{-3}$, in other words dimensions of a volume $[P(k)]=L^3$. It is therefore also useful to define the dimensionless power spectrum
\begin{equation}\label{eq:Dimensionless Power Spectrum}
    \mathcal{P}(k)\equiv k^3\,P(k)\,.
\end{equation}

Conversely, the correlation function is determined in terms of the power spectrum as
\begin{align}
    \xi_\rho(|\mathbf{x}-\mathbf{x}'|)&=V\int\frac{d^3k}{(2\pi)^3}\frac{d^3k'}{(2\pi)^3}\langle\tilde\delta(\mathbf{k})\tilde\delta(\mathbf{k}')\rangle\,e^{i\mathbf{k}\cdot\mathbf{x}-i\mathbf{k}'\mathbf{x}'}\,,\nonumber\\
    &=\int\frac{d^3k}{(2\pi)^3}\,P_\rho(k)\,e^{i\mathbf{k}\cdot(\mathbf{x}-\mathbf{x}')}\,,
\end{align}
and in particular the variance reads
\begin{equation}
    \xi(0)=\sigma^2=\frac{1}{2\pi^2}\int_0^\infty\frac{dk}{k}\,\mathcal{P}(k)\,.
\end{equation}
As discussed, this variance is however of no particular interest, and we should instead consider the variance of smoothed-out overdensities defined in Eq.~\eqref{eq:DefSigmaRFirst}, which in analogy to the variance above can be written in terms of the dimensionless powerspectrum as
\begin{equation}\label{eq:def_sigmaR_2}
			\boxed{\sigma_R^2= \int\frac{d k}{k}\mathcal{P}_m(k)\tilde W^2(kR)\, ,}
\end{equation}
where the Fourier transform of the tophat function is given by
\begin{equation}
    \tilde W(kR) \equiv \frac{3j_1(kR)}{kR}=\frac{3\left[\sin(kR)-kR\cos(kR)\right]}{(kR)^3}
\end{equation}
with $j_1$ a spherical Bessel function. Note that the additional factor introduced by the window function $\tilde W^2(kR)$ is unity for $kR\ll 1$, while it tends to zero for $kR\gg 1$. Thus, as expected, the smoothing effectively cuts off small scales of $k>1/R$ and effectively sums up the dimensionless powerspectrum for all large scales. However, since no structure is expected on the largest scales, $\sigma_R$ can be viewed as a measure of the clumpiness of a given variable around scale $R$.

\paragraph{Perturbation Initial Conditions: Curvature Powerspectrum.}

 As mentioned, in the scalar sector, it is sufficient to provide the initial conditions for a single scalar perturbation because of the adiabatic assumption. We therefore have to specify the initial statistical distribution of one quantity that, from the considerations above, can for instance be given in the form of the powerspectra in Eqs.~\eqref{eq:Power Spectrum} and \eqref{eq:Dimensionless Power Spectrum}. While it would be possible to assign the initial conditions to either the gravitational potential $\phi_\text{in}$ or the density contrast $\delta_\text{in}$, it is custom in cosmology to provide the random initial conditions in terms of a combined gauge-invariant variable called the \textit{curvature perturbations}\footnote{The nomenclature hinges on the relation of $\mathcal{R}$ with the three-dimensional curvature of surfaces in the comoving frame.} $\mathcal{R}_\text{in}$ (see e.g. \cite{Gorbunov:2011zzc,maggiore2018gravitationalV2,dodelson2020modern}). 
 
 This variable has the very convenient property that its momentum mode is independent of time outside of the horizon both in RD and MD, hence $\tilde{\mathcal{R}}_\text{in}(\mathbf{k})=\tilde{\mathcal{R}}(\mathbf{k},a)$. Therefore, the curvature perturbations provide the ideal initial conditions for momentum modes entering the Hubble radius one by one and start their cosmic evolution.
 In fact, the existence of a conserved scalar perturbation on super Hubble scales regardless of the presence of constituents can be viewed as a definition of adiabatic perturbation solutions \cite{Weinberg2008Cosmology}.


 Within $\Lambda$CDM, the dimensionless power spectrum of curvature perturbations is parameterized by a powerlaw around a \textit{pivot scale} $k_p$ of the form
\begin{equation}\label{eq:InitialCurvaturePowerSectrum Dimensionless}
	\boxed{\mathcal{P}_\mathcal{R}(k)= A_s\left(\frac{k}{k_p}\right)^{n_s-1}} \,,\qquad k_p=0.05\,\text{Mpc}^{-1}\,.
\end{equation}
Thus the random initial conditions of scalar perturbations are determined by the small amplitude $A_s$ and the \textit{scalar spectral index} $n_s$ which measures the departure from a perfectly scale-independent initial spectrum. As we will see below, $\Lambda$CDM favors a nearly scale-independent powerspectrum but requires a small spectral tilt to the red, hence $n_s-1<0$. The choice of reverence value $k_p$ is a matter of convention. Note, however, that both the amplitude $A_s$ and the spectral index $n_s$ depend on the pivot scale. 

For every scalar perturbation quantity, for example the scalar potential modes $\tilde\phi(\mathbf{k},a)$, we can therefore factor out the random initial conditions by defining the deterministic variable
\begin{equation}\label{eq:Deterministic Variable}
    \tilde\phi_\text{d}(k,a)\equiv \tilde\phi(\mathbf{k},a)/\mathcal{R}(\mathbf{k})\,.
\end{equation}
These variable capture the deterministic evolution after Hubble radius reentry that only depends on the modulus $k$ and time $a$. 

With the assumptions of Gaussian initial conditions of adiabatic perturbations, the two numbers $A_s$ and $n_s$, together with the knowledge of the super-Hubble conservation of the associated curvature perturbations are in principle all we need in order to specify the random initial conditions of our cosmological model. Hence, the precise mechanism of initial conditions is in that sense not important to do cosmology. However, we still want to quickly mention two concrete possibilities that also address the horizon problem introduced above, starting with the most popular and most developed paradigm. 

\paragraph{Mechanisms for Initial Conditions: Inflation and Bouncing Cosmologies.}

\textit{Inflation} is the proposition that before the expected radiation dominating era of the FLRW evolution, the universe started off with an exponential expansion driven by the potential energy of a primordial scalar field \cite{Guth:1980zm,Starobinsky:1980te,Sato:1980yn,Mukhanov:1981xt} (see also \cite{Liddle:2000cg,Tsujikawa:2003jp,Cheung:2007st,Weinberg:2008hq,Weinberg2008Cosmology,Gorbunov:2011zzc,Rubakov:2017xzr,Vazquez:2018qdg,maggiore2018gravitationalV2,dodelson2020modern}). Such a scalar field called \textit{inflaton} is not known, but could very well be associated to a beyond GR mode. An early accelerated expansion solves first of all the horizon problem, because an initially causally connected region of comoving space is stretched far beyond the Hubble radius, making up our entire observable universe. Thus, widely spread regions that at the time of the CMB could naively not have been in causal contact according to a standard evolution actually originate from a same small patch eventually contained in a single particle horizon. Note, however, that inflation also requires a further mechanism that transforms the energy of the inflaton into the energy content of a standard cosmology, including matter, in a process called \textit{reheating}, the details of which are entirely unclear. 

Moreover, while the process of inflation essentially completely empties the observable universe, the quantum fluctuations of the inflaton that dominates the energy density at the end of inflation causes the initial density perturbations that can be observed in the CMB and eventually grew to the structure of the universe observed today. More precisely, the fluctuations can be viewed to cause an end of inflation at slightly different times in different regions. In an exponential expansion, such tiny fluctuations are however stretched to macroscopic scales. An important point is that the associated curvature perturbations at the relevant scales today have exited the Hubble radius by the end of inflation and are thus conserved until their reentry, regardless of the unknown details of reheating happening in between.
In the simplest inflatory models, these fluctuations are generically adiabatic, Gaussian and nearly scale-invariant. Remarkably, these properties of fluctuations were predicted by inflation before their observation in the CMB. 

Another interesting option for a mechanism of initial conditions are for instance \textit{bouncing cosmologies}, with the philosophically and theoretically appealing possibility that our current expanding universe emerged from a contracting universe with a (possibly quantum) turnover point at large densities (see \cite{Battefeld:2014uga,Brandenberger:2016vhg,Ijjas:2018qbo} for a review). Both options come with their conceptual advantages and disadvantages \cite{Brandenberger:1999sw,Battefeld:2014uga,Brandenberger:2016vhg}, although inflation is by far the most widely accepted and developed paradigm. Interestingly, future measurements of primordial gravitational waves might be able to discriminate between different scenarios of the beginning of our universe.

\paragraph{The Flatness Problem.}
Finally, we want to mention another puzzle concerning initial conditions that an early universe scenario might want to address, namely the fact that today we observe an almost perfectly spatially flat or in other words Euclidean universe. The reason this is hard to explain for a universe that started in a radiation dominated phase followed by matter domination can be seen as follows: Based on such a standard evolution of the universe, an initial small curvature contribution one might expect with random initial conditions, will eventually dominate due to its $\sim a^{-2}$ scaling given in Eq.~\eqref{eq:FriedmannEquationGen}, compared to relativistic and non-relativistic matter scaling as $\sim a^{-4}$, respectively $\sim a^{-3}$. More precisely, we can rewrite the Friedmann equation [Eq.~\eqref{eq:FriedmannEquationGen}] as
\begin{equation}
    \frac{k}{a^2H^2}=\Omega(t)-1\,,
\end{equation}
where we have defined the time dependent total energy density quantity\footnote{Note the difference to the time independent density parameters introduced in Eq.~\eqref{eq:DefDensityParameters}, which represent the energy densities today.}
\begin{equation}
    \Omega(t)\equiv \frac{\kappa_0\rho(t)}{3H^2(t)}\,.
\end{equation}
A vanishing curvature $k=0$ therefore requires that $\Omega(t)=1$ for all times. However, over a radiation and matter dominated era, for which $a\propto t^\alpha$ with\footnote{Recall that $\alpha=1/2$ and $\alpha=2/3$ for radiation, respectively matter domination.} $\alpha<1$ (see Eq.~\eqref{eq:GenSolutionScaleFactor}) and hence $H=\alpha/t$, such that
\begin{equation}
    \frac{k}{a^2H^2}\propto k\,t^2{1-\alpha}\,,
\end{equation}
is growing over time. Observing a nearly zero curvature today therefore naively requires very special and fine-tuned initial conditions for the spacial curvature. 

Yet, inflation offers an elegant solution to this ``flatness problem'', since an initial period of accelerated expansion with constant energy density would naturally wash away any initial spacial curvature with $\sim a^{-2}$ \cite{Guth:1980zm}. In other words, as explained above, according to inflation the universe today is but a tiny patch of the initial spacetime, small enough so that spacial curvature is negligible. Note that this process also washes out all radiation and non-relativistic matter, that has to be recreated during the period of reheating. On the other hand, bouncing cosmologies do not all come with a mechanism to address the flatness problem, although a subset of models do \cite{Brandenberger:2016vhg}.

\subsection{The Cosmic Microwave Background}\label{sSec:CMB}

Next, we want to analyze the cosmic microwave background radiation more closely and especially describe its perturbations based on the initial conditions described above. As discussed, the CMB radiation is characterized by a temperature averaged over the sky today
\begin{equation}
    \boxed{T_0=\frac{1}{4\pi}\int_{S^2} d\Omega\, T_0(\theta,\phi)\,,}
\end{equation} 
that is perturbed by tiny anisotropies that depend on the direction in the sky 
\begin{equation}\label{eq:DefTemperaturFluctuations}
    \boxed{\delta T_0(\theta,\phi)=T_0(\theta,\phi)- T_0\,.}
\end{equation}
For brevity it will be useful to indicate the dependence on the polar angles by the direction of the unit vector $\mathbf{n}= \mathbf{n}(\theta,\phi)$ [Eq.~\eqref{eq:Def Direction of Propagation n}] already employed in previous chapters. Moreover, instead of a Fourier transform adequate for observables over space, it is useful to expand the temperature scalar fluctuation in terms of spherical harmonics in order to separate different angular scales
\begin{equation}
    \frac{\delta T_0(\mathbf{n})}{T_0}=\sum_{l=1}^\infty \sum_{m=-l}^l a_{lm} Y_{lm}( n)\,.
\end{equation}
Note that the sum starts at $l=1$, since the constant mode $l=0$ corresponds to the angular average that vanishes by the definition in Eq.~\eqref{eq:DefTemperaturFluctuations}.
Conversely, the perturbation coefficients are related to the temperature fluctuations through
\begin{equation}\label{eq:alm in terms of temp}
    a_{lm}=\int_{S^2} d\Omega \frac{\delta T_0(\mathbf{n})}{T_0}\,Y^*_{lm}(\mathbf{n})\,.
\end{equation}
And since $\delta T_0$ is real, the perturbation coefficients satisfy
\begin{equation}
    a^*_{lm}=(-1)^m\,a_{l-m}\,,
\end{equation}
because of the relation
\begin{equation}
    Y^*_{lm}=(-1)^m \,Y_{l-m}\,.
\end{equation}

Now, since the temperature fluctuations $\delta T_0(\mathbf{n})$ and the associated harmonic transform $a_{lm}$ have a stochastic origin, we are again interested in the corresponding averaged quantities, in particular the two-point correlation function $\langle\delta T_0(\mathbf{n})\delta T_0(\mathbf{n}')\rangle$, where, as discussed, in cosmology, the formal ensemble average $\langle ...\rangle$ is replaced by an average over space. Moreover, the cosmological principle imposes a rotational and translational invariance of all averaged quantities. In particular, the mean value $\langle\delta T_0(\mathbf{n})\rangle$ is independent of the angles (and the position) and thus since by definition
\begin{equation}
     \int_{S^2} d\Omega\, \delta T_0(\theta,\phi)=0
\end{equation}
its average imposes 
\begin{equation}
    \int_{S^2} d\Omega\, \langle\delta T_0(\mathbf{n})\rangle=0 \quad\Rightarrow\quad \langle\delta T_0(\mathbf{n})\rangle=0\,.
\end{equation}
Just as for the powerspectrum discussed above, the information of the two-point function (and for a Gaussian distribution of all n-point functions) is conveniently captured within the correlation function of the harmonic modes $a_{lm}$. And again, rotational invariance implies that
\begin{equation}
    \langle a_{lm}a^*_{lm}\rangle= C_l \,\delta_{ll'}\delta_{mm'}\,,
\end{equation}
for some function $C_l$ called \textit{angular powerspectrum} that is real and positive and only depends on $l$. Hence, the different $a_{lm}$ are uncorrelated and the distribution is completely characterized by the variance that is related to the temperature fluctuations through
\begin{equation}\label{eq:Theoretical Cl}
    C_l=\frac{1}{2l+1}\sum_{m=-l}^l\langle|a_{lm}|^2\rangle=\frac{1}{4\pi}\int_{S^2}d\Omega d\Omega'\,P_l (\mathbf{n}\cdot \mathbf{n}')\,\frac{\langle \delta T_0(\mathbf{n})\delta T_0(\mathbf{n}')\rangle}{T_0^2}\,,
\end{equation}
where we have used the identity of the Legendre polynomials
 \begin{equation}
     P_l (\mathbf{n}\cdot \mathbf{n}')=\frac{4\pi}{2l+1}\sum_{m=-l}^l Y^*_{l m}(\mathbf{n})Y_{lm}(\mathbf{n}')\,.
 \end{equation}

 Note that the angle $\theta$ on the sky is approximately inversely proportional to the multipole moment $l\sim\pi/\theta$. This follows from the number of nodes within the Legendre polynomial $P_l(\cos\theta)$, which for instance for the dipole $l=1$ has two nodes and therefore separates the circle of $\theta$ in two parts of angular distance $\pi$. The quadrupole with $l=2$ then has four nodes separated approximately by $\pi/2$ and so and so forth. It is therefore useful to introduce the \textit{characteristic angular scale} $\theta_l$ through
\begin{equation}\label{eq:relTheta and l}
    \boxed{\theta_l=\frac{\pi}{l}\,,}
\end{equation}
which reflects this approximate relation. 

 \paragraph{The Observed Angular Powerspectrum.}

 However, in the case of the CMB, observationally it is not possible to compensate the lack of access to multiple ``realizations'' of the universe through an average over space, as we can only observe the CMB at one particular location over the sphere. The best we can do is to count on the integration over the sphere to provide a good enough statistical probe, which in the case of $a_{lm}$ reduces to a sum over $m$. In other words, the quantity that is actually observed in the CMB is \cite{Weinberg2008Cosmology,maggiore2018gravitationalV2}
 \begin{equation}
    C_l^\text{obs}=\frac{1}{2l+1}\sum_{m=-l}^l\,|a_{lm}|^2=\frac{1}{4\pi}\int_{S^2}d\Omega d\Omega'\,P_l (\mathbf{n}\cdot \mathbf{n}')\,\frac{ \delta T_0(\mathbf{n})\delta T_0(\mathbf{n}')}{T_0^2}\,.
\end{equation}
This introduces an intrinsic uncertainty especially limiting for low $l$, since for a given $l$ only maximally $2l+1$ coefficients can be sampled. More precisely, the fractional mean square error is known as the \textit{cosmic variance} and is for a Gaussian distribution given by \cite{Weinberg2008Cosmology}
\begin{equation}
    \left\langle\left(\frac{C_l-C_l^\text{obs}}{C_l}\right)^2\right\rangle=\frac{2}{2l+1}\,,
\end{equation}
which fortunately decreases fast enough for large $l$.

\begin{figure}[H]
\centering
\includegraphics[scale=0.35]{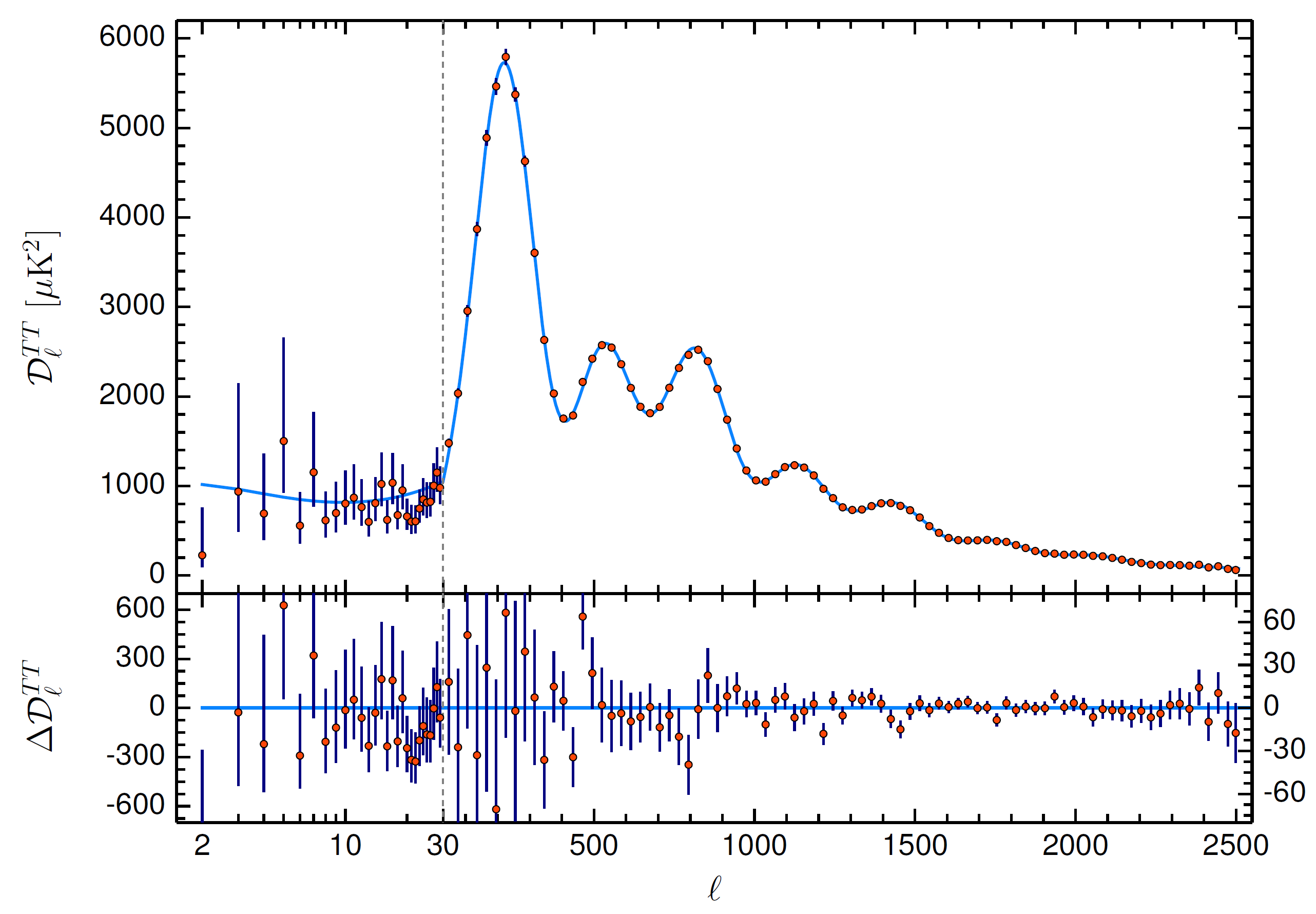}
\caption{\small (Upper panel) The anisotropies in the CMB given by the temperature powerspectrum as measured by the Planck satellite (points) compared to the best-fit $\Lambda$CDM (blue line). The variance of the temperature fluctuations as a function of scale $\mathcal{D}_l^{TT}=l(l+1)C_l^{TT}T^2_0/(2\pi)$ is given in terms of multipole moments $l$ associated to the angular scale. Note the change in $y$ axis al $l=30$. (Lower panel) Difference between the data and the best-fit model. (Figure taken from \textit{Planck Collaboration, N. Aghanim et al., (2018)} \cite{Planck:2018vyg}.)}
\label{fig:PlackCMB}
\end{figure}

The observed angular power spectrum is depicted in Figure~\ref{fig:PlackCMB} showing data points of the latest CMB anisotropy measurements by the Planck satellite \cite{Planck:2018vyg}. Plotted is the conventional combination
\begin{equation}
    \mathcal{D}_l^{TT}\equiv l(l+1)C_l^{TT}T^2_0/(2\pi)\,,
\end{equation}
where the $TT$ superscript indicates that it is the angular powerspectrum associated to the temperature two-point function.\footnote{Not to be confused with the $TT$-gauge introduced in earlier chapters.} Observe that the plot starts at $l=2$, since the $l=1$ dipole mode is used to compute the particular velocity of the earth compared to the CMB rest frame that defines the background FLRW coordinates. Moreover, the cosmic variance is clearly visible in Figure~\ref{fig:PlackCMB} with increasing error bars at low $l$. 

 \paragraph{The Computed Angular Powerspectrum.} To be able to compare the observed CMB anisotropies with a cosmological model, we need a few further considerations and manipulations.
 First of all, note that as opposed to the observation, theoretically we are not limited to our position in the universe and can compute the spacial averages. Thus, in theory, we can work with the statistically accurate definition in Eq.~\eqref{eq:Theoretical Cl}. In particular, we would of course like to relate the stochastic properties of the temperature fluctuation variable to the initial powerspectrum of fluctuations introduced in Section~\ref{sSec:Initial Conditions}. Since this initial power spectrum is by definition provide in Fourier space, to do this, we first perform a Fourier transform of the temperature fluctuations and simply evaluate it today $t=t_0$ at our location at the origin $\mathbf{x}=0$
 \begin{equation}\label{eq:FourierTransform dT}
     \delta T_0(\mathbf{x}=0,\mathbf{n})=\sqrt{V}\int\frac{d^3k}{(2\pi)^3}\,\delta \tilde T_0(\mathbf{k},\mathbf{n})\,.
 \end{equation}
 
 Now it turns out that the Fourier transform $\delta \tilde T_0(\mathbf{k},n)$ only depends on the angular direction $\mathbf{n}$ through the combination $\mathbf{n}\cdot  \hat k\equiv \mu$, with $\hat k\equiv \mathbf{k}/k$ that can be parameterized by an angle $\theta$ for a suitably rotated coordinate system such that $\mathbf{n}\cdot \hat k=\cos(\theta)$. This implies that the Fourier transform $\delta \tilde T_0$ can further be expanded in terms of Legendre polynomials that form a complete set of functions over the range of $\mu$. Hence,
 \begin{equation}
     \delta \tilde T_0(\mathbf{k},\mu)=\sum_{l=0}^\infty (2l+1)\,i^l\,\delta \tilde T_l(\mathbf{k})\,P_l(\mu)
 \end{equation}
 Plugging the expanded Fourier transform in Eq.~\eqref{eq:FourierTransform dT} into the expressions in Eqs.~\eqref{eq:alm in terms of temp} and \eqref{eq:Theoretical Cl}, while using the identity
 \begin{equation}
     \int_{S^2} d\Omega\, Y^*_{lm}(\mathbf{n})P_{l'}(\hat k\cdot n)=\delta_{ll'}\frac{4\pi}{2l+1}Y^*_{lm}(\hat k)\,,
 \end{equation}
 one obtains (see e.g. \cite{maggiore2018gravitationalV2})
 \begin{equation}\label{eq:Cl intermediate}
     C_l=4\pi V\int \frac{d^3k}{(2\pi)^3}\frac{d^3k'}{(2\pi)^3}\,P_l(\hat k\cdot\hat k')\,\frac{\langle \delta \tilde T_l(\mathbf{k})\delta \tilde T_l^*(\mathbf{k}')\rangle}{T_0^2}\,.
 \end{equation}

 Finally, this is the form for which the angular power spectrum can be related to the stochastic initial conditions provided by the curvature powerspectrum $\mathcal{P}_\mathcal{R}(k)$ defined in Eq.~\eqref{eq:InitialCurvaturePowerSectrum Dimensionless} by defining deterministic Fourier modes as in Eq.~\eqref{eq:Deterministic Variable}. More precisely, we define 
 \begin{equation}
      \delta \tilde T^d_{l}(k)\equiv  \delta \tilde T_l(\mathbf{k})/\mathcal{R}(\mathbf{k})\,,
 \end{equation}
 where we simply divide out the stochastic initial conditions and are left with an observable that underwent a deterministic evolution from Hubble entry until today. Plugging this back into Eq.~\eqref{eq:Cl intermediate} while using $P_l(1)=1$, the angular powerspectrum becomes
 \begin{equation}
     \boxed{C_l=4\pi \int_0^\infty \frac{dk}{k} \frac{|\delta \tilde T^d_{l}(k)|^2}{T_0^2}\,\mathcal{P}_\mathcal{R}(k)\,.}
 \end{equation}
 Hence, as expected, the CMB angular powerspectrum depends on the stochastic initial conditions, as well as on the modulus of the moments $\delta \tilde T^d_{l}(k)$ that were created during the history of the universe in a given cosmological model.

 \paragraph{CMB Constraints on the $\Lambda$CDM Parameters.}

 Given a cosmological model with initial conditions parameterized by the two numbers $A_s$ and  $n_s$ characterizing the primordial powerspectrum $\mathcal{P}_\mathcal{R}(k)$ together with additional free parameters describing the cosmic evolution that can compute the moments $\delta \tilde T^d_{l}(k)$, one can therefore constrain the parameter space of the model by fitting the computed $C_l$ to the observed $C_l^\text{obs}$. Concretely, the $\Lambda$CDM model, assuming an Euclidean universe and a minimal value of neutrino masses in Eq.~\eqref{eq:NeutrinoMasses}, has only six free parameters, which can be chosen to be the set
 \begin{equation}\label{eq:LCDM parameters}
     \boxed{\Lambda\text{CDM parameters}\;=\{h,\omega_m,\omega_b,\tau_\text{rei},n_s,A_s\}\,,}
 \end{equation}
 where $\tau_\text{rei}$, the only parameter not introduced so far, is the optical depth due to reionization, that controls amount of scattering of CMB photons off free electrons in the cosmic gas after it had been reionized, as discussed further below (see also \cite{Weinberg2008Cosmology,dodelson2020modern}).\footnote{Most of the gas in the universe observed today is ionized up to redshift $z\sim 6$ \cite{dodelson2020modern}.} Moreover, we have defined the physical density parameters of each species
 \begin{equation}\label{eq:DensityParametersSmall Physical}
     \omega_\lambda\equiv \Omega_\lambda h^2\,.
 \end{equation}
 Such a redefinition of parameters is useful, as the $\omega_\lambda$ do no longer explicitly depend on the Hubble constant $h$, a dependence which was introduced in $\Omega_\lambda$ through the critical density $\rho_c=\rho_0=3H_0^2/\kappa_0$. The density parameters in Eq.~\eqref{eq:DensityParametersSmall Physical} are therefore much closer to the physical densities $\rho_\lambda$ and do not carry an unnecessary uncertainty due to the imperfect knowledge of the Hubble constant. In fact, the CMB puts stronger constraints on the physical density parameters \cite{dodelson2020modern}. 
 Moreover, defining density parameters without explicit dependence on $H_0$ will be crucial in the subsequent Chapter~\ref{Sec:CsomoTensions}.
 In terms of the physical density parameters $\omega_\lambda$, the Hubble function in Eq.~\eqref{eq:HubbleParameterLCDMSecond} becomes
\begin{align}\label{eq:HubbleParameterLCDMThird}
		H^2_\text{\tiny $\Lambda$CDM} = C_H^2 \,\Big(\omega_m (1+z)^{3} + \omega_r (1+z)^{4} + \omega_\Lambda\Big)\,,
\end{align}	
where we defined the factor
\begin{equation}
    C_H\equiv 100\ \text{km}\,\text{s}^{-1}\,\text{Mpc}^{-1}\,,
\end{equation}
that arises due to the definition of $h$ in Eq.~\eqref{eq:Def little h}. Moreover, the no-curvature constraint on the density parameters in Eq.~\eqref{eq:ConstraintOnDensityParametersFirst} now reads
\begin{equation}\label{eq:ConstraintOnDensityParametersSecond}
		\omega_\Lambda = h^2-\omega_m-\omega_r\, .
\end{equation}

 Thus, computing $C_l(h,\omega_m,\omega_b,\tau_\text{rei},n_s,A_s)$ for the standard $\Lambda$CDM model by numerically solving the Boltzmann-Einstein system of equations while varying the parameters, the result can be fitted against the observed angular powerspectrum $C_l^\text{obs}$. The result is also found in Figure~\ref{fig:PlackCMB}. Foremost, the perfect fit confirms that $\Lambda$CDM is a cosmological model that can indeed capture the production of anisotropies in the CMB. However, beyond that, the CMB is such a powerful probe, that, together with additional information within the polarization anisotropies due to Compton scattering before decoupling (see e.g. \cite{Weinberg2008Cosmology,dodelson2020modern}), it allows to constrain all free $\Lambda$CDM parameters to a high degree of precision \cite{Planck:2018vyg}. In other words, assuming the $\Lambda$CDM cosmological model, the CMB observation basically fixes all cosmological parameters. Within that model, one can then compute derived quantities such as other density parameters or the amount of large scale clustering of matter. The Table~\ref{table:PlankBestFit} shows the Planck best-fit values from \cite{Planck:2018vyg} of the six $\Lambda$CDM parameters that we will use throughout this work as a baseline cosmology, together with a subset of derived quantities.

\begin{table}[H]
\centering
\begin{tabular}{p{6.4cm} p{2cm} p{3cm}}
 \hline\hline
 Parameter& Symbol & Planck best-fit \\
 \hline
 Hubble constant & $h$   & $0.673$   \\
 Matter density parameter & $\Omega_m$ / $\omega_m$   & $0.316$ / $0.143$   \\
 Baryon density parameter & $\Omega_b$ / $\omega_b$   & $0.0494$ / $0.0224$   \\
 Optical depth due to reionization & $\tau_\text{rei}$   & $0.0543$   \\
 Scalar spectral density & $n_s$   & $0.966$   \\
 Scalar power spectrum amplitude & $A_s$   & $2.10 \times 10^{-9}$   \\
 \hline
 CC density parameter & $\Omega_\Lambda$ / $\omega_\Lambda$   & $0.684$ / $0.310$   \\
 Radiation density parameter & $\omega_r$   & $4.25\times 10^{-5}$ \\
 Redshift at decoupling & $z_*$   & $1090$   \\
 Matter clustering amplitude & $\sigma_8$   & $0.81$   \\
 \hline
\end{tabular}
\caption{\small The values of cosmological parameters used in this work. The six parameters in the upper part define the primary free parameters of the Euclidean $\Lambda$CDM cosmological standard model, as inferred by the Planck best-fit values in \cite{Planck:2018vyg}. The other parameters are derived quantities in the model.}
\label{table:PlankBestFit}
\end{table}

 \paragraph{The Features of the Anisotropic CMB.} 
 It is worth understanding the features of the angular powerspectrum in Figure~\ref{fig:PlackCMB} in more detail and qualitatively consider the various effects that go into the computation of the angular power spectrum.
The primary source of the temperature fluctuations in the CMB observed today are dominated by the effects that lead to a release of an anisotropic radiation at the time of recombination, usually denoted as $z_*$. The leading effects are \cite{Weinberg2008Cosmology,dodelson2020modern}:
\begin{enumerate}[(1)]
    \item Intrinsic fluctuations in the temperature of the equilibrium primordial plasma at $z_*$.
    \item The gravitational red- and blueshift due to fluctuations in the scalar gravitational potential at $z_*$ known as \textit{Sachs Wolfe} (SW) effect.
    \item A Doppler effect caused by velocity fluctuations in the primordial baryon-photon plasma at $z_*$.
\end{enumerate}
Yet, between this time of last scattering and today, the freely propagating CMB photons were further perturbed through various effects ranging from gravitational lensing to Compton scattering by electrons within hot gas of clusters of galaxies known as the \textit{Sunyaev-Zel'dovich} (SZ) effect. The two dominant such ``propagation'' effects are, however
\begin{enumerate}[(1)]
\setcounter{enumi}{3}
    \item Thomson scattering of photons by free electrons produced during the reionization of the cosmic gas that wash out the anisotropies, characterized by the optical depth $\tau_\text{rei}$.\footnote{Note however, that by the time of reionization, the cosmic gas diffused by a lot, such that only a small fraction of CMB photons scattered off free electrons again.}
    \item The gravitational red- and blueshift due to time-varying\footnote{Note that time-dependence is necessary, since otherwise the energy of the photons falling in and climbing out of the potentials would not be altered.} fluctuations in the scalar gravitational potential between $z_*$ and today, known as \textit{integrated Sachs Wolfe} (ISW) effect.
\end{enumerate}
Since, the gravitational potentials remain constant during matter dominationas, a fact that we will discuss in more detail below, the effect can be split into an \textit{early ISW} effect, cause by the time-variation that creeps in from RD that was not far away from decoupling and the \textit{late ISW} due to the presence of late-time dark energy.\footnote{Moreover the CMB photons all fell into the gravitational potential that we are sitting in now. However, this effect simply contributes to the monopole and does not cause any further anisotropy.} While the late ISW only affects the largest scales at $l\lesssim 30$ the early ISW adds coherently to the intrinsic temperature fluctuations \cite{dodelson2020modern}.

These effects combine to produce a angular powerspectrum in Figure~\ref{fig:PlackCMB}. Most notable is the oscillatory feature that can be traced back to pressure or sound waves in the primordial baryon-photon fluid. Since the pressure waves with finite sound speed $c_s$
only had time to travel a comoving distance given by the sound horizon defined in Eq.~\eqref{eq:SoundHorizonSecond z}, the location of the first acoustic peak $l_A$ captures this scale within the angular powerspectrum. The scale in terms of $l$ is translated into an approximate angular scale on the sky known as \textit{acoustic scale} through Eq.~\eqref{eq:relTheta and l}
\begin{equation}\label{eq:RelationAngularScaleto l}
    \boxed{\theta_A=\frac{\pi}{l_A}\,.}
\end{equation}
The variance $\sqrt{C_{l}}$ of the harmonic modes $a_{lm}$ with corresponding harmonic parameter $l=l_A$ peaks on these scales as the anisotropies show the largest deviations from their zero mean. Furthermore, the shape of the full spectrum shows three distinctive regimes, namely:
\begin{enumerate}[(i)]
    \item A nearly constant plateau for $l\lesssim 30$ on scales much larger than the sound horizon at decoupling. These extremely large-scale modes did not cross the Hubble radius before decoupling, and therefore offer a particularly direct observational probe of the initial conditions. However, these scales are plagued by the intrinsic uncertainty of the cosmic variance.
    \item A harmonic sequence between $30\lesssim l\lesssim 1500$ given by a series of acoustic peaks and troughs of damped wave within the primordial baryon-photon plasma. The first peak at around $l\sim 180$ sets the scale for the maximum distance that the sound waves could have traveled starting at $a_\text{in}$.
    \item A photon diffusion tale for $1500\lesssim $, where fluctuations are erased due to the scattering of photons around $z_*$ with a wavelength smaller than the mean free path.
\end{enumerate}

\paragraph{CMB Distance Priors.}

The most interesting feature in the CMB is the first and biggest acoustic peak whose location is determined by the acoustic scale $\theta_A$, which more precisely defines a scale that depends on the time of decoupling $\theta^*_A=\theta_A(z_*)$.
As discussed in Sec.~\ref{sSec:HomIsoUniverse}, a small angle is determined by a standard ruler of known proper size $s$ through the relation in Eq.~\eqref{eq:AngularDiameterDistanceDefAngle} involving the angular diameter distance in Eq.~\eqref{eq:AngularDistanceSecond}. In this case, the proper distance of the standard ruler provided by the sound horizon is given by $s(z_*)=r_s(z_*)/(1+z_*)$, such that the acoustic scale is given by
\begin{equation}\label{eq:AcousticScaleDef}
    \boxed{\theta^*_A=\frac{s(z_*)}{d_A(z_*)}=\frac{r^*_s}{(1+z_*)d^*_A}=\frac{r^*_s}{d^*_C}\,,}
\end{equation}
where the sound horizon of the baryon-photon fluid reads
\begin{equation}
		r_\text{s}(z) =\int^\infty_z\,\frac{d z'}{H(z')}\,c_s(z)\,,
\end{equation}
with sound speed defined in Eq.~\eqref{eq:SoundspeedPrimordialFluid}
\begin{equation}\label{eq:SoundspeedPrimordialFluidSecond}
    c_s(z)\equiv\frac{1}{\sqrt{3(1+R_{br}(z))}}\,, \qquad R_{br}(z)\frac{3}{4}\frac{\omega_b}{\omega_r(1+z)}\,,
\end{equation}
and where $d_C$ is the comoving distance given in Eq.~\eqref{eq:ComovingDistanceThird}
\begin{equation}
		d_C(z) =\int^z_0\,\frac{d z'}{H(z')}\,.
\end{equation}

The acoustic scale is therefore a very important indicator of the CMB anisotropy measurement that defines a fundamental distance ration representing a fundamental pillar of the shape of the angular powerspectrum. This scale is also known as the first CMB \textit{distance prior} that represents a particularly well measured quantity of the CMB \cite{WMAP:2008lyn,Chen:2018dbv}.

It is useful to introduce a second distance prior called \textit{shift parameter} $R^*$ that instead of the location of the peaks mainly controls the heights of the peaks, influencing the CMB temperature spectrum along the line-of-sight \cite{WMAP:2008lyn,Chen:2018dbv}
\begin{equation}\label{eq:ShiftParameter}
   \boxed{R^* \equiv (1+z_*)d_A(z_*)\sqrt{\Omega_m H_0^2}=C_H\,d^*_C\sqrt{\omega_m}\,.}
\end{equation}
By its definition, changing the shift parameter has a similar effect on the CMB powerspectrum as varying the total matter density $\omega_m$, which modulates the overall amplitude and the relative heights of the peaks due to a modified ration $\omega_{cdm}/\omega_m$ at fixed $\omega_b$ \cite{dodelson2020modern}. The slightly obscure physical meaning of the shift parameter is historic in origin. It can be viewed as a proxy for the distance ratio between the angular diameter distance and the Hubble radius at decoupling in the case of a purely matter dominated universe $H(Z_*)\sim H_0\sqrt{\Omega_m}(1+z_*)^{3/2}$, which is a good approximation at decoupling
\begin{equation}
    \frac{d_A(z_*)}{\ell_H(Z_*)}=d_A(z_*)H(z_*)\simeq \sqrt{1+z_*} R^*\,.
\end{equation}
We will however bow to the convention in the literature and in the following use the shift parameter in Eq.~\eqref{eq:ShiftParameter}. What matters, is that the two combinations of distance priors are only very weakly correlated and are very well constrained by the CMB data.
Their latest values using the Planck 2018 release can be found in \cite{Chen:2018dbv}.

 \subsection{Large Scale Structure}\label{sSec:LargeScaleStructure}

The small perturbations in the cosmos at the time of the CMB eventually grow to form the web of dark matter and galaxy clusters observed today. Studying the distribution of gravitating matter on large scales represent a second main cosmological observation, that is however slightly different from the model-matching of the CMB data discussed above. To understand these local LSS measurements, however, we first need to develop the necessary theoretical background.

\begin{figure}[H]
\centering
\includegraphics[scale=0.7]{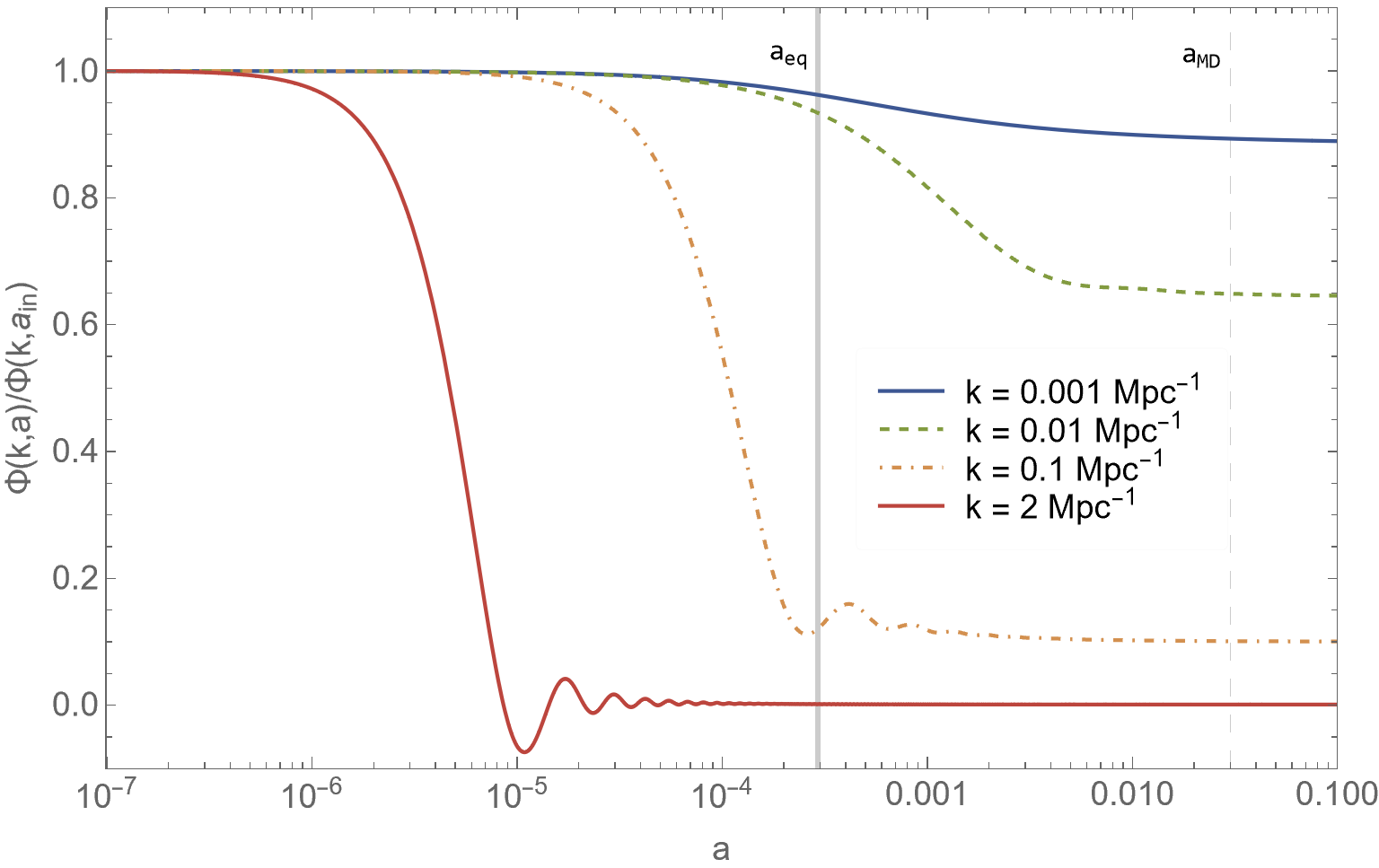}
\caption{\small The linear $\Lambda$CDM evolution of the Fourier modes $\tilde \phi_\text{d}(k,a)$ of the gravitational potential normalized to the value of the potential at $a_{in}$ for modes of different wavenumber. The mode with $k=0.001$ Mpc$^{-1}$ (blue solid) remains outside of the Hubble radius, while the mode $k=0.01$ Mpc$^{-1}$ (green dashed) enters during radiation/matter equality at $a_\text{eq}$. The two remaining modes (orange dot-dashed and red solid) enter during radiation domination. The plot was obtained by numerically solving the cosmological Einstein-Boltzmann system of equations (see e.g \cite{dodelson2020modern}) while neglecting any baryonic effects.}
\label{fig:PlotPhisNoBaryons}
\end{figure}

\paragraph{A Set of Late Time Assumptions.}
 In order to describe the growth of matter perturbations at late times on the relevant scales of the still linear large scale structures, it is necessary to understand the evolution of the scalar metric perturbation associated to the gravitational potential from its initial conditions to the present time. As discussed, we will work in the conformal Newtonian gauge. Leaving all details to the numerous existing reviews \cite{Coles:1995bd,Liddle:2000cg,Mukhanov:2005sc,Weinberg2008Cosmology,Gorbunov:2011zzc,maggiore2018gravitationalV2,dodelson2020modern}, we want to focus on the strictly relevant and therefore make several assumptions for the late universe. By late universe, we mean times deep in matter-domination after decoupling characterized by some\footnote{A good value to choose is for instance $a_{MD}=0.03$ (see Fig.~\ref{fig:PlotPhisNoBaryons}).} $a_{MD}$ all the way to the present $a_{MD}<a<1$. 
 
 First of all, we are primarily interested in solutions on the sub-Hubble scales that as introduced corresponds to the large-$k$ limit $k\gg a H$, which is satisfied for all LSS observations \cite{dodelson2020modern}. Moreover, at late times, we will assume that baryons can indeed be described together with cold dark matter, hence assume a zero baryon pressure, which is an excellent approximation after decoupling. Together with the assumption of negligible neutrino masses, this implies that the matter perturbations can be described by a total perturbation $\delta\rho_m=\delta\rho_b+\delta\rho_{cdm}$. Furthermore, we assume that radiation is completely negligible. Together with the $\Lambda$CDM assumption of a CC that has no perturbations, this implies that cold matter is the dominant clustering component. In fact, our considerations will be valid even in a more general dark energy scenario, as long as matter remains the dominant clustering component. Note, however, that this does not imply that matter must be the dominant energy component, which today is of course not true with dark energy dominating the energy budget. In summary, we therefore make the following assumptions
 \begin{equation}\label{eq:LateTimeAssumptions}
 \text{Assumptions for } a>a_{MD}:\;
		\left\{\begin{array}{l}
		\displaystyle
			k\gg a H\,,\\
			\displaystyle
			\rho_m\tilde\delta_m(\mathbf{k},a)=\rho_b\tilde\delta_b(\mathbf{k},a)+\rho_{cdm}\tilde\delta_{cdm}(\mathbf{k},a)\,,\\
			\displaystyle
			\tilde\delta_m\text{ dominant clustering component.}
		\end{array}\right.
	\end{equation}

Finally, at late times the flat $\Lambda$CDM model with initially six free parameters given Eq.~\eqref{eq:LCDM parameters} is essentially governed by the two remaining parameters $h$ and $\omega_m$, with a background expansion governed by the Hubble function of the form
 \begin{align}\label{eq:HubbleParameterLCDMFourthLateTime}
		H^2_\text{\tiny $\Lambda$CDM} = C_H^2 \,\Big(\omega_m (1+z)^{3} + \omega_\Lambda\Big)\,,
\end{align}	
where the CC density parameter is determined through
\begin{equation}\label{eq:ConstraintOnDensityParametersThirdLateTimes}
		\omega_\Lambda = h^2-\omega_m\, .
\end{equation}

\paragraph{The Linear Matter Power Spectrum.} 
 Again, we further assume that any scalar part in the anisotropic stress tensor vanishes, such that the scalar metric perturbations are governed by a single variable $\phi(\mathbf{x},a)$ describing the gravitational potential above an FLRW background, whose conventions we define by requiring that under the late time assumptions in Eq.~\eqref{eq:LateTimeAssumptions} the potential satisfies the Poisson equation
 \begin{equation}
     \nabla^2\phi(\mathbf{x},a)=-\frac{\kappa_0}{2}\,a^2\,\delta\rho_m(\mathbf{x},a)\,,
 \end{equation}
 where we recall the definition $\kappa_0\equiv 8\pi G_0$ with $G_0$ the (bare) Newton constant. In Fourier space, this becomes
 \begin{equation}
     k^2\tilde\phi(\mathbf{k},a)=-\frac{\kappa_0}{2}\,\rho_m(a) a^2\,\tilde\delta_m(\mathbf{k},a)\,,
 \end{equation}
 that we can rewrite through $\rho_m=\Omega_m\rho_c/a^3$, with $\rho_c=\rho_0$ given by Eq.~\eqref{eq:CriticalDensityToday} as 
 \begin{equation}\label{eq:MatterOverdensitiesToPhi}
    \boxed{\tilde\delta_m(\mathbf{k},a)=\frac{2 k^2 a}{3\Omega_m H_0^2} \,\tilde\phi(\mathbf{k},a)}\,,\qquad a>a_{MD}\,.
 \end{equation}
 Thus, in order to understand the late time matter overdensities, we need to describe how the gravitational potential $\tilde\phi(\mathbf{k},a)$ evolved starting from its initial conditions set at some $a_\text{in}$ deep in the radiation domination era that we can effectively set to zero $a_\text{in}\rightarrow 0$. Conventionally, these initial conditions are given by the stochastic variable of curvature perturbations $\tilde{\mathcal{R}}(\mathbf{k})$ introduced in Sec.~\ref{sSec:Initial Conditions}. And as discussed, the deterministic variable that captures the subsequent evolution is simply given by 
 \begin{equation}
     \tilde\phi_\text{d}(k,a)=\tilde\phi(\mathbf{k},a)/\tilde{\mathcal{R}}(\mathbf{k})\,.
 \end{equation}

 It turns out that if we are interested in the late universe only, hence $a_{MD}<a<1$, the evolution of $ \tilde\phi_\text{d}$ can be separated into a $k$-dependent \textit{transfer function} $T(k)$ that effectively captures the evolution from $a_\text{in}$ to $a_{MD}$ and a time dependent \textit{growth factor} $D_+(a)$ describing the $k$ independent evolution after $a_{MD}$. The notation $D_+$ will become clear below. 
 
 This is possible, because at the linear level, each $k$ mode evolves individually, which justifies the use of an effective transfer function $T(k)$ from $a_\text{in}$ to $a_{MD}$. During matter domination, the potentials at all scales remain constant for a while, such that one can define a time $a_{MD}$ at which the evolution at all scales becomes the same. The evolution of different modes of the gravitational potential $\tilde\phi$ up until $a_{MD}$ is shown in Fig.~\ref{fig:PlotPhisNoBaryons}. These plots were obtained by numerically solving the Einstein-Boltzmann system of equations. 
 
 After $a_{MD}$, however, all relevant modes are well inside the horizon and the evolution remains scale-independent in the large $k$ limit, described through a single function $D_+$, which in general has to be determined numerically as well, but which admits an analytic $\Lambda$CDM solutions in the late time regime of Eq.~\eqref{eq:LateTimeAssumptions}. 

 Thus, effectively, the late time evolution between $a_{MD}$ and today of each $\tilde\phi_d(k)$ is simply governed by a growth factor $D_+(a)$ and its initial value at $a_{MD}$ is determined by the transfer function $T(k)$. The precise relation further depends on normalization conventions. Namely, since at $a_{MD}$, the gravitational potential outside the Hubble radius is related to the curvature potential as \cite{dodelson2020modern} 
 \begin{equation}
     \tilde\phi_\text{superH}(\mathbf{k},a_{MD})=\frac{3}{5}\tilde{\mathcal{R}}(\mathbf{k})\,,
 \end{equation}
 while the growth factor $D_+(a)$ is defined such that $D_+(a_{MD})=a_{MD}$, a convenient convention for the effective evolution of the gravitational potential reads
 \begin{equation}\label{eq:PhiRelTo T and D}
     \tilde\phi(\mathbf{k},a)=\frac{3}{5}\tilde{\mathcal{R}}(\mathbf{k})\,T(k)\,\frac{D_+(a)}{a}\,,\qquad a>a_{MD}\,.
 \end{equation}
The associated deterministic functions then become
  \begin{equation}
     \tilde\phi_\text{d}(k,a)=\frac{3}{5}\,T(k)\,\frac{D_+(a)}{a}\,,\qquad a>a_{MD}\,.
 \end{equation}
Note that the convention in Eq.~\eqref{eq:PhiRelTo T and D} sets $T(0)=1$, which simply states that for modes that only entered the Hubble radius after $a_{MD}$ there was no prior evolution of the deterministic variable.

 Plugging this expression into Eq.~\eqref{eq:MatterOverdensitiesToPhi} one obtains
 \begin{equation}\label{eq:MatterOverdensitiesTocurvaturePert}
    \boxed{\tilde\delta_m(\mathbf{k},a)=\frac{2 k^2}{5\omega_m C_H^2} \,\tilde{\mathcal{R}}(\mathbf{k})\,T(k)\,D_+(a)}\,,\qquad a>a_{MD}\,,
 \end{equation}
 such that the approximate form of the linear matter power spectrum in terms of the growth factor and the transfer function becomes (see also \cite{dodelson2020modern})
\begin{equation}\label{eq:LinearMatterPowerSpec}
			P_m(k,a) = \frac{8\pi^3}{25}\frac{k}{\omega_m^2C_H^4}\,T^2(k)\,D^2_+(a)\,\mathcal{P}_\mathcal{R}(k)\propto k^{n_s}\,T^2(k)\,D^2_+(a)\ ,
\end{equation}
where we recall the definition of parametrization of the primordial dimensionless power spectrum of curvature perturbations [Eq.~\eqref{eq:InitialCurvaturePowerSectrum Dimensionless}]
\begin{equation}
			\mathcal{P}_\mathcal{R}(k)= A_s\left(\frac{k}{k_p}\right)^{n_s-1},\qquad
				k_p=0.05\,\text{Mpc}^{-1}\ .
\end{equation}
Therefore, to obtain the matter power spectrum that can be related to observations from given initial conditions parametrized by $A_s$ and $n_s$, we effectively need to know the transfer function $T(k)$ and the growth factor $D_+(a)$. It will also be convenient to define the associated dimensionless matter powerspectrum 
\begin{equation}\label{eq:LinearMatterPowerSpecSecond}
			\boxed{\mathcal{P}_m(k,a)\equiv \frac{k^3}{2\pi^2}P_m(k,a) = \frac{4}{25}\frac{k^4}{\omega_m^2C_H^4}\,T^2(k)\,D^2_+(a)\,\mathcal{P}_\mathcal{R}(k)\,.}
\end{equation}

\begin{figure}[H]
\centering
\includegraphics[scale=0.5]{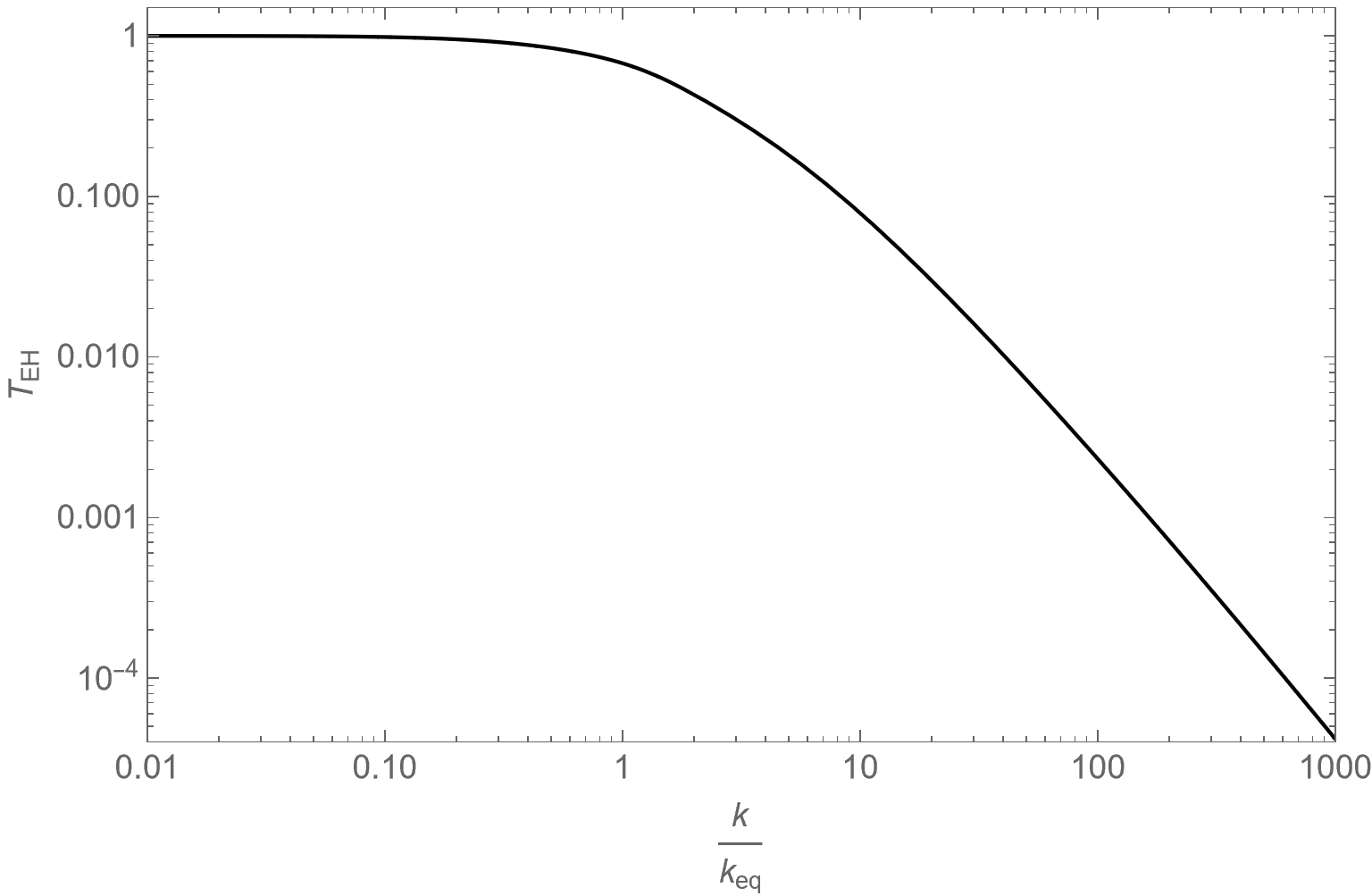}
\caption{\small The Eisenstein-Hu transfer function $T_\text{EH}(k)$ \cite{Eisenstein:1997ik} measured in dimensionless units of $k/k_\text{eq}$ for the parameters in Table~\ref{table:PlankBestFit} and with $k_\text{eq}=0.073\, \omega_m\,\text{Mpc}^{-1}$.}
\label{fig:EHT}
\end{figure}

\paragraph{The Transfer function.} 
In principle, $T(k)$ has to be evaluated numerically by integrating the evolution equation for each $k$ between $a_\text{in}$ and $a_{MD}$. However, the result can compactly be summarized in an analytic fitting formula that we will explicitly introduce below. But first, we want to understand the qualitative form of it. As already mentioned, for the largest scales which enter the horizon during or after MD, the transfer function is unity, hence $T(0)=1$. During RD, on the other hand, structure grows more slowly, in other words, the potentials decay and the transfer function for modes entering the Hubble radius in radiation domination should be suppressed more and more as modes enter earlier before radiation/matter equality (see Fig.~\ref{fig:EHT}). Translated to the linear matter power spectrum in Eq.~\eqref{eq:LinearMatterPowerSpec}, we therefore have $P_m\propto k^{n_s}$, with $n_s\simeq 0.97$ on large scales, while it should be a decreasing function of $k$ on small scales with a turnover at a scale $k_\text{eq}$ characterizing modes that enter the Hubble radius at radiation/matter equality. Interestingly, measuring the turnover point of the matter power spectrum therefore allows to constrain the amount of matter in the universe. 

One thing we did not yet take into account is that before decoupling, dark matter and baryons can absolutely not be described together, as the baryons are tightly coupled to photons. As already mentioned, within this primordial plasma the baryon overdensities do not grow inside the Hubble radius, since also the photon perturbations do not grow due to pressure. This means that only a fraction $\omega_{cdm}/\omega_m$ of the total matter contributed to the collapse of perturbations before decoupling, that still grew in RD, although slower compared to matter domination. The amount of baryons in the universe therefore leads to a further suppression of small scales that were already sub-Hubble before recombination.

Lastly, the pressure present in the primordial plasma results in oscillation in the transfer function (not present in Fig.~\ref{fig:EHT}) that are associated to sound waves driven by the gravitational potential perturbations. These oscillations in turn translate into oscillations of the matter power spectrum that are known as \textit{baryon acoustic oscillations} (BAO). Interestingly, these features, depending on the known scale of the sound horizon at decoupling, can be detected in the clustering of galaxies and can serve as a ``standard ruler'' of the late time universe.

As mentioned, for practical purposes, it is very useful to have an analytic fitting formula for the $\Lambda$CDM transfer function. We will present here the Eisenstein-Hu fitting formula \cite{Eisenstein:1997ik} that takes into account the baryonic suppression at small scales that proves important for an accurate computation of the clustering amplitude. This will in particular be relevant for the application in Chapter~\ref{Sec:CsomoTensions}. Mainly adopting the notation of the original work \cite{Eisenstein:1997ik} the transfer function is given by
\begin{equation}\label{eq:EisensteinHuFittingFormula}
    T_\text{EH}(k)\equiv T(q(k)))=\frac{\ln(2\text{e} + 1.8\, q)}{\ln(2\text{e} + 1.8 \,q)+\left[14.2 + \frac{731}{1+62.5\,q}\right]q^2}\,,
\end{equation}
where 
\begin{equation}
    q(k) = \frac{k}{\text{Mpc}^{-1}}\frac{T_{2.7}^2}{\Gamma_\text{eff}(k)}\,,
\end{equation}
with $T_{2.7}=1.00944$ the temperature of the CMB in $2.7\ \text{K}$ units \cite{Fixsen:2009ug} and where
\begin{align}
    \Gamma_\text{eff}(k) &= \omega_m\left(\alpha_\Gamma + \frac{1-\alpha_\Gamma}{1+(0.43ks)^4}\right)\,,\\
    \alpha_\Gamma &= 1-0.328\ln(431\omega_m)\frac{\omega_b}{\omega_m}
					+ 0.38\ln(22.3\omega_m)\left(\frac{\omega_b}{\omega_m}\right)^2\,,\\
     s &= \frac{44.5 \ln(9.83/\omega_m)}{\sqrt{1+10(\omega_b)^{3/4}}}\,\text{Mpc}\,.
\end{align}
The fitting formula is plotted in Fig.~\ref{fig:EHT} for the values of the $\Lambda$CDM parameters in Table~\ref{table:PlankBestFit} and with $k_\text{eq}=0.073\, \omega_m\,\text{Mpc}^{-1}$. Note that in the definition of the transfer function we measure $k$ in $\text{Mpc}^{-1}$ units, instead of the frequently employed $\text{Mpc}^{-1}\,h$, a detail which will also become important in the subsequent Chapter~\ref{Sec:CsomoTensions}.

 \paragraph{The Growth Factor.}
 As discussed, after decoupling at late times, the time evolution of sub-Hubble matter perturbations is entirely governed by the growth factor. In order to determine this function, one has to solve the evolution equation of matter oversensitive for $a>a_{MD}$, which in $\Lambda$CDM with the late-time assumptions in Eq.~\ref{eq:LateTimeAssumptions} is given by (see e.g. \cite{dodelson2020modern})
 \begin{equation}
     \frac{d^2\delta_m(a)}{da^2}+\frac{d\ln(a^3 H(a))}{da}\,\frac{d\delta_m(a)}{da}-\frac{3\Omega_m H^2_0}{2a^5 H^2(a)}\,\delta_m(a)=0\,.
 \end{equation}
 Through Eq.~\eqref{eq:MatterOverdensitiesTocurvaturePert}, this linear growth equation translates into a differential equation for $D(a)$ of the form
 \begin{equation}\label{eq:LinearGrowthEq}
     \boxed{D''(a)+\left(\frac{3}{a}+\frac{H'(a)}{H(a)}\right)\,D'(a)-F(a)\,D(a)=0\,,}
 \end{equation}
 where we defined
 \begin{equation}\label{eq:FFunctionGrowthFactor}
     F(a)\equiv \frac{3\Omega_m H_0^2}{2a^5 H^2(a)}=\frac{3\omega_m C_H^2}{2a^5 H^2(a)}\,.
 \end{equation}

 It is important to note here, that the linear growth equation [Eq.~\eqref{eq:LinearGrowthEq}] is valid for general expansion histories $H(a)$ that can be different from $\Lambda$CDM, in particular for a more general equation of state of dark energy, as long as the late-time assumptions described above are satisfied. Generically, the equation has to be solved numerically. But if we assume a solution, let's call it $D_2(a)$, one can construct another solution of the form
 \begin{equation}\label{eq:SecondSol}
     D_1(a)=D_2(a)\int_{0}^adx_a\frac{H_0}{x_a^3 \,D_2^2(x_a)\,H(x_a)}\,.
 \end{equation}
 
For a $\Lambda$CDM universe, where matter and a cosmological constant $\Lambda$ with $w=-1$ are the	dominant late-time energy components\footnote{In fact, this is also true in the presence of a curvature component.} it turns out that
\begin{equation}
    D_2(a)=\frac{H(a)}{H_0}\,,
\end{equation}
is a solution. Note, however, that this is a decaying solution $D_2(a)=D_-(a) \propto H(a)$. A second solution $D_1(a)$ can then be obtained through Eq.~\eqref{eq:SecondSol}.
Together with the initial condition $D(a_{MD})=a_{MD}$ at which $H(a_{MD})=H_0\,\Omega_m^{1/2}\,a_{MD}^{-3/2}$, this gives an analytic expression for the growing solution $D_1=D_+$
\begin{equation}\label{eq:analytic_Da}
			\boxed{D_+(a) =\frac{5\Omega_m}{2}\frac{H(a)}{H_0}I(a)\,,}
\end{equation}
where, for later use, we have defined here the dimensionless integral
\begin{equation}
    I(a)\equiv \int^a_0 d x_a\frac{H_0^3}{\big(x_aH(x_a)\big)^3}\,.
\end{equation}
As the name suggests, the growth factor $D_+$ in Eq.~\eqref{eq:analytic_Da} will at late-times dominate over the decaying solution $D_-$.

It will also be useful to define the linear \textit{growth rate}
\begin{equation}\label{eq:LinearGrowthRate}
			\boxed{f\equiv \frac{d\ln{D_+}}{d\ln a}=a\frac{D_+'(a)}{D_+(a)}\,.}
\end{equation}
In $\Lambda$CDM the growth rate is well approximated by the expression
		\begin{equation}\label{eq:growthrateEstimate}
			f(a)\simeq\left(\frac{\Omega_m H_0^2a^{-3}}{H^2}\right)^{0.55}= \left(\frac{\omega_m C_H^2a^{-3}}{H^2}\right)^{0.55}.
		\end{equation}

At this point, a word on the validity of the linear perturbation approximation used above is in order. While as discussed, the matter powerspectrum $P_m$ has a turnover at the scale $k_\text{eq}$, the associated dimensionless powerspectrum $\mathcal{P}_m\sim k^3 P_m$ defined in Eq.~\eqref{eq:LinearMatterPowerSpecSecond} remains an increasing function of $k$ although the increase is damped for $k>k_\text{eq}$. It is $\mathcal{P}_m$ representing the dimensionless variance of the Fourier modes of each $k$ that can be used as an indicator for whether our linear approximation is still valid. Whenever $\mathcal{P}_m\gtrsim 1$ indicates non-linear perturbations,
which for $\Lambda$CDM at $a=1$ is given by the scale $k_{NL}\sim 0.25 \,h \text{ Mpc}^{-1}$ \cite{maggiore2018gravitationalV2,dodelson2020modern}. But on larger scales, the linearity assumptions still hold with $\mathcal{P}_m\ll 1$. Knowing that after $a_{MD}$ the powerspectra did not change their shape due to the $k$ independent evolution through the growth factor, but simply gained in amplitude on all scales, this implies that at earlier times $a_{MD}<a<1$ the non-linearity scale $k_{NL}$ is pushed to higher and higher scales. In other words, at earlier times, only the perturbations on the smallest scales behaved non-linearly, while $k_{NL}$ decreases over time.  

\paragraph{Observing the Large Scale Structure.} 

For observations in position space, it is useful to characterize the amplitude of matter fluctuations on a particular scale through $\sigma_R$, the variance of smoothed-out matter overdensities in spheres of comoving radius $R$ introduced in Eq.~\eqref{eq:DefSigmaRFirst}. Being interested on the large scale structure, of which we have a good analytic understanding, the scale $R$ can be chosen such that it still corresponds to the linear modes. However, we should also not choose $R$ too large, where structure fades out to the homogeneous background. It is therefore custom to choose
\begin{equation}
    R=8\ h^{-1}\text{Mpc}\,,
\end{equation} 
for which the associated clustering amplitude $\sigma_8\lesssim 1$ remains below the non-linearity scale. On smaller scales with $\sigma_8\gtrsim 1$ the linear approximation breaks down, since in that case, any given volume of radius R is likely to have a density that is very different from the cosmic mean.

However, when trying to measure $\sigma_8$ through the direct observation of the matter content in the universe, immediately two major problems arise:
\begin{enumerate}[(1)]
    \item The clustering amplitude $\sigma_8$ characterized the distribution of the total matter that is dominated by dark matter. Thus, by observing the visible matter for instance by mapping all galaxies in function of redshift up to a given distance, we actually only observe a small fraction of the clustering matter in the universe. This problem is known as the fundamental bias.
    \item Moreover, there is also a fundamental uncertainty in the redshift information due to the unknown peculiar velocities of the galaxies, called redshift space distortions (RSD).
\end{enumerate}
There are two main approaches of addressing these difficulties and conducting LSS survey measurements of matter clustering that are currently done by several collaborations \cite{DES:2017myr, DES:2021wwk, KiDS:2020suj, Heymans:2020gsg,DES:2022vuu}.

The first is to anyway observe the galaxy distributions, but to use the galaxy maps as tracers of the underlying dark matter distribution. In doing so, the redshift space distortions, hence the second problem listed above, can actually be turned into information on dark matter \cite{Jackson:1971sky,Kaiser:1987qv,Hamilton:1997zq,Percival:2008sh,Yoo:2008tj,dodelson2020modern}. This is because on large scales, the peculiar velocities of galaxies are not random, but are correlated with the overdense regions exerting a strong gravitational pull. And by the equivalence principle, dark matter is affected by the gravitational potentials in exactly the same way. The continuity equation relates the velocities of matter to the change in matter densities that, because at late times the time evolution is exclusively governed by the growth factor $D_+(a)$, is proportional to the growth rate $f$ introduced in Eq.~\eqref{eq:LinearGrowthRate}
\begin{equation}
    \delta'_m(a)=\frac{D'_+(a)}{D_+(a)}\delta_m(a)=\frac{f}{a}\delta_m(a)\,.
\end{equation}
This is the heuristic reason why LSS surveys based on RSD measurements are fundamentally sensitive to the combination $f\sigma_8$.

The second option is to use lensed light as a measurement of the total matter clusters. In particular, weak gravitational lensing induces distortions in the shapes of distant galaxies. While of course the initial shape of a galaxy is not known, it is possible to learn about the large scale matter structure by monitoring correlations between the galaxy shape distortions known as \textit{cosmic shear} measurements \cite{Refregier:2003ct,Bartelmann:2010fz,Kilbinger:2014cea, Mandelbaum:2017jpr,Yoo:2008tj,dodelson2020modern}. Such weak lensing surveys mainly constrain the particular combination
\begin{equation}\label{eq:S8}
    S_8\equiv \sigma_8\,\sqrt{\frac{\Omega_m}{0.3}}=\frac{\sigma_8}{h}\sqrt{\frac{\omega_m}{0.3}}\,.
\end{equation}
Observe that through the $\Lambda$CDM estimate of the growth rate in Eq.~\eqref{eq:growthrateEstimate} of $f_0\sim(\Omega_m)^{0.55}$, $S_8$ is very closely related to the combination $f\sigma_8$ targeted by galaxy surveys.

\subsection{Dark Energy and the Cosmological Constant.}\label{sSec:CCDarkEnergy}

Finally, early time measurements from the CMB, together with BAO constraints as well as direct local observations, all indicate that the current expansion of the universe is accelerating $\ddot a>0$ (see e.g. \cite{Weinberg2008Cosmology,dodelson2020modern}).
The most direct evidence for an accelerated late time expansion comes from Type Ia supernovae measurements \cite{SupernovaSearchTeam:1998fmf,Perlmutter:1999jt} that can be used as standard candles to measure the luminosity distance $d_L$ defined in Eq.~\eqref{eq:LuminosityDistanceSecond} that depends on the Hubble flow $H(z)$. An accelerated expansion implies that at earlier times, hence larger distances, the change in scale factor was smaller in the past $\dot a(t)< \dot a(t_0)$ for some $t<t_0$, instead of the expected $\dot a(t)> \dot a(t_0)$ in a pure matter and radiation dominated universe. A direct measurement of an accelerated expansion therefore requires the measurement of a change in scale factor over time.  At fixed $H_0$, hence fixed $\dot a(t_0)$, this implies a longer lifespan of the universe. Indeed, in a universe fully dominated by matter $\Omega_m\sim 1$, the expected expansion age would be less than the estimated age of the oldest observed stars. Moreover, a dark energy dominated phase at late times influences the growth of structure in that the gravitational potential starts decaying again, slowing down the growth of matter perturbations as captured by the growth factor $D_+(a)$ introduced in Sec.~\ref{sSec:LargeScaleStructure}. Through this effect, probes of the local large scale matter structure also provide evidence for a cosmological model with $\Omega\simeq 0.7$ also at the level of the perturbations (see e.g. \cite{dodelson2020modern}).

While observationally, an accelerated expansion therefore resides on solid grounds, the theoretical description thereof remains a mystery. As discussed in Sec.~\ref{sSec:HomIsoUniverse} an accelerated expansion requires a cosmic fluid with equation of state $w<-1/3$ called dark energy, that dominates the current energy budget of the universe. Within the $\Lambda$CDM model, dark energy and the associated accelerated expansion is described through the simplest and most natural option in GR, namely a cosmological constant. As already mentioned in Sec.~\ref{sSec:HomIsoUniverse} introducing a positive cosmological constant term $\Lambda>0$ in the gravitational Lagrangian as in Eq.~\eqref{eq:EinsteinHilbertAction} can be viewed as introducing a dark energy matter component with energy momentum tensor of the form [Eq.~\eqref{eq:EMTforCC}]
\begin{equation}
    T^{\mu\nu}_{\Lambda}=-\frac{\Lambda}{\kappa_0} g^{\mu\nu}\,.
\end{equation}
This energy-momentum tensor is covariantly conserved in any metric theory of gravity due to the metricity of the Levi-Civita connection and therefore, the fluid with constant positive energy density $\rho>0$ satisfies the energy conservation equation [Eq.~\eqref{eq:ConservationOfMatterCosmology}]
\begin{equation}\label{eq:ConservationOfMatterCosmologySecondDE}
    \dot\rho+3H(\rho+p)=0\,,
\end{equation}
with 
\begin{equation}
    \rho=-T_{\Lambda\,0}^0=\text{const.}\,,\quad \rho=-p \quad\Leftrightarrow \quad w = -1\,.
\end{equation}
Thus, since $w<-1/3$, eventually, as the CC component becomes dominant, the universe will transit into a state of accelerated expansion and according to the Friedmann equation [Eq.~\eqref{eq:FriedmannEquationGen}] the Hubble parameter asymptotes towards a constant value of
\begin{equation}
    H^2=\frac{\Lambda}{3}\,.
\end{equation}

This concludes the introduction of the $\Lambda$CDM cosmological standard model. In the next section, we will explore the possibility that current tensions in the cosmological observations might require a departure from this model and ask whether two of the most promising tensions can be solved consistently. This investigation will be heavily based on the concepts introduced above.

\newpage
\thispagestyle{plain} 
\mbox{}

\chapter{Cosmological Tensions Guiding the Path Beyond $\Lambda$CDM}\label{Sec:CsomoTensions}

A ``tension'' is a term that is commonly used to describe an apparent discrepancy between two distinct measurements or inferences of a physical observable. In a sense, the theories of physics and in particular cosmology, are decisively guided through the advent of such tensions. While many apparent tensions eventually turned out to be due to simple statistical fluctuations, errors in the analysis or poorly modelled systematics, some resolutions required a fundamental revision of the underlying theoretical description. Indeed, also many cosmological anomalies have proven in the past to hold the power to guide the path towards novel physical understanding, such that their study may result in much more interesting consequences than the precise value of the observational parameter they arise from.

Interestingly, the current data interpreted through the cosmological standard model described in Chapter~\ref{Sec:CosmoNutshell} still exhibits numerous such tensions (see \cite{Zhao:2017cud,Riess:2019qba,Knox:2019rjx,DiValentino:2020vvd,DiValentino:2020zio,DiValentino:2021izs,Perivolaropoulos:2021jda,Abdalla:2022yfr,Peebles:2022akh,Hu:2023jqc} and references therein) which therefore provide an opportunity for new discovery. In the following, we will focus on two of today's most significant such discrepancies between different measurements of cosmological observables, and propose a largely model agnostic approach to draw first conclusions on the guiding principles a consistent resolution of both these tensions might provide.

\section{The $H_0$ and $\sigma_8$ Tensions}\label{sSec:TheH0andSigma8Tensions}

The two arguably most prominent cosmological tensions (see \cite{Riess:2019qba, Knox:2019rjx, DiValentino:2020vvd, DiValentino:2020zio,DiValentino:2021izs, Perivolaropoulos:2021jda,Abdalla:2022yfr,Hu:2023jqc} and references therein) are discrepancies in the values of the Hubble 
constant $H_0$ given in Eq.~\eqref{eq:HubbleConstant} and the clustering amplitude $\sigma_8$ defined in Eq.~\eqref{eq:DefSigmaRFirst} as the variance of smoothed out matter overdensities, with a smoothing scale of $R=8\,h$ Mpc$^{-1}$. Note that these two observables do not play a mere supporting role in the history of the universe, but are central objects in cosmology, determining the current expansion rate and the large scale distribution of matter. Any viable cosmological model should be able to provide coherent values of these observables across different types of measurements. Therefore, if the tensions are real and not due to some systematic measurement errors, the $H_0$ and $\sigma_8$ tensions of $\Lambda$CDM would provide clear evidence for a missing piece in the current standard cosmological model, necessitating a departure in one way or another. Excitingly, the tensions could also indicate a first observation of beyond GR physics.

\subsection{Observational Evidence}

Both tensions arise primarily between the Planck values \cite{Planck:2018vyg} inferred from the CMB in comparison to local direct measurements of the expansion rate \cite{Riess:2019cxk,Riess:2020fzl,Riess:2021jrx,Pesce:2020xfe,Wong:2019kwg} and large scale structure 
(LSS) surveys \cite{DES:2017myr, DES:2021wwk, KiDS:2020suj, Heymans:2020gsg} respectively. Concretely, if the theoretical angular powerspectrum, together with information in the polarization of the CMB, computed within the $\Lambda$CDM model with six free parameters in Eq.~\eqref{eq:LCDM parameters} is fitted against the observed angular powerspectrum as explained in Sec.~\ref{sSec:CMB}, the inferred value of the current expansion rate is about $h\sim 0.67$. This value lies however between\footnote{Here $\sigma$ denotes the standard deviation of the distribution given by the difference between estimates of two independent measurements in units of uncertainty typically taken to be the posterior errors of the two experiments. A certain number times $\sigma$ then quantifies the probability of rejecting the hypothesis that the two measurements are still the same, assuming a Gaussian distribution. See \cite{Abdalla:2022yfr} for subtleties regarding the interpretation of such a quantification of tensions.} $4.5\sigma$ to $6.3\sigma$ below direct measurements that lie around $h\sim 0.74$ depending on different measurement combinations \cite{Riess:2019qba,Abdalla:2022yfr}. Hence, schematically,
\begin{equation}
    \text{Hubble tension:}\qquad h^\text{Planck} \;<\; h^\text{local}\,.
\end{equation}
The most significant direct and largely model independent measurements of $H_0$ are based on the same type of supernova observations used as standard candles to infer the accelerated expansion of the universe as described in Sec.~\ref{sSec:CCDarkEnergy}. In contrast to acceleration measurements, however, the determination of $H_0$ requires absolute distance measurements which need a so-called \textit{distance ladder} calibration that starts off by nearby geometric distance measurements based on the parallax given by the motion of the earth. The most prominent intermediate piece within the distance ladder towards the supernovae is based on pulsating stars called Cepheid variables that exhibit a strong relationship between their luminosity and the pulsation period used by the SH0ES collaboration \cite{Riess:2019cxk,Riess:2020fzl,Riess:2021jrx}. An overview of different measurement methods as well as the associated systematics can be found in \cite{Abdalla:2022yfr}. Here, we do not want to go into the details of the observations but simply want to assume the reality of the tension that now persisted over many years supported by multiple independent measurements.

On the other hand, computing $\sigma_8$ within $\Lambda$CDM model given by the Plank best fit values results in $\sigma_8\sim 0.81$ which lies $2-3\sigma$ above the locally measured value \cite{Abdalla:2022yfr}, primarily consisting of weak lensing measurements and galaxy cluster counts as described in Sec.~\ref{sSec:LargeScaleStructure}
\begin{equation}
    \sigma_8\text{ tension:}\qquad \sigma_8^\text{Planck} \;>\; \sigma_8^\text{local}\,.
\end{equation}
Such late time measurements of the clustering amplitude are computed assuming the standard $\Lambda$CDM model. However, note that compared to the CMB observations, LSS probes only depend on the low redshift cosmology. Again, we refer to \cite{Abdalla:2022yfr} for a complete account of measurement methods and systematics. Often, the tension is also quantified in terms of $S_8$ defined in Eq.~\eqref{eq:S8}. While the $\sigma_8$ tension is still of less significance compared to the $H_0$ tension and the associated debate on its reality is still roaring, we will take the multiple independent local measurements pointing towards a discrepancy in comparison to the Planck data as enough evidence to require that any proposed alternative cosmological model should at least not worsen the present $\sigma_8$ tension.

\subsection{Solving the $H_0$ and $\sigma_8$ Tension.} 

As discussed, direct measurements of $H_0$ that only lightly depend on the very late time expansion history can to a large extent be considered to be model independent. The opposite is the case for the inference through the CMB powerspectrum, that fit the data with the free parameters of a given model. Assuming that neither the early time nor late time observations are flawed in an unexpected way, a solution to the $H_0$ tension therefore likely demands a modification of the cosmological model that shifts the preferred value of $H_0$ to the larger value measured at late times when fitting the new model to the CMB powerspectrum for instance. The question is of course how a model needs to be modified in order to sill fit the CMB data while exhibiting a positive shift in the value of $H_0$.

\begin{figure}[H]
\centering
\includegraphics[scale=0.55]{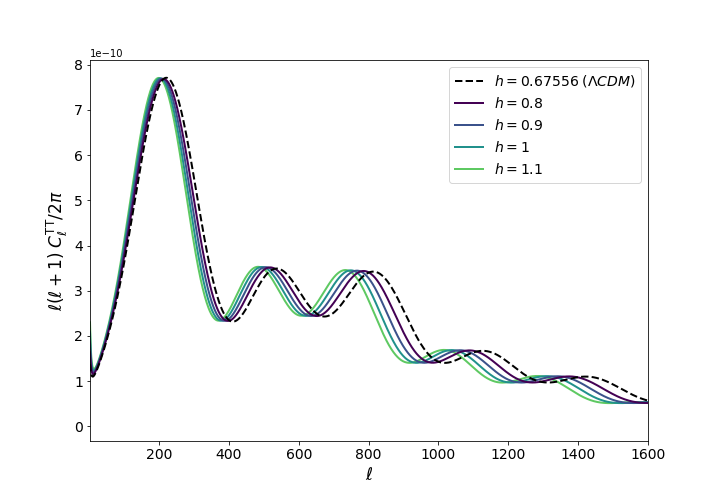}
\caption{\small The theoretical powerspectrum of anisotropies $C_l^{TT}$ in the CMB as a function of the multipole moment $l$ of spherical harmonics computed for the $\Lambda$CDM model with parameters given in Table~\ref{table:PlankBestFit} but for different values of $h$ at fixed $\omega_m$. The angular powerspectrum was computed with the Cosmic Linear Anisotropy Solving System ({\fontfamily{qcr}\selectfont class}) code \cite{lesgourgues2011cosmic,Blas:2011rf}.}
\label{fig:Cl_h}
\end{figure}

\paragraph{Early- and Late-Time Solutions to the Hubble Tension.}
It is instructive to consider what happens to the CMB powerspectrum when simply shifting the value of $h$ while keeping all other parameters fixed as depicted in Fig.~\ref{fig:Cl_h}. If we would naively increase the value of $h$ to become compatible with the direct measurements of $H_0$, evidently the main effect on the CMB is a shift of the acoustic peaks to the left. Therefore, what sets the value of $h$ when fitting the $\Lambda$CDM model to the CMB data is mainly the position of the peaks. As discussed in Sec.~\ref{sSec:CMB}, the position of the peaks is captured by the distance prior given by the angular scale [Eq.~\eqref{eq:AcousticScaleDef}]
\begin{equation}\label{eq:AngularScaleCosmicTenstions}
    \theta^*_A =\frac{r_s(z_*)}{d_C(z_*)}
\end{equation}
where $z_*$ is the redshift at decoupling. Indeed, this angular scale has the geometric interpretation of determining the location of the first peak by providing the angular scale set by the maximum distance the primordial plasma sound waves could have travelled as determined by the comoving sound horizon [Eq.~\eqref{eq:SoundHorizonSecond z}]
compared to the comoving distance to us [Eq.~\eqref{eq:ComovingDistanceThird}]
at decoupling. And since the multipole moment $l$ is inversely proportional to the angular separation [Eq.~\eqref{eq:RelationAngularScaleto l}], a shift of the peak to the left observed in Fig.~\ref{fig:Cl_h} implies that a naive change of the $\Lambda$CDM parameter $h$ induces an increase in the angular scale $\theta^*_A$.

Thinking this chain of reasoning in reverse, any alternative cosmological model that reduces the angular scale $\theta^*_A$ observed in the CMB has the chance of solving the $H_0$ tension. This is because when determining the preferred cosmological parameters of such a model in a fit to the CMB, a decrease in angular scale will very likely be compensated by an increase in the preferred value of the Hubble constant parameter $h$. Hence, very crudely, solutions to the Hubble tension can be divided into two classes: \textit{early-time solutions} that decrease the sound horizon in Eq.~\eqref{eq:AngularScaleCosmicTenstions} mainly affecting physics before decoupling, and \textit{late-time solutions} that lead to an increase in the cosmic distance to the CMB that depends on the expansion history after decoupling.

Over the last six years, the community put great efforts towards developing models beyond $\Lambda$CDM that alleviate the Hubble tension, as for instance in
\cite{Renk:2017rzu,Bolejko:2017fos,deFelice:2017paw,Poulin:2018cxd,Amendola:1999er,DiValentino:2019ffd,Smith:2019ihp,Alcaniz:2019kah,Frusciante:2019puu,DeFelice:2020sdq,Heisenberg:2020xak,Zumalacarregui:2020cjh,Gomez-Valent:2020mqn, Ballesteros:2020sik, Jimenez:2020bgw, DiValentino:2020naf,Banerjee:2020xcn, Braglia:2020auw, Braglia:2020iik,Krishnan:2021dyb,Jedamzik:2020krr, Teng:2021cvy, Ballardini:2021evv} (see also \cite{Riess:2019qba, Knox:2019rjx, DiValentino:2020vvd, DiValentino:2020zio,DiValentino:2021izs, Perivolaropoulos:2021jda,Abdalla:2022yfr}). The proposals include a wide range of other ideas such as the introduction of primordial magnetic fields modifying the recombination history \cite{Jedamzik:2020krr}, employing departures from isotropy or homogeneity \cite{DiValentino:2021izs}, or considering spacial curvature \cite{Bolejko:2017fos}, all the way to considering new interactions in the dark sector \cite{Amendola:1999er,DiValentino:2019ffd}. The most popular scenarios are however dynamical dark energy models, that modify the background expansion history either by replacing the late-time $\Lambda$ driven accelerated expansion or by introducing a short period of accelerated expansion before decoupling. We want to note that despite the multitude of different ideas, due to the large dependence of both the sound horizon in Eq.~\eqref{eq:SoundHorizonTensions} and the comoving distance in Eq.~\eqref{eq:ComovingDistanceTensions} on the background expansion $H(z)$, most models, but in particular late-time solutions, mainly rely on modifications of the expansion history $H(z)$ in order to reconcile late and early time measurements.

\paragraph{Adding the $\sigma_8$ Tension.} However, a general observation is that almost all proposals for solving the Hubble tension in fact worsen the $\sigma_8$ tension. In other words, within the modified cosmological models together with the newly inferred cosmological parameters, the computed value of $\sigma_8$ lies even higher compared to the local measurements \cite{Renk:2017rzu,Frusciante:2019puu,deFelice:2017paw,DeFelice:2020sdq,Heisenberg:2020xak}. While the $\sigma_8$ tension is not as significant yet as the $H_0$ tension, one can argue that any viable solution to the Hubble tension should at least not worsen the $\sigma_8$ discrepancy. Viewed individually,  alternative cosmological models are in fact able to alleviate the $\sigma_8$ tension, where a selection of solutions are found in \cite{Lambiase:2018ows, 
Keeley:2019esp, DiValentino:2019ffd, Jedamzik:2020zmd, Clark:2021hlo, SolaPeracaula:2021gxi, Alestas:2021xes, Nunes:2021ipq, Schoneberg:2021qvd, Alestas:2021luu, Ye:2021iwa}. The crux seems therefore to lie in the simultaneous resolution of both tensions.

This observation lies at the heart of the motivation for the subsequent work, in which we want to understand if and how it is possible to consistently solve both the $H_0$ and the $\sigma_8$ tension in alternative cosmologies. A major goal in answering this question is to remain as model independent as possible in order to account for the plethora of possible $\Lambda$CDM departures, none of which stands out as a clear favorite theory yet. This led to develop a method to study small deviations from a given cosmological model that we will present in the next section. This method will then be applied to the particular case of the $H_0$ and $\sigma_8$ tensions above.

\section{A Model Agnostic Method to Study Alternative Cosmologies}\label{sSec:ModelIndepMethod}

In this section, we present a largely model independent approach to formulate constraints on small departures of cosmological models. Since this method will in the following mainly be applied to late time departures from $\Lambda$CDM, the presentation will for brevity focus on that specific context. However, we want to stress that our method is in principle applicable in a much broader context and is in particular a priori not tied to late time considerations or the $\Lambda$CDM model as baseline cosmology.

\paragraph{Late-Time assumptions.} The starting point of the method is a standard model that depends on a finite set of parameters, of which we would like to study small corrections and understand their implications on the parameter space. Let's therefore for concreteness consider a $\Lambda$CDM cosmology, which at the background level can effectively be described by Eq.~\eqref{eq:HubbleParameterLCDMThird}. At late times, the radiation is negligible. Moreover, we will generally consider the $\Lambda$CDM parameters $\tau_\text{rei}$ and $A_\text{s}$, i.e. the optical depth to reionization and the amplitude of the spectrum of scalar perturbations as fixed.\footnote{In general, modifications of cosmological perturbations, for instance within clustering dark energy models, may lead to a modification of the integrated Sachs-Wolfe (ISW) effect \cite{Sachs:1967er} that might affect the determination of $\tau_\text{reio}$ and $A_\text{s}$. However, we will leave the study of such effects for future work.} Thus, at late times, the $\Lambda$CDM cosmological model is essentially governed by two free and dimensionless parameters, the Hubble constant $h$ and the matter abundance 
$\omega_m$ through Eq.~\eqref{eq:HubbleParameterLCDMFourthLateTime}
\begin{equation}
	H_\text{\tiny $\Lambda$CDM}^2(h,\omega_m)= C_H^2 \left(\omega_m (1+z)^{3}+ \omega_\Lambda\right)\,,
\end{equation}
where
\begin{equation}
	\omega_\Lambda=h^2-\omega_m\,.
\end{equation}
Recall that the factor $C_H\equiv 100 \ \text{km}\,\text{s}^{-1}\,\text{Mpc}^{-1}$ arises from the definition of the dimensionless parameter $h$ in $H_0\equiv 100 \ h \ \text{km}\,\text{s}^{-1}\,\text{Mpc}^{-1}$.

\paragraph{Departure from the Base Cosmology.}
Alternative cosmological models can then be characterized by small variations of the expansion history at fixed values of all the cosmological parameters, that we will denote by $\delta H(z)$, together with variations in other observables, such as for instance the gravitational constant $G_\text{eff}=G_0+\delta G(z)$. However, for simplicity, the exposition of the method will first focus on $\delta H(z)$, and only address the general case towards the end of this section. Thus, small changes in the cosmological model will result in a deformation of the background evolution
\begin{equation}
	H(h, \omega_m) = H_\text{\tiny $\Lambda$CDM}(h,\omega_m) + \delta H(z)\,.
\end{equation}
At this stage, $\delta H(z)$ is a completely arbitrary function (see Fig.~\ref{fig:PlotDers} for a concrete example). A restriction to late-time modifications, then merely imposes the constraint of the form $\delta H(z) = 0$ for e.g. $z>300$, while we will also generally require $\delta H(0) = 0$ for basic observational consistency

\paragraph{Variation of the Cosmological Parameters.}
A generic deviation from $\Lambda$CDM at fixed parameters will however also modify the observationally preferred values of the $\Lambda$CDM parameters, in this case $H_0$ and $\omega_m$. This generic fact can be captured through the notion of a total variation $\Delta H$ that includes a variation of the model parameters. Working at first order in deviations, the true, observationally preferred Hubble parameter in the alternative cosmology therefore takes the general form
\begin{equation}
	H(H_0+\delta H_0, \omega_m+\delta\omega_m) = H_\text{\tiny $\Lambda$CDM}(H_0,\omega_m) + \Delta H\ .
\end{equation}
where the dimensionless total variation in the Hubble parameter reads
\begin{equation}\label{eq:DeltaH}
	\boxed{\frac{\Delta H(z)}{H(z)} = \frac{H_0^2}{H^2(z)}\frac{\delta h}{h} + m(z)\,\frac{\delta \omega_m}{\omega_m}+  \frac{\delta H(z)}{H(z)}\ ,}
\end{equation}
with 
\begin{equation}
    m(z)\equiv \frac{\omega_m C_H^2}{2H^2}\left((1+z)^{3}-1\right)\,,
\end{equation}
and where we denote $H_\text{\tiny $\Lambda$CDM}$ simply by $H$, since we are working to first order. This is the true variation of the expansion history induced by a small deformation from the cosmological model that takes into account an altered determination of the cosmological parameters themselves.

At the background level, the total variation of the Hubble function in Eq.~\eqref{eq:DeltaH} can then be propagated to any other cosmological observable $\mathcal{O}(z)$. Indeed, the total variation of any such observable can be written as
\begin{equation}\label{eq:VarGen}
	\boxed{\frac{\Delta\mathcal{O}(z)}{\mathcal{O}(z)} = I_\mathcal{O}(z)\frac{\delta h}{h}+ J_\mathcal{O}(z)\frac{\delta\omega_m}{\omega_m}+ \int^\infty_0\frac{d x_z}{1+x_z}R_\mathcal{O}(x_z, z)\frac{\delta H(x_z)}{H(x_z)}\ ,}
\end{equation}
where the functions $I_\mathcal{O}(z)$, $J_\mathcal{O}(z)$ and $R_\mathcal{O}(x_z, z)$ are functions of the cosmological parameters and the expansion history $H(z)$. In fact, at late times, we will be able to provide definite analytic expressions of all quantities. 

\paragraph{Connecting Model Departures with the Shift in Parameters.}
The important question at this stage is how the initial modification of the Hubble parameter $\delta H(z)$ is related to the modifications of the cosmological parameters $\delta h$ and $\delta\omega_m$. Knowing this relation is what we are ultimately interested in. The crucial point of the proposed method is that it is enough to choose in this case two very well measured anchor observables $A^*_i\equiv A_i(z_*)$ at some redshift $z_*$ (which for us will be the redshift at decoupling), whose value should absolutely not change in the alternative cosmological model and demand that the variation of these anchor observables vanish $\Delta A^*_i\overset{!}{=}0$. Imposing this minimal observational consistency is what allows to formulate a \textit{response function} that captures the effect of the general modified expansion history $\delta H(z)$ on the variation in the inferred $\Lambda$CDM parameters $\delta h$ and $\delta \omega_m$.

Indeed, the system of equations
\begin{equation}\label{eq:AcnhorConstraints}
        \frac{\Delta A^*_i}{A^*_i} = I^*_{A_i}\frac{\delta h}{h}+J^*_{A_i}\frac{\delta \omega_m}{\omega_m}+  \int^\infty_0\frac{d x_z}{1+x_z}R^*_{A_i}(x_z)\frac{\delta H(x_z)}{H(x_z)} \overset{!}{=} 0\,,
\end{equation}
for $i=1,2$ can readily be solved to give
\begin{align}
        \frac{\delta h}{h}&=\int^\infty_0\frac{d x_z}{1+x_z}\,\frac{1}{D^*_{12}}\left(J_{A_1}^*R^*_{A_2}(x_z)-J_{A_2}^*R^*_{A_1}(x_z)\right)\frac{\delta H(x_z)}{H(x_z)}\,\nonumber\\
        &\equiv \int^\infty_0\frac{d x_z}{1+x_z}\mathcal{R}_{h}(x_z)\frac{\delta H(x_z)}{H(x_z)}\,,\label{eq:ResH0Full}\\
        \frac{\delta \omega_m}{\omega_m}&=\int^\infty_0\frac{d x_z}{1+x_z}\,\frac{1}{D^*_{12}}\left(I_{A_2}^*R^*_{A_1}(x_z)-I_{A_1}^*R^*_{A_2}(x_z)\right)\frac{\delta H(x_z)}{H(x_z)}\,,\nonumber\\
        &\equiv \int^\infty_0\frac{d x_z}{1+x_z}\mathcal{R}_{\omega_m}(x_z)\frac{\delta H(x_z)}{H(x_z)} \,,\label{eq:ResomegamFull}
\end{align}
with
\begin{equation}\label{eq:ResDFull}
    D^*_{12}\equiv J_{A_2}^*I_{A_1}^*-J_{A_1}^*I_{A_2}^*\,.
\end{equation}
This provides the response functions $\mathcal{R}_{h}$ and $\mathcal{R}_{\omega_m}$ that very generically and once and for all (up to the choice of anchor observables) describe how a generic modification of the expansion history $\delta H$ results in a variation of the Hubble constant and the matter abundances.
Combining Eqs.~\eqref{eq:ResH0Full} and \eqref{eq:ResomegamFull} with the general total variation of observables $\mathcal{O}$ in Eq.~\eqref{eq:VarGen}, the above results then allow for the computation of the response function of any cosmological quantity
\begin{equation}\label{eq:RespGenO}
	\boxed{\frac{\Delta\mathcal{O}(z)}{\mathcal{O}(z)} =\int^\infty_0\frac{d x_z}{1+x_z}\, \mathcal{R}_\mathcal{O}(x_z, z)\frac{\delta H(x_z)}{H(x_z)}\,,}
\end{equation}
where the response function is explicitly given by
\begin{equation}\label{eq:def_response_g}
		\mathcal{R}_\mathcal{O}(x_z, z) \equiv I_\mathcal{O}(z)\mathcal{R}_h(x_z)
			+ J_\mathcal{O}(z)\mathcal{R}_{\omega_m}(x_z) + R_\mathcal{O}(x_z, z)\,.
\end{equation}

In practice, when applying the method to the $H_0$ and $\sigma_8$ tensions we will be able to effectively neglect any variations on the matter abundances and thus set $\delta\omega_m=0$, an assumption which we will explicitly justify below. In this case, it is sufficient to choose a single observable, for which we impose at a specific time $z_*$ the constraint in Eq.~\eqref{eq:AcnhorConstraints} with $\delta \omega_m=0$ to obtain the simple relation
\begin{align}\label{eq:Resph}
	\boxed{\frac{\delta h}{h} = -\int\frac{d x_z}{1+x_z}\frac{R^*_{A}(x_z)}{I^*_{A}}\frac{\delta H(x_z)}{H(x_z)} = \int\frac{d x_z}{1+x_z}\mathcal{R}_{h}(x_z)\frac{\delta H(x_z)}{H(x_z)}\,.}
\end{align}
Of course, agreeing with just one observable as imposed by Eq.~\eqref{eq:AcnhorConstraints} is by far enough for a $\Lambda$CDM departure to be viable. However, this simple method already allows the derivation of general conditions that \textit{any} model must
at least satisfy in order not to be immediately ruled out by observations. As we will describe, such necessary conditions can already impose stringent analytic constraints on the allowed modifications at the level of the expansion history. For instance, by computing the response function of the observable $\sigma_8$, we will be able to formulate necessary constraints on the functional form of $\delta H$ in order to achieve desired modifications in both the Hubble constant and the clustering amplitude. These conditions crucially depend on the functional form of the response functions $\mathcal{R}_{h}$ and $\mathcal{R}_{\sigma_8}$, which we want to stress, capture once and for all the response of any completely arbitrary background modification $\delta H$ that could arise in any imaginable way.

\paragraph{Generalization to Perturbations.}
Yet, in some cases the effects of an alternative model might certainly not only be restricted to a modification of the background, but will for instance generally also affect cosmological perturbations that will influence other quantities $Q_i(z)$ of the base model. At the linear level, such additional deviations affecting a particular observable $\mathcal{O}$ can simply be captured by adding a sum of the form
\begin{equation}\label{eq:VarAll}
	\sum_i \int^\infty_0\frac{d x_z}{1+x_z}\mathcal{Q}_{i\mathcal{O}}(x_z)\frac{\delta Q_i(x_z)}{Q_i(x_z)}\,,
\end{equation}
to the total variation in Eq.~\eqref{eq:VarGen}, which defines the additional response functions $\mathcal{Q}_{i\mathcal{O}}(x_z)$. 
As generality increases, however, constraining $\delta H$ as well as $\delta Q_i$ will require an increasing number of anchor conditions that reflect multiple observational constraints.

\section{$H_0$ and $\sigma_8$ Tension Constraints on Late Time Cosmologies}\label{sSec:H0 and sigma8 Constraints on LateTimeSolutions}

We will now apply the method introduced above to the specific case of the $H_0$ and $\sigma_8$ tensions introduced in Sec.~\ref{sSec:TheH0andSigma8Tensions} and try to provide guidelines for generic models beyond the standard $\Lambda$CDM to simultaneously solve both tensions. We will in a first step focus on deviations from the background spacetime expansion history, which will already provide noteworthy insights. Subsequently, a first step towards adding the layer of perturbations will be made.

\subsection{Modifying the Background Expansion}\label{ssSec:dH}

A broad class of proposed solutions to the $H_0$ tension modify the $\Lambda$CDM 
background without introducing significant deviations in the perturbations, i.e.
without introducing new clustering species or modifying quantities like the 
gravitational coupling $G$. Indeed, the main effect of most models to alleviate the Hubble tension lies on the background expansion, such that we will in a first approximation neglect any effects at the level of cosmological perturbations.

\paragraph{Fixing the Observational Anchors.}

The first and crucial task in the method described in Sec.~\ref{sSec:ModelIndepMethod} above is to determine the response functions $\mathcal{R}_h$ and $\mathcal{R}_{\omega_m}$ of the free  base cosmology parameters defined in Eqs.~\eqref{eq:ResH0Full} and \eqref{eq:ResomegamFull}. These response functions carry the information on how a generic modification of the background expansion history $\delta H$ translates into a shift in the observationally preferred values of $H_0$ and $\omega_m$. For this, we need to choose two anchor observables $A_{1,2}$ whose values at some particular redshift we do not want to alter.

A very natural choice to place the observational anchors is the CMB representing the bedrock of modern cosmology and in particular its associated distance priors $\theta_A^*=\theta_A(z_*)$ and $R^*=R(z_*)$ defined in Eqs.~\eqref{eq:AcousticScaleDef} and \eqref{eq:ShiftParameter}, where $z_*\simeq 1090$ is here the redshift at decoupling. 
For convenience, we reproduce here their definitions
\begin{align}
		\theta^*_A &= \frac{r_\text{s}(z_*)}{d_C(z_*)}\, ,\\
		R^* &= C_H\,d_C(z_*)\sqrt{\omega_m}\, ,
\end{align}
with the comoving sound horizon [Eq.~\eqref{eq:SoundHorizonSecond z}]
\begin{equation}\label{eq:SoundHorizonTensions}
		r_\text{s}(z) =\int^\infty_z\,\frac{d z'}{H(z')}\,c_s(z)\,,
\end{equation}
and the comoving distance [Eq.~\eqref{eq:ComovingDistanceThird}]
\begin{equation}\label{eq:ComovingDistanceTensions}
		d_C(z) =\int^z_0\,\frac{d z'}{H(z')}\,.
\end{equation}

The choice of the distance priors as observational anchor points is justified in that they each constrain an orthogonal fundamental aspect of the CMB angular powerspectrum, namely the location and the relative heights of the acoustic peaks respectively. Moreover, as discussed, the distance priors are particularly well measured observables of the CMB. Of course, as already discussed, the effects of modified cosmologies are by no means restricted to variations in these two quantities only, but demanding that $\theta^*_A$ and $R^*$ remain approximately fixed can be viewed as a minimal requirement to not directly be excluded from observational constraints. We therefore impose 
\begin{equation}\label{eq:CMBDistancePriorsConstrants}
\begin{split}
		\frac{\Delta \theta^*_A}{\theta^*_A} &= \frac{\Delta r_\text{s}^*}{r_\text{s}^*} - \frac{\Delta d_C^*}{d_C^*}\overset{!}{=} 0\,,\\
		\frac{\Delta R^*}{R^*} &= \frac{\Delta d_C^*}{d_C^*} + \frac{\delta\omega_m}{2\omega_m}\overset{!}{=} 0\,.
  \end{split}
\end{equation}

Hence, in order to bring these anchor constraints into the form of Eq.~\eqref{eq:AcnhorConstraints}, we need to compute the variation of the comoving distance and the comoving sound horizon\footnote{Note that by their definition, the comoving, the luminosity and the angular diameter distances $d_C$, $d_L$ and $d_A$ in fact all share the same total variation.}
	\begin{equation}
		\left\{\begin{array}{l}
		\displaystyle
			I_{d_C}(z)  = -\frac{1}{d_C(z)}\int^z_0 d x_z\frac{H_0^2}{H^3(x_z)}\\[8pt]
			\displaystyle
			J_{d_C}(z) = -\frac{1}{d_C(z)}\int^z_0 d x_z\frac{H_0^2}{H^3(x_z)}m(x_z)\\[8pt]
			\displaystyle
			R_{d_C}(x_z,z) = -(1+x_z)\frac{\Theta(z-x_z)}{d_C(z)H(x_z)}
		\end{array}\right.
	\end{equation}
and 
	\begin{equation}
		\left\{\begin{array}{l}
		\displaystyle
			I_{r_\text{s}}(z) = -\frac{1}{r_\text{s}(z)}\int^\infty_z d x_z\frac{H_0^2}{H^3(x_z)}c_\text{s}(x_z)\\[8pt]
			\displaystyle
			J_{r_\text{s}}(z) = -\frac{1}{r_\text{s}(z)}\int^\infty_z d x_z\frac{H_0^2}{H^3(x_z)}m(x_z)c_\text{s}(x_z)\\[8pt]
			\displaystyle
			R_{r_\text{s}}(x_z,z) = -\frac{(1+x_z)c_\text{s}(x_z)}{r_\text{s}(z)}\frac{\Theta(x_z-z)}{H(x_z)}
		\end{array}\right.
	\end{equation}
 where $\Theta$ is the Heaviside step function. Thus
 \begin{align}
        \frac{\Delta d^*_C}{d^*_C} &= I^*_{d_C}\frac{\delta h}{h}+J^*_{d_C}\frac{\delta \omega_m}{\omega_m}+  \int^\infty_0\frac{d x_z}{1+x_z}R^*_{d_C}(x_z)\frac{\delta H(x_z)}{H(x_z)} \,,\\
        \frac{\Delta r^*_\text{s}}{r^*_\text{s}} &= I^*_{r_\text{s}}\frac{\delta h}{h}+J^*_{r_\text{s}}\frac{\delta \omega_m}{\omega_m}+  \int^\infty_0\frac{d x_z}{1+x_z}R^*_{r_\text{s}}(x_z)\frac{\delta H(x_z)}{H(x_z)}\,.
 \end{align}

 Using these results together with Eq.~\eqref{eq:CMBDistancePriorsConstrants}, we immediately obtain the response functions $\mathcal{R}_h$ and $\mathcal{R}_{\omega_m}$ through Eqs.~\eqref{eq:ResH0Full}, \eqref{eq:ResomegamFull} and \eqref{eq:ResDFull} with the replacements
 \begin{align}
     I^*_{A_1} &\rightarrow I^*_{r_s}-I^*_{d_A}\,,& J^*_{A_1} &\rightarrow J^*_{r_s}-J^*_{d_A}\,,& R^*_{A_1}&\rightarrow R^*_{r_s}-R^*_{d_A}\,,\\
     I^*_{A_2} &\rightarrow I^*_{d_A}\,,& J^*_{A_2} &\rightarrow J^*_{d_A}+\frac{1}{2}\,,& R^*_{A_2}&\rightarrow R^*_{d_A}\,.
 \end{align}
The resulting response functions $\mathcal{R}_h$ and $\mathcal{R}_{\omega_m}$ capture the variations in the observationally preferred late-time $\Lambda$CDM parameters induced by changes in the Hubble function $\delta H(z)$ produced by a completely generic alternative cosmological model through
\begin{align}
	\frac{\delta h}{h} &=\int^\infty_0\frac{d x_z}{1+x_z}\, \mathcal{R}_h(x_z)\frac{\delta H(x_z)}{H(x_z)}\,,\label{eq:RespGenH0}\\
 \frac{\delta\omega_m}{\omega_m} &=\int^\infty_0\frac{d x_z}{1+x_z}\, \mathcal{R}_{\omega_m}(x_z)\frac{\delta H(x_z)}{H(x_z)}\,.\label{eq:RespGenomegam}
\end{align}
The notion of ``observationally preferred'' is in this case provided by the CMB priors that are left unchanged, such that all modified cosmologies remain roughly compatible with the CMB. These general and analytic response functions can now be used in order to quickly estimate the effect on the cosmological parameter space of every specific model based on their impact on the expansion history.

\begin{figure}
\centering
\includegraphics[scale=0.55]{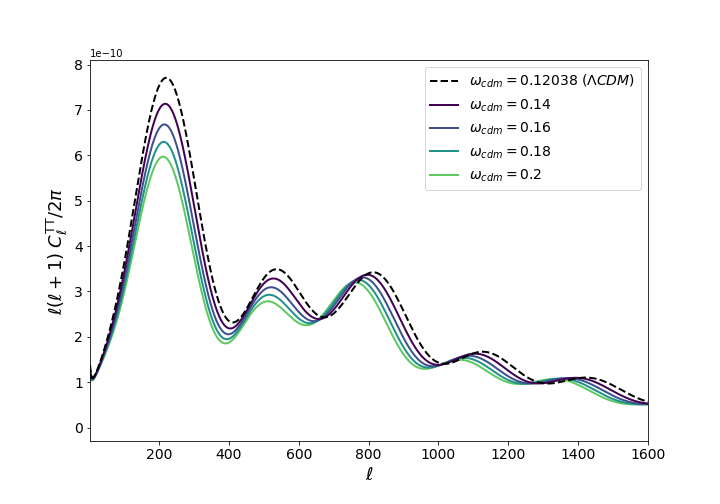}
\caption{\small The theoretical powerspectrum of anisotropies $C_l^{TT}$ in the CMB as a function of the multipole moment $l$ of spherical harmonics computed for the $\Lambda$CDM model with parameters given in Table~\ref{table:PlankBestFit} but for different values of $\omega_{cdm}$ at fixed $h$ and $\omega_b$. The angular powerspectrum was computed with the {\fontfamily{qcr}\selectfont class} code \cite{lesgourgues2011cosmic,Blas:2011rf}.}
\label{fig:Cl_w0}
\end{figure}

While the expressions derived above are general in principle, the neglect of variation of the additional $\Lambda$CDM parameters $\omega_b$, $\tau_\text{rei}$, $n_s$ and $A_s$ compared to $h$ and $\omega_m$ is only valid at late cosmic times. Hence, in the following we will also restrict ourselves to late-time modifications, and choose specifically $\delta H(z)=0$ for $z>300$. Within this range, however, the response function of $\omega_m$ remains very close to zero $\mathcal{R}_{\omega_m}\simeq 0$. This means that we can actually also keep the cosmological parameter $\omega_m$ fixed for late-time modifications. Indeed, a comparison of Fig.~\ref{fig:Cl_h} showing the variation of the angular powerspectrum with respect to the parameter $h$ with Fig.~\ref{fig:Cl_w0} in which the density parameter of cold dark matter and thus effectively $\omega_m$ is varied, confirms that both variations are rather orthogonal. The variation of $h$ mainly affects the location of the acoustic peaks characterized by the angular scale, whereas a variation of $\omega_m$ mostly results in a modification of the relative heights of the peaks constrained by the shift parameter. The result, that $\mathcal{R}_{\omega_m}$ remains negligible, therefore implies that a deviation in the late-time expansion history mainly only results in a shift of the acoustic peaks. 

\begin{figure}[H]
    \centering
    \includegraphics[width=0.7\textwidth]{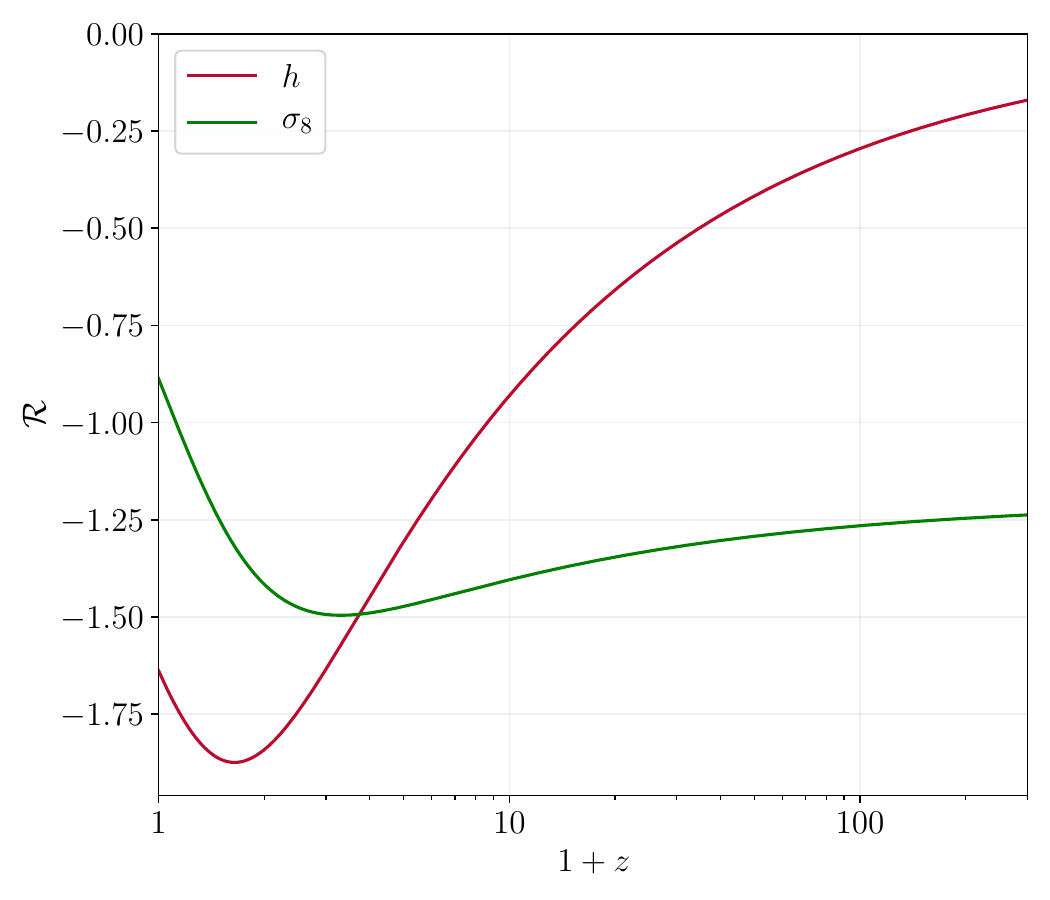}
    \caption{\small The response functions $\mathcal{R}_{h}(z)$ and $\mathcal{R}_{\sigma_8}(z, 0)$ as defined in Eqs.~\eqref{eq:Resph} and \eqref{eq:Respsigma8} respectively. Both responses remain strictly negative over the entire rage $0<z<300$ in which the expansion history is modified. (Figure taken from \textit{L. Heisenberg, H. Villarrubia-Rojo, J. Zosso, (2022)} \cite{Heisenberg:2022gqk}.)}
    \label{fig:responses}
\end{figure}

Thus, effectively at late times we are left with one free $\Lambda$CDM parameter to vary, the Hubble constant $h$. In this case, following Eq.~\eqref{eq:Resph} while imposing a single observational constraint on the CMB angular scale $A=\theta^*$ in Eq.~\eqref{eq:Resph}, the response function of the Hubble constant has the simple form
\begin{equation}\label{eq:ResponseH0Simp}
	\mathcal{R}_h(z) \simeq -\frac{R^*_{d_A}(z)}{I^*_{d_A}}\, .
\end{equation}
The resulting response function is depicted in Figure \ref{fig:responses}. The most notable feature is that the response function remains strictly negative over the entire redshift range $0<z<300$. This allows us to draw the first, very general conclusion: In order to increase the value of $H_0$, hence $\delta h>0$, to alleviate the $H_0$ tension at the level of the background, any alternative model at least needs to satisfy $\delta H(z)<0$ for some $z$ in the late-time range. Observer that this very clean result is well in line with the intuition that solving the Hubble tension at late times requires an increase in the comoving distance to decoupling $d_C(z_*)$ compared to $\Lambda$CDM. In Sec.~\ref{sSec:LessonsOnDynamicalDE} we will further analyze this result in the specific but obvious application to dynamical dark energy and beyond GR theories.

\paragraph{The $\sigma_8$ Response Function.}
As described in Sec.~\ref{sSec:ModelIndepMethod} the result of the response function of $H_0$ in Eq.~\eqref{eq:ResponseH0Simp}, together with the verification that fixing $\omega_m$ is an excellent approximation at late times, now allow the computation of the response function defined in Eqs.~\eqref{eq:RespGenO} and \eqref{eq:def_response_g} of any other cosmological quantity for which the total variation of the form in Eq.~\eqref{eq:VarGen} has been computed. In order to attempt an answer to the question on whether and how it is possible for cosmological departures to solve both the $H_0$ and $\sigma_8$ tensions, we will apply this method to compute the response function of the clustering amplitude $\sigma_8$. 

Hence, the next task in our approach is to find an analytic formula for $\sigma_8$ whose total variation we can compute. To achieve this, we need to resort to the various concepts and observables introduced back in Sec.~\ref{sSec:LCDM}. First of all, recall that the clustering amplitude $\sigma_8$ is defined as the variance of smoothed out overdensities, which can be written in terms of a weighted integral over the matter powerspectrum [Eq.~\eqref{eq:def_sigmaR_2}]
\begin{equation}
			\sigma_8^2(a)= \int_0^\infty\frac{d k}{k}\,\mathcal{P}_m(k,a)\,\tilde W^2(kR)\, ,
\end{equation}
where we measure the scales $k$ in terms of Mpc$^{-1}$, with
\begin{equation}
    R= 8 \,h^{-1}\text{ Mpc}^{-1}\,,
\end{equation}
and where $\tilde W$ is the Fourier transform of the tophat function
\begin{equation}
    \tilde W(x) = \frac{3j_1(x)}{x}\,.
\end{equation}
At late-times, the matter powerspectrum can be written in terms of the primordial curvature powerspectrum $\mathcal{P}_\mathcal{R}$ defined in Eq.~\eqref{eq:InitialCurvaturePowerSectrum Dimensionless} and a subsequent evolution characterized by the $k$ dependent transfer function and the time dependent growth factor [Eq.~\eqref{eq:LinearMatterPowerSpecSecond}]
\begin{equation}
			\mathcal{P}_m(k,a) =\frac{4}{25}\frac{k^4}{\omega_m^2C_H^4}\,T^2(k)\,D^2(a)\,\mathcal{P}_\mathcal{R}(k)\, ,
\end{equation}
where we recall that
\begin{equation}
	\mathcal{P}_\mathcal{R}(k)= A_s\left(\frac{k}{k_p}\right)^{n_s-1},\qquad k_p=0.05\,\text{Mpc}^{-1}\,.
\end{equation}

For the transfer function, we can employ the analytic Eisenstein-Hu fitting formula \cite{Eisenstein:1997ik} $T_{EH}(k)$ introduced in Eq.~\eqref{eq:EisensteinHuFittingFormula} that importantly takes into account the small scale baryonic suppression as discussed in Sec.~\ref{sSec:LargeScaleStructure}. On the other hand, the growth factor $D(a)$ is defined as the growing solution to the linear growth equation [Eq.~\eqref{eq:LinearGrowthEq}] that is valid as long as the late-time assumptions listed in Eq.~\eqref{eq:LateTimeAssumptions} are satisfied. Gathering all together, $\sigma_8$ in $\Lambda$CDM is therefore given by
\begin{equation}
			\sigma_8^2(a)= \frac{4}{25}\,\frac{D^2(a)}{\omega_m^2}\,\mathcal{I}\, ,
\end{equation}
where
\begin{equation}\label{eq:DefIofsigm8}
    \mathcal{I}\equiv \int_0^\infty\frac{d k}{k}\,\left(\frac{k}{C_H}\right)^4\,T_{EH}^2(k)\,\mathcal{P}_\mathcal{R}(k)\,\tilde W^2(kR)\,.
\end{equation}
Given this expression, the total variation of $\sigma_8$ at fixed $\omega_m$ can be written as
\begin{equation}
    \boxed{\frac{\Delta\sigma_8}{\sigma_8}=\frac{\Delta D}{D}+\frac{1}{2}\frac{\Delta\mathcal{I}}{\mathcal{I}}\,,}
\end{equation}
where
\begin{align}
\frac{\Delta\mathcal{I}}{\mathcal{I}}=&\;\frac{2}{\mathcal{I}} \,\bigg[\int_0^\infty\frac{d k}{k}\,\left(\frac{k}{C_H}\right)^4\,T_{EH}^2(k)\,\mathcal{P}_\mathcal{R}(k)\,\tilde W(kR)\,\Delta\tilde W(kR) \bigg]\,,\\
=& -\frac{2}{\mathcal{I}}\int^\infty_0\frac{d k}{k}T^2(k)\mathcal{P}_\mathcal{R}(k)\left(\frac{k}{C_H}\right)^4\,\tilde W(kR)\,kR\,\tilde W'(kR)\; \frac{\delta h}{h}\,.
\end{align}
In this final expression, we can easily read off the value of $I_\mathcal{I}$, whereas $R_\mathcal{I}=0$.

Here, the only non-straightforward variation is the variation of the growth factor $D(a)$. Defined as the growing solution of the linear growth equation [Eq.~\eqref{eq:LinearGrowthEq}], its total variation is determined by solving the associated inhomogeneous equation
\begin{equation}
    \frac{d^2}{da}\Delta D(a)+\frac{d\log(a^3 H(a)}{da}\frac{d}{da}\Delta D(a)-F(a)\Delta D(a)=g(a)\,,
\end{equation}
where $g(a)$ is the inhomogeneous correction that reads
\begin{equation}
    g(a)\equiv -\frac{d}{da}\left(\frac{\Delta H(a)}{H(a)}\right)\frac{dD_+(a)}{da}-2F(a)D_+(a)\frac{\Delta H(a)}{H(a)}\,.
\end{equation}
Given two solutions $D_1$ and $D_2$ of the homogeneous equation  in Section~\ref{sSec:LargeScaleStructure}, one can construct a solution to the inhomogeneous equation for any $g(a)$ through the Wronskian method by
\begin{equation}
    \Delta D(a)=D_2(a)\int_0^adx_a\frac{D_1(x_a)g(x_a)}{W(x_a)}-D_1(a)\int_0^adx_a\frac{D_2(x_a)g(x_a)}{W(x_a)}\,,
\end{equation}
where in this case, the Wronskian reads
\begin{equation}
    W(a)=D_1(a)D_2'(a)-D_2(a)D_1'(a)=-\frac{H_0}{a^3H(a)}\,.
\end{equation}
Thus, the total variation of the growth factor becomes
\begin{equation}
		\boxed{\Delta D (a)= \frac{H(a)}{H_0} \int^a_0d x_a \frac{x_a^3H^2(x_a)}{H_0^2}\Big(I(a)-I(x_a)\Big) g(x_a)\ ,}\label{eq:DeltaD}\\
\end{equation}

This provides all the ingredients for the total variation of the observable $\sigma_8$. However, in order to determine the associated response function, that in the absence of $\mathcal{R}_{\omega_m}$ is entirely determined through $\mathcal{R}_{h}$ in Eq.~\eqref{eq:def_response_g} we further need to perform a change of variables to redshift space and determine
\begin{subequations}
\begin{align}
    I_{\sigma_8}(z)&=I_D(z)+\frac{1}{2}I_\mathcal{I}\,,\\
    R_{\sigma_8}(x_z,z)&=R_D(x_z,z)\,.
\end{align}
\end{subequations}
 We provide the full expressions of $I_D(z)$ and $R_D(z)$ in Appendix~\ref{sec:app_formulaeVariations}. In the same Appendix~\ref{sec:app_formulaeVariations} we gather for convenience all full analytic expressions relevant in this chapter. Finally, the response function of $\sigma_8$ today is given by the combination
\begin{equation}\label{eq:Respsigma8}
	\boxed{\mathcal{R}_{\sigma_8}(z, 0) = I_{\sigma_8}(0)\mathcal{R}_{H_0}(z) + R_{\sigma_8}(z, 0)\,.}
\end{equation}
The result, which we want to emphasize again is computed once and for all, is plotted in Fig.~\ref{fig:responses}. Hence, for a given background $\Lambda$CDM departure $\delta H(z)$ the corresponding shift in $\sigma_8$ can be computed analytically through Eq.~\eqref{eq:RespGenO}
\begin{equation}\label{eq:RespGensigma8}
	\frac{\Delta\sigma_8(z)}{\sigma_8(z)} =\int^\infty_0\frac{d x_z}{1+x_z}\, \mathcal{R}_{\sigma_8}(x_z, z)\frac{\delta H(x_z)}{H(x_z)}\,.
\end{equation}

Just as the response function of the Hubble constant $\mathcal{R}_{h}$, also $\mathcal{R}_{\sigma_8}$ remains negative on the entire redshift range $0<z<300$. This functional form of the response functions immediately allow drawing simple conclusions with substantial impact. As we already concluded from the shape of $\mathcal{R}_{h}$, solving the Hubble tension, hence $\delta h> 0$ at least requires $\delta H(z)<0$ for some $z$. Similarly, in order to alleviate the $\sigma_8$ tension, which requires $\Delta\sigma_8<0$
, is only possible if $\delta H(z)>0$ at some $z$. Thus, the exact opposite of what is required from the $H_0$ tension. This very simple observation within the proposed language of response functions is at the root of the observation that most Hubble tension solutions mess up the $\sigma_8$ inference.
In other words, in order to address both tensions and thus increase the value of the Hubble constant while simultaneously decreasing the clustering amplitude demands more sophisticated models and at the very least necessarily requires $\delta H(z)$ to change sign. This general result can readily be used to rule out specific models proposed in the 
literature. Again, in Sec.~\ref{sSec:LessonsOnDynamicalDE} we will analyze this result in the context of dark energy and beyond GR theories.

Furthermore, it is interesting to also compute the response functions of the combinations of observables $f\sigma_8$ with $f$ the growth rate given in Eq.~\eqref{eq:FFunctionGrowthFactor} and $S_8$ defined in Eq.~\eqref{eq:S8}, that are more directly targeted by RSD and weak lensing surveys, respectively. The variation of the growth rate is given by
\begin{equation}
    \Delta f = \frac{d}{d\log a}\frac{\Delta D}{D}\,,
\end{equation}
while the total variation of $S_8$ (at fixed $\omega_m$) reads
\begin{equation}
    \frac{\Delta S_8}{S_8}(z) = \frac{\Delta\sigma_8}{\sigma_8}(z) - \frac{\delta h}{h}\,,
\end{equation}
and therefore
\begin{align}
    \mathcal{R}_{f\sigma_8}(x_z,z) &= \mathcal{R}_{f}(x_z,z) + \mathcal{R}_{\sigma_8}(x_z,z)\,,\label{eq:Respfsigma8} \\
    \mathcal{R}_{S_8}(x_z,z) &= \mathcal{R}_{\sigma_8}(x_z,z) - \mathcal{R}_h(x_z)\,.\label{eq:RespS8} 
\end{align}
The explicit formula for $\mathcal{R}_{f}(x_z,z)=I_f(z)\mathcal{R}(x_z)+R_f(x_z,z)$ is again provided in Appendix~\ref{sec:app_formulaeVariations}.
		
\begin{figure}
\centering
\includegraphics[scale=0.6]{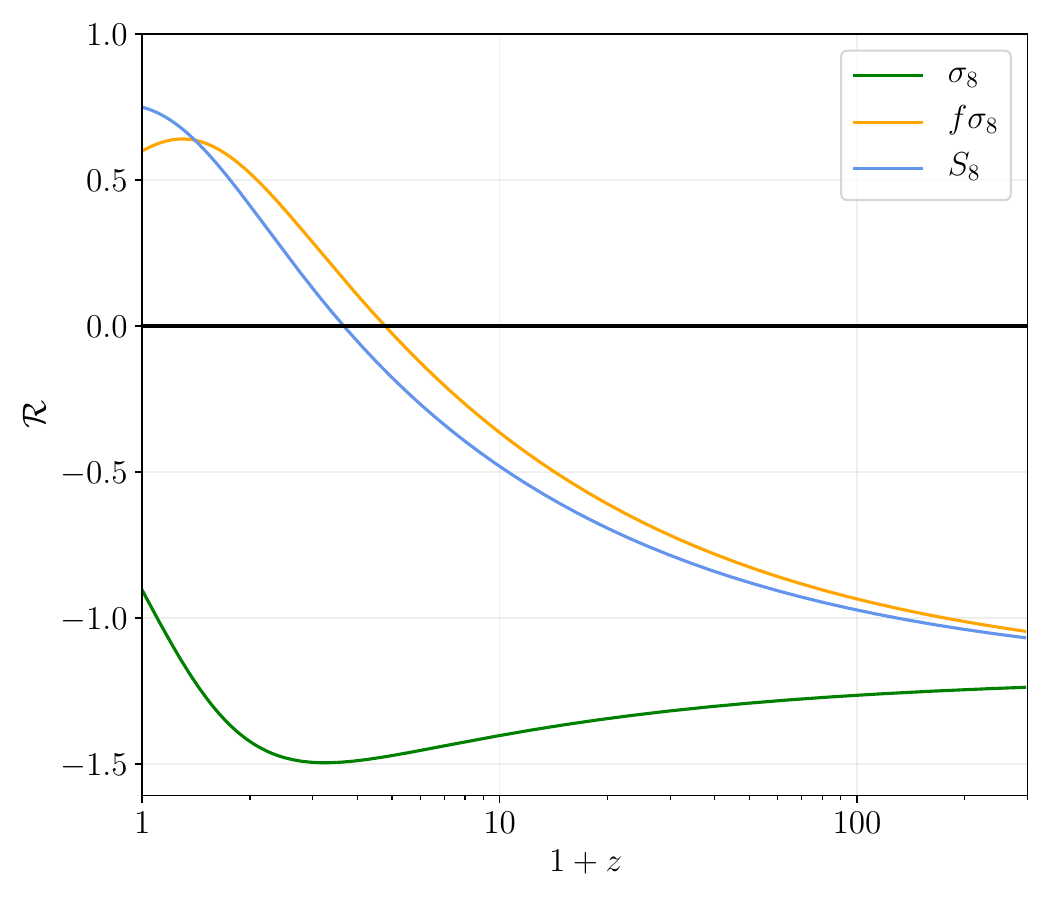}
\caption{\label{fig:responsesS8} \small The response functions $\mathcal{R}_{\sigma_8}(z,0)$, $\mathcal{R}_{f\sigma_8}(z, 0)$ and $\mathcal{R}_{S_8}(z, 0)$ as given in Eqs.~\eqref{eq:Respsigma8}, Eqs.~\eqref{eq:Respfsigma8} and Eqs.~\eqref{eq:RespS8} respectively. (Figure taken from \textit{L. Heisenberg, H. Villarrubia-Rojo, J. Zosso, (2022)} \cite{Heisenberg:2022gqk}.)}
\end{figure}

The comparison of the response functions for $\sigma_8$, $f\sigma_8$ and $S_8$ is depicted in Fig.~\ref{fig:responsesS8}. Observe that while $\mathcal{R}_{\sigma_8}$ remains entirely negative, the responses for $S_8$ and $f\sigma_8$ at very late-times change their sign. This feature could be noteworthy, if with the results of upcoming LSS surveys, we find ourselves in a situation where the clustering amplitude tension is clearly present in 
$S_8$ and $f\sigma_8$, but not in $\sigma_8$, or vice versa. This structural difference in their response functions might in this case be regarded as a potential explanation, and could provide us with hints about the shape of $\delta H(z)$.

\subsection{Beyond the expansion history: $G_\text{eff}$}\label{sSec:dG}

As discussed above, at the level of the homogeneous and isotropic background, the $H_0$ and $\sigma_8$ tensions require diametrically opposite deviations from $\Lambda$CDM. One may wonder whether a more natural solution could therefore lie at the level of the cosmological perturbations. Indeed, any consistent late-time dark energy model (see Sec.~\ref{sSec:LessonsOnDynamicalDE}) must include perturbations as well. In realistic scenarios, however, the perturbation level is in most cases not independent of the background behavior and their observational impact is very different. In fact, the main driving force for the values of $\sigma_8$ and $H_0$ mostly lies in the evolution of the FLRW background, which justified the study in the preceding sections.

Yet, in particular for DE models for example through clustering of perturbations and beyond GR theories in general, one typical mechanism that can significantly affect the matter growth and therefore $\sigma_8$ is an effective change in Newtons constant (see e.g. \cite{Heisenberg:2020xak}). Note that while a modification in $G$ strictly speaking also affects the background equation governing the evolution of the Hubble function, such a change is effectively degenerate with $\delta H$ and therefore already contained in the analysis above. 

We therefore define $G_\text{eff}(a)=G+\delta G(a)$ to first order in the sub-Hubble regime and analyze its effect on $\sigma_8$. Observe that, for simplicity, we consider here only scale invariant modifications. Moreover, we will further require that the assumptions underlying the evolution equation of the growth factor [Eq.~\eqref{eq:LinearGrowthEq}] still hold. In this case, the effect of changing the effective gravitational coupling on $\sigma_8$ is precisely characterized through the modification of said evolution equation
\begin{equation}
	D''(a) + \frac{d\log (a^3H)}{da}D'(a) - \frac{G_\text{eff}(a)}{G_0}F(a)D(a) = 0\, .
\end{equation}
Hence
\begin{equation}
    \frac{d^2}{da^2}\delta_G D(a) + \frac{d\log (a^3H)}{da}\frac{d}{da}\delta_G D(a)
        - F(a)\delta_G D(a) = F(a)D(a)\frac{\delta G(a)}{G_0}\ ,
\end{equation}
where in this case $\delta_G D$ stands for a variation keeping all the cosmological parameters and $H(z)$ fixed. The particular solution is this time given by
\begin{align}\label{eq:deltaGofD}
    \boxed{\frac{\delta_G D}{D} = \frac{H(a)}{D(a)H_0}\int^a_0d x_a \frac{x_a^3H^2(x_a)}{H_0^2}\Big(I(a)-I(x_a)\Big)F(x_a)D(x_a)\frac{\delta G(x_a)}{G_0}\, .}
\end{align}

As discussed in Sec.~\ref{sSec:ModelIndepMethod}, going beyond the background evolution induces additional small deviations from $\Lambda$CDM, parametrized by additional functions $\delta Q_i(a)$, which in this case is given by $\delta G(a)$. At the linear level, this additional deviation affects a particular observable through an additive factor given by an associated response function [Eq.~\eqref{eq:VarAll}].
Therefore, the total variation of the growth factor including $\delta_G D$ reads
\begin{equation}
			\frac{\Delta D(z)}{D(z)}\bigg\lvert_\text{full} = \int^{\infty}_0\frac{d x_z}{1+x_z}\left(
				\mathcal{R}_D(x_z, z)\frac{\delta H(x_z)}{H(x_z)}
				+\mathcal{G}_D(x_z, z)\frac{\delta G(x_z)}{G}\right)\,,
\end{equation}
where the additional response function $\mathcal{G}_D$ can be read off from Eq.~\eqref{eq:deltaGofD} to be
\begin{align}
        \frac{\delta_G D(z)}{D(z)}\equiv \int^\infty_0 \frac{d x_z}{1+x_z} \mathcal{G}_{D}(x_z, z)\frac{\delta G(x_z)}{G_0}\,,
\end{align}
with
\begin{equation}\label{eq:G_eff_response}
    \boxed{\mathcal{G}_{D}(x_z, z) = \frac{H^3(z) D(x_z)}{H_0^3 D(z)}\Big(I(z)-I(x_z)\Big)F(x_z)\frac{\Theta(x_z-z)}{(1+x_z)^4}\, .}
\end{equation}

\begin{figure}
\centering
\includegraphics[scale=0.6]{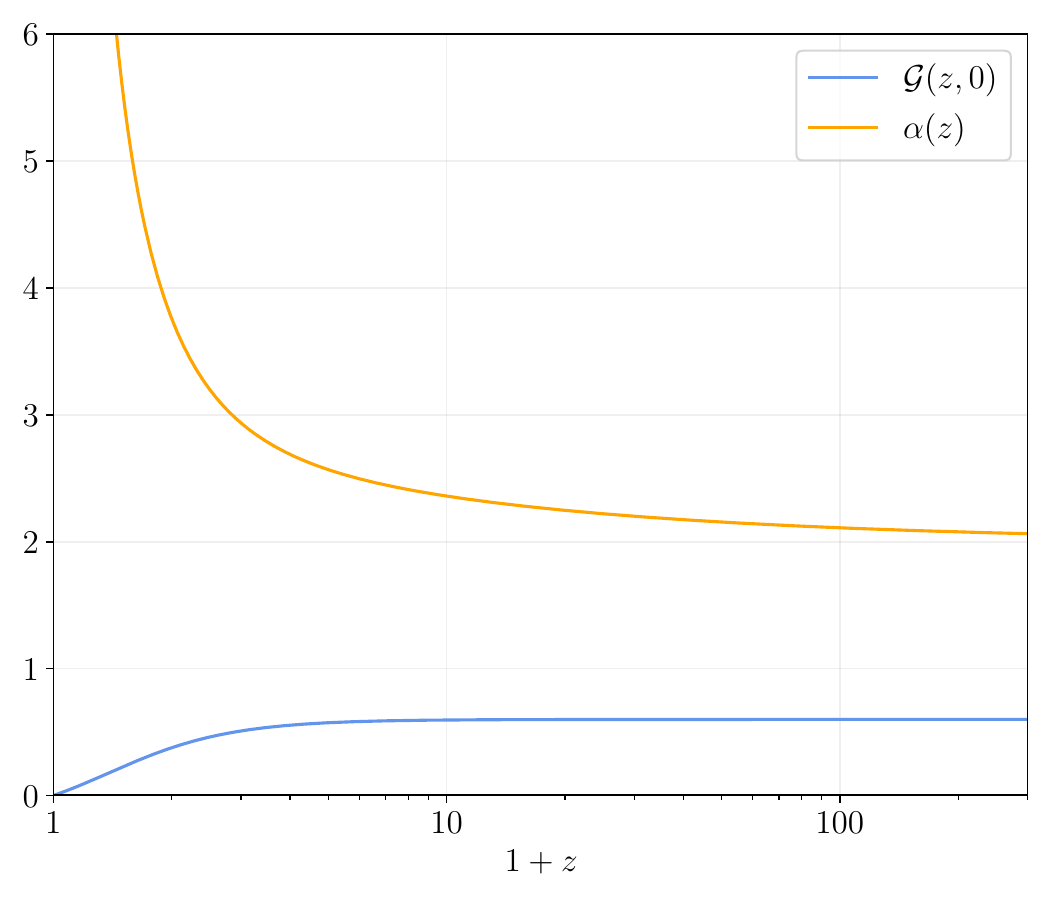}
\caption{\small{The response function $\mathcal{G}(z, 0)$ and the function $\alpha(z)$ as defined in Eqs.~\eqref{eq:G_eff_response} and \eqref{eq:alpha} respectively. Both functions remain strictly positive over the entire rage $0<z<300$ in which the expansion history is modified. (Figure taken from \textit{L. Heisenberg, H. Villarrubia-Rojo, J. Zosso, (2022)} \cite{Heisenberg:2022gqk}.)}}
\label{fig:responsesG}
\end{figure}

Translated to $\sigma_8$ that linearly depends on the growth factor, the additional variation of the growth factor directly affects the clustering amplitude and therefore
\begin{equation}
    \mathcal{G}_{\sigma_8}(x_z,z)=\mathcal{G}_{D}(x_z,z)\,.
\end{equation}
In summary, the total variation of $\sigma_8$ reads
\begin{equation}\label{eq:RespGenWithG}
\boxed{\frac{\Delta\sigma_8(z)}{\sigma_8(z)}\bigg\lvert_\text{full} =\int^\infty_0\frac{d x_z}{1+x_z}\, \mathcal{R}_{\sigma_8}(x_z,z)\frac{\delta H(z)}{H(z)}+\int^\infty_0\frac{d x_z}{1+x_z}\, \mathcal{G}_{\sigma_8}(x_z,z)\frac{\delta G(z)}{G_0}\,,}
\end{equation}
where the response function $\mathcal{G}_{\sigma_8}(z,0)$ today is again 
presented visually in Fig.~\ref{fig:responsesG}.

Including two free functions $\delta H(z)$ and $\delta G(z)$ render the results more general but also implies that within our minimal approach of enforcing a single observational anchor point given by the CMB acoustic scale, it is not possible to derive similarly strong necessary conditions on the two functional form of the free functions $\delta H$ and $\delta G$. While it would be possible to simply add additional observational constraints or even switch to a more quantitative study, we choose here to proceed by restrict ourselves to the case in which $\delta H(z)<z$ for all $0<z<300$. In other words, we want to solve the Hubble tension at the level of the background in a simple way, as it is realized in many physically relevant models, and analyze whether it is possible to reconcile the associated unwanted increase in $\sigma_8$ at the level of the perturbations.  

Indeed, naively including $\delta G(z)$ we have enough freedom to increase $H_0$ while reducing $\sigma_8$ by reducing the effective strength
of gravity enough, i.e. $\delta G(z)<0$. In our approach, we can turn this intuition into a precise statement of a necessary condition. Namely, the results above, in particular the fact that $\mathcal{G}_{\sigma_8}$ remains strictly positive, imply that in order to reduce the value of $\sigma_8$, while increasing $H_0$ it must be that
\begin{equation}\label{eq:G_eff_bound}
    \frac{\delta G(z)}{G_0} < \alpha(z)\frac{\delta H(z)}{H(z)}<0\qquad
        \text{for some }z\ ,
\end{equation}
where we have defined the strictly positive function
\begin{equation}\label{eq:alpha}
    \alpha(x_z)\equiv
-\mathcal{R}_{\sigma_8}(x_z, 0)/\mathcal{G}_{\sigma_8}(x_z, 0)\,,
\end{equation} 
plotted in Fig.~\ref{fig:responsesG}.

In the next section, we want to analyze the results above in the context of a more concrete scenario of late-time dark energy models characterized by a general equation of state $w(t)$. This will in particular draw a connection to beyond GR theories, which most often are used as concrete theories behind the more phenomenological DE models.


\section{Lessons on Dynamical Dark Energy}\label{sSec:LessonsOnDynamicalDE}

This last section will apply the in principle totally model independent study of arbitrary deviations $\delta H$, to a specific class of theories, namely the popular late-time dark energy models. This will in particular provide an answer to the question \cite{Heisenberg:2022gqk}: 
``Can the $H_0$ and $\sigma_8$ tensions be simultaneously relieved, modifying only the dark energy equation of state $w(z)$ at late times?'' As already discussed, a motivation for such a special focus on late-time dynamical dark energy models, is that most beyond GR models introduced in Sec.~\ref{ssSec: A Exact Theories} employed in the context of cosmology effectively introduce a dynamical dark energy component that modifies the background evolution at different cosmological epochs. Such late-time DE models were studied and motivated long before the advent of the current Hubble tension, based on entirely different motivations.
Moreover, also from an observational side current cosmological data, which is beginning to disfavor a pure $\Lambda$CDM model, could in particular necessitate a time varying late time equation of state \cite{Zhao:2017cud}.

\paragraph{Dynamical Dark Energy Beyond the Cosmological Constant.}
While a cosmological constant at the root of the current accelerated expansion nevertheless performs rather well in comparison to observations, there are deep theoretical issues in determining the origin of a constant energy density that does not dilute as the universe expands. Quantum mechanics would in principle provide a very compelling explanation in terms of the vacuum energy of quantum fields coupling to a quantum effective field theory approach to gravity. However, taking such a gravitation of quantum matter seriously leads to the biggest discrepancy in theoretical physics of  50 to 120 orders of magnitude between the expected and observed value of $\Lambda$. More precisely, the extremely low value of $\Lambda$ compared to the theoretical expectation requires a very unnatural \textit{fine-tuning} of parameters. In simple terms, this means that we have absolutely no idea why the CC should admit its measured value. This most apparent lack of understanding is known as the CC problem that we already mentioned a few times and that we will discuss in more detail in Sec.~\ref{sSec: The CC Problem}. Together with the inflation paradigm that requires a dynamical accelerated expansion that cannot be due to a fixed constant energy density simply because the acceleration has to end, the CC problem represents a big motivation to study models of dynamical dark energy.

Except for the value $w=-1$ of a cosmological constant discussed in Sec.~\ref{sSec:CCDarkEnergy} and in contrast to radiation and non-relativistic matter, there is no particular reason why the equation of state of dark energy should be constant \cite{Weinberg2008Cosmology}. Thus, at the level of the homogeneous and isotropic background, the energy density of generic dark energy is given by Eq.~\eqref{eq:GeneralRhoNotConstOmega}, with the constraint that it should reduce to the value of the density parameter $\Omega_\Lambda$ today, such that in terms of redshift
\begin{equation}\label{eq:rhoDEz_w}
		\Omega_\text{DE}(z)\equiv\frac{\rho_\text{DE}(z)}{\rho_c} =\Omega_\Lambda \,\exp\left(3\int^z_0\frac{d z'}{1+z'}\,\big[1+w(z')\big]\right)\,.
\end{equation}

\paragraph{Dynamical Dark Energy and the Hubble Tension.}
In order to solve the $H_0$ tension, it is known that such a phenomenological description of a dark energy fluid requires a value of the equation of state of $w <-1$ (see e.g. \cite{Planck:2018vyg}), called \textit{phantom} equation of state. With the results derived in Sec.~\ref{ssSec:dH} we can understand this statement very cleanly. For this, we need to translate the generic modification from the $\Lambda$CDM expansion history at fixed cosmological parameters
\begin{equation}\label{eq:DefLCDMDeviationH}
    \delta H(z)= H(z)-H_\text{\tiny $\Lambda$CDM}(z)
\end{equation}
to a deviation in the equation of state. Writing 
\begin{equation}\label{eq:HubbleModModel}
		H^2(h, \omega_m) = H^2_{\text{\tiny $\Lambda$CDM}}(h, \omega_m) + H_0^2\delta\Omega\ .
\end{equation}
where 
\begin{equation}
\Omega_\text{DE}(z)\equiv\Omega_\Lambda + \delta\Omega(z) \,,
\end{equation}
with $\delta\Omega=0$ at $z=0$, we have at first order the relation
\begin{equation}
		\frac{\delta H(z)}{H(z)} = \frac{H_0^2}{2H^2(z)}\delta\Omega(z)\, .
\end{equation}
Observe that while $\delta H/H$ needs to be small by assumption of linear perturbations, $\delta\Omega(z)$ can still become substantial. For a dynamical dark energy model, we would therefore require that the still completely generic function $\delta\Omega(z)$ is such that
 \begin{equation}
		\Omega_\text{DE}(z) > 0\,.
\end{equation}

\begin{figure}
\centering
\includegraphics[scale=0.5]{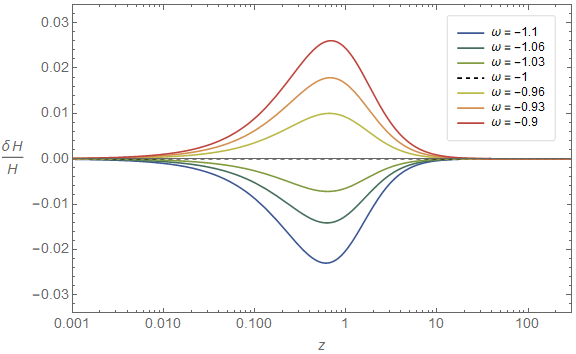}
\caption{\small The deviation from a $\Lambda$CDM expansion history $\delta H(z)/H(z)$ for a Hubble function of the form in Eq.~\eqref{eq:HubbleModModel} for a dark energy model with constant equation of state $w$ given in Eq.~\eqref{eq:OmegaConstw} for different values of the equation of state.}
\label{fig:PlotDers}
\end{figure}
Through Eq.~\eqref{eq:rhoDEz_w} we can now relate variation of the Hubble function to the equation of state of dark energy $w(z)$
	\begin{equation}\label{eq:Omegaz_w}
		\delta \Omega(z) = \Omega_\Lambda \left\{\exp\left(3\int^z_0\big(1+w(z)\big)\frac{d z}{1+z}\right)-1\right\}\, .
	\end{equation}
 and hence
 \begin{equation}\label{eq:Omegaz_wSecond}
		\boxed{\frac{\delta H(z)}{H(z)}=  \frac{H_0^2\Omega_\Lambda}{2H^2(z)} \left\{\exp\left(3\int^z_0\big(1+w(z)\big)\frac{d z}{1+z}\right)-1\right\}\, .}
	\end{equation}

To get an intuition, consider first for simplicity a constant equation of state, for which
 \begin{equation}\label{eq:OmegaConstw}
		\delta \Omega(z) = \Omega_\Lambda \left\{(1+z)^{3(1+w)}-1\right\}\,,
	\end{equation}
 and hence the sign of $1+w$ is directly correlated with the sign of $\delta\Omega$ and thus of $\delta H$. The associated deviations of the expansion history for a constant DE equation of state $w$ compared to the CC value are plotted in Fig.~\ref{fig:PlotDers}. At each $z$ at very late-times in DE domination, $H(z)$ satisfying $\dot H>0$ lies below the $\Lambda$CDM value of a smaller accelerated expansion with $\dot H = \text{const.}$ because $H_0$ is kept fixed. However, note that indeed, the deviations are well confined within the interval $0<z<300$, assumed in Sec.~\ref{ssSec:dH}. Projecting back in time, eventually, matter and radiation dominate again and the expansion resumes to be a $\Lambda$CDM expansion. This is in contrast to what one would obtain if instead of modifying the equation of state one would directly change the cosmological parameter $\Omega_\Lambda$ which would affect the entire expansion history.

 Now for a more general equation of state $w(z)$ one can still conclude that the necessary condition for solving the Hubble tension derived in Sec.~\ref{ssSec:dH}, which reads $\delta H(z_1)<0$ at some $z_1$, directly requires $w(z_2)<-1$ at some $z_2$, not necessarily equal to $z_1$ though. In other words, in order to have the slightest chance of solving the Hubble tension, a phantom equation of state is required.

\paragraph{Dynamical Dark Energy and the $\sigma_8$ Tension.}

On the other hand, as derived in Sec.~\ref{ssSec:dH}, the strictly negative response function $\mathcal{R}_{\sigma_8}$ implies that alleviating the $\sigma_8$ tension at least requires $\delta H(z)>0$ at some time $z$. This again directly translates to a condition on DE equation of state of the form $w(z)>-1$ for some $z$, which is again exactly opposite to what a solution of the Hubble tension requires. This result can also be understood intuitively, since a phantom-like evolution of dark energy in general extends the matter-dominated phase and therefore boosts the matter growth. This is exactly opposite to what is required to alleviate the $\sigma_8$ tension. A coherent DE model that solves the Hubble tension but also simultaneously addresses the $\sigma_8$ therefore requires $\delta H(z)$ as well as $1+w(z)$ to change sign. Such a crossing of the $\Lambda$CDM value is known as crossing of the \textit{phantom divide}. 

\begin{figure}[H]
\centering
\includegraphics[scale=0.5]{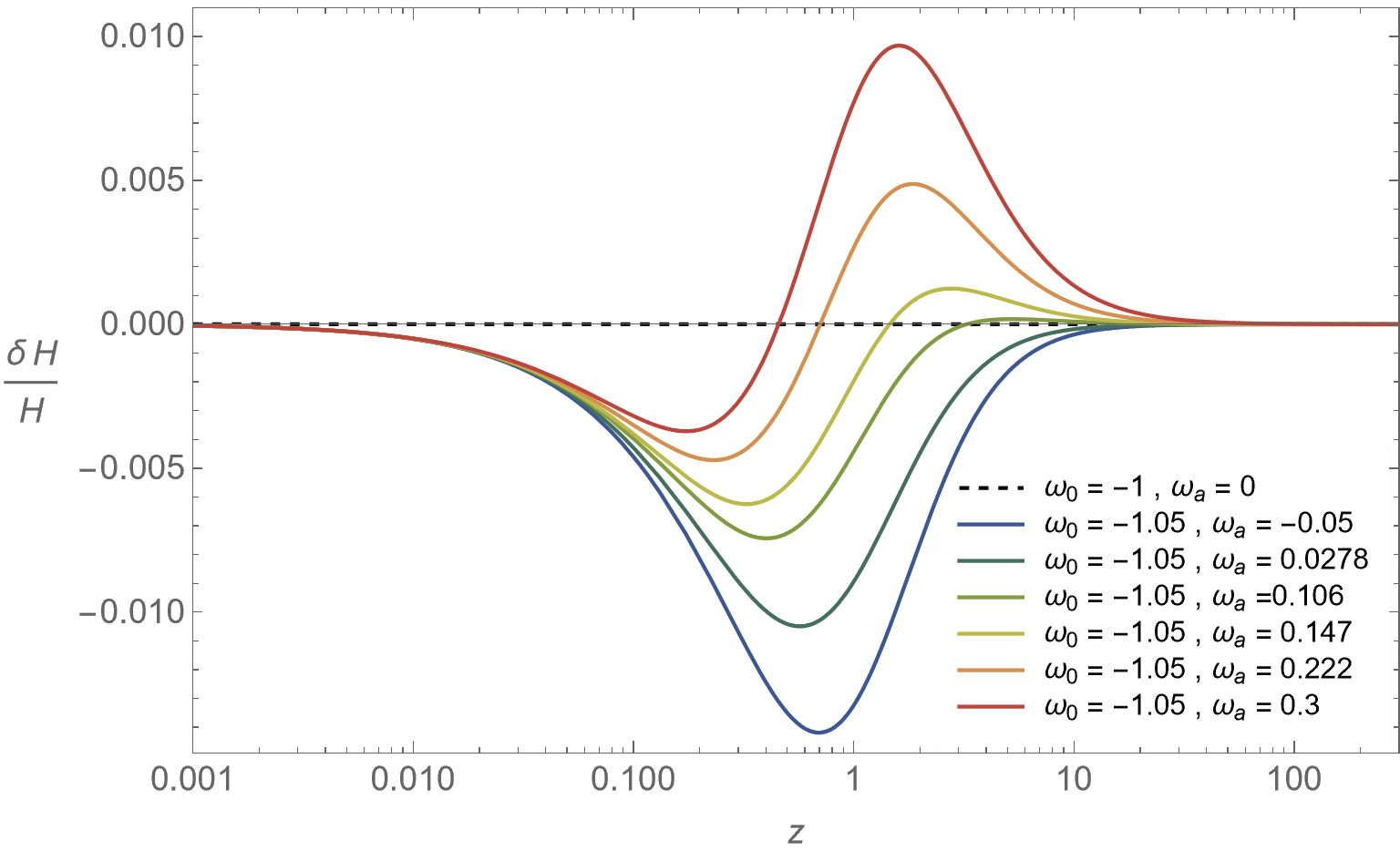}
\caption{\small The deviation from a $\Lambda$CDM expansion history $\delta H(z)/H(z)$ for a Hubble function of the form in Eq.~\eqref{eq:HubbleModModel} for a dark energy model with a CPL \cite{Chevallier:2000qy, Linder:2002et} a parameterization $w=w_0 + w_a(1-a)$ of the equation of state for different values of the parameters. We choose a fixed parameter $w_0=-1.05$ to obtain a deformation $\delta H(z)$ that changes sign at late times.}
\label{fig:PlotDersGen}
\end{figure}

As a simple example of a DE model that includes enough freedom to model a crossing of $w=-1$ we consider a so-called CPL \cite{Chevallier:2000qy, Linder:2002et} parameterization
\begin{equation}
    w(a)=w_0 + w_a(1-a)\,.
\end{equation}
Choosing for concreteness a fixed parameter $w_0=-1.05$, one can indeed model a phantom crossing and hence a deformation $\delta H(z)$ that changes sign at late times as shown in Fig.~\ref{fig:PlotDersGen}. Given such an explicit model with a known functional form of $\delta H(z)$ our method introduced in Secs.~\ref{sSec:ModelIndepMethod} and \ref{sSec:H0 and sigma8 Constraints on LateTimeSolutions} above,  allows the direct computation of the induced shifts in the cosmological observables $H_0$ and $\sigma_8$. This is achieved by simply plugging the departure from the $\Lambda$CDM expansion history $\delta H(z)$ into the response formulas in Eqs.~\eqref{eq:RespGenH0} and \eqref{eq:RespGensigma8} of $H_0$ and $\sigma_8$ respectively. This simple example is also an opportunity to test our analytical approach against a numerical calculation within {\fontfamily{qcr}\selectfont class}, keeping fixed the acoustic scale $\theta^*_A$ and $\omega_m$. As shown in Table~\ref{tab:w0wa} the analytic results show a very satisfactory overall performance. In particular, we want to point out the case with $w_a=0.174$, for which the shift in $H_0$ and $\sigma_8$ indeed both have the right sign. On a quantitative level, however, it seems that our example of a CPL equation of state is not able to provide a big enough shift in the parameters to fully resolve the tensions.

 \begin{table}    
\begin{center}
        \begin{tabular}{r|r@{\hskip 12pt}rcr@{\hskip 12pt}r}
            				         & \multicolumn{2}{c}{$100\times\delta h/h$} 
                                     & 
                                     & \multicolumn{2}{c}{$100\times \Delta \sigma_8/\sigma_8$}\\[4pt]
                              $w_a$  & {\fontfamily{qcr}\selectfont class}
                                     & Analytical
                                     & \phantom{aaa}
                                     & {\fontfamily{qcr}\selectfont class}
                                     & Analytical\\
	    	\hline\hline
	    	$-0.05   $&$   2.81   $&$    2.61    $&&$    2.24    $&$     2.08 $\\
			$ 0.03   $&$   1.89   $&$    1.80    $&&$    1.47    $&$     1.39 $\\
			$ 0.11   $&$   0.946  $&$    0.93    $&&$    0.66    $&$     0.64 $\\
			$ 0.174  $&$   0.095  $&$    0.093   $&&$   -0.089   $&$    -0.092$\\
			$ 0.22   $&$  -0.52   $&$   -0.53    $&&$   -0.64    $&$    -0.65 $\\
			$ 0.3    $&$  -1.54   $&$   -1.62    $&&$   -1.56    $&$    -1.64 $
	    \end{tabular}
\end{center}
	    \caption{Comparison of our analytical results with the full computation
	    	in {\fontfamily{qcr}\selectfont class}, keeping fixed $\theta_*$ and $\omega_m$, for a
	    	dark energy model with an equation of state $w(a)=-1.05+w_a(1-a)$. Notice
	    	that in this case $\delta H(z)$ changes sign, as shown in the right pannel in
	    	Fig.~\ref{fig:PlotDersGen}. Even though they are too small to relieve the tensions, 
	    	this example shows that when $\delta H(z)$ changes sign it is possible to 
	    	increase $h$ while reducing $\sigma_8$.}
	    \label{tab:w0wa}
	\end{table}

\paragraph{Implication of a Phantom Equation of State.}
Within a phenomenological description of a dark energy background fluid through its $\rho$ and $p$, a phantom equation of state $w <-1$ may appear as simply exploring a further range of parameter space. Yet, phantom energy as the name suggests may come with rather unusual underlying physics. For instance, together with the cosmological constraint of $\rho>0$ on the energy density of a homogeneous and isotropic energy-momentum tensor Eq.~\eqref{eq:EnergyMomentumHomogeneousIsotropic}, all so-called classical or \textit{pointwise energy conditions} on the energy momentum tensor exclude the phantom regime, hence imply $w\geq -1$ \cite{Carroll:2003st,Kontou:2020bta}. Moreover, note that for $w<-1$ the energy density increases in an expanding universe. Because of this behavior, a constant phantom equation of state exhibits a future curvature singularity at finite time where the scale factor diverges, leading to a so-called \textit{phantom energy disaster} \cite{Carroll:2003st,Weinberg2008Cosmology}. It is interesting to consider what a phantom equation of state entails on more realistic effective models describing dynamical dark energy. As we will see, the necessity of phantom energy could even point towards a departure from GR on cosmological scales.

\paragraph{Dynamical Dark Energy and Additional Degrees of Freedom.} 

The most straightforward way to model a dynamical dark energy fluid is by introducing new degrees of freedom that modify the Einstein equations to allow for de-Sitter like solutions without the introduction of a cosmological constant (see e.g. \cite{Copeland:2006wr,Durrer:2007re,Durrer:2008in,Tsujikawa:2010zza,Li:2012dt,Mortonson:2013zfa,Joyce:2016vqv,DeFelice:2016yws,Bahamonde:2017ize,Brax:2017idh,Tawfik:2019dda,Frusciante:2019xia} for a topic specific review). Yet, the additional fields driving the accelerated expansion do not necessarily need to be associated to a non-minimal field introduced in Sec.~\ref{sSec:Metric Theories} intrinsic to the gravity sector.

Indeed, the simplest example for a concrete model representing an additional degree of freedom that drives an accelerated expansion, also in the case of inflation, is to consider a minimally coupled scalar field with a potential energy called \textit{quintessence} with action (see e.g. \cite{Weinberg2008Cosmology})
\begin{equation}\label{ActionQuintessence}
    S^{\myst{Quint}}=\frac{1}{2\kappa_0}\int \dd^4 x\sqrt{-g}\,R-\int \dd^4 x\sqrt{-g}\left(\frac{1}{2}g^{\mu\nu}\nabla_\mu\Phi\nabla_\nu\Phi+V(\Phi)\right)\,.
\end{equation}
The associated covariantly conserved energy momentum tensor of the scalar field reads
\begin{equation}
    T^\Phi_{\mu\nu}=\nabla_\mu\Phi\nabla_\nu\Phi-g_{\mu\nu}\left(\frac{1}{2}\nabla_\mu\Phi\nabla^\mu\Phi+V(\Phi)\right)\,.
\end{equation}
On an FLRW background on which the scalar field is only allowed to carry a time dependence, the energy density and pressure of the scalar field read
\begin{align}
    \rho_\Phi=\frac{1}{2}\dot\Phi^2+V\,,\\
    p_\Phi=\frac{1}{2}\dot\Phi^2-V\,,
\end{align}
and thus the associated equation of state becomes
\begin{equation}\label{eq:OmegaQuint}
    w=\frac{\frac{1}{2}\dot\Phi^2+V}{\frac{1}{2}\dot\Phi^2-V}=-1+\frac{\dot\Phi^2}{V}+\mathcal{O}\left(\dot\Phi^4\right)\,,
\end{equation}
where we expanded for $\dot\Phi^2\ll V$. Moreover, the covariant energy conservation equation [Eq.~\eqref{eq:ConservationOfMatterCosmologySecondDE}] becomes
\begin{equation}
    \ddot\Phi+3H\dot\Phi+V'(\Phi)=0\,.
\end{equation}
The shape and size of the potential can then be arranged to obtain the desired dominant late time behavior. Here, we assume that $V>0$, in order to obtain a solution close to de-Sitter space of a positive cosmological constant. Hence, quintessence is not phantom since its equation of state satisfies
\begin{equation}
    w \geq -1\,.
\end{equation}

More sophisticated models of dynamical DE are therefore required to describe a phantom equation of state necessary for the resolution of the $H_0$ tension. However, in the following, we want to discuss a series of arguments of why it could be more natural to promote the new DE degrees of freedom to non-minimal fields in metric theories beyond GR. This will smoothly lead to a first assessment of the implications of our general results in Sec.~\ref{sSec:H0 and sigma8 Constraints on LateTimeSolutions}.

\paragraph{Reasons to consider Beyond GR Theories.} 

While as mentioned, the additional propagating DOFs driving a beyond $\Lambda$CDM late time acceleration with possible solutions to the $H_0$ tension do not necessarily need to be associated to a non-minimal field, no known matter field can drive the accelerated expansion. In this sense, the introduction of extra (in this case matter) degrees of freedom seems rather ad-hoc. 
The question whether such additional DOFs could instead be incorporated in metric theories of gravity as alternatives to GR seems to reside on valid grounds. Recall that for cosmological purposes it is accurate to talk about alternative theories of gravity, as one requires large departures from GR on cosmological scales such that mere perturbations on top of GR are not sufficient.
Indeed, as discussed in detail, new degrees of freedom that mainly interact gravitationally, very naturally arise in beyond GR theories such that it could be regarded as reasonable to consider alternative theories of gravity introduced in Sec.~\ref{ssSec: A Exact Theories} to represent an underlying mechanism for beyond $\Lambda$CDM models. Associated is of course the hope to eventually come across a resolution of the more fundamental questions within the theories of gravity, including the CC problem. 

Moreover, an additional compelling argument for beyond GR theories could be the fact that the cosmological tensions calling for a departure of $\Lambda$CDM actually prefer a more negative equation of state $w \leq -1$, implying $\dot H>0$, opposite to what quintessence like models are able to cover. 
Equation~\eqref{eq:OmegaQuint} indicates that for the minimal scalar theory considered above a phantom equation of state with $V>0$ would require a wrong sign of the kinetic term, resulting in an untenable Ostrogradsky kinetic instability (recall Sec.~\ref{sSec:OstrogradskyTheorem}) at the level of the interactions that are present if $w\neq -1$. This instability is related to the fact that in a cosmological context with $\rho>0$ all the various pointwise energy-conditions \cite{Kontou:2020bta} on the energy momentum tensor of the associated fluid imply that $w\geq -1$. While this does not per se guarantee that generically a phantom equation of state implies perturbative instabilities, also because $p<-\rho$ does not directly imply $\delta p<-\rho$ at the level of the perturbations, a violation of the energy conditions is generically not desirable, since for most explicit examples of low energy scalar theories a violation entails an instability of the theory \cite{Dubovsky:2005xd,Carroll:2003st,Ludwick:2017tox}.

Other than assuming nontrivial interactions between dark energy and dark matter, the key for stable periods of phantom behavior is in fact the introduction of higher-order derivative self interactions that become important at higher energy scales \cite{Nicolis:2009qm,Creminelli:2010ba,DeFelice:2016yws,Kobayashi:2019hrl}. And such higher-order derivative self indications precisely arise in beyond GR theories such as Horndeski theories or Generalized Proca discussed in Sec.~\ref{ssSec: A Exact Theories}. In other words, beyond GR theories are able to accommodate violations of the energy conditions without immediately being ruled out by the presence of perturbative instabilities. In fact, non-minimal fields per definition do not possess a well-defined covariantly conserved energy-momentum tensor in the first place. Thus, it could be that the requirement of energy conservation of minimally coupled matter fields is too restrictive in order to describe the observed dynamics of the late time expansion, indicating that indeed a non-minimal coupling might be required.


\paragraph{Constraining Metric Theories Beyond GR.} 

In this context, the general guiding principles for a consistent resolution of both the $H_0$ and $\sigma_8$ tensions in Sec.~\ref{sSec:H0 and sigma8 Constraints on LateTimeSolutions} can be used to discuss first implications on the theory space beyond GR.

First of all, recall that when concentrating on the pure background modifications of a resulting cosmological model, we were able to conclude that solving both the $H_0$ and $\sigma_8$ simultaneously inevitably requires a change of sign of $\delta H(z)$ translating into a crossing of the phantom divide $w=-1$. It turns out, however, that while a pure phantom equation of state could be consistently described through the introduction of non-linear interactions as mentioned above, a consistent phantom crossing seems even harder to obtain. At least several concrete examples indicate that instabilities and divergences may occur in the transition towards and out of the phantom phase, and a consistent crossing might only be obtained with the aid of multiple scalar degrees of freedom in the effective fluid (see \cite{Kobayashi:2019hrl,Li:2010hm} and references therein). Taking this remark seriously would mean that a consistent resolution of the cosmological tensions would require a multi-scalar generalization of the metric theories introduced in Sec.~\ref{ssSec: A Exact Theories}.

On the other hand, single scalar field metric theories such as Horndeski theory [Eq.~\eqref{eq:ActionHorndeski}] or generalized Proca [Eq.~\eqref{eq:ActionGenProca}] may still address both tensions if modifications on the perturbative level are able to compensate for the worsening of the $\sigma_8$ tension induced through a pure phantom effective equation of state. However, as concrete studies show for instance in the case of a generalized Proca model \cite{DeFelice:2020sdq,Heisenberg:2020xak}, adding the perturbation level typically does not provide the necessary freedom to consistently solve both tensions. In the contrary, it is predominately observed that perturbations of the late DE models increase the preferred value of $\sigma_8$ even more. In our approach, such statements could be made without a full-fledged Boltzmann code based computation, simply through our considerations in Sec.~\ref{sSec:dG} above. Namely, it is known that luminal Horndeski and GP models [Eqs.~\eqref{eq:Hsurv} and \eqref{eq:GenProcaLuminal}] considered in Sec.~\ref{ssSec:Speed Of Gravity Beyond GR} with a phantom equation of state do not have the freedom at the perturbation level to actually reduce the effective gravitational constant \cite{Amendola:2017orw}.\footnote{This is at least true in the linear sub-Hubble regime within the quasi-static approximation.} While already at the intuitive level one can conclude that therefore such models are not able to address both the $H_0$ and $\sigma_8$ tensions simultaneously, such a reasoning can be put on firm footing through the constraint in Eq.~\eqref{eq:G_eff_bound}. In fact, such a behavior of increasing the gravitational coupling instead of decreasing it is generally expected for any model exhibiting DE clustering of additional dark energy perturbation components.

Observe that in the above argument we used the additional powerful constraint on the gravitational wave speed discussed in Sec.~\ref{ssSec:PropagationSpeedConstraints}, which precisely highly constrains higher derivative theories necessary for a stable phantom energy behavior. In general, since the results derived in Sec.~\ref{sSec:H0 and sigma8 Constraints on LateTimeSolutions} only rest upon a very minimal set of observable constraints, they can be combined with other existing constraints to further the ability of theoretical guidance. In particular, the surviving Horndeski and GP models with luminal propagation also tend to fall short due to stringent constraints on the sign of the ISW effect \cite{Renk:2017rzu,Noller:2018wyv,Nakamura:2018oyy}. Moreover, BAO measurements providing a second handle on the sound horizon scale at low redshift generally put tight constraints on late-time $\Lambda$CDM modifications \cite{Benevento:2020fev,Jedamzik:2020zmd}. An associated combined study with our work will however be left for the future.

\section{Summary and Outlook}

The methodology developed in this work allowed us to identify a broad class of alternative cosmological models that can \textit{not} solve both the $H_0$ and $\sigma_8$ tensions. Focusing on late DE models, with general equation of state $w(z)$ governing the background evolution and a modified gravitational coupling $G_\text{eff}=G+\delta G(z)$ at the level of perturbations, we derived a set of necessary conditions that must be met to simultaneously alleviate both the $H_0$ and the $\sigma_8$ tensions, i.e. $\delta H_0>0$ and $\Delta\sigma_8<0$, namely
\begin{enumerate}[i)]
	\item Solving the $H_0$ tension $\Rightarrow$ $\delta H(z)<0$ at some $z$ (DE: $w(z)<-1$).
	\item If $G_\text{eff}=G_0$:\\[3pt]
		Solving the $H_0$ and $\sigma_8$ tensions $\Rightarrow$ $\delta H(z)$ changes sign at some $z$ \\(DE: $w(z)$ crosses the phantom divide $w=-1$).
	\item If $G_\text{eff}=G_0+\delta G$ and $\delta H<0$ (DE: $w(z)<-1$):\\[3pt]
		Solving the $H_0$ and $\sigma_8$ tensions $\Rightarrow$
		$\displaystyle \frac{\delta G}{G}<\alpha(z)\frac{\delta H}{H}<0$ at some $z$, with $\alpha(z)>0$.		
\end{enumerate}

The necessary conditions above represent the first results obtained with the general method presented in Sec.~\ref{sSec:ModelIndepMethod}. Quite generally, a first conclusion from the above discussion could be that despite the large amount of proposed beyond $\Lambda$CDM models proposed to solve the current cosmological tensions, luckily in a sense, it is actually not at all trivial to do so consistently. In this context, the necessary conditions derived here can be viewed as guiding principles towards the direction of a viable extension of the current standard cosmology.

In the future, it would be of great interest to apply similar considerations to early-time solutions to the Hubble tension as well. Such a study would however require the implementation of the methodology to a linear Boltzmann solver such as {\fontfamily{qcr}\selectfont class} \cite{lesgourgues2011cosmic,Blas:2011rf}, due to the lack of analytic formulas.
In this context, a new period of accelerated expansion before decoupling due to dynamical dark energy can also in principle address the Hubble tension by increasing $\delta H$ and therefore reducing the sound horizon (see e.g. \cite{Kamionkowski:2022pkx,Poulin:2023lkg} for a review). While so-called \textit{early dark energy} models lack the motivation in connection to the CC problem, they are still rather well embedded in beyond GR theories and can in particular be expected to arise within string theory. However, also early time solutions generically seem to fall short when probed against additional cosmological data and in particular also generically worsen the $\sigma_8$ tension \cite{Jedamzik:2020zmd,Hill:2020osr} with the potential of providing hints towards building successful models beyond $\Lambda$CDM.


\part{Theoretical Consistency at the Quantum Level}\label{Part: Quantum Gravity}

\small
\noindent
\emph{\ul{Personal Contribution and References}}\\ 
\footnotesize 
\textit{Chapter~\ref{Sec.Quantum Stability} is based on \textbf{L. Heisenberg, J. Noller, J. Zosso, 2020} \cite{Heisenberg:2020cyi}, \textbf{L. Heisenberg, J. Zosso, 2021} \cite{Heisenberg:2020jtr} and \textbf{C. de Rham, L. Heisenberg, A. Kumar, J. Zosso, 2022} \cite{deRham:2021yhr}. Parts of the following treatment are also inspired from \cite{WaldBook,Feynman:1996kb,Weinberg:1995mt,Burgess:2003jk,maggiore2008gravitational,zee_quantum_2010,Donoghue:2012zc,zee2013einstein,carroll2019spacetime}.}
\normalsize

\vspace{5mm}

\noindent
\textbf{Summary of Part IV}\\ 
\noindent
While large portions of the quantum realm of gravity theories remains a mystery, the first contacts between the two worlds still represents a valuable ground for performing internal consistency tests on the purely theoretical level. In particular, in the case of general relativity there exists a well established quantum effective field theory formulation that not only identifies GR as the unique theory of massless a spin 2 graviton, but also serves as a computational tool in particular based on powerful amplitude methods. It can therefore be argued that quantum consistency within an effective field theory approach represents an additional viability criteria for alternative theories of gravity that might have a purely theory based constraining power.

Starting with a concise review of the quantum EFT of GR, we introduce the notion of quantum stability that we will subsequently analyze for selected models of metric theories of gravity of the Horndeski class. In doing so, we correct previous beliefs by giving pertinent dimensional arguments backed up through extensively double-checked explicit one-loop computations. Finally, based on the Isaacson approach to describing dynamical degrees of freedom of metric theories, we will close with a speculation on a possible alternative approach to quantization. 


\chapter{GR and Quantum Physics}\label{Sec:Gravity and Quantum Physics}

Formulating a consistent theory of \textit{quantum gravity} is a long-standing and major problem of theoretical physics. The unification of today's main pillars of fundamental physics, quantum field theory and gravity theory in the form of GR, has proven to be a most refractory problem, withstanding decades of research (see e.g. \cite{Rayski:1978jda,WaldBook,Isham:1992ms,Kiefer:2004xyv,Rickles:2006ee,Strominger:2009aj,Anderson:2010xm,Lindesay:2013iba,zee2013einstein,Ashtekar:2014ife,Giddings:2022jda}). The issue even already begins at the question of what is meant by a ``quantum theory of gravity'' and whether it is indeed necessary to find a quantum generalization of the gravitational interaction. 

In any case, there exists a list of unsolved conceptual problems that need to be addressed in one way or the other. As concerns the current description of gravity, the mere fact that matter, as a source of the gravitational field, seems to behave quantum mechanically leads to a general expectation for the necessity of a unified description. In particular, the quantum mechanical superposition principle for matter that sources a gravitational field can be used to argue in favor for a quantum behavior of the gravitational field \cite{DeWitt:1957obj,DiMauro:2021mcu}. Moreover, also the singularity theorems of GR \cite{Penrose:1964wq,Hawking:1967ju} indicating a breakdown of the current gravity theory in understanding the very early universe or the final stages of black-hole evolution are read as hints for the necessity of a high-energy quantum generalization \cite{Kiefer:2004xyv}. On the other hand, also quantum field theory will eventually need to deal with gravity, at the very least at the Planck scale, where a concentration of energy sufficiently localized is expected to collapse into a black hole \cite{zee2013einstein}. Even more fundamentally, the notion of a manifold at the basis of a description of spacetime seems to unavoidably require a quantum mechanical update. Indeed, if viewed as a collection of ``events'', as discussed at the beginning of Sec.~\ref{Part: EFT of Gravity}, a spacetime defined through quantum events can fundamentally only be resolved up to quantum uncertainties. 

Certainly, the lack of empirical probes in the regime of quantum gravity represents one of the major reasons for the current lack of a unifying framework. But also deep unsolved conceptual issues arise that challenge our very understanding of the interpretation of physical experiments, indicating that a theory of quantum gravity would require a radical change in the mathematical and conceptual framework.\footnote{Note that in the light of the so called ``measurement problem'' (see e.g. \cite{Hodgson:1993,Schlosshauer:2003zy}), this is partially already the case for quantum mechanics alone.} For instance, the current understanding of quantum field theory crucially relies on the existence of a classical spacetime, providing a notion of time and space on which local observables can be defined, together with a sense of causality. Within classical metric theories of gravity, the physical metric precisely provides such a notion of a universal spacetime that can be probed independently of an observer or a particular measurement device. Yet, as soon as one naively transfers the fundamental necessity of quantum fluctuations to the gravitational field itself, it immediately invokes a series of theoretical issues that are at present seem unclear on how to resolve (see also \cite{Isham:1992ms,Kiefer:2004xyv,Anderson:2010xm,zee2013einstein}).

Having said that, this does not mean that no progress has been made in considering an overlap between gravity and the quantum world. Leaving aside the attempts of formulating concrete theories of quantum gravity \cite{Green:1987sp,Green:1987mn,Polchinski:1998rq,Polchinski:1998rr,Weinberg:2000cr,Zwiebach:2004tj,Mukhi:2011zz,Rovelli:1997yv,Gambini:2011zz,Ashtekar:2017yom,Ashtekar:2021kfp} and conjectured transdimensional relationships between the two worlds \cite{Maldacena:1997re,Polchinski:2010hw,Hubeny:2014bla,Penedones:2016voo} that exerted a substantial influence on today's physics practice, there are two distinct areas where the first approaches of the gravitational and quantum worlds can be explored very humbly by still assuming a well-defined notion of background spacetime.

On the one hand, one can consider QFTs of matter fields on a fixed curved background spacetime, thus investigating the non-localizable effects beyond the local Minkowski approximation (see e.g. \cite{misner_gravitation_1973,Birrell:1982ix,WaldBook,Fulling:1989nb,Wald:1995yp,Ford:1997hb,Mukhanov:2007zz,Parker:2009uva,carroll2019spacetime} for a review). This approach in particular famously lead to Hawking's result of radiating black holes that is still the subject of extended debates on black hole entropy and black hole evaporation 
\cite{Bekenstein:1972tm,Hawking:1974rv,Hawking:1976ra,Unruh:1976db,Page:1993wv,Jacobson:1993vj,Wald:1993nt,Bekenstein:1994bc,Strominger:1994tn,Wald:1995yp,Lowe:1999pk,Ryu:2006bv,Page:2013dx,zee2013einstein,Polchinski:2016hrw,Wallace:2017wzs,Wallace:2017yfi,Wallace:2017tfa,Wall:2018ydq,Raju:2020smc,Penington:2019npb,Almheiri:2019psf,Almheiri:2019hni,Almheiri:2019qdq,Almheiri:2020cfm,Renner:2021qbe}.

On the other, within the framework of quantum effective field theories \cite{Weinberg:1978kz,Gasser:1983yg,Gasser:1984gg,Polchinski:1992ed,Georgi:1993hh,Weinberg:1995mt,Burgess:2007pt, Davidson:2020gsx} already mentioned back in Sec.~\ref{sSec:Why and how go beyond GR} it is possible to formulate a low energy quantum theory of gravity by viewing the gravitational field as an ordinary massless and gauge symmetric spin 2 field on Minkowski spacetime \cite{Feynman:1963ax,Weinberg:1964ew,Weinberg:1965rz,DeWitt:1967ub,Deser:1969wk,BOULWARE1975,PhysRev.96.1683,tHooft:1974toh,Donoghue:1993eb,Donoghue:1994dn,Dunbar:1994bn,Donoghue:1995cz,Feynman:1996kb,Weinberg:1995mt,Bjerrum-Bohr:2002gqz,Khriplovich:2002bt,Burgess:2003jk,maggiore2008gravitational,zee_quantum_2010,Donoghue:2012zc,zee2013einstein,PetrovKopeikinLompayTekin+2017}. It is this approach that we now want to analyze in more detail.


\section{Quantization of GR as a Field Theory}\label{sSec:Quantization of Gravity}

The standard perturbative quantization procedures that are applicable on relativistic field theories, in particular the standard canonical quantization procedure, crucially rely on the existence of a classical background spacetime that provides a well-defined reference system \cite{Dirac:1925jy,Dirac:1927dy,Schwinger:1951xk,Dirac:1958xyv,Weinberg:1995mt,Srednicki:2007qs,Folland:2008zz,Schwartz:2014sze}. This is precisely what prevents a straightforward application of the known quantization schemes to gravity, as it still appears to be unclear how to deal with a quantized reference system (see however e.g. \cite{Aharonov:1984zz,Toller:1996ki,Poulin:2006ryq,Giacomini:2017zju,Frauchiger_2018}. Thus, a possible approach to quantizing a gravitational field with the currently well understood methods is to consider an ordinary relativistic field defined on a Minkowski spacetime and try to describe the gravitational interaction with the same QFT formalism that worked so well for all other fundamental forces of nature that are mediated by the exchange of bosonic field excitations. From this point of view, there would be no fundamental difference between the gravitational field and the force carrier fields of the standard model of particle physics. Or in the words of Feynman: let's just assume ``that gravitation is a new field, number 31'' \cite{Feynman:1996kb}.

\paragraph{GR as a Field Theory.} Postulating therefore a fixed reference Minkowski spacetime, one can try to formulate a consistent QFT, or more precisely a quantum EFT, of gravity. The first question is, what the integer spin of the mediating bosonic field should be that couples to the energy-momentum tensor $T_{\mu\nu}$ of matter fields. First, the long range nature of the gravitational force indicates that the field should be (at least close to) massless. Moreover, while spins $\geq 3$ are ruled out from the start (see e.g. \cite{maggiore2008gravitational,Schwartz:2014sze}), also a spin 1 field is quickly rejected, simply due to the fact that gravity is never repulsive.\footnote{Furthermore, at first order it would actually even already be impossible to consistently couple a massless gauge invariant vector field to a symmetric energy-momentum tensor \cite{maggiore2008gravitational}.} And while a coupling between a scalar field and the trace of $T_{\mu\nu}$ is perfectly consistent, a spin 0 boson is ruled out by the experimental fact that electromagnetic waves, described by a traceless energy-momentum tensor, also feel the gravitational force. In conclusion, gravity should be mediated by perturbations, the \textit{gravitons}, of a bosonic, massless spin 2 field (see also \cite{Feynman:1996kb,maggiore2008gravitational,zee_quantum_2010}).

Under this assumption, and the fact that on a Lorentz symmetric background a massless spin 2 field can only be described by a two-index, symmetric Lorentz field $h_{\mu\nu}$ if the theory comes with a gauge symmetry of the form
\begin{equation}\label{eq:LinearGaugeSymmTrans}
    h_{\mu\nu}\rightarrow h_{\mu\nu}+2\partial_{(\mu}\xi_{\nu)}\,,
\end{equation}
actually completely determines the form of the free second order action to be (see e.g. \cite{Feynman:1996kb,maggiore2008gravitational,zee_quantum_2010})
\begin{align}\label{ActionGR2nd Part IV}
    _{\mys{(2)}}S^{\myst{GR}}_G=\frac{-M_\text{P}^2}{8}\int d^4x\,\Big[\partial_\mu h_{\alpha\beta}\partial^\mu h^{\alpha\beta}-\partial_\mu h^t\partial^\mu h^t+2\partial_\mu h^{\mu\nu}\partial_\nu h^t -2\partial_\mu h^{\mu\nu}\partial_\alpha h\ud{\alpha}{\nu}\Big]\,,
\end{align}
Here, we have chosen the dimensionality factor
\begin{equation}
    M_\text{P}=\frac{1}{\kappa_0}=\frac{1}{8\pi G}
\end{equation}
to match the expression of the leading order perturbative action in GR in Eq.~\eqref{ActionGR2nd}.\footnote{Note also that we momentarily still stick here to our convention of treating every field as dimensionless.} Of course, the fact that we recover the perturbative linearized action of GR is already a first hint that we are on the right track. Note however the crucial difference between the statements here in Eq.~\eqref{ActionGR2nd Part IV} and the action in Eq.~\eqref{ActionGR2nd}. Back in Sec.~\ref{ssSec:Local Wave Equation in GR} we were working in perturbation theory around a known solution to the Einstein equations and split every field content into a clearly separable high- and low-frequency part, where $h_{\mu\nu}$ denoted the high-frequency metric perturbation. This Isaacson split was what allowed us to locally consider a chart in which the low-frequency metric reduced to the Minkowski form. Here, on the other hand, Eq.~\eqref{ActionGR2nd Part IV} represents the action of a free spin 2 field $h_{\mu\nu}$ on a Minkowski metric in a global Minkowski chart.

\paragraph{Quantization and Non-Linearities.} This free field theory can then be quantized, under the usual subtleties of a gauge invariant field. Indeed, in a QFT, it is decisive to only quantize the physical propagating degrees of freedom of a theory. 
At the linear level it would be sufficient to introduce an additional gauge fixing term, that would simultaneously also allow determining the graviton propagator (see e.g. \cite{maggiore2008gravitational,zee_quantum_2010}). However, in a more rigorous treatment the consistent quantization of the full field theory of GR \cite{Feynman:1963ax,DeWitt:1967ub,tHooft:1974toh,Donoghue:1995cz} requires the methods of Faddeev and Popov \cite{Faddeev:1967fc} (see also \cite{WaldBook,Peskin:1995ev,Weinberg:1996kr,Srednicki:2007qs,zee_quantum_2010,Schwartz:2014sze,Flory:2012nk}).
This is because the theory of a spin 2 graviton describing the gravitational force is only consistent as a non-linear theory. In the following, we will for completeness offer a short summary on such a bottom-up approach on the gravitational qEFT action. 

So far we did not explicitly introduce any interactions, but only considered a second order action of the gravity field that does not talk to the action of pure matter fields
\begin{equation}
    S=\phantom{}_{\mys{(2)}}S^{\myst{GR}}_G[h]+ S^{\myst{pure}}_\text{m}[\Psi_m]\,.
\end{equation}
In other words, the matter action currently lacks any gravity and to describe it, the spin 2 field should of course couple to matter fields. More precisely, the graviton should couple to the total energy momentum tensor of matter fields $T_{\mu\nu}$, that in a Minkowski chart satisfies the on-shell conservation law (recall Secs.~\ref{sSec: Theories of Minkowski Spacetime} and \ref{sSec:Covariant Consrevation})
\begin{equation}\label{eq:ConservationEq}
    \partial_\mu T^{\mu\nu}=0\,.
\end{equation}
This energy momentum tensor serves as a source of the graviton field in the equations of motion by introducing the obvious coupling term to matter
\begin{equation}\label{eq:Interaction term}
    S^{\myst{int}}\sim \int d^4 x \,h_{\mu\nu} T^{\mu\nu}\,.
\end{equation}
Note however, that introducing this term, hence an interaction of matter fields with the new gravity field, also has implications on the action of the matter fields and in particular on the conservation equation in Eq.~\eqref{eq:ConservationEq}, since the conservation of $T_{\mu\nu}$ crucially depends on the equations of motion of the matter fields.\footnote{In other words, by including the interaction term in Eq.~\eqref{eq:Interaction term}, the matter energy-momentum is no longer conserved as the system of matter fields can lose energy to the new field.} Indeed, through Noethers' theorem (see Appendix.~\ref{sApp: Noethers Theorem}) the invariance of the background spacetime under time translations only ensures that the total energy-momentum tensor of the system, including the one of the new field $h_{\mu\nu}$ is conserved. 

Thus, consistency requires that the spin 2 field $h_{\mu\nu}$ that itself carries energy and momentum inevitably needs to represent a non-linear field theory. On the other hand, introducing non-linear terms inevitably also requires a generalization of the gauge symmetry transformation in Eq.~\eqref{eq:LinearGaugeSymmTrans}, rendering the graviton field a non-abelian gauge field. The resulting iterative procedure of finding appropriate higher order matter couplings [Eq.~\eqref{eq:Interaction term}] and ensuring higher order gauge invariance, that we will only sketch here, naturally leads to the formulation of a total matter and gravitational action of the form (see \cite{Gupta:1954zz,PhysRev.98.1118,Ogievetsky:1965zcd,Deser:1969wk,BOULWARE1975,Feynman:1963ax,maggiore2008gravitational,Deser:2009fq,Padmanabhan:2004xk})
\begin{equation}
    S=S^{\myst{GR}}_G[h]+ S_\text{m}[h, \Psi_m]\,.
\end{equation}
The pure action of the graviton field is corrected by a series of non-linear self-interacting terms that gradually involve more powers of the fields but keep two powers of derivative operators (omitting all index contraction structures)
\begin{equation}\label{eq:Low EFT GR}
    \boxed{S^{\myst{GR}}_G[h]\sim M_\text{P}^2\int d^4 x\left[(\partial h)^2+h^2\partial^2h+h^3\partial^2 h+...\right]\,.}
\end{equation}
On the other hand, also the total matter action including the interaction with the graviton field will receive an infinite series of correcting terms that can be written as
\begin{equation}\label{eq:expansion Matter Int}
    S^{\myst{GR}}_\text{m}[h,\Psi_m]\sim S^{\myst{pure}}_\text{m}[\Psi_m]+\int d^4x \,\left[h \,\frac{\delta S_\text{m}}{\delta \bar{g}} \big\lvert_{\bar g_{\mu\nu}=\eta_{\mu\nu}}+\,h^2 \,\frac{\delta^2 S_\text{m}}{\delta \bar{g}^2} \big\lvert_{\bar g_{\mu\nu}=\eta_{\mu\nu}}+...\right]\,,
\end{equation}
where $\bar{g}_{\mu\nu}$ represents an auxiliary field that temporally replaced the Minkowski metric $\eta_{\mu\nu}$. Note that the leading order interaction term above simply corresponds to $S^{\myst{int}}[h,\Psi_m]$ in Eq.~\eqref{eq:Interaction term}.

Remarkably,\footnote{But as it seems only up to hindsight's from the geometric approach \cite{Padmanabhan:2004xk,maggiore2008gravitational}.} these two series can be resummed by realizing that a coordinate gauge symmetric version of Eq.~\eqref{eq:expansion Matter Int} corresponds to a Taylor expansion of the initial matter action with the replacements 
\begin{equation}\label{eq:DefMetric}
    \eta_{\mu\nu}\rightarrow \eta_{\mu\nu}+h_{\mu\nu}\,,\quad\text{with}\quad   g_{\mu\nu}\equiv \eta_{\mu\nu}+h_{\mu\nu}\,,
\end{equation}
and
\begin{equation}
\partial_\mu\rightarrow \nabla_\mu\,,
\end{equation}
where $\nabla_\mu$ represents the covariant derivative with respect to the Levi-Civita connection defined in Eq.~\eqref{eq:DefCovariantDerivativeD}, such that \cite{Deser:1969wk,BOULWARE1975,Deser:2009fq,Padmanabhan:2004xk}
\begin{equation}
    S^{\myst{GR}}_\text{m}[g,\Psi_m]=\int d^4x \sqrt{-g}\,L^{\myst{pure}}_\text{m}[g,\Psi_\text{m}]\,.
\end{equation}
Moreover, the purely gravitational action upon the identification in Eq.~\eqref{eq:DefMetric} and up to subtleties on boundary terms that require external input \cite{,maggiore2008gravitational,Padmanabhan:2004xk}, can be shown to recover the full Einstein-Hilbert term in Eq.~\eqref{eq:EinsteinHilbertAction} \cite{Deser:1969wk,BOULWARE1975,Feynman:1963ax,Deser:2009fq}
\begin{equation}
    S^{\myst{GR}}_G\sim M_\text{P}^2\int d^4 x\left[(\partial h)^2+h^2\partial^2h+h^3\partial^2 h+...\right]\sim \frac{M_\text{P}^2}{2}\int d^4x \sqrt{-g} R\,,
\end{equation}
where $R$ corresponds to the Ricci scalar with respect to the Levi-Civita tensor defined in Eq.~\eqref{eq:RicciScalar App}.

\paragraph{Recovering Classical Results.} What is more important than the resummation of the infinite series, is however that to lower order in the non-linear expansion in $h$ we obtain a well-defined effective QFT that can recover classical results of GR.\footnote{In fact, in practice it is much more efficient to actually start from Einsteins theory of gravity and rewrite it in terms of a quantum EFT by expanding all the expressions about flat Minkowski space.} Indeed, as is well known, in such a qEFT formulation one can in particular recover the form of the Newtonian potential in the non-relativistic limit (see e.g. 
\cite{Donoghue:1993eb,Donoghue:1994dn,Akhundov:1996jd,Bjerrum-Bohr:2002gqz,Khriplovich:2002bt,maggiore2008gravitational,zee2013einstein})
\begin{equation}\label{eq:classical Newtonian Potential}
    V(r)=-\frac{G m_1 m_2}{r}\,.
\end{equation}
 In contrast to a geometric GR result however, the potential above arises in this context as consequence of the virtual exchanges of gravitons at the classical tree level at lowest order through the Feynman diagram in Fig.~\ref{GravityClassical}. 
 Such ``classical'' lowest order tree level effects for instance also including Compton scattering of gravitons with massive particles \cite{Feynman:1963ax}. 
 
 Due to the non-linearity of the theory as well as its relativistic nature, these are however inevitably corrected through higher order tree level diagrams, both by considering higher order diagrams in the non-linear expansion in Eq.~\eqref{eq:Low EFT GR}, as well as higher order tree-level graphs with additional virtual external legs (see \cite{Feynman:1963ax,maggiore2008gravitational,zee_quantum_2010,Helling2012}). Indeed, the infinite interaction series in Eq.~\eqref{eq:Low EFT GR} capture the non-linearity of GR and account for relativistic effects, for instance in the Newtonian potential \cite{Weinberg1972}.

From this perspective, it therefore seems that one can indeed describe gravity as a quantum field theory of gravitons. This result, together with the confirmed existence of gravitational waves and the clear analogy to the electromagnetic waves, forms the natural justification of the general belief that the concept of gravitons is viable, although the observation of individual on-shell gravitons with any conceivable experiment seems well out of reach \cite{Feynman:1963ax,zee2013einstein}.

\begin{figure}
\begin{center}
\begin{fmffile}{GravityClassical}
\begin{fmfgraph*}(80,60)
    \fmfleft{i1,i2}
    \fmfright{o1,o2}
    \fmf{plain_arrow,tension=3}{i1,v1,i2}
    \fmf{plain_arrow,tension=3}{o1,v2,o2}
    \fmf{photon}{v1,v2}
     \fmfdot{v1,v2}
\end{fmfgraph*}
\end{fmffile}
\end{center}
\caption{\small{Feynman diagram of a virtual exchange of gravitons between two massive sources represented by two massive scalar particles. This contribution recovers the Newtonian potential in the non-relativistic limit.}}
\label{GravityClassical} 
\end{figure}

We also want to seize the opportunity to remark at this point that, although technically possible, in this framework it does not make much sense to replace the classical Minkowski reference metric with a curved background metric about which one quantizes a field $h_{\mu\nu}$. Indeed, given the above interpretation of graviton exchange as being at the root of the Newtonian gravitational force, a curved background metric, hence a quantum EFT of gravitons on a curved background would in some sense overcount the effects of the gravitational field.


\section{Quantum Stability of GR}\label{sSec: Radiative Stability of GR}

\paragraph{The Non-Renormalizability of GR.} 

As a quantum theory, however, it is of course not enough to simply recover classical results, because inevitably quantum loop corrections are generated. In this context, it is important to realize that the interaction terms of the quantum EFT of GR are in the terminology of the renormalization group structure ``irrelevant operators'' (a terminology further explained below). In order to see this, it is important, however, that we canonically normalize the graviton field $h$, since in a perturbative field theory is always crucial to compare interaction terms with the leading kinetic term. We therefore perform the rescaling
\begin{equation}
    h\rightarrow \frac{h}{M_\text{P}}\,,
\end{equation}
such that from now on $[h]=E^1$ and Eq.~\eqref{eq:Low EFT GR} can be written as
\begin{equation}\label{eq:expansion GR action}
     S^{\myst{GR}}_G[h]\sim\int d^4 x\left[(\partial h)^2+\frac{h^2\partial^2h}{\MPl}+\frac{h^3\partial^2 h}{\MPl^2}+...\right]\sim \int d^4 x\,\sum_{i=0}^{\infty}\,(\partial h)^2\,\alpha_\text{cl}^i\,,
\end{equation}
where we identified a classical expansion parameter
\begin{equation}\label{eq:alphaCl GR}
    \boxed{\alpha^{\myst{GR}}_\text{cl}\equiv \frac{h}{M_\text{P}}\,.}
\end{equation}
The irrelevant nature of the interactions of gravitons can now be understood by examining for instance the dimension of the coupling of the first interaction term with $i=1$ in Eq.~\eqref{eq:expansion GR action}
\begin{equation}
   (\partial h)^2 \alpha_\text{cl}\sim \lambda \,(\partial h)^2 h \,,
\end{equation}
where $[\lambda]=E^{-1}$, since $\lambda \sim 1/M_\text{P}$. 

In the early days of QFT, this simple fact was viewed as one of the main problems of unifying gravity with quantum physics, as GR is therefore a so called \textit{non-renormalizable} theory (see \cite{Shomer:2007vq} for a nice review of this statement). From that, nowadays arguably outdated \cite{Cao1993,Weinberg:1995mt,Burgess:2006bm,Burgess:2007pt,zee_quantum_2010} viewpoint, renormalization was viewed as a procedure to eliminate divergences in loop computations by (very schematically) first regularizing divergent integrals by an arbitrary energy (or momentum) cutoff $\Lambda$\footnote{Not to be confused with the cosmological constant of course.} that ultimately was sent to $\Lambda\rightarrow\infty$ in order to obtain a theory that is valid on all scales. Within such a renormalization-procedure, by simple dimensional reasons any higher order correction from a ``non-renormalizable'' interaction with energy dimensions $[\lambda]< 1$ of the coupling necessarily still involve a power of $\Lambda$ in the numerator and thus blow up in the high-energy limit (see e.g. \cite{zee_quantum_2010}). Or in more accurate terms, one would need an infinite amount of counter terms to cancel the divergences in the high-energy limit. This was taken as a reason to promote renormalizability as a ``principle of nature'' and disregard any non-renormalizable theory, including the above quantum theory of spin 2 gravitons.

\paragraph{Wilsonian Renormalization and Quantum Effective Field Theory.} 

From a modern perspective on the other hand, as we already discussed back in Sec.~\ref{sSec:Why and how go beyond GR}, any theory describing physics at some accessible energy scale, in particular the known relativistic field theories, are to be understood as effective field theories, in the sense that on a very fundamental level they are low energy approximations to a known or unknown framework of smaller-scale (UV) details. Such a UV completion might even require a departure from the field theory perspective all together (see e.g. \cite{Weinberg:1995mt}). However, and crucially, knowledge of the UV theory is not required for a description of lower scales of energy. Compared to previous attempts of formulating an ultimate theory of physics, this approach incorporates in sense a more humble viewpoint, adapted to a realistic practice of physics.

Such an effective (field) theory perspective was in particular solidified through a deeper understanding of renormalization pioneered by Wilson among others \cite{Wilson:1973jj}. Within the framework of quantum field theory, any low-energy effective description given by a renormalizable set of interactions inevitably includes an infinite number of non-renormalizable interactions, that are however suppressed by a certain energy scale (see \cite{Weinberg:1978kz,Gasser:1983yg,Gasser:1984gg,Arzt:1992wz,Burgess:1992gx,Polchinski:1992ed,Cao1993,Georgi:1993hh,Donoghue:1994dn,Weinberg:1995mt,Burgess:2003jk,Burgess:2006bm,Burgess:2007pt,Weinberg:2008hq,zee_quantum_2010,Donoghue:2012zc,Endlich:2017tqa,Davidson:2020gsx}).\footnote{While the possibility for a field theory to flow to a non-trivial fixed point given by a scale invariant conformal field theory in the renormalization group flow represents a theoretically appealing escape from an infinite tower of successive EFTs, such a scenario seems realistically unlikely for an ultimate theory of physics. In particular this seems not to be the case for the standard model of particle physics, in which for instance the quartic Higgs couplings is in the absence of miracles expected to eventually hit a Landau pole \cite{Burgess:2007pt}, not to mention the completely unknown awaiting at the Planck scale.} These correcting terms announce the effects of a higher order theory whose low-energy influence can however entirely be described in terms of the degrees of freedom at low-energies. Thus, from that perspective, a regularizing cutoff $\Lambda$ is not arbitrary but captures an upper bound of a validity of a theory given by the energy scale at which the quantum corrections start to dominate over the original theory. 

In practice, the computation of EFT quantum corrections of a certain theory requires the identification of a ``lowest order'' action, in most cases corresponding to a field theory description of known classical physics, that serves as a starting point for computing the propagators and vertex Feynman rules (see e.g. \cite{Donoghue:1995cz,Peskin:1995ev,Weinberg:1995mt,Srednicki:2007qs,Schwartz:2014sze}). This action corresponds to the analogue of theories of type (A) that we described in Chapter~\ref{Sec:The Theory Space Beyond GR} to represent the principal part of a theory that in particular defines the number and type of dynamical degrees of freedom and describes classical physics, for instance in terms of tree-level interactions (see e.g. \cite{Helling2012}). Based on such a principal part, the quantum corrections are computed through loop contributions whose UV divergences require a regularization that usually leads to a renormalization of the classical operators but also inevitably generates a tower of new operators corresponding to all possible symmetry respecting interaction terms that can be written down in an energy expansion. When included in the final action, such quantum correcting operators must however be treated as pure perturbations in the sense discussed in Chapter~\ref{Sec:The Theory Space Beyond GR} corresponding to type (B) corrections to a principal part. Most importantly, quantum corrections are imperatively required to not introduce additional DOFs into a theory, which in practice results in the discussed additional constraints that need to be imposed when considering the effects of such additional operators (recall the discussion in Sec.~\ref{sSec:OstrogradskyTheorem} and see \cite{Eliezer:1989cr,Simon:1990PhysRevD41,Simon:1990jn,Yunes:2013dva}). 

In computing observable quantum corrections through regularization of divergences, it is however important to realize that only logarithmic divergences are relevant. Only the logarithmic divergences have a universal behavior that is completely determined by the low energy theory while power-law divergences, for instance quadratic or quartic, are generally UV sensitive. Fortunately, at low enough energy scales, such stronger divergences are completely unobservable compared to the log-divergent corrections \cite{Arzt:1992wz}. In this context, it is therefore advisable to employ dimensional regularization together with the minimal subtraction (MS) scheme \cite{Weinberg:1973xwm,tHooft:1973mfk} or the more practical modification known as $\overline{\text{MS}}$ that also absorbs an additional universal constant (see \cite{Weinberg:1995mt,Peskin:1995ev,Schwartz:2014sze}), which exclusively captures these physical logarithmic divergences. 
In particular, it was found that the use of a cutoff regularization comes with considerable disadvantages that can lead to erroneous conclusions depending on the choice of field variables \cite{Burgess:1992gx}. Given the absence of an algorithm to identify the in this context ``right'' variables, it is therefore recommended to relinquish on the use of hard cutoffs when estimating the size of new quantum correcting EFT operators and instead use dimensional regularization, which is insensitive to the choice of employed variables.

\paragraph{Quantum Stability of Irrelevant principal Parts.}
From the Wilsonian viewpoint on renormalization discussed above, renormalizable interactions are special in the sense that at sufficiently low energies below the cutoff, the renormalizable interactions, known as \textit{relevant operators}, will therefore dominate over the non-renormalizable interaction terms correspondingly called \textit{irrelevant operators}. This is one of the reasons that very often the principal part of a theory is given by a renormalizable theory with a set of relevant interactions. 

However, it is very well possible that irrelevant interactions represent a key feature of a classical field theory and need to be included in the principal part of the theory. This is obviously the case for the qEFT of gravity considered here. Similarly, this also applies to the Horndeski type theories considered in Sec.~\ref{ssSec: A Exact Theories}, that involve non-trivial higher-order derivative self interactions. Recall, however, that a consistent inclusion of such operators was only possible due to a very special structure of these interactions that ensured an absence of any Ostrogradsky instability (recall the Theorem~\ref{Thm:OstrogradskyTheorem}). Viewed as a quantum EFT, for such theories it is therefore important to ensure that new operators generated through quantum corrections do not spoil this classical fine-tuned structure of irrelevant terms. 

Naively, this seems hard to achieve if one wants to use such a theory on energy scales at which the classical irrelevant operators become important, because at such high-energy scales also the irrelevant operators generated through loop corrections are expected to become large. Thus, it is imperative to ensure that such a quantum detuning of classical operators is absent, in such a way that there exist a parametrically large regime in which the irrelevant interactions of the principal part may unfold their influence, while quantum corrections remain well under control. Such a property is known as \textit{quantum stability} or \textit{radiative stability} of a theory. 
This can generally be achieved either through a large enough suppression of renormalizing quantum corrections beyond the cutoff of the EFT or if none of the quantum corrections are of the form of the classical operators at all, in which case one talks about a \textit{non-renormalization} of classical structures.

\paragraph{Non-Renormalization and Radiative Stability of GR.}
Thus, in this language the quantum EFT of spin 2 gravitons represents a theory, whose classical principal part crucially relies on irrelevant operators. In the following, we will therefore discuss the quantum stability of GR, which will in particular also introduce relevant techniques employed in the subsequent Chapter~\ref{Sec.Quantum Stability}. It is important to stress, however, that in discussing radiative stability, we will exclusively focus here on the purely gravitational part of the action as it is often the custom. In Sec.~\ref{sSec: The CC Problem} below, we will come back to the important question of including also the interactions with matter fields.

While loop corrections within the quantum EFT of GR have been computed explicitly under tremendous efforts \cite{tHooft:1974toh,Goroff:1985sz}, we will here be able to draw stringent qualitative results by purely resorting to a power-counting argument based on dimensional analysis as well as Lorentz invariance. More precisely, we will be able to predict the schematic structure of all quantum induced operators, hence the quantum corrections to the qEFT of GR, that arise by virtue of the regularization of observationally relevant logarithmic divergences in the loop computations (recall the discussion above). The power of this argument even allows for general conclusions at presumably all orders in loops and external legs. 

Let's however first of all concentrate on one-loop corrections, more precisely $1$PI one-loop diagrams. The starting point of the power-counting method is the realization that the general schematic structure of the classical GR Lagrangian in terms of graviton fields that we identified in Eq.~\eqref{eq:expansion GR action} implies that each vertex of the associated Feynman rules introduces as many factors of $1/\MPl$ as there are external legs in a particular loop graph. Concretely, the cubic vertex, corresponding to $i=1$ in Eq.~\eqref{eq:expansion GR action}, only has one power of $1/\MPl$ and in a given loop graph can only contribute with one external legs, as to form a one-loop graph each vertex has to contribute exactly two internal legs. Similarly, the quartic vertex with $i=2$ always has two external legs ect. Moreover, each external leg will contribute one power of the graviton field variable $h$ to the associated quantum correction. One can therefore establish the general rule that the quantum induced operators at one-loop will precisely be given by an expansion in $\alpha_\text{cl}=h/\MPl$ [Eq.~\eqref{eq:alphaCl GR}] with an increasing number of external legs in the graphs. However, since this expansion parameter is dimensionless, dimensional analysis then requires that in order to form a quantum corrective operator in the Lagrange density of the theory, there are four powers of energy dimensions missing, which, due to the lack of any other energy scale in the theory must be provided by derivative operators on the fields. Thus, very generally, the schematic form of the one-loop quantum corrections to GR must be
\begin{equation}
    L^\text{q}_{\myst{1-loop}}\sim \partial^4 \left(\frac{h}{\MPl}\right)^j\,,\quad j\geq 2\,.
\end{equation}
This result immediately implies that all one-loop corrections are strictly distinct from the classical operators of GR, simply due to the presence of four powers of derivatives, a statement verified by the explicit computations.

Pressing on, it in now possible to extend the above power-counting arguments to higher loop orders as well. This is done by noting that at a fixed number of external legs, any higher-loop diagram necessarily involves more powers of $1/\MPl$ than its lower-loop counterpart. Either, an additional internal vertex is introduced, or an existing vertex is replaced by a vertex with more legs, such that it can contribute additional internal lines. However, again on purely dimensional grounds, these additional powers of $1/\MPl$ can only be balanced out by extra derivative operators. Furthermore, Lorentz invariance and the fact that the graviton field $h_{\mu\nu}$ always carries two Lorentz indices, requires the additional derivative operators to always come in pairs. In summary, the quantum corrections to GR to all orders therefore have the generic form
\begin{equation}
    L^\text{q}\sim \partial^4 \left(\frac{\pd^2}{\MPl^2}\right)^k\left(\frac{h}{\MPl}\right)^j\sim (\pd h)^2 \left(\frac{\pd^2}{\MPl^2}\right)^{1+k}\left(\frac{h}{\MPl}\right)^{j-2} \,,\quad j\geq 2\,,\;k\geq 0\,.
\end{equation}
The last step is a simple rearranging of terms in order to facilitate an order comparison to the all important kinetic term of the theory. 

On top of the classical expansion parameter in Eq.~\eqref{eq:alphaCl GR} that captures the degree of classical non-linearities, we can therefore identify a quantum expansion parameter 
\begin{equation}\label{eq:alphaQ GR}
    \boxed{\alpha^{\myst{GR}}_\text{q}\equiv \frac{\pd^2}{M_\text{P}^2}\,,}
\end{equation}
that is unique to the quantum loop expansion. The full quantum EFT action of GR can therefore schematically be written as 
\begin{equation}\label{eq:expansion GR action 2s}
     \boxed{S^{\myst{GR}}_G[h]\sim \int d^4 x\,(\partial h)^2\,\left[\alpha_\text{cl}^i+\alpha_\text{q}^{1+k}\alpha_\text{cl}^{j}\right]\,,\quad i,j,k\geq 0\,.}
\end{equation}

While Eq.~\eqref{eq:expansion GR action 2s} would obviously not be enough to perform explicit computations, these gross outlines of the form of the action already entail rather important results of the qEFT of gravity. First and foremost, the simple fact that the series of the quantum induced operators inevitably includes at least one power of the quantum expansion parameter $\alpha^{\myst{GR}}_\text{q}$ directly implies that none of the classical operators are directly corrected through quantum operators. This proves a so-called non-renormalization of the classical graviton action that comes with the important conclusion that there might exist a regime in which $\alpha_\text{cl}\sim 1$, hence $h\sim \MPl$, where classical non-linear terms of gravity become important, while quantum corrections still remain under control with $\alpha_\text{q}\ll 1$. This is a non-trivial statement, since from the point of view of the quantum EFT, one could naively expect that in a regime in which the irrelevant interactions of GR start to dominate, the equally irrelevant quantum corrections might take over as well, thus indicating the breakdown of the EFT expansion. For a gravity theory, it is however of imminent importance that non-linear effects can be described without the interference of any quantum corrections. Note also that while our arguments purely focused on the gravitational action, similar considerations would identify the same expansion parameters in Eq.~\eqref{eq:expansion GR action 2s} for the interactions with matter fields.

\paragraph{Schwarzschild Solution as a Qualitative Example.}

\begin{figure}
\centering
\includegraphics[scale=0.45]{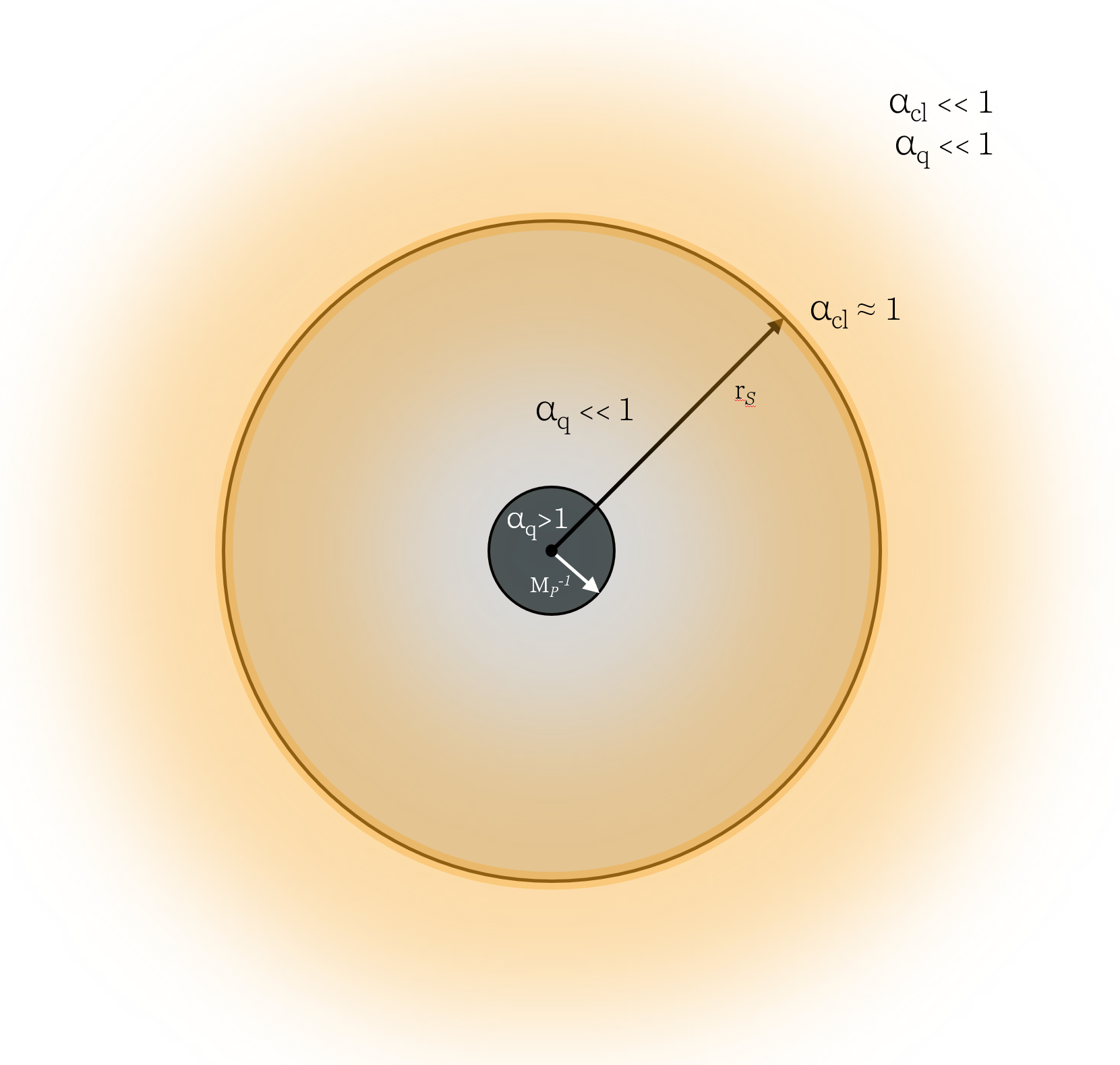}
\caption{\small{Schematic illustration of the quantum stability of GR: For solutions with $r_\text{S}\gg \MPl^{-1}$, there exists a parametrically large regime around the scale $r\sim r_\text{S}$, where classical non-linearities are large $\alpha^{\myst{GR}}_\text{cl}\sim r_\text{S}/r \sim 1 $, while quantum corrections remain suppressed $\alpha^{\myst{GR}}_\text{q}\sim 1/(\MPl^2 r^2)\ll 1$.}}
\label{fig:VainshteinGR}
\end{figure}

It is instructive to estimate the size of the classical and quantum operators for a known solution.
For the gravitational field created by localized matter of mass $M$, hence a Schwarzschild solution with Schwarzschild radius 
\begin{equation}
    r_\text{S}=2GM\sim \frac{M}{\MPl^2}\,,
\end{equation}
we expect that 
\begin{equation}\label{eq:Schwarzschild Solution}
    h\sim M_\text{P}\frac{r_\text{S}}{r}\sim \frac{M}{M_\text{P} r}\,.
\end{equation}
In this case, the order of the classical and quantum expansion parameters in Eqs. \eqref{eq:alphaCl GR} and \eqref{eq:alphaQ GR} become
\begin{equation}\label{eq:ExpansionParamsGR S}
    \alpha^{\myst{GR}}_\text{cl}=\frac{h}{M_\text{P}}\sim\frac{r_\text{S}}{r}\sim \frac{G M}{r}\,,\quad \alpha^{\myst{GR}}_\text{q}=\frac{\partial^2}{M_\text{P}^2}\sim\frac{1}{\MPl^2 r^2}\sim \frac{G}{r^2}\,.
\end{equation}
Thus, as nicely illustrated in Fig.~\ref{fig:VainshteinGR} we recover the intuitive result that as long as $r_\text{S}\gg \MPl^{-1}$, there exist a parametrically large regime on scales $r\sim r_\text{S}$, in which classical non-linear effects can be studied while quantum effects remain well under control $\alpha^{\myst{GR}}_\text{q}\ll \alpha^{\myst{GR}}_\text{cl}\sim 1 $. Only as the Planck scale is approached, quantum corrections might dominate, indicating the breakdown of the EFT. Well above the scale of the Schwarzschild radius, also the classical non-linearities become negligible $\alpha^{\myst{GR}}_\text{cl}\ll 1$ and the theory is well approximated by the linear terms.

\paragraph{Recovering the Type (B) Perturbative EFT of GR}

Furthermore, the general structure of the quantum correcting operators in Eq.~\eqref{eq:expansion GR action 2s}, together with the expectation to recover a full non-linear gauge diffeomorphic symmetry, strongly suggests that, just as for the classical Einstein-Hilbert term, the series can be resummed to yield an expansion in all possible contractions of curvature invariants. For instance, the one-loop terms ($k=0$) with four powers of derivatives should naturally restructure to 
\begin{equation}
    (\partial h)^2 \alpha_\text{q} \sum_{j=0}^\infty\alpha^j_\text{cl}\sim (\partial h)^2\frac{\partial^2}{\MPl^2}\sum_{j=0}^\infty \left(\frac{h}{\MPl}\right)^j\sim R^2\,,
\end{equation}
upon the identification 
\begin{equation}
    g_{\mu\nu}=\eta_{\mu\nu}+\frac{h_{\mu\nu}}{\MPl}\,.
\end{equation}
Here, the $R^2$ terms contain squares of the Ricci scalar as well as the Ricci tensor, while contractions between the Riemann tensor can up to boundary terms be recast in terms of Ricci tensors and Ricci scalars by using the fact that the Gauss-Bonnet scalar is purely topological. Two loop terms ($k=1$) then correspond to curvatures cubed corrections
\begin{equation}
     (\partial h)^2 \alpha^2_\text{q} \sum_{j=0}^\infty\alpha^j_\text{cl}\sim(\partial h)^2\left(\frac{\partial^2}{\MPl^2}\right)^2\sum_{j=0}^\infty \left(\frac{h}{\MPl}\right)^j\sim \frac{R^3}{\MPl^2}\,,
\end{equation}
and so and so forth. At one and two loops, these identifications were indeed confirmed through explicit computations \cite{tHooft:1974toh,Goroff:1985sz,Donoghue:1995cz}.

Presumably, the counterterm structure of the quantum EFT of gravitons with corresponding quantum correcting operators therefore precisely recovers the perturbative EFT of GR discussed in Sec.~\ref{ssSec: B Perturbative Theories}. The principal part is played in the quantum context by the classical operators that are used to compute the propagator as well as the initial vertices. The classical equations are then imposed as constraints on the additional quantum corrective operators so as to ensure that they do not lead to additional propagating DOFs. Indeed, observe that in a quantum EFT, any quantum induced operator merely serves as a perturbative correcting term and should by no means be considered as a genuine operator of the theory, since otherwise the nature of the theory would fundamentally change.

Observe that upon imposing the vacuum equations of motion, any quantum correction at one-loop can in fact be disregarded, because they simply vanish. The first non-trivial effect comes from two-loop divergences, thus recovering the theory in Eq.~\eqref{eq:QEFT GR}. These statements are but a rephrasing of the early-day findings that the qEFT of GR is still ``renormalizable'' at the one loop level \cite{tHooft:1974toh,Kalmykov:1998cv}. This is however no longer true as soon as interaction with matter fields are considered, which is confirmed by the non-trivial one-loop quantum correction that was obtained for the Newtonian potential between two energy sources.

\paragraph{Quantum Corrections to the Gravitational Potential.}
A standard application of the quantum EFT of GR is the computation of the quantum corrections to the gravitational potential through graviton loops. Note, however, that such a computation of course crucially relies on the interaction vertices of the graviton with matter particles, which we did not take into account in the discussion of quantum stability above. However, graviton loop corrections based on the well known interaction vertices of the principal part (see e.g. \cite{Bjerrum-Bohr:2002gqz}) are still characterized by the same expansion parameters. 

The final result of the one loop corrections to the Newtonian potential in Eq. \eqref{eq:classical Newtonian Potential} reads
\cite{Donoghue:1993eb,Donoghue:1994dn,Dunbar:1994bn,Akhundov:1996jd,Bjerrum-Bohr:2002gqz,Khriplovich:2002bt,Donoghue:2012zc,zee2013einstein}
\begin{equation}\label{eq:GravPotentialCorrections}
    V(r)=-\frac{G m_1 m_2}{r}\left(1+a\,\frac{G(m_1+m_2)}{ r c^2}+b\,\frac{G\hbar}{ r^2 c^3}+...\right)\,.
\end{equation}
where a careful examination of subtleties in defining the notion of a potential, as well as a correction of apparent computation errors determines the exact values of the constants $a$ and $b$ to be \cite{Bjerrum-Bohr:2002gqz,Donoghue:2012zc,zee2013einstein}
\begin{equation}
    a=3\,,\qquad b= \frac{41}{10\pi} \,.
\end{equation}
We have restored here units of $c$ and $\hbar$ to render obvious the nature of the relativistic and quantum corrections.

Interestingly, but perhaps not entirely surprising, comparing the result of the gravitational potential in Eq. \eqref{eq:GravPotentialCorrections} with the form of the classical and quantum expansion factors $\alpha_\text{cl}$ and $\alpha_\text{q}$ in Eq.~\eqref{eq:ExpansionParamsGR S}, shows that the form of the non-relativistic corrections precisely correspond to the classical non-linearity expansion parameter, while the quantum correction is up to dimensionality factors entirely governed by the quantum loop expansion parameter.

The main conclusion that can be drawn from such an assessment of the quantum corrections to GR is that they remain unobservably small. Numerically, the magnitude of the quantum corrections is of the order of $10^{-40}$ at a distance of one Fermi \cite{Donoghue:2012zc}. This however also implies that the theory of gravitons seems extremely robust against quantum corrections. Or in the words of Donoghue: ``The gravitational quantum correction is the smallest perturbative correction of all our fundamental theories. So instead of general relativity being the worst quantum theory as is normally advertised, perhaps it should be considered the best!'' \cite{Donoghue:2012zc}.
While it is of course also unfortunate that gravitational quantum corrections of the qEFT of GR seem tremendously out of empirical reach, the mere understanding of why it is possible to compute quantum corrections is remarkable. This is only possible due to EFT structure of physical phenomenon that allow for a computation of low-energy results without the knowledge of all small scale details.

\paragraph{Intimate Connection to Non-Abelian Vector Theories.}

Finally, we want to mention a rather interesting consequence of a perturbative effective field theory viewpoint of GR, namely its ``secret'' close connection to non-abelian gauge theories at the level of scattering amplitudes. While Yang-Mills theory and the perturbative EFT of GR both share the status of the unique low-energy effective theory of a massless spin 1 and spin 2 field, respectively, at first sight the two theories remain still rather different. For instance, their structure of the internal symmetries are not at all comparable. Moreover, the non-linear interactions of gauge theories remain renormalizable while the ones of the gravitons do not. 

However, as already mentioned, the distinction between renormalizable and non-renormalizable theories seems not very fundamental. And indeed, there appears to exist a fundamental connection between the two theories at the level of the amplitudes. Namely, under what is known as the \textit{color-kinematics duality}, the amplitudes of gravitons can be written as the products of Yang-Mills gluon amplitudes \cite{Bern:2008qj,Bern:2010ue} (see also \cite{zee_quantum_2010,Bern:2019prr}). Such a map between scattering amplitudes of gluons and gravitons falls under the concept of \textit{double copy} that applies to a large variety of gauge and gravity theories.\footnote{Originally, such an intimate relation between gauge theories and gravity was already hinted through the (Kawai, Lewellen, and Tye) KLT relations in string theory \cite{Kawai:1985xq}, indicating that the closed string spectrum could naturally be described as two copies of open string spectra.} 

Interestingly, given that based on EFT techniques post Newtonian and post Minkowskian corrections to analytic computations of gravitational waveforms can be computed via scattering amplitudes of gravitons see \cite{Porto:2016pyg,Levi:2018nxp,Cheung:2018wkq}, the gluon graviton double copy opens the possibility to map the concrete computation of GW waveform models to gluon scattering amplitudes. Applying the powerful tools of amplitude computations it was in fact possible to complement and extend state-of-the-art computations in particular by providing results valid up to all orders in the velocity expansion (see \cite{Buonanno:2022pgc,Adamo:2022dcm} for a review).



\newpage
\thispagestyle{plain} 
\mbox{}

\chapter{Quantum Stability of Metric Theories}\label{Sec.Quantum Stability}

With the analysis of GR as a quantum Effective theory in the previous Chapter and in particular the statement about its quantum stability, the question immediately arises whether more general metric theories of gravity also enjoy such a stability. In other words, quantum stability can be viewed as a diagnostic tool to probe the intrinsic theoretical consistency of models beyond GR. 

In this context, theories with higher-order derivative self-interactions are particularly interesting. As we discovered explicitly in Chapter~\ref{Sec:The Theory Space Beyond GR} it is possible to construct metric theories with interaction terms that contain more than two derivative operators that nevertheless do not suffer from Ostrogradsky instabilities, such as the Horndeski type theories we defined in Eqs.~\eqref{eq:ActionHorndeski}, \eqref{eq:ActionVectorHorndeski}, \eqref{eq:ActionSVH} and \eqref{eq:ActionGenProca}. Recall that these theories could therefore be considered as ``exact'' theories (of type (A)) that could serve as a principal part for perturbative expansions (of type (B)). In the present quantum EFT context, such a distinction now translates into the statement that the exact theories (A) form consistent classical field theories whose quantum corrections can be investigated, effectively giving rise to the corresponding perturbative expansion (B).

Recall that theories with non-linear derivative interactions are particularly interesting, as they introduce a Vainshtein screening mechanism into the theory, as discussed in Sec.~\ref{ssSec:Screening}. However, this mechanism crucially relies on a regime in which the non-linear interactions become large near a massive source, such that the kinetic term of perturbations gets enhanced significantly, which in turn weakens their non-minimal coupling to the physical metric. 

Asking for theoretical consistency at the quantum EFT level, this immediately leads to the question whether these classical interactions are stable under quantum corrections. This is because the Vainshtein mechanism precisely relies on scales for which non-linear interactions, hence irrelevant interactions usually suppressed by the cutoff of the EFT, are large compared to the kinetic term. As already discussed, one could naively expect that in the high-energy regime, such an EFT is not protected against equally irrelevant quantum corrections. More precisely, one could generally expect that the generated quantum corrections also become large at high energies and therefore lead to a breakdown of the EFT. Yet, as we have already explicitly shown in the case of GR in Sec.~\ref{sSec: Radiative Stability of GR}, there is the possibility that the EFT is organized in such a way that there exist a parametrically large regime in which classical non-linearities dominate, while quantum effects are still under control. This can in particular be achieved if quantum corrections fundamentally differ from their classical principal parts, a statement known as non-renormalization.


For simplicity, we will however mostly restrict our attention to the quantum stability of field theories of non-minimal fields without taking into account the coupling between the non-minimal fields and the gravitons of the physical metric. In other words, we will analyze the quantum stability of field theories with higher-order derivative self interactions on the flat Minkowski spacetime assumed in this quantum EFT formalism and neglect any coupling to the quantum fields of the physical metric. This could be justified by the simple observation that if we consider a quantum EFT of a non-minimal field that is governed by a strong-coupling scale $\Lambda \ll M_\text{P}$, then any interaction vertex with a graviton will inevitably be suppressed by an additional factor of $1/M_\text{P}$, and so will their quantum corrections. A more in-depth analysis of the possible consequences of graviton couplings are however left for future work.

\section{Flat-Space Galileon Theories}\label{sSec:FlatSpaceGalileons}

As a warm-up for considering quantum stability of theories with higher-order derivative self-interactions, we should first discuss the flat-space Galileon theories \cite{Nicolis:2008in,Deffayet:2009wt,Deffayet:2009mn}, that also provide the historical context in which these aspects were first discussed. In Sec.~\ref{ssSec: A Exact Theories} we already mentioned Galileon theories when introducing Horndeski gravity. While the Galileon interactions naturally arise as the zero-helicity mode in different contexts of massive gravity theories, and in particular also in five-dimensional braneworld models \cite{Dvali:2000hr}, they can be viewed as a local version of the Horndeski action Eq.~\eqref{eq:ActionHorndeski} with the additional requirement of a name providing invariance under the Galilean transformation
\begin{equation}\label{eq:GalileonSymmetry}
    \pi\rightarrow\pi+c+x_\mu b^\mu\,,
\end{equation}
with constants $c$ and $b^\mu$. 

\paragraph{The Galileon Lagrangians.} Indeed, the action of the Galileons is given by a sum of Lagrange densities that precisely correspond to the ones in Eq.~\eqref{eq:ActionHorndeski} with $g_{\mu\nu}\rightarrow\eta_{\mu\nu}$ and trivial prefactors of the generalized functionals $G_i$ \cite{Nicolis:2008in,Heisenberg:2018vsk}
\begin{subequations}\label{eq:ActionGalileons}
\begin{align}
        L^{\myst{Gal}}_2=&\,X\,,\\
        L^{\myst{Gal}}_3=&\,\frac{1}{\Lambda^3}\,X\,\Box\pi\,,\\
        L^{\myst{Gal}}_4=&\,\frac{1}{\Lambda^6}\,X\left[(\Box\pi)^2-\Pi^{\mu\nu}\Pi_{\mu\nu}\right]\,,\\
        L^{\myst{Gal}}_5=&\,\frac{1}{\Lambda^9}\,X\Big[(\Box\pi)^3-3\,\Box\pi\,\Pi^{\mu\nu}\Pi_{\mu\nu}+2\,\Pi_{\mu\nu}\Pi^{\nu\lambda}\Pi\du{\lambda}{\mu}\Big]\,,
\end{align}
\end{subequations}
where we bow to conventions and write $\Phi=\pi/\Lambda$, with $\Lambda$ the natural mass scale of the EFT, as well as $X\equiv - ({1}/{2}) \partial_\mu\pi\partial^\mu\pi$ and $\Pi_{\mu\nu}\equiv\partial_{\mu}\partial_{\nu}\pi$.
Just as Horndeski gravity, this theory also preserves equations of motion with only two derivatives per field, despite the presence of the non-trivial derivative interactions, rendering it stable under Ostrogradski instabilities (recall Sec.~\ref{sSec:OstrogradskyTheorem}). As such, the Galileon theory therefore represents a viable classical flat space theory. This special feature of the theory can in fact be understood by noting that these terms can be recast in terms of contractions with the antisymmetric Levi-Civita tensor that naturally kills the appearance of higher-order terms in derivatives per fields in the equations of motion
\begin{subequations}\label{eq:ActionGalileons2}
\begin{align}
        L^{\myst{Gal}}_2=&\,\pi\, \epsilon^{\mu\nu\rho\sigma}\epsilon\ud{\alpha}{\nu\rho\sigma}\Pi_{\mu\alpha}\,,\\
        L^{\myst{Gal}}_3=&\,\frac{\pi}{\Lambda^3}\, \epsilon^{\mu\nu\rho\sigma}\epsilon\ud{\alpha\beta}{\rho\sigma}\Pi_{\mu\alpha}\Pi_{\nu\beta}\,,\\
        L^{\myst{Gal}}_4=&\,\frac{\pi}{\Lambda^6}\, \epsilon^{\mu\nu\rho\sigma}\epsilon\ud{\alpha\beta\gamma}{\sigma}\Pi_{\mu\alpha}\Pi_{\nu\beta}\Pi_{\rho\gamma}\,,\\
        L^{\myst{Gal}}_5=&\,\frac{\pi}{\Lambda^9}\,\epsilon^{\mu\nu\rho\sigma}\epsilon\ud{\alpha\beta\gamma\delta}{}\Pi_{\mu\alpha}\Pi_{\nu\beta}\Pi_{\rho\gamma}\Pi_{\sigma\delta}\,.
\end{align}
\end{subequations}
From this perspective, it is also clear why there cannot be any $L_6$ term, as there are simply no further indices of Levi-Civita tensors to be contracted with.
Observe also that, interestingly, the use of the Levi-Civita tensor already played a crucial role in restricting the equations of motion to second order in sGB and ddR gravity introduced in Sec.~\ref{ssSec: A Exact Theories}.

\paragraph{Galileon Non-Renormalization.}

Exactly as for the quantum EFT of GR, quantum contributions will automatically generate all sorts of higher order derivative terms as corrections to the classical operators in Eqs.~\eqref{eq:ActionGalileons}. It is well known that also the Galileon operators, that from a renormalization group perspective correspond to irrelevant operators of the EFTs, will however not be renormalized as the quantum corrections that are also of an irrelevant nature are fundamentally distinct (recall the equivalent statements for the qEFT of GR in Sec.~\ref{sSec: Radiative Stability of GR} above). More precisely, all terms generated by quantum loops have more derivatives per fields compared to the nonlinear Galileon interactions, a statement known as the \textit{Galileon non-renormalization theorem} \cite{Luty:2003vm,Nicolis:2004qq,Burgess:2006bm,Hinterbichler:2010xn,dePaulaNetto:2012hm,Rham2013,Heisenberg:2014raa,Goon:2016ihr} (see however comments below).

This statement, that has been verified by explicit calculations to various loop orders both using Feynman diagrams, as well as background field methods, can readily be understood already at the level of power-counting in dimensional analysis that we now want to expose in a particularly enlightening manner (see also \cite{Luty:2003vm,Nicolis:2004qq,Nicolis:2008in,Hinterbichler:2010xn,Deffayet:2015rzg}). In the same schematic notation employed for GR in the previous chapter, the classical principal part of the Galileon Lagrangian in Eqs.~\eqref{eq:ActionGalileons} can be written as
\be\label{eq:Galileon Classical Schematic}
L^{\myst{Gal}}\sim (\partial \pi)^2 + (\partial \pi)^2 \left(\frac{\partial^2\pi}{\Lambda^3}\right)^l+ \,, \quad 3\geq l\geq1\,.
\ee

Lets for simplicity start by analyzing the counterterms generated by the $L^{\myst{Gal}}_{3}$ interaction, hence corresponding to $l=1$ in the above expansion, that ultimately give rise to quantum correcting operators in the EFT. Just as for GR, the schematic form of this contribution at one loop is completely fixed by Lorentz invariance and the fact that the theory only involves one energy scale $\Lambda$. In detail, in parallel to the arguments given in the case of GR in Sec.~\ref{sSec: Radiative Stability of GR}, the fact that each $L^{\myst{Gal}}_{3}$ comes with a factor $1/\Lambda^3$ fixes the number of derivatives per external field for a given number of vertices, introducing a dimensionless combination $\partial \pi /\Lambda^3$ for each vertex insertion. Moreover, since there are no other scales in the theory, each such correcting quantum operator needs to involve an additional prefactor of the form $\partial^4$. Thus, since a loop contribution at least requires two insertions of $L^{\myst{Gal}}_{3}$ vertices, the associated one-loop quantum corrections are of the generic schematic form
\be\label{eq:Fgal OneLoop}
L^{\myst{q}}_3\sim\partial^{4}\left(\frac{\partial^2\pi}{\Lambda^3}\right)^m  \,, \qquad m\geq2\,.
\ee
For example, the one-loop corrections to the propagator are given by $m=2$, hence two $\mathcal{L}_{3}$ insertions, while corrections to the $3$-pt vertex involve three $\mathcal{L}_{3}$ vertices, thus $m=3$. 

Similar arguments can be used for the $L^{\myst{Gal}}_{4}$ and $L^{\myst{Gal}}_{5}$ interaction terms, as well as corrections that are made up of mixed terms. Observe, however, that since dimensionally, these are just multiples of the $L^{\myst{Gal}}_{3}$ vertex given by the $l=2$ and $l=3$ terms in Eq.~\eqref{eq:Galileon Classical Schematic}, any one-loop quantum correction of the classical Galileon action takes on the form in Eq.~\eqref{eq:Fgal OneLoop}. For $L^{\myst{Gal}}_{4}$, in this case, $m=2$ simply would correspond to a single insertion of the interaction as a correction to the propagator for instance. 

Furthermore, in this language also higher-loop corrections can easily be incorporated. Such internal higher loop insertions generally introduce additional powers of $\Lambda$ in the denominator, that on dimensional grounds need to be compensated by additional powers of derivatives, coming in pairs to ensure Lorentz invariance of the action. Recall that this is because in order to add a loop to a diagram while keeping the number of external legs fixed necessarily requires the inclusion of an additional vertex (or the replacement of an existing vertex with a vertex that has additional legs).

Thus, in summary, all quantum corrections of Galileon theory are dimensionally bound to the following schematic form
\be\label{eq:Fgal}
L^{\myst{q}}\sim\partial^{4}\left(\frac{\partial^2}{\Lambda^2}\right)^{n}\left(\frac{\partial^2\pi}{\Lambda^3}\right)^m \,, \qquad n\geq0\,,\;m\geq2\,.
\ee
In order to properly understand the statement of the non-renormalization theorem, however, a further manipulation is needed. Namely, just as explained in Sec.~\ref{sSec: Radiative Stability of GR}, we should rearrange the schematic form of the quantum corrections in Eq.~\eqref{eq:Fgal} to allow for a direct comparison with the kinetic term. Such a rewriting yields
\be\label{eq:Fgal second}
L^{\myst{q}}\sim \left(\frac{\partial^{2}}{\Lambda^{2}}\right)^{3+n}(\partial \pi)^2\left(\frac{\partial^2\pi}{\Lambda^3}\right)^{m-2}\sim (\partial \pi)^2\,(\alpha_\text{q})^{3+n} \,(\alpha_\text{cl})^{m-2} \,, \quad n\geq0\,,\;m\geq2\,,
\ee
which, allows for the identification of a classical and a quantum expansion parameter 
\begin{eqnarray}\label{eq:Classical And Quantum Expansion Op}
    \boxed{\alpha_\text{cl}\equiv \frac{\partial^2\pi}{\Lambda^3}\,,\quad  \alpha_\text{q}\equiv \frac{\partial^2}{\Lambda^2}\,,}
\end{eqnarray}
respectively (recall the treatment of GR). Indeed, observe that the parameter $\alpha_\text{cl}$
precisely governs the expansion of the classical Lagrangian in Eq.~\eqref{eq:Galileon Classical Schematic}, while $\alpha_\text{q}$
is only present in the quantum corrections.

Thus, the full quantum corrected Galileon Lagrangian has the generic form
\be\label{eq:GalileonLagrangian Final Schematic}
\boxed{L^{\myst{Gal}}\sim (\partial \pi)^2 \left[\left(\alpha_\text{cl}\right)^l+ (\alpha_\text{q})^{3+n} \,(\alpha_\text{cl})^{m-2}\right]\,, \quad 3\geq l\geq0\,,\;n\geq0\,,\;m\geq2\,.}
\ee
Quantum corrections force the introduction of all possible correcting operators also at higher orders in derivatives. In other words, the principal part given by the Galileon action is supplemented by additional operators that generate the perturbative Langrangian of what we defined to represent a type (B) theory in Chapter~\ref{Sec:The Theory Space Beyond GR}. Recall, however, that it is crucial to view such additional quantum correcting interactions as pure perturbations to the original action that do not alter the number of propagating degrees of freedom of the principal part. 

Furthermore, in Eq.~\eqref{eq:GalileonLagrangian Final Schematic}, the non-renormalization is manifest through the additional factor of $(\alpha_\text{q})^{3}$ that is only present in the quantum correcting terms. 
At this point, it should however be noted that this statement of the Galileon non-renormalization entirely focuses on the log-divergent terms relevant for the low-energy EFT, that are captured through our power-counting arguments. Indeed, only the logarythmic divergences play a relevant role in determining a low-energy EFT (recall the discussion in Sec.~\ref{sSec: Radiative Stability of GR}) while stronger divergences can generally depend on details of the UV completion. However, for Galileon theories it is possible to formulate a stronger statement, hence a stronger Galileon non-renormalization theorem \cite{Nicolis:2004qq,Hinterbichler:2010xn} that is intimately related to the Galileon symmetry in Eq.~\eqref{eq:GalileonSymmetry} and in principle includes all type of corrections that might become important at even higher energy scales, in particular also including possible loops of other heavy fields coupled to the Galileon field in a Galileon invariant way \cite{Goon:2016ihr}. In the following, we will however refer to Galileon non-renormalization as the statement in Eq.~\eqref{eq:GalileonLagrangian Final Schematic} relevant for the assessment of the radiative stability of the EFT structure, that also corresponds to the notion of quantum stability employed in the case of GR in Sec.~\ref{sSec: Radiative Stability of GR}. However, crucially, such statements should be viewed as purely applying to the gravitational sector governed by massless fields (see Sec.~\ref{sSec: The CC Problem} below).

Crucially, low-energy non-renormalization manifest in the form of the Eq.~\eqref{eq:GalileonLagrangian Final Schematic} allows for the existence of a regime below the UV cutoff, in which classical non-linearities grow large $\alpha_\text{cl}\sim 1$ ($\pd^2\pi\sim\Lambda^3$) compared to the kinetic term, while quantum corrections still remain under control $\alpha_\text{q}\ll 1$ ($\pd^2 \ll \Lambda^2$). 
Note that in this non-linear regime, higher loop corrections, hence larger $n$, are naturally suppressed by additional factors of $\alpha_{\text{q}}$. In fact, the potentially worrisome expansion is not the expansion in loops, but rather the expansion in external legs of the quantum corrections, or in other words the expansion at large integer $m$. Simply based on the expansions in Eq.~\eqref{eq:GalileonLagrangian Final Schematic} in the regime $\alpha_\text{cl}\gg 1$, while $\alpha_\text{q}\ll 1$, one would need to conclude, that the EFT breaks down as terms with higher numbers of background fields ($m$) at fixed numbers of loops ($n$) are considered, regardless of whether $\alpha_{\text{q}}\ll 1$ or not. However, at a closer look this issue is cured precisely by the observation that in the non-linear regime the tree level kinetic term gets enhanced such that upon canonical normalization the local cutoff gets effectively shifted towards the UV. Thus, as long as the classical contributions do not lead to ghost instabilities as is the case by construction, the quantum fluctuations are rather further suppressed on such scales in contrast to what one could have expected \cite{Nicolis:2004qq} (see also similar arguments in the case of massive gravity \cite{deRham:2013qqa}).

\paragraph{Relation to Vainshtein Screening.}

Recall that the existence of the two distinct expansion parameters in the EFT Lagrangian in Eq.~\eqref{eq:GalileonLagrangian Final Schematic} is particularly important, as it allows for regions where classical non-linearities induce a Vainshtein screening mechanism, while quantum corrections are still under control. As already discussed, this is interesting for the application to cosmology of Horndeski theory as the covarantized version of the Galileon theories. In this context, the Vainshtein screening can in local dense regions effectively suppress the coupling of the non-minimal field to the physical metric. On large scales, beyond the Vainshtein radius, both classical and quantum derivative self interactions become negligible, such that the scalar degree of freedom can be used as an extension of classical gravity. Thus, in these theories, the Vainshtein radius precisely plays the role of the strong coupling scale, at which $\alpha_\text{cl}\sim 1$, and in this respect therefore corresponds to the Schwarzschild radius of the qEFT of GR. 

It is instructive to consider an explicit perturbative example of a localized, spherically symmetric source to illustrate these points (see e.g. \cite{Nicolis:2004qq,Nicolis:2008in,Babichev:2009jt,Hinterbichler:2010xn}) although here we will remain in our purely schematic analysis. These considerations will strikingly parallel the considerations of the Schwarzschild solution in the case of GR discussed in Sec.~\ref{sSec: Radiative Stability of GR}.

To start, it should be noted, that such concrete examples are for computational reasons usually exclusively treated in the so-called Einstein frame, where the kinetic terms of the redefined Lorentz field $\hat{h}_{\mu\nu}$ and $\pi$ are disentangled. With such a redefinition of the physical metric to an effective metric, the scalar field gains an apparent direct coupling to the energy-momentum tensor of matter, which render calculations straightforward. A redefinition of the physical metric to decouple the fields is however only necessary in the first place if particular non-trivial non-minimal couplings are present. One should therefore always recall that strictly speaking it is more accurate to think about non-trivial scalar field solutions as a product of the non-minimal couplings, rather than their unphysical apparent coupling to matter in the Einstein frame.

That being said, the solution for an isolated source of mass $M$ of a non-minimally coupled Galileon field depends on the type of Galileon interactions considered, as well as on the precise regime in terms of radial coordinate $r$, as governed by the Vainshtein radius, that typically has the following form
\begin{equation}
    r_\text{V}\sim \frac{1}{\Lambda}\left(\frac{M}{M_\text{P}}\right)^a\,,\quad 0<a<1\,.
\end{equation}
Recall that the Vainshtein radius is determined as the length scale below which the kinetic mixing of the physical metric and the non-minimal field (or equivalently the Einstein-frame matter coupling) is effectively suppressed, such that the solution essentially recover the GR result. For our purposes it is however enough to know that the solution for the scalar field has the following schematic form with different powers of $b$ for different regimes and interactions
\begin{equation}
    \pi\sim \Lambda^3 r_\text{V}^2\left(\frac{r_\text{V}}{r}\right)^b\,,\quad b>-1\,.
\end{equation}

Plugging this solution into our definitions of the classical and quantum expansion parameters, we have
\begin{equation}
    \alpha_\text{cl}= \frac{\partial^2\pi}{\Lambda^3}\sim \left(\frac{r_\text{V}}{r}\right)^{2+b}\,,
\end{equation}
while
\begin{equation}
    \alpha_\text{q}=\frac{\partial^2}{\Lambda^2}\sim \frac{1}{\Lambda^2 r^2}\,.
\end{equation}
Thus, indeed, the Vainshtein radius plays the role of the strong coupling scale, since $\alpha_\text{cl}\sim 1$ as $r\rightarrow r_\text{V}$. Furthermore, as long as there is a clear separation of scales $r_\text{V}\gg \Lambda^{-1}$, which is assured whenever $M\gg \MPl$ there indeed exists a desired parametrically large regime in which classical non-linearities are large, while quantum corrections remain well suppressed. The Picture is therefore very much like the one in GR presented in Fig.~\ref{fig:VainshteinGR} with the replacements $r_\text{S}\rightarrow r_\text{V}$ and $M_\text{P}^{-1}\rightarrow \Lambda^{-1}$. 
A concrete example on cosmological scales would be the case where $r_\text{V}\sim H_0^{-1}$ and $\pi \sim \MPl$, such that on these scales $\partial \sim H_0$ and one has that $\partial^2 \pi \sim \MPl H_0^2 = \Lambda^3$ while $\partial^2/\Lambda^2 \sim 1/\MPl \ll 1$ (see e.g. \cite{Noller:2018eht}).

\section{Horndeski Under The Quantum Loupe}\label{sSec:sec_HornSurv}

\small{\textit{Parts of this section are taken over from the original work \cite{Heisenberg:2020cyi} of the author. Based on this remark, we will refrain from introducing explicit quotation marks to indicate direct citations.}}
\\

\normalsize
\noindent
The quantum stability of Galileon theories through the non-renormalization discussed above can be viewed as vital for the viability of the EFT structure of the associated metric theories of gravity. However, Galileon interactions on curved spacetimes, hence Horndeski gravity, are fundamentally different form the pure Galileon interactions in that they lose their invariance under the Galileon symmetry. In particular, the dependence of the Horndeski Lagrangian's on the arbitrary functionals $G_i(\pi)$ of the scalar field explicitly break the invariance under the Galileon symmetry in Eq.~\ref{eq:GalileonSymmetry} that crucially rely on the high derivative structure. And since the strong version of the Galileon non-renormalization theorem crucially depends on this Galileon symmetry, it not clear a priori if the quantum stability properties of the Galileons is in fact also present within its covariant generalization. In fact, generally quite the opposite was assumed.

Here we want to show, however, that a Galileon-like EFT can be radiatively stable even in the presence of operators that explicitly break the Galileon symmetry, implying that the non-renormalization of the classical EFT operators does not depend on the presence of the Galileon symmetry but is a more general result of higher order derivative EFTs. To exemplify this statement, we will focus on a very particular classical and flat-space theory that we choose to be of the form
\begin{equation}
S^{\myst{sH}}=\int \mathrm{d}^{4}x\left(L^{\myst{sH}}_{2}+L^{\myst{sH}}_{3}+L^{\myst{sH}}_{4}\right),\label{Gact}
\end{equation}
where the individual Lagrangian pieces are
\begin{subequations}\label{eq:Action sH}
\begin{align}
L^{\myst{sH}}_{2}={}&\tilde{c}_2 \pi\epsilon^{\mu\nu\rho\sigma}\epsilon\ud{\alpha}{\nu\rho\sigma}\,\partial_{\mu}\partial_{\alpha} \pi,\label{L1}\\
L^{\myst{sH}}_{3}={}&\frac{\tilde{c}_3}{\Lambda^3} \pi\epsilon^{\mu\nu\rho\sigma}\epsilon\ud{\alpha\beta}{\rho\sigma}\,\partial_{\mu}\partial_{\alpha} \pi\,\partial_{\nu}\partial_{\beta}\pi,\\
L^{\myst{sH}}_{4}={}&\frac{\tilde{c}_4}{\tilde{\Lambda}^2} \pi^3\epsilon^{\mu\nu\rho\sigma}\epsilon\ud{\alpha}{\nu\rho\sigma}\,\partial_{\mu}\partial_{\alpha} \pi\,.
\end{align}
\end{subequations}
Writing the Lagrangians in terms of their explicit antisymmetric structure will become useful later on. Note that $L^{\myst{sH}}_{2}$ simply corresponds to a kinetic term, while $L^{\myst{sH}}_{3}$ is first the standard Galileon term. It is $L^{\myst{sH}}_{4}$ that represents the interesting Galileon symmetry breaking interaction. In the following, we will explicitly compute all one-loop quantum corrections of this theory in order to prove their non-renormalization.

\subsection{Motivation Through Luminal Horndeski Action}

While it is interesting to analyze the quantum stability of the EFT in Eqs.~\eqref{eq:Action sH} based on their symmetry breaking properties, the Lagrangian actually enjoys a further motivation. Namely, particular choices of the functionals in the action of the luminal Horndeski theory identified back in Eq.~\eqref{eq:Hsurv} naturally reduce to Eq.~\eqref{eq:Action sH} for the pure scalar sector on Minkowski spacetime. For convenience, we reproduce here the Lagrangian of the luminal Horndeski gravity
\begin{equation}\label{eq:HsurvSecond}
L^{\myst{H}}_{\myst{luminal}}=G_2(\pi,X)-G_3(\pi,X)\Box\pi+G_4(\pi)\,R\,.
\end{equation}

A particularly interesting model in the context of linear cosmological perturbations identified in \cite{Noller:2018eht} is obtained through the choices
\begin{equation} \label{Gdefs}
G_2=X,\qquad G_3=\frac{c_3}{\Lambda^3}X \qquad \text{and}  \qquad G_4=\frac{\MPl^2}{2}\left( 1+\frac{c_4 \pi^2}{\MPl\tilde{\Lambda}}\right) \,.
\end{equation}
However, in this formulation the kinetic terms of the scalar field will be coupled to the Lorentz tensor perturbations $h_{\mu\nu}$ of the physical metric through the non-minimal coupling given by $G_4$. Thus, the pure scalar sector of the dynamical fields should be analyzed by disentangling this apparent kinetic mixing, which can conveniently be done by transforming the full theory to the Einstein frame. In such a formulation, this model can explicitly be written as
\begin{equation} \label{exampleTheory}
S^{\myst{sH}}_{\myst{luminal}}=\int \mathrm{d}^{4}x \sqrt{-g} \left[ \frac12\MPl^2R+X\left(1-\frac{c_3}{\Lambda^3} \Box \pi+\frac{c_4^2 \pi^2}{\tilde{\Lambda}^2}\right)\right]\,,
\end{equation}
up to leading order in $1/\MPl$ and where we have absorbed a numerical ${\cal O}(1)$ factor into $\tilde\Lambda$. The flat-space, pure scalar part of this action then precisely reduces to Eqs.~\eqref{eq:Action sH}. 

Note that such an explicit transformation to the Einstein frame introduces a direct coupling of the Horndeski scalar to the standard matter fields. However, such a coupling to matter fields only arises indirectly due to the non-minimal interaction between the scalar and the physical metric, such that these couplings will effectively be suppressed at the same level as the explicit couplings between the scalar and the metric perturbations. In this work, we will therefore fully concentrate on the scalar Horndeski interactions and ignore any further interaction terms.

\subsection{A First Look at Non-Renormalization}\label{Non-renormalization}

To assess the quantum stability of the Galileon breaking theory in Eqs.~\eqref{eq:Action sH} we will start by offering a powercounting argument based on dimensional analysis. To achieve this, we can now harness the particularly simple statement of the non-renormalization of Galileon theories in the previous section, which will allow us to very quickly draw the most important conclusions.

Let's therefore start by writing the Lagrangian in the schematic form of Eq.~\eqref{eq:Galileon Classical Schematic} in order to identify the new classical expansion parameter of the Galileon breaking interaction
\be\label{eq:Galileon Classical Schematic Second}
L^{\myst{sH}}\sim (\partial \pi)^2 + (\partial \pi)^2 \left(\frac{\partial^2\pi}{\Lambda^3}\right)+ (\partial \pi)^2 \left(\frac{\pi^2}{\tilde \Lambda^2}\right) \,.
\ee
We immediately recognize the two classical dimensionless parameters.
First focusing on counterterms at one-loop giving rise to quantum correcting operators in the EFT, we will now analyze each of the three distinctive contributions separately.
\begin{enumerate}
\item
\textbf{Pure Galileon $L^{\myst{sH}}_{3}$ insertions:} 
These contributions will lead to the quantum corrections already identified in Eq.~\eqref{eq:Fgal OneLoop} of the schematic form
\be\label{eq:Fgal OneLoop Repeat}
\partial^{4}\left(\frac{\partial^2\pi}{\Lambda^3}\right)^m \sim (\partial \pi)^2 (\alpha_\text{q})^3(\alpha_\text{cl})^{m-2} \,, \quad n\geq0\,,\;m\geq2\,.
\ee
where the quantum expansion operator was identified in Eq.~\eqref{eq:Classical And Quantum Expansion Op}. Thus, all terms generated through quantum loops have more derivatives per field than the classical cubic Galileon interaction. Since we are not interested in repeating well known computations of Galileon theories, this sector will in the following not be relevant for us.
\item
\textbf{Pure non-Galileon $L^{\myst{sH}}_{4}$ insertions:} 
The $L^{\myst{sH}}_{4}$ term, provides the interesting interaction vertex whose implications for radiative stability we would like to study. Note that this new interaction is irrelevant as well and will only be of importance as soon as $\pi^2\sim\tilde{\Lambda}^2$, hence $\alpha_{\tilde{\text{cl}}}\sim 1$.
Based on dimensional analysis, the generated quantum correcting operators will be of the schematic form 
\be\label{eq:Fnew}
\pd^4\left(\frac{\pi}{\tilde{\Lambda}}\right)^{2o}\sim (\partial \pi)^2\,\alpha_{\tilde{\text{q}}} \,(\alpha_{\tilde{\text{cl}}})^{o-1}\,,\;o\geq 1\,,
\ee
where we have identified the two new classical and quantum expansion parameters
\be\label{eq: New Classical And Quantum Expansion Op}
\boxed{\alpha_{\tilde{\text{cl}}}=\frac{\pi^2}{\tilde{\Lambda}^2}\,,\quad \text{and}\quad \alpha_{\tilde{\text{q}}}=\frac{\partial^2}{\tilde{\Lambda}^2}}
\ee

In a sense, this simple dimensional analysis can be regarded as our main result, as is shows that also the quantum corrections of the Galileon symmetry breaking interaction will not renormalize any of the classical operators. Although the difference between classical and quantum operators is in this case more subtle as these quantum corrections also generate operators with fewer derivatives per field as the Galileon interactions, the clear distinction through the presence of the quantum expansion factor $\alpha_{\tilde{\text{q}}}$ ensures that on the relevant scale $\pi^2\sim\tilde{\Lambda}^2$, one should expect to find a parametrically large regime, for which the generated quantum interactions remain suppressed as long as $\pd^2 \ll \tilde{\Lambda}^2$.
\item
\textbf{Mixing of $L^{\myst{sH}}_{3}$ and $L^{\myst{sH}}_{4}$ insertions:} Finally, those counterterms induced by mixed vertices from both the Galileon and non-Galileon interactions $L^{\myst{sH}}_{3}$ and $L^{\myst{sH}}_{4}$ generate counterterms which at one-loop take the form
\be\label{eq:Fmix}
\pd^4\left(\frac{\pi^2}{\tilde{\Lambda}^2}\right)^o\left(\frac{\pd^2\pi}{\Lambda^3}\right)^m\sim (\partial \pi)^2 \alpha_{\tilde{\text{q}}} \,(\alpha_{\tilde{\text{cl}}})^{o-1}(\alpha_{\text{cl}})^m\,,\quad o,m\geq 1\,.
\ee
Again, these will not generate any operators of the same form as the classical initial interactions in Eq.~\eqref{eq:Galileon Classical Schematic Second}. On scales for which both of these classical higher order self-interactions become relevant, there again exists a regime in which the quantum contributions are suppressed by the parameter $\alpha_{\tilde{\text{q}}}$.
\end{enumerate}

With the same arguments as we have employed in the Galileon case in Sec.~\ref{sSec:FlatSpaceGalileons}, higher loop contributions will merely introduce further factors of $\alpha_{\text{q},\tilde{\text{q}}}$. Thus, the final quantum corrected EFT takes on the final form
\be\label{eq: sH Final EFT action}
\boxed{L^{\myst{sH}}\sim (\partial \pi)^2\left[1 +\alpha_{\text{cl}}+\alpha_{\tilde{\text{cl}}}+\alpha_{\tilde{\text{q}}}^{1+p}\,\alpha_{\text{q}}\,\alpha_{\tilde{\text{cl}}}^{o-1}\,\alpha_{\text{cl}}^m\right]\,,\quad  m,n,p\geq 0\,,\; o\geq 1}
\ee
showing that indeed none of the classical operators will be renormalized and the classical Lagrangian terms are protected against quantum corrections. Moreover, also the potentially worrisome expansion in high numbers of external legs of the quantum induced vertices is cured through the same arguments as in the Galileon case. In summary, the above analysis suggests that the general EFT organization of the considered theory remains healthy on all of the relevant scales below the true UV cutoff.

Furthermore, as long as $\alpha_{\text{q},\tilde{\text{q}}}\ll 1$ the principle classical operators may become important, inducing interesting non-linear behavior, including a Vainshtein screening as for the pure Galileon theories, while all quantum corrections remain suppressed. However, in the present case it might be more subtle to find regimes in which both classical expansion parameters become large at the same time due to the fact that the quantum operators generated by $L^{\myst{sH}}_{4}$ can have fewer derivatives per fields as the classical $L^{\myst{sH}}_{3}$. This is however still possible as long as there exists a further hierarchy between the two energy scales in the EFT, namely $\Lambda\ll \tilde\Lambda\ll\MPl$.

The aim of the remainder of this section is now to substantiate the above claims through the explicit computation of the divergent parts of the one-loop quantum corrections for the flat space model that ultimately determine the form of the quantum correcting operators. We will star by presenting an explicit Schwinger-DeWitt calculation of the divergent one loop effective action up to the fourth order in background fields. This computation will be subsequently generalized through a geometrical interpretation of the second order differential operator that will allow us to resum the contributions of all $n$-point functions into a single expression, providing a closed algorithm for the calculation of one-loop counterterms to any order. These results are then double-checked against direct evaluations within the Feynman diagrammatic momentum space method.

\subsection{Explicit One-Loop Computations}\label{sSec:Explicit one loop computations luminal H}

This subsection is devoted to the explicit computation of all counter term structures up to the 4-point function at one-loop of the theory in Eqs.~\eqref{eq:Action sH}, with comments on higher order results. As mentioned, we will do so by three independent methods, all focusing on the log-divergent parts of the radiative corrections in dimensional regularization with the minimal subtraction $\overline{\mathrm{MS}}$ scheme in mind (recall the discussion in Sec.~\ref{sSec: Radiative Stability of GR}). Moreover, for simplicity, we will momentarily also switch to Euclidean space, hence $\eta_{\mu\nu}\rightarrow \delta_{\mu\nu}$ which for a Minkowski background can be obtained through a standard analytic continuation. Thus, in this section we will also not distinguish between Lorentz and spacial indices such that Greek and Latin indices can be used interchangeably.

These methods were already applied in the context of the scalar Galileon model. The expansion of the one-loop effective action in terms of universal functional traces was used in \cite{dePaulaNetto:2012hm}, in order to calculate the correction to the two point function, and was also confirmed by Feynman diagrammatic methods. The divergent part of the on-shell one-loop $4$-point function was investigated in detail in \cite{Kampf2014} and further generalized in \cite{Heisenberg:2019udf} beyond the on-shell limit up to one-loop $5$-point correlation functions. And in \cite{Heisenberg:2019wjv}, these results were extended to arbitrary $n$-point functions by the geometrical formulation mentioned above.

The use of three different methods provides a powerful crosscheck of the results. Moreover, this also allowed to show that the three methods provide off-shell results that can be compared one-to-one, contrary to previous belief in the literature where it was thought that only on-shell results could be compared (see \cite{Heisenberg:2019udf,Heisenberg:2019wjv}).

\subsubsection{\ul{Schwinger-DeWitt One-Loop Effective Action}}

The combination of the background field method with gauge preserving heat-kernel techniques provides efficient ways of computing the logarithmically divergent part of the effective action \cite{Schwinger1961,DeWitt1964,Atiyah1973,Abbott1982a,Barvinsky1985}. We will first perturbatively expand the effective action in traces expressed in a universal functional form, whose values are readily calculable by means of the generalized Schwinger-DeWitt formalism.


The bedrock of the computation of the one-loop effective action is the background field method. 
For this, the Galileon field is split into its classical background and small quantum fluctuation
\begin{align}\label{fieldSplit}
\pi(x)= \bar{\pi}(x)+\delta \pi(x)\,.
\end{align}
The Euclidean one-loop effective action is then given by
\begin{align}\label{EffA}
\Gamma_{1}=\frac{1}{2}\text{Tr}\log F(\partial)\,,
\end{align}
where the general form of the scalar second order differential operator reads
\begin{align}
F(\partial^x)\delta(x,x')=\left.\frac{\delta^2 S[ \pi]}{\delta \pi(x)\delta \pi(x')}\right|_{\pi=\bar{\pi}}\,.\label{SecOrdOp}
\end{align}
The one-loop counterterms induced by the action Eq.~\eqref{Gact} up to a given order in fields and derivatives can be obtained by calculating the logarithmic divergent part of the one-loop effective action Eq.~\eqref{EffA} in the background field approach using the generalized Schwinger-DeWitt technique \cite{Barvinsky1985}.  

\paragraph{Fundamental operators and expansions.}
This method starts by splitting the scalar second order differential operator Eq.~\eqref{SecOrdOp} into its principal part $\Delta$ and the subleading, background field dependent perturbation $Y=Y( \bar{\pi})$
\be
F(\partial)=\Delta+Y\, ,
\ee
with
\begin{align}
\Delta={}&-\delta^{a b}\pd_a\pd_b\,,\\
Y={}&\tilde{c}_3\,\frac{12}{\Lambda^3}\left( \Delta  \bar{\pi}\Delta-\pd^a\pd^b  \bar{\pi}\pd_a\pd_b\right)-\tilde{c}_4\,\frac{36}{\tilde{\Lambda}^2}\left( \bar{\pi}^2\Delta+ \bar{\pi}\Delta \bar{\pi}\right)\, , \label{Y}
\end{align}
where we have canonically normalized by setting 
\begin{equation}
    \tilde{c}_2=-\frac{1}{12}
\end{equation}
Note that Eq.~\eqref{Y} contains the contributions proportional to $\tilde{c}_3$ and $\tilde{c}_4$.

The splitting Eq.~\eqref{split} together with an expansion of the logarithm in Eq.~\eqref{EffA} leads to
\begin{align}
\label{logExp}
\frac{1}{2}\Tr \ln F(\nabla)&=\frac{1}{2}\Tr \ln\left[ \Delta\right]+\frac{1}{2}\Tr\left[Y\frac{1}{\Delta}\right]-\frac{1}{4}\Tr\left[Y\frac{1}{\Delta}Y\frac{1}{\Delta}\right]\nonumber\\
&+\frac{1}{6}\Tr\left[Y\frac{1}{\Delta}Y\frac{1}{\Delta}Y\frac{1}{\Delta}\right]+\mathcal{O}(Y^4)\, ,
\end{align}
where $\frac{1}{\Delta}$ denotes the inverse of the principal operator.

The method now consists of transforming the expansion above into a sum of terms proportional to universal functional traces whose divergent part can readily be evaluated. In flat spacetime, the only non-vanishing universal functional traces in dimensional regularization with $d=4-2\epsilon$\footnote{Note that we have not carried around the various factors of $d$ arising when converting the Levi-Civita structure in the Lagrangian Eq.~\eqref{L1} to contractions of the metric tensor, since the divergent part at one loop is blind to the extra $\epsilon$ terms. Moreover, the theory could have been defined from the start without explicit use of any Levi-Civita symbol.} have the form
\be
\label{UFT 2s}
\Tr\;\mathcal{Y}^{\mu_{\scaleto{1\mathstrut}{4pt}}...\mu_{\scaleto{2n-4\mathstrut}{4pt}}}(\bar{\pi})\,\pd_{\mu_{\scaleto{1\mathstrut}{4pt}}}...\pd_{\mu_{\scaleto{2n-4\mathstrut}{4pt}}}\,\frac{1}{\Delta^n}\bigg\rvert_{\text{div}}=\,\frac{(-1)^n}{16 \bar{\pi}^2\,\epsilon}\,\int\mathrm{d}^{4}x\,\mathcal{Y}^{\mu_{\scaleto{1\mathstrut}{4pt}}...\mu_{\scaleto{2n-4\mathstrut}{4pt}}}(\bar{\pi})\,\frac{\delta^{(n-2)}_{\mu_{\scaleto{1\mathstrut}{4pt}} ... \mu_{\scaleto{2n-4\mathstrut}{4pt}}}}{2^{n-2}\,(n-1)!}\, ,
\ee
where $n\geq 2$ and  $\delta^{(n-2)}_{\mu_{\scaleto{1\mathstrut}{4pt}} ... \mu_{\scaleto{2n-4\mathstrut}{4pt}}}$ is the totally symmetrized product of $n-2$ metrics. Observe that the background field dependent piece $\mathcal{Y}(\bar{\pi})$ just goes along the ride, regardless of its specific form. 

Any term appearing in the expansion Eq.~\eqref{logExp} can be cast into the specific form appearing on the left-hand side of Eq.~\eqref{UFT 2s} by commuting all the operators $\frac{1}{\Delta}$ to the right. Note that
\be \label{Com1 2s}
\scaleto{\left[\frac{1}{\Delta}\,,Y\right]=-\,\frac{1}{\Delta}\,[\Delta\,,Y]\,\frac{1}{\Delta}\mathstrut}{24pt}\, ,
\ee
where each commutation increases the number $n$ of inverse operators $\frac{1}{\Delta}$ as well as the number of derivatives acting on the background operator Y
\be
 [\Delta\,,Y]=(\Delta Y)-2(\pd^\alpha Y)\pd_\alpha \, .
\ee
Given that one is only interested in counterterms up to a given order in the fields as well as a given order in derivatives applied to them, the procedure above is efficient in the sense that the log expansion Eq.~\eqref{logExp} will be cut off by the maximum number of background fields one is interested in, while the iterative commutation of operators Eq.~\eqref{Com1 2s} will eventually hit the threshold of derivatives applied on the background fields, such that all traces indeed can take on a universal functional form Eq.~\eqref{UFT 2s}.


\paragraph{Results up to the 4-Point Function.}

Here, we will compute the logarithmic divergent part of the one-loop effective action up to four background fields, that is the 4-point function contributions, acted on by a maximum of ten derivatives, which translates into a limitation to ten external momenta. 

First of all, note that from Eq.~\eqref{UFT 2s} it follows that the linear terms $\Tr\left[Y\frac{1}{\Delta}\right]$ with $n=1$ remain finite in dimensional regularization and can thus be disregarded. This directly implies that the 1-point tadpole contribution and the 2-point contribution proportional to $\tilde{c}_4$ do not contribute.

The next term in the log expansion in Eq.~\eqref{logExp} $\sim Y^2$ give rise to an already known, pure Galilean contribution to the 2-point function\footnote{see eg. \cite{dePaulaNetto:2012hm,Heisenberg:2019udf,Heisenberg:2019wjv}} and new contributions to the 3- and 4-point functions proportional to $\tilde{c}_3 \tilde{c}_4$ and $\tilde{c}_4^2$ respectively:
\begin{align}
\Gamma_{1,3}^{\rm div} &\supset - \frac{54}{16 \pi^2\epsilon}\,\frac{\tilde{c}_3\tilde{c}_4}{\Lambda^3\tilde{\Lambda}^2}\,\int \mathrm{d}^4x \,  \bar{\pi}\,\Delta  \bar{\pi}\,\Delta^2  \bar{\pi}\, ,\label{resultc3c4} \\
\Gamma_{1,4}^{\rm div} &\supset - \frac{324}{16 \pi^2\epsilon}\,\frac{\tilde{c}_4^2}{\tilde{\Lambda}^4}\,\int \mathrm{d}^4x \,  \bar{\pi}^2\,(\Delta  \bar{\pi})^2\label{resultc4c4}\, .
\end{align}
The concise form of the above results can be obtained by performing several tuned integrations by parts and the equivalence to more basic results can conveniently be checked by going into momentum space which eliminates this freedom of representation.

In the same spirit, the $\sim Y^3$ term in Eq.~\eqref{logExp} will yield a known contribution $\sim \tilde{c}_3^3$ to the 3-point function and a novel mixed contribution $\sim \tilde{c}_3^2\tilde{c}_4$ to the 4-point function, while other contributions will depend on more than four background fields. The next order will then merely contribute to the 4-point function via a pure Galileon contribution $G(\tilde{c}_3^4)$, which we are not interested in here. The final results up to the fourth order in background fields read
\begin{IEEEeqnarray}{rCl}\label{finalResults}
\Gamma_{1,2}^{\rm div} &=\frac{-1}{16 \pi^2\epsilon}\,\int \mathrm{d}^4x &\, \frac{9}{4}\,\frac{\tilde{c}_3^2}{\Lambda^6}\,  \bar{\pi}\,\Delta^4  \bar{\pi}\,,\nonumber\\
\Gamma_{1,3}^{\rm div} &=\frac{1}{16 \pi^2\epsilon}\,\int \mathrm{d}^4x \, &\left[\frac{\tilde{c}_3^3}{\Lambda^9}\left\{\tfrac{63}{4} \Delta{} \bar{\pi} (\Delta^2{} \bar{\pi})^2+ \tfrac{9}{2}(\Delta{} \bar{\pi})^2 \Delta^3{} \bar{\pi}  - \tfrac{9}{2}  \bar{\pi}\Delta^2{} \bar{\pi}\Delta^3{} \bar{\pi}\right.\right.\nonumber\\
&&\left.\left.- \tfrac{9}{4} \bar{\pi}  \Delta{} \bar{\pi} \Delta^4{} \bar{\pi} + \tfrac{27}{8}  \bar{\pi}^2 \Delta^5{} \bar{\pi} \right\}-54\,\frac{\tilde{c}_3\tilde{c}_4}{\Lambda^3\tilde{\Lambda}^2}  \bar{\pi}\,\Delta  \bar{\pi}\,\Delta^2  \bar{\pi}\right]\nonumber\, , \\
\Gamma_{1,4}^{\rm div}& = \frac{-1}{16 \pi^2\epsilon}\,\int \mathrm{d}^4x&\left[G(\tilde{c}_3^4)+\frac{\tilde{c}_3^2\tilde{c}_4}{\Lambda^6\tilde{\Lambda}^2}\{\tfrac{657}{10} (\Delta{} \bar{\pi})^4 -  \tfrac{2727}{2} \bar{\pi}(\Delta{} \bar{\pi})^2 \Delta^2{} \bar{\pi}  + \tfrac{1134}{5}\bar{\pi}^2 (\Delta^2{} \bar{\pi})^2  \right.\nonumber\\
&&\left.
-\tfrac{666}{5}\bar{\pi}^2 \Delta{} \bar{\pi} \Delta^3{} \bar{\pi} +\bar{\pi}^3 \tfrac{114}{5} \Delta^4{} \bar{\pi}  +  \tfrac{3969}{5} \bar{\pi}\Delta{} \bar{\pi}   \partial_{a}\Delta{} \bar{\pi}\partial^{a}\Delta{} \bar{\pi}\right.\nonumber\\
&&\left.
+ 432 \Delta{} \bar{\pi}\Delta^2{} \bar{\pi}\partial_{a} \bar{\pi}\partial^{a} \bar{\pi} + \tfrac{2592}{5} \Delta{} \bar{\pi}\partial_{b}\partial_{a}\Delta{} \bar{\pi}\partial^{a} \bar{\pi}\partial^{b} \bar{\pi} \right.\nonumber\\
&&\left.
+\tfrac{216}{5}(\bar{\pi} \Delta^2{} \bar{\pi}-  2 (\Delta{} \bar{\pi})^2) \,\partial_{b}\partial_{a} \bar{\pi}\partial^{b}\partial^{a} \bar{\pi}+\tfrac{468}{5} \bar{\pi}\Delta{} \bar{\pi} \partial_{c}\partial_{b}\partial_{a} \bar{\pi} \partial^{c}\partial^{b}\partial^{a} \bar{\pi}\right.\nonumber\\
&&\left.
+\tfrac{1152}{5} \Delta{} \bar{\pi}\partial^{b}\partial^{a} \bar{\pi} \partial_{c}\partial_{b} \bar{\pi} \partial^{c}\partial_{a} \bar{\pi}\}+324\,\frac{\tilde{c}_4^2}{\tilde{\Lambda}^4}\,  \bar{\pi}^2\,(\Delta  \bar{\pi})^2\right]\,.
\end{IEEEeqnarray}
The contributions coming from purely Galileon interactions coincide with the known results in the literature (see for instance \cite{Heisenberg:2019wjv}). We see exactly that our dimensional analysis performed in Sec.~\ref{Non-renormalization} is directly reflected in the individual counterterms generated at one loop. For instance, the three point function of the pure Galileon interactions proportional to $\tilde{c}_3^3$ generates an operator involving 10 derivatives, compared to the classical $\mathcal{L}_{3}$ Lagrangian with 4 derivatives. This counterterm is suppressed as long as $ \alpha_{\text{q}}=\frac{\partial^2}{\Lambda^2}\ll1$ and the large number of derivatives generated is at the heart of the well-known non-renormalization theorem of the Galileon. Interestingly, we also see this non-renormalization property for the pure symmetry breaking and mixed contributions calculated above, as already anticipated by the dimensional analysis in Sec.~\ref{Non-renormalization}. Explicitly, the correction to the four point function originating from the symmetry breaking interaction proportional to $\tilde{c}_4^2$ yields a contribution with four derivatives applied on the four fields, while the classical $\mathcal{L}_{4}$ Lagrangian only involves two. Hence, the non-renormalization holds and the generated counterterms remain suppressed, assuming $\alpha_{\tilde{\text{q}}}=\frac{\partial^2}{\tilde{\Lambda}^2}\ll1$ in this case.
The same is true for the mixed counterterms, i.e. proportional to combined powers of $\tilde{c}_3$ and $\tilde{c}_4$. The contribution in $\Gamma_{1,3}^{\rm div}$ proportional to $\tilde{c}_3\tilde{c}_4$ and the one proportional to $\tilde{c}_3^2\tilde{c}_4$ in $\Gamma_{1,4}^{\rm div}$ also give rise to counterterms involving two more derivatives, as compared to the classical Lagrangian. This can be viewed as a remnant of the pure Galileon non-renormalization theorem.

Summarizing, we conclude that our specific, cosmologically relevant Horndeski survival model shares a non-trivial non-renormalization theorem even in the presence of symmetry breaking operators.


\subsubsection{\ul{Closed Algorithm: Geometrical Resummation}}

We will now proceed and present a closed algorithm for the calculation of the divergent one-loop effective action to any order. On the one hand, this will give a non-trivial check of the above results and on the other it will allow us to have access to arbitrary higher order terms.
This can be done by interpreting the background scalar field contribution to the second-order fluctuation term as an effective inverse metric that enables one to define geometrical objects which bring the fluctuation operator into the form of a minimal second-order operator, necessary for the utilization of the original Schwinger-DeWitt technique. In this way, the divergent part of the one-loop effective action of all n-point functions are resummed in a single expression, from which individual contributions can be directly retrieved by expanding the curvature invariants in terms of the effective metric. 

Similar to the previous computation, we split the Galileon field into it's background and perturbation part as in Eq.~\eqref{fieldSplit}. The one-loop effective action is again given by $\Gamma_{1}=\frac{1}{2}\text{Tr}\log F(\partial)$. 
This time we represent the scalar second order differential operator as
\begin{align}\label{DiffOpGeom}
F(\partial^x)\delta(x,x')=\left.\frac{\delta^2 S[ \bar{\pi}]}{\delta \bar{\pi}(x)\delta \bar{\pi}(x')}\right|_{\pi=\bar{\pi}}=\left(-M^{\mu\nu}\partial_{\mu}^{x}\partial_{\nu}^{x}+\Gamma^{\nu}\partial_{\nu}^{x}+P\right)\delta(x,x')\,.
\end{align}
For the theory at hand in Eq.~\eqref{Gact} the explicit contributions are
\begin{align}
&M^{\mu\nu}={}-\left(2\tilde{c}_2\,\varepsilon^{\mu\alpha\rho\sigma}{{\varepsilon}^{\nu}}_{\alpha\rho\sigma}+6\frac{\tilde{c}_3}{\Lambda^3}{\varepsilon}^{\mu\alpha\rho\sigma}{{\varepsilon}^{\nu\beta}}_{\rho\sigma}\,\partial_{\alpha}\partial_{\beta} \bar{\pi} +6\color{black} \bar{\pi}^2\frac{\tilde{c}_4}{\tilde{\Lambda}^2} \varepsilon^{\mu\alpha\rho\sigma}{{\varepsilon}^{\nu}}_{\alpha\rho\sigma}\right)\,,\label{expSOO} \\
&\Gamma^{\nu}={}0 \, ,\nonumber\\
&P={} 6 \bar{\pi}\frac{\tilde{c}_4}{\tilde{\Lambda}^2}\varepsilon^{\mu\alpha\rho\sigma}{{\varepsilon}^{\nu}}_{\alpha\rho\sigma}\partial_{\mu}\partial_{\nu} \bar{\pi}\, .\label{expSOOP}
\end{align}
Recall that we are working in Euclidean space, so $\varepsilon_{\mu\nu\rho\sigma}\varepsilon^{\mu\nu\rho\sigma} = d!$ (where $d=4$ for us) and setting $c_2 = -1/12$ canonically normalizes the kinetic term.
Note that in the absence of the Galileon symmetry breaking interaction $\mathcal{L}_{4}$, only the symmetric tensor $M^{\mu\nu}$ would contribute, which has been discussed in detail in \cite{Heisenberg:2019wjv}.

The algorithm starts by identifying the symmetric tensor $M^{\mu\nu}$ as the inverse of an effective metric $M_{\mu\nu}$, such that 
\be\label{defeffM}
M_{\mu\rho}M^{\rho\nu}=\delta^{\mu}_{\nu}\,,
\ee
assuming that the effective metric is non-degenerate $M\equiv \det [M_{\mu\nu}]\neq 0$. The effective metric $M_{\mu\nu}$ then allows the definition of a corresponding metric compatible covariant derivative $\nabla^{M}_{\mu}$ with associated connection
\begin{equation}
    \Gamma^{\rho}_{\mu\nu}(M)=\frac{M^{\rho\sigma}}{2}\left(\partial_{\mu}M_{\sigma\nu}+\partial_{\nu}M_{\sigma\mu}-\partial_{\sigma}M_{\mu\nu}\right)\,,
\end{equation}
such that $\nabla^{M}_{\rho}M_{\mu\nu}=0$. This provides us with an effective Laplacian
\begin{equation}
    \Delta_{M}\equiv -M^{\mu\nu}\nabla^{M}_{\mu}\nabla^{M}_{\nu}\,,
\end{equation}
with which we can reformulate the first term in Eq.~\eqref{DiffOpGeom}
\begin{equation}
    -M^{\mu\nu}\partial_{\mu}\partial_{\nu}=\Delta_{M}-M^{\mu\nu}\Gamma^{\rho}_{\mu\nu}(M)\nabla^{M}_{\rho}\,.
\end{equation}

Thus, the operator Eq.~\eqref{DiffOpGeom} can be rewritten in terms of quantities defined through the effective metric as
\begin{align}\label{Op2}
F(\nabla^{M})=\Delta_{M}-2L^{\rho}\nabla_{\rho}^{M}+P\,,
\end{align} 
where $L^{\rho}$ is defined to be
\begin{align}
L^{\rho}\equiv\frac{1}{2}M^{\mu\nu}\Gamma^{\rho}_{\mu\nu}(M)\,.
\end{align} 

Finally, by redefining the covariant derivative 
\begin{equation}
    \mathcal{D}_{\mu}\equiv \nabla^{M}_{\mu}+M_{\mu\nu}L^{\nu}\,,
\end{equation}
the second-order fluctuation operator in Eq.~\eqref{SecOrdOp} can be brought into a minimal second order form
\begin{align}\label{Opmin}
F(\mathcal{D})=-\mathcal{D}_{\mu}\mathcal{D}^{\mu}+U\,,
\end{align}
where all the linear terms have been absorbed by the potential part
\begin{align}\label{pot}
U\equiv\nabla_{\nu}^{M}L^{\nu}+L_{\nu}L^{\nu}+P\,.
\end{align}

Using heat-kernel techniques, the one-loop divergences of the effective action Eq.~\eqref{EffA} can then be expressed in a closed form in terms of geometrical curvature invariants of the effective metric $M_{\mu\nu}$ and the potential $U$ \cite{Heisenberg:2019wjv}
\begin{IEEEeqnarray}{rCl}\label{OneLoopGgen}
\Gamma_{1}^{\mathrm{div}}&=-\frac{\chi(\mathcal{M})}{180\varepsilon}-\frac{1}{32 \bar{\pi}^2\varepsilon}&\int_{\mathcal{M}}\mathrm{d}^4x\,\sqrt{M}\,\left\{\frac{1}{60}M^{\mu\rho}M^{\nu\sigma}R_{\mu\nu}(M)R_{\rho\sigma}(M)\right.\nonumber\\
&&\left.\;+\frac{1}{120}R^2(M)-\frac{1}{6}R(M)U+\frac{1}{2}U^2\right\}\,,
\end{IEEEeqnarray}
where $\chi(\mathcal{M})=\frac{1}{32 \pi^2}\int_{\mathcal{M}}\mathrm{d}^4x\,\sqrt{M}\,\mathcal{G}(M)$ is the Euler characteristic of $\mathcal{M}$ in $d=4$ dimensions in terms of the Gauss-Bonnet scalar [Eq.~\eqref{eq:GBScalar}]
\begin{equation}
    \mathcal{G}(M)=R_{\mu\nu\rho\sigma}(M)R^{\mu\nu\rho\sigma}(M)-4R_{\mu\nu}(M)R^{\mu\nu}(M)+R^2(M)\,.
\end{equation}
However, since the effective metric is symmetric and metric compatible, the Gauss-Bonnet term can be discarded in four dimensions (we have explicitly checked that all resulting $\bar \pi$ interactions are indeed total derivatives, as expected) and we are thus left with
\begin{align}\label{OneLoopGour}
\Gamma_{1}^{\mathrm{div}}=-\frac{1}{32 \pi^2\varepsilon}\int_{\mathcal{M}}\mathrm{d}^4x\,\sqrt{M}\left\{\frac{R_{\mu\nu}R^{\mu\nu}}{60}
+\frac{R^2}{120}-\frac{RU}{6}+\frac{U^2}{2}\right\}\,.
\end{align}

In order to extract one-loop counterterms from the full resummed result in Eq.~\eqref{OneLoopGour}, one simply plugs in the effective metric and its inverse and expands up to the desired order of background fields $ \bar{\pi}$. The explicit expressions of the effective metric and its determinant up to $4^{\text{th}}$ order in the field $ \bar{\pi}$ can be found in the Appendix \ref{explicityGeom}. In this way, all geometrical objects defined above can be expanded in the number of background fields such that the method provides a closed algorithm for the calculation of all the one-loop counterterms of the theory.
Doing so to the required orders in $\bar\pi$, we indeed precisely recover all expressions in Eq.~\eqref{finalResults}.

However, the same can be obtained by resorting to metric perturbation tools without ever needing to perturbatively invert the effective inverse metric $M^{\mu\nu}$. We refer the reader to section 4 in \cite{Heisenberg:2019wjv} for more details. First of all, we expand the effective metric employed in the geometrized formulation to a desired order $n$
\be\label{Mexp}
M_{\mu\nu}=\delta_{\mu\nu}+\sum_{l=1}^n h^{\scaleto{(l)\mathstrut}{6pt}}_{\mu\nu}\,,
\ee
where $\delta_{\mu\nu}$ is the leading term corresponding to a vanishing background field. Using this generic expansion, one can thus apply standard perturbation methods in order to calculate up to the n$^{\text{th}}$ variation of Eq.~\eqref{OneLoopGour} with respect to the inverse effective metric $M^{\mu\nu}$
\be\label{Gexp}
\sum_{l=0}^n\frac{1}{l!}\,\delta^l\Gamma_{1}^{\mathrm{div}}\biggr\rvert_{\scaleto{M^{\mu\nu}=\delta^{\mu\nu}\mathstrut}{6pt}}\,.
\ee

The connection to a specific theory is then done by interpreting the expression of the effective inverse metric in Eq.~\eqref{expSOO} as well as a perturbative expansion in $ \bar{\pi}$
\be\label{Minvexp}
M^{\mu\nu}=\delta^{\mu\nu}+\sum_{l=1}^n \frac{1}{l!}H_{l}^{\mu\nu}\,.
\ee
For the theory at hand, the series stops at the second order and the explicit expressions are
\begin{IEEEeqnarray}{rCl}\label{Hdef}
H_1^{\mu\nu}&=&12\frac{c_3}{\Lambda^3}\,\left[\partial^\mu\partial^\nu \bar{\pi}-\delta^{\mu\nu}\,\Delta \bar{\pi}\right]\nonumber\\
H_2^{\mu\nu}&=&-72\frac{c_4}{\Lambda^2}\,\delta^{\mu\nu}\, \bar{\pi}^2 \\
H_{l>2}^{\mu\nu}&=&0\,.\nonumber
\end{IEEEeqnarray}
In that way, the series in Eq.~\eqref{Gexp} can make contact with the specific theory at hand by relating the two expansions Eqs.~\eqref{Mexp} and \eqref{Minvexp} to each other at each order. For example, the first two relations are
\begin{IEEEeqnarray}{rCl}\label{Hhrel}
h^{\scaleto{(1)\mathstrut}{6pt}\,\mu\nu}&=&-H_1^{\mu\nu}\nonumber\\
h^{\scaleto{(2)\mathstrut}{6pt}\,\mu\nu}&=&2H_1^{\mu\rho}H_{1\,\rho}^{\nu}-H_2^{\mu\nu}\,.
\end{IEEEeqnarray}

This allows us to obtain the divergent one-loop contributions to the effective action of any order $\Gamma_{1,i}^{\mathrm{div}}$ by inserting Eq.~\eqref{Hdef} into the expansion in Eq.~\eqref{Gexp} and extracting the term with the desired number $i$ of background fields $ \bar{\pi}$. Note that the potential term in Eq.~\eqref{pot} already contributes at the lowest order in the metric expansion Eq.~\eqref{Gexp} with two background fields through the operator $P$ [Eq.~\eqref{expSOOP}]. In doing so, we again recover exactly the same results [Eq.~\eqref{finalResults}] of the previous section.


\subsubsection{\ul{Feynman Diagram Computation}}

Additionally, we offer here a discussion of the individual one-loop Feynman diagrams, which will give direct access to the $\overline{\text{MS}}$-counterterms.
We will then make the link between the previous one-loop effective action computations and the explicit Feynman diagram results, which closes the picture and serves as a complementary check of our calculations. 

We will therefore now calculate the divergent part of the one-loop $1$PI Feynman diagrams up to four external legs. 
Each diagram represents a contribution to the reduced matrix element $\mathcal{A}$ in the perturbative expansion of the S-matrix:
\be\label{Smatrix}
\bra{k_{\text{out}}}\mathcal{S}\ket{k_{\text{in}}}\bigg\rvert_{1\rm PI}=1+(2 \pi)^4\,\delta^4(k_{\text{out}}-k_{\text{in}})\, \mathcal{A}\;.
\ee
The reduced matrix element is calculated by summing over all possible Wick contractions of the form 
\begin{subequations}
\begin{align}
& \contraction{}{\pi}{(x)}{\pi}
 \pi(x)  \bar{\pi}(y)=D(x-y)\\
& \bcontraction{}{\pi}{(x)}{ket{k}}
 \pi(x)\ket{k} =\mr{e}^{-ikx} \\
& \bcontraction{}{ket{k}}{}{\pi}
\bra{k} \pi(x) = \mr{e}^{ikx} \, ,
\end{align}
\end{subequations}
where
\be\label{eq:ScalarPropagator}
D(x-y)=\int \frac{d^4p}{(2 \pi)^4}\,\mr{e}^{ip(x-y)}\; \, \frac{1}{p^2}
\ee
is the propagator of the massless scalar field with implicit Feynman-prescription.

Following the $\overline{\text{MS}}$-scheme, the one-loop counterterms can then be inferred from the UV divergence of the $1$PI diagrams, which we will again extract using a dimensional regularization procedure with $d=4-2\epsilon$. We are thus after the log-divergent part of the one-loop $1$PI diagrams with $n$ external legs $\mathcal{A}_n^{\text{div}}$ which will be a function of the external momenta $k_i$, $ i=1,..,n-1$, since the overall delta-function $\delta^4(k_{\text{out}}-k_{\text{in}})$ always allows expressing one momentum $k_n$ in terms of the others. For consistency, we will treat all momenta as incoming throughout this section.
In the following, we will calculate all the divergent one-loop off-shell contributions up to four external legs.

\paragraph{1-Point Function.} 
Since, we have only the interaction in $L^{\myst{sH}}_{3}$ with three legs, there is only one 1-point diagram at 1-loop order, namely the tadpole as shown in Fig. \ref{Feynman_diagrams_1} below.
\begin{figure}[H]
	\centering
    \includegraphics[scale=0.2]{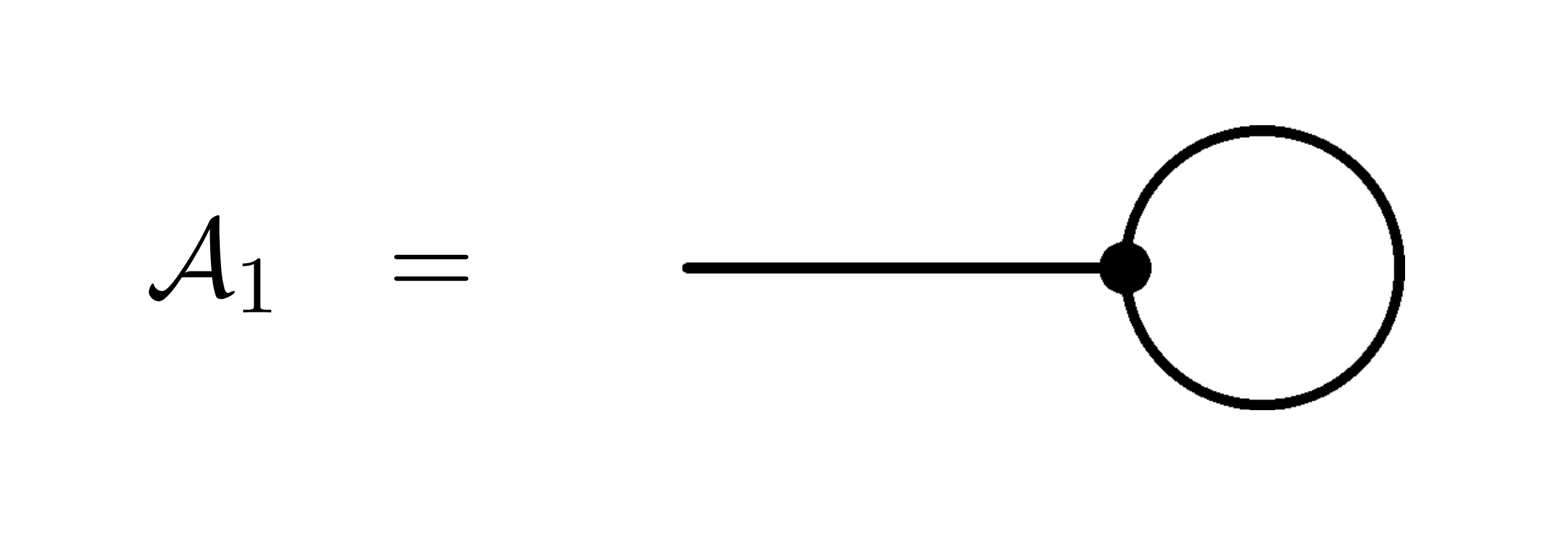}
	\caption{\small The tadpole contribution coming from $L^{\myst{sH}}_{3}$.}
	\label{Feynman_diagrams_1}
\end{figure}
\noindent However, due to the antisymmetric structure of the interaction and the number of derivatives per field involved, the tadpole contribution vanishes identically.
\begin{align}
\mathcal{A}_{1}={}&0\,.
\label{eq:1ptmatter}
\end{align}

\begin{figure}[H]
	\centering
    \includegraphics[scale=0.3]{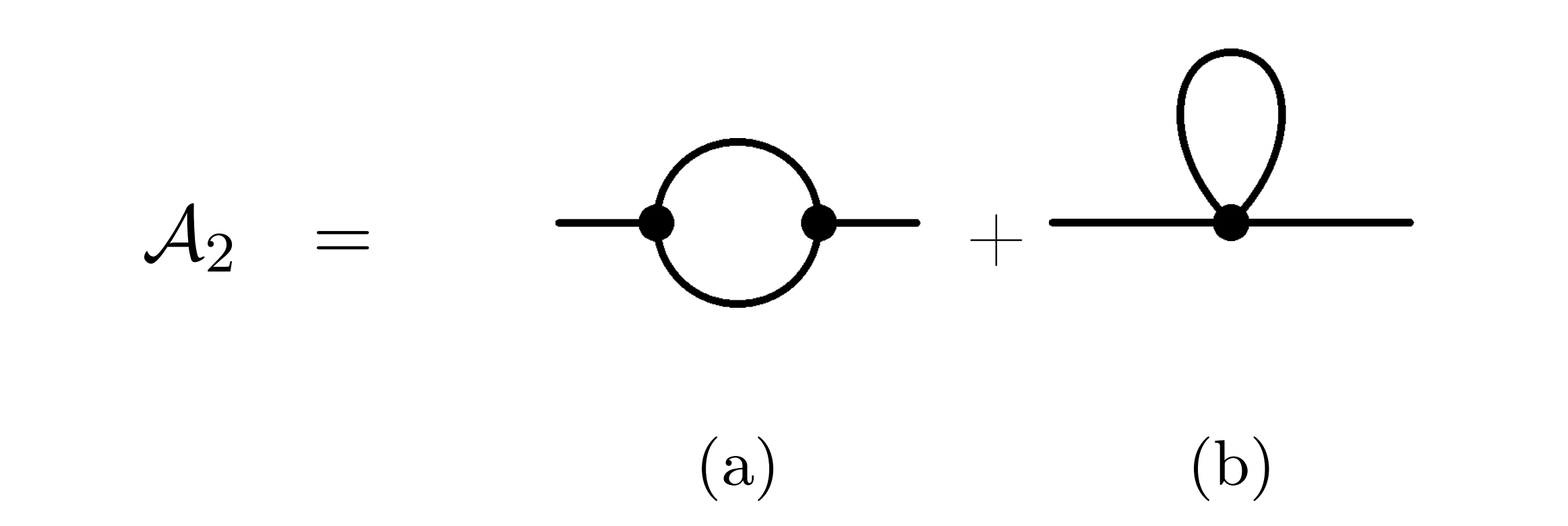}
\caption{\small{One-loop contributions to the 2-point function originating from (a) the cubic Galileon interaction $L^{\myst{sH}}_{3}$ and (b) the Galileon breaking interaction $L^{\myst{sH}}_{4}$.}}
\label{Feynman_diagrams_2pt}
\end{figure}

\paragraph{2-Point Function.} At 1-loop order, there are only the two diagrams shown in Fig. \ref{Feynman_diagrams_2pt} that contribute to the 2-point function.
The first diagram is the standard Galileon diagram coming from the $L^{\myst{sH}}_{3}$ interactions
\begin{align}
\mathcal{A}_{2a}^{\rm div}=& -9\,\frac {\tilde{c}_3^2}{\Lambda^6} \int  \frac{d^d p}{(2 \pi)^d} \frac{9 (p\cdot k_1)^4-16 p^2k_1^2(p\cdot k_1)^2+9p^4k_1^4}{p^2\; (p-k_1)^2}\biggr\rvert_{\rm div}   \nonumber\\
=&-\frac{1}{16 \pi^2\epsilon}\,\frac{\tilde{c}_3^2}{\Lambda^6}\;\frac{9}{4}p^8 \,,
\end{align}
On the other hand the tadpole type diagram \ref{Feynman_diagrams_2pt}(b) simply gives zero in dimensional regularization
\begin{align}
\mathcal{A}_{2b}=& \;36\,\frac {\tilde{c}_4}{\tilde{\Lambda}^2}\int  \frac{d^d p}{(2 \pi)^d} \frac{p^2+k_1^2}{p^2} =0  \,.
\end{align}

\noindent Only starting from the 3-point function onward, things will get interesting.

\begin{figure}[H]
	\centering
    \includegraphics[scale=0.3]{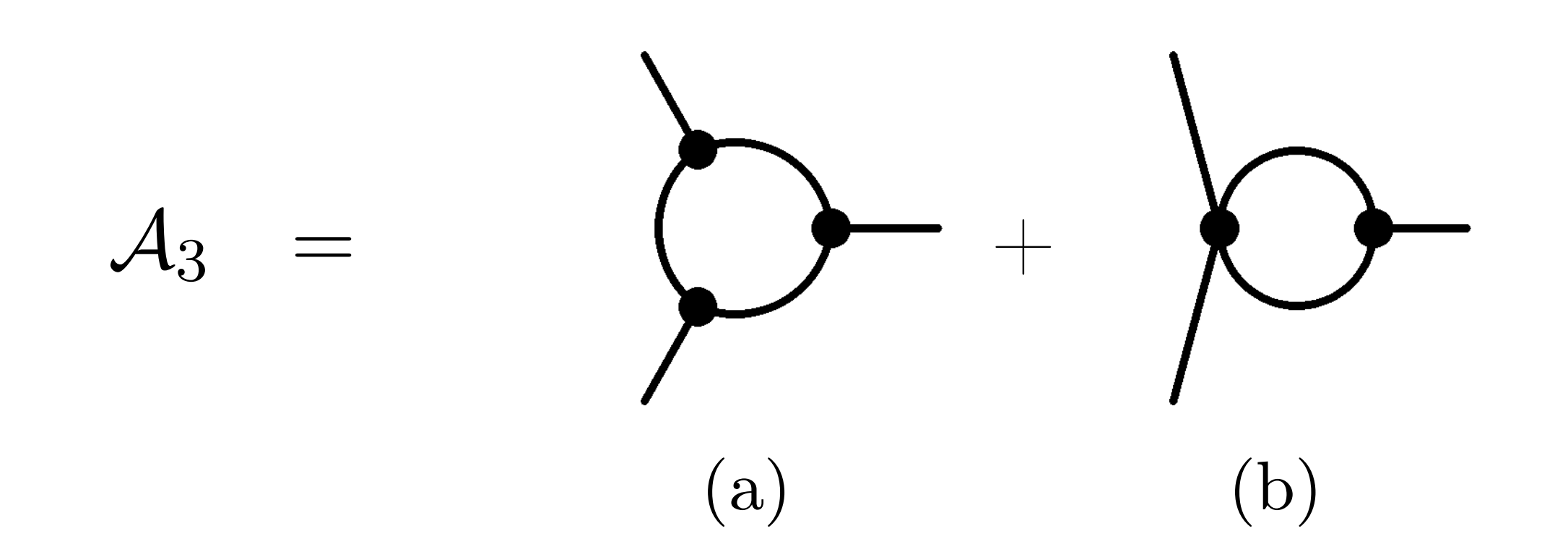}
\caption{\small{One-loop contributions to the 3-point function originating from (a) solely the cubic Galileon interaction and (b) a mixing between the Galileon and the Galileon breaking interaction.}}
\label{Feynman_diagrams_3pt}
\end{figure}

\paragraph{3-Point Function.} There are again only two distinct $1$PI contributions with three external legs at one-loop order as shown in Fig. \ref{Feynman_diagrams_3pt}.

The first diagram [Fig.~\ref{Feynman_diagrams_3pt} (a)] arises purely from the Galileon interaction and is therefore of no particular interest to us. The relevant contribution will come from the diagram [Fig.~\ref{Feynman_diagrams_3pt} (b)]. There are three distinct channels which need to be considered. For each of these channels at fixed vertices, there are a priori $72$ different ways of contracting in the S-matrix expansion in Eq.~\eqref{Smatrix} or in other words $3!4!$ different ways of distributing the $L^{\myst{sH}}_{3}$ and $L^{\myst{sH}}_{4}$ insertions over the legs in Fig~\ref{Feynman_diagrams_3pt} (b) divided by the symmetry factor of two. Note that for vertices with a different number of legs, there is no additional vertex exchange factor which could cancel the $1/2!$ in the exponential expansion in Eq.~\eqref{Smatrix}. The final result reads

\begin{IEEEeqnarray}{rCl}\label{Fresultc3c4}
\mathcal{A}_{3b}^{\rm div} =& -\frac{54}{16 \pi^2\epsilon}\,\frac{\tilde{c}_3\tilde{c}_4}{\Lambda^3\tilde{\Lambda}^2}\left[\right.&\left.2k_1^6+5 k_1^4k_2^2+6 k_1^4k_{12}+4 k_1^2k_{12}^2+8k_1^2k_2^2k_{12}
\right.\nonumber \\
&&\left. 
+4 k_2^2k_{12}^2+6 k_{2}^4k_{12}+5k_1^2 k_2^4+2k_2^6\right]\,
\end{IEEEeqnarray}
where we denote $k_{ij}\equiv k_i\cdot k_j$. Note that the result is symmetric under the exchange of momenta $k_1\leftrightarrow k_2$, as it should be.

\begin{figure}[H]
	\centering
    \includegraphics[scale=0.38]{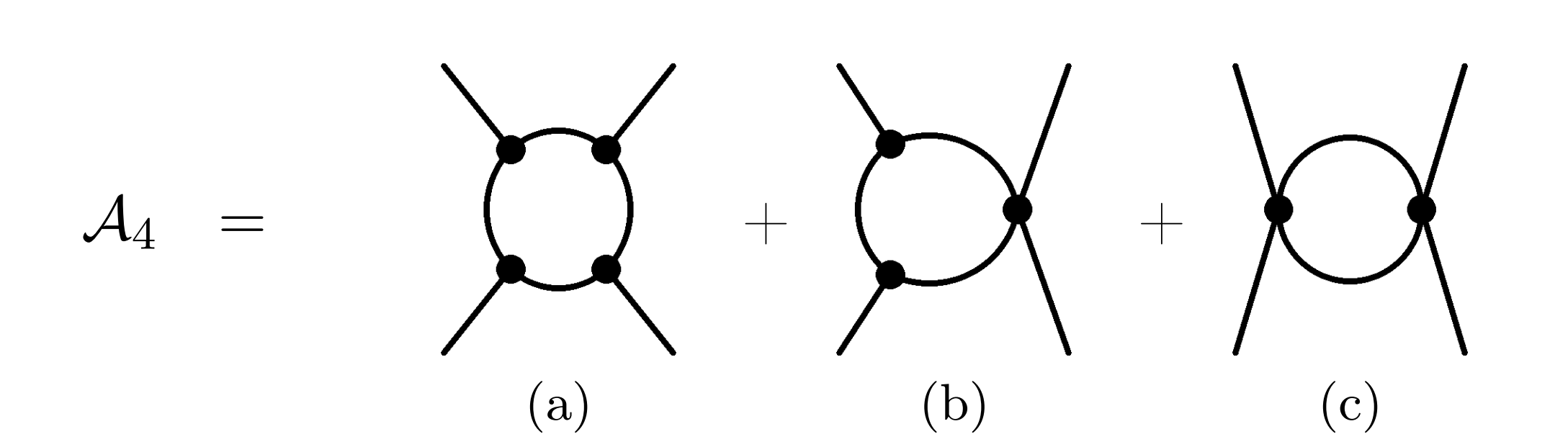}
\caption{\small One-loop graphs of the 4-point function coming from (a) four insertions of $L^{\myst{sH}}_{3}$, (b) the mixing between $L^{\myst{sH}}_{3}$ and $L^{\myst{sH}}_{4}$ and (c) two insertions of $L^{\myst{sH}}_{4}$.}
\label{Feynman_diagrams_4pt}
\end{figure}

\paragraph{4-Point Function.} At one-loop with four external legs, there are three distinct contributions, but one is again a pure Galileon result coming solely from $L^{\myst{sH}}_{3}$. The two interesting diagrams involving $L^{\myst{sH}}_{4}$ are depicted in Fig.~\ref{Feynman_diagrams_4pt} (b) and (c).

The first diagram [Fig.~\ref{Feynman_diagrams_4pt} (b)] comes in a total of six different channels. Let's also quickly go through the combinatorics: For each of the six channels and fixed vertices there are $3!^24!=864$ different ways of distributing the insertions over the legs, since the symmetry factor of the diagram in Fig.~\ref{Feynman_diagrams_4pt} (b) is one. Vertex exchange of the two $\mathcal{L}_{3}$ insertions then introduces an additional factor of $2!$. Added up, the log-divergent part is calculated to be
\begin{IEEEeqnarray}{rCl}\label{Fresultc3c3c4}
\mathcal{A}_{4b}^{\rm div} &=-&\frac{648}{16 \pi^2\epsilon}\,\frac{\tilde{c}_3^2\tilde{c}_4}{\Lambda^6\tilde{\Lambda}^2}\nonumber\\
&\times&\Big[ k_i^8-\tfrac{9}{2} k_i^6k_{j}^2+4 k_i^6k_{ij}+6 k_i^4k_{ij}^2+\tfrac{44}{3} k_i^4 k_{ij}k_{il}+\tfrac{16}{3} k_i^4 k_{ij}k_{jl}-\tfrac{2}{3} k_i^4 k_{jl}^2 \nonumber \\
&&
-\tfrac{27}{2} k_i^4k_j^2 k_{ij}-\tfrac{73}{6} k_i^4k_j^2 k_{il}-\tfrac{83}{6} k_i^4k_j^2 k_{jl}-\tfrac{26}{3} k_i^4k_j^4-19 k_i^4k_j^2k_l^2 \nonumber \\
&&
+4 k_i^2 k_{ij}^3+\tfrac{52}{3} k_i^2 k_{ij}^2k_{il}+\tfrac{13}{3} k_i^2 k_{ij}^2k_{jl}+\tfrac{29}{3} k_i^2 k_{ij}k_{jl}^2+\tfrac{56}{3} k_i^2 k_{ij}k_{il}k_{jl} \nonumber \\
&&
-\tfrac{29}{3} k_i^2k_j^2 k_{ij}^2-\tfrac{38}{2} k_i^2k_j^2 k_{il}^2-\tfrac{52}{3} k_i^2k_j^2 k_{ij}k_{il}-\tfrac{70}{3} k_i^2k_j^2 k_{il}k_{il}-\tfrac{109}{3} k_i^2k_j^2k_l^2 k_{ij} \nonumber \\
&&
+\tfrac{10}{3}  k_{ij}^4+ 14 k_{ij}^3k_{il}+\tfrac{64}{3} k_{ij}^2k_{il}^2+\tfrac{112}{3} k_{ij}^2k_{il}k_{jl}
\Big]
\end{IEEEeqnarray}
where this should be read as a sum over all $i,j,l=1,2,3$ but $i\neq j\neq l$. This of course reflects again the symmetry under exchange of momenta $k_1$, $k_2$ and $k_3$.

The contribution in Fig.~\ref{Feynman_diagrams_4pt} (c) with two insertions of $L^{\myst{sH}}_{4}$ on the other hand comes in three different channels, each of which gets a contribution of $4!^2/2=288$ due to the symmetry factor of the diagram together with an additional vertex exchange contribution $2!$ which here just cancels the expansion coefficient $1/2!$ at second order. The total gives rise to a divergent part of the form 
\begin{IEEEeqnarray}{rCl}\label{Fresultc4c4}
\mathcal{A}_{4c}^{\rm div} =& -\frac{1296}{16 \pi^2\epsilon}\,\frac{\tilde{c}_4^2}{\tilde{\Lambda}^4}\left[\right.&\left.k_i^4+2 k_i^2k_{jl}+3 k_i^2k_j^2\right]\,.
\end{IEEEeqnarray}
where again $i,j,l=1,2,3$ but $i\neq j\neq l$.

\paragraph{Cross-Check Against Effective Action Computations.}

The divergent part of the one-loop effective action of both the Schwinger-DeWitt and the geometrical computation can be compared to the off-shell Feynman diagram results by making use of the generating functional property of the effective action for one-loop $1$PI $n$-point correlation functions. Thus, Fourier-transformed functional derivatives of $ \Gamma_{1}^{\rm div}$ with respect to the background fields should coincide with the one-loop calculations of the diagrammatic method. 

The effective action is a $1$PI generating functional in the sense that repeatedly applying functional derivatives with respect to the background field yields $1$PI correlation functions
\be
\frac{\delta^n\Gamma[\bar{\pi}]}{\delta\bar{\pi}(x_1)...\delta\bar{\pi}(x_n)}\biggr\rvert_{\bar{\pi}=\langle\pi\rangle} =\langle \pi(x_1)...\pi(x_n) \rangle_{1\rm PI}\,.
\ee
The $1$PI correlation functions in turn are given by the sum of all connected $1$PI diagrams with $n$ external points by the usual cancellation of exponentiated disconnected diagrams in the numerator and denominator. Thus, Fourier transformed functional derivatives of our divergent one-loop effective action results in Eq.~\eqref{finalResults} at vanishing mean field should coincide with the corresponding divergent off-shell results of the $1$PI diagrams calculated above.

As an explicit example, consider the 3-point effective action contribution [Eq.~\eqref{resultc3c4}] proportional to $\tilde{c}_3\tilde{c}_4$
\be
\Gamma_{1,3}^{\rm div}(\tilde{c}_3\tilde{c}_4) = - \frac{54}{16 \pi^2\epsilon}\,\frac{\tilde{c}_3\tilde{c}_4}{\Lambda^3\tilde{\Lambda}^2}\,\int d^4x \,  \bar{\pi}\,\Delta  \bar{\pi}\,\Delta^2  \bar{\pi}\,.
\ee 
Taking three functional derivatives of this expression gives
\begin{align}
\frac{\delta^3\Gamma_{1,3}^{\rm div}(\tilde{c}_3\tilde{c}_4)}{\delta\bar{\pi}(x_1)\delta\bar{\pi}(x_2)\delta\bar{\pi}(x_3)} = \frac{54}{16 \pi^2\epsilon}&\,\frac{\tilde{c}_3\tilde{c}_4}{\Lambda^3\tilde{\Lambda}^2}\,\int d^4x \Big[  \delta^4(x-x_1)\,\left(\pd_{x}^2 \, \delta^4(x-x_2)\right)\nonumber\\
&\times \left(\pd_{x}^4 \delta^4(x-x_3)\right)+ 5\, \text{perm.}\Big]\,.
\end{align}
Fourier transforming results in replacing each delta function by an exponential dependence on the corresponding incoming momentum 
\begin{equation}
    \int d^4x_j \,\mathrm{e}^{ik_jx_j}\delta(x-x_j)=\mathrm{e}^{ik_jx}\,,
\end{equation}
yielding the overall momentum conservation 
\begin{equation}
    \int d^4x \,\mathrm{e}^{i(k_1+k_2+k_3)x}=(2\pi)^4\delta^4(k_1+k_2+k_3)\,,
\end{equation}
which matches the factor in the definition of the reduced matrix element in Eq.~\eqref{Smatrix}. The various derivatives $\pd_x^2$ which applied on the exponential factors exactly reproduce the momentum structure in Eq.~\eqref{Fresultc3c4} upon expressing $k_3$ in terms of the other momenta, showing the equivalence of the results. The same can be done for each term of our results in Eq.~\eqref{finalResults}, all matching their corresponding Feynman diagram calculation.

Thus, have explicitly checked that indeed all the effective action calculations in Eq.~\eqref{finalResults} match their corresponding off-shell $\overline{\text{MS}}$-counterterm calculated through Feynman diagram techniques. In particular, the results in Eqs.~\eqref{resultc3c4} and \eqref{resultc4c4} perfectly agree with Eqs.~\eqref{Fresultc3c4} and \eqref{Fresultc4c4} respectively upon performing the transformations. The expression proportional to $\sim \tilde{c}_3^2\tilde{c}_4$ in Eq.~\eqref{finalResults} matches also perfectly well the calculation of Eq.~\eqref{Fresultc3c3c4}.

The comparison and matching of the different methods provides a powerful check of our results, as they rely on fundamentally distinct concepts. Especially the agreement between the Feynman calculations and computations relying on the effective action can hardly be a coincidence, as the only common ground is the input of the Lagrangian.


\section{Quantum Stability of Generalized Proca}\label{Sec:GenProca Quantum Stability}
\small{\textit{Parts of this section are taken over from the original work \cite{Heisenberg:2020jtr,deRham:2021yhr} of the author. Based on this remark, we will refrain from introducing explicit quotation marks to indicate direct citations.}}
\\

\normalsize
\noindent
Finally, we also want to analyze the quantum stability of generalized Proca theory, whose associated covariant gravity theory we introduced back in Eq.~\eqref{eq:ActionGenProca}. Again, we will however focus on the flat-space GP theory and leave an explicit analysis of the couplings to the graviton for future studies. Yet, similar arguments given in the case of Horndeski theory above suggests that such couplings would not significantly alter the EFT structure, as each mixed vertex comes with a heavy plank mass suppression.

The generalized Proca theories, also known as vector Galileons, possess an intimate relation to scalar Galileons. At high energies way above the vector mass, the longitudinal polarization dominates, such that the theory acquires a Galileon symmetry and half of the generalized Proca interactions reduce to pure scalar Galileon terms. 

Despite this fact, the quantum stability analysis of GP provides a very interesting addition to the Galileon theory and its Horndeski generalization discussed above, as it will involve entirely new features. In particular, a second energy scale, the mass of the vector, will play an essential role. In fact, a first look at the radiative stability of GP was already taken in \cite{Charmchi:2015ggf}. This work computed the one-loop quantum corrections to the 2-point function and found that the classical operators are in fact renormalized, in contrast to the scalar cases discussed above, suggesting that GP might not be radiatively stable.

Nevertheless, our analysis will show that even though there is indeed no GP non-renormalization, the corresponding corrections remain largely suppressed in such a way as the mass of henceforth introduced ghost degrees of freedom reside safely above the EFT cutoff. What is more, we are able to prove by direct calculation at one-loop, substantiated by detailed all order arguments in the decoupling limit, that the GP EFT precisely reorganizes itself such that potentially worrisome terms that would blow up in the high energy limit are all cancelled. We therefore claim radiative stability of the generalized Proca interactions, in the sense that the EFT remains perfectly healthy below its cutoff scale. 

What is more, at the strong coupling scale well above the vector mass, in the so-called decoupling limit, the theory in fact reorganized itself again with a clear hierarchy between classical and quantum operators. This conclusion is ultimately enabled by an explicit identification of global classical and quantum expansion parameters of the theory, while showing that the decoupling limit indeed represents the appropriate limit to analyze the quantum induced operators that are most relevant on scales close to the cutoff of the theory. by the explicit identification of global classical and quantum expansion parameters of the theory.

Through this thorough analysis of the behavior of generalized Proca theories under loop corrections therefore substantially extend and in particular also correct the earlier results in \cite{Charmchi:2015ggf}. Moreover, these results are especially noteworthy with possible gravitational and cosmological applications in mind, as the hierarchy between classical and quantum non-linearities allow for regimes in which the former dominate while the EFT description is still protected against quantum detuning. Including gravity and matter fields, this allows for the possibility for the presence of a natural Vainshtein screening in dense regions, while the additional vector field serves as a generalization of gravity on cosmological scales.

In the following, we will start in Sec.~\ref{GPm} by introducing the flat-space generalized Proca model and analyze its structure in more detail, while in particular also introduce the notion of a decoupling limit in Sec.~\ref{ssSec:Decoupling Limit}. In a subsequent section [Sec.~\ref{OneLoop}] we first offer explicit calculations of one-loop UV logarithmic divergences of Feynman diagrams up to four external legs, cross-checked by the entirely independent heat kernel method we already employed in the scalar theory case above. Finally, in Sec.~\ref{QuantumStability} decoupling limit arguments then allow for in depth understanding of the explicitly computed results, and also pave the way for a complete generalization to all orders.

Bowing to conventions in the literature, we will also momentarily employ a mostly minus metric-sign convention $(+,-,-,-)$ in this section.


\subsection{Flat-Space Generalized Proca Theory}
\label{GPm}

The most general action of a local and Lorentz-invariant massive vector field theory on Minkowski spacetime with second order equations of motion and three propagating degrees of freedom is restricted to the following structure \cite{Heisenberg:2014rta,Jimenez:2016isa}
\begin{equation}
    S^{\myst{GP}}=\int d^4x\sum_{i=2}^6\,L^{\myst{GP}}_i
\end{equation}
where
\begin{IEEEeqnarray}{rCl}\label{Lagrangians}
L^{\myst{GP}}_2 &=&  \Lambda_2^4\, f_2 \left(\frac{mA_\mu}{\Lambda_2^2},\frac{F_{\mu\nu}}{\Lambda_2^2},\frac{\tilde{F}_{\mu\nu}}{\Lambda_2^2}\right)\, ,\nonumber \\
 L^{\myst{GP}}_3 &=& \frac{-\Lambda_2^2}{6} \, f_3 \left(Z\right) \, \epsilon^{\mu\nu\rho\sigma}\epsilon\ud{\alpha}{\nu\rho\sigma} \, \pd_\mu A_\alpha\,, \nonumber\\
L^{\myst{GP}}_4 &=&  \frac{-1}{2}\;\epsilon^{\mu\nu\rho\sigma}\epsilon\ud{\alpha\beta}{\rho\sigma} \left(\, f_4 \left(Z\right)\pd_\mu A_\alpha \, \pd_\nu A_\beta + \tilde{f}_4 \left(Z\right) \pd_\mu A_\nu \, \pd_\alpha A_\beta \right), \\
L^{\myst{GP}}_5 &=& \frac{-1}{\Lambda_2^2}\, \epsilon^{\mu\nu\rho\sigma}\epsilon\ud{\alpha\beta\gamma}{\sigma} \left(f_5 \left(Z\right)\pd_\mu A_\alpha \, \pd_\nu A_\beta \,\pd_\rho A_\gamma + \tilde{f}_5 \left(Z\right) \pd_\mu A_\nu\, \pd_\alpha A_\beta \,\pd_\rho A_\gamma \right),\nonumber \\
L^{\myst{GP}}_6 &=& \frac{-1}{\Lambda_2^4} \, \epsilon^{\mu\nu\rho\sigma}\epsilon\ud{\alpha\beta\gamma\delta}{} \left(f_6 \left(Z\right)\pd_\mu A_\alpha\, \pd_\nu A_\beta\, \pd_\rho A_\gamma\, \pd_\sigma A_\delta + \tilde{f}_6 \left(Z\right) \pd_\mu A_\nu\, \pd_\alpha A_\beta\, \pd_\rho A_\gamma\, \pd_\sigma A_\delta \right),\nonumber
\end{IEEEeqnarray}
where the vector field $A_\mu$ here is dimensionful of mass dimension one, and where we now have defined the dimensionless combination
\begin{equation}
   Z\equiv \frac{m^2A^2}{\Lambda_2^4}\,.
\end{equation}
There thus exist two natural classical energy scales of the theory given by the mass $m$ and the interaction scale $\Lambda_2$ which controls the interactions expanded in the number of fields $n$ through factors of $1/\Lambda_2^{2n-4}$. The dimensionless combination $m/\Lambda_2$ can be viewed in some sense as a coupling constant, generally assumed to be small, hence
\begin{equation}\label{eq:hierachy scales}
    m\ll \Lambda_2\,.
\end{equation}
Moreover, the numerical factors appearing in the definitions of $L^{\myst{GP}}_3$ and $L^{\myst{GP}}_4$ are pure convenience.

As it was already the case for the covariant version of Generalized Proca theory, the first Lagrangian term $L^{\myst{GP}}_2$ contains all possible vector potential contributions including the mass term, as well as kinetic and interaction terms constructed out of the building blocks $A_\mu$, its field strength $F_{\mu\nu}$ and the dual $\tilde{F}^{\mu\nu}\equiv\frac{1}{2}\epsilon^{\mu\nu\rho\sigma}F_{\rho\sigma}$. By construction, these terms do not give rise to any dynamics for the temporal component $A_0$. On the other hand, $L^{\myst{GP}}_{3,.., 6}$ represent the derivative self-interactions which despite their higher number of derivatives do not increase the number of propagating degrees of freedom of the theory, which therefore remains at three. Thus, the theory is stable under Ostrogradski instabilities (recall Sec.~\ref{sSec:OstrogradskyTheorem}) and remains free of ghosts \cite{Heisenberg:2014rta}. It is interesting to note that this property is again ensured by their construction via two Levi-Civita tensors, which at the level of the equations of motion only allows for at most second order terms restricted to the very specific gauge invariant schematic form $\sim\pd F$. This follows from the fact that two derivatives applied to the same field can only enter through \begin{equation}
    \epsilon^{\mu\nu\rho\sigma}\epsilon\ud{\alpha\beta}{\rho\sigma}\pd_\mu\pd_\alpha A_\beta\sim\pd_\mu\pd^\mu A^\nu-\pd_\mu\pd^\nu A^\mu=\pd_\mu F^{\mu\nu}\,,
\end{equation}
while all other derivative terms remain first order.

Being interested in quantum corrections which potentially renormalize the given classical structure, we will for simplicity only perform explicit calculations based on a particularly interesting subset of the most general construction above and choose a minimal model with standard canonically normalized kinetic and mass terms and where
\be
\label{FunctionChoice}
 f_{3,4}(Z)=c_{3,4}\,Z \; ,\quad \tilde{f}_4(Z)=\tilde{c}_{4}\,Z  \; ,\quad f_{5,6}(Z)=c_{5,6} \; ,\quad \tilde{f}_{5,6}(Z)=\tilde{c}_{5,6}\, ,
\ee
with constants $c_i$.
With this choice, the terms proportional to $c_5$ and $c_6$ are total derivatives and effectively drop out of the analysis. Up to total derivatives, the remaining terms can be written as
\begin{subequations}\label{action}
\begin{align}
& L^{\myst{GP}}_{2}=-\frac{1}{4}F^2+\frac{1}{2}m^2A^2\,,\\
& L^{\myst{GP}}_3 = \frac{m^2}{\Lambda_2^2} \,c_3\, A^2 \pd\cdot A\,, \\
& L^{\myst{GP}}_4 = \frac{m^2}{\Lambda_2^4}\,A^2\,\left(c_4\left[(\pd\cdot A)^2-\pd_\mu A_\nu\pd^\nu A^\mu\right]+\tilde{c}_4\,F^2\right)\,, \\
& L^{\myst{GP}}_5 =  -\frac{1}{\Lambda_2^2}\,\tilde{c}_5\, \epsilon^{\mu\nu\rho\sigma}\epsilon\ud{\alpha\beta\gamma}{\sigma} \pd_\mu A_\nu\, \pd_\alpha A_\beta \,\pd_\rho A_\gamma \, \\
&  L^{\myst{GP}}_6 = - \frac{1}{\Lambda_2^4}\,\tilde{c}_6\,\epsilon^{\mu\nu\rho\sigma}\epsilon\ud{\alpha\beta\gamma\delta}{} \pd_\mu A_\nu\, \pd_\alpha A_\beta\, \pd_\rho A_\gamma\, \pd_\sigma A_\delta \,.
\end{align}
\end{subequations}
Note that the operator proportional to $\tilde{c}_4$ is actually a higher order $L^{\myst{GP}}_{2}$ term. We will nevertheless keep it in order to explicitly see what happens with this class of terms.

This theory is fundamentally different from the scalar theories analyzed in the previous sections in that it involves two energy scales $\Lambda_2$ and $m$, where in particular the presence of the mass scale will complicate the power-counting in dimensional analysis, that indeed can lead to the naive conclusion that the theory might not be radiatively stable as claimed in the work of Charmchi \textit{et al.} \cite{Charmchi:2015ggf}. This is ultimately related to the fact that the vector propagator of the theory reads
\be\label{propagator}
D_{\mu\nu}(x-y)=\int \frac{\mr{d}^4p}{(2\pi)^4}\,\mr{e}^{ip(x-y)}\; i\, \frac{-\eta_{\mu\nu}+{\scriptstyle \frac{p_\mu p_\nu}{m^2}}}{p^2-m^2}
\ee
with implicit Feynman-prescription, which at high energies behaves as $\sim\frac{1}{m^2}$ compared to for instance the scalar propagator in Eq.~\eqref{eq:ScalarPropagator}.

\subsection{Decoupling Limit}
\label{ssSec:Decoupling Limit}

While it would be of course possible to analyze the quantum stability of the theory in Eq.~\eqref{action} though direct computations, which we will in fact explicitly carry out in Sec.~\ref{OneLoop}, a qualitative understanding of the radiative behavior of this multiscale theory in fact necessitates the introduction of the concept of a \textit{decoupling limit} (DL), which we will now want to introduce. The decoupling limit ultimately corresponds to a well-defined limit of energy scales well above the mass of the vector field. 

That this limit must exist within the validity of the EFT is essential, and is guaranteed by the assumed hierarchy between the energy scales in Eq.~\eqref{eq:hierachy scales}. The necessity for a careful definition of a decoupling limit is already apparent from the high energy behavior of the vector propagator in Eq.~\eqref{propagator} that is not well-defined in the limit $m\rightarrow 0$. Moreover, a naive massless limit in the GP action [Eq.\eqref{action}] is ill-defined, as it introduces a discontinuity in terms of degrees of freedom. This is in close analogy to the relationship between the quantum EFTs of a massive and a massless spin 2 field. And just as for massive theories of gravity, a proper decoupling limit can therefore only be introduced by reformulating the theory in terms of a faithful description with an explicit scalar field through a mechanism that goes back to the work of St\"uckelberg \cite{Stueckelberg:1900zz,GREEN1991462,Siegel:1993sk,Arkani-Hamed:2002bjr,Ruegg:2003ps}, which we already introduced. In fact, in this context, a St\"uckelberg reformulation can be viewed as an explicit reintroduction of the eaten Goldstone boson. 

Indeed, recall that when introducing GP gravity [Eq.~\eqref{eq:ActionGenProca}] in Sec.~\ref{ssSec: A Exact Theories} we already noted that it is useful to rewrite a massive vector theory in a more faithful way by introducing a redundancy in the form of an additional scalar field $\phi$ through the replacement [Eq.~\eqref{eq:StuckelbergReplacementA}]\footnote{Recall that this replacement is not a conventional change of field variables and neither a decomposition of $A_\mu$ into transverse and longitudinal degrees of freedom. It is a way to introduce redundancy in the description.}
\be\label{Stuckelberg}
A_\mu\rightarrow A_\mu+\tfrac{1}{m}\partial_\mu\phi \,.
\ee
where the mass scale is fixed by canonically normalizing the kinetic term of the scalar field. The specific form of the replacement in Eq.~\eqref{Stuckelberg} is such that gauge invariant terms remain untouched, such that the new theory is effectively obtained by making the replacements [Eq.~\eqref{eq:EffectiveStuckelbergRep}]
\be\label{Stuckelberg2}
A_\mu\rightarrow \frac{1}{m}D_\mu\phi \,,\quad F\rightarrow F \quad\text{and} \quad\tilde{F}\rightarrow \tilde{F}
\ee
where we have defined the covariant derivative $D_\mu\phi\equiv\pd_\mu\phi+m A_\mu$. As already discussed, the resulting theory is invariant under the simultaneous gauge transformation
\be\label{gaugeS}
\phi\rightarrow\phi+m\,\alpha\, ,\quad A_\mu\rightarrow A_\mu-\pd_\mu\alpha\,,
\ee
while a unitary gauge choice $\alpha=-\frac{\phi}{m}$, setting $\phi=0$ recovers the original theory in Eq.~\eqref{action}, showing that the new theory is indeed equivalent and also only propagates three degrees of freedom. 

In order to identify the correct massless limit of a theory in terms of its decoupling limit, one first has to identify its lowest strong coupling scale $\Lambda_3$. This scale needs to be separated from the vector mass $m$ by a parametrically large gap, essential for the healthiness of an EFT and is in fact a requirement for a decoupling limit to be well-defined. The lowest strong coupling scale of the theory is, in this case, found by looking considering the pure scalar sector. For instance, the $2\rightarrow 2$ tree-level amplitude $\mathcal{A}_{\scriptstyle2\rightarrow 2}$ induced by the operator of the schematic form 
\begin{equation}
    L^{\myst{GP}}_4 \sim\frac{m^2}{\Lambda_2^4}\frac{1}{m^4}(\pd\phi)^2(\pd^2\phi)^2
\end{equation}
that goes like 
\begin{equation}
    \mathcal{A}_{\scriptstyle2\rightarrow 2}\sim\frac{E^6}{\Lambda_2^4m^2}\,,
\end{equation}
such that at energies above the scale
\be
\Lambda_3\equiv \left(\Lambda_2^2m\right)^{\frac{1}{3}}\,,
\ee
the theory becomes strongly interacting.

Thus, indeed as long as Eq.~\eqref{eq:hierachy scales} is satisfied, this new scale is separated from the vector mass $m$ by a parametrically large gap as required. The decoupling limit is then defined as the zero mass limit, while $\Lambda_3$ is kept fix
\be\label{DL}
m\rightarrow 0 \quad\text{and}\quad \Lambda_2\rightarrow \infty\,,\;\;\text{while}\quad \Lambda_3\equiv\left(\Lambda_2^2m\right)^{\frac{1}{3}}=\text{constant}\,.
\ee
This limit can therefore be viewed as a zoom towards the high-energy regime of the strong coupling scale $\Lambda_3$ of the theory that lies however below the cutoff.

Observe that in the DL, the vector field in the Stückelberg replacement in Eq.~\eqref{Stuckelberg} becomes entirely negligible and only the vector fields that can be written in terms of field strengths $F$ and $\tilde{F}$ survive. In other words, the decoupling limit can be taken directly in the original theory in Eq.~\eqref{action} through the replacements
\be\label{StuckelbergDL}
A_\mu\rightarrow \frac{1}{m}\pd_\mu\phi \,,\quad F\rightarrow F \quad\text{and} \quad\tilde{F}\rightarrow \tilde{F}\,.
\ee
This implies that the coupled gauge symmetry in Eq.~\eqref{gaugeS} is broken apart and only a non-trivial gauge transformation of $A_\mu$ prevails, while the scalar field merely retains an independent global shift symmetry
\be\label{gaugeDL}
\phi\rightarrow\phi+c\, ,\quad A_\mu\rightarrow A_\mu-\pd_\mu\alpha\,.
\ee

Most importantly, in the DL the descriptions of the degrees of freedom is therefore entirely faithful (recall the Definition~\ref{DefFaithfulRep}), with $A_\mu$ exclusively propagating the two vector transverse modes, where the gauge symmetry in Eq.~\eqref{gaugeDL} ensures the absence of ghost instabilities, while the scalar $\phi$ describes the longitudinal scalar DOF. This crucial decoupling of the vector DOFs from the Goldstone boson is what explains the name of the decoupling limit. Moreover, we already want to note that the decoupling limit is also particularly useful when analyzing the quantum stability of an EFT, as it focuses on the high energy behavior right at the relevant scale, while ignoring all others.

In the following, our usage of the decoupling limit will be twofold: In Sec.~\ref{OneLoop} we will first offer explicit calculations of the most important counterterms at one-loop of GP in unitary gauge [Eq.~\eqref{action}] and employ the decoupling limit in order to identify potentially worrisome quantum corrections. In a second step in Sec.~\ref{QuantumStability}, the decoupling limit will be taken from the beginning at the level of the action, which will allow for an explanation of the conclusions drawn on the basis of the explicit calculations at one-loop. This matching provides explicit evidence for the commutation of the two operations of taking the decoupling limit and calculating quantum corrections, which in turn gives additional support for the final thorough hierarchy classification of terms in the EFT performed as a decoupling limit analysis. It should be emphasized at this point, however, that the decoupling limit is just a very convenient tool we choose to employ, but the entire analysis could very well have been done without ever mentioning the St\"uckelberg mechanism.


\subsection{One-Loop Computations in the Unitary Gauge}
\label{OneLoop}

Let's therefore consider the minimal GP theory in the unitary gauge with Lagrangians given in Eq.~\eqref{action} and analyze its radiative stability. After discussing a general power-counting that identify potentially worrisome terms, explicit calculations of the logarithmic divergent part of the $1$PI Feynman diagrams up to four external legs are presented. These computations are again crosschecked with a perturbative calculation based on the generalized Schwinger-DeWitt method developed in \cite{BARVINSKY19851} that we also already employed in the scalar theory case in Sec.~\ref{sSec:Explicit one loop computations luminal H}. In doing so, we in particular correct and extend the previous work of Charmchi \textit{et al.} \cite{Charmchi:2015ggf}. 

\subsubsection{\ul{Feynman Diagram Calculation}}

In this subsection, we compute the quantum behavior of the generalized Proca model in the unitary gauge at the one-loop level using standard Feynman diagram techniques that we already employed for the luminal Horndeski theory in Sec.~\ref{sSec:Explicit one loop computations luminal H}. Each diagram represents a contribution to the reduced matrix element $\mathcal{A}_{1\text{PI}}$ in the perturbative expansion of the S-matrix that is again given by [Eq.~\eqref{Smatrix}]
\be\label{SmatrixGP}
\bra{k_{\text{out}}}\mathcal{S}\ket{k_{\text{in}}}\bigg\rvert_{1\text{PI}}=1+(2\pi)^4\,\delta^4(k_{\text{out}}-k_{\text{in}})\,i \mathcal{A}_{1\text{PI}} \;.
\ee
The reduced matrix element is calculated by summing over all possible Wick contractions of the form
\begin{subequations}\label{Rules}
\begin{align}
& \contraction{}{A}{_\mu(x)}{A}
A_\mu(x) A_\nu(y)=D_{\mu\nu}(x-y)\\
& \bcontraction{}{A^\mu}{(x)}{ket{k,\epsilon}}
A^\mu(x)\ket{k,\epsilon} = \epsilon^\mu_k\; \mr{e}^{-ikx} \\
& \bcontraction{}{ket{k,\epsilon}}{}{A}
\bra{k,\epsilon}A^\mu(x) = \epsilon^{*\mu}_k\; \mr{e}^{ikx} \, ,
\end{align} 
\end{subequations}
where the propagator $D_{\mu\nu}(x-y)$ was defined in \eqref{propagator} and $\epsilon^{\mu}_k$ denotes the associated polarization vector.

Moreover, again following the $\overline{\text{MS}}$-scheme, the one-loop counterterms can be inferred from the UV divergence of the $1$PI diagrams which we will extract using dimensional regularization. We are thus after the log-divergent part of the one-loop $1$PI diagrams with $n$ external legs $\mathcal{A}_n^{\text{div}}$ which will be a function of the external momenta $k_i$, $\scriptstyle i=1,..,n-1$, since the overall delta-function $\delta^4(k_{\text{out}}-k_{\text{in}})$ always allows expressing one momentum $k_N$ in terms of the others. As before, we will also treat all momenta as incoming, such that the overall delta-function translates to $\sum_{i=1}^n k_i=0$.

Our choices of functionals in Eq.~\eqref{FunctionChoice} defining the minimal GP model in Eq.~\eqref{action} only allow for vertices with up to four legs whose value depend on the derivative structure of the insertion that translates in Fourier space into a dependence on the momenta which run on each leg. 
In the following, we will therefore calculate explicit divergent off-shell contributions with up to four external legs and comment on their implications.

\paragraph{2-Point Function.}

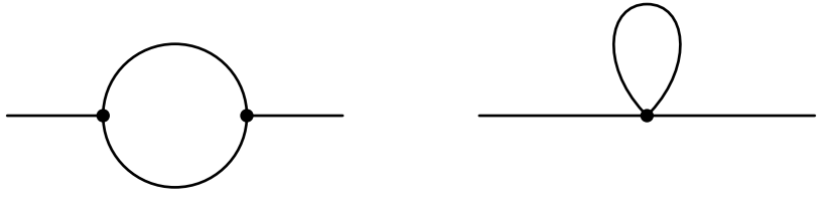
\begin{figure}[H]
\begin{center}
\begin{fmffile}{loops2pf}
\begin{fmfgraph*}(125,65)
     \fmfleft{i}
     \fmfright{o}
     \fmf{plain,tension=3}{i,v1}
     \fmf{plain,left=1}{v1,v2}
     \fmf{plain,left=1}{v2,v1}
     \fmf{plain,tension=3}{v2,o}
     \fmfdot{v1,v2}
    \end{fmfgraph*}
    \qquad \qquad
  \begin{fmfgraph*}(125,65)
 	\fmfleft{i}
     \fmfright{o}
     \fmf{plain,tension=3}{i,v1}
     \fmf{plain}{v1,v1}
     \fmf{plain,tension=3}{v1,o}
     \fmfdot{v1}
\end{fmfgraph*}
\end{fmffile}
\end{center}
\caption{\small{Two distinct one-loop $1$PI diagrams, giving rise to corrections of the two point function. The first diagram represents contributions from the three possible combinations out of $L^{\myst{GP}}_{3}$ and $L^{\myst{GP}}_{5}$ and the second one separate contributions from $L^{\myst{GP}}_{4}$ and $L^{\myst{GP}}_{6}$. Each diagram comes with a symmetry factor of $2$.}}
\label{loopdia} 
\end{figure}

Within our minimal generalized Proca model [Eq.~\eqref{action}], the perturbative renormalization procedure of the two-point function at one-loop requires the calculation of only two distinct $1$PI diagram structures depicted in Fig.\ref{loopdia} coming from $L^{\myst{GP}}_{3,5}$ and $L^{\myst{GP}}_{4,6}$ respectively.

It is instructive to first carry out a power-counting analysis in dimensional regularization of the expected structure of matrix elements. Note that the unusual behavior of the high-energy of the vector propagator in Eq.~\eqref{propagator} considerably modifies such a dimensional analysis in comparison to what we were able to do in the scalar case in Secs.~\ref{sSec:FlatSpaceGalileons} and \ref{sSec:sec_HornSurv}. However, using the power in cutoff $\Lambda_2$ of each vertex in Eq.~\eqref{action} and the fact that the propagator can maximally go like $1/m^2$, together with Lorentz invariance, one can expect the following general structure of the matrix elements
\small
\begin{subequations}
\begin{align}\label{2ptDim}
\mathcal{A}_{\scriptstyle L_3L_3}&\sim\,\frac{m^4}{\Lambda_2^4}\left(m^2+k^2+\frac{k^4}{m^2}+\frac{k^6}{m^4}\right)\,, \\
\mathcal{A}_{\scriptstyle L_3L_5,\,L_4}&\sim\,\frac{m^2}{\Lambda_2^4}\left(m^4+m^2k^2+k^4+\frac{k^6}{m^2}+\frac{k^8}{m^4}\right)\,, \\
\mathcal{A}_{\scriptstyle L_5L_5,\,L_6}&\sim\,\frac{1}{\Lambda_2^4}\left(m^6+m^4k^2+m^2k^4+k^6+\frac{k^8}{m^2}+\frac{k^{10}}{m^4}\right)\,,
\end{align}
\end{subequations}
\normalsize
where $k$ stands for external momenta and the series presumably stops at $1/m^2$ or $1/m^4$, depending on how many propagators are involved in the loop. 

From these schematic expressions, we can conclude two things. First of all, in this massive vector theory, classical operators are renormalized in contrast to the previous theories that we analyzed. For example, from the terms proportional to $k^2$ we expect an explicit detuning of the gauge invariant kinetic combination by the introduction of a quantum correction of the form
\be\label{Odetuning}
\sim\frac{m^4}{\Lambda_2^4}(\pd_\mu A^\mu)^2\,.
\ee
However, it is important to realize that naturally this term is heavily suppressed by the additional factor of $m^4/\Lambda_2^4$ (recall the assumption in Eq.~\eqref{eq:hierachy scales}) and treated as a perturbation it does by definition not alter the number of propagating degrees of freedom of the principal part. Moreover, non-renormalization is foremost a desired property to ensure that there exist a parametrically large high-energy regime for which classical non-linearities can become significant while quantum corrections remain under control. In the high energy limit way above the vector mass, the quantum correction in Eq.~\eqref{Odetuning} remains well suppressed, such that it does not pose any threat to the viability of the EFT structure. In fact, this is a generic result, and all quantum induced operators that are of the same structure as the classical terms are always suppressed my enough factors of $m/\Lambda_4$ so as to play a negligible role in the high-energy limit.

Yet, there still are rather dangerous terms in the expansion in Eq.~\eqref{2ptDim}, namely the induced operators that exhibit powers of the vector mass $m$ in the denominator. Such terms arise because of the aforementioned modified high energy behavior of the vector propagator in Eq.~\eqref{propagator}. Such an enhancement can be enough that at energies in the strong coupling regime close to the cutoff, these quantum terms would dominate over the classical structure and thus destabilize the EFT description. This can explicit be shown by taking the decoupling limit of such terms.
For instance, consider a quantum correction induced by a $\sim k^8$ contribution in $\mathcal{A}_{\scriptstyle L_3L_5,\,L_4}$ or $\mathcal{A}_{\scriptstyle L_5L_5,\,L_6}$ and take the decoupling limit [Eq.~\eqref{DL}] towards the high-energy strong coupling scale
\be\label{DimExpectk8DL}
\sim\frac{\pd^8}{\Lambda_2^4m^2}\,A^2\xrightarrow[]{\text{DL}}\frac{\pd^6}{\Lambda_3^6}\frac{\pd^2}{m^2}\left(\pd\phi\right)^2\rightarrow \infty\,.
\ee
Hence, in the decoupling limit, this term technically blows up, and the quantum correction is out of control compared to the classical kinetic term. We want to stress here that at this point, the decoupling limit is merely a convenient tool in order to investigate the stability of the EFT description by comparing classical and quantum induced operators. Without any knowledge of the St\"uckelberg mechanism \eqref{Stuckelberg} one would reach exactly the same conclusion by power-counting, for instance in cutoff-regularization.

Thus, at first sight, generalized Proca theory might indeed be radiatively unstable, as was previously claimed. In the following we will however show by explicit calculation that despite this bad expectation based on dimensional analysis, the one-loop corrections precisely organize themselves in such a way that their quantum stability is not spoiled by excessive mass terms in the denominator.

In order to evaluate the contributions explicitly one has to perform the usual sum over all possible Wick-contractions which a priory gives $3!^2/2=18$ and $4!/2=12$ possible contractions for each diagram respectively without counting the vertex exchange factor which as usual cancels the prefactor of the exponential expansion. Considering all possible combinations of vertices, summing up all diagrams and following a standard dimensional regularization procedure with $d=4+2\epsilon$\footnote{Note that at one loop the divergent part is blind to the extra factors of $d$ in the Levi-Civita contractions, such that we will disregard them.}, the divergent part of the reduced matrix element up to two powers of momenta reads
\begin{IEEEeqnarray}{rCl}\label{2ptFeyn}
\mathcal{A}_2^{\text{div}}&=\frac{\epsilon^{\alpha}_{\scriptstyle k}\,\epsilon^{\beta}_{\scriptstyle -k}}{16\pi^2\epsilon\,\Lambda_2^4}&\left[\,k^2 \eta_{\alpha\beta}\,m^4\left(-3\,c^2_3+6\,\tilde{c}_4-4\,c_3\tilde{c}_5 +2\,\tilde{c}^2_5\right)+ \eta_{\alpha\beta}\,m^6\left(-3\,c^2_3+6\,\tilde{c}_4\right)\right.\nonumber\\
&&\left.\,
+ k_\alpha k_\beta\,m^4\left(12\,c^2_3-6\,\tilde{c}_4+16\,c_3\tilde{c}_5 +\tfrac{11}{2}\,\tilde{c}^2_5\right)\right. \nonumber\\
&&\left.\,
+k^2 k_\alpha k_\beta\,m^2\left(-3\,c_3^2-2\,c_3\tilde{c}_5+\tfrac{2}{3}\,\tilde{c}_5^2\right)-k^4\eta_{\alpha\beta}\,\frac{m^2}{\Lambda_2^4}\,\frac{19}{6}\,\tilde{c}_5^2\right. \nonumber \\
&&\left.\,
+k^4 k_\alpha k_\beta \left(\tfrac{1}{2}\,c_3^2-\tfrac{13}{12}\,\tilde{c}_5^2\right)+k^6\eta_{\alpha\beta}\,\frac{4}{3}\,\tilde{c}_5^2\right. \nonumber \\
&&\left.\,
+k^6\,\frac{1}{m^2}\,\frac{1}{6}\,\tilde{c}_5^2\left(k_\alpha k_\beta-k^2\eta_{\alpha\beta}\right)
 \right] 
\end{IEEEeqnarray}
where $k_1=-k_2=k$. Note in particular the different relative factors compared to Eq. $(3.13)$ in \cite{Charmchi:2015ggf}. In particular, when trying to reproduce the results of \cite{Charmchi:2015ggf} we already obtain discrepancies in earlier steps, for instance Eq. (3.9). Given that our computation in Eq.~\eqref{2ptFeyn} is confirmed by an entirely independent method (see Eq.~\eqref{finalResultsG2} below) we are rather confident about the correctness of our results.

Observe that $L^{\myst{GP}}_6$ and the term in $L^{\myst{GP}}_4$ proportional to $c_4$ do not contribute at all. Moreover, the counterterm induced by $\tilde{c}_4$ preserves the ghost free structure of the kinetic and mass term $\left(\Box+m^2\right)\,\eta_{\alpha\beta}-\pd_\alpha \pd_\beta$ as could have been expected by the structure of the operator, while the contributions from $L^{\myst{GP}}_3$ and $L^{\myst{GP}}_5$ introduce the anticipated detuning which leads to an operator in  Eq.~\eqref{Odetuning}. 

However, as discussed above, only the terms involving powers of the mass scale in the denominator, with a corresponding power of external momenta equal or higher than eight, are troublesome. Let's thus focus on the last line in Eq.~\eqref{2ptFeyn}. First of all, note that terms with momenta to a power of ten are absent, even though they technically would have been allowed on dimensional grounds. Hence, the structure of the generalized Proca model is precisely such, that these dangerous corrections are canceled. However, there is still a contribution $\sim k^8$. But remarkably, the theory only allows for this contribution to induce a counterterm with the specific gauge preserving combination $\Box\eta_{\alpha\beta}-\pd_\alpha \pd_\beta$. This cures the EFT structure as can be seen in the decoupling limit, where, compared to  Eq.~\eqref{DimExpectk8DL}, we now have
\be\label{DimExpectk6F2DL}
\sim\frac{\pd^6}{\Lambda_2^4m^2}\,F^2\xrightarrow[]{\text{DL}}\frac{\pd^6}{\Lambda_3^6}F^2\,,
\ee
perfectly fitting into the hierarchy between classical and quantum terms.

At this point, the cancellations observed above magically seem to rescue the EFT. In order to better understand these nice properties of the EFT, we will change gears in the next section [Sec.~\ref{QuantumStability}] and perform a thorough decoupling limit analysis, which will allow us to extrapolate quantum stability of the generalized Proca theory in its full generality.
But first, let's also explicitly calculate the higher point one-loop contributions.

\paragraph{3-Point Funciton.}

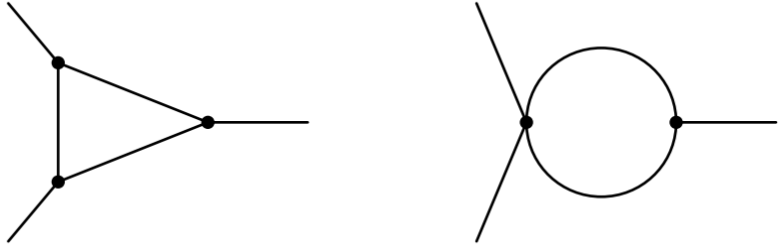
\begin{figure}[H]
\begin{center}
\begin{fmffile}{loops3pf}
\begin{fmfgraph*}(125,90)
     \fmfleft{i1,i2}
     \fmfright{o}
     \fmf{plain,tension=3}{i1,v1}
     \fmf{plain,tension=3}{i2,v2}
     \fmf{plain}{v1,v2}
     \fmf{plain}{v1,v3}
     \fmf{plain}{v2,v3}
     \fmf{plain,tension=3}{v3,o}
     \fmfdot{v1,v2,v3}
    \end{fmfgraph*}
    \qquad \qquad
  \begin{fmfgraph*}(125,90)
 	\fmfleft{i1,i2}
     \fmfright{o}
      \fmf{plain,tension=3}{i1,v1}
     \fmf{plain,tension=3}{i2,v1}
     \fmf{plain,left=1}{v1,v2}
     \fmf{plain,left=1}{v2,v1}
     \fmf{plain,tension=3}{v2,o}
     \fmfdot{v1,v2}
\end{fmfgraph*}
\end{fmffile}
\end{center}
\caption{\small{Two distinct one-loop $1$PI diagrams, giving rise to corrections of the three point function. The diagrams represent contributions from the four possible combinations out of $L^{\myst{GP}}_{3}$ and $L^{\myst{GP}}_{5}$ and contributions from the mixing of even and odd numbered interaction terms respectively.}}
\label{loop3pf} 
\end{figure}

With three external legs, there exist as well two distinct $1$PI one-loop diagram structures represented in Fig.\ref{loop3pf}. The first diagram receives contributions from combinations out of $L^{\myst{GP}}_{3}$ and $L^{\myst{GP}}_{5}$ and the second one pairs $L^{\myst{GP}}_{4,6}$ with $L^{\myst{GP}}_{3,5}$.

Again, we can take a look at what awaits us by invoking dimensional analysis together with Lorentz invariance. Note that there are now three indices of external polarization vectors to be contracted.
\small
\begin{subequations}
\begin{align}\label{3ptDim}
\mathcal{A}_{\scriptstyle L_3^3}&\sim\,\frac{m^6}{ \Lambda_2^6}\left(k+\frac{k^3}{m^2}+\frac{k^5}{m^4}+\frac{k^7}{m^6}\right)\,, \\
\mathcal{A}_{\scriptstyle L_3^2 L_5,\, L_3 L_4}&\sim\,\frac{m^4}{\Lambda_2^6}\left(m^2k+k^3+\frac{k^5}{m^2}+\frac{k^7}{m^4}+\frac{k^9}{m^6}\right)\,, \\
\mathcal{A}_{\scriptstyle L_3 L_5^2,\, L_3 L_6,\, L_5 L_4}&\sim\,\frac{m^2}{\Lambda_2^6}\left(m^4k+m^3k^3+k^5+\frac{k^7}{m^2}+\frac{k^9}{m^4}+\frac{k^{11}}{m^6}\right)\,, \\
\mathcal{A}_{\scriptstyle L_5^3,\, L_5 L_6}&\sim\,\frac{1}{\Lambda_2^6}\left(m^6k+m^4k^3+m^3k^5+k^7+\frac{k^9}{m^2}+\frac{k^{11}}{m^4}+\frac{k^{13}}{m^6}\right)\,,
\end{align}
\end{subequations}
\normalsize
where again $k$ denote external momenta and the series stops at $1/m^4$ or $1/m^6$ depending on how many propagators are involved. Hence, by the same arguments as above, we should give special attention to the $ L_3 L_5^2$, $ L_5^3$ and $ L_5 L_6$ contributions containing external momenta to the power 11 or higher, as they potentially destabilize the EFT structure.

In order to calculate the diagrams explicitly, let's quickly go through the combinatorics. The first diagram gets four different contributions from combinations out of $ L_{3}$ and $ L_{5}$. For each of these, there are $3!$ possible ways of exchanging the vertices, which for the $ L_{3}^3$ contribution for example simply cancels the prefactor of the expansion of the exponential. But for the combinations $ L_{3}^2 L_5$ and $ L_{3} L_5^2$ a redistribution of vertices leads to three distinct results, hence each of these come only with a vertex exchange factor $2!$. After that, there remains $3!^3$ possible Wick-contractions for each diagram, as the symmetry factor is 1. Note that not all of these contractions are independent, of course. 

For the second diagram, there are three distinct channels which need to be considered. For each of these channels at fixed vertices, there are a priori $72$ different ways of contracting in the S-matrix expansion in Eq.~\eqref{SmatrixGP} or in other words $3!4!$ different ways of distributing the insertions over the legs divided by the symmetry factor of two. Note that for vertices with a different number of legs, there is no additional vertex exchange factor which could cancel the $1/2!$ in the exponential expansion in Eq.~\eqref{SmatrixGP}.

Due to a fast-growing complexity of the off-shell expressions, we will only explicitly show here the lowest order momentum result and leave the remaining part in a schematic sum of contributions $M^{\mys{(i)}}_{3}(c_j)$ where $i$ denotes the power of external momenta involved, while the arguments ($c_j$) indicate which diagrams contribute at the given order. In order not to clutter the schematic expansion, we will further leave the argument in a general form whenever all possible contributions are involved. The detailed expressions can be found in an ancillary file of the publication \cite{Heisenberg:2020jtr}.
\begin{IEEEeqnarray}{rCl}\label{3ptFeyn}
\mathcal{A}_3^{\text{div}}&=\frac{m^6}{16\pi^2\epsilon\,\Lambda_2^6}&\left[M^{\scaleto{(1)\mathstrut}{6pt}}_{3}(\scaleto{c_3^3,c_3^2\tilde{c}_5,c_3(c_4+\tilde{c}_4),\tilde{c}_5\tilde{c}_4\mathstrut}{9pt})+\frac{1}{m^2}M^{\scaleto{(3)\mathstrut}{6pt}}_{3}(\scaleto{c_j\mathstrut}{9pt})+\frac{1}{m^4}M^{\scaleto{(5)\mathstrut}{6pt}}_{3}(\scaleto{c_j\mathstrut}{9pt})+\frac{1}{m^6}M^{\scaleto{(7)\mathstrut}{6pt}}_{3}(\scaleto{c_j\mathstrut}{9pt})\right.\nonumber\\
&&\left.\,
+\frac{1}{m^8}M^{\scaleto{(9)\mathstrut}{6pt}}_{3}(\scaleto{c_3\tilde{c}_5^2,\tilde{c}_5^3,c_3\tilde{c}_6,\tilde{c}_5\tilde{c}_6\mathstrut}{9pt})\right]\,.
\end{IEEEeqnarray}
with the explicit leading order expression
\begin{IEEEeqnarray}{rCl}
M^{\scaleto{(1)\mathstrut}{6pt}}_{3}&=-&2i \bigl(2 c_{3}{}^3 + c_{3}{} (c_{4}{} - 2 \tilde{c}_{4}{}) + 9 c_{3}{}^2 \tilde{c}_{5}{} - 9  \tilde{c}_{5}{}\tilde{c}_{4}{}\bigr)\nonumber\\
&&\times \bigl(\epsilon_{23}{} \epsilon k_{11}{} + \epsilon_{13}{} \epsilon k_{22}{} -  \epsilon_{12}{} (\epsilon k_{31}{} + \epsilon k_{32}{})\bigr)\,,
\end{IEEEeqnarray}
where we denote $\epsilon_{ij}\equiv\epsilon_{k_i}\cdot\epsilon_{k_j}$ and $ \epsilon k_{ij}\equiv \epsilon_{k_i}\cdot k_j$.

Regardless of the precise form of the contributions, the important result from Eq.~\eqref{3ptFeyn} is really the simple absence of any contribution going like $\sim k^{13}$ or $\sim k^{11}$ which would destabilize the EFT.\footnote{Note that with an odd number of external fields we do not have any marginal contribution which should preserve gauge invariance.} Hence, again, even though dangerous contributions would technically be allowed in Eq.~\eqref{3ptDim} the calculated series stops at a healthy order.

At this point, one could conclude that all quantum corrections which renormalize the given classical structure involving gauge breaking operators, although being heavily suppressed, come from diagrams involving either $ L^{\myst{GP}}_3$ of $ L^{\myst{GP}}_5$. In other words, upon a restriction of the generalized Proca model to the even numbered terms $ L^{\myst{GP}}_2$, $ L^{\myst{GP}}_4$ and $ L^{\myst{GP}}_6$ by choosing $c_3=\tilde{c}_5=0$, the only one-loop correction so far is a gauge preserving operator proportional to $\tilde{c}_4$. Moreover, this choice is technically natural, since with only $ L^{\myst{GP}}_{4,6}$ insertions no diagrams with an odd number of external legs can be constructed. However, these properties are lost as soon as corrections to higher point functions are taken into account, as it will become clear through the 4-point example below.

\paragraph{4-Point Function.}

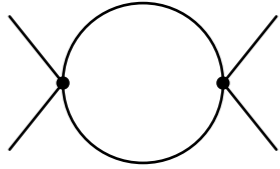
\begin{figure}[H]
\begin{center}
\begin{fmffile}{loops4pf}
\begin{fmfgraph*}(125,50)
     \fmfleft{i1,i2}
     \fmfright{o1,o2}
     \fmf{plain,tension=3}{i1,v1}
      \fmf{plain,tension=3}{i2,v1}
     \fmf{plain,left=1}{v1,v2}
     \fmf{plain,left=1}{v2,v1}
     \fmf{plain,tension=3}{v2,o1}
     \fmf{plain,tension=3}{v2,o2}
     \fmfdot{v1,v2}
    \end{fmfgraph*}
\end{fmffile}
\end{center}
\caption{\small{A $1$PI diagram, giving rise to corrections of the four point function. The diagram represents contributions from the three possible combinations out of $L^{\myst{GP}}_{4}$ and $L^{\myst{GP}}_{6}$.}}
\label{4ptloopdia} 
\end{figure}

We thus also calculate corrections to the four point function, but restrict ourselves by simplicity to the contributions of the diagram in Fig. \ref{4ptloopdia}. The symmetry factor of the diagram is two, such that for each of the three distinct channels there are at first sight $4!4!/2=288$ possible Wick contractions. Again, vertex exchange cancels the $2!$ in the exponential expansion.
\noindent
After adding all the contractions together and going through the dimensional regularization procedure, the schematic form of the result reads
\begin{IEEEeqnarray}{rCl}\label{4ptFeyn}
\mathcal{A}_4^{\text{div}}&=\frac{m^8}{16\pi^2\epsilon\,\Lambda_2^8}&\left[M^{\scaleto{(0)\mathstrut}{6pt}}_{4}(\scaleto{\tilde{c}_4^2\mathstrut}{9pt})+\frac{1}{m^2}M^{\scaleto{(2)\mathstrut}{6pt}}_{4}(\scaleto{(c_4+\tilde{c}_4)^2,\tilde{c}_4\tilde{c}_6\mathstrut}{9pt})+\frac{1}{m^4}M^{\scaleto{(4)\mathstrut}{6pt}}_{4}(\scaleto{c_j^2\mathstrut}{9pt})+\frac{1}{m^6}M^{\scaleto{(6)\mathstrut}{6pt}}_{4}(\scaleto{\tilde{c}_j^2\mathstrut}{9pt})\right.\nonumber \\
&&\left.\,
+\frac{1}{m^8}M^{\scaleto{(8)\mathstrut}{6pt}}_{4}(\scaleto{\tilde{c}_j^2\mathstrut}{9pt})+\frac{1}{m^{10}}M^{\scaleto{(10)\mathstrut}{6pt}}_{4}(\scaleto{(c_4+\tilde{c}_4)\tilde{c}_6,\tilde{c}_6^2\mathstrut}{9pt})+\frac{1}{m^{12}}M^{\scaleto{(12)\mathstrut}{6pt}}_{4}(\scaleto{\tilde{c}_6^2\mathstrut}{9pt})\right]\,,
\end{IEEEeqnarray}
with the same notation as above and details in the ancillary file of \cite{Heisenberg:2020jtr}. Thus, also the classical terms $L^{\myst{GP}}_4$ and $L^{\myst{GP}}_6$ get renormalized. In particular, $M^{\scaleto{(2)\mathstrut}{6pt}}_{4}$ generates operators of the form\footnote{As can be conveniently seen in the position space result in Eq.~\eqref{finalResultsG4} of the next subsection.} $A^\mu A^\nu\left(\pd_\mu A_\alpha \pd_\nu A^\alpha+\pd_\alpha A_\mu \pd^\alpha A_\nu\right)$ which destroy the classical ghost-free tuning. Yet again, these operators come with heavy suppression such that the associated ghost degree of freedom will have a mass way above the cutoff. Only the contribution involving twelve external momenta is potentially worrisome, because it naively diverges in the decoupling limit. Yet, again, we expect the corresponding counterterm to preserves gauge invariance\footnote{Due to the immense complexity of the expression resulting from the Feynman diagram calculation at this order, there is no use in trying to explicitly show this statement in the present unitary gauge calculation.} such that the actual decoupling limit is of the form
\be\label{DimExpectk6F2DL 2s}
\sim\frac{\pd^8}{\Lambda_2^8m^4}\,F^4\xrightarrow[]{\text{DL}}\frac{\pd^8}{\Lambda_3^8}\frac{F^2}{\Lambda_3^4}\,F^2\,.
\ee
as will become clear in section \ref{QuantumStability} when performing the decoupling limit analysis.

\subsubsection{\ul{Schwinger-DeWitt One-Loop Effective Action}}

As already mentioned, as a complementary check we compute one-loop counterterms using an alternative, effective action based method which combines background field and generalized Schwinger-DeWitt techniques \cite{BARVINSKY19851} that we already employed in Sec~\ref{sSec:Explicit one loop computations luminal H}. 

\paragraph{The Schwinger-DeWitt Method for Massive Vector Fields.}
First, however, we need to generalize the covariant Schwinger-DeWitt method that we already introduced in Sec~\ref{sSec:Explicit one loop computations luminal H} to the present case of a massive vector theory.\footnote{See \cite{Ruf:2018vzq} for a specific application of the method to a similar theory.} The starting point is the one-loop effective action [Eq.~\eqref{EffA}]
\be
\label{effAGP}
\Gamma_{1}=\frac{i}{2}\Tr \ln \hat{\mathcal{F}}\, ,
\ee
whose divergent part we want to calculate after a split of the field $A_\mu\rightarrow\bar{A}_\mu+B_\mu$ into background and quantum parts. Here, $\hat{\mathcal{F}}$ denotes the bilinear form of the action in Eq.~\eqref{action}
\be
\label{split}
S^{(2)}=-\frac{1}{2}\int d^4 x\, B\, \hat{\mathcal{F}}\, B\, , \quad \hat{\mathcal{F}}=\hat{D}_2+\hat{P}\, .
\ee
This action can be decomposed into it's principal part $[\hat{D}_2]_{\mu\nu}=\mathcal(\Box+m^2)\eta_{\mu\nu}-\pd_\mu\pd_\nu$ and the subleading perturbations $\hat{P}=\sum_{i=3}^6\,\hat{D}_i(\bar{A})$ depending on the background field, which originate from the interaction terms $L^{\myst{GP}}_{3,..6}$.

The decomposition in Eq.~\eqref{split} together with an expansion of the logarithm in Eq.~\eqref{effAGP} leads to
\be
\label{traceExp}
\Tr \ln \hat{\mathcal{F}}=\Tr \ln \hat{D}_2+\Tr\left[\hat{\mathcal{P}}\hat{D}_2^{-1}\right]-\frac{1}{2}\Tr\left[\hat{P}\hat{D}_2^{-1}\hat{P}\hat{D}_2^{-1}\right]+\mathcal{O}(\hat{P}^3)\, ,
\ee
where the principal operator can be inverted to give
\be
[\hat{D}_2^{-1}]^{\mu\nu}=\frac{1}{\Box+m^2}\left(\eta^{\mu\nu}+\frac{\pd^\mu\pd^\nu}{m^2}\right).
\ee

The trick is now to transform the expansion in Eq.~\eqref{traceExp} above into a sum of terms proportional to universal functional traces whose divergent part can be evaluated by resorting to Schwinger-DeWitt techniques \cite{BARVINSKY19851}. In flat space-time, the only non-vanishing universal functional traces in dimensional regularization with $d=4-2\epsilon$ are
\small
\be
\label{UFT}
\Tr\;\mathcal{P}^{\mu_{\scaleto{1\mathstrut}{4pt}} ... \mu_{\scaleto{2N\mathstrut}{4pt}}}(\bar{A})\,\pd_{\mu_{\scaleto{1\mathstrut}{4pt}}}...\pd_{\mu_{\scaleto{2N\mathstrut}{4pt}}}\,\frac{1}{(\Box+m^2)^n}\bigg\rvert_{\text{div}}=
\,\frac{i}{16\pi^2\,\epsilon}\,\int\mathrm{d}^{4}x\,\mathcal{P}^{\mu_{\scaleto{1\mathstrut}{4pt}} ... \mu_{\scaleto{2N\mathstrut}{4pt}}}(\bar{A})\,\frac{(-1)^{n}\,m^{2l}}{2^{N}\,l!(n-1)!}\,\eta^{(N)}_{\mu_{\scaleto{1\mathstrut}{4pt}} ... \mu_{\scaleto{2N\mathstrut}{4pt}}}\, ,
\ee
\normalsize
where $2N=2n-4+2l$, $N\geq1$, $n\geq1$, $l=0,1,2,...$ and $\eta^{(n-2+l)}_{\mu_{\scaleto{1\mathstrut}{4pt}} ... \mu_{\scaleto{2n-4+2l\mathstrut}{4pt}}}$ is the totally symmetrized product of $n-2+l$ metrics. Note that the background field dependent piece $\mathcal{P}(\bar{A})$ just goes along the ride, regardless of its precise structure. 

The terms appearing in the expansion in Eq.~\eqref{traceExp} are cast into the specific form appearing on the left-hand side of Eq.~\eqref{UFT} by commuting all the operators $1/(\Box+m^2)$ to the right. This procedure is efficient, as each commutation decreases the number of partial derivatives in the numerator of Eq.~\eqref{UFT} compared to the factors of $1/(\Box+m^2)$ and increases the number of derivatives on the background operator $\hat{P}$:
\be\label{Com1}
\left[\frac{1}{\Box+m^2}\,,\hat{P}\right]=-\,\frac{1}{\Box+m^2}\,[\Box\,,\hat{P}]\,\frac{1}{\Box+m^2}\,,
\ee
where
\begin{equation}
    [\Box\,,\hat{P}]=(\Box\hat{P})+2(\pd^\alpha\hat{P})\pd_\alpha
\end{equation}
This means that the while the log expansion in Eq.~\eqref{traceExp} will be cut off by the maximum number of background fields one is interested in, the iterative commutation of operators Eq.~\eqref{Com1} will constantly increase the number of derivatives applied on the background fields, which thus allows for the computation of counterterms up to any desired but fixed order in derivatives as well as in the fields.

Note that in contrast to the massless case, the expansion in factors of $\frac{m^2}{\Box}$ measured by the integer $l$ allows for divergent contributions of the linear terms in Eq.~\eqref{traceExp} with $n=1$. However, tadpole contributions arising from the interaction terms $L^{\myst{GP}}_{3}$ and $L^{\myst{GP}}_{5}$ are immediately ruled out by the odd number of derivative factors. Thus, the linear terms will only provide potential corrections to the two point function via contributions from $L_{4}$ and $L_{6}$. 
The next terms in the log expansion of Eq.~\eqref{traceExp} $\sim \hat{P}^2$ give rise to contribution to the 2-point function originating in the interactions $L^{\myst{GP}}_{3}$ and $L^{\myst{GP}}_{5}$ and contributions to the 3- and 4-point functions by a suitable mixing of all interaction terms.
As concerns higher point results, we won't need terms in the log expansion of  Eq.~\eqref{traceExp} higher than $\sim \hat{P}^3$, as these cover all cases considered through Feynman calculations carried out above.

\paragraph{Results up to 4-Point.}
For the full logarithmically divergent one-loop contribution to the 2-point effective action up to a power of four derivatives we find
\begin{IEEEeqnarray}{rCl}\label{finalResultsG2}
\Gamma_{1,2}^{\rm div} &=\frac{m^4}{16 \pi^2\epsilon\,\Lambda_2^4}\,\int \mathrm{d}^4x &\, \left[\;\left(-\frac{3}{2}c_3^2+3\,\tilde{c}_4-2\,c_3\tilde{c}_5+\tilde{c}_5^2\right)\pd_\mu \bar{A}_\nu\pd^\mu \bar{A}^\nu\right.\nonumber\\
&&\left.
+\left(-\frac{3}{2}c_3^2+3\,\tilde{c}_4\,\right)m^2\bar{A}_\mu\bar{A}^\mu\phantom{\tfrac{\tilde{c}_5^2}{\Lambda^4}}\right.\nonumber\\
&&\left.
+\left(\;\;\,6\,c_3^2-3\,\tilde{c}_4+8\,c_3\tilde{c}_5+\frac{11}{4}\tilde{c}_5^2\right)(\pd_\mu \bar{A}^\mu)^2 \right.\nonumber\\
&&\left. 
-\left(\;\;\,\frac{3}{2}c_3^2+c_3\tilde{c}_5-\frac{1}{3}\tilde{c}_5^2\right)\frac{1}{m^2}\pd_\mu\pd_\nu \bar{A}^\nu\Box\bar{A}^\mu\right.\nonumber\\
&&\left. 
-\;\;\;\,\frac{19}{12}\, \tilde{c}_5^2\,\frac{1}{m^2}\Box \bar{A}_\mu\Box\bar{A}^\mu \right].
\end{IEEEeqnarray}
Additionally, we present here a selection of the most relevant 3 and 4-point leading order results up to three powers of external momenta

\begin{IEEEeqnarray}{rCl}\label{finalResultsH}
\Gamma_{1,3}^{\rm div} &\supset\frac{1}{16 \pi^2\epsilon}\frac{m^6}{\Lambda_2^6}\int \mathrm{d}^4x \,& \bigg[9\,\tilde{c}_4\tilde{c}_5\,\bar{A}^2\pd_\alpha\bar{A}^\alpha+\frac{1}{12}\tilde{c}_5^3\frac{1}{m^2}\Big\{11\left(\pd_\alpha\bar{A}^\alpha\right)^3\nonumber\\
&&
+51\,\pd_\alpha\bar{A}^\alpha\left(\pd_\mu\bar{A}_\nu\pd^\mu\bar{A}^\nu+\pd_\mu\bar{A}_\nu\pd^\nu\bar{A}^\mu\right)\nonumber\\
&&
-6\,\pd^\mu\bar{A}^\nu\pd_\alpha\bar{A}_\mu\pd^\alpha\bar{A}_\nu-2\,\pd^\mu\bar{A}^\nu\pd_\nu\bar{A}_\alpha\pd^\alpha\bar{A}_\mu\Big\} \bigg]
\label{finalResultsG3}\\
\Gamma_{1,4}^{\rm div} &\supset\frac{1}{16 \pi^2\epsilon}\frac{m^6}{\Lambda_2^8}\int \mathrm{d}^4x \,&\bigg[9\,\tilde{c}_4^2\,m^2\big(\bar{A}^2\big)^2\nonumber\\
&&
-\left(2\,c_4^2+16\,c_4\tilde{c}_4+20\,\tilde{c}_4^2+3\,\tilde{c}_4\tilde{c}_6\right)\bar{A}^2\, \pd_\alpha\bar{A}_\beta\pd^\beta\bar{A}^\alpha\nonumber\\
&&
-\left(2c_4^2-10c_4\tilde{c}_4-2\tilde{c}_4^2-3\tilde{c}_4\tilde{c}_6\right)\bar{A}^2\left(\pd_\alpha\bar{A}^\alpha\right)^2\nonumber\\
&&
+2\left(c_4^2+3c_4\tilde{c}_4+9\tilde{c}_4^2\right)\bar{A}^2\,\pd_\alpha\bar{A}_\beta\pd^\alpha\bar{A}^\beta\nonumber\\
&&
-2\left(c_4^2+3c_4\tilde{c}_4+3\tilde{c}_4^2\right)\bar{A}^\mu\bar{A}^\nu\pd_\mu\bar{A}_\alpha\pd_\nu\bar{A}^\alpha\nonumber\\
&&
+4\left(4c_4^2+\tilde{c}_4^2\right)\bar{A}^\mu\bar{A}^\nu\pd_\mu\bar{A}_\nu\pd_\alpha\bar{A}^\alpha\nonumber\\
&&
-2\left(2c_4^2-3c_4\tilde{c}_4+5\tilde{c}_4^2\right)\bar{A}^\mu\bar{A}^\nu\pd_\alpha\bar{A}_\mu\pd^\alpha\bar{A}_\nu\bigg].\label{finalResultsG4}
\end{IEEEeqnarray}

\paragraph{Cross-Check Against Feynman Diagram Computations.}
In order to relate these results to the Feynman diagram calculations carried out above, recall that the effective action is a generating functional of $1$PI correlation functions
\be
\frac{\delta^n\Gamma[\bar{\pi}]}{\delta\bar{A}^{\mu_1}(x_1)...\delta\bar{A}^{\mu_n}(x_n)}\biggr\rvert_{\bar{A}=\langle A\rangle} =\langle A_{\mu_1}(x_1)...A_{\mu_n}(x_n) \rangle_{1\rm PI}\,.
\ee
The $1$PI correlation functions in turn are given by the sum of all $1$PI diagrams with $n$ external points. Thus, Fourier transformed functional derivatives of divergent one-loop effective action results at vanishing mean field should coincide with the corresponding divergent off-shell results of the $1$PI Feynman diagrams. We explicitly checked this for all the expressions above. For instance, for the 2-point result in Eq.~\eqref{finalResultsG2} it can be seen by eye that it precisely matches the momentum space calculation in Eq.~\eqref{2ptFeyn} as the conversion merely introduces a factor of $1/2$. 

These results therefore serve as highly nontrivial checks of the Feynman diagram based momentum space calculations above, as the only common ground of the two methods is the input of the Lagrangian. Due to exceeding computational cost for results at high orders in derivatives, we restricted ourselves to the computation of terms involving a maximum of four derivatives acting on the background fields, which translates into a limitation to four powers of external momenta. However, since Feynman diagram calculations are not structured in an expansion of external momenta\footnote{Meaning that the integration of a given Feynman loop-diagram expression directly gives the result to all orders of external momenta.} the matching of the results at low powers of momenta gives very strong support for the validity of the entire momentum-space calculation. This provides us with confidence for the correctness of our results, especially in comparison with the previous computations in \cite{Charmchi:2015ggf}.


\subsection{Decoupling Limit Analysis}
\label{QuantumStability}

Starting from the above one-loop results, we will now intend a complete radiative stability analysis of the Generalized Proca EFT. To this end we will leave the unitary gauge employed in the previous section and instead take the decoupling limit already at the level of the Lagrangian. In a first step, this will allow us to shed light on the observed cancellations in the unitary gauge calculations by rederiving the most important aspects of the results still within our minimal generalized Proca model Eq.~\eqref{action}. At the same time, this confirms that just as within massive gravity theories (see e.g. \cite{Hinterbichler:2011tt,deRham:2013qqa}) taking the decoupling limit and computing quantum corrections are two operations which commute. 

Based on this knowledge, we will in a second step establish healthiness of the full generalized Proca theory in Eq.~\eqref{Lagrangians} under quantum corrections at all orders by showing that all these models admit a well-defined decoupling limit where classical and quantum operators are structured in a sound hierarchy. In particular, we will identify the correct classical and quantum expansion parameters of the theory. 

\subsubsection{\ul{Reinterpretation of the Unitary Gauge Results}}

\paragraph{Decoupling Limit Action.}
As discussed in Sec.~\ref{ssSec:Decoupling Limit}, rewriting the generalized Proca model in Eq.~\eqref{action} by introducing a St\"uckelberg field $\phi$ allows one to take a smooth $m\rightarrow 0$ limit without loosing any degrees of freedom, known as the decoupling limit. In practice, this limit can be taken by carrying out the replacements in Eq.~\eqref{StuckelbergDL} within the unitary gauge action [Eq.~\eqref{action}] which yields
\begin{equation}\label{actionDL}
\begin{split} 
& L^{\myst{GP}}_{2}=-\frac{1}{4}F^2+\frac{1}{2}(\pd\phi)^2\,,\\
& L^{\myst{GP}}_3 = \frac{1}{\Lambda_3^3} \,c_3\, (\pd\phi)^2 \Box \phi\,, \\
& L^{\myst{GP}}_4 = \frac{1}{\Lambda_3^6}\,(\pd\phi)^2\,c_4\,\left[(\Box \phi)^2-\left(\pd_\mu \pd_\nu\phi\right)^2\right]\,, \\
& L^{\myst{GP}}_5 = -\frac{1}{\Lambda_3^3}\,\tilde{c}_5\,\tilde{F}^{\mu\alpha}\tilde{F}^{\nu}_{\;\alpha} \pd_\mu \pd_\nu\phi\,, \\
&  L^{\myst{GP}}_6 = -\frac{1}{\Lambda_3^6}\,\tilde{c}_6\,\tilde{F}^{\mu\alpha}\tilde{F}^{\nu\beta} \pd_\mu \pd_\nu\phi\pd_\alpha \pd_\beta\phi\,.
\end{split}
\end{equation}
Hence, the mass term of the vector field and the term proportional to $\tilde{c}_4$ vanish, while the terms proportional to $c_3$ and $c_4$ reduce to pure scalar Galileon interactions that we introduced in Eq.~\eqref{eq:ActionGalileons}.

Recall that in the DL the transverse modes, that are described by a massless and gauge invariant vector field $A_\mu$, are decoupled in a symmetry sense from the longitudinal scalar mode $\phi$. Moreover, for an interacting theory, the decoupling limit works as a high energy limit way above the vector mass and right at the lowest strong coupling scale $\Lambda_3=(\Lambda_2^2m)^{\frac{1}{3}}$.
This makes contact with the Goldstone boson equivalence theorem: At rest, all three polarizations are equivalent, but at higher energies, the transverse polarizations and the rapidly moving longitudinal polarization are clearly distinguished. 

In the decoupling limit, the analysis of the radiative stability of the EFT is thus reduced to an analysis of the behavior of the Goldstone. The only operators which survive this high energy limit are the least suppressed ones and thus the decoupling limit puts focus on the operators with the poorest behavior, showing its particular usefulness for analyzing the quantum stability of an EFT.
In the light of this discussion, it is perhaps less surprising that it turns out that the structure of the generalized Proca EFT bares many similarities with equally self-interacting non-abelian $SU(2)$ spin $1$ fields endowed with a mass term and thus the hierarchy structure of the weak sector of the Standard Model EFT.

\paragraph{Special Structure of High Momenta Operators.}
We will now show how to infer the explicitly computed general structure of highest order operators in the unitary gauge directly from the decoupling limit. This means that in general, from the healthiness of the EFT in the decoupling limit, we can predict the qualitative aspect of the cancellations of leading order terms in the unitary gauge observed in the previous section [Sec.~\ref{OneLoop}].

The crucial thing to note is that in the DL the propagators of the fields regain their usual $1/p^2$ high energy behavior. This is manifest from the explicit DL Lagrangian in Eq.~\eqref{actionDL}, but can also be concluded in the following alternative way. Namely, in the full but Stückelberged action, instead of the unitary gauge choice one could for instance choose the gauge $\pd_\mu A^\mu+m\phi=0$. Implementing this new gauge in a Fadeev-Popov procedure, one obtains the following propagators of $A_\mu$ and $\phi$ \cite{Hinterbichler:2011tt}
\be
\frac{-i\,\eta_{\mu\nu}}{p^2+m^2}\quad\text{and}\quad \frac{-i}{p^2+m^2}\,,
\ee
which at high energies behave as $\sim \frac{1}{p^2}$ compared to the $\sim\frac{1}{m^2}$ behavior of the unitary gauge propagator in Eq.~\eqref{propagator}. These propagators then nicely translate into the propagators of a massless, gauge invariant vector field in the Lorenz gauge and a massless scalar field respectively.

This recovery of the usual behavior of propagators is what allows us now to employ the time-tested methods of power-counting of Secs.~\ref{sSec:FlatSpaceGalileons} and \ref{sSec:sec_HornSurv}. Indeed, since the $L^{\myst{GP}}_3$ and $L^{\myst{GP}}_4$ Lagrangians precisely reduce to scalar Galileon terms in the decoupling limit, their quantum induced operators at one loop will coincide with the analysis carried out in Sec.~\ref{sSec:FlatSpaceGalileons}, namely [Eq.~\eqref{eq:Fgal second}]
\be\label{eq:GalileonLagrangian Oneloop Schematic Second}
L^{\text{q}}\sim\partial^{4}\left(\frac{\partial^2\pi}{\Lambda_3^3}\right)^i\sim (\partial \pi)^2  (\alpha_\text{q})^{3} \,(\alpha_\text{cl})^{i-2}\,, \quad i\geq2\,,
\ee
where the classical and quantum expansion parameters are given by [Eq.~\eqref{eq:Classical And Quantum Expansion Op}]
\begin{eqnarray}\label{eq:Classical And Quantum Expansion Op second}
    \alpha_\text{cl}= \frac{\partial^2\pi}{\Lambda_3^3}\,,\quad  \alpha_\text{q}= \frac{\partial^2}{\Lambda_3^2}\,.
\end{eqnarray}

From these results, one can now readily translate back to the unitary gauge results and obtain their highest order terms in momenta (or derivatives). More precisely, we can recover all highest order quantum corrections generated exclusively through $L^{\myst{GP}}_3$ and the $c_4$ term in $L^{\myst{GP}}_4$ via a replacement $\pd\phi\rightarrow m\,A$. Indeed, from Eq.~\eqref{eq:GalileonLagrangian Oneloop Schematic Second} the corresponding highest order quantum correction in the unitary gauge schematically read
\be\label{FL3L4}
L^{\text{q}}_{L_3,L_4(c_4)}\sim m^i\frac{\pd^{4+i}}{\Lambda_3^{3i}}\,A^i\sim \frac{\pd^{4+i}}{\Lambda_2^{2i}}\,A^i   \,, \quad i\geq2\,.
\ee
For instance, the two-point correction, corresponding to $i=2$, precisely matches the highest order result of the $c_3^2$ term in the explicit result in Eq.~\eqref{2ptFeyn}.
Moreover, these considerations also directly explain the absence of high momenta power contributions proportional to $c_4$ in Eq.~\eqref{2ptFeyn}, because $L^{\myst{GP}}_{4}$ in the DL does not contribute to the one-loop correction with two external legs. In fact, the result in Eq.~\eqref{FL3L4} matches precisely with all the explicit unitary gauge results (at all numbers of external legs $i$ considered) [Eqs.~\eqref{3ptFeyn} and \eqref{4ptFeyn}].

Recall, however, that only quantum corrections involving $L^{\myst{GP}}_5$ and $L^{\myst{GP}}_6$ required non-trivial cancellations of the leading order estimates in order to remain healthy. These cancellations can also readily be explained from the point of view of the decoupling limit. 

To see this, let's first focus on the one loop corrections to the propagator. $L^{\myst{GP}}_5$ in the decoupling limit is an interaction term between the massless vector and the Goldstone [Eq.~\eqref{actionDL}]. This directly implies that in the high energy limit no one-loop diagram can be formed between the two terms $L^{\myst{GP}}_3L^{\myst{GP}}_5$ and hence, terms proportional to $c_3\tilde{c}_5$ have no impact close to the cutoff scale $\Lambda_3$ and remain highly suppressed. This is in perfect agreement with the obtained results in Eq.~\eqref{2ptFeyn}. 

\begin{figure}[H]
\begin{center}
\begin{fmffile}{loops2pfDL}
\begin{fmfgraph*}(125,50)
     \fmfleft{i}
     \fmfright{o}
     \fmf{plain,tension=3}{i,v1}
     \fmf{photon,left=1}{v1,v2}
     \fmf{photon,left=1}{v2,v1}
     \fmf{plain,tension=3}{v2,o}
     \fmfdot{v1,v2}
    \end{fmfgraph*}
    \qquad \qquad
  \begin{fmfgraph*}(125,50)
     \fmfleft{i}
     \fmfright{o}
     \fmf{photon,tension=3}{i,v1}
     \fmf{plain,left=1}{v1,v2}
     \fmf{photon,left=1}{v2,v1}
     \fmf{photon,tension=3}{v2,o}
     \fmfdot{v1,v2}
    \end{fmfgraph*}
\end{fmffile}
\end{center}
\caption{\small{Two distinct one-loop $L^{\myst{GP}}_{5}$ diagram contributions in the decoupling limit, giving rise to corrections of the two point function. Solid lines represent scalar legs or propagators. Each external leg comes with two derivatives applied on the field $\sim\pd^2\phi$. Wiggled lines correspond to gauge preserving vector legs $\sim F$ or corresponding propagators. In the decoupling limit, propagators have a good $\sim 1/p^2$ high energy behavior.}}
\label{2ptloopdiaDL} 
\end{figure}
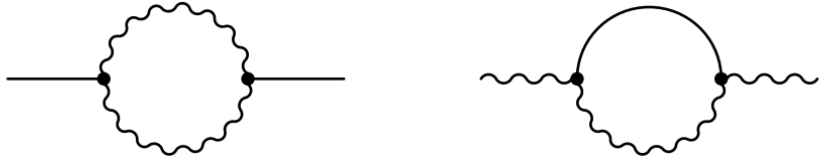

However, two distinct diagrams can be formed with two $L^{\myst{GP}}_5$ insertions depicted in Fig.~\ref{2ptloopdiaDL}, where straight lines denote scalar legs $\sim\pd^2\phi$ or scalar propagators that now behave as $\sim 1/p^2$, while the wiggled lines correspond to massless vector legs $\sim F$ or propagators $\sim 1/p^2$.

\noindent
The first diagram induces a counterterm proportional to $(\pd^2\phi)^2$ since, just as in the pure Galileon case, the external legs carry two derivatives per scalar field. This is a contribution of the same order as in Eq.~\eqref{eq:GalileonLagrangian Oneloop Schematic Second}. On the other hand, because of the two vector legs, the contribution from the second diagram is bound to give rise to a gauge invariant operator, such that in total we have
\be\label{ctDL1}
L^{\text{q}}_{L_5L_5}\sim \tilde{c}_5^2\left[\frac{\pd^4}{\Lambda_3^6}\left(\pd^2\phi\right)^2+\frac{\pd^6}{\Lambda_3^6}\left(F\right)^2\right]\,.
\ee
This nicely explains why there could not be a contribution proportional to $\sim k^{10}$ in the one-loop two-point corrections in the unitary gauge result in Eq.~\eqref{2ptFeyn}, simply from the fact that terms of this order cannot be formed in the decoupling limit. Moreover, again translating back to the unitary gauge through $\pd\phi\rightarrow m\,A$ leads to
\be
L^{\text{q}}_{L_5L_5}\sim \tilde{c}_5^2\left[\frac{m^2\pd^4}{\Lambda_3^6}\left(\pd A\right)^2+\frac{\pd^6}{\Lambda_3^6}\left(F\right)^2\right]\sim \tilde{c}_5^2\left[\frac{\pd^4}{\Lambda_2^4}\left(\pd A\right)^2+\frac{\pd^6}{\Lambda_2^4m^2}\left(F\right)^2\right] \,,
\ee
which is qualitatively in perfect agreement with the $F^2$ structure obtained in the last line of Eq.~\eqref{2ptFeyn}.

\begin{figure}[H]
\begin{center}
\begin{fmffile}{loops4pfDL}
\begin{fmfgraph*}(100,40)
     \fmfleft{i1,i2}
     \fmfright{o1,o2}
     \fmf{plain,tension=3}{i1,v1}
      \fmf{plain,tension=3}{i2,v1}
     \fmf{photon,left=1}{v1,v2}
     \fmf{photon,left=1}{v2,v1}
     \fmf{plain,tension=3}{v2,o1}
     \fmf{plain,tension=3}{v2,o2}
     \fmfdot{v1,v2}
    \end{fmfgraph*}
    \qquad
  \begin{fmfgraph*}(100,40)
     \fmfleft{i1,i2}
     \fmfright{o1,o2}
     \fmf{photon,tension=3}{i1,v1}
      \fmf{plain,tension=3}{i2,v1}
     \fmf{photon,left=1}{v1,v2}
     \fmf{plain,left=1}{v2,v1}
     \fmf{photon,tension=3}{v2,o1}
     \fmf{plain,tension=3}{v2,o2}
     \fmfdot{v1,v2}
    \end{fmfgraph*}
    \qquad
  \begin{fmfgraph*}(100,40)
     \fmfleft{i1,i2}
     \fmfright{o1,o2}
     \fmf{photon,tension=3}{i1,v1}
      \fmf{photon,tension=3}{i2,v1}
     \fmf{plain,left=1}{v1,v2}
     \fmf{plain,left=1}{v2,v1}
     \fmf{photon,tension=3}{v2,o1}
     \fmf{photon,tension=3}{v2,o2}
     \fmfdot{v1,v2}
    \end{fmfgraph*}
\end{fmffile}
\end{center}
\caption{\small{Three distinct one-loop $L^{\myst{GP}}_{6}$ diagram contributions in the decoupling limit, giving rise to corrections of the four point function. Solid lines represent scalar legs or propagators. Each external leg comes with two derivatives applied on the external field $\sim\pd^2\phi$. Wiggled lines correspond to gauge preserving vector legs $\sim F$ or corresponding propagators. In the decoupling limit, propagators have a good $\sim 1/p^2$ high energy behavior.}}
\label{4ptloopdiaDL} 
\end{figure}
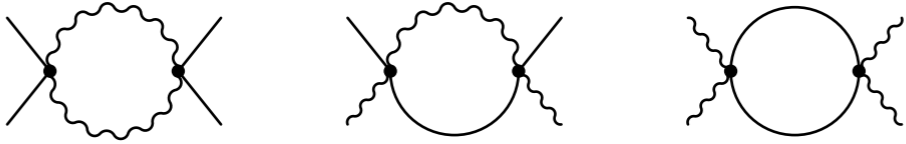

Similar for higher point functions. For example, $\tilde{c}_5^3$ diagrams at most generate counterterms of the form \small$\sim\tfrac{1}{\Lambda_3^9}\pd^6(\pd^2\phi)F^2$\normalsize, while it is not possible to form contributions going like \small$\sim\tfrac{1}{\Lambda_3^9}\pd^8F^3$\normalsize. This is explicitly confirmed by the 3-point calculation in Eq.~\eqref{3ptFeyn}. In a similar manner, the 4-point results can be understood. For instance, the $\tilde{c}_6^2$ high energy contributions come from the three diagrams in Fig.~\ref{4ptloopdiaDL} in the decoupling limit. The schematic form of the corresponding counterterms is
\be\label{ctDL2}
L^{\text{q}}_{L_6L_6}\sim \tilde{c}_6^2\left[\frac{\pd^4}{\Lambda_3^{12}}\left(\pd^2\phi\right)^4+\frac{\pd^6}{\Lambda_3^{12}}\left(\pd^2\phi\right)^2(F)^2+\frac{\pd^8}{\Lambda_3^{12}}\left(F\right)^4\right]\,.
\ee
In unitary gauge this corresponds to
\be\label{UnitaryGaugeL5}
L^{\text{q}}_{L_6L_6}\sim \tilde{c}_6^2\left[\frac{m^4\pd^4}{\Lambda_3^{12}}\left(\pd A\right)^4+\frac{m^2\pd^6}{\Lambda_3^{12}}\left(\pd A\right)^2\left(F\right)^2+\frac{\pd^9}{\Lambda_3^{12}}\left(F\right)^4\right]\,,
\ee
which for instance shows that $M^{\scaleto{(12)\mathstrut}{6pt}}_{4}$ in \eqref{4ptFeyn} indeed possesses a gauge invariant structure.

\subsubsection{\ul{Quantum Stability of Generalized Proca Theories}}

The results above confirm that the decoupling limit is the ideal tool to analyze the quantum stability of GP type theories, as it allows for a pertinent understanding of the higher energy behavior in simple power-counting terms. In other words, the regime at high energy that is most important to quantum stability can viably be analyzed using dimensional arguments. Thus, generalizing the decoupling limit arguments above to all higher point functions, as well as higher loop orders, indicate that the EFT structure is stable under all possible quantum corrections from an EFT point of view. We will now precisely provide such a general argument that will in fact go beyond the specific minimal model chosen in Eq.~\eqref{action}, but includes the full spectrum of generalized Proca EFT's in Eq.~\eqref{Lagrangians}. 

First, note that up to total derivatives, a generic generalized Proca theory with classical action [Eq.~\eqref{Lagrangians}] can be expanded in a sum of terms with a schematic form\footnote{We assume here implicitly, that the generic functions $f$ in [Eq.~\eqref{Lagrangians}] are smooth, such that they admit a Taylor series expansion.}
\be\label{SchematicL}
L^{\myst{GP}}\sim \left(F^2+m^2 A^2\right)\left(\frac{mA}{\Lambda_2^2}\right)^{2a_1}\left(\frac{F}{\Lambda_2^2}\right)^{a_2}\left(\frac{\pd A}{\Lambda_2^2}\right)^{a_3}\,,\quad a_{1,2,3}\geq0\,,\;a_3\leq 4\, ,
\ee
where of course all suppressed Lorentz indices need to be contracted, and the classical structure is such that the theory only propagates the required three degrees of freedom. The first two operators in Eq.~\eqref{SchematicL} take care of the dimension, while combinations of the dimensionless factors with powers $a_i$ can be divided into three classes of terms:
\begin{itemize}
\item Terms coming from expanding the function $f_2$ in terms of its arguments, which includes the kinetic and mass term. In Eq.~\eqref{SchematicL}, these terms correspond to operators with $a_3=0$, where the kinetic and mass term are the ones given by $a_1=a_2=a_3=0$. 

Note that on top of the basic $L^{\myst{GP}}_2$ terms, in the specific minimal model [Eq.~\eqref{action}] employed in the major part of this work also the term proportional to $\tilde{c}_4$ correspond to this class.
\item Terms resulting from the expansion of the functions $f_i$, $i=3,..,6$. These terms correspond to Galileon like contributions. They do not include any powers of $a_2$ and are proportional to $m^2A^2$ only. 

In Eq.~\eqref{action} these correspond to the $c_3$ and $c_4$ terms.
\item Terms coming from expanding the functions $\tilde{f}_i$, $i=4,5,6$. These terms correspond to genuinely new vector derivative self interactions with $a_2=0$ as well  but proportional to $F^2$. 

In Eq.~\eqref{action} these correspond to the terms proportional to $\tilde{c}_5$ and $\tilde{c}_6$.
\end{itemize}
This justifies the specific choice of our model in Eq.~\eqref{FunctionChoice} in retrospective, since we cover all interesting cases. 

The St\"uckelberg trick described in Sec.~\ref{ssSec:Decoupling Limit} can now be applied to each classical term in Eq.~\eqref{SchematicL} which reformulates the theory using gauge redundancy. This alternative formulation of a generic generalized Proca term then allows performing the decoupling limit [Eq.~\eqref{DL}] on each individual operator. We obtain that indeed the limit is well-defined, and the classical Lagrangian reduces to
\be\label{SchematicLDL}
L^{\myst{GP}}_{\text{DL}}\sim \left(F^2+(\pd\phi)^2\right)\left(\frac{\pd^2\phi}{\Lambda_3^3}\right)^{a_3}\,,\quad 3\geq a_{3}\geq0\,,
\ee
where terms proportional to $F^2$ are limited to $a_{3}\leq 2$. Moreover, all the operators proportional to $f_6(x)$ in Eq.~\eqref{Lagrangians} are actually total derivatives \cite{Heisenberg:2014rta,Jimenez:2016isa,Heisenberg:2018vsk,Jimenez:2019hpl}, such that in particular also the operator which would lead to a $a_{3}=4$ decoupling limit contribution has vanishing equations of motion. This nicely reflects the fact that for the terms involving only the scalar field $\phi$, the individual classical operators only remain ghost-free in the specific scalar Galileon form.

Including now loop contributions, the specific form of Eq.~\eqref{SchematicLDL} implies that each vertex in the decoupling limit comes at least with a factor of $1/\Lambda_3^3$. This means that in dimensional regularization and only considering $1$PI diagrams at one loop, where each vertex at least includes one external leg while two legs are contributing to the loop, there are only two distinct schematic building blocks for quantum induced operators \small$\frac{\pd F}{\Lambda_3^3}\;$\normalsize and \small$\frac{\pd^2\phi}{\Lambda_3^3}\;$\normalsize. Therefore, a general one loop counterterm in the decoupling limit has the generic form\footnote{Once again this schematic form is fixed through Lorentz invariance, power-counting and the well-behaved propagators in the decoupling limit}
\be\label{fullDL}
L^{\text{q}}_{\text{DL}}\sim \pd^4\left(\frac{\pd F}{\Lambda_3^{3}}\right)^{2 b_2}\left(\frac{\pd^2\phi}{\Lambda_3^{3}}\right)^{b_3}\sim
\begin{cases}
F^2\left(\frac{\pd^2}{\Lambda_3^2}\right)^{2+b_2}\left(\frac{F^2}{\Lambda_3^4}\right)^{b_2-1}\left(\frac{\pd^2\phi}{\Lambda_3^3}\right)^{b_3}&,\quad b_2\geq 1\\
\\
(\pd\phi)^2\left(\frac{\pd^2}{\Lambda_3^2}\right)^{3+b_2}\left(\frac{F^2}{\Lambda_3^4}\right)^{b_2}\left(\frac{\pd^2\phi}{\Lambda_3^3}\right)^{b_3-2}&,\quad b_3\geq 2
\end{cases} 
\ee
where $2 b_2+b_3=N\geq 2$ with $N$ the number of external fields and $b_{2,3}\geq0$ positive integers.\footnote{The two possible reformulations in Eq.~\eqref{fullDL} result from comparing the operator to the two kinetic terms $F^2$ and $(\pd\phi)^2$ of the theory. For most $b_i$ values, either one can be employed. Only for $b_1=0$ the upper one looses its sense, while the lower one is not valid whenever $b_3<2$.} For example, the cases $\{b_2=0,b_3=2\}$, $\{b_2=1,b_3=0\}$ correspond to the counterterms we already encountered in Eq.~\eqref{ctDL1}, while $\{b_2=0,b_3=4\}$, $\{b_2=1,b_3=2\}$ and $\{b_2=2,b_3=0\}$ are covered in Eq.~\eqref{ctDL2}. Thus, on top of the two expansion parameters $\alpha_{\text{cl}}$ and $\alpha_{\text{q}}$ defined in the pure scalar Galileon context in Eq.~\eqref{eq: New Classical And Quantum Expansion Op}, we identify a second quantum expansion parameter 
\be
\alpha_{\tilde{\text{q}}}=\frac{F^2}{\Lambda_3^4}\,.
\ee
One can easily generalize this analysis to higher loops. Each additional loop comes with an increase in factors of $1/\Lambda_3^3$ compared to the same diagram without the additional loop. This is because in order to add a loop to a diagram while keeping the number of external legs fixed necessarily requires the inclusion of an additional vertex or the addition of legs to existing vertices. Now, since the number of external legs remains the same in this comparison, to match dimensions these factors can only be compensated with additional powers of derivatives. Thus, higher loops will merely introduce additional factors of $\alpha_{\text{q}}$.\footnote{Actually, Lorentz invariance requires the additional factor to be $\pd^6/\Lambda_3^6$.}

From here on, the analysis exactly parallels the one employed for the consolidation of radiative stability of various derivative self-interacting theories such as scalar Galileons \cite{Luty:2003vm,Nicolis:2004qq,Burgess:2006bm,Hinterbichler:2010xn,Hinterbichler:2011tt,deRham:2012ew,Goon:2016ihr,HorndeskiSurvivals}: The complete EFT Lagrangian can be written as an expansion in the three parameters $\alpha_{\text{cl}}$, $\alpha_{\text{q}}$ and $\alpha_{\tilde{\text{q}}}$ equivalent to the Galileon case discussed in Sec.~\ref{sSec:FlatSpaceGalileons}
\be\label{eq:FinalExpansionGP}
\boxed{L^{\myst{GP}}_{\text{DL}}\sim\left(F^2+(\pd\phi)^2\right)\,\alpha_{\text{cl}}^{a_3}+\left(F^2+(\pd\phi)^2\right)\,\alpha_{\text{q}}^{2+n}\alpha_{\tilde{\text{q}}}^l\,\alpha_{\text{cl}}^m}
\ee
where $ 3\geq a_3\geq0\,,\;l,n,m\geq0$. As expected, the quantum induced operators are distinct from their classical counterpart to all orders due to the presence of the quantum parameters $\alpha_{\text{q},\tilde{\text{q}}}$. More precisely, every loop operator inevitably carries a non-zero power of these quantum parameters. This marks a clear separation between classical and quantum terms and implies non-renormalization of classical terms in the DL. Just as in the scalar Galileon case, there exists a regime below the energy scale $\Lambda_3$ where quantum contributions are heavily suppressed $\alpha_{\text{q},\tilde{\text{q}}}\ll 1$, while classical non-linear terms, although equally non-renormalizable, are important compared to the kinetic term $\alpha_{\text{cl}}\sim\mathcal{O}(1)$. Hence, in the decoupling limit, the theory is stable under quantum corrections.

However, the beauty of the decoupling limit analysis is that this statement in fact directly implies the quantum stability of the whole theory. More precisely, the clear separation between classical and quantum terms in Eq.~\eqref{eq:FinalExpansionGP} is of course only valid in the decoupling limit. In the unitary gauge computation, we explicitly encountered quantum corrections of the same form
as classical operators, leading to a potential detuning and introduction of ghost instabilities. However, since the decoupling limit analysis focuses on the relevant high-energy scale $\Lambda_3$ quantum stability in the DL immediately also implies that all operators which did not survive the decoupling limit are in fact further suppressed by factors of $m/\Lambda_3$ in comparison to operators present in the DL such that they have no impact on the viability of the EFT.
Hence, the generalized Proca EFT [Eq.~\eqref{Lagrangians}] does not lose its key properties when including quantum corrections in their full generality and the effective description is theoretically viable.

Moreover, the commutativity of decoupling limit and quantum correction calculations allows translating the expansion in Eq.~\eqref{fullDL} back to the unitary gauge, from which one can infer the cancellation of dangerous leading order terms of loop corrections in generic unitary gauge calculations. In particular, this gives access to the least suppressed quantum corrections in the original formulation
\be
L^{\text{q}}\sim \pd^4\left(\frac{\pd F}{\Lambda_3^{3}}\right)^{2b_2}\left(\frac{m\,\pd A}{\Lambda_3^{3}}\right)^{b_3}\,,
\ee
which are the ones with $b_3=0$, hence, the ones which preserve gauge invariance. Other contributions with non-zero $b_3$ and operators which do not survive the decoupling limit are further suppressed by factors of $m/\Lambda_3$.

Finally, we want to mention that the same arguments presented above also apply to an even broader class of theories \cite{deRham:2021yhr} also known as \textit{Proca nuevo} \cite{deRham:2020yet}, for which the constraints for Ostrogradsky stability relies on an  infinite tower of massive vector interactions. We already briefly talked about this theory back in Sec.~\ref{ssSec: A Exact Theories}.


\newpage
\thispagestyle{plain} 
\mbox{}


\chapter{Challenges of the Quantum EFT of Gravity}\label{Sec:Challenges of the Quantum EFT of Gravity}

In this final chapter of this monograph, we want to draw the attention to some challenges that the quantum effective field theory of gravity introduced above has to face and end with a speculation on a possible alternative approach of constructing a gravity theory describing quantized gravitons. Despite its various successes in particular as a computational tool, the quantum EFT of GR that we discussed in Chapter~\ref{Sec:Gravity and Quantum Physics} seems flawed in various aspects, including presumably the biggest open puzzle of modern physics.

In its core, the standard approach of perturbatively quantizing a relativistic field theory of gravity relies on the assumption of the presence of an underlying classical and Lorentz invariant Minkowski spacetime, on top of which a self-interacting massless spin-2 field is considered. The true dynamical variables of the spin-2 gauge field can then be quantized through standard techniques. As already insisted on, such standard perturbative quantization schemes necessitate the existence of a classical background spacetime to provide a well-defined reference system. The resulting predictive quantum effective field theory of gravity is able to describe the principles of the gravitational force through virtual exchanges of gravitons, while on-shell gravitons provide an alternative particle picture to the phenomenon of gravitational waves.

However, such a departure from a geometrical interpretation of gravitation as a manifestation of space-time curvature does not come without its drawbacks. First of all, such a field theoretic description at best obscures the notion of causality a generic and dynamical spacetime metric on a manifold would define, in particular in connection with gravitational horizons. In the field theoretic approach, the only classical structure at hand that could provide a well understood causal structure is the underlying Minkowski metric. As already remarked, in the same way as matter field QFTs can be considered on a curved background spacetime one could technically also quantize a spin-2 field on a more general spacetime background, which would however not make much sense if one wants to account for the gravitational force through graviton exchange (we will come back to this point in Sec.~\ref{sSec:LetGravityBeGravity} below).

More generally, squeezing the theory of gravity into the corset of established field theoretic methods and simply treating it as an additional force carrying quantum field of nature seems to undo some of the most important deep insights into the phenomenon of gravitation that lay at the basis of the formulation of general relativity. As discussed in Sec.~\ref{Sec:The Generalization to Gravity} the postulation of the Einstein equivalence principle leads to a fundamental distinction between the gravitational force and other forces of nature, as there are by principle no true localizable effects of gravity. This is the crucial property that allows for an interpretation of gravitation as a manifestation of spacetime curvature, whose presence can only be determined through sufficiently non-local experiments that probe the inhomogeneity of the gravitational field, in particular through tidal forces on geodesic motion. As such, the natural reference system of relativistic theories on Minkowski spacetime given by inertial observers, is to be replaced by local freely falling frames, in which the gravitational force is fundamentally absent for local events. Alongside comes the impossibility of defining a local notion of energy-momentum for the gravitational field, with all its implications on the conservation of energy and momentum.

Given that a quantum field theoretic perspective on GR seems not entirely self-consistent \cite{Padmanabhan:2004xk} and often requires input from its geometric formulation, while still relying on concepts such as the energy-momentum tensor of gravitons (through Noethers' theorem on the translation symmetry of the Minkowski reference), at the very least leaves a feeling of unease. Along the same lines one could also challenge the wide-spread practice of obtaining graviton Feynman rules by directly perturbing geometric curvature tensor fields, since from the viewpoint of a theory on a manifold with generic metric the initial split between the classical background metric and the dynamical quantum field is in general not well-defined.

Foremost, however, it appears that a quantum field theory of gravitons in the form it was introduced in Chapter~\ref{Sec:Gravity and Quantum Physics} is experimentally ruled out by the simple observation that quantum vacuum energy does evidently not gravitate as predicted by this framework, a statement we want to elaborate on in the next subsection.



\section{The Cosmological Constant Problem}\label{sSec: The CC Problem}

The \textit{cosmological constant problem} is a longstanding puzzle that in particular challenges the quantum EFT picture of GR \cite{Weinberg:1988cp,Martin:2012bt,Burgess:2013ara,Polchinski:2006gy,zee_quantum_2010,zee2013einstein}, that we already mentioned on several occasions, for instance in Sec.~\ref{sSec:LessonsOnDynamicalDE}.  Technically, there exist two facets of the puzzle. On the one hand there is the so called ``old'' CC problem that deals with the apparent complete absence of gravitating vacuum energy, and therefore dates before the observed accelerated expansion of the universe \cite{SupernovaSearchTeam:1998fmf,SupernovaCosmologyProject:1998vns} that suggest a non-zero net value of the cosmological constant in the cosmological standard model (recall Part.~\ref{Part: Cosmological Testing Ground}). A solution to the ``new'' cosmological constant problem would then not only need to explain the old CC problem in terms of quantum instabilities, but also explain the origin of the observed accelerated expansion of the universe. As already mentioned, solving the new CC puzzle serves in fact as a prime motivation to look at generalizations to GR on cosmological scales. We will however first assume GR and briefly comment on more general metric theories of gravity at a later stage.

First of all, it is important to realize that for GR as a metric theory of gravity in purely classical terms, one can argue that there is in fact no real CC ``problem''. In Chapter~\ref{Sec:General Relativity} we derived the equations of motion of the unique metric theory of gravity that only involves the physical metric $g_{\mu\nu}$ as a gravitational field to be of the form [Eq.~\eqref{eq:EinsteinFieldEquations}]
\begin{equation}\label{eq:EinsteinFieldEquations CC}
    R_{\mu\nu}-\frac{1}{2}g_{\mu\nu}\,R+\Lambda g_{\mu\nu}=\kappa_0\,T_{\mu\nu}\,,
\end{equation}
with $\Lambda$ a constant that represents the infamous CC and $T_{\mu\nu}$ is the total matter energy-momentum tensor. 

Now, many important applications of GR, including the black hole solutions, are so-called vacuum equations that are solved outside any matter source $T_{\mu\nu}=0$. For asymptotically flat solutions, it is then further imposed that $\Lambda=0$. In order to properly discuss the vacuum equations of GR, we should however first answer the question of what a vacuum actually means in terms of matter fields, a question that is often not touched upon, since absolute vacuum energy does not affect non-gravitational physics in any way. In other words, only differences in energies have a physically observable effect. Very generally, a vacuum should therefore be defined as a zero point energy density that is locally constant and the same for every inertial observer. In this case, Lorentz invariance imposes the following form for the energy momentum tensor of vacuum for a theory on Minkowski spacetime in Minkowski coordinates \cite{Sakharov:1967pk}
\begin{equation}\label{eq:EMT Vacuum}
    T^{\myst{vac}}_{\mu\nu}=-\rho_{\myst{vac}}\,\eta_{\mu\nu}\,.
\end{equation}
Here $\rho_{\myst{vac}}$ represents a constant, such that the conservation equation $\pd^\mu T^{\myst{vac}}_{\mu\nu}=0$ is satisfied. Generally, the vacuum energy-momentum tensor can be viewed as a minimum energy in which all derivative operator terms are neglected. Due to its unobservability $\rho_{\myst{vac}}$ is then often shifted to zero, very much like the minimum of a potential can be adjusted arbitrarily. For now, it is however instructive to leave the absolute value of the vacuum energy as an arbitrary constant. In this case, minimal and universal coupling in a metric theory then ensures that such a vacuum energy simply gravitates as
\begin{equation}\label{eq:EMT Vacuum gen}
    \boxed{T^{\myst{vac}}_{\mu\nu}=-\rho_{\myst{vac}}\,g_{\mu\nu}\,.}
\end{equation}

We can now come back to the Einstein equations [Eq.~\eqref{eq:EinsteinFieldEquations CC}] and observe that the concept of vacuum solutions is in fact not altered even for a non-trivial value of $\rho_{\myst{vac}}$. This is because the presence of such a term simply results in a redefinition of an effective cosmological constant
\begin{equation}\label{eq:EFF CC}
    \Lambda_{\myst{eff}}\equiv \Lambda + \kappa_0 \rho_{\myst{vac}}\,,
\end{equation}
such that the vacuum Einstein equations again read
\begin{equation}
     R_{\mu\nu}-\frac{1}{2}g_{\mu\nu}\,R+\Lambda_{\myst{eff}}\, g_{\mu\nu}=0\,,
\end{equation}
where now, boundary conditions impose a value on $\Lambda_{\myst{eff}}$ instead of the bare value $\Lambda$. Of course, even classically it would be nice to understand the measured value of the CC in the context of cosmology (see Part~\ref{Part: Cosmological Testing Ground}) from first principles, but fixing it by hand does not represent a major problem.

The real puzzle arises when quantum physics is considered. First, quantum field theory predicts a value for the energy density of empty space that is enormously larger than the cosmologically observed constant through the zero-point energy induced by vacuum fluctuations. Heuristically, this can be understood through the analogy of viewing quantum field theory as an infinite set of harmonic oscillators, each of which comes with its own zero-point energy. Crude estimates of the minimal value of the vacuum energy density contributed by known matter fields lie so many orders of magnitude above the observed value, that either an extremely unnatural fine-tuning between the expected $\rho_{\myst{vac}}$ and the bare CC in Eq.~\eqref{eq:EFF CC} is required, or we do not yet understand how quantum vacuum energy gravitates.

While this last point is hard to address in a purely classical theory of GR, the quantum EFT of GR considered in Chapter~\ref{Sec:Gravity and Quantum Physics} precisely gives an answer to this question, namely, quantum matter gravitates though interactions with the graviton field. However, it seems that this answer is clearly the wrong one. This is not only because the predicted value of $\rho_{\myst{vac}}$ is enormous, but because it is completely out of control. In other words, the CC term obtains large quantum corrections due to the coupling of the graviton field with massive particles of the standard model such as the Higgs field, let alone any hypothetical fields that might be discovered beyond the energy scales that were currently probed.

\begin{figure}
\begin{center}
\begin{fmffile}{loopsGraviton}
\begin{fmfgraph*}(45,30)
    \fmfleft{i}
    \fmfright{o}
    \fmf{photon,tension=2}{i,v1}
    \fmf{plain,left=1}{v1,o,v1}
     \fmfdot{v1}
    \end{fmfgraph*}
    \quad
  \begin{fmfgraph*}(35,30)
    \fmfleft{i1,i2}
    \fmfright{o}
    \fmf{photon,tension=2}{i1,v1}
    \fmf{photon,tension=2}{i2,v1}
    \fmf{plain,left=1}{v1,o,v1}
     \fmfdot{v1}
    \end{fmfgraph*}
    \quad
  \begin{fmfgraph*}(60,30)
     \fmfleft{i}
     \fmfright{o}
     \fmf{photon,tension=2}{i,v1}
     \fmf{plain,left=1}{v1,v2}
     \fmf{plain,left=1}{v2,v1}
     \fmf{photon,tension=2}{v2,o}
     \fmfdot{v1,v2}
    \end{fmfgraph*}
     \quad
  \begin{fmfgraph*}(30,30)
    \fmfleft{i1,i2,i3}
    \fmfright{o}
    \fmf{photon,tension=2}{i1,v1}
    \fmf{photon,tension=2}{i2,v1}
    \fmf{photon,tension=2}{i3,v1}
    \fmf{plain,left=1}{v1,o,v1}
     \fmfdot{v1}
    \end{fmfgraph*}
       \quad
  \begin{fmfgraph*}(50,30)
     \fmfleft{i1,i2}
     \fmfright{o}
     \fmf{photon,tension=2}{i1,v1}
     \fmf{photon,tension=2}{i2,v1}
     \fmf{plain,left=1}{v1,v2}
     \fmf{plain,left=1}{v2,v1}
     \fmf{photon,tension=2}{v2,o}
     \fmfdot{v1,v2}
    \end{fmfgraph*}
     \quad
  \begin{fmfgraph*}(45,30)
     \fmfleft{i1,i2}
     \fmfright{o}
     \fmf{photon,tension=2}{i1,v1}
     \fmf{photon,tension=2}{i2,v2}
     \fmf{plain,left=0.5}{v1,v2}
     \fmf{plain,left=0.7}{v2,v3}
     \fmf{plain,left=0.7}{v3,v1}
     \fmf{photon,tension=2}{v3,o}
     \fmfdot{v1,v2,v3}
    \end{fmfgraph*}
\end{fmffile}
\end{center}
\caption{\small{All one-loop $1$PI contributions through matter loops up to three external graviton legs that lead to quantum corrections of the cosmological constant term of GR.}}
\label{GravitonLoopsCC} 
\end{figure}
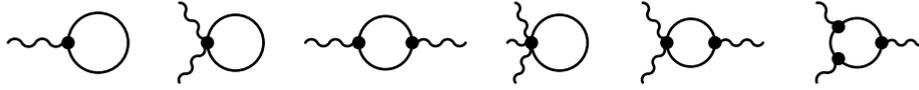

Indeed, the statement of the non-renormalization properties of GR in Sec.~\ref{sSec: Radiative Stability of GR} only hold in the pure gravity sector. The reason these previous arguments no longer go through is because of the additional mass scale $M$ carried by matter fields that dominates the powercounting of the loop corrections at low energies. Essentially, this additional scale implies that the quantum corrections to the graviton action coming from matter loops are not bound to involve a higher power of derivatives. Concretely, at one-loop the quantum corrections are expected to provide an expansion of the form
\begin{equation}
    (M^4+M^2\pd^2+\pd^4)\sum_{i=0}^{\infty}\left(\frac{h}{\MPl}\right)^i\,.
\end{equation}
As dictated by the full non-linear gauge symmetry, the first set of terms $\sim M^4$ that dominate at the lowest energy scales will precisely resum to give a non-trivial correction to the classical CC term
\begin{equation}
    M^4\sum_{i=0}^{\infty}\left(\frac{h}{\MPl}\right)^i\sim \sqrt{-g} M^4\,,
\end{equation}
upon the usual identification
\begin{equation}
    g_{\mu\nu}=\eta_{\mu\nu}+\frac{h_{\mu\nu}}{\MPl}\,.
\end{equation}
This can readily be verified through direct computations of the log-divergences of the one-loop $1$PI diagrams, the first few of which are depicted in Fig.~\ref{GravitonLoopsCC}.

These quantum corrections could be attributed to either the CC term in the gravitational action
\begin{equation}
    \sim -\MPl^2 \int d^4x \sqrt{-g} \left[2\Lambda +\frac{M^4}{\MPl^2}\right]
\end{equation}
or simply as a vacuum energy contribution to the matter Lagrangian. Either way, and despite the Plank suppression through the gravitational coupling, the associated quantum corrections are out of control \cite{Weinberg:1988cp,Martin:2012bt,Burgess:2013ara,zee_quantum_2010,zee2013einstein}. Every massive matter field is expected to correct the CC in such a way, destabilizing a classical initial value and rendering a cancellation at astronomical orders of magnitude untenable, as any new field beyond the standard model, as well as any higher loop corrections would again destroy such a fragile fine-tuning.\footnote{This is the case despite the fact that fermionic fields contribute to the CC with an opposite sign. However, in the absence of supersymmetry, the associated cancellation is way too weak.} In this respect, the CC problem is very close to the so called ``hierarchy problem'' of the Higgs mass, with the difference that the instability is already provided by known fields of the standard model of particle physics.

On the other hand, the terms scaling like $M^2\pd^2$ will renormalize the Ricci scalar, wile the last term corresponds to the curvature squared terms that we already encountered. Higher loop contributions would then involve additional factors of $M^2/\MPl^2$ or $\pd^2/\MPl^2$. Observe that the difference to the case of the generalized Proca theory discussed in Sec.~\ref{Sec:GenProca Quantum Stability} is that the quantum EFT of GR already at the classical level is given by an infinite sum of terms, such that low energy quantum corrections that at first sight are well suppressed by the cutoff of the EFT become large upon the resummation of terms. Also, the problem really arises at low energies in the IR, since in the high-energy limit above the mass $M$, quantum stability of GR is recovered. 

Moreover, we want to note that the CC problem in the quantum EFT of gravity is very robust. In particular, as noted in  \cite{Polchinski:2006gy} (see also \cite{zee2013einstein}), any clever mechanism that would cancel the unwanted Feynman diagrams in Fig.~\ref{GravitonLoopsCC} is ruled out if the interactions of the graviton with matter are to explain the force of gravity. This is because, for instance, connecting the first diagram in Fig.~\ref{GravitonLoopsCC} with two photon legs to the propagator of some atomic nucleus, gives a contribution to the gravitational mass of the associated nucleus. However, the exact same diagram with the graviton removed would give a contribution to the inertial mass of the nucleus. Thus, by the equivalence principle, it seems not possible to simply delete the graviton interaction with matter loops without serious issues. The root of the problem could here be identified in the fact that the quantum EFT of gravity requires the gravitational interaction to be local, such that one cannot mess with the vertices that couple the graviton to matter fields \cite{zee2013einstein}. After all, these vertices are at the core of explaining gravitation through exchanges of virtual gravitons that represent the initial purpose of the construction of the quantum EFT of GR.

In summary, it thus seems that the recipe of the quantum theory of gravitons on how matter, and in particular a quantum vacuum energy, gravitates is not the correct one.

\section{Let Gravity be Gravity: A Speculation on a Physical Quantization}\label{sSec:LetGravityBeGravity}

In the light of the rather serious challenges of the standard approach to a quantum EFT of GR outlined above, we would like to end this work by speculating about an alternative possibility of describing quantized gravitons within GR. The goal here will of course not be to propose a theory of quantum gravity, but to advertise an appropriate use of known quantization techniques in the context of geometric gravity theories based on non-trivial spacetime structures associated to a physical metric. Indeed, the considerations above indicate that a local description of the gravitational interaction through standard quantum field theoretic tools is not viable. Vice versa, this suggests that one should perhaps not abandon the theoretic and equivalence principle based successes of a reinterpretation of freely falling frames as the true local inertial frames.
This would mean that one should insist on a treatment of gravitation that is fundamentally distinct from the one of the other fundamental interactions. 

Concretely, instead of building up a theory of quantized gravitons from scratch, one could try to apply the time-tested methods of quantizing dynamical degrees of freedom within geometric formulations of gravity, restricted to situations in which perturbative quantization schemes are well-defined. In GR as a theory on curved spacetime with dynamical metric, the main problem in such an approach seems finding an appropriate definition of dynamical degrees of freedom.\footnote{Also the path integral approach on quantizing GR ultimately seems to fail due to the fundamental difficulty of isolating its true dynamical degrees of freedom.}
However, in Sec.~\ref{ssSec:Local Wave Equation in GR} we did precisely that, namely identify the dynamical DOFs of GR described as high-frequency perturbations well separated from a slowly varying but otherwise arbitrary physical background metric. Recall in particular that it was the assumption of the Isaacson split between parametrically separated high- and low-frequency contributions that enabled an unambiguous identification of the propagating degrees of freedom as solutions to a wave equation, up to standard gauge symmetry redundancies. Moreover, this parametric separation assures the existence of an extended patch on which the arbitrary low-frequency metric can be described as a Minkowski chart, while the high-frequency metric perturbations enjoy all defining properties of a spin-2 Lorentz field. In particular, also their energy-momentum content back-reacting on the slowly-varying background is now well-defined.
Considering such a freely falling frame of low-frequency quantities seems in fact the only possibility for a local description of Lorentzian matter physics while keeping a non-trivial, localized dynamics of gravitational DOFs. This is because without the presence of a clear separation of scales, a restriction to the total freely falling frame by definition washes out any non-trivial gravitational effects.

On the other hand, assuming an Isaacson split, the slowly-varying part of the physical metric can locally serve as a well-defined Minkowski reference on which a standard quantization of the spin 2 high-frequency Lorentz fields in addition to the matter fields should be possible. This would allow for a description of the particle analogue of gravitational waves through quantized gravitons. Note, however, that such an approach does not try to describe the bulk of gravitation in terms of a quantum field theory, but restricts a quantization to the high-frequency dynamical degrees of freedom, the consequences of which to our knowledge remain entirely unexplored. Such a take on quantizing the propagating DOFs of GR would leave a clear distinction between gravity and other fundamental interactions, and therefore evade all the challenges discussed above that a full quantum EFT of gravitons has to face. In particular, such a distinction could provide a new interesting handle on the CC problem, because the prescription of how quantum matter interacts with gravitons is fundamentally altered. More precisely, within such a perturbative Isaacson approach, the background equations that predominately set the cosmological constant on a classical level would be independent from the quantized high-frequency graviton and matter fields, whose energy-momentum content may only perturb the background.

In summary, taking the necessity of an Isaacson split in defining gravitational DOFs seriously could allow for a well-defined quantization while keeping intact the concepts of a spacetime as an independent stage for quantum physics. In this context, beyond the low-frequency freely falling frame discussed above, it would then actually also make sense to talk about a quantum EFTs on curved spacetime, including the description of quantized gravitons on a curved background spacetime. Curiously, today's technical possibilities of quantization procedures on curved background spacetimes precisely overlap with the definition of a low-frequency Isaacson background, which in the language of non-trivial background quantization is known as the \textit{nice slice} criterion \cite{Polchinski:1995ta} that is satisfied as soon as the background fields vary sufficiently slowly with respect to a given time-variable.


%% file: IntoAndOutro/1.Appendix.tex
\part*{APPENDICES}\addcontentsline{toc}{part}{Appendices}

\begin{appendices}

\newpage
\thispagestyle{plain} 
\mbox{}

\chapter{Gauge Freedom and Symmetries in Physics}\label{App: Symmetires in Physics}

\noindent
\footnotesize
\emph{\ul{References}}\\ 
\textit{The following is based on the reviews in 
\cite{Trautman_1963,Kosmann-Schwarzbach2011}.}

\vspace{5mm}

\noindent
The notion of symmetry plays an essential role in modern physics. Symmetries provide a guideline for constructing theories and have profound implications due to their close connection to conservation laws and mathematical identities. Moreover, a single physical system can always be described in multiple ways, however, our choice of description should of course not alter the physical content. Such a redundancy in the description is known as gauge freedom and also plays a key role in theories of physics. It is therefore worth taking the time to properly introduce these notions and state the connection between the concept of symmetries and gauge freedom, as well as the close relationship between symmetries and conservation laws known as the Noether theorems.

\section{Gauge Freedom}\label{sApp: Gauge Freedom}

Let's therefore consider a theory of a physical system described by an action integral over some domain $\Omega$
\begin{equation}
    S=\int_\Omega \mathcal{L}(x; u(x)) d^nx\,,
\end{equation}
where
\begin{equation}
\mathcal{L}(x; u(x))=\mathcal{L}\left(x_i; u_A, \partial_i u_A, \partial_i\partial_j u_A\right)\,,
\end{equation}
is the Lagrange density that depends on the $n$ independent variables $x_i$, $i=1,...,n$ and $N$ dynamical variables $u_A$, $A=1,...,N$ and their derivatives, and where $\partial_i\equiv \partial /\partial x_i$. For simplicity, we assume only first and second derivatives to appear. The set of functions $u_A(x)$ completely describe some physical system and the corresponding equations of motion are obtained by varying the action
\begin{equation}\label{eq:initialEOMs}
    \frac{\delta S}{\delta u_A(x)}=\mathcal{E}^A(x ; u(x))=0\,,
\end{equation}
where it is assumed that the dynamical variables are independent.

As already mentioned, a given physical system or situation can always be described in different ways. This includes for instance different choices of frames of references, often encoded in the choice of variables $x_i$, or alternative choices for the dynamical functions $u_A$. Such a freedom or redundancy of description is generally called a ``gauge freedom''. More precisely, we will define \ul{gauge transformations} to denote all transformations 
\begin{equation}\label{eq:GaugeTransformation}
\begin{aligned}
     x_i &\rightarrow x'_i =f_i(x)\,,\\
    u_A(x)& \rightarrow u'_A(x')=F_A(x;u)\,,
    \end{aligned}
\end{equation}
such that \cite{Trautman_1963}
\begin{equation}\label{eq:GaugeTransformationCondition}
    S'=\int_{\Omega'} \mathcal{L}'(x'; u') d^nx'=\int_\Omega \left[\mathcal{L}(x; u) -\partial_i Q^i(x; u) \right]d^nx\,,
\end{equation}
under the assumption that all dynamical variables $u_A$ and their derivatives vanish at the boundary of $\Omega$.
Here, $\Omega'=f(\Omega)$ denotes the image of $\Omega$ under Eq.~\eqref{eq:GaugeTransformation} and the functions $Q$ are arbitrary as long as they vanish upon the vanishing of the dynamical variables and their derivatives. The condition in Eq.~\eqref{eq:GaugeTransformationCondition} ensures that the new equations of motion
\begin{equation}\label{eq:EOMSNoetherA}
    \frac{\delta S'}{\delta u'_A(x')}=\mathcal{E}'^A(x' ; u')=0\,,
\end{equation}
are equivalent in physical content, since $\delta W'=0$ if and only if $\delta W=0$. Note that these new equations of motion must not in general be of the same form as the initial relations in Eq.~\eqref{eq:initialEOMs}, since also the Lagrange density has a different form.

\section{Symmetries}\label{sApp: Symmetries}

Very generally, one talks about a symmetry of a physical system if a feature of the system remains unchanged under some transformation. We will however more specifically be interested in continuous symmetries of theories or models described by an action, depending on a set of variables, and the associated governing equations of motion. In this case, a symmetry is a transformation of the variables that leaves the form of the equations of motion invariant.

Thus, in this sense, we define a \ul{symmetry} of the action, to be a gauge transformation in Eq.~\eqref{eq:GaugeTransformation} that leaves the form of the equations of motion invariant
\begin{equation}
   \mathcal{E}^A(x;u)\rightarrow \mathcal{E}'^A(x';u')= \mathcal{E}^A(x';u')\,.
\end{equation}
In terms of the action of the theory or the physical system, this is the case if for a certain choice of function $Q$, the form of the Lagrangian density is unaltered
\begin{equation}\label{eq:SymmetryConditionLagrangianA}
     \mathcal{L}'(x';u')= \mathcal{L}(x';u')\,.
\end{equation}

For continuous symmetries, we further distinguish between \ul{global symmetries} that form a $p$-dimensional Lie group $G_p$, with $p$ a finite integer that characterizes the number of continuous group parameters $\alpha_a$, $a=1,...,p$, and \ul{local symmetries}, that form an infinite dimensional group $G_{\infty q}$, where the transformations depend on $q$ arbitrary functions of the variables $x_i$, hence, $\alpha_a(x)$, $a=1,...,q$. For both types of symmetries, the transformations [Eq.~\eqref{eq:GaugeTransformation}] can more precisely be written as
\begin{equation}\label{eq:GaugeTransformationLieGroup pre}
\begin{aligned}
     x_i &\rightarrow x'_i=f_i(x;\alpha) \,,\\
    u_A(x)&\rightarrow u'_A(x')=F_A(x;u;\alpha)\,,
    \end{aligned}
\end{equation}
where the functions $f$ and $F$ also depend on the continuous group parameters  $\alpha$ and their derivatives. The transformations can then be characterized through the infinitesimal transformations to first order in $\alpha$, 
\begin{equation}\label{eq:GaugeTransformationLieGroup}
\begin{aligned}
     x_i &\rightarrow x'_i= x_i + \delta x_i \,,\\
    u_A(x)&\rightarrow u'_A(x')= u_A(x) + \delta  u_A(x) \,.
    \end{aligned}
\end{equation}

Since in the action, the variables $x_i$ are dummy variables because they are being integrated over, it is useful to define the \ul{total variation}\footnote{In fact, this variation corresponds to minus the usual definition of the generalized ``Lie derivative'' along the direction of the vector field generating the transformation.} of the dynamical functions
\begin{equation}
    \bar{\delta}u_A\equiv u'_A(x)-u_A(x)=\delta u_A(x) -\delta x^i\partial_i u_A(x)\,.
\end{equation}
For a finite Lie group $G_p$ with constant parameters $\alpha_a$ we can write
\begin{equation}\label{eq:GaugeTransformationLieGroup 2s}
\begin{aligned}
     x_i &\rightarrow x'_i=f_i(x;0) + \partial_a f_i(x;0) \alpha^a  \,,\\
    u_A(x)&\rightarrow u'_A(x')=F_A(x;u;0)+\partial_a F_A(x;u;0)\alpha^a \,,
    \end{aligned}
\end{equation}
up to order $\mathcal{O}(\alpha^2)$ where we assumed that the identity is given by $\alpha_a=0$ and therefore define 
\begin{equation}\label{eq:bardeltaForGlobalSymmetryA}
    \bar{\delta}u_A=\left(\partial_a F_A(x;u;0)-\partial_a f^i(x;0)\partial_i u_A\right) \alpha^a\equiv \gamma_{Aa}(x)\,\alpha^a\,.
\end{equation}
Similarly, for a local symmetry and by assuming that only first derivatives of the group parameters enter in Eq.~\eqref{eq:GaugeTransformationLieGroup} , we can write
\begin{equation}
    \bar{\delta}u_A=\gamma_{Aa}(x)\,\alpha^a-\Delta^i_{Aa}\,\partial_i\,.
\end{equation}

\section{The Noether Theorems}\label{sApp: Noethers Theorem}

\begin{theorem}\label{Thm:NoetherTheorem} \textbf{The Noether Theorems} \cite{Noether:1918zz,Trautman_1963,konopleva1981gauge,Kosmann-Schwarzbach2011}
\begin{enumerate}[(I)]
    \item If the action integral $S$ is symmetric with respect to a p-dimensional Lie group $G_p$, then there exist $p$ linearly independent conserved on-shell currents constructed out of Lagrangian derivatives.
    \item If the action integral $S$ is symmetric with respect to an infinite group $G_{\infty q}$, then there exist $q$ \ul{Bianchi identity relations} between the Lagrangian derivatives. 
\end{enumerate}
Moreover, the converse of these statements are also true.
\end{theorem}

We will refrain from providing a rigorous proof of the above theorems, but instead sketch the most important steps and elaborate on these results, and refer for more details to \cite{Noether:1918zz,Trautman_1963,konopleva1981gauge,Kosmann-Schwarzbach2011}. Together with the gauge condition Eq.~\eqref{eq:GaugeTransformationCondition} and the arbitrariness of $\Omega$, the symmetry condition on the Lagrangian density in \eqref{eq:SymmetryConditionLagrangianA} implies that
\begin{equation}
    \bar{\delta}\mathcal{L}+\partial_i\bar{\delta}Q^i+\partial_i(\mathcal{L}\delta x^i)=0\,,
\end{equation}
where we denote
\begin{equation}
    \bar{\delta}F\equiv F(x;u')-F(x;u)\,,
\end{equation}
for any functional $F(x;u)$. Using the definitions of $\bar{\delta}\mathcal{L}$ and of the equations of motion [Eq.~\eqref{eq:EOMSNoetherA}], this condition can be rewritten as \cite{Trautman_1963}
\begin{equation}
    \mathcal{E}^A\bar{\delta}u_A+\partial_i\bar{\delta} J^i=0\,,
\end{equation}
where the ``current'' $J^i$ depends on the Lagrange density and its derivatives, as well as on the functionals $Q$. 

For the first Noether theorem $(I)$, one can show that $\bar{\delta} J^i=\alpha^aJ^i_a$, such that together with Eq.~\eqref{eq:bardeltaForGlobalSymmetryA}, we obtain
\begin{equation}
    \mathcal{E}^A\gamma_{Aa}+\partial_i J^i_a=0\,.
\end{equation}
Therefore, indeed the global symmetry provides $p$ conservation laws $\partial_i J^i_a=0$, provided that the equations of motion are satisfies (on-shell condition). Note, however, that these currents are not unique and depend on an arbitrary choice of the functionals $Q$. On the other hand, if the symmetry is local (second Noether theorem $(II)$), one obtains equations that depend linearly on the $\alpha_a(x)$ and their derivatives. The resulting identities that are linear and homogeneous in the equations of motion are called ``Bianchi identities'' and are given by \cite{Trautman_1963}
\begin{equation}
    \mathcal{E}^A\gamma_{Aa}+\partial_i(\mathcal{E}^A\Delta^i_{Aa})=0\,.
\end{equation}

\section{Gauge Symmetries and Proper Symmetries}\label{sApp: Gauge Symmetries and Proper Symmetries}

Global and local symmetries have thus rather different implications on the theory. In fact, local symmetries are closely tied to the underlying gauge freedom and can therefore mainly still be viewed as a redundancy in the description of a physical theory and are thus referred to as \ul{gauge symmetries} as opposed to the global \ul{proper symmetries}. 

This is because in most cases, the gauge freedom of a theory can be promoted to a gauge symmetry, by adding additional structure to the theory. Indeed, since physical quantities should not depend on any gauge freedom, it should be possible to find a corresponding explicitly invariant formulation of the theory under the corresponding local symmetry. 
For instance, a theory in Minkowski spacetime, formulated in terms of a preferred Minkowski frame, can always be turned into a coordinate transformation invariant description by promoting the metric to a general (auxiliary) tensor field and by introducing a covariant derivative (see Chapter~\ref{Sec:Theories of Gravity} for more details). In this way, the general \textit{gauge transformation} of changing coordinates is promoted to a \textit{gauge symmetry}. Indeed, even Newtonian gravity admits a coordinate invariant formulation \cite{misner_gravitation_1973}.

In practice, one generic strategy to promote a gauge redundancy to an associated gauge symmetric formulation is given by the so called St\"ukelberg trick \cite{Stueckelberg:1900zz,GREEN1991462,Siegel:1993sk,Arkani-Hamed:2002bjr,Ruegg:2003ps}.\footnote{The St\"ukelberg trick is in particular known for constructing gauge invariant formulations of massive theories.} Instead of directly promoting a gauge freedom to a symmetry, however, such a St\"ukelberg construction typically starts from an existing global symmetry that is based on the gauge transformation in question and promotes the entire gauge freedom to a local gauge symmetry. In particular, a coordinate invariant formulation of SR can be viewed as being based on the global Poincaré symmetry of Minkowski spacetime. In fact, as proven by Utiyama \cite{Utiyama:1956sy,konopleva1981gauge,Aldaya:2021zqm}, any global symmetry based on gauge transformations can be promoted to a local gauge symmetry under a so-called minimal coupling approach that relies on the St\"ukelberg trick.

Thus, while a gauge symmetric formulation can always be constructed, the in a sense ``true'' symmetry of a theory (or a particular solution of a theory) is given by the global or proper symmetries. Foremost, only the finite symmetries associated to $G_p$ lead to so called \ul{proper conservation laws} of part (I) of Noethers' Theorem~\ref{Thm:NoetherTheorem} above that imply stringent physical consequences on a theory \cite{Noether:1918zz,Trautman_1963,konopleva1981gauge,Kosmann-Schwarzbach2011}. However, gauge symmetries still play an important role in theory building\footnote{For instance the Standard Model of particle physics is heavily structured by the bosonic gauge groups.}, such that in particular the type of gauge freedom that a theory possesses has an important impact.

At this point, it should be noted that a priori a finite symmetry group $G_p$ is only proper if it is not a direct subgroup of an infinite dimensional symmetry group $G_{\infty q}$.\footnote{In other words, a proper global symmetry that cannot be extended to an infinite dimensional group $G_{\infty q}$ without the introduction of any auxiliary fields.} Indeed, any general group $G_{\infty q}$ always contains a finite subgroup $G_p$, however, the associated conservation laws are not on the same footing \cite{Noether:1918zz,Trautman_1963,konopleva1981gauge,Kosmann-Schwarzbach2011}.

A good example for the above discussion is the existence of a general coordinate invariant formulation of a generic theory with a metric on a manifold. Invariance of the action under all coordinate transformations necessarily implies the invariance under specific coordinate transformations, such as time translations. However, such an invariance as a subgroup of an infinite dimensional symmetry group does not imply the existence of a globally conserved quantity that one can call energy, as it would be possible in a theory like SR that admits a proper symmetry under time translations. Note, on the other hand, that this does not prevent us from formulating a generally covariant formulation of SR by promoting the metric to an auxiliary field that also transforms under coordinate transformations.

In certain special cases, a local symmetry can however contain a proper global symmetry as a subgroup. This is the case for a formulation of a theory that is invariant under a $U(1)$ gauge transformation, such as electrodynamics described by a vector potential $A_\mu$ that is symmetric under the gauge transformation $ A_\mu\rightarrow   A_\mu+\partial_\mu \Lambda$.
In that case, the global symmetry connected to charge conservation is automatically present. This can be traced back to the simple fact that the global symmetry of charge conservation is associated to a constant $\Lambda$, such that 
the total variation (or ``Lie derivative'') of the gauge field identically vanishes $\bar{\delta}A_\mu=\partial_\mu\Lambda\equiv 0$.

In contrast, for a generic coordinate transformation invariant theory with a metric, the total variation corresponding to for instance the subgroup of time translations, which in this case corresponds to the Lie derivative of the metric that we will define in Sec.~\ref{sApp:DiffsAndLieDer} below, does not vanish. This is only the case for special metric solutions such as the Minkowski metric.

\newpage
\thispagestyle{plain} 
\mbox{}
\chapter{Differential Geometry in a Nutshell}\label{App:DiffGeo}

\noindent
\emph{\ul{References}}\\ 
\textit{The following is partially based on the reviews in 
\cite{misner_gravitation_1973,WaldBook,Blau2017,carroll2019spacetime,Renner2020}.}

\section{Manifold, Curves and Tangent Spaces}\label{sApp:ManifoldCurvesTangent}

In simple terms, a differentiable, four-dimensional (topological) \ul{manifold} $\M$ is a smooth set of points that locally behaves as $\mathbb{R}^4$. More formally, for any open subset $U$ of $\M$ we can define a \ul{chart} through a bijective map $\phi : U\rightarrow \mathbb{R}^4$, whose image is open in $\mathbb{R}^4$. A chart map therefore labels every point $p\in U$ by a unique tuple of numbers $\phi(p)=(x^0(p),...,x^3(p))$, where each component of the map $x^\mu$, with $\mu=0,1,2,3$, are functions from $U$ to $\mathbb{R}$, and therefore corresponds to a coordinate system. Moreover, for smoothness, we require that any chart transition map $\phi'\circ \phi^{-1}: \mathbb{R}^4\rightarrow \mathbb{R}^4$ is $C^\infty$. The union of all charts should of course cover all $\M$.

Without adding additional structure, we can define functions $f : \M\rightarrow \mathbb{R}$ on the manifold, as well as \ul{curves} $\gamma :\mathbb{R} \rightarrow  \M$. And combining these two notions, we can talk about a \ul{tangent vectors} $\underline v_{\gamma,p} : C^\infty(\M)\rightarrow \mathbb{R} $, which are defined as $\underline v_{\gamma,p}(f)\equiv \frac{d(f\circ \gamma)}{d\lambda}\lvert_{\lambda_0}$, that is, the directional derivative of a function along a curve with $\gamma(\lambda_0)=p$. The space of all tangent vectors at each point defines a vector space called \ul{tangent space} $T_p$, while the corresponding dual space $T^*_p$ is known as \ul{cotangent space}. The collection of all tangent spaces will be denoted by $T_\M$.

A given chart map $\phi$ then provides us with the associated representations in terms of coordinates we need, in order to do physics and explicit calculations. For any function $f$ on the manifold, there exists a coordinate representation $F : \mathbb{R}^4\rightarrow \mathbb{R}$, given by $F=f\circ \phi^{-1}$ such that we can define $\partial_\mu f(p)\equiv \frac{\partial F(\phi(p))}{\partial x^\mu}$. Likewise, curves on the manifold $\gamma :\mathbb{R} \rightarrow  \M$ are represented by $\Gamma :\mathbb{R} \rightarrow  \mathbb{R}^4$, where $\Gamma(\lambda)=(\phi \circ \gamma)(\lambda)$. The $\mu$th component of the map $\Gamma(\lambda)$ will be denoted by $x^\mu(\lambda)$. In the main text, curves are simply denoted by their components. This notation is convenient, as the coordinates $x^\mu$ used for the representation of the curves is explicit. Similarly, functions along a curve $f\circ\gamma:\mathbb R\rightarrow\mathbb R$ will also be denoted as $f(\lambda)$.

\paragraph{Coordinate Induced Basis.}
Given a coordinate system $x^\mu$, it is particularly useful to connect the basis of the tangent vector space to the choice of coordinates by the use of a so called \ul{coordinate induced basis}. This basis is obtained by noting that 
\begin{equation}\label{eq:Tangent Vector Derivation A}
    \underline v_{\gamma}(f)=   \frac{df(\lambda)}{d\lambda}=\frac{d(f \circ \gamma)}{d\lambda}=\frac{d(f\circ\phi^{-1} \circ\phi \circ \gamma)}{d\lambda}=\frac{d(\phi \circ \gamma)^\mu}{d\lambda}\frac{\partial F}{\partial x^\mu}=\dot x^\mu\partial_\mu f
\end{equation}
where we defined $\dot x^\mu(\lambda)\equiv \frac{dx^\mu(\lambda)}{d\lambda}= \frac{d(\phi \circ \gamma)^\mu(\lambda)}{d\lambda}$. Since this is true for any function $f$, we can promote this expression and write any tangent vector in terms of a differential operator
\begin{equation}\label{eq:Tangent Vector A}
    \underline v_\gamma =\dot x^\mu\underline \partial_\mu\,,
\end{equation}
where $\dot x^\mu$ are the components of the vector. Hence, the vectors $\{\underline{\partial}_\mu\}$ form a basis of the tangent space. Given that expression, the coordinate induced dual basis is provided by the gradient of the coordinate functions $\{\underline{\dd x}^\mu\}$, such that
\begin{equation}
    \underline{\dd x}^\mu(\underline{\partial}_\nu)=\underline{\partial}_\nu(x^\mu)=\frac{\partial x^\mu}{\partial x^\nu}=\delta\ud{\mu}{\nu}\,.
\end{equation}
Applied to a scalar function we have $\underline{\dd f}=\partial_\mu f \underline{\dd x}^\mu$ and hence in summary
\begin{equation}
    \underline{\dd f}\left(\underline{v}_\gamma\right)=\dot x^\mu\partial_\mu f=\underline{v}_\gamma(f)=\underline{v}_\gamma(\underline{\dd f})\,,
\end{equation}
where in the last equality, we identified the dual of the cotangent space $(T_p^*)^*$ with the tangent space itself $T_p^*$ through the isomorphism that maps any element $\underline V$ to $[\underline V] : T_p^*\rightarrow \mathbb{R}$, with $[\underline V](\underline{\omega})\equiv \underline{\omega}(\underline V)$.

\paragraph{Change of Coordinates.}
Physics should of course not depend on our choice of coordinate system. In particular, the tangent vectors of Eq.~\eqref{eq:Tangent Vector A} are physical quantities that are invariant under a change of coordinates, as it is evident from their definition. However, a new coordinate system $x'^\mu$ defines a new set of coordinate induced tangent space basis $\{\underline{\partial}'_\mu\}$. Hence, since the basis vectors of the tangent space depend on the coordinates, also the associated components of vectors will transform under a change of coordinates. Formally, a change of coordinates is given by a chart transition map $\phi'\circ \phi^{-1}$, where we have introduced a new chart $\phi'(p)=(x'^0(p),...,x'^3(p))$. Henceforth and in the main text, such a change of coordinates will simply be denoted by $x^\mu\rightarrow x'^\mu=x'^\mu(x)$, as the $\mu$th component of the map $\phi'\circ \phi^{-1}$. In order to relate the two basis $\{\underline{\partial}_\mu\}$ and $\{\underline{\partial}'_\mu\}$ at the same point on the manifold, it is useful to consider the action on a generic function $f$
\begin{align*}
    \underline{\partial}_\mu(f)&=\partial_\mu f = \frac{\partial F}{\partial x^\mu}= \frac{\partial(f\circ\phi'^{-1}\circ\phi'\circ\phi^{-1})}{\partial x^\mu}=\frac{\partial(\phi'\circ\phi^{-1})^\nu}{\partial x^\mu}\frac{\partial(f\circ\phi'^{-1})}{\partial x'^\nu}\\
    &=\frac{\partial x'^\nu}{\partial x^\mu}\,\partial'_\mu f=\frac{\partial x'^\nu}{\partial x^\mu}\,\underline{\partial}'_\mu(f)\,,
\end{align*}
and hence
\begin{equation}
    \underline{\partial}_\mu=\frac{\partial x'^\nu}{\partial x^\mu}\,\underline{\partial}'_\nu\,.
\end{equation}
Thus, by demanding $\underline v_\gamma=\dot x'^\mu\underline{\partial}'_\mu=\dot x^\mu\underline{\partial}_\mu=\dot x^\mu\frac{\partial x'^\nu}{\partial x^\mu}\,\underline{\partial}'_\nu$, we identify the transformation rule for the components of a tangent vector as
\begin{equation}\label{eq:VecTransformationA}
    \dot x'^\mu=\frac{\partial x'^\mu}{\partial x^\nu}\dot x^\nu\,.
\end{equation}

The transformation rule under a change of coordinates of the components of a generic vector $\underline V=V^\mu\underline\partial_\mu$, where $\underline V : T^*_p\rightarrow \mathbb{R}$, is therefore given by
\begin{equation}
    V'^\mu=\frac{\partial x'^\mu}{\partial x^\nu}\,V^\nu\,.
\end{equation}
And consequently, the components of a covector $\underline \omega=\omega_\mu\underline{\dd x}^\mu$, where $\underline \omega : T_p\rightarrow \mathbb{R}$, transform by the inverse of Eq.~\eqref{eq:VecTransformationA}, namely
\begin{equation}
    \omega'_\mu=\frac{\partial x^\nu}{\partial x'^\mu}\,\omega_\nu\,.
\end{equation}
More generally, the components of an $\binom{r}{s}$-tensor
\begin{equation}
    \underline T=T\ud{\mu_1 ... \mu_r}{\nu_1 ... \nu_s}\,\underline\partial_{\mu_1}\otimes ...\otimes \underline\partial_{\mu_r}\otimes \underline{\dd x}^{\nu_1}\otimes...\otimes \underline{\dd x}^{\nu_s}\,,
\end{equation}
defined as a multilinear map from a collection of $r$ covectors and $s$ vectors to $\mathbb{R}$, transform as
\begin{equation}
    T'\phantom{}\ud{\mu_1 ... \mu_r}{\nu_1 ... \nu_s}=\frac{\partial x'^{\mu_1}}{\partial x^{\alpha_1}}...\frac{\partial x'^{\mu_r}}{\partial x^{\alpha_r}}\frac{\partial x^{\beta_1}}{\partial x'^{\nu_1}}...\frac{\partial x^{\beta_s}}{\partial x'^{\nu_s}}\, T\ud{\alpha_1 ... \alpha_r}{\beta_1 ... \beta_s}\,.
\end{equation}
Finally, we define \ul{vector fields} $\underline V : \M \rightarrow T_\M$ as functions that produce a vector in $T_p$ at each point $p$, with an obvious generalization to \ul{tensor fields.}

Below, as well as in the main text, we mostly talk about the coordinate representation of various objects and generically do not distinguish between objects on the manifold and their coordinate representations. In particular, we will mostly only consider components of tensor fields and frequently simply refer to them as tensors. Of course, physics itself does not depend on our choice of specific coordinates (or tangent space basis) and thus only coordinate independent objects serve as well-defined physically measurable quantities. However, in practice, there is no harm in working in a specific coordinate representation. Indeed, any quantity constructed out of different tensor components, whose indices are fully contracted, defines a scalar that is invariant under coordinate transformations. Moreover, if tensor components are zero in one chart, they also vanish in any other chart.

\section{Diffeomorphisms and Lie-Derivative}\label{sApp:DiffsAndLieDer}

A \ul{diffeomorphism} is defined as an invertible smooth map $\Phi : \M\rightarrow \N$ between two manifolds $\M$ and $\N$. If such a diffeomorphism exists, then we actually identify $\M$ with $\N$. A diffeomorphism (or more generally any map between manifolds) allows the definition of a \ul{pushforward} $\Phi_* : T_p \rightarrow T_{\Phi(p)}$ on vectors $\underline V\in T_p$ given by $(\Phi_* \underline V)(f)=\underline V(f\circ \Phi)$. Such a push-forward defines a natural way to ``carry along'' vectors at $p\in\M$ to vectors at $\Phi(p)\in\M$.
Similarly, we define a natural \ul{pullback} $\Phi^* : T^*_{\Phi(p)} \rightarrow T^*_p$ on covectors $\underline\omega\in T_p^*$, defined through $(\Phi^*\underline\omega)(\underline V)=\underline\omega(\Phi_* \underline V)$. Since a diffeomorphism is invertible, and $\Phi_*=(\Phi^{-1})^*$, we can more generally define the pullback of an $\binom{r}{s}$-tensor $T$ as
\begin{equation}\label{eq:pullbackGen A}
    (\Phi^*\underline T)(\omega^{(1)},...,\omega^{(r)},V^{(1)},...,V^{(s)})=\underline T((\Phi^{-1})^*\omega^{(1)},...,(\Phi^{-1})^*\omega^{(r)},\Phi_*V^{(1)},...,\Phi_*V^{(s)})\,,
\end{equation}
whereas the pushforward is given by
\begin{equation}\label{eq:pushforwradGen A}
    (\Phi_*\underline T)(\omega^{(1)},...,\omega^{(r)},V^{(1)},...,V^{(s)})=\underline T(\Phi^*\omega^{(1)},...,\Phi^*\omega^{(r)},(\Phi^{-1})_*V^{(1)},...,(\Phi^{-1})_*V^{(s)})\,.
\end{equation}

\paragraph{Chart Representation and Relation to Change of Coordinates.} Since we want to work in specific charts, we should ask the question of how the components of a vector $\underline V=V^\mu\underline{\partial}_\mu$ at $T_p$ in some chart $\phi$ with components $x^\mu$ change in the new tangent space $T_{\Phi(p)}$ after we perform a pushforward $\Phi_*$. For simplicity, we assume, here, that the chart $\phi$ includes both points $p$ and $\Phi(p)$. The answer is
\begin{align*}
    (\Phi_*V)^\mu \underline\partial_\mu (f)&=(\Phi_*V)^\mu\frac{\partial(f\circ\phi^{-1})}{\partial x^\mu}= V^\mu \underline\partial_\mu (f\circ \Phi)=V^\mu\frac{\partial(f\circ\phi^{-1}\circ\phi\circ \Phi\circ\phi^{-1})}{\partial x^\mu}\\
    &=V^\mu\frac{\partial(\phi\circ \Phi\circ\phi^{-1})^\nu}{\partial x^\mu}\frac{\partial (f\circ\phi^{-1})}{\partial x^\nu}=V^\mu\frac{\partial(\phi\circ \Phi\circ\phi^{-1})^\nu}{\partial x^\mu} \underline\partial_\nu (f)\,.
\end{align*}
In other words, the components of the vector $\Phi_*\underline{V}$ at the point $\Phi(p)$ in the chart $\phi=(x^0,...,x^3)$ are given by
\begin{equation}\label{eq:Coefficients of push forward A}
    (\Phi_*V)^\mu(x_{\Phi(p)})=\frac{\partial(\phi\circ \Phi\circ\phi^{-1})^\mu}{\partial x^\nu}V^\nu(x_p)\,.
\end{equation}
Note that a pushforward corresponds to an ``active'' transformation of vectors and tensors by mapping every point on the manifold to a different point through $\Phi$ and dragging along the tensors of the corresponding tangent spaces. However, such an active transformation is actually equivalent to a ``passive'' coordinate transformation. Indeed, the components of the vector $\Phi_*\underline{V}$ at $\Phi(p)$ in the coordinate system $x^\mu$ given in Eq.~\eqref{eq:Coefficients of push forward A} are precisely the components $V'^\mu$ of $\underline V$ at $p$ in a chart $\phi'=(x'^0,...,x'^3)$ where the new chart is given by the pullback of the original coordinates $\phi'\equiv \Phi^*\phi=\phi\circ \Phi$, such that
\begin{equation}\label{eq:Coefficients of coord transf A}
    V'^\mu(x'_p)=\frac{\partial x'^\mu}{\partial x^\nu}V^\nu(x_p)=\frac{\partial(\phi\circ \Phi\circ\phi^{-1})^\mu}{\partial x^\nu}V^\nu(x_p)\,.
\end{equation}
Comparing to Eq.~\eqref{eq:Coefficients of push forward A}, we indeed have
\begin{equation}\label{eq:EquivalenceDiffCordA}
    (\Phi_*V)^\mu(x_{\Phi(p)})=V'^\mu(x'_p)\,,
\end{equation}
where we identify $x_{\Phi(p)}\leftrightarrow x'_p$. Through Eq.~\eqref{eq:pushforwradGen A}, this correspondence can be extended to an arbitrary tensor
\begin{equation}\label{eq:EquivalenceDiffCordA aT}
    (\Phi_* T)\ud{\mu_1...\mu_r}{\nu_1...\nu_s}(x_{\Phi(p)})=T'\phantom{}\ud{\mu_1...\mu_r}{\nu_1...\nu_s}(x'_p)\,.
\end{equation}

A diffeomorphism $\Phi$ therefore provides both an ``active'' transformation through the pushforward, as well as a ``passive'' operation, in which the effect of $\Phi$ is given by inducing a natural change of coordinates $x^\mu\rightarrow x'^\mu(x)$. More precisely, in the active point of view, we talk about how tensors change under transformations between different points on the manifold in one single chart (for simplicity). On the other hand, in the passive point of view, the tensors at the point $p$ are left untouched, and instead we evaluate these tensor components in a new coordinate induced basis given by the pullback of the coordinates at the point $\Phi(p)$. By Eq.~\eqref{eq:EquivalenceDiffCordA} these two formulations are however equivalent in practice.

Equivalently, through $\Phi^*=\big(\Phi^{-1}\big)_*$, one can also compare the pullback of a vector
\begin{equation}\label{eq:Coefficients of pull back A}
    (\Phi^*V)^\mu(x_{p})=\frac{\partial(\phi\circ \Phi^{-1}\circ\phi^{-1})^\mu}{\partial x^\nu}V^\nu(x_{\Phi(p)})\,,
\end{equation}
with the components of a coordinate transformation
\begin{equation}\label{eq:Coefficients of coord transf 2 A}
    \tilde{V}^\mu(x_p)=\frac{\partial \tilde{x}^\mu}{\partial x^\nu}V^\nu(x'_p)\,,
\end{equation}
where $x'_p$ is the inverse coordinate transformation from $\tilde x^\mu(x)$. For $\tilde \phi=\Phi_* \phi=\big(\Phi^{-1}\big)^*\phi=\phi\circ \Phi^{-1}$ the pushforward of the chart $\phi$ (such that $\tilde \phi\circ \phi$ is the inverse of the chart transition map $\phi'\circ\phi^{-1}=\phi\circ\Phi\circ\phi^{-1}$), the two expressions on the right-hand-side in Eqs.\eqref{eq:Coefficients of pull back A} and \eqref{eq:Coefficients of coord transf 2 A} are equal after identifying $x_{\Phi(p)}\leftrightarrow x'_p$ and therefore
\begin{equation}
    (\Phi^*V)^\mu(x_{p})=\tilde{V}^\mu(x_p)\,.
\end{equation}
And again, we can use Eq.~\eqref{eq:pullbackGen A} to generalize to 
\begin{equation}\label{eq:EquivalenceDiffCord2A}
    (\Phi^*T)\ud{\mu_1...\mu_r}{\nu_1...\nu_s}(x_{p})=\tilde{T}\ud{\mu_1...\mu_r}{\nu_1...\nu_s}(x_p)\,.
\end{equation}

Note that the pullback map allows us to compare tensors at different spacetime points that are related through a diffeomorphism $\Phi$ in the sense that we can compare $\underline T(\Phi(p))$ to $\underline T(p)$, by pulling back and compare $\underline \Phi^*T$ to $\underline T$ at $p$. This provides a notion of symmetry, in the sense that if $\underline \Phi^*T=\underline T$, then the Tensor does not change under $\Phi$. Equivalently, the symmetry condition can also be written as $\underline \Phi_*T=\underline T$, where now we compare at $\Phi(p)$ instead. In Section \ref{sApp: Spacetime Gaugefreedom and symmetries} below, we will further analyze the notion of spacetime symmetry and the connection to the concept of symmetry introduced in Appendix \ref{App: Symmetires in Physics}.

\paragraph{Flows and Lie Derivative.}
It will be very useful to formulate an infinitesimal version of diffeomorphisms, in particular in connection to statements of symmetry (see Sec.~\ref{sApp: Spacetime Gaugefreedom and symmetries}). Such an infinitesimal notion is provided by the concept of \ul{flows}, defined as a one-parameter family of diffeomorphisms $\Phi_\lambda:\mathbb{R}\times\M\rightarrow\M$, such that $\Phi_\lambda\circ \Phi_{\tau}=\Phi_{\lambda+\tau}$. This property in particular implies that $\Phi_0$ is the identity and $\Phi_{-\lambda}=\Phi_\lambda^{-1}$. At any point $p\in\M$ a flow defines a curve $\gamma: \mathbb{R}\rightarrow \M$ through $p$, by $\gamma(\lambda)=\Phi_\lambda(p)$, with $p=\gamma(0)$. Thus, a flow also defines a tangent vector at each point $p\in\M$ (see Eq.~\eqref{eq:Tangent Vector A}) and therefore a vector field $\underline v:\M\rightarrow T_\M$. Conversely, it can be shown \cite{WaldBook} that any vector field $\underline v$ uniquely defines a flow $\Phi^v_\lambda$, through the \ul{integral curves} of the vector field, that are the solutions to
\begin{equation}\label{eq:integral curves A}
    \frac{dx^\mu(\lambda)}{d\lambda}=v^\mu(x(\lambda))\,,
\end{equation}
where $x^\mu(\lambda)$ are the components of the integral curves in a given chart.

Infinitesimally, the diffeomorphism generated by the vector field $\underline v$ is given by moving any point $p\in\M$ with coordinates $x^\mu_p$ to a \ul{neighbouring point} $q=\Phi(p)\in\M$ with coordinates 
\begin{equation}\label{eq:active inf transf A}
    x^\mu_q=x^\mu_p+dx^\mu(x_p)\,,
\end{equation}
where $dx^\mu\equiv \Delta \lambda v^\mu=\Delta \lambda \dot x^\mu$ are the components of the \ul{infinitesimal tangent vector}
\begin{equation}
    \underline{d v} = dx^\mu\underline{\partial}_\mu= \Delta\lambda \,\dot x^\mu \underline{\partial}_\mu\,.
\end{equation}
Hence, infinitesimal tangent vectors, unlike regular tangent vectors, connect the two neighboring points $p$ and $q$ corresponding to the values $x^\mu(0)$ and $x^\mu(\Delta\lambda)$ of the integral curve. For instance, the infinitesimal difference $df$ in a scalar $f(\lambda)$ along a curve $x^\mu(\lambda)$ between two neighboring points reads
\begin{equation}
    df=dx^\mu \partial_\mu f =\underline{\dd f}(\underline{dv})=\underline{dv}(f)=\underline{dv}(\underline{\dd f})\,.
\end{equation}

Equation~\eqref{eq:integral curves A} is of course just the reverse of the definition of a tangent vector in Eq.~\eqref{eq:Tangent Vector Derivation A} through the directional derivative 
\begin{equation}\label{eq:DirectionalDerivativeA}
    \frac{df(\lambda)}{d\lambda}=\lim_{\Delta \lambda\rightarrow 0}\left(\frac{f(\lambda+\Delta \lambda)-f(\lambda)}{\Delta \lambda}\right)=\dot x^\nu\partial_\nu f\,.
\end{equation}
Note, in particular, that the object $\partial_\nu f$ defines the components of the covector given by the differential $\underline{\dd f}=\partial_\mu f \underline{\dd x}^\mu$. As we will now explore, the notion of flows will allow us to define a directional derivative of a vector and generally of tensors. Indeed, the notion of directional derivative of a vector is a subtle issue. A naive definition of the directional derivative of the components of a vector field $X^\mu(\lambda)$ along a curve could be
\begin{equation}
    \frac{dX^\mu(\lambda)}{d\lambda}=\lim_{\Delta \lambda\rightarrow 0}\left(\frac{X^\mu(\lambda+\Delta \lambda)-X^\mu(\lambda)}{\Delta \lambda}\right)=\dot x^\nu\partial_\nu X^\mu\,.
\end{equation}
However, note that this expression compares the components of the vector field at two different points, hence one compares the components with respect to two in principle different tangent space basis. This is not particularly useful, and indeed, the object $\partial_\nu X^\mu$ does not transform as a tensor. In order to generalize the concept of directional derivatives of tensors, we therefore need to specify how we want to relate different tangent spaces to each other.

Given a diffeomorphism $\Phi$, the pullback and pushforward of that diffeomorphism precisely provide us with a notion of relation between tangent spaces. In particular, the pullback of a flow now allows comparing a tensor field at two different but infinitesimally close points $p$ and $\Phi_{\Delta\lambda}(p)$, in the limit where $\Delta\lambda\rightarrow 0$. Thus, a flow generated by a vector field $\underline v$ defines a generalized directional derivative of an $\binom{r}{s}$-tensor field $\underline T$ along the vector field $\underline v$, known as \ul{Lie derivative}
\begin{equation}\label{eq:LieDerivativeDefA}
    \mathcal{L}_v\underline T=\lim_{\Delta\lambda\rightarrow 0}\frac{\big(\Phi^v_{\Delta\lambda}\big)^*\underline{T}- \underline T}{\Delta\lambda}=\frac{d}{d\lambda}\big(\Phi^v_{\lambda}\big)^*\underline T\,,
\end{equation}
where both $\underline T$ and $\big(\Phi^v_{\Delta\lambda}\big)^*\underline{T}$ are tensors at $p$. Alternatively, one can also define
\begin{equation}\label{eq:LieDerivativeDef2A}
    \mathcal{L}_v\underline T=\lim_{\Delta\lambda\rightarrow 0}\frac{\underline T-\big(\Phi^v_{\Delta\lambda}\big)_*\underline{T} }{\Delta\lambda}=-\frac{d}{d\lambda}\big(\Phi^v_{\lambda}\big)_*\underline T\,,
\end{equation}
where this time we compare the expressions at $\Phi^v_{\Delta\lambda}(p)$.
This derivative measures how a $\binom{r}{s}$-tensor changes as we \ul{Lie-drag} it along the integral curves of a given vector field $\underline v$. Moreover, $\mathcal{L}_v\underline{T}$ defines a new $\binom{r}{s}$-tensor at $p$. Clearly, it generalizes the directional derivative of a function in Eq.~\eqref{eq:DirectionalDerivativeA}, since for functions we recover $\mathcal{L}_vf=v(f)$.
In terms of components of the tensor and $q={\Phi^v_{\Delta\lambda}(p)}$ we have
\begin{equation}\label{eq:LieDerivativeDefCompA}
    \mathcal{L}_vT\ud{\mu_1...\mu_r}{\nu_1...\nu_s}=\lim_{\Delta\lambda\rightarrow 0}\left(\frac{\big(\Phi^v_{\Delta\lambda}\big)^*\left[T\ud{\mu_1...\mu_r}{\nu_1...\nu_s}(x_q)\right]-T\ud{\mu_1...\mu_r}{\nu_1...\nu_s}(x_p)}{\Delta\lambda}\right)\,,
\end{equation}
or equivalently
\begin{equation}\label{eq:LieDerivativeDefComp2A}
    \mathcal{L}_vT\ud{\mu_1...\mu_r}{\nu_1...\nu_s}=\lim_{\Delta\lambda\rightarrow 0}\left(\frac{T\ud{\mu_1...\mu_r}{\nu_1...\nu_s}(x_q\big)-\big(\Phi^v_{\Delta\lambda}\big)_*\left[T\ud{\mu_1...\mu_r}{\nu_1...\nu_s}(x_p)\right]}{\Delta\lambda}\right)\,.
\end{equation}

In the same way as active diffeomorphisms are related to coordinate transformations, Lie derivatives are related to infinitesimal coordinate transformations. It is in particular useful to also define the Lie derivative in terms of the passive picture of coordinate transformations, in order to derive the expressions in a given chart. In the passive picture, we do not use the diffeomorphism $\Phi$ infinitesimally generated by a vector field through Eq.~\ref{eq:passive inf transf A} to push forward tensors to different points on the manifold as in Eq.~\eqref{eq:LieDerivativeDefComp2A}, but rather define a chart transition map $\phi'\circ\phi^{-1}=\phi\circ\Phi\circ\phi^{-1}$, which gives rise to an infinitesimal coordinate transformation that relabels each point on the manifold
\begin{equation}\label{eq:passive inf transf A}
    x'^\mu_p=x^\mu_p+dx^\mu(x_p)\,.
\end{equation}
Using Eq.~\eqref{eq:EquivalenceDiffCordA aT}, we can then rewrite Eq.~\eqref{eq:LieDerivativeDefComp2A} in terms of passive coordinate transformations
\begin{equation}\label{eq:LieDerivativeDefCompCoords2A}
    \mathcal{L}_vT\ud{\mu_1...\mu_r}{\nu_1...\nu_s}=\lim_{\Delta\lambda\rightarrow 0}\left(\frac{T\ud{\mu_1...\mu_r}{\nu_1...\nu_s}(x'_p)-T'\phantom{}\ud{\mu_1...\mu_r}{\nu_1...\nu_s}(x'_p)}{\Delta\lambda}\right)\,.
\end{equation}
where we replaced $x_q\rightarrow x'_p$. Note that $\big(\Phi^v_{\Delta\lambda}\big)_*\left[T\ud{\mu_1...\mu_r}{\nu_1...\nu_s}(x_p)\right]$ is an expression evaluated at $x_q$ after performing the pushforward.

For a vector $X^\mu$, we can readily compute the expression in Eq.~\eqref{eq:LieDerivativeDefCompCoords2A}, using 
\begin{equation}
    X'^\mu(x')=\frac{\partial x'^\mu}{\partial x^\alpha} X^\alpha(x)=(\delta^\mu_\alpha+\partial_\alpha dx^\mu)X^\alpha(x)+\mathcal{O}(\Delta\lambda^2)\,,
\end{equation}
and $X^\mu(x')=X^\mu(x)+dx^\mu\partial_\mu X^\mu(x)+\mathcal{O}(\Delta\lambda^2)$, with $dx^\mu=\Delta\lambda \,v^\mu$ in order to obtain
\begin{equation}\label{eq:CompOfLieDerVEcA}
    \mathcal{L}_v X^\mu=v^\lambda\partial_\lambda X^\mu-X^\lambda\partial_\lambda v^\mu\,. 
\end{equation}
Similarly, for a covector $\omega_\mu$ we get
\begin{equation}
    \mathcal{L}_v \omega_\mu=v^\lambda\partial_\lambda \omega_\mu+\omega_\lambda\partial_\mu v^\lambda\,. 
\end{equation}
The obvious generalization to an arbitrary tensor is therefore
\begin{equation}\label{eq:LieDerCompA}
    \mathcal{L}_v T^{\mu...}{}_{\nu...}=v^\lambda\partial_\lambda T^{\mu...}{}_{\nu...}-\partial_\lambda v^\mu T^{\lambda...}{}_{\nu...}-...+\partial_\nu v^\lambda T^{\mu...}{}_{\lambda...}+...\,.
\end{equation}

For completeness, we also rewrite the formula in Eq.~\eqref{eq:LieDerivativeDefCompA} in terms of the passive picture by using Eq.~\eqref{eq:EquivalenceDiffCord2A}
\begin{equation}\label{eq:LieDerivativeDefCompCoordsA}
    \mathcal{L}_vT\ud{\mu_1...\mu_r}{\nu_1...\nu_s}=\lim_{\Delta\lambda\rightarrow 0}\left(\frac{\tilde T\ud{\mu_1...\mu_r}{\nu_1...\nu_s}(x_p)-T\ud{\mu_1...\mu_r}{\nu_1...\nu_s}(x_p)}{\Delta\lambda}\right)\,,
\end{equation}
where the coordinate change is given by $\tilde \phi\circ\phi^{-1}=\phi\circ\Phi^{-1}\circ \phi^{-1}$, such that 
\begin{equation}
    \tilde x_p=x_p-dx^\mu(x_p)\,,
\end{equation}
and where
\begin{equation}\label{eq:DefTildeTransform}
    \tilde T\ud{\mu_1...\mu_r}{\nu_1...\nu_s}(x_p)=\frac{\partial\tilde x^{\mu_1}}{\partial x^{\alpha_1}}...\frac{\partial\tilde x^{\mu_r}}{\partial x^{\alpha_r}}\frac{\partial x^{\beta_1}}{\partial \tilde x^{\nu_1}}...\frac{\partial x^{\beta_s}}{\partial \tilde x^{\nu_s}}T\ud{\alpha_1...\alpha_r}{\beta_1...\beta_s}(x'_p)\,.
\end{equation}

\paragraph{Lie Bracket.}

The Lie derivative of a vector field $ \mathcal{L}_v\underline{X}$ defines a commutator known as \ul{Lie Bracket}
\begin{equation}
    \mathcal{L}_v\underline{X}\equiv [\underline v,\underline X]\,,
\end{equation}
that satisfies
\begin{equation}
    [\underline v,\underline X]=-[\underline X,\underline v]\,,
\end{equation}
as well as the Jacobi identity. To see this, one can explicitly compute the commutator for the vector fields $\underline v=v^\mu\underline{\partial}_\mu$ and $\underline X=X^\mu\underline{\partial}_\mu$ in a given chart induced basis and compare with Eq.~\eqref{eq:CompOfLieDerVEcA}
\begin{equation}\label{eq:LieDerivativeVectorsDef}
    [v^\mu\underline{\partial}_\mu,X^\nu\underline{\partial}_\nu]=\left(v^\lambda\partial_\lambda X^\mu-X^\lambda \partial_\lambda v^\mu\right)\underline{\partial}_\mu=\mathcal{L}_vX^\mu\underline{\partial}_\mu\equiv[\underline v,\underline X]^\mu \underline{\partial}_\mu
\end{equation}

From this expression, it is also clear that the basis vectors of a coordinate induced basis commute
\begin{equation}
    [\underline{\partial}_\mu,\underline{\partial}_\nu]=0\,.
\end{equation}
In fact, locally also the converse is true: If the Lie bracket of two vector fields vanishes, there locally exists a chart in which the two vectors are the basis vectors of that chart.

\section{Connection and Curvature}\label{sApp: connection and curvature}

Instead of relying on flows in order to compare tensors at different points, and therefore different, unrelated tangent spaces, on the manifold, we can introduce an additional structure on the manifold that uniquely connects any two tangent spaces $T_p$ and $T_q$. This again allows to generalize the directional derivative defined in Eq.~\eqref{eq:DirectionalDerivativeA}. Moreover, in GR, the additional structure can be associated to the experiment of moving a gyroscope in spacetime and monitor its direction.

\paragraph{Affine Connection and Parallel Transport.}
This additional structure is mathematically captured by the notion of an \ul{affine connection} that we define by associating to any curve $\gamma$ on the manifold a \ul{parallel transport} $\tau_\gamma(u,\lambda): T_q \rightarrow T_p$, where $q=\gamma(\lambda)$ and $p=\gamma(u)$, such that $\tau_\gamma(\lambda,\lambda)$ is the identity and $\tau_\gamma(v,u)\circ \tau_\gamma(u,\lambda)=\tau_\gamma(v,\lambda)$.  Moreover, in the limit $t\rightarrow u$ the map to first order only depends on the direction $\dot x_\gamma^\beta (u)$ of the curve $\gamma$ at $u$. In other words, there exist \ul{connection coefficients} $\,\DGamma^{\,\mu}_{\,\alpha\beta}(\gamma(\lambda))$ such that
\begin{equation}
    \frac{d}{d\lambda}\tau_\gamma(u,\lambda)^\mu_\alpha\big\lvert_{\lambda=u}=\,\DGamma^{\,\mu}_{\,\alpha\beta}(p)\,\dot x_\gamma^\beta (u)\,.
\end{equation}
A connection is therefore completely determined by its connection coefficients. The parallel transport is then extended to $\binom{r}{s}$-tensor through the requirements $\tau_\gamma(u,\lambda)(c)=c$ for $c\in \mathbb{R}$, linearity of tensor products and $\tau_\gamma(u,\lambda)(\Tr[T])=\Tr[\tau_\gamma(u,\lambda)(T)]$.

Through this map that connects different tangent spaces, we thus define a generalization of the directional derivative in Eq.~\eqref{eq:DirectionalDerivativeA} of a vector field $X^\mu$ along a curve $x^\mu(\lambda)$ at $u=0$
\begin{align}
    \frac{\DD X^\mu}{d\lambda}\equiv&\lim_{\Delta\lambda\rightarrow 0}\frac{\tau(0,\Delta\lambda)^\mu_\alpha X^\alpha(x_q)-X^\mu(x_p)}{\Delta\lambda}=\frac{d}{d\lambda}\tau(0,\lambda)^\mu_\alpha X^\alpha\big\lvert_{\lambda=0} \,,\nonumber\\
    =&\,v^\beta\left(\partial_\beta X^{\mu}+\,\DGamma^{\,\mu}_{\,\alpha\beta} X^\alpha\right)\,,\label{eq:CovDerSimpA}
\end{align}
where $v^\beta=\dot x^\beta$. Note that unlike the components of the Lie derivative in Eq.~\eqref{eq:CompOfLieDerVEcA}, this so called \ul{covariant derivative} only depends on the direction of the curve $v^\beta$ at the single point $p$ and is therefore independent of the choice of the curve. We can therefore more formally define the covariant derivative of a $\binom{r}{s}$-tensor field $\underline{T}$ in the direction of a vector field $\underline{V}$ and associated to a connection as (compare to the Lie derivative in Eq.~\eqref{eq:LieDerivativeDefA})
\begin{align}
    \DGrad_V \underline{T}\equiv \lim_{\Delta\lambda\rightarrow 0}\frac{\tau(0,\Delta\lambda)\underline{T}-\underline{T}}{\Delta\lambda}\,,
\end{align}
all evaluated at $p$. This defines again a $\binom{r}{s}$-tensor, since like for the Lie derivative, the difference between two tensors on the same tangent space $T_p$ is again a tensor on that tangent space. 

However, because the covariant derivative is independent of the choice of the curve, the vector field in the expression $\DGrad_V \underline{T}$ can also be regarded as an additional argument in order to define a new $\binom{r}{s+1}$-tensor. This is the reason why the covariant derivative generalizes the object $\partial_\nu X^\mu$ to the expression inside the round brackets in Eq.~\eqref{eq:CovDerSimpA} 
\begin{align}
    \DGrad_{\underline{\partial}_{\nu}} X^\mu = \DGrad_\nu X^\mu \equiv \partial_\nu X^\mu+\,\DGamma^{\,\mu}_{\,\nu\alpha}X^\alpha\,,
\end{align}
that are the  proper components of a $\binom{1}{1}$-tensor. Similarly, the components of the covariant derivative of a covector give rise to a $\binom{0}{2}$-tensor whose components read
\begin{align}
    \DGrad_\nu \omega_\mu\equiv \partial_\nu \omega_\mu-\,\DGamma^{\,\alpha}_{\,\nu\mu}\omega_\alpha\,,
\end{align}
with an obvious generalization to arbitrary tensors.

\paragraph{Torsion and Curvature.}
Note that the connection coefficients themselves are not components of a tensor. This simply follows from the quotient theorem and the observation that $\partial_\nu V^\mu$ is not a tensor. However, two tensorial objects, the field strengths of the connection, can be constructed out of them, namely the \ul{torsion}
\begin{equation}\label{eq:DefTorsion}
    \mathcal{T}\ud{\mu}{\alpha\beta}\equiv 2\,\DGamma^{\,\mu}_{\,[\alpha\beta]}\,,
\end{equation}
which is just the antisymmetric part of the connection coefficients, as well as the \ul{curvature tensor}
\begin{equation}\label{eq:Riemann TensorA}
    \mathcal{R}\ud{\mu}{\nu\rho\sigma}\equiv\,\DGamma^{\,\mu}_{\,\nu\sigma,\rho}-\,\DGamma^{\,\mu}_{\,\nu\rho,\sigma}+\,\DGamma^{\,\mu}_{\,\lambda\rho}\;\DGamma^{\,\lambda}_{\,\nu\sigma}+\,\DGamma^{\,\mu}_{\,\lambda\sigma}\;\DGamma^{\,\lambda}_{\nu\rho}\,.
\end{equation}
In relation to their geometric meaning, these two objects naturally appear in the commutator of the covariant derivative acting on a function $f$ and a vector $V^\mu$
\begin{equation}\label{eq:Torsion Gen Appendix}
    [\DGrad_\rho,\DGrad_\sigma]f= - \mathcal{T}\ud{\lambda}{\rho\sigma}\DGrad_\lambda f\,,
\end{equation}
and (see e.g. \cite{carroll2019spacetime})
\begin{equation}\label{eq:Riemann Gen Appendix}
    [\DGrad_\rho,\DGrad_\sigma]V^\mu=\mathcal{R}\ud{\mu}{\nu\rho\sigma}V^\nu - \mathcal{T}\ud{\lambda}{\rho\sigma}\DGrad_\lambda V^\mu\,.
\end{equation}
However, note that at this stage there is no unique notion of connection and covariant derivative and without any relation to physical experiments, the choice is a priory arbitrary.

\section{Metric and Pseudo-Riemannian Geometry}\label{sApp: Metric and Riemannian G}

\paragraph{Metric and Non-Metricity.}
In order to define physical distances, in space and time, a manifold must also be equipped with a \ul{metric}. In a metric theory of gravity, without surprise, the central object capturing gravitation will be the metric. It is defined by a symmetric $\binom{0}{2}$-tensor
\begin{equation}
    \underline{g}=g_{\mu\nu} \,\underline{\dd x}^\mu\otimes \underline{\dd x}^\nu\,,
\end{equation}
which satisfies the following additional properties: The metric is assumed to be non-degenerate, and hence invertible, in the sense that there exists a tensor $g^{\mu\nu}$ such that $g_{\mu\nu}g^{\nu\rho}=\delta^\nu_\rho$, where $\delta^\mu_\nu$ is the unit matrix. The metric is also required to have a definite signature, in our case $(-,+,+,+)$, that is, at every point in $\mathcal{M}$, $g_{\mu\nu}$ has one negative and three positive Eigenvalues when viewing the components as a matrix. Therefore, the \ul{determinant of the metric} components $g\equiv \det g_{\mu\nu}$ always satisfies $g<0$. Furthermore, at every point, the metric defines a raising and lowering map which connects the tangent and cotangent spaces by allowing to raise and lower indices. 

Most importantly, the metric determines a magnitude in terms of a physical norm of vector fields $\underline{V}=v^\mu\underline{\partial}_\mu$ at each point through
\begin{equation}\label{eq:Norm of VectorA}
    g(\underline{V},\underline{V})=g_{\mu\nu}v^\mu v^\nu\,.
\end{equation}
As such, a metric therefore also introduces a notion of distance between two neighboring points. This infinitesimal distance $ds$, termed \ul{line element}, is therefore given by
\begin{equation}
    ds^2\equiv g(\underline{dX},\underline{dX})= g_{\mu\nu} dx^\mu dx^\nu\,.
\end{equation}
Interestingly, through the so called \ul{polarization identity}
\begin{equation}
    g(\underline{V},\underline{W})=\frac{1}{4}\left[g(\underline{V}+\underline{W},\underline{V}+\underline{W})-g(\underline{V}-\underline{W},\underline{V}-\underline{W})\right]\,,
\end{equation}
the full metric can be reconstructed from the knowledge of the spacetime-distance between neighboring points.

With the metric at hand, an additional tensorial object can be constructed out of a general connection $\,\DGamma^{\,\mu}_{\,\alpha\beta}$, namely the \ul{non-metricity}
\begin{equation}\label{eq:DefNonMetricity}
    \mathcal{Q}_{\alpha\mu\nu}\equiv \DGrad_\alpha g_{\mu\nu}=\partial_\alpha g_{\mu\nu}-2\,\DGamma^{\,\lambda}_{\,\alpha(\mu}\, g_{\nu)\lambda}\,.
\end{equation}

\paragraph{Levi-Civita Connection.}
A differential manifold with a metric as defined above is known as a \ul{pseudo-Riemannian manifold}. According to the fundamental theorem of Riemannian geometry \cite{levi1917nozione,zbMATH06520113,zbMATH00052737}, there exists a unique connection, called the \ul{Levi-Civita connection}, with associated \ul{Christoffel symbols} $\Gamma^\mu_{\alpha\beta}$ and covariant derivative $\nabla_\mu$, such that
\begin{align}
    \Gamma^\mu_{[\alpha\beta]}& =0\,,\label{eq:NoTorsionA}\\
    \nabla_\alpha g_{\mu\nu}& =0\,.\label{eq:NoNonMetricityA}
\end{align}
Note that this precisely correspond to the conditions that the connection is free of torsion and non-metricity
\begin{align}
    \mathcal{T}\ud{\mu}{\alpha\beta}& \overset{!}{=}0\,,\\
    \mathcal{Q}_{\alpha\mu\nu}& \overset{!}{=}0\,.
\end{align}
The connection coefficients of the Levi-Civita connection are completely determined by the metric
\begin{equation}\label{eq:Christoffel Appendix}
    \Gamma^\alpha_{\ \mu\nu}=\frac12g^{\alpha\beta}(\partial_\mu g_{\nu\beta}+\partial_\nu g_{\mu\beta}-\partial_\beta g_{\mu\nu})\,.
\end{equation}
In fact, it follows that any connection with coefficients $\,\DGamma^{\,\mu}_{\,\alpha\beta}$ admits the following decomposition 
\begin{equation}
    \,\DGamma^{\,\mu}_{\,\alpha\beta}=\Gamma^{\mu}_{\alpha\beta}+\mathcal{K}\ud{\mu}{\alpha\beta}+\mathcal{L}\ud{\mu}{\alpha\beta}\,,
\end{equation}
where the \ul{contortion} and \ul{disformation} pieces are completely determined by the torsion and the non-metricity respectively
\begin{equation}
   \mathcal{K}\ud{\mu}{\alpha\beta}\equiv \frac{1}{2}\,\mathcal{T}\ud{\mu}{\alpha\beta}+\mathcal{T}^{\;\;\mu}_{(\alpha\;\,\beta)}\,,\quad \mathcal{L}\ud{\mu}{\alpha\beta}\equiv \frac{1}{2}\,\mathcal{Q}\ud{\mu}{\alpha\beta}-\mathcal{Q}^{\;\;\;\mu}_{(\alpha\;\,\beta)}\,.
\end{equation}

\paragraph{Riemann Curvature with Levi-Civita Connection.}
Since the Christoffel symbols are completely determined by the metric, so is the associated \ul{Riemann curvature tensor}\footnote{For simplicity, we will reserve the terminology "Riemann curvature tensor" for the curvature tensor associated to the Levi-Civita connection.}
\begin{equation}\label{eq:Riemann Tensor A}
    R\ud{\mu}{\nu\rho\sigma}=\Gamma^\mu_{\nu\sigma,\rho}-\Gamma^\mu_{\nu\rho,\sigma}+\Gamma^\mu_{\lambda\rho}\Gamma^\lambda_{\nu\sigma}+\Gamma^\mu_{\lambda\sigma}\Gamma^\lambda_{\nu\rho}\,.
\end{equation}
Moreover, the Riemann tensor fulfills several symmetry properties and identities
\begin{align}
    R_{\alpha\beta\mu\nu}&=-R_{\alpha\beta\nu\mu}=-R_{\beta\alpha\mu\nu}=R_{\mu\nu\alpha\beta}\, ,\\
    R_{[\alpha\beta\mu\nu]}&=0\, ,\\
    R^\alpha_{\ [\beta\mu\nu]}&=0\, ,\\
    \nabla_{[\rho}R^\alpha_{\ |\beta|\mu\nu]}&=0\, ,
\end{align}
where the last two are known as the 1$^\text{st}$ and $2^\text{nd}$ \ul{Bianchi identities}. The Riemann tensor has only one independent contraction known as the \ul{Ricci tensor}
\begin{equation}
    R_{\mu\nu}\equiv R^\lambda_{\ \mu\lambda\nu}\,,
\end{equation}
which can be further contracted to define the \ul{Ricci scalar}
\begin{equation}\label{eq:RicciScalar App}
    R\equiv g^{\mu\nu}R_{\mu\nu}\,.
\end{equation}

As explained in the main text, the Riemann tensor $R\ud{\alpha}{\beta\mu\nu}$ provides a measure of how ``curved'' the spacetime determined by the metric is. Moreover, the parallel transport associated to the Levi-Civita connection can be related to the physical operation of monitoring the direction of a gyroscope moving in spacetime without applied forces.

\paragraph{Lie Derivative on a torsion-free Pseudo-Riemannian Manifold.}
For a torsion free covariant derivative $\nabla_\mu$, the expression for the components of the Lie derivative in Eq.~\eqref{eq:LieDerCompA} can be rewritten as \cite{carroll2019spacetime}
\begin{equation}\label{eq:FormulaLieDerivativeGeneral}
    \mathcal{L}_V T^{\mu...}{}_{\nu...}=V^\lambda\nabla_\lambda T^{\mu...}{}_{\nu...}-\nabla_\lambda V^\mu T^{\lambda...}{}_{\nu...}-...+\nabla_\nu V^\lambda T^{\mu...}{}_{\lambda...}+...\,,
\end{equation}
which is explicitly covariant. In particular, the Lie derivative along $V^\mu$ of the metric $g_{\mu\nu}$ reads
\begin{equation}
    \mathcal{L}_v g_{\mu\nu}=v^\alpha\nabla_\alpha g_{\mu\nu}+(\nabla_\mu v^\alpha)g_{\alpha\nu}+(\nabla_\nu v^\alpha)g_{\mu\alpha}\,,
\end{equation}
which reduced to
\begin{equation}\label{eq:LieDerivativeOfMetricA}
    \mathcal{L}_v g_{\mu\nu}=2\nabla_{(\mu}v_{\nu)}\,,
\end{equation}
for a covariant derivative that is also metric compatible. Note also that the Lie derivative does not commute with the raising and lowering of indices, such that is it important to specify whether we apply it on a vector component or the corresponding covector.

\section{Normal Coordinates}\label{sApp: Normal Coordinates}

\paragraph{Riemann Normal Coordinates.}
At every point $p\in \M$ of a pseudo-Riemannian manifold $\M$ with Levi-Civita connection, we can construct geodesic, or \ul{Riemann normal coordinates} $\{y^\mu\}$ for which (see e.g. \cite{misner_gravitation_1973,zbMATH00052737,zbMATH06520113,carroll2019spacetime,Renner2020})
\begin{equation}\label{eq:RiemannNormalCoordinates}
    g_{\mu\nu}(y)|_p=\eta_{\mu\nu}\quad\text{and} \quad g_{\mu\nu,\alpha}(y)|_p=0\,,
\end{equation}
where $\eta_{\mu\nu}$ are the components of the Minkowski metric in inertial coordinates 
\begin{equation}
    \eta_{\mu\nu}\equiv\text{diag}(-1,+1,+1,+1)\,.
\end{equation}
Hence, up to first order, the components of the metric correspond to the Minkowski metric. Note that the second condition in Eq.~\eqref{eq:RiemannNormalCoordinates} implies, that the Christoffel symbols also vanish at $p$
\begin{equation}\label{eq:Condition on Connection}
    \Gamma^\rho_{\mu\nu}(y)|_p=0\,.
\end{equation}
However, the derivatives of the Christoffel symbols at $p$ or equivalently the second derivatives of the metric do not vanish in general. Indeed, if we choose the point $p$ to be the origin of the coordinate system, the components of the metric in Riemann normal coordinates can in general only be constructed such that the metric adopts the Minkowski form up to second order in an expansion around the origin
\begin{equation}
    g_{\mu\nu}(y)=\eta_{\mu\nu}+\mathcal{O}\left(\frac{y^2}{D^2}\right)\,,
\end{equation}
where $D^{-2}\sim |R_{\mu\nu\rho\sigma}|$ defines a scale of length (or time) associated to the spacetime curvature.

Normal coordinates should clearly be distinguished from the concept of \textit{flatness} in Definition~\ref{DefFlatness}. It is also important to note that Riemann normal coordinates can explicitly only be constructed if torsion and non-metricity vanishes \cite{Hartley:1995dg,carroll2019spacetime,Renner2020}. More precisely, for a general connection to satisfy the definition of Riemann normal coordinates, both the conditions in Eqs.~\eqref{eq:RiemannNormalCoordinates} and \eqref{eq:Condition on Connection} on the metric and the connection 
need to be separately satisfied as they are a priori independent of each other, hence
\begin{align}\label{eq:conditionsNormal Coords}
    g_{\mu\nu,\alpha}(y)|_p=0\,,\qquad \DGamma^{\,\mu}_{\,\alpha\beta}(y)|_p=0\,.
\end{align}
However, since both the torsion and the non-metricity are proper tensor fields, no coordinate transformation will be able to transform them away. Conversely, if the components of torsion and non-metricity vanish in some coordinate system, they will be zero in any coordinate system. On the other hand, if there exists a coordinate chart for which at some point the connection vanishes $\,\DGamma^{\,\mu}_{\,\alpha\beta}=0$, then through Eq.~\eqref{eq:DefTorsion} also torsion vanishes, while if both conditions in Eq.~\eqref{eq:conditionsNormal Coords} are satisfied, also non-metricity defined in Eq.~\eqref{eq:DefNonMetricity} will be zero.

\paragraph{Fermi Normal Coordinates.} The Riemann normal coordinates defined above correspond to a local inertial frame at a single point in spacetime, and therefore also at a single moment in the proper time of an observer. However, for a physical observer on a timelike geodesic, one can actually construct Riemann normal coordinates at every instant in proper time $\tau$ along the geodesic. The resulting chart scalars $\{t,y^i\}$ are called \ul{Fermi normal coordinates} and corresponds to what is known as a \ul{freely falling frame} that defines local inertial frames along the entire geodesic.

The construction of Fermi normal coordinates can intuitively be understood as follows \cite{maggiore2008gravitational,zee2013einstein}: Along the geodesic $y^\mu(\tau)$ one chooses $t=\tau$ and $y^i=0$, hence $y^\mu(\tau)=(\tau,0)$. Moreover, it is possible to construct an orthonormal coordinate system (using gyroscopes) that one can parallel transport along the entire geodesic, such that
\begin{equation}
    g_{\mu\nu}(y)|_\gamma = \eta_{\mu\nu}\,,
\end{equation}
where here we indicate that an expression is evaluated on a point on the geodesic $y^\mu(\tau)$ by $|_\gamma$.
Furthermore, it is possible to construct the coordinate system such that also
\begin{equation}
    \Gamma^\rho_{\mu\nu}(y)|_\gamma=0\,,
\end{equation}
along the entire geodesic. That this is possible is not evident, but can be shown by explicitly constructing a coordinate system for which the components of the metric read \cite{1963JMP,zee2013einstein}
\begin{equation}\label{FermiNormalCoords}
g_{\mu\nu}(y)=
\eta_{\mu\nu}-
    \left(\begin{array}{c|c c c} 
    	R_{0k0l}|_\gamma\, y^ky^l &  &\frac{2}{3}R_{0kjl}|_\gamma\, y^ky^l&\phantom{0} \\
    	\hline 
    	 &  &&\\
\frac{2}{3}R_{0kjl}|_\gamma\, y^ky^l&  &\frac{1}{3}R_{ikjl}|_\gamma\, y^ky^l& \\
 \phantom{0} &  & & \\
    \end{array}\right)+\mathcal{O}\left(\frac{y^3}{D^3}\right)\,,
\end{equation}
where again $D^{-2}\sim |R_{\mu\nu\rho\sigma}|$.

\section{Spacetime Gauge Freedom and Symmetries}\label{sApp: Spacetime Gaugefreedom and symmetries}

\paragraph{Gauge Freedom and Gauge Symmetry.}
In theories of gravity that are described through a differentiable manifold, a metric and a set of additional fields $(\M,\underline{g},\underline{\Psi})$, the freedom of performing diffeomorphisms $\Phi: \M\rightarrow \M$ represents a gauge freedom in the sense defined in Appendix~\ref{App: Symmetires in Physics}. More precisely, the solutions $(\M,\underline{g},\underline{\Psi})$ and $(\M,\underline{\Phi^*g},\underline{\Phi^*\Psi})$ represent the same physical spacetime \cite{WaldBook}. In most formulations of gravity theories, this gauge freedom of coordinate transformations is promoted to a gauge symmetry of the theory. In that case, the action of the theory is invariant under 

\paragraph{Global Symmetries.} 
On the other hand, a diffeomorphism is called a \ul{symmetry transformation} of a tensor field $\underline T$, if 
\begin{equation}
    \underline \Phi_*T=\underline T\,.
\end{equation}
That is, a spacetime symmetry is associated to the invariance of an underlying structure on the manifold. Most importantly, if the metric of a spacetime (defined in Sec.~\ref{sApp: Metric and Riemannian G} below) is left invariant
\begin{equation}\label{eq:IsometryA}
    (\Phi^*g)_{\mu\nu}=g_{\mu\nu}\,,
\end{equation}
the corresponding diffeomorphism is termed an \ul{isometry}. For a diffeomorphism invariant theory, this notion of symmetry coincides with the concept of global or proper symmetry introduced in Appendix~\ref{App: Symmetires in Physics}. This is because the action of a coordinate invariant theory is a scalar under general coordinate transformations. The Lagrangian density for instance contains terms of the form $g_{\mu\nu}\Psi^{\mu\nu}$, where $\Psi^{\mu\nu}$ is any tensor constructed out of the additional fields. By construction, any such coordinate scalar expressions are invariant under generic coordinate transformations
\begin{equation}
    g'_{\mu\nu}\Psi'^{\mu\nu}=  \frac{\partial x^\alpha}{\partial x'^\mu}\frac{\partial x^\beta}{\partial x'^\nu}\,g_{\alpha\beta}\,\frac{\partial x'^\mu}{\partial x^\rho}\frac{\partial x'^\nu}{\partial x^\sigma}\Psi^{\rho\sigma}= g_{\alpha\beta}\Psi^{\alpha\beta}\,.
\end{equation}
Suppose that for a given solution for the metric $\underline{g}$, there exists a particular diffeomorphism that leaves this metric invariant, hence, is a symmetry transformation [Eq.~\eqref{eq:IsometryA}]. This implies, that there exists a coordinate system in which the metric has a specific form $\mathcal{G}_{\mu\nu}$, in which the action of the coordinate transformation associated to the diffeomorphism reads
\begin{equation}
    \mathcal{G}'_{\mu\nu}=\Lambda\ud{\alpha}{\mu}\Lambda\ud{\beta}{\nu} \mathcal{G}_{\alpha\beta}=\mathcal{G}_{\mu\nu}\,.
\end{equation}
Then, the theory formulated in this coordinate system, in which we now treat the metric as an auxiliary field, is symmetric under the global transformation
\begin{equation}
   \Psi'_{\mu}=\Lambda\ud{\alpha}{\mu}\Psi_\alpha\,,
\end{equation}
applied to all the other fields, since any term in the action still remains invariant. For instance 
\begin{equation}
    \mathcal{G}_{\mu\nu}\Psi'^{\mu\nu}= \mathcal{G}_{\mu\nu}\,\Lambda\ud{\mu}{\alpha}\Lambda\ud{\nu}{\beta}\Psi^{\alpha\beta}= \mathcal{G}_{\alpha\beta}\Psi^{\alpha\beta}\,.
\end{equation}

Lie derivative provides us with an infinitesimal notion of spacetime symmetry, in the sense that 
\begin{equation}
    \mathcal{L}_vT\ud{\mu_1...\mu_r}{\nu_1...\nu_s}=0
\end{equation}
everywhere if and only if for all $\lambda$, $\Phi_\lambda$ is a symmetry transformation of $T\ud{\mu_1...\mu_r}{\nu_1...\nu_s}$.  
In particular, a vector field $\underline{V}$ that generates an isometry is called a \ul{Killing vector field} and satisfies (see Eq.~\eqref{eq:LieDerivativeOfMetricA})
\begin{equation}
    \mathcal{L}_V g_{\mu\nu}=2\nabla_{(\mu}V_{\nu)}=0\ .
\end{equation}
where we have used the torsion-freeness and metricity of the connection. Note that the Lie derivative is closely related to the total variation $\bar{delta}$ that we introduced in Appendix~\ref{App: Symmetires in Physics}, in the sense that $\bar{\delta} T\ud{\mu_1...\mu_r}{\nu_1...\nu_s}=0$ for the transformation generated by the vector field $\underline{V}$ if and only if $\mathcal{L}_vT\ud{\mu_1...\mu_r}{\nu_1...\nu_s}=0$.

\chapter{Proofs and Derivations}\label{App:Proofs}


\section{Proof of Eq.~\eqref{eq:condition deviation}}\label{sApp:SpacialGeodesicDeviation}
\textit{A variant of this proof can also be found in \cite{Blau2017}.}
\\

\noindent
Consider a two-parameter family of timelike geodesics $x^\mu(\tau,z)$, with infinitesimal tangent vector
\begin{equation}
    d x^\mu=\frac{\partial}{\partial\tau}x^\mu(\tau,z)\,d\tau=\dot x^\mu(\tau,z)\,d\tau\,.
\end{equation}
and infinitesimal spacelike deviation vector
\begin{equation}
    \delta x^\mu=\frac{\partial}{\partial z}x^\mu(\tau,z)\,dz\,.
\end{equation}
In this Appendix we want to prove Eq.~\eqref{eq:condition deviation} that states that
\begin{equation}\label{eq:condition deviation init Proof}
    dx_\mu \delta x^\mu = \text{const.}\,.
\end{equation}

\begin{proof}
First of all, recall that by definition the tangent of the geodesic and the deviation vector, as two natural basis vectors for the two-parameter family of geodesics, satisfy Eq.~\eqref{eq:LieDerivativeVectorsDef}
\begin{equation}\label{eq:conditionLieBracket Proof}
    [\underline{d x},\underline{\delta x}]^\mu=0=dx^\lambda\nabla_\lambda \delta x^\mu-\delta x^\lambda \nabla_\lambda dx^\mu\,.
\end{equation}
Moreover, the tangent vector of a geodesic is by definition parallel transported along it [Eq.~\eqref{eq:ParallelTransportOfVector}]
\begin{equation}\label{eq:ParallelTransportOfVector Proof}
    \frac{D \,dx^\mu}{d\tau}=d x^\nu \nabla_\nu dx^{\mu}=0\,.
\end{equation}

In the same language, the condition in Eq.~\eqref{eq:conditionLieBracket Proof} therefore implies that
\begin{equation}
     \frac{D \,\delta x^\mu}{d\tau}=d x^\nu \nabla_\nu \delta x^{\mu}=\delta x^\nu \nabla_\nu dx^{\mu}\,.
\end{equation}
Contracting this expression with the tangent vector of the geodesic, we thus have the identity
\begin{equation}\label{eq:transverse condition Proof}
    dx_\mu \frac{D \,\delta x^\mu}{d\tau}=\frac{1}{2}\delta x^\nu \nabla_\nu (dx_\mu dx^{\mu})=0\,,
\end{equation}
where the expression vanishes for an affinely parameterized geodesic with $dx_\mu dx^{\mu}= C =$ const., such that $\nabla_\nu C=\partial_\mu C=0$. In other words, the parallel transport of the deviation vector along the geodesic is transverse to the geodesic tangent vector.

This further implies that
\begin{equation}
    \frac{d}{d\tau}(dx_\mu \delta x^\mu)=\delta x_\mu\frac{D}{d\tau}dx^\mu+d x_\mu\frac{D}{d\tau} \delta x^\mu=0\,.
\end{equation}
Here, the first term vanishes due to Eq.~\eqref{eq:ParallelTransportOfVector Proof} while the second one due to Eq.~\eqref{eq:transverse condition Proof}. Hence, indeed the contraction between the infinitesimal tangent vector of the geodesic and the infinitesimal deviation vector remains a constant along the geodesic
\begin{equation}\label{eq:condition deviation Proof}
    dx_\mu \delta x^\mu = \text{const.}\,,
\end{equation}
which is what we wanted to show.
\end{proof}

Observe that this implies that the component of the deviation vector in the direction of the geodesic does not contain any interesting information and can be set to zero without loss of generality. Indeed, according to Eq.~\eqref{eq:condition deviation Proof} a vector $\delta x^\mu$ that lies parallel to the tangent of the geodesic can only be a deviation vector if $\delta x^\mu= C\, \dot x^\mu$ for $C$ a constant and therefore does not contain useful information on the geodesic deviation. Thus, without loss of information, one can choose a deviation vector that satisfies
\begin{equation}\label{eq:condition deviation 2 Proof}
    dx_\mu \delta x^\mu = 0\,.
\end{equation}

\section{Proof of Eq.~\eqref{eq:RelBlanchet}}\label{DerivationEq}
\textit{This proof is taken over from the original work of the author in \cite{Heisenberg:2023prj} and is based on private communications with Prof. Blanchet. An alternative version of a similar proof can also be found in \cite{BlanchetPaper}.}
\\

\noindent
In this appendix, we provide a derivation of the relation [Eq.~\eqref{eq:RelBlanchet}]
\begin{equation}\label{eq:RelBlanchetA}
   \left[\frac{n'_in'_j}{1-\vec{n}'\cdot\vec{n}}\right]^\text{TT}=\sum_{l=2}^\infty \frac{2(2l+1)!!}{(l+2)!}\left[n_{L-2}n'_{\langle ijL-2\rangle}\right]^\text{TT}\,,
\end{equation}
where a superscript TT implies a contraction of free indices with the TT projector with respect to $\vec{n}$, which we defined in Eq.~\eqref{eq:Projectors}.

\begin{proof} First, note that due to the TT projection, the left-hand side of Eq.~\eqref{eq:RelBlanchetA} vanishes for $\vec{n}'=\vec{n}$, while this is also true for the right-hand side. Thus, in the following, we consider the case $\vec{n}'\neq\vec{n}$, which implies $|\vec{n}'\cdot\vec{n}|<1$ such that we can expand the left-hand side in terms of a geometric series
    \begin{align}
        \frac{n'_in'_j}{1-\vec{n}'\cdot\vec{n}}&=n'_in'_j\sum_{l=2}^\infty(\vec{n}'\cdot\vec{n})^{l-2}=\sum_{l=2}^\infty n_{L-2}n'_{ijL-2}\,,
    \end{align}
where we use the notation introduced in Sec.~\ref{App:TTM Expansion}. The remaining task is to rewrite $n'_{ijL-2}$ in terms of its STF part $n'_{\langle ijL-2\rangle}$, by using the formula [see e.g. Eq.~(A21a) in \cite{Blanchet:1985sp}]
    \begin{equation}
        n_L=\sum_{k=0}^{[l/2]}a_k^l\delta_{\{2K}n'_{\langle L-2K\rangle\}}\,,
    \end{equation}
where 
    \begin{equation}
        a_k^l\equiv\frac{(2l-4k+1)!!}{(2l-2k+1)!!}\,,
    \end{equation}
and where $[l/2]$ selects the integer part of $l/2$. Moreover, the operator $\{...\}$ on tensor indices $A_{\{L\}}$ denotes the sum $\sum_{\sigma\in S} A_{i_{\sigma(1)}...i_{\sigma(l)}}$, where $S$ is the smallest set of permutations of $(1...l)$, which makes $A_{\{L\}}$ fully symmetric in $L$. For instance,  \begin{equation}\label{eq:ExampleExpansion}
        \delta_{\{ab}n'_{ij\}}=\,\delta_{ab}n'_{ij}+\delta_{ai}n'_{bj}+\delta_{aj}n'_{bi}+\delta_{bi}n'_{aj}+\delta_{bj}n'_{ai}+\delta_{ij}n'_{ab}\,.
    \end{equation}

With this in hand,
    \begin{align}
        \frac{n'_in'_j}{1-\vec{n}'\cdot\vec{n}}=\sum_{l=2}^\infty n_{L-2}\sum_{k=0}^{[\frac{l-2}{2}]}a_k^l\delta_{\{2K}n'_{\langle ijL-2-2K\rangle\}}\,.
    \end{align}
Notice, however, that due to the TT projection with respect to $\vec{n}$ on the free indices $ij$ and the contraction of the remaining indices with $n_{L-2}$, any term which involves one of the indices $i$ or $j$ within the Kronecker delta, hence terms containing $\delta_{ij}$, $\delta_{ia}$ or $\delta_{ja}$ for any index $a$, will vanish. For example, for the term $l=4$ and $k=1$, written out in Eq.~\eqref{eq:ExampleExpansion}, only the first term will survive. Thus,
    \begin{align}
        \left[\frac{n'_in'_j}{1-\vec{n}'\cdot\vec{n}}\right]^\text{TT}&=\sum_{l=2}^\infty n_{L-2}\sum_{k=0}^{[\frac{l-2}{2}]}\left[\delta_{2K}n'_{\langle ij L-2-2K\rangle}\right]^\text{TT} a_k^l\,b_k^l\,,\notag\\
        &=\sum_{l=2}^\infty\sum_{k=0}^{[\frac{l-2}{2}]}a_k^l\,b_k^l\left[n_{L-2-2K}n'_{\langle ij L-2-2K\rangle}\right]^\text{TT}\,,
    \end{align}
where 
\begin{equation}
    b_k^l\equiv\frac{(l-2)!}{2^kk!(l-2-2k)!}\,,
\end{equation}
is the number of terms within the sum $\delta_{\{2K}n'_{L-2-2K\}}$.

We can now rearrange the sum over positive integers $l$ and $k$ as
\begin{align}
    \sum_{l=2}^\infty\sum_{k=0}^{[\frac{l-2}{2}]}&=\sum_{l,k}
    [2\leq l\leq \infty] [0\leq k\leq \frac{l-2}{2}]=\sum_{l,k}
    [0\leq k\leq \frac{l-2}{2}\leq \infty]\notag\\
    &=\sum_{p,k}
    [0\leq k\leq \frac{p+2k-2}{2}\leq\infty]=\sum_{p,k}[2\leq p\leq \infty][0\leq k\leq \infty]\,,
\end{align}
where we defined $p\equiv l-2k$ and, in this context, the angular brackets $[B]$ of a Boolean expression denote Inverson brackets,\footnote{Inverson Brackets are defined as $[B]=\begin{cases}1\,&\text{if $B$ is true}\\0\,,&\text{otherwise}\end{cases}$.} such that
\begin{align}
       \left[\frac{n'_in'_j}{1-\vec{n}'\cdot\vec{n}}\right]^\text{TT}&=\sum_{p=2}^{\infty}\sum_{k=0}^\infty a_k^{p+2k}\,b_k^{p+2k}\left[n_{P-2}n'_{\langle ij P-2\rangle}\right]^\text{TT}\,,\notag\\
        &=\sum_{l=2}^{\infty}\underset{\equiv S_l}{\underbrace{\sum_{k=0}^\infty a_k^{l+2k}\,b_k^{l+2k}}}\left[n_{L-2}n'_{\langle ij L-2\rangle}\right]^\text{TT}\,,
\end{align}
where in the last step we simply relabeled $p\rightarrow l$. The sum over $k$ can indeed be resummed explicitly to give
\begin{align}
    S_l=\sum_{k=0}^\infty \frac{(2l+1)!!}{(2l+2k+1)!!}\, \frac{(l+2k-2)!}{2^kk!(l-2)!}=\frac{2^{2+l}\Gamma(l+\frac{3}{2})}{\sqrt{\pi}\Gamma(l+3)}=\frac{2(2l+1)!!}{(l+2)!}\,,
\end{align}
where in the last step we used the identities of gamma functions
\begin{equation}
    \Gamma(l+3)=(l+2)!\,,
\end{equation}
and
\begin{equation}
    \Gamma(l+\frac{3}{2})=\frac{\sqrt{\pi}(2l+1)!!}{2^{1+l}}\,.
\end{equation}
Comparing to Eq.~\eqref{eq:RelBlanchetA}, this concludes the proof.
\end{proof}

\section{Proof of Lemma~\ref{LemmaP}}\label{App:Proof of Lemma}
\textit{This proof is adapted from the original work of the author in \cite{Heisenberg:2023prj}. Parts of the steps are based on explicit computations in \cite{Stein:2010pn}.}
\\

\noindent
In this section, we present the proof of Lemma~\ref{LemmaP}. 
\begin{proof}
    The energy-momentum tensor appearing on the left-hand side of Eq.~\eqref{eq:EqLemma} of the Lemma was defined in Eq.~\eqref{eq:SecondOrderVariationEMTensor}, and therefore, the lemma is proven if we can show that in the limit to null infinity 
    \begin{equation}
        \Bigg\langle\frac{\delta \phantom{}_{\mys{(2)}}S^{\myst{flat}}_\text{eff}}{\delta \eta^{\mu\nu}}\Bigg\rangle = \Bigg\langle\frac{\delta \phantom{}_{\mys{(2)}}S_\text{eff}}{\delta  g^{L\mu\nu}}\Bigg\rangle\,,
    \end{equation}
    where recall here that the background metric must be treated as an independent field.  
    
    Since we only consider local and covariant theories with Levi-Civita connection, a variation of the second-order action $\phantom{}_{\mys{(2)}}S_\text{eff}$ with respect to the independent background metric $ g^L_{\mu\nu}$ will only arise from polynomial contributions of the background metric, the associated covariant derivatives $\nabla_\mu$ through its Christoffel symbols $\Gamma\ud{\mu}{\alpha\beta}$, and contractions of background curvature invariants $R_{\mu\nu\rho\sigma}[g^L]$, the last two of which vanish for a flat metric by definition. 
    To prove Lemma~\ref{LemmaP}, we therefore need to show that any term in the \textit{non-flat} perturbed action $\phantom{}_{\mys{(2)}}S_\text{eff}$ that involves background curvature or connection operators does not contribute to the effective stress-energy tensor $\phantom{}_{\mys(2)}t_{\mu\nu}$ at null infinity.
    
    For any term in the perturbed action that contains curvature or connection quantities, a variation of a curvature or connection coefficient with respect to the background metric $g_L^{\mu\nu}$ can be written as \cite{Stein:2010pn}
    \begin{equation}
        \delta_{ g^L} \left[\phantom{}_{\mys{(2)}}S_\text{eff}\right]\supset\int\dd^4 x\sqrt{- g^L} \; \nabla_\sigma P\ud{\sigma}{\mu\nu} \; \delta  g_L^{\mu\nu}
    \end{equation}
    for some tensor $P\ud{\sigma}{\mu\nu}$ upon integration by parts. Such a contribution vanishes
    \begin{equation}
        \phantom{}_{\mys{(2)}}t_{\mu\nu}\supset-2\Big\langle\nabla_\sigma P\ud{\sigma}{\mu\nu}\Big\rangle=0
    \end{equation}
    due to property (II) of the average (see Sec.~\ref{ssSec:IsaacsonInGR}).
    
    On the other hand, if the variation of such a term is hitting a background metric, the resulting term will still contain either a curvature or a connection operator. Therefore, any such contribution will vanish in the limit to null infinity.
\end{proof}

\section{Proof of Theorem~\ref{Theorem1}}\label{App:ProofOfMemoryTheorem}
\textit{This proof is adapted from the original work of the author in \cite{Heisenberg:2023prj}.}
\\

\noindent
In this section, we present the proof of Theorem~\ref{Theorem1}. 

\begin{proof}
     Let  $S_\text{eff}=\phantom{}_{\mys(0)}S_\text{eff}+\phantom{}_{\mys(2)}S_\text{eff}$ be the effective action of a dynamical metric theory (see Definition \ref{DefMetricTheory}), where here we have ignored the linear term because it will not contribute to the equations of motion of the perturbations at leading order [Eq.~\eqref{eq:EOMIISTh}], or to the leading-order memory-evolution equation [Eq.~\eqref{eq:EOMISTh}].
   In the following, every equality or proportionality is to be understood as a statement in the limit to null infinity. Moreover, proportionalities are relations up to scalar functions, which may depend on the Minkowski background $\bar\eta=\{\eta_{\mu\nu},\bar\Psi\}$ at null infinity, as well as the dimensionful bare Newton's constant $\kappa_0$. 

    With this in mind, let us separately consider the left-hand side and right-hand side of the leading-order, low-frequency equation [Eq.~\eqref{eq:EOMISTh}], starting with the former. The left-hand side of Eq.~\eqref{eq:EOMISTh} arises to leading order and in the limit to null infinity from $\delta\phantom{}_{\mys(0)}S_\text{eff}/\delta  g_L^{\mu\nu}$ and a subsequent split in Eq.~\eqref{eq:IsaacsonSplitGen2 Thm} into the exact Minkowski solution $\bar{\eta}=\{\eta_{\mu\nu},\bar \Psi\}$ that satisfies the vacuum equations and low-frequency perturbations $\delta g^L_{\mu\nu}$ and $\delta\Psi^L$, such that the resulting operator splits into a homogeneous piece $_{\mys{(0)}}\mathcal{G}_{\mu\nu}[\bar{\eta}]=0$ and an inhomogeneous part $_{\mys{(1)}}\mathcal{G}_{\mu\nu}[\delta g^L,\delta\Psi^L]$ [see Eq.~\eqref{eq:EOMIS2}].
    
    Note that this operator $_{\mys{(1)}}\mathcal{G}_{\mu\nu}[\delta g^L,\delta\Psi^L]$ has the same functional form as the operator with the low-frequency perturbations replaced by the high-frequency ones, which is the operator of the propagation equation [Eq.~\eqref{eq:EOMIISTh}] in the limit to null infinity. Therefore, assumption (iii) also dictates the functional form of the inhomogeneous piece $_{\mys{(1)}}\mathcal{G}_{\mu\nu}[\delta g^L,\delta\Psi^L]$. 
    More precisely, by assumption (iii), the operator takes the form of a massless wave operator $_{\mys{(1)}}\mathcal{G}_{\mu\nu}[\delta g^H,\delta\Psi^H]\propto\Box \hat{h}^H_{\mu\nu}$ upon a field redefinition $\hat{h}_{\mu\nu}^H=W(\delta g^H,\delta \Psi^H)$ and $\hat{\psi}^H=V(\Psi^H)$ and after imposing the conditions in Eq.~\eqref{eq:ThLorenzGauge}. Thus, there exists a redefinition $\delta \hat h^L_{\mu\nu}=\hat{W}(\delta g^L_{\mu\nu},\delta \Psi^L)$, together with a suitable choice of $\xi_L^{\mu}$ in a coordinate transformation $x^\mu\rightarrow x^\mu+\xi_L^{\mu}$, such that the remaining leading-order operator reduces to a decoupled wave operator $\Box\delta \hat h^L_{\mu\nu}$.
    The corresponding gauge choice is the Lorenz gauge on the rescaled, low-frequency metric perturbation, $\partial^\mu\delta \hat h^L_{\mu\nu}=0$. The existence of this gauge follows from the conservation of the energy-momentum tensor $\partial^\mu \phantom{}_{\mys{(2)}}t_{\mu\nu}=0$ (see below), which in turn can be established from property (II) of the average (see Sec.~\ref{ssSec:IsaacsonInGR}). If the total energy-momentum tensor is moreover traceless, it is possible to impose tracelessness on $\delta \hat h^L_{\mu\nu}$, and, in this case, Eq.~\eqref{eq:ProofLHS} follows directly by choosing $\delta \hat h^L_{\mu\nu}=W(\delta h^L_{\mu\nu},\delta \psi^L)$. If the total energy momentum tensor is not trace-less, we can perform a further redefinition to a trace-reversed variable $\delta \hat{\bar{h}}^L_{\mu\nu}\equiv \delta \hat h^L_{\mu\nu}-\frac{1}{2}\eta_{\mu\nu}\delta \hat h^{Lt}$, which yields Eq.~\eqref{eq:ProofLHS} in terms of this new variable.
    Thus, in summary, the left-hand side of Eq.~\eqref{eq:EOMISTh} is given by the inhomogeneous contribution
    \begin{equation}\label{eq:ProofLHS}
    \phantom{}_{\mys{(1)}} \mathcal{G}_{\mu\nu}[\eta_0,\{\delta h^L,\delta\Psi^L\}] \propto \Box\delta \hat h^L_{\mu\nu}\,.
    \end{equation}
     
    The right-hand side of Eq.~\eqref{eq:EOMISTh} only contributes to the inhomogeneous equation and is, by definition, the energy-momentum (pseudo)tensor, which in the limit to null infinity reduces to $\phantom{}_{\mys{(2)}} t_{\mu \nu} \to t_{\mu\nu}[\eta_0,\{\hat h^H,\hat\Psi^H\}]$. In this limit, we can then write
     \begin{equation}\label{eq:ProofRHS}
        \big\langle \phantom{}_{\mys{(2)}}\mathcal{G}_{\mu\nu}\big\rangle=2\kappa_0\,A(\bar\eta)\,\phantom{}_{\mys{(2)}}t_{\mu\nu}[\eta_0,\{\hat h^H,\hat\Psi^H\}]\,,
    \end{equation}
    where the proportionality can only depend on a scalar function of the Minkowski background $\bar\eta$, while dimensional analysis requires the presence of a factor of $\kappa_0$.
    Combining Eqs.~\eqref{eq:ProofLHS} and \eqref{eq:ProofRHS} establishes that the memory-evolution equation can be written as
    \begin{equation}\label{eq:MemoryEq Proof}
        \Box\delta \hat h^L_{\mu\nu}=-2\kappa_0 \bar{A}\phantom{}_{\mys{(2)}}t_{\mu\nu}[\hat h^H,\hat\psi^H]\,.
    \end{equation}

Will now establish the properties (a)--(c) of the flat energy-momentum (pseudo)tensor. Property (a), the conservation of the energy-momentum (pseudo)tensor $\partial^\mu\phantom{}_{\mys{(2)}}t_{\mu\nu}=0$,  was already touched upon when describing the left-hand side of Eq.~\eqref{eq:EOMISTh}, but let us be more explicit here. This (pseudo)tensor is conserved because it is defined in terms of the average of the variation of the second-order effective action with respect to the background metric [Eq.~\eqref{eq:SecondOrderVariationEMTensor}]. When we take the divergence of this quantity, the derivative commutes with the average symbol and then hits the variation itself, which is a rank-2 (pseudo)tensor. Property (II) of the average procedure (see Sec.~\ref{ssSec:IsaacsonInGR}), however, guarantees that the average of the divergence of a tensor vanishes up to boundary terms, which are higher order in the ratio of the frequency scales. Therefore, the divergence of the (pseudo)tensor also vanishes up to these boundary terms, which establishes property (a). 

Let us now focus on property (b), which tells us that the energy-momentum (pseudo)tensor near null infinity can be decomposed into two terms: a standard GR term and a term that depends on the perturbations of the additional gravitational fields. In order to establish this property, we must first massage the second-order effective action near null infinity. In the limit to null infinity,
    \begin{align}
       0=\phantom{}_{\mys{(1)}}\mathcal{G}_{\mu\nu}[\delta g^H,\delta \Psi^H]&\propto\left[\frac{\delta \phantom{}_{\mys(2)}S^{\mys{flat}}_\text{eff}}{\delta \delta g^{H\mu\nu}}\right]^H\,,\\
        0=\phantom{}_{\mys{(1)}}\mathcal{J}[\delta g^H,\delta\Psi^H]&\propto\left[\frac{\delta \phantom{}_{\mys(2)}S^{\mys{flat}}_\text{eff}}{\delta \delta \Psi^{H}}\right]^H\,,
    \end{align}
    and hence, the leading-order propagation equations [Es.~\eqref{eq:EOMIISTh}] only depend on the $\phantom{}_{\mys(2)}S^{\mys{flat}}_\text{eff}$ of Eq.~\eqref{eq:flatSA}. 
    Assumption (iii) of the theorem then implies that there exists a set of redefined perturbation variables $\{\hat{h}_{\mu\nu}^H,\hat{\psi}^H\}$ for which the effective action $\phantom{}_{\mys(2)}S^{\mys{flat}}_\text{eff}$ can be written as
     \begin{align}\label{ProofActionGenn2nd}
    \phantom{}_{\mys(2)}S^{\mys{flat}}_\text{eff}\propto&\int\dd^4x\sqrt{-\eta}\,\bigg[\hat{h}^{H\mu\nu}\mathcal{E}^{\alpha\beta}_{\mu\nu}\hat{h}^H_{\alpha\beta}+\mathcal{J}[\hat\psi^H]\bigg]\,.
    \end{align}
    The first term is the usual Fierz-Pauli operator with indices contracted with the independent background metric $\eta_{\mu\nu}$. The second term corresponds schematically to the sum of terms of the high-frequency perturbations of the additional gravitational fields that lead to the corresponding equations of motion. Generally, for ghost-free theories, these operators are also expected to involve only up to two derivative operators. 
    Up to integrations by parts, this is the unique action that will lead to a set of decoupled massless wave equations for the leading-order, high-frequency tensor perturbations upon imposing the adequate gauge choices of Eq.~\eqref{eq:ThLorenzGauge}. 
    
    With this in hand, we can now establish property (b).  Lemma \ref{LemmaP} and the perturbed generalized field equations imply that, in the limit to null infinity,
     \begin{equation}\label{eq:RHSP}
    \Big\langle \phantom{}_{\mys{(2)}}\mathcal{G}_{\mu\nu}[\hat h^H,\hat\psi^H]\Big\rangle \propto \Bigg\langle\frac{\delta \phantom{}_{\mys{(2)}}S^{\mys{flat}}_\text{eff}}{\delta \eta^{\mu\nu}}\Bigg\rangle \propto \phantom{}_{\mys{(2)}}t_{\mu \nu}\,.
     \end{equation}
    In other words, in this limit, the effective stress-energy (pseudo)tensor of radiation, and therefore, also the right-hand side of the memory equation [Eq.~\eqref{eq:EOMISTh}], only requires knowledge of the flat, second-order effective action $\phantom{}_{\mys(2)}S^{\mys{flat}}_\text{eff}$ as well. Since $\phantom{}_{\mys(2)}S^{\mys{flat}}_\text{eff}$ can be written in the form of Eq.~\eqref{ProofActionGenn2nd}, it can be split into a term that only depends on the metric perturbation $h^H_{\mu\nu}$ and another set of terms, each of which only depends on the field perturbation $\psi^H$ of every additional field $\Psi$. At this point, we want to mention that if the second order action for the extra non-minimal fields only involve up to two derivative operators, then so will the corresponding energy-momentum tensor. Each of these terms can be treated separately when computing the radiative energy-momentum (pseudo)tensor, and hence
     \begin{equation}\label{eq:ProofRHS2}
         \phantom{}_{\mys{(2)}}t_{\mu\nu}=\phantom{}_{\mys{(2)}}t^{\myst{GR}}_{\mu\nu}[\hat h^H]+\phantom{}_{\mys{(2)}}t^{\hat{\psi}}_{\mu\nu}[\hat \psi^H] \,.
    \end{equation}
   Moreover, since the metric perturbation piece in the flat, effective second-order action in Eq.~\eqref{ProofActionGenn2nd} admits the same form as in GR, the first term is proportional to the radiative energy-momentum tensor of GR. After imposing the gauge of Eq.~\eqref{eq:ThLorenzGauge}, the latter reads
    \begin{equation}
        \phantom{}_{\mys{(2)}}t^{\myst{GR}}_{\mu\nu}\propto\Big\langle \partial_\mu \hat h^H_{\alpha\beta}\partial_\nu \hat h^{H\alpha\beta}\Big\rangle\,,
    \end{equation}
thus establishing property (b). 

    Finally, let us consider property (c), which states that the energy-momentum (pseudo)tensor is invariant under infinitesimal coordinate transformations. The assumptions on asymptotic flatness of the peacetime, together with the split in Eq.~\eqref{eq:ProofRHS2} and the assumption that $\hat{\psi}^H$ does not depend on $h^H_{\mu\nu}$,  directly implies that $\phantom{}_{\mys{(2)}}t_{\mu\nu}$ is invariant under infinitesimal diffeomorphisms  
    \begin{equation}
        x^\mu\rightarrow x^\mu+\xi_H^{\mu}\,.
    \end{equation} 
    This is because the assumption of a local-Lorentz preserving background with $\bar\Psi=0$  for fields that are not a scalar and $\partial_\alpha\bar\Psi=0$ for scalars implies that the Lie derivative of the first-order perturbations with respect to a vector field generating such a diffeomorphism will vanish. Therefore, only the metric perturbations $h^H_{\mu\nu}$ transform under the gauge shift. Moreover, since the $\hat\psi^H$ perturbations are independent of $h^H_{\mu\nu}$, the coordinate gauge transformation only affects the term $\phantom{}_{\mys{(2)}}t^{\myst{GR}}_{\mu\nu}$ , which is invariant upon integration by parts and after imposing the Lorenz gauge Eq.~\eqref{eq:ThLorenzGauge}.

    We are now ready to perform the final step of the proof. Namely, since properties (a)--(c) of the energy-momentum tensor $\phantom{}_{\mys{(2)}}t_{\mu\nu}$ establish that it corresponds to a well-behaved asymptotic energy momentum tensor of massless fields, it will necessarily be of the form [Eq.~\eqref{eq:General Form asymptotic EMT Thm}]
   \begin{equation}\label{eq:General Form asymptotic EMT Proof}
    \phantom{}_{\mys{(2)}}t_{\mu\nu}(u,r,\Omega)= \phantom{}_{\mys{(2)}}t_{00}(u,r,\Omega)\,l_\mu l_\nu=\frac{1}{r^2}\,F(u,\Omega)\,l_\mu l_\nu
    \end{equation}
    for some function $F(u,\Omega)$ and where
    \begin{equation}
    l_\mu=-\nabla_\mu t+\,\nabla_\mu r\,,\quad  u= t-r\,,
    \end{equation}
    as demonstrated in Sec.~\ref{sSec:UnifiedTreatmentof Null and Ordinary Memory}. Thus, since in the derivation of Eq.~\eqref{eq:DispMemoryGR} the energy momentum tensor, up to the special form in Eq.~\eqref{eq:General Form asymptotic EMT Proof} only played the role of a spectator, one can carry out the exact same steps as in Sec.~\ref{sSec:UnifiedTreatmentof Null and Ordinary Memory} to solve Eq.~\eqref{eq:MemoryEq Proof} and arrive at 
    \begin{align}\label{NonLinDispMemory Proof}
    \delta \hat h_{L,ij}^\text{TT}(u,r,\Omega)=\,\frac{\kappa_0 \bar{A}}{2\pi r} \int_{-\infty}^u du'\int_{S^2} d^2\Omega'\,\left(r^2 \phantom{}_{\mys{(2)}}t_{00}(u',r,\Omega')\right)\,\left[\frac{n'_in'_j}{1-\mathbf{n}'\cdot\mathbf{n}}\right]^\text{TT}\,.
\end{align}

    Having established  Eq.~\eqref{NonLinDispMemory Thm} together with all the required properties (a)--(c) of the radiative energy-momentum tensor $\phantom{}_{\mys{(2)}}t_{\mu\nu}$, we have therefore proven the theorem.
\end{proof}

\section{Derivation of Eq.~\eqref{NonLinDispMemoryGen2}}\label{app:ExampleNullMemoryKForm}
\textit{This derivation is taken over from the original work of the author in \cite{Heisenberg:2023prj}.}
\\

We present in this appendix a derivation of the tensor null memory for the particular, but fairly general, class of gauge-Invariant $k$-form metric theories given in Eq.~\eqref{NonLinDispMemoryGen2}. Consider therefore an arbitrary number of additional dynamical $k$-form connection fields that we schematically denote as $\Psi$, with an associated, Abelian, gauge symmetry for $k>1$ (for $k=0$, the field simply corresponds to a scalar field). The action in Eq.~\eqref{eq:ActionMetricTheory} is thus constructed out of curvature invariants of the metric, as well as field strengths $\mathscr{F}\equiv d\Psi$ that are invariant under Abelian gauge transformations
\begin{equation}
    \Psi\rightarrow \Psi + d\Lambda\,,
\end{equation}
where $\Lambda$ are arbitrary $(k-1)$-forms and $d$ is the exterior derivative. These dynamical $k$-form fields are assumed to describe $N$ additional propagating gravitational degrees of freedom.
Note that in a local chart-induced basis, a $k$-form field reads
\begin{equation}
    \Psi=\frac{1}{k!}\Psi_{\mu_1\mu_2...\mu_k}\,dx^{\mu_1}\wedge dx^{\mu_2}\wedge...\wedge dx^{\mu_k}\equiv\frac{1}{k!}\Psi_{K}\,\dd X^{K}\,.
\end{equation}

We now consider theories for which the assumptions of Theorem~\ref{Theorem1} hold. In particular, the leading-order propagation equations of the theory reduce to a set of decoupled wave equations for the leading-order waves $\Psi^H_K$ upon imposing the analog of the Lorentz gauge 
\begin{equation}\label{eq:ThLorenzGaugeKForm}
\partial^{\mu_i}\hat{\Psi}_K^H=0\,,
\end{equation}
on each of the field perturbations. The assumptions of the theorem are rather natural assumptions because, the theory is covariant, locally Lorentz invariant and massless. Moreover, any term in the action that would lead to higher-order derivatives in the equations of motion of the leading-order waves does not contribute to leading order in the small-coupling approximation. An additional assumption here is that, in the limit to null infinity, there is no coupling between different fields at the level of the leading-order, high-frequency equation.

Following the proof of Theorem~\ref{Theorem1}, the flat, effective, second-order action of Eq.~\eqref{ProofActionGenn2nd} can thus be written as
\begin{align}\label{ProofActionGenn2ndKForm}
    \phantom{}_{\mys(2)}S^{\mys{flat}}_\text{eff}=&\frac{-1}{2\kappa_\text{eff}}\int\dd^4x\,\sqrt{-\eta}\bigg[\hat{h}^{\mu\nu}_H\mathcal{E}^{\alpha\beta}_{\mu\nu}\hat{h}^H_{\alpha\beta}+\sum_{\Psi}\frac{1}{2q}\eta^{\mu_1\nu_1}...\eta^{\mu_{q}\nu_{q}}\hat{\mathscr{F}}^H_{\mu_1...\mu_{q}}\hat{\mathscr{F}}^H_{\nu_1...\nu_{q}}\bigg]\,,
\end{align}
where $q\equiv k+1$, $\kappa_\text{eff}=\kappa_0A(\bar\eta)$ and $\hat{\mathscr{F}}^H=d\hat{\Psi}^H$ are the field strength of the canonically normalized $k$-form perturbations $\hat{\Psi}^H$. The first term in Eq.~\eqref{ProofActionGenn2ndKForm} is again the usual Fierz-Pauli operator with indices contracted with the independent background metric $\eta_{\mu\nu}$, while the second term corresponds to the sum of kinetic terms of all the additional gravitational, gauge-invariant, $k$-form field perturbations.

With this explicit form of the flat effective action at hand, we can now explicitly compute the resulting effective stress-energy tensor for the additional leading-order waves (see also \cite{Navarro:2012hv} for a study of the properties of the energy-momentum tensor of $k$-forms)
    \begin{align}
    \begin{split}
        \phantom{}_{\mys(2)}t_{\mu\nu}^{\hat{\Psi}}=\frac{1}{2\kappa_\text{eff}}\sum_{\Psi}\bigg\langle&\hat{\mathscr{F}}^H_{\mu K}\hat{\mathscr{F}}_{\nu}^{HK}-\frac{1}{2q}\eta_{\mu\nu}\hat{\mathscr{F}}^H_{Q}\hat{\mathscr{F}}^{HQ}\bigg\rangle\,.
     \end{split}
    \end{align}
Imposing the equations of motion [Eq.~\eqref{eq:ThWaveEq}], as well as the Lorenz gauge [Eq.~\eqref{eq:ThLorenzGaugeKForm}], while recalling that the averaging allows for integrations by parts, the only surviving terms are
    \begin{align}\label{eq:IndexStP}
        \phantom{}_{\mys(2)}t_{\mu\nu}^{\hat{\Psi}}=\frac{1}{2\kappa_\text{eff}}\sum_{\Psi}\bigg\langle\partial_\mu\hat{\Psi}^H_K\partial_\nu\hat{\Psi}^{KK}\bigg\rangle\,.
    \end{align}
Including the Fierz-Pauli result, the total energy-momentum tensor therefore reads
\begin{align}\label{eq:EMTotKForm}
        \phantom{}_{\mys(2)}t_{\mu\nu}=\frac{1}{2\kappa_\text{eff}}\bigg\langle&\frac{1}{2}\partial_\mu\hat{h}^H_{\alpha\beta}\partial_\nu\hat{h}^{H\alpha\beta}+\sum_{\Psi}\partial_\mu\hat{\Psi}^H_K\partial_\nu\hat{\Psi}^{HK}\bigg\rangle\,.
    \end{align}
This (pseudo)energy-momentum tensor is indeed conserved (and traceless), as well as gauge invariant (both under linearized diffeomorphisms and gauge transformations of the $k$-form fields, upon imposing a Lorenz gauge).

The above also means that the (pseudo)energy-momentum tensor can be written in terms of the radiative modes of each field, which are the solutions to the wave equations that we can expand in terms of a polarization basis. For the metric perturbation, we have the usual TT tensor modes of the physical metric
    \begin{equation}
        \hat{h}^{H\text{TT}}_{\mu\nu}=h^{H\text{TT}}_{\mu\nu}=\epsilon^+_{\mu\nu} \,\hat h_++\epsilon^\times_{\mu\nu} \,\hat h_\times\,,
    \end{equation}
where
    \begin{equation}
        \epsilon^+_{\mu\nu} \epsilon^{+\mu\nu}=\epsilon^\times_{\mu\nu} \epsilon^{\times\mu\nu}=2\,,\quad\epsilon^+_{\mu\nu} \epsilon^{\times\mu\nu}=0\,.
    \end{equation}
For the additional gravitational fields, $\Psi$, by assumption they describe $N$ additional radiative modes $\psi_\lambda$, where $\lambda=1,..,N$. These can be written in some polarization basis as\footnote{A practical way of completely fixing the gauge is the use of spacetime light-cone coordinates, as exemplified for instance in \cite{Heisenberg:2019akx} for 2-forms. These polarizations should, however, not be confused with the gravitational polarizations of the physical metric defined in Sec.~\ref{sSec:GWPolGen}.} 
    \begin{equation}
        \hat{\Psi}^p_K=\sum_{\text{P}}\epsilon^\text{P}_K \,\hat\psi_\text{P}\,,\quad\text{where}\quad
        \epsilon^\text{P}_K \epsilon^{\text{P}'K}=\delta^{\text{PP}'}\,.
    \end{equation}
    
These solutions to the wave equation in the limit to null infinity take the form (cf. Eq.~\eqref{eq:OutgoingPlaneWave})
    \begin{align}\label{eq:OutgoingPlaneWaveKForm}
    &\left(\hat{h}_+\,,\;\hat{h}_\times\,,\;\hat{\psi}_\lambda\right) \sim \frac{1}{r} \left[f_+(u,\theta,\phi),f_\times(u,\theta,\phi), f_\lambda(u,\theta,\phi) \right]\,.
    \end{align}
    for some real functions $f_{+,\times,\lambda}$. Together with the form of the energy-momentum tensor Eq.~\eqref{eq:EMTotKForm} this behavior near null infinity implies that the radiative energy-momentum tensor can indeed be written as 
    \begin{equation}\label{eq:StressEnergyGen}
        \phantom{}_{\mys(2)}t_{\mu\nu}=\frac{1}{2\kappa_\text{eff}}\bigg\langle | \dot{\hat{h}}_+|^2+| \dot{\hat{h}}_\times|^2+\sum_{\lambda=1}^N|\dot{\hat{\psi}}_\lambda|^2\bigg\rangle \,l_\mu l_\nu\,,
    \end{equation}
    where $l_\mu=-\nabla_\mu t+\nabla_\mu r$ and the falloffs of Eq.~\eqref{eq:OutgoingPlaneWaveKForm} impose the scaling $t_{\mu\nu}\sim r^{-2}$.

    Following Sec.~\ref{sSec:UnifiedTreatmentof Null and Ordinary Memory}, the tensor null-memory formula therefore takes the form
\begin{equation}\label{NonLinDispMemoryGen2App}
     \delta h_H^{l m}(u,r)=\frac{1}{r} \sqrt{\frac{(l-2)!}{(l+2)!}}\int_{S^2} \dd^2 \Omega'\,\bar{Y}^{l m}(\Omega')\,\int_{-\infty}^{u}\dd u'\,r^2\bigg\langle | \dot{\hat{h}}_+|^2+| \dot{\hat{h}}_\times|^2+\sum_{\lambda=1}^N|\dot{\hat{\psi}}_\lambda|^2\bigg\rangle\,.
\end{equation}
Note that the $\hat{\psi}$ here are the canonically normalized modes. The coupling constants of a specific theory will then enter the memory formula by transforming back to the physical modes.

The example presented here could be readily generalized by the inclusion of non-Abelian, 1-form gauge fields with a simple and compact but otherwise arbitrary gauge group and field strength $d\mathrm{F}=d\Psi+\Psi\wedge \Psi$. Such theories are a natural generalization of the SVT theory considered in this work, but would not change the result significantly, as the background solution requires a vanishing vector field in this case. Moreover, just as for massless Abelian fields, no higher-order self-interaction terms exist in $d=4$, which would still lead to second-order equations of motion \cite{Deffayet:2010zh}. Similarly, in the same way as Abelian 1-forms can be generalized to non-Abelian gauge groups, $k$-forms have non-Abelian generalizations, which typically require the use of gerbes (see e.g. \cite{Strobl:2016aph,Breen:2001ie}). We conjecture, that also in such a general framework an adapted version of the memory equation [Eq.~\eqref{NonLinDispMemoryGen2}] should hold.

\chapter{Analytic Formulas}\label{App:Formulas}


\section{Transverse Traceless Multipole Expansion}\label{App:TTM Expansion}

For convenience, we gather in this appendix a collection of definitions and formulas from the unifying review by Thorne \cite{Thorne:1980ru} for different multipole expansions used in this work. In particular, we will touch upon spin-weighted spherical harmonics, pure-spin transverse-traceless tensor harmonics, as well as STF tensors, all tied to the $SO(3)$ rotation group and the irreducible representations thereof. The notation for STF tensors will, however, be adapted to a more contemporary custom (see for instance \cite{poisson2014gravity,Blanchet:2013haa,maggiore2008gravitational}). For more details and derivations, we refer the reader to \cite{Thorne:1980ru}, while part of the treatment can also be found in \cite{Kidder:2007rt,maggiore2008gravitational,Creighton:2011zz} and in the Appendix of \cite{Nichols:2017rqr}.


\subsubsection{\ul{Spin-Weighted Spherical Harmonics}}


Spin-weighted spherical harmonics can be constructed from ordinary spherical harmonics through applications of angular derivative operators
\begin{equation}
 \phantom{}_{\mys{s}}Y_{lm}(\theta,\phi)=
\begin{cases}
			\sqrt{\frac{(l-s)!}{(l+s)!}}\eth^s Y_{lm}(\theta,\phi)\,, & l\geq s\geq 0\\
            (-1)^s \sqrt{\frac{(l+s)!}{(l-s)!}}\bar{\eth}^{-s} Y_{lm}(\theta,\phi)\,, & 0> s\geq -l\\
            0\,, & |s|>l\\
\end{cases}
\end{equation}
 where the angular derivative operator $\eth$ and its complex conjugate $\bar{\eth}$ are defined through their action on functions $f_s$ of spin-weight $s$
 \begin{align}\label{eth}
	\eth f_s(\theta,\phi) &\equiv -\sin^s\theta\left(\partial_\theta+i\,\csc\theta\right)(f_s \sin^{-s}\theta)\,,\\
	\bar\eth f_s(\theta,\phi) &\equiv -\sin^{-s}\theta(\partial_\theta-i\,\csc\theta)\left(f_s \sin^{s}\theta\right)\,.
\end{align}
Here, the spin-wight $s$ of a function $f_s$ on the sphere, parameterized through a polar and an azimuthal angle $(\theta,\phi)$ is defined through its transformation under the $U(1)$ gauge freedom of angle $\psi$ on the sphere (see e.g. \cite{DAmbrosio:2022clk})
\begin{equation}\label{eq:SpinWeightTransfos App}
    f_s(\theta,\phi)\rightarrow f'_s(\theta,\phi)=f_s(\theta,\phi)\,e^{is\psi}.
\end{equation}
The application of the operator $\eth$ on a function $f_s$ defines a quantity with spin-weight $s+1$, while $\bar{\eth}$ lowers the spin-weight by one unit.
Explicitly, the spin-weighted spherical harmonics can be written as
\begin{align}
    \phantom{}_{\mys{s}}Y_{lm}(\theta,\phi)=&(-1)^m\,e^{i\,m\phi}\,\sqrt{\frac{(l+m)!(l-m)!(2l+1)}{4\pi(l+s)!(l-s)!}}\,\sin^{2l}\frac{\theta}{2}\notag\\
    &\times\sum_{q=0}^{l-s} \binom{l-s}{q} \binom{l+s}{q+s-m}\,(-1)^{l-s-q}\,\cot^{2q+s-m}\frac{\theta}{2}\,.
\end{align}

Useful identities include:\\
\noindent
\textbf{Orthogonality and Completeness.}
The orthogonality and completeness relations are inherited from the ordinary spherical harmonics
\begin{align}
    \int_{S^2}\dd\Omega \,\phantom{}_{\mys s}Y_{lm}(\theta,\phi) \phantom{}_{\mys s}Y^*_{l'm'}(\theta,\phi)&=\delta_{ll'}\delta_{mm'}\,,\label{OrthogonalitySWSH}\\
     \sum_{l=0}^\infty \sum_{m=-l}^l \,\phantom{}_{\mys s}Y_{lm}(\theta,\phi) \phantom{}_{\mys s}Y^*_{lm}(\theta',\phi')&=\delta(\phi'-\phi)\delta(\cos\theta'-\cos\theta)\,.
\end{align}

\noindent
\textbf{Raising and Lowering of Spinweight.}
\begin{align}
    \sqrt{(l-s)(l+s+1)}\,_{\mys{s+1}}Y_{lm}&=\eth\,_{\mys s}Y_{lm}\,,\label{eq:ASWSHid1}\\
     -\sqrt{(l+s)(l-s+1)}\,_{\mys{s-1}}Y_{lm}&=\bar{\eth}\,_{\mys s}Y_{lm}\,.
\end{align}

\noindent
\textbf{Complex Conjugation and Angle Shifts.}
Complex conjugation is given by
\begin{equation}
    (-1)^{s+m}\, \phantom{}_{\mys{-s}}Y^*_{l-m}(\theta,\phi)=\phantom{}_{\mys s}Y_{lm}(\theta,\phi)\,,\label{CCSWSH}
\end{equation}
while shifts of $\pi$ in the angular arguments give rise to the following identities
\begin{align}
    \phantom{}_{\mys s}Y_{lm}(\pi-\theta,\phi)&=(-1)^{l+m}\, \phantom{}_{\mys{-s}}Y_{lm}(\theta,\phi) \label{thshift}\\
    \phantom{}_{\mys s}Y_{lm}(\theta,\phi+\pi)&=(-1)^{m}\, \phantom{}_{\mys s}Y_{lm}(\theta,\phi)\label{phshift}\,.
\end{align}
Combining Eq.~\eqref{CCSWSH} with Eq.~\eqref{thshift}, we have
\begin{equation}\label{CCSWSHandthshift}
    \phantom{}_{\mys s}Y_{lm}(\pi-\theta,\phi)=(-1)^{s+l}\, \phantom{}_{\mys s}Y^*_{l-m}(\theta,\phi)\,.
\end{equation}

\noindent
\textbf{Triple Integral.}
The integral of three spin weighted-spherical harmonics satisfying $s_1+s_2+s_3=0$ is given by
\begin{equation}\label{SWSHTrippleInt}
    \int_{S^2}\dd^2\Omega\,_{\mys{s_1}}Y^{l_1m_1}\,_{\mys{s_2} }Y^{l_2m_2}\,_{\mys{s_3}}Y^{l_3m_3}
    = \sqrt{\frac{\prod_{i=1}^3(2l_i+1)}{4\pi}} \footnotesize{
\begin{pmatrix}
l_1 & l_2 & l_3\\
m_1 & m_2 & m_3
\end{pmatrix}
\begin{pmatrix}
l_1 & l_2 & l_3\\
-s_1 & -s_2 & -s_3
\end{pmatrix}}\,,
\end{equation}
where the round brackets define Wigner $3-j$ symbols.
In particular, the integral therefore only gives non-zero values for $m_1+m_2+m_3=0$ and $|l_1-l_2|\leq l_3\leq l_1+l_2$. If the condition $s_1+s_2+s_3=0$ is not satisfied, the relation is not valid and one should resort to formulas in terms of gamma functions, as for instance presented in the appendix of \cite{Favata:2008yd}. 
\\

\noindent
\textbf{Relation to Wigner-D Matrices.}
The SWSHs are related to Wigner-D rotation matrices as follows\footnote{We assume the conventions for Wigner-D matrices $\mathfrak{D}^l_{m'm}(\psi,\theta,\phi)$ used in \textsc{Mathematica}.}
\begin{equation}
     \phantom{}_{\mys s}Y_{lm}(\theta,\phi)=(-1)^s\sqrt{\frac{2l+1}{4\pi}}\,\mathfrak{D}^l_{-sm}(0,\theta,\phi)
\end{equation}
Using the properties of the Wigner-D matrices as representations of the rotation group, we can write
\begin{equation}
    \phantom{}_{\mys s}Y_{lm}(R_1R_2)=\sum_{m'} \phantom{}_{\mys s}Y_{lm'}(R_2)\mathfrak{D}^l_{m'm}(R_1)\,,
\end{equation}
and thus, the SWSHs transform under rotations as
\begin{equation}\label{rotSWSH}
    \phantom{}_{\mys s}Y_{lm}(\theta',\phi')=\sum_{m'}\phantom{}_{\mys s}Y_{lm'}(\theta,\phi)\mathfrak{D}^l_{m'm}(R)\,e^{is\psi}\,,
\end{equation}
where the extra factor of $e^{is\psi}$ capture the spin-weighted transformations defined in Eq.~\eqref{eq:SpinWeightTransfos App}.



\subsubsection{\ul{Pure Spin Transverse-Traceless Tensor Harmonics}}

The electric- and magnetic-parity pure-spin TT harmonics can be constructed from ordinary spherical harmonics through
\begin{align}
    T^{\scriptscriptstyle{E2}\,\scriptstyle{l m
    }}_{ij}&\equiv r^2\sqrt{\frac{2(l-2)!}{(l+2)!}}\,\perp_{ijab}\partial_a\partial_b\,Y_{lm}\,,\\
    T^{\scriptscriptstyle{B2}\,\scriptstyle{l m
    }}_{ij}&\equiv r\sqrt{\frac{2(l-2)!}{(l+2)!}}\,\perp_{ijab}\epsilon_{cd(a}\partial_{b)}r\,n_d\partial_c\,Y_{lm}\,.
\end{align}
where $n_i$ is a unit radial vector and the TT projector $\perp_{ijab}$ is defined in Eq.~\eqref{eq:Projectors}.
They satisfy
\begin{align}
    (-1)^m\bar{T}^{\scriptscriptstyle{P}\,\scriptstyle{l -m
    }}_{ij}(\theta,\phi)&=T^{\scriptscriptstyle{P}\,\scriptstyle{l m
    }}_{ij}(\theta,\phi)\,,\\
    \int_{S^2}\dd\Omega \,T^{\scriptscriptstyle{P}\,\scriptstyle{l m
    }}_{ij}(\theta,\phi)\,T^{\scriptscriptstyle{P'}\,\scriptstyle{l' m'
    }}_{ij}(\theta,\phi)&=\delta_{PP'}\delta_{ll'}\delta_{mm'}\,.
\end{align}



\subsubsection{\ul{STF Tensors}}

Denoting a multi-index of $l$ spatial indices collectively as $L\equiv i_1i_2...i_l$, the projection of a tensor component $A_L$ onto its STF part will be referred to as $A_{\langle L\rangle}$. Note that $B_{L}A_{\langle L\rangle}=B_{\langle L\rangle}A_{\langle L\rangle}$, where a sum over $l$ is implicit. Moreover, $n_L$ stands for the tensor product of $l$ radial vectors $n_i$. An expansion in terms of STF tensors $n_{\langle L\rangle}=n_{\langle L\rangle}(\theta,\phi)$ is equivalent to an expansion in spherical harmonics
\begin{align}\label{eq:AYlmToNL}
    n_{\langle L\rangle}&=\frac{4\pi\,l!}{(2l+1)!!}\sum_{m=-l}^l\,\mathcal{Y}_{\langle L\rangle}^{lm}\, Y_{lm}\,,\\
    Y_{lm}&=\mathcal{Y}_L^{lm}\,n_L=\mathcal{Y}_L^{lm}\,n_{\langle L\rangle}\,,
\end{align}
where $\mathcal{Y}_L^{lm}$ are angle-independent, STF tensors connecting the two basis. They satisfy
\begin{align}
    (-1)^m\bar{\mathcal{Y}}^{l-m}_L&=\mathcal{Y}^{lm}_L\,\\
    \bar{\mathcal{Y}}_L^{lm}\mathcal{Y}_L^{lm'}&=\frac{(2l+1)!!}{4\pi\,l!}\,\delta_{mm'}\,.
\end{align}




\subsubsection{\ul{Relations Between Harmonics}}

The pure-spin TT harmonic tensors are related to spin weight $s=\pm 2$ spin-weighted spherical harmonics through
\begin{align}
    T^{\mys{E2}\,\mys{l m
    }}_{ij}&=\frac{1}{\sqrt{2}}\left(\phantom{}_{\mys{-2}}Y_{lm}\,m_im_j+\phantom{}_{\mys{2}}Y_{lm}\,\bar{m}_i\bar{m}_j\right)\,,\\
    T^{\mys{B2}\,\mys{l m
    }}_{ij}&=-\frac{i}{\sqrt{2}}\left(\phantom{}_{\mys{-2}}Y_{lm}\,m_im_j-\phantom{}_{\mys{2}}Y_{lm}\,\bar{m}_i\bar{m}_j\right)\,,
\end{align}
while the relation to the STF basis is given by
\begin{align}
    T^{\mys{E2}\,\mys{l m}}_{ij}&=\perp_{ijab}\sqrt{\frac{2(l-1)l}{(l+1)(l+2)}}\,\mathcal{Y}_{abL-2}^{l m}\,n_{L-2}\,,\\
    T^{\mys{B2}\,\mys{l m}}_{ij}&=\perp_{ijab}\sqrt{\frac{2(l-1)l}{(l+1)(l+2)}}\,\epsilon_{cd(a}\,\mathcal{Y}_{b)dL-2}^{l m}\,n_{cL-2}\,.
\end{align}



\subsubsection{\ul{Expansion of a Rank-2, TT Tensor}}

Using the definitions above, we will now relate the expansions of an arbitrary, rank-2, TT tensor $H_{ij}^\text{TT}(u,r,\theta,\phi)$, in terms of spin-weighted spherical harmonics, pure-spin TT tensor harmonics, and STF tensors. 


\noindent \textbf{Spin-Weighted Spherical Harmonic Expansion.}\label{Ap:SWSHExpansion}
Using the transverse null vector of spin-weight $s=-1$
\begin{equation}
    \bar m^i\equiv\frac{1}{\sqrt{2}}(u^i-iv^i)\,,
\end{equation}
we can define the spinweight $s=-2$ scalar quantity
\begin{equation}\label{eq:ADefHs2}
    H\equiv H_{ij}^\text{TT}\, \bar m^i\, \bar m^j=H_{ij}\, \bar m^i\, \bar m^j\,,
\end{equation}
which we expand in terms of spin-weighted spherical harmonics as
\begin{equation}\label{eq:AExpansionHlm}
    H=\sum_{l=2}^\infty\sum_{m=-l}^l\,H^{lm}\, \phantom{}_{\scriptscriptstyle-2}Y_{lm}\,,
\end{equation}
where
\begin{align}
    H^{lm}(u,r)=\int_{S^2}\dd^2\Omega\,\phantom{}_{\scriptscriptstyle-2}Y^*_{lm}\, H\,.
\end{align}
\noindent \textbf{TT Tensor Expansion.}
The general expansion in terms of TT tensor harmonics reads
\begin{align}
   H_{ij}^\text{TT}=\sum_{l=2}^\infty\sum_{m=-l}^l \left[ U^{l m}\,T^{\scriptscriptstyle{E2}\,\scriptstyle{l m
    }}_{ij}+V^{l m}\,T^{\scriptscriptstyle{B2}\,\scriptstyle{l m
    }}_{ij}\right]\,,
\end{align}
where 
\begin{align}
    U^{l m}(u,r)&=\int_{S^2}\dd^2\Omega \,T^{\mys{E2}\,\scriptstyle{l m
    }}_{ij}\,H^\text{TT}_{ij}\,,\label{eq:AUlmCalc}\\
    V^{l m}(u,r)&=\int_{S^2}\dd^2\Omega \,T^{\mys{B2}\,\mys{l m
    }}_{ij}\,H^\text{TT}_{ij}\,.\label{eq:AVlmCalc}\\
\end{align}
The expansion coefficients $U^{l m}$ and $V^{l m}$ are the electric- and magnetic-parity multipole moments, respectively, also commonly known as mass and current multipole moments. In the literature, it is common to explicitly factor out the $r$ dependence of mass and current multipole moments, which we will not do here. Since $H^\text{TT}_{ij}$ is real, they satisfy
\begin{align}
    \bar{U}^{l m}=(-1)^m U^{l -m}\,,\quad \bar{V}^{l m}=(-1)^m V^{l -m}\,.
\end{align}
\noindent \textbf{STF Expansion.}
The corresponding STF multipole expansion is given by
\begin{align}\label{eq:AExpansionULVL}
\begin{split}
   H_{ij}^\text{TT} =4\perp_{ijab}\sum_{l=2}^\infty &\frac{1}{l !}\Big[\,U_{klL-2}\,n_{L-2}\\
   &+\frac{2l}{l+1}\epsilon_{cd(a}\, V_{b)cL-2}\,n_{dL-2}\Big]\,,
\end{split}
\end{align}
where
\begin{align}
    U_{ijL-2}&=\frac{1}{4\pi}\frac{l(l-1)(2l+1)!!}{2(l+1)(l+2)}\int\dd^2\Omega\, n_{L-2}\,H_{ij}^\text{TT} \,,\\
    V_{ijL-2}&=\frac{1}{4\pi}\frac{(l-1)(2l+1)!!}{4(l+2)}\int\dd^2\Omega\, n_{aL-2}\,\epsilon_{iab}H_{bj}^\text{TT} \,.
\end{align}

\noindent \textbf{Relations between Expansion Coefficients.}
The multipole moments in a spin-weighted basis are related to the electric- and magnetic-parity moments as
\begin{equation}\label{eq:AHlmToUlmVlm}
    H^{lm}=\frac{1}{\sqrt{2}}\left[U^{lm}-iV^{lm}\right]\,,
\end{equation}
which can be inverted to give
\begin{align}
    U^{lm}&=\frac{1}{\sqrt{2}}\left[H^{lm}+(-1)^m \,\bar{H}^{l-m}\right]\,,\label{eq:AUVlmToHlm}\\
    V^{lm}&=\frac{i}{\sqrt{2}}\left[H^{lm}-(-1)^m \,\bar{H}^{l-m}\right] \,.
\end{align}
On the other hand, the relation between an STF and a TT tensor basis is given by
\begin{align}
  U^{l m}&=\frac{16\pi}{(2l+1)!!}\sqrt{\frac{(l+1)(l+2)}{2(l-1)l}}\,\bar{\mathcal{Y}}_{L}^{l m}\,U_L\,,\\
  V^{l m}&=\frac{-32\pi l}{(l+1)(2l+1)!!}\sqrt{\frac{(l+1)(l+2)}{2(l-1)l}}\,\bar{\mathcal{Y}}_{L}^{l m}\,V_L\,,\label{eq:AUVlmToULVL}\\
    U_L&=\frac{l!}{4}\sqrt{\frac{2(l-1)l}{(l+1)(l+2)}}\sum_{m=-l}^l\, \mathcal{Y}_L^{l m}\,U^{l m}\,,\\
    V_L&=\frac{-(l+1)!}{8l}\sqrt{\frac{2(l-1)l}{(l+1)(l+2)}}\sum_{m=-l}^l\, \mathcal{Y}_L^{l m}\,V^{l m}\,.
\end{align}

\newpage
\section{Total Variation and Response Functions}\label{sec:app_formulaeVariations}
	This appendix gathers the analytical expressions for the total variation of all relevant quantities used Chapter~\ref{Sec:CsomoTensions}. This Appendix is adapted from the Appendix A in \cite{Heisenberg:2022gqk}.
	Unless otherwise stated, every function inside the integrals depends on the integration 
	variable $x_z$.
	Comoving, luminosity and angular diameter distance:
	\begin{equation}
		\left\{\begin{array}{l}
		\displaystyle
			I_{d_C}(z) = I_{d_L}(z) = I_{d_A}(z) = -\frac{1}{\chi(z)}\int^z_0 d x_z\frac{H_0^2}{H^3}\\[8pt]
			\displaystyle
			J_{d_C}(z) = J_{d_L}(z) = J_{d_A}(z) = -\frac{1}{\chi(z)}\int^z_0 d x_z\frac{H_0^2}{H^3}m(x_z)\\[8pt]
			\displaystyle
			R_{d_C}(x_z,z) = R_{d_A}(x_z,z) = R_{d_A}(x_z,z) = -(1+x_z)\frac{\Theta(z-x_z)}{\chi(z)H(x_z)}
		\end{array}\right.
	\end{equation}
	Comoving sound horizon:
	\begin{equation}
		\left\{\begin{array}{l}
		\displaystyle
			I_{r_\text{s}}(z) = -\frac{1}{r_\text{s}(z)}\int^\infty_z d x_z\frac{H_0^2}{H^3}c_\text{s}(x_z)\\[8pt]
			\displaystyle
			J_{r_\text{s}}(z) = -\frac{1}{r_\text{s}(z)}\int^\infty_z d x_z\frac{H_0^2}{H^3}m(x_z)c_\text{s}(x_z)\\[8pt]
			\displaystyle
			R_{r_\text{s}}(x_z,z) = -\frac{(1+x_z)c_\text{s}(x_z)}{r_\text{s}(z)}\frac{\Theta(x_z-z)}{H(x_z)}
		\end{array}\right.
	\end{equation}
	Integral defined in \eqref{eq:DefIofsigm8}:
	\begin{equation}
		\left\{\begin{array}{l}
			\displaystyle
			I_{\mathcal{I}} = -\frac{2}{\mathcal{I}}\int^\infty_0\frac{ d k}{k}T^2(k)\mathcal{P}_\mathcal{R}(k)\left(\frac{k}{C_H}\right)^4kR\,W(kR)W'(kR)\\[8pt]
			\displaystyle
			J_{\mathcal{I}} = \frac{2\omega_m}{\mathcal{I}_k}\int^\infty_0\frac{ d k}{k}T(k)\frac{\partial T(k)}{\partial\omega_m}\mathcal{P}_\mathcal{R}(k)\left(\frac{k}{C_H}\right)^4W^2(kR)\\[8pt]
			\displaystyle
			R_{\mathcal{I}} = 0
		\end{array}\right.
	\end{equation}
	Growth factor:
	\begin{equation}
		\left\{\begin{array}{l}
			 \scriptstyle
			I_{D}(z) = -\frac{H(z)}{H_0D(z)}\int^\infty_z\frac{ d x_z}{1+x_z}\frac{H_0^2D}{H^2}
				\Bigg\{\frac{H_0}{H}f + \frac{N(z, x_z)F}{1+x_z}
				\left(1+\frac{(1+x_z)^3}{F}\frac{d\log H}{dx_z}f\right)\Bigg\}\\
			\scriptstyle
			J_{D}(z) = \frac{1}{5} + \frac{H(z)}{H_0D(z)}\int^\infty_z\frac{ d x_z}{(1+x_z)^2}N(z,x_z)FD \\
				\scriptstyle \qquad\qquad
					- \frac{H(z)}{H_0D(z)}\int^\infty_z\frac{ d x_z}{1+x_z}mD
					\Bigg\{\frac{H_0}{H}f + \frac{N(z,x_z)F}{1+x_z}\left(1+\frac{(1+x_z)^3}{F}\frac{d\log H}{dx_z}f\right)\Bigg\}\\
			\scriptstyle
			R_{D}(x_z,z) = -\frac{H(z)D(x_z)}{H_0D(z)}
				\Bigg\{\frac{H_0}{H(x_z)}f(x_z) + \frac{N(z, x_z)F(x_z)}{1+x_z}
				\left(1+\frac{(1+x_z)^3}{F(x_z)}\frac{d\log H}{d x_z}f(x_z)\right)\Bigg\}
				\Theta(x_z-z)
		\end{array}\right.
	\end{equation}
	where
	\begin{equation}
		N(z, x_z)\equiv \frac{H^2(x_z)}{(1+x_z)^3H_0^2}\Big(I(z)-I(x_z)\Big)\ ,\qquad
		I(z)\equiv \int^\infty_z d x_z(1+x_z)\frac{H_0^3}{H^3(x_z)}\ .
	\end{equation}
	Linear growth rate:
	\begin{equation}
		\left\{\begin{array}{l}
			\scriptstyle
			I_{f}(z) = -I_D(z)\left(1+\frac{1+z}{f}\frac{d\log H}{dz}\right) - \frac{H_0^2}{H^2(z)}\\
        \scriptstyle\qquad\qquad
				-\frac{(1+z)^2H_0^2}{f(z)H^2(z)D(z)}\int^\infty_z\frac{ d x_z}{(1+x_z)^5} FD\left(1+\frac{(1+x_z)^3}{F}\frac{d\log H}{dx_z}f\right)\\
            \scriptstyle
			J_{f}(z) = -\left(J_D(z)-\frac{1}{5}\right)\left(1+\frac{1+z}{f}\frac{d\log H}{dz}\right) - m(z)\\
            \scriptstyle \qquad\qquad
				+\frac{(1+z)^2}{f(z)H^2(z)D(z)}\int^\infty_z\frac{ d x_z}{(1+x_z)^5} H^2FD
					\Big\{1-m\left(1+\frac{(1+x_z)^3}{F}\frac{d\log H}{dx_z}f\right)\Big\}\\
            \scriptstyle
			R_{f}(x_z,z) = -R_D(x_z, z)\left\{1+\frac{1+z}{f}\left(\frac{d\log H}{dz}-\frac{\delta(x_z-z)}{\Theta(x_z-z)}\right)\right\}\\
             \scriptstyle\qquad\qquad\qquad
				-\frac{1}{(1+z)f(z)}\frac{(1+z)^3H^2(x_z)D(x_z)}{(1+x_z)^3H^2(z)D(z)}\frac{F(x_z)}{1+x_z}\left(1+\frac{(1+x_z)^3}{F(x_z)}\frac{d\log H}{dx_z}f(x_z)\right)\Theta(x_z-z)
		\end{array}\right.
	\end{equation}
	Supernova absolute magnitude:
	\begin{equation}
		\left\{\begin{array}{l}
			\displaystyle
			I_{M} = -\frac{5}{M}-\frac{5}{M\sum_{ij}(C^{-1})_{ij}}\sum_{ij}(C^{-1})_{ij}I_\chi(z_j)\\[8pt]
			\displaystyle
			J_{M} = -\frac{5}{M\sum_{ij}(C^{-1})_{ij}}\sum_{ij}(C^{-1})_{ij}J_\chi(z_j)\\[8pt]
			\displaystyle
			R_{M}(x_z) = -\frac{5}{M\sum_{ij}(C^{-1})_{ij}}\sum_{ij}(C^{-1})_{ij}R_\chi(x_z, z_j)
		\end{array}\right.
	\end{equation}
	As discussed in the main text around Eq. \eqref{eq:def_response_g}, the response function
	for any cosmological quantity $\mathcal{O}$ can be computed as
	\begin{equation}
		\mathcal{R}_\mathcal{O}(x_z, z) \equiv I_\mathcal{O}(z)\mathcal{R}_h(x_z)
			+ J_\mathcal{O}(z)\mathcal{R}_{\omega_m}(x_z) + R_\mathcal{O}(x_z, z)\, .
	\end{equation} 
	All the response functions relevant for this work can be derived from the formulas above.
	These include:
	\begin{subequations}
	\begin{align}
		\mathcal{R}_{\sigma_8} &= \mathcal{R}_D - \mathcal{R}_{\omega_m} + \frac{1}{2}\mathcal{R}_{\mathcal{I}_k}\ ,\\
		\mathcal{R}_{S_8} &= \mathcal{R}_{\sigma_8} - \mathcal{R}_h + \frac{1}{2}\mathcal{R}_{\omega_m}\ ,\\
		\mathcal{R}_{f\sigma_8} &= \mathcal{R}_{f} + \mathcal{R}_{\sigma_8}\ .
	\end{align}
	\end{subequations}

\section{Explicit geometrical objects}\label{explicityGeom}

In this appendix, we provide the explicit expressions of the effective metric and its determinant up to $4^{\text{th}}$ order in the field $ \bar{\pi}$ of Sec.~\ref{sSec:Explicit one loop computations luminal H}.

Expanding the Levi-Civita symbols and canonically normalizing by setting $\tilde{c}_2=-\frac{1}{12}$, the effective inverse metric in Eq.~\eqref{expSOO} reads
\begin{align}\label{Minv}
M^{\mu\nu}={}&\delta^{\mu\nu} + 12 \frac{ \tilde{c}_3}{\Lambda^3}\,\left[\delta^{\mu \nu } \Delta \bar{\pi} + \partial^{\nu }\partial^{\mu }  \bar{\pi}\right]-  36\frac{\tilde{c}_4}{\tilde{\Lambda}^2}\,\delta^{\mu\nu}   \bar{\pi}^2 \,.
\end{align}
By treating this as a perturbation of the Euclidean metric in powers of the field $\bar{\pi}$ as $M^{\mu\nu}=\delta^{\mu\nu}+M_{\bar{\pi}}^{\mu\nu}$, this expression can be perturbatively inverted in a Neuman series schematically of the form $(I+M_{\bar{\pi}})^{-1}=\sum_{n=0}^\infty \,(-1)^n M^n_{\bar{\pi}}$ which yields
\begin{IEEEeqnarray}{rCl}\label{Meff}
M_{\mu\nu} &=&\delta_{\mu\nu} - 12\frac{\tilde{c}_3}{\Lambda^3}\left[ \delta_{ab} \Delta \bar{\pi} +  \partial_{b}\partial_{a}  \bar{\pi}\right] +36 \frac{ \tilde{c}_4}{\tilde{\Lambda}^2}\delta_{\mu\nu}   \bar{\pi}^2+144\frac{\tilde{c}_3^2}{\Lambda^6}\left[\delta_{\mu \nu } (\Delta\bar{\pi})^2 + \partial_{a}\partial_{\nu }  \bar{\pi} \partial_{\mu }\partial^{a}  \bar{\pi} \right.
\nonumber \\ 
&&  
\left.+ 2 \Delta \bar{\pi} \partial_{\nu }\partial_{\mu }  \bar{\pi}\right] -  864 \frac{ \tilde{c}_3 \tilde{c}_4 }{\Lambda^3\tilde{\Lambda}^2} \left[\delta_{\mu \nu } \Delta \bar{\pi}   \bar{\pi}^2 +   \bar{\pi}^2 \partial_{\nu}\partial_{\mu }  \bar{\pi}\right]-1728 \frac{ \tilde{c}_3^3 }{\Lambda^9}\left[\delta_{\mu \nu } (\Delta\bar{\pi})^3\right. 
\nonumber \\ 
&& 
\left. + 3 \Delta \bar{\pi} \partial_{a}\partial_{\nu }  \bar{\pi} \partial_{\mu }\partial^{a}  \bar{\pi} + \partial_{b}\partial_{a}  \bar{\pi} \partial_{\mu }\partial^{a}  \bar{\pi} \partial_{\nu }\partial^{b}  \bar{\pi} + 3 (\Delta\bar{\pi})^2 \partial_{\nu }\partial_{\mu }  \bar{\pi} \right] +1296 \frac{ \tilde{c}_4^2}{\tilde{\Lambda}^4}\delta_{\mu\nu}   \bar{\pi}^4
\nonumber \\ 
&& 
+ 15552\frac{ \tilde{c}_3^2 \tilde{c}_4}{\Lambda^6\tilde{\Lambda}^2} \left[ \delta_{\mu \nu } (\Delta\bar{\pi})^2   \bar{\pi}^2 +   \bar{\pi}^2 \partial_{a}\partial_{\nu }  \bar{\pi} \partial_{\mu }\partial^{a}  \bar{\pi} + 2 \Delta \bar{\pi}  \bar{\pi}^2 \partial_{\nu }\partial_{\mu } \bar{\pi} \right]
\nonumber \\ 
&& 
 +20736\frac{ \tilde{c}_3^4 }{\Lambda^{12}}\left[ \delta_{\mu \nu } (\Delta \bar{\pi})^4 + 6 (\Delta \bar{\pi})^2 \partial_{a}\partial_{\nu } \bar{\pi} \partial_{\mu }\partial^{a} \bar{\pi} + 4 \Delta  \bar{\pi} \partial_{b}\partial_{a} \bar{\pi} \partial_{\mu }\partial^{a} \bar{\pi} \partial_{\nu }\partial^{b} \bar{\pi}  \right.
\nonumber \\ 
&& \left.+ \partial_{c}\partial_{b} \bar{\pi} \partial^{c}\partial_{a} \bar{\pi} \partial_{\mu }\partial^{a} \bar{\pi} \partial_{\nu }\partial^{b} \bar{\pi} + 4 (\Delta \bar{\pi})^3 \partial_{\nu }\partial_{\mu } \bar{\pi}\right]
+ {\cal O}( \bar{\pi}^5)\,.
\end{IEEEeqnarray}
where we explicitly show the expansion up to $4^{\text{th}}$ order in the field, such that Eq.~\eqref{defeffM} is satisfied up to fourth order. The effective determinant is then defined as
\begin{align}
M \equiv \frac{1}{4!}\epsilon^{\mu_1 \mu_2 \mu_3 \mu_4}\epsilon^{\nu_1 \nu_2 \nu_3 \nu_4} M_{\mu_1 \nu_1}M_{\mu_2 \nu_2}M_{\mu_3 \nu_3}M_{\mu_4 \nu_4},
\end{align}
Explicitly we find
\begin{IEEEeqnarray}{rCl}\label{detM}
M&=& 1 - 36 \frac{\tilde{c}_3}{\Lambda^3}\Delta  \bar{\pi}+144 \frac{\tilde{c}_4}{\tilde{\Lambda}^2}   \bar{\pi}^2 + 72 \frac{ \tilde{c}_3^2}{\Lambda^6}\left[11 (\Delta \bar{\pi})^2 + \partial_{b}\partial_{a} \bar{\pi} \partial^{b}\partial^{a} \bar{\pi} \right] -  6480 \frac{\tilde{c}_3 \tilde{c}_4}{\Lambda^3\tilde{\Lambda}^2} \Delta  \bar{\pi}  \bar{\pi}^2
\nonumber \\ 
&& 
- 288\frac{\tilde{c}_3^3}{\Lambda^9}\left[47 (\Delta \bar{\pi})^3 + 15 \Delta  \bar{\pi} \partial_{b}\partial_{a} \bar{\pi} \partial^{b}\partial^{a} \bar{\pi} + 2 \partial^{b}\partial^{a} \bar{\pi} \partial_{c}\partial_{b} \bar{\pi} \partial^{c}\partial_{a} \bar{\pi}\right]+12960 \frac{ \tilde{c}_4^2 }{\tilde{\Lambda}^4} \bar{\pi}^4
\nonumber \\ 
&&
+ 15552\frac{ \tilde{c}_3^2 \tilde{c}_4}{\Lambda^6\tilde{\Lambda}^2} \left[11 (\Delta \bar{\pi})^2  \bar{\pi}^2 +  \bar{\pi}^2 \partial_{b}\partial_{a} \bar{\pi} \
\partial^{b}\partial^{a} \bar{\pi}\right] +2592\frac{ \tilde{c}_3^4 }{\Lambda^{12}}\left[75 (\Delta \bar{\pi})^4\right.
 \nonumber \\
 && 
 \left.+ 58 (\Delta \bar{\pi})^2 \partial_{b}\partial_{a} \bar{\pi} \partial^{b}\partial^{a} \bar{\pi} + 16 \Delta  \bar{\pi} \partial^{b}\partial^{a} \bar{\pi} \partial_{c}\partial_{b} \bar{\pi} \partial^{c}\partial_{a} \bar{\pi} + 2 \partial^{b}\partial^{a} \bar{\pi} \partial^{c}\partial_{a} \bar{\pi} \partial_{d}\partial_{c} \bar{\pi} \partial^{d}\partial_{b} \bar{\pi} \phantom{\frac{1}{2}} \right.
 \nonumber \\
 && 
\left.+ \partial_{b}\partial_{a} \bar{\pi} \partial^{b}\partial^{a} \bar{\pi} \partial_{d}\partial_{c} \bar{\pi} \partial^{d}\partial^{c} \bar{\pi} \right]+ {\cal O}( \bar{\pi}^5)\,.
\end{IEEEeqnarray}

The effective metric [Eq.~\eqref{Meff}] and its inverse [Eq.~\eqref{Minv}] can then be used in order to determine the various geometrical objects defined in Sec.~\ref{sSec:Explicit one loop computations luminal H}. For example, the associated Christoffel symbols up to third order are given by
\begin{IEEEeqnarray}{rCl}
 \Gamma^{\rho}_{\mu\nu}(M)&=&\frac{M^{\rho\sigma}}{2}\left(\partial_{\mu}M_{\sigma\nu}+\partial_{\nu}M_{\sigma\mu}-\partial_{\sigma}M_{\mu\nu}\right)
 \nonumber \\ 
 &=&\scaleto{ 
 -6 \frac{ \tilde{c}_3}{\Lambda^3} \left[\delta_{\nu }{}^{\rho } \partial_{\mu }\Delta \bar{\pi} + \delta_{\mu }{}^{\rho } \partial_{\nu }\Delta \bar{\pi} -  \delta_{\mu \nu } \partial^{\rho }\Delta \bar{\pi} + \partial^{\rho }\partial_{\nu }\partial_{\mu }\bar{\pi}\right]+36\frac{ \tilde{c}_4 }{\tilde{\Lambda}^2} \left[\delta_{\nu }{}^{\rho } \bar{\pi} \partial_{\mu }\bar{\pi}+ \delta_{\mu }{}^{\rho } \bar{\pi} \partial_{\nu }\bar{\pi} -  \delta_{\mu \nu } \bar{\pi} \partial^{\rho }\bar{\pi}\right]
\mathstrut}{19pt}\nonumber \\ 
&& \scaleto{ 
+72\frac{c_{3}{}^2}{\Lambda^6} \left[\delta_{\nu }{}^{\rho } \Delta \bar{\pi} \partial_{\mu }\Delta \bar{\pi} + \delta_{\mu }{}^{\rho } \Delta \bar{\pi} \partial_{\nu }\Delta \bar{\pi} -  \delta_{\mu \nu } \Delta \bar{\pi} \partial^{\rho }\Delta \bar{\pi} - 2 \partial_{\nu }\partial_{\mu }\bar{\pi} \partial^{\rho }\Delta \bar{\pi}+ \delta_{\mu \nu } \partial^{a}\Delta \bar{\pi} \partial^{\rho }\partial_{a}\bar{\pi}  \right.
\mathstrut}{19pt}\nonumber \\ 
&& \scaleto{  
\left. + \partial_{a}\partial_{\nu }\partial_{\mu }\bar{\pi} \partial^{\rho }\partial^{a}\bar{\pi}+ \partial_{\nu }\Delta \bar{\pi} \partial^{\rho }\partial_{\mu }\bar{\pi} + \partial_{\mu }\Delta \bar{\pi} \partial^{\rho }\partial_{\nu }\bar{\pi} + \Delta \bar{\pi} \partial^{\rho }\partial_{\nu }\partial_{\mu }\bar{\pi}\right]-216 \frac{c_{3}{} c_{4}{} }{\Lambda^3\tilde{\Lambda}^2}\left[\delta_{\nu }{}^{\rho } \Delta \bar{\pi} \partial_{\mu }\Delta \bar{\pi}\right.
\mathstrut}{19pt}\nonumber \\ 
&& \scaleto{ 
+ \delta_{\mu }{}^{\rho } \Delta \bar{\pi} \partial_{\nu }\Delta \bar{\pi} -  \delta_{\mu \nu } \Delta \bar{\pi} \partial^{\rho }\Delta \bar{\pi} - 2 \partial_{\nu }\partial_{\mu }\bar{\pi} \partial^{\rho }\Delta \bar{\pi}+ \delta_{\mu \nu } \partial^{a}\Delta \bar{\pi} \partial^{\rho }\partial_{a}\bar{\pi} + \partial_{a}\partial_{\nu }\partial_{\mu }\bar{\pi} \partial^{\rho }\partial^{a}\bar{\pi} \phantom{\frac{1}{2}}
 \mathstrut}{19pt}\nonumber \\ 
&& \scaleto{  
\left. + \partial_{\nu }\Delta \bar{\pi} \partial^{\rho }\partial_{\mu }\bar{\pi} + \partial_{\mu }\Delta \bar{\pi} \partial^{\rho }\partial_{\nu }\bar{\pi}+ \Delta \bar{\pi} \partial^{\rho }\partial_{\nu }\partial_{\mu }\bar{\pi}  \right] - \frac{c_{3}{}^3 }{\Lambda^9}\left[\delta_{\nu }{}^{\rho } \Delta \bar{\pi}^2 \partial_{\mu }\Delta \bar{\pi} + \delta_{\mu }{}^{\rho } \Delta \bar{\pi}^2 \partial_{\nu }\Delta \bar{\pi}\right.
\mathstrut}{19pt} \nonumber \\ 
&&\scaleto{ 
-  \delta_{\mu \nu } \Delta \bar{\pi}^2 \partial^{\rho }\Delta \bar{\pi}- 3 \partial_{a}\partial_{\nu }\bar{\pi} \partial_{\mu }\partial^{a}\bar{\pi} \partial^{\rho }\Delta \bar{\pi} - 4 \Delta \bar{\pi} \partial_{\nu }\partial_{\mu }\bar{\pi} \partial^{\rho }\Delta \bar{\pi} + 2 \delta_{\mu \nu } \Delta \bar{\pi} \partial^{a}\Delta \bar{\pi} \partial^{\rho }\partial_{a}\bar{\pi} \phantom{\frac{1}{2}}
 \mathstrut}{19pt}\nonumber \\ 
&& \scaleto{ 
+ \partial_{\mu }\partial^{a}\bar{\pi} \partial_{\nu }\Delta \bar{\pi} \partial^{\rho }\partial_{a}\bar{\pi} + \partial_{\mu }\Delta \bar{\pi} \partial_{\nu }\partial^{a}\bar{\pi} \partial^{\rho }\partial_{a}\bar{\pi}+ 2 \partial^{a}\Delta \bar{\pi} \partial_{\nu }\partial_{\mu }\bar{\pi} \partial^{\rho }\partial_{a}\bar{\pi} + 2 \Delta \bar{\pi} \partial_{a}\partial_{\nu }\partial_{\mu }\bar{\pi} \partial^{\rho }\partial^{a}\bar{\pi} \phantom{\frac{1}{2}}
  \mathstrut}{19pt}\nonumber \\ 
&& \scaleto{ 
-  \partial_{\mu }\partial^{a}\bar{\pi} \partial_{\nu }\partial^{b}\bar{\pi} \partial^{\rho }\partial_{b}\partial_{a}\bar{\pi} + \partial_{b}\partial_{a}\partial_{\nu }\bar{\pi} \partial_{\mu }\partial^{a}\bar{\pi} \partial^{\rho }\partial^{b}\bar{\pi}+ \partial_{b}\partial_{a}\partial_{\mu }\bar{\pi} \partial_{\nu }\partial^{a}\bar{\pi} \partial^{\rho }\partial^{b}\bar{\pi} \phantom{\frac{1}{2}}
 \mathstrut}{19pt}\nonumber \\ 
&& \scaleto{
 \left.+ 2 \Delta \bar{\pi} \partial_{\nu }\Delta \bar{\pi} \partial^{\rho }\partial_{\mu }\bar{\pi}+ 2 \Delta \bar{\pi} \partial_{\mu }\Delta \bar{\pi} \partial^{\rho }\partial_{\nu }\bar{\pi} + \Delta \bar{\pi}^2 \partial^{\rho }\partial_{\nu }\partial_{\mu }\bar{\pi} \right]
+ {\cal O}( \bar{\pi}^4)\,.\phantom{\frac{1}{2}}\mathstrut}{19pt}
\end{IEEEeqnarray}

In practice, however, a convenient way to proceed is to expand all the objects in Eq.~\eqref{OneLoopGour} in terms of the effective metric $M_{\mu\nu}$, it's inverse $M^{\mu\nu}$ and the determinant $M$ and plug in the expressions in Eqs.~\eqref{Meff}, \eqref{Minv} and \eqref{detM} respectively, in order to obtain the desired divergent one-loop effective action expressions in terms of the scalar background field. Alternatively, one can also expand Eq.~\eqref{OneLoopGour} in a perturbation series of the inverse metric and match it to Eq.~\eqref{Minv} without ever referring to the explicit expression of the effective metric, as explained in the main text in Sec.~\ref{sSec:Explicit one loop computations luminal H}.

\end{appendices}

